\newcommand{\sides}{twoside}
\newcommand{\userefs}{true}

\documentclass[a4paper,11pt,\sides]{thesis_style}
\usepackage{fancyhdr,bm}
\usepackage{mathrsfs,amsmath,amssymb,cancel,rotating}
\usepackage{ifthen,setspace,graphicx,verbatim,subfigure,booktabs}
\usepackage{multirow}
\usepackage{color}
\usepackage[usenames,dvipsnames]{xcolor}
\usepackage{captcont}
\usepackage[pagebackref=true]{hyperref}

\newcommand{\lpage}{\textbf{\textit{\thepage}}\ \ \textbar}
\newcommand{\rpage}{\textbar\ \ \textbf{\textit{\thepage}}}
\newcommand{\thesistitle}{Investigation of the transfer and dissipation of energy in isotropic turbulence}

\newcommand{\frun}[1]{\texttt{#1}}
\newcommand{\drun}[1]{\texttt{#1}}

\renewcommand{\vec}[1]{\boldsymbol{\mathbf{#1}}}
\newcommand{\ord}[1]{O(#1)}
\newcommand{\vmod}[1]{\lvert \vec{#1} \rvert}

\newcommand{\kmax}{k_{\textrm{max}}} 

\newcommand{\kcut}{\Lambda} 
\newcommand{\pimax}{\varepsilon_T}
\newcommand{\ktop}{k_{\textrm{top}}} 
\renewcommand{\geq}{\geqslant}
\renewcommand{\leq}{\leqslant}
\newcommand{\re}[1]{\textrm{Re} \left[ #1 \right]}
\newcommand{\dns}{\textup{\textbf{DNS}$2012$}}
\newcommand{\addref}{[\textbf{reference}]}

\newcommand{\epsw}{\varepsilon_W}
\newcommand{\eddie}{\texttt{eddie}}

\newcommand{\etal}{\textit{et al.}}
\newcommand{\Ceps}{C_\varepsilon}
\newcommand{\unitM}{\mathbb{I}}
\newcommand{\tr}{\textrm{tr}}
\newcommand{\realpart}{\textrm{Re}}
\newcommand{\CPiVAL}{0.47}
\newcommand{\CLVAL}{19.1}

\newcommand{\fvect}[1]{\hat{#1}}
\newcommand{\high}{+}
\newcommand{\low}{-}
\newcommand{\measure}[2]{\frac{d^d #1\ d #2}{(2\pi)^{d+1}}}
\newcommand{\fmeasure}[2]{d\hat{#1}}
\newcommand{\order}[1]{O(#1)}

\hypersetup{
    linktoc=page,
    pdftitle={\thesistitle},    
    pdfauthor={Samuel R. Yoffe},     
    pdfcreator={Samuel R. Yoffe}   
}
\ifthenelse{\equal{\userefs}{true}}{
\hypersetup{
    colorlinks=true,       
    linkcolor=red,          
    citecolor=blue,        
    urlcolor=OliveGreen           
}
}{
\hypersetup{
    colorlinks=true,       
    linkcolor=black,          
    citecolor=black,        
    urlcolor=black           
}
}

\renewcommand*{\backref}[1]{}
\renewcommand*{\backrefalt}[4]{%
  \ifcase #1 %
    \relax
  \or
$\langle\langle$~Cited~on~page~#2.~$\rangle\rangle$%
  \else
$\langle\langle$~Cited~on~pages~#2.~$\rangle\rangle$%
  \fi%
}

\begin{document}

\pagestyle{empty}
\pagestyle{empty}

\begin{center}
\LARGE
\onehalfspacing 
\textbf{\thesistitle}

\vspace{2cm}
\includegraphics[width=0.35\linewidth]{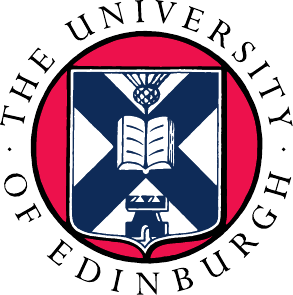}

\vspace{2cm}

\singlespacing

\Large
\textit{Samuel Robert Yoffe}

\vfill

\normalsize
A thesis submitted in fulfilment of the requirements\\
for the degree of Doctor of Philosophy\\
to the\\
University \textit{of} Edinburgh\\
\ \\
March, 2012

\end{center}

\frontmatter

\onehalfspacing

\renewcommand{\chaptermark}[1]{\markboth{#1}{}}
\renewcommand{\headheight}{28pt}

\renewcommand{\headrulewidth}{0pt}
\renewcommand{\footrulewidth}{0pt}

\fancypagestyle{plain}{
 \fancyhf{} 
 \renewcommand{\headrulewidth}{0pt}
 \ifthenelse{\equal{\sides}{oneside}}
 {
   \fancyfoot[R]{\rpage}
 }
 {
   \fancyfoot[RO]{\rpage}
   \fancyfoot[LE]{\lpage}
 }
}

\pagestyle{fancy}
\fancyhf{}
\renewcommand{\chaptermark}[1]{\markboth{#1}{}}
\ifthenelse{\equal{\sides}{oneside}}
 {
   \fancyfoot[R]{\rpage}
 }
 {
   \fancyfoot[RO]{\rpage}
   \fancyfoot[LE]{\lpage}
 }

\chapter*{}
\thispagestyle{empty}
 \vspace{5cm}
\begin{flushright}
 \Large\textit{For my dad.}
\end{flushright}

\chapter{Abstract}

\noindent Numerical simulation is becoming increasingly used to support theoretical effort into understanding the turbulence problem. We develop theoretical ideas related to the transfer and dissipation of energy, which clarify long-standing issues with the energy balance in isotropic turbulence. These ideas are supported by results from large scale numerical simulations.

Due to the large number of degrees of freedom required to capture all the interacting scales of motion, the increase in computational power available has only recently allowed flows of interest to be realised. A parallel pseudospectral code for the direct numerical simulation (DNS) of isotropic turbulence has been developed. Some discussion is given on the challenges and choices involved. The DNS code has been extensively benchmarked by reproducing well established results from literature.

The DNS code has been used to conduct a series of runs for freely-decaying turbulence. Decay was performed from a Gaussian random field as well as an evolved velocity field obtained from forced simulation. Since the initial condition does not describe developed turbulence, we are required to determine when the field can be considered to be evolved and measurements are characteristic of decaying turbulence. We explore the use of power-law decay of the total energy and compare with the use of dynamic quantities such as the peak dissipation rate, maximum transport power and velocity derivative skewness. We then show how this choice of evolved time affects the measurement of statistics. In doing so, it is found that the Taylor dissipation surrogate, $u^3/L$, is a better surrogate for the maximum inertial flux than dissipation.

Stationary turbulence has also been investigated, where we ensure that the energy input rate remains constant for all runs and variation is only introduced by modifying the fluid viscosity (and lattice size). We present results for Reynolds numbers up to $R_\lambda = 335$ on a $1024^3$ lattice. Using different methods of vortex identification, the persistence of intermittent structure in an ensemble average is considered and shown to be reduced as the ensemble size increases. The longitudinal structure functions are computed for smaller lattices directly from an ensemble of realisations of the real-space velocity field. From these, we consider the generalised structure functions and investigate their scaling exponents using direct analysis and extended self-similarity (ESS), finding results consistent with the literature. An exploitation of the pseudospectral technique is used to calculate second- and third-order structure functions from the energy and transfer spectra, with a comparison presented to the real-space calculation. An alternative to ESS is discussed, with the second-order exponent found to approach 2/3.

The dissipation anomaly is then considered for both forced and free-decay. Using different choices of the evolved time for a decaying simulation, we show how the behaviour of the dimensionless dissipation coefficient is affected. The K\'arm\'an-Howarth equation (KHE) is studied and a derivation of a work term presented using a transformation of the Lin equation. The balance of energy represented by the KHE is then investigated using the pseudospectral method mentioned above. The consequences of this new input term for the structure functions are discussed. Based on the KHE, we develop a model for the behaviour of the dimensionless dissipation coefficient that predicts $\Ceps = \Ceps(\infty) + C_L/R_L$. DNS data is used to fit the model. We find $\Ceps(\infty) = 0.47$ and $C_L = 19.1$ for forced turbulence, with excellent agreement to the data.

Theoretical methods based on the renormalization group and statistical closures are still being developed to study turbulence. The dynamic RG procedure used by Forster, Nelson and Stephen (FNS) is considered in some detail and a disagreement in the literature over the method and results is resolved here. An additional constraint on the loop momentum is shown to cause a correction to the viscosity increment such that all methods of evaluation lead to the original result found by FNS. The application of statistical closure and renormalized perturbation theory is discussed and a new two-time model probability density functional presented. This has been shown to be self-consistent to second order and to reproduce the two-time covariance equation of the local energy transfer (LET) theory. Future direction of this work is discussed.

\chapter{Declaration}

I declare that this thesis was composed by myself and that, except where explicitly stated otherwise in the text, the work contained therein is my own or was carried out in collaboration with Professors W. D. McComb and A. Berera, and Dr. M. Salewski.

\vspace{1em}
\noindent Work from section \ref{sec:Taylor_surr} was published in McComb, Berera, Salewski and Yoffe \cite{McComb:2010p250} and the original formulation of the model developed in section \ref{sec:model_DA} was presented in the \emph{arXiv} preprint McComb, Berera, Salewski and Yoffe \cite{McComb:2010p1601}. The analysis of chapter \ref{sec:RG} appeared in Berera and Yoffe \cite{Berera:2010p789}.

\vspace{1in}
\begin{flushright}
 \emph{S. R. Yoffe}\\
 March, 2012
\end{flushright}

\chapter{Acknowledgements}

First and foremost, I wish to extend my gratitude to my supervisors, Professors W. David McComb and Arjun Berera. Without their continued support I would never have completed this thesis. I wish to thank Prof. McComb for his patient guidance and motivation towards research. I thank Prof. Berera for sharing his knowledge and enthusiasm with me, as well as for being approachable with any problems I had. I have learnt a lot from working with both of them.

Particular thanks are due to Dr. Matthew Salewski, for his friendship and many stimulating discussions on the topic of turbulence.

I cannot describe how indebted I am to my wonderful girlfriend, Amanda, whose love and encouragement will always motivate me to achieve all that I can. I could not have written this thesis without her support; in particular, my peculiar working hours and erratic behaviour towards the end could not have been easy to deal with!

Of course, I would never have made it this far without the love and support of my family, particularly my mum and brother, Joe. Their interest (a fa\c{c}ade though it may have been!) in my work and pride at my achievements has always been an inspiration.

I could also have not made it through without the many friends I have made along the way. I particularly wish to thank my colleagues and flatmates Gavin and Liam, as living and working with them was a privilege. I also thank Eoin for the many jamming sessions and encouraging the creation of the physics dept. football team, \emph{the Feynmen}. 

When I joined the particle theory group, I was instantly made to feel welcome and included, for which I owe additional thanks to Erik, Claudia, Simone, Thomas and Brian. I extend my thanks and best wishes to the entire PPT corridor and the students and post-docs I got to share lunch, coffee and/or (several) pints with.

I would like to thank Jane Patterson for her kindness and ensuring my PhD career ran smoothly.

I gratefully acknowledge the generosity and support of the \emph{Edinburgh Compute and Data Facility}. My funding was provided by the STFC, to whom I am eternally grateful for this opportunity.

\singlespacing
\tableofcontents
\newpage

\listoffigures
\listoftables

\mainmatter
\onehalfspacing

\renewcommand{\chaptermark}[1]{\markboth{#1}{}}
\ifthenelse{\equal{\sides}{oneside}}
 {
   \renewcommand{\headrulewidth}{0.5pt}
   \fancyfoot[R]{\rpage}
   \fancyhead[R]{\textit{\nouppercase{\rightmark}}}
 }
 {
   \renewcommand{\headrulewidth}{0.5pt}
   \fancyhead[RO]{\textit{\nouppercase{\rightmark}}}
   \fancyhead[LE]{\textit{\chaptername\ \thechapter\ \ ---\ \ \leftmark}}
 }


\chapter{Introduction to fluid turbulence}

\vspace{-1in}
\begin{center}
 \setlength\fboxsep{0px}
 \setlength\fboxrule{0.5pt}
 \fbox{
  \hspace{-3.625px}\includegraphics[width=0.8\textwidth,height=0.5512\textwidth]{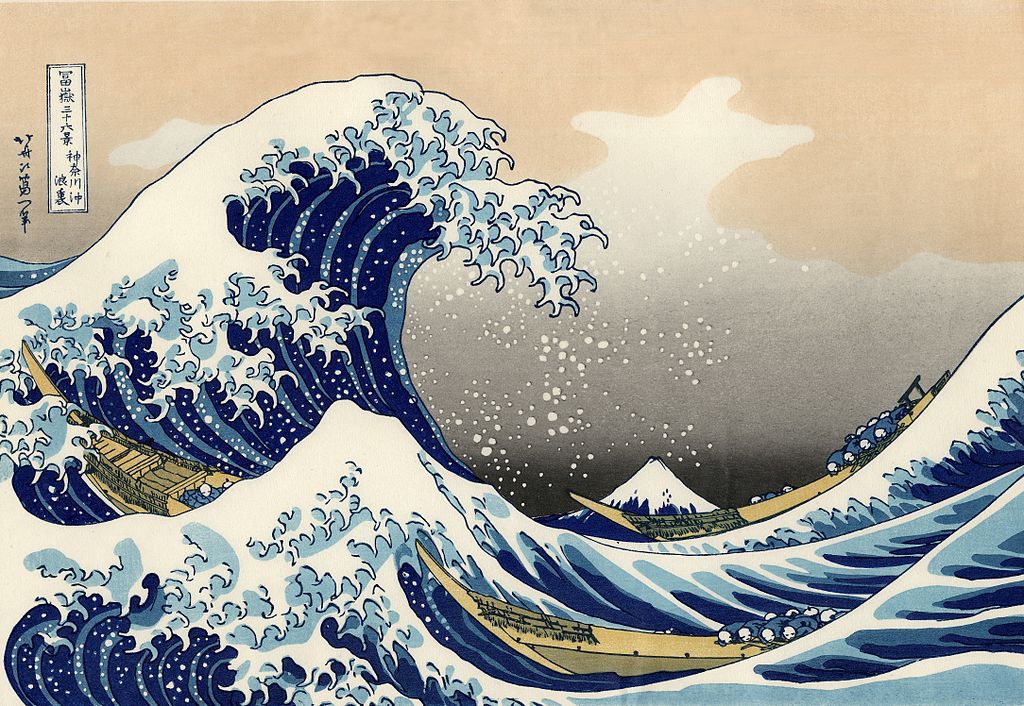}\hspace{-3.65px}
 }
\end{center}
\vspace{-2em}
\begin{flushright}
 {\small \emph{The Great Wave off Kanagawa}; Hokusai, circa 1830.\hspace{0.6in}\ }
\end{flushright}

\newpage
\section{Introduction}
Turbulence is ubiquitous; it can be seen all around us as we go about our everyday lives. From the wind that drags a piece of litter along the floor to raging rivers. It exists at many length-scales, from the blood in your veins to interstellar gas clouds. Indeed, the majority of flows of interest in engineering applications are also turbulent, whether it be the flow over a new suspension bridge or through the pipelines that provide us with water or gas. Laminar flow is very much the exception, not the rule.

Despite this, turbulence remains one of the unsolved problems of classical mechanics. This is due to the non-linear equation of motion which governs its behaviour. This leads to the chaotic motion which has intrigued scientists and inspired artists for centuries (for example, the famous Japanese woodblock print produced around 1830, presented above, or the sketches of Leonardo da Vinci). While it is difficult to provide a precise definition of turbulence, we can say that it is characterised by fluctuations on a large range of interacting scales and is accompanied by large amounts of energy dissipation. Therefore, turbulence cannot sustain itself and, unless driven by an external force, is a decaying phenomenon.

Real flows are not homogeneous or isotropic due to boundary conditions and constraints on the flow. However, away from any boundaries and in a frame moving with the mean flow, a range of scales significantly smaller than the size of the system may be considered to be homogeneous and locally isotropic. A study of the properties of turbulence which is isotropic and homogeneous is therefore not necessarily a pointless endeavour. The simplifications that arise from assuming these (statistical) properties allow the fundamental properties of turbulence to be investigated, free of any complications of interaction with boundaries or the mean flow itself.

Homogeneous, isotropic turbulence (HIT) was first realised in a laboratory in the 1930s using a grid placed in a wind tunnel. This is clearly not homogeneous in the streamwise direction, but by transforming to a frame moving with the mean flow this is equivalent to homogeneous turbulence which is decaying in time. Unfortunately, we cannot generate stationary HIT in a lab. With the increase in available computing power, it has now become practical to perform numerical simulations of large-scale flows. Direct numerical simulation of all the scales of turbulent motion has shown excellent agreement with experimental data and offers a route to study stationary (albeit artificially maintained) HIT.

Since the equations of motion are mathematically intractable, much effort has been spent on the construction of statistical descriptions of turbulence, with varying success. The bedrock problem of statistical physics, an infinite hierarchy of unclosed equations, has to be treated carefully and cleverly, and remains an active area of research today.

\section{The equations of fluid motion}
Consider a fluid of density $\rho(\vec{x},t)$ in three-dimensions. By considering the flux of mass into (or out of) a test volume due to a flow of velocity $\vec{U}(\vec{x},t)$ through its surface, we can write an expression for the time rate of change of the mass contained. Since mass must be conserved, we have
\begin{equation}
 \label{eq:continuity_gen}
 \frac{\partial \rho}{\partial t} + \vec{\nabla}\cdot(\rho \vec{U}) = 0
\end{equation}
at all points $\vec{x}$. This is known as the \emph{continuity equation}. We can use the density to define the convective derivative, which is the Lagrangian time rate of change moving with a fluid expressed in Eulerian (fixed) coordinates \cite{Patterson:1983}. This is achieved by considering the density at two successive times with a small separation,
\begin{align}
 \delta \rho(\vec{x},t) &= \rho(\vec{x}+\vec{U}\delta t, t + \delta t) - \rho(\vec{x},t) \nonumber \\
 &= \delta t \left( \frac{\partial}{\partial t} + \vec{U}\cdot\vec{\nabla} \right) \rho(\vec{x},t) \nonumber \\
 &= \delta t \frac{D \rho(\vec{x},t)}{D t} \ .
\end{align}
If the fluid is incompressible, there cannot be any variation of density as the fluid moves and we must satisfy
\begin{equation}
 \frac{D \rho(\vec{x},t)}{Dt} = 0 \ ,
\end{equation}
which, for the continuity equation \eqref{eq:continuity_gen} to hold, requires that the velocity field be solenoidal (or divergenceless),
\begin{equation}
 \label{eq:continuity}
 \vec{\nabla}\cdot\vec{U} = 0 \ .
\end{equation}
This is referred to as the \emph{incompressibility condition}, and for an incompressible flow is equivalent to equation \eqref{eq:continuity_gen}.

\subsection{The Navier-Stokes equations}
The convective derivative can also be used to find the acceleration of a volume of fluid in our Eulerian coordinates \cite{Patterson:1983},
\begin{equation}
 \vec{A}(\vec{x},t) = \frac{D \vec{U}(\vec{x},t)}{Dt} \ ,
\end{equation}
from which Newton's second law gives us the rate of change of momentum
\begin{equation}
 \rho \frac{D U_\alpha(\vec{x},t)}{Dt} = \rho F_\alpha(\vec{x},t) + \frac{\partial}{\partial x_\beta} \sigma_{\alpha\beta} \ ,
\end{equation}
where $\vec{F}$ is the external body force (density) acting on the fluid and $\sigma_{\alpha\beta}$ the stress tensor. Note that, for the velocity field to remain solenoidal, the external force must also satisfy $\vec{\nabla}\cdot\vec{F} = 0$. The Greek tensor indices $\alpha, \beta = 1,2,3$ label the three mutually-orthogonal components of our flow and we employ the Einstein summation convention, by which repeated indices are summed. The stress tensor may be decomposed into an isotropic (normal) stress and a \emph{deviatoric} component \cite{Patterson:1983},
\begin{equation}
 \sigma_{\alpha\beta} = -P \delta_{\alpha\beta} + 2\mu_0 \big( S_{\alpha\beta} - \tfrac{1}{3}\tr{S}\delta_{\alpha\beta} \big) \ ,
\end{equation}
where $\delta_{\alpha\beta}$ is the Kronecker-$\delta$ and the deviatoric part (second term) has been expressed for a Newtonian fluid in terms of the molecular viscosity, $\mu_0$, and the (symmetric) rate-of-strain tensor,
\begin{equation}
 S_{\alpha\beta} = \frac{1}{2} \left( \frac{\partial U_\alpha}{\partial x_\beta} + \frac{\partial U_\beta}{\partial x_\alpha} \right) \ .
\end{equation}
For an incompressible fluid, $P$ becomes the thermodynamic pressure and the trace of the rate-of-strain tensor vanishes,
\begin{equation}
 \tr{S} = \vec{\nabla}\cdot\vec{U} = 0 \ ,
\end{equation}
which leads us to the \emph{Navier-Stokes equation}
\begin{equation}
 \rho\left( \frac{\partial U_\alpha(\vec{x},t)}{\partial t} + U_\beta(\vec{x},t) \frac{\partial U_\alpha(\vec{x},t)}{\partial x_\beta} \right) = -\frac{\partial P(\vec{x},t)}{\partial x_\alpha} + \mu_0 \nabla^2 U_\alpha(\vec{x},t) + \rho F_\alpha(\vec{x},t) \ .
\end{equation}
Note that the continuity equation has been used to eliminate the term generated when the derivative acts on the second term of the strain rate tensor. By considering the density to be constant, we define the \emph{kinematic} viscosity $\nu_0 = \mu_0/\rho$ and write the Navier-Stokes equation (hereafter NSE) in vector form as
\begin{equation}
 \label{eq:NSE_x}
 \frac{\partial \vec{U}}{\partial t} + \big(\vec{U}\cdot\vec{\nabla}\big) \vec{U} = -\frac{1}{\rho} \vec{\nabla} P + \nu_0 \nabla^2 \vec{U} + \vec{F} \ .
\end{equation}
Together with initial and boundary conditions, this equation is believed to completely describe the flow of an incompressible Newtonian fluid, including both laminar and turbulent regimes.

The non-linear term $\big(\vec{U}\cdot\vec{\nabla}\big) \vec{U}$ present in the NSE is responsible for the difficulty in its solution. It couples together a wide range of scales, allowing them to exchange energy with one another.

In the absence of the viscous term, this equation is referred to as the \emph{Euler} equation. It should be noted that the derivation of the Navier-Stokes equation relies on the fluid being treated as a continuum. Since turbulence generates smaller and smaller scales, it has been suggested that the equations may not be suitable for the study of turbulence if scales comparable to the molecular mean free path are excited \cite{thesis:apquinn,davidson:2004-book}. It is the viscous term which comes to the rescue, suppressing the small scales and preventing this from occurring. Also, it is this dissipative nature of turbulence that prevents the use of classical variational principle approaches \cite{McComb:2012}.

The non-linearity does, however, ensure that small enough scales are always created such that viscosity is important and the energy is lost. Turbulence is therefore a decaying phenomenon, characterised by a large amount of energy dissipation. As such, for turbulence to be sustained it requires an input of energy. We shall see that, in the case of a mean flow, this energy can be taken from the mean flow itself.

\subsection{The Reynolds equation}\label{subsec:Reynolds_equation}

Since turbulence is a property of fluid flow rather than the fluid itself \cite{TennekesLumley:1972},
we consider decomposing the flow into its mean and fluctuating parts
\begin{equation}
 \label{eq:Reynolds_decomp}
 \vec{U} = \overline{\vec{U}} + \vec{u} \ , \qquad\qquad \langle \vec{u} \rangle = 0 \ ,
\end{equation}
where the fluctuating part must average to zero. The angle brackets $\langle \cdots \rangle$ denote an ensemble average and an overline represents a mean value. This is called \emph{Reynolds decomposition}. The incompressibility condition then reads
\begin{equation}
 \vec{\nabla}\cdot \overline{\vec{U}} + \vec{\nabla}\cdot \vec{u} = 0 \ .
\end{equation}
Averaging this equation, we see that $\vec{\nabla}\cdot \overline{\vec{U}} = 0$ and we must have
\begin{equation}
 \vec{\nabla}\cdot \vec{u}  = 0\ ;
\end{equation}
in other words, the mean flow and fluctuations are separately incompressible.

We may construct the Reynolds equation for the mean flow by inserting the decomposition in equation \eqref{eq:Reynolds_decomp} into equation \eqref{eq:NSE_x} and averaging to find
\begin{equation}
 \label{eq:Reynolds_equation}
 \frac{\partial \overline{U}_\alpha}{\partial t} + \overline{U}_\beta \partial_\beta \overline{U}_\alpha = -\frac{1}{\rho} \partial_\alpha \overline{P} + \nu_0 \nabla^2 \overline{U}_\alpha - \partial_\beta \langle u_\alpha u_\beta \rangle + \overline{F}_\alpha \ .
\end{equation}
The tensor $-\rho \langle{\vec{u}\otimes\vec{u}}\rangle$ is known as the \emph{Reynolds stress} and the mean flow must do work against it, thus energy is removed from the mean flow by the fluctuations \cite{TennekesLumley:1972}. This `production method' feeds the turbulence with energy. By trying to relate $\langle u_\alpha u_\beta \rangle$ to the mean rate-of-strain tensor, $\overline{S}_{\alpha\beta}$, this equation is used as a model in numerical investigations of turbulence (see RANS in section \ref{subsec:DNS}).

A similar equation for the fluctuations $\vec{u}$ can be found \cite{McComb:1990-book} by inserting the decomposition in equation \eqref{eq:Reynolds_decomp} into the Navier-Stokes equations \eqref{eq:NSE_x} and subtracting equation \eqref{eq:Reynolds_equation},
\begin{equation}
 \frac{\partial u_\alpha}{\partial t} + u_\beta \partial_\beta u_\alpha + u_\beta \partial_\beta \overline{U}_\alpha + \overline{U}_\beta \partial_\beta u_\alpha = -\frac{1}{\rho} \partial_\alpha p + \nu_0 \nabla^2 u_\alpha + \partial_\beta \langle u_\alpha u_\beta \rangle + f_\alpha \ ,
\end{equation}
which can be seen to include the convection of the fluctuations by the mean flow and the Reynolds stress now as a production term.

\section{Homogeneity and isotropy}
As mentioned in the introduction at the beginning of this chapter, homogeneity and isotropy introduce a great deal of simplifications to the statistical study of turbulence. It should be borne in mind that these are statistical properties of the probability distribution of the velocity field, not of an instantaneous snapshot of the velocity field. (Likewise, the concept of stationarity applies to average values.)
\begin{description}
 \item[Homogeneity:]{A consequence of translation invariance of the probability distribution. This prevents absolute positions from affecting a measurement or result; instead, only relative separations $\vec{r} = \vec{x}'-\vec{x}$ can be involved.}

 \item[Isotropy:]{A consequence of the invariance of the probability distribution under rotations of the coordinate system. This requires that there be no favoured direction in the system, and statistical properties become a function of only the scalar separation $r = \lvert\vec{r}\rvert$ between points.}
\end{description}

We now consider a homogeneous, isotropic system, which will be the case for the rest of this thesis. Homogeneity is broken by boundary conditions, so our system must fill all of space. Since there cannot be any preferred direction for isotropy to be valid, there must be no mean flow, $\overline{\vec{U}} = 0$. We are therefore limited to studying only the velocity fluctuations, $\vec{u}$, which must satisfy
\begin{equation}
 \label{eq:NSE_x_iso}
 \frac{\partial u_\alpha(\vec{x},t)}{\partial t} + u_\beta(\vec{x},t) \frac{\partial u_\alpha(\vec{x},t)}{\partial x_\beta} = -\frac{1}{\rho} \frac{\partial p(\vec{x},t)}{\partial x_\alpha} + \nu_0 \nabla^2 u_\alpha(\vec{x},t) + f_\alpha(\vec{x},t) \ ,
\end{equation}
supplemented by the incompressibility condition, $\partial_\alpha u_\alpha(\vec{x},t) = 0$ (and $\partial_\alpha f_\alpha(\vec{x},t) = 0$). In the absence of a mean flow, the fluctuations have no production term, such as that provided by shear flows, and require an artificial source of energy injection to be maintained.

\subsection{Correlations of the velocity field}
Direct solution of the Navier-Stokes equations has so far proven to be unsuccessful and this has led many to study turbulence as a statistical problem. The general behaviour of a flow can then be investigated and these techniques have been used in many applications in, for example, engineering, along with research on the fundamental processes of turbulence.

A key concept in the statistical study of any problem involving many degrees of freedom is the correlation of the field with itself at other positions in space or time. We therefore define the correlation tensor
\begin{equation}
 \label{eq:1:corr_tensor}
 C_{\alpha\beta}(\vec{x},\vec{x}';t,t') = \langle u_\alpha(\vec{x},t) u_\beta(\vec{x}',t') \rangle
\end{equation}
as being the correlation of the velocity field at points $(\vec{x},t)$ and $(\vec{x}',t')$. This is also called the \emph{second-order moment} of the velocity field. The third-order correlation tensor (or moment) may be defined in a similar way,
\begin{equation}
 C_{\alpha\beta\gamma}(\vec{x},\vec{x}',\vec{x}'';t,t',t'') = \langle u_\alpha(\vec{x},t) u_\beta(\vec{x}',t') u_\gamma(\vec{x}'',t'') \rangle \ ,
\end{equation}
and so on for moments of higher orders. Note that the positions $\{\vec{x},\vec{x}',\cdots\}$ and times $\{t,t',\cdots\}$ need not be unique.

Using the constraint of homogeneity, the correlation tensor given in equation \eqref{eq:1:corr_tensor} can be written
\begin{align}
 C_{\alpha\beta}(\vec{x},\vec{x}';t,t') &= \langle u_\alpha(\vec{x},t) u_\beta(\vec{x}+\vec{r},t') \rangle \nonumber \\
 &= \langle u_\alpha(\vec{0},t) u_\beta(\vec{0}+\vec{r},t') \rangle \nonumber \\
 &= C_{\alpha\beta}(\vec{r};t,t') \ ,
\end{align}
where $\vec{r} = \vec{x}'-\vec{x}$ is the relative separation vector. For the correlation tensor to be isotropic, we must be able to express it in terms of invariant tensors,
\begin{equation}
 \label{eq:C_iso_expansion}
 C_{\alpha\beta}(\vec{r};t,t') = A(r;t,t') \delta_{\alpha\beta} + B(r;t,t') \frac{r_\alpha r_\beta}{r^2} \ ,
\end{equation}
where the coefficients are functions of the scalar separation $r$ only.

The second-order moment is associated with the kinetic energy (density) of the fluctuations, which is defined as
\begin{align}
 E(t) &= \tfrac{1}{2} \langle u_\alpha(\vec{x},t) u_\alpha(\vec{x},t) \rangle \\
 &= \tfrac{1}{2} \big( \langle u_x^2 \rangle + \langle u_y^2 \rangle + \langle u_z^2 \rangle \big) \nonumber \ .
\end{align}
When the system is isotropic, we expect that the average velocity correlation in each direction is the same, such that $\langle u_x^2 \rangle = \langle u_y^2 \rangle = \langle u_z^2 \rangle = u^2$ and we have
\begin{align}
E(t) &= \tfrac{3}{2} u^2 \ .
\end{align}
This defines the root-mean-square (rms) velocity, $u(t) = \sqrt{2 E(t)/3}$.

\subsection{Longitudinal and transverse correlations}\label{subsec:long_trans_corr}
\begin{figure}[tb!]
 \begin{center}
  \subfigure[Orientation of the correlations]{
   \includegraphics[width=0.4\textwidth]{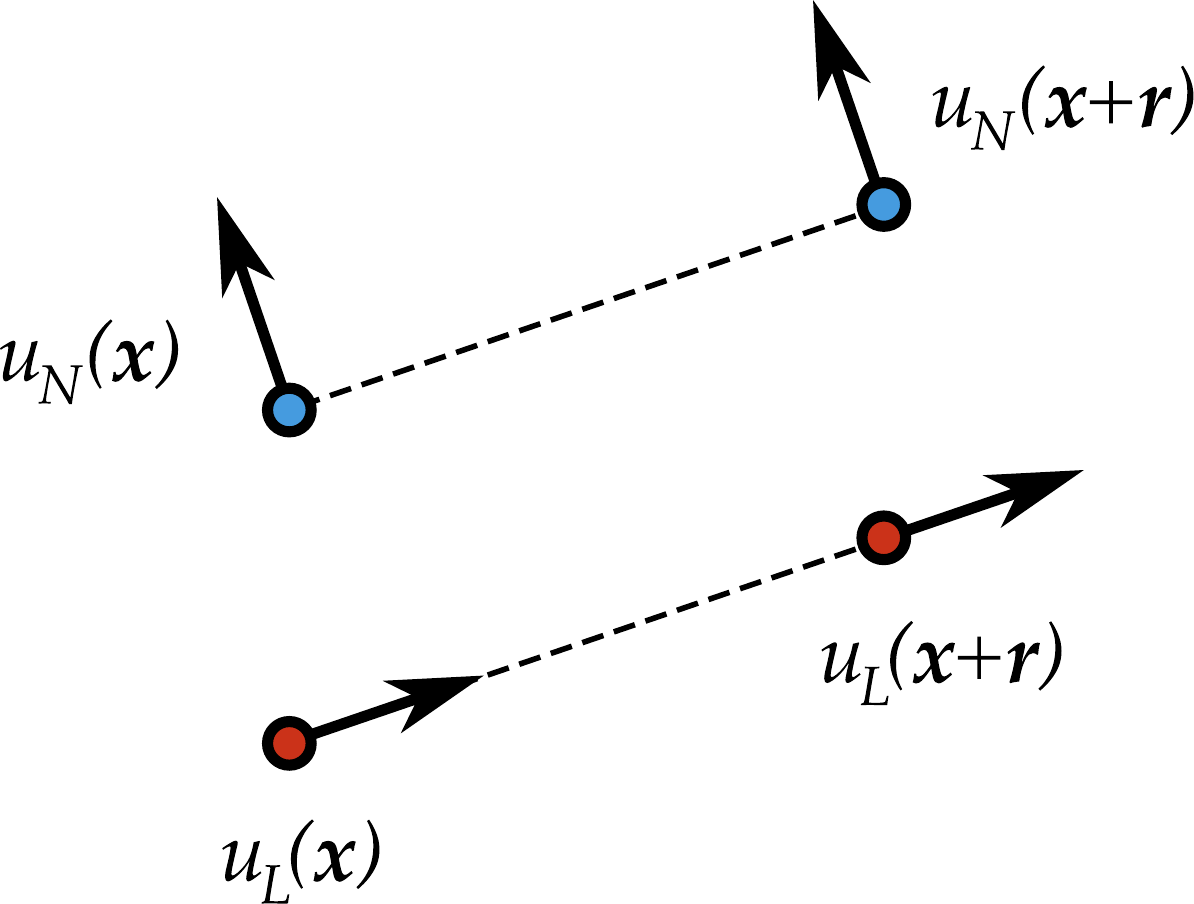}
  }\hspace{2em}
  \subfigure[Variation of the correlation functions]{
   \includegraphics[width=0.5\textwidth]{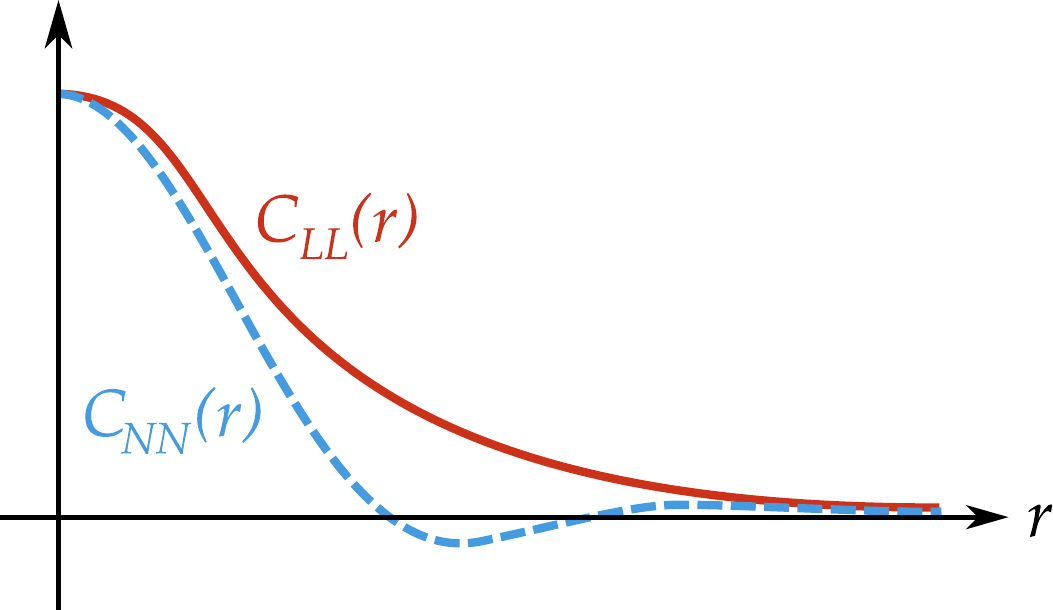}
  }
 \end{center}
 \caption{The orientation and variation of the second-order, two-point single-time longitudinal and transverse correlation functions.}
 \label{fig:long_trans_corr2}
\end{figure}

For the purpose of this section, we specialise to the two-point, single-time case and as such drop the time argument. Looking once again at the isotropic correlation tensor, we see that the separation vector $\vec{r}$ has actually broken isotropy by introducing a sense of direction. It therefore seems natural to consider the correlations parallel and perpendicular to this induced direction,
\begin{align}
 C_{LL}(r) &= \langle u_L(\vec{x}) u_L(\vec{x}+\vec{r}) \rangle \nonumber \\
 C_{NN}(r) &= \langle u_N(\vec{x}) u_N(\vec{x}+\vec{r}) \rangle \ ,
\end{align}
where $u_L = \vec{u}\cdot\vec{\hat{r}}$ is the component of the velocity in the direction $\vec{\hat{r}} = \vec{r}/r$ and $u_N$ is any normal component. The orientation and variation of the two isotropic correlation functions is sketched in figure \ref{fig:long_trans_corr2}. It can be shown that the isotropic expansion of the correlation tensor in equation \eqref{eq:C_iso_expansion} can be written in terms of these longitudinal and transverse correlation functions as \cite{Batchelor:1953-book,MoninYaglom:vol2}
\begin{equation}
 \label{eq:R_iso_exp}
 C_{\alpha\beta}(\vec{r}) = C_{NN}(r) \delta_{\alpha\beta} + \Big( C_{LL}(r) - C_{NN}(r) \Big) \frac{r_\alpha r_\beta}{r^2} \ .
\end{equation}
Moreover, the continuity equation may be used to show that $\partial C_{\alpha\beta}(\vec{r})/\partial r_\alpha = \partial C_{\alpha\beta}(\vec{r})/\partial r_\beta = 0$, which requires the relationship between the longitudinal and transverse components to be
\begin{equation}
 \label{eq:CNN_exp}
 C_{NN}(r) = \left( 1 + \frac{r}{2} \frac{\partial}{\partial r} \right) C_{LL}(r) \ .
\end{equation}

A similar analysis may be performed for the two-point, third-order moment
\begin{align}
 C_{\alpha\beta,\gamma}(\vec{r}) &= \langle u_\alpha(\vec{x}) u_\beta(\vec{x}) u_\gamma(\vec{x}+\vec{r}) \rangle \nonumber \\
 &= \Big[ C_{LL,L}(r) - 2C_{LN,N}(r) - C_{NN,L}(r) \Big] \frac{r_\alpha r_\beta r_\gamma}{r^3} + C_{NN,L}(r) \frac{r_\gamma}{r}\delta_{\alpha\beta} \nonumber \\
 &\qquad + C_{LN,N}(r) \left[ \frac{r_\alpha}{r} \delta_{\beta\gamma} + \frac{r_\beta}{r} \delta_{\alpha\gamma} \right] \ .
\end{align}
The comma is used to indicate which indices correspond to velocities evaluated at $\vec{x}+\vec{r}$, and each can be a longitudinal or transverse component.
The longitudinal and transverse correlations must ensure that the continuity equation $\partial C_{\alpha\beta,\gamma}(\vec{r}) / \partial r_\gamma = 0$ is satisfied and therefore have the relationships \cite{Batchelor:1953-book,MoninYaglom:vol2}
\begin{align}
 C_{LN,N}(r) &= \frac{1}{2} \left( 1 + \frac{r}{2} \frac{\partial}{\partial r} \right) C_{LL,L}(r) \nonumber \\
 C_{NN,L}(r) &= -\frac{1}{2} C_{LL,L}(r) \ .
\end{align}

\subsection{Scales of turbulent motion}\label{subsec:lengthscales}
Turbulence is a problem of many length-scales. The longitudinal correlation function can be used to establish two characteristic scales which describe the correlations. To do this, we consider the dimensionless correlation function
\begin{equation}
 f(r) = \frac{C_{LL}(r)}{u^2} \ ,
\end{equation}
where $u$ is the rms velocity. At $r = 0$, the correlation is just $\langle u^2_L \rangle$. Since the system is isotropic, we expect the energy to be distributed equally in each direction so $\langle u^2_L \rangle = u^2$. In which case, the dimensionless correlation function starts from unity at $r = 0$ and decays to zero as $r$ is increased. The first length-scale we define is by considering $f(r)$ to be exponentially decaying,
\begin{equation}
 f(r) \sim \exp \left( -\frac{r}{L} \right) \ .
\end{equation}
This defines the \emph{integral length-scale}, $L$. Note that this form does not respect the symmetry of $C_{LL}(r)$ which is even in $r$, and hence so is $f(r)$. This requires that the gradient be zero at $r = 0$, which is not the case for the exponential. Since this exercise is merely to introduce the integral scale and its physical interpretation, we continue regardless. The integral (length)scale is a measure of the large scale correlated fluctuations and can be equivalently evaluated as
\begin{equation}
 L = \int_0^\infty dr\ f(r) \ .
\end{equation}

The second length-scale we can define is found by Taylor expanding the dimensionless correlation function close to $r = 0$,
\begin{align}
 f(r) &= f(0) + r\left.\frac{\partial f}{\partial r}\right\vert_{r = 0} + \frac{r^2}{2} \left. \frac{\partial^2 f}{\partial r^2}\right\vert_{r = 0} + \cdots \nonumber \\
 \label{eq:lambda}
 &\simeq 1 - \frac{r^2}{2\lambda^2} \ ,
\end{align}
where the first derivative vanishes because the function is even and we define the \emph{Taylor microscale} as
\begin{equation}
 \frac{1}{\lambda^2} = - \left. \frac{\partial^2 f(r)}{\partial r^2}\right\vert_{r = 0} \ .
\end{equation}
This was actually defined in Taylor \cite{Taylor:1935p308} through the transverse correlation function, $g(r) = C_{NN}(r)/u^2$, using an osculating parabola $g(r) = 1 - r^2/\lambda^2$ such that $g(\lambda) = 0$. The second-order isotropic relation in equation \eqref{eq:CNN_exp} is then used to obtain $f(r)$ as found in equation \eqref{eq:lambda}. For isotropic turbulence, this was found by Taylor \cite{Taylor:1935p308} (see also Batchelor \cite{Batchelor:1953-book}) to be equivalent to
\begin{equation}
 \lambda^2 = \frac{15\nu_0 u^2}{\varepsilon} \ .
\end{equation}
Due to its definition in the proximity of $r = 0$, the Taylor microscale can be regarded as a `small' scale. But it is not necessarily the smallest scale that can be generated. In fact, it cannot be a dissipation scale because it involves the rms velocity in its definition, where the dissipation scales should be insensitive to the large scale motion \cite{TennekesLumley:1972}.

\subsection{Reynolds number}\label{subsec:1:reynolds_number}
Whilst studying the transition from laminar to turbulent flow, Reynolds noticed that the onset occurred at a critical value of a dimensionless parameter, now known as the Reynolds number. This is defined as
\begin{equation}
 Re = \frac{Ul}{\nu_0} \ ,
\end{equation}
where $U,l$ are characteristic velocity and length-scales, respectively, such as the centre-line velocity and radius of the pipe in pipe flow. Above the critical value, the flow was always seen to be turbulent.

For our purposes, we define two Reynolds numbers using the rms velocity and corresponding to the two length-scales of the previous section: the \emph{integral Reynolds number} and \emph{Taylor-Reynolds number},
\begin{equation}
 R_L = \frac{u L}{\nu_0} \ , \qquad\qquad R_\lambda = \frac{u \lambda}{\nu_0} \ .
\end{equation}

The Reynolds number can be seen as measuring the comparative strength of the inertial and viscous forces at work. The inertial term has dimension $[u]^2[L]^{-1}$ and the viscous term $[\nu_0][u][L]^{-2}$. If we take our characteristic scales as representative, then the Reynolds number can be seen as the ratio
\begin{equation}
 R_L = \frac{u^2/L}{\nu_0 u /L^2} \ ;
\end{equation}
that is, it quantifies the relationship between inertial forces driving the turbulence and viscous forces suppressing the small scales.

\section{The spectral representation of turbulence}
Spectral methods have enjoyed great success in the analysis of differential equations. The essential step is expanding the velocity field as a linear superposition of a set of orthonormal basis functions, whose form depends on the geometry and symmetry of the problem. Further discussion can be found in section \ref{subsec:2:spectral_methods}. We now consider studying our isotropic turbulence in a spectral representation.

\subsection{Spectral equation of motion}
To begin, we consider our system to be contained in a box of side $L$ with periodic boundary conditions (which allow us to maintain isotropy). Since our system is periodic, we may expand the velocity field in a Fourier series
\begin{equation}
 u_\alpha(\vec{x},t) = \sum_{\vec{k}} u_\alpha(\vec{k},t) e^{i\vec{k}\cdot\vec{x}} \ ,
\end{equation}
which has the spectrum
\begin{equation}
 u_\alpha(\vec{k},t) = \left(\frac{1}{L} \right)^3 \int d^3x\ u_\alpha(\vec{x},t) e^{-i\vec{k}\cdot\vec{x}} \ .
\end{equation}
The momentum in this finite system is quantised since all waves have to `fit in the box' and can take the values
\begin{equation}
 k_\alpha = \frac{2\pi}{L} n_\alpha \ , \qquad\qquad n_\alpha \in \mathbb{Z} \ .
\end{equation}
The components $n_\alpha$ are known as the \emph{wavenumbers} and $\vec{n}$ the \emph{wavevector}. With the choice $L = 2\pi$ we see the momentum and wavevector coincide.

As we take the infinite system limit $L \to \infty$, the Fourier series goes over to the Fourier transform,
\begin{align}
 u_\alpha(\vec{x},t) &= \int d^3k\ u_\alpha(\vec{k},t) e^{i\vec{k}\cdot\vec{x}} \ , \qquad u_\alpha(\vec{k},t) &= \left(\frac{1}{2\pi} \right)^3 \int d^3x\ u_\alpha(\vec{x},t) e^{-i\vec{k}\cdot\vec{x}} \ .
\end{align}
We now take the Fourier transform of the Navier-Stokes equation as given in equation \eqref{eq:NSE_x_iso}, noting that differentiation in configuration space becomes multiplication, since
\begin{equation}
 \frac{\partial^m}{\partial x_\alpha^m} \int d^3k\ u_\beta(\vec{k},t) e^{i\vec{k}\cdot\vec{x}} = \int d^3k\ (ik_\alpha)^m u_\beta(\vec{k},t) e^{i\vec{k}\cdot\vec{x}} \ .
\end{equation}
By setting $m = 1, \alpha = \beta$, we see that the continuity equation becomes equivalent to the requirement that the velocity coefficient be orthogonal to its wavevector,
\begin{equation}
 \vec{k}\cdot\vec{u}(\vec{k},t) = 0 \ .
\end{equation}

The non-linear term requires a little more attention when performing the Fourier transform (where we use $\mathcal{F}\big[ \cdots \big]$ to denote the Fourier transform):
\begin{align}
 \mathcal{F} \left[ u_\beta \frac{\partial u_\alpha}{\partial x_\beta} \right] &= \left(\frac{1}{2\pi}\right)^3 \int d^3x\ \left[ u_\beta(\vec{x},t) \frac{\partial u_\alpha(\vec{x},t)}{\partial x_\beta} \right] e^{-i\vec{k}\cdot\vec{x}} \\
 &= \left(\frac{1}{2\pi}\right)^3 \int d^3x \int d^3j \int d^3q\ u_\beta(\vec{j},t)e^{i\vec{j}\cdot\vec{x}} (iq_\beta) u_\alpha(\vec{q},t) e^{i\vec{q}\cdot\vec{x}} e^{-i\vec{k}\cdot\vec{x}} \nonumber \\
 &= i\int d^3j \int d^3q\ q_\beta\ u_\beta(\vec{j},t) u_\alpha(\vec{q},t) \left\{\left(\frac{1}{2\pi}\right)^3 \int d^3x\ e^{i\vec{x}\cdot(\vec{j}+\vec{q}-\vec{k})} \right\} \nonumber \ .
\end{align}
The term in the braces is nothing but the definition of the Dirac $\delta$-function, which we insert and perform the integral over $\vec{q}$ to obtain
\begin{align}
 \label{eq:do_delta_int}
 \mathcal{F} \left[ u_\beta(\vec{x},t) \frac{\partial u_\alpha(\vec{x},t)}{\partial x_\beta} \right] 
  &= i\int d^3j \int d^3q\ q_\beta\ u_\beta(\vec{j},t) u_\alpha(\vec{q},t)\ \delta(\vec{j} + \vec{q} - \vec{k}) \nonumber \\
  &= i\int d^3j\ (k_\beta - j_\beta)\ u_\beta(\vec{j},t) u_\alpha(\vec{k}-\vec{j},t) \nonumber \\
  &= ik_\beta \int d^3j\ u_\beta(\vec{j},t) u_\alpha(\vec{k}-\vec{j},t) \ ,
\end{align}
where the last line used the incompressibility condition. The equation of motion becomes
\begin{align}
 \frac{\partial u_\alpha(\vec{k},t)}{\partial t} + ik_\beta \int d^3j\ u_\beta(\vec{j},t) u_\alpha(\vec{k}-\vec{j},t) = \frac{k_\alpha}{i\rho} p(\vec{k},t) + \nu_0 (ik_\beta)^2 u_\alpha(\vec{k},t) + f_\alpha(\vec{k},t) \ .
\end{align}
As the field is incompressible, the pressure term can be eliminated by multiplying by $k_\alpha$ (and summing)
\begin{align}
 i k_\alpha k_\beta \int d^3j\ u_\beta(\vec{j},t) u_\alpha(\vec{k}-\vec{j},t) = -\frac{ik^2}{\rho} p(\vec{k},t) \ ,
\end{align}
which, along with a relabelling of the dummy indices $\beta,\alpha \to \gamma,\beta$, is rearranged to
\begin{align}
 -\frac{i}{\rho} p(\vec{k},t) = i \frac{k_\beta k_\gamma}{k^2} \int d^3j\ u_\gamma(\vec{j},t) u_\beta(\vec{k}-\vec{j},t)
\end{align}
and substituted back into the transformed equation to give
\begin{align}
 \left( \frac{\partial}{\partial t} + \nu_0 k^2 \right) u_\alpha(\vec{k},t) &= -i k_\gamma \left( \delta_{\alpha\beta} - \frac{k_\alpha k_\beta}{k^2} \right) \int d^3j\ u_\gamma(\vec{j},t) u_\beta(\vec{k}-\vec{j},t) + f_\alpha(\vec{k},t) \nonumber \\
 \label{eq:NSE_k_p1}
 &= -ik_\gamma P_{\alpha\beta}(\vec{k}) \int d^3j\ u_\gamma(\vec{j},t) u_\beta(\vec{k}-\vec{j},t) + f_\alpha(\vec{k},t) \ ,
\end{align}
where the \emph{projection operator} has been defined as
\begin{equation}
 \label{eq:def:projection}
 P_{\alpha\beta}(\vec{k}) = \delta_{\alpha\beta} - \frac{k_\alpha k_\beta}{k^2} \ .
\end{equation}
This operator has the properties that
\begin{align}
 P_{\alpha\beta}(\vec{k}) k_\alpha = 0 \ , \qquad P_{\alpha\beta}(\vec{k}) u_\alpha(\vec{k},t) = 0  \qquad \textrm{and} \qquad \tr{P} = P_{\alpha\alpha}(\vec{k}) = 2 \ .
\end{align}
It can be seen to ensure that the velocity field remains solenoidal, since the projection operator subtracts from the velocity field any divergence which is present.

Returning to equation \eqref{eq:NSE_k_p1}, we see that we are free to exchange the dummy indices $\beta,\gamma$. Also, since the integration is over all space, we may relabel $\vec{j} \to \vec{k}-\vec{j}$ (with unit Jacobian). This allows us to define the symmetric (vertex) operator
\begin{equation}
 \label{eq:def:vertex}
 M_{\alpha\beta\gamma}(\vec{k}) = \frac{1}{2i} \Big( k_\gamma P_{\alpha\beta}(\vec{k}) + k_\beta P_{\alpha\gamma}(\vec{k}) \Big)
\end{equation}
and write the Navier-Stokes equation in Fourier space as
\begin{equation}
 \label{eq:NSE}
 \left( \frac{\partial}{\partial t} + \nu_0 k^2 \right) u_\alpha(\vec{k},t) = M_{\alpha\beta\gamma}(\vec{k}) \int d^3j\ u_\beta(\vec{j},t) u_\gamma(\vec{k}-\vec{j},t) + f_\alpha(\vec{k},t) \ .
\end{equation}
This will be the starting point for most of our work with the Navier-Stokes equation. The vertex operator satisfies the following relations
\begin{align}
 P_{\alpha\rho}(\vec{k}) M_{\rho\beta\gamma}(\vec{k})& = M_{\alpha\beta\gamma}(\vec{k})\ , \nonumber \\
 M_{\alpha\beta\gamma}(\vec{k}) &= M_{\alpha\gamma\beta}(\vec{k})\ , \quad\text{and} \nonumber \\
 M_{\alpha\beta\gamma}(-\vec{k}) &= - M_{\alpha\beta\gamma}(\vec{k}) \ .
\end{align}
Note that the force must also be solenoidal ($\vec{k}\cdot\vec{f}(\vec{k},t) = 0$) to ensure that the velocity remains so.

\subsection{Energy balance and the energy cascade}
We start by considering the correlation in Fourier space between two modes with momenta $\vec{k},\vec{k}'$. By inserting the Fourier transformation, we find that this is
\begin{align}
 \langle u_\alpha(\vec{k},t) u_\beta(\vec{k}',t') \rangle &= \left(\frac{1}{2\pi} \right)^6 \int d^3x \int d^3x'\ \langle u_\alpha(\vec{x},t) u_\beta(\vec{x}',t') \rangle e^{-i(\vec{k}\cdot\vec{x} + \vec{k}'\cdot\vec{x}')} \nonumber \\
 &= \left(\frac{1}{2\pi} \right)^6 \int d^3x \int d^3r\ C_{\alpha\beta}(\vec{r};t,t')\ e^{-i(\vec{k}+\vec{k}')\cdot\vec{x}}\ e^{-i\vec{k}'\cdot\vec{r}} \nonumber \\
 &= \delta(\vec{k}+\vec{k}') \left(\frac{1}{2\pi} \right)^3 \int d^3r\ C_{\alpha\beta}(\vec{r};t,t') \ e^{-i\vec{k}'\cdot\vec{r}} \nonumber \\
 &= \delta(\vec{k}+\vec{k}')\ C_{\alpha\beta}(\vec{k},\vec{k}';t,t') \ ,
\end{align}
where the second line used the homogeneity of the system to write $\vec{x}'=\vec{x}+\vec{r}$ and the penultimate line used the integral definition of the $\delta$-function. Thus the Fourier transform of the two-point correlation tensor is
\begin{equation}
 C_{\alpha\beta}(\vec{k};t,t') = \langle u_\alpha(\vec{k},t) u_\beta(-\vec{k},t') \rangle \ .
\end{equation}
Homogeneity has imposed unimodal coupling of the velocity field. This is because the field was real in configuration space and as such the Fourier transform possesses \emph{Hermitian symmetry}, that is
\begin{equation}
 u_\alpha(-\vec{k},t) = u^*_\alpha(\vec{k},t) \ ,
\end{equation}
with the star representing complex conjugation. Isotropy further restricts the form of the correlation tensor. In Fourier space, we must have
\begin{equation}
 C_{\alpha\beta}(\vec{k};t,t') = P_{\alpha\beta}(\vec{k}) C(k;t,t') \ .
\end{equation}

\subsubsection{The energy spectrum}
With the Fourier decomposition of the velocity field, we are expanding in a superposition of (standing) waves. With these waves, one can associate an amount of energy to their oscillation, which can be interpreted as the energy contained in motions of a certain length-scale. The energy spectrum therefore gives a scalar representation of how the energy is distributed among modes or, equivalently, length-scales. In configuration-space, the total energy of the system is given by the average kinetic energy,
\begin{equation}
 E(t) = \tfrac{1}{2} \langle u_\alpha(\vec{x},t) u_\alpha(\vec{x},t) \rangle \ ,
\end{equation}
where the density has been taken to be unity or $E$ is defined as the kinetic energy per unit mass.
This can be written in terms of the Fourier-components as
\begin{align}
 E(t) &= \tfrac{1}{2} \int d^3k\ \langle u_\alpha(\vec{k},t) u_\alpha(-\vec{k},t) \rangle \nonumber \\
 &= \int dk\ k^2 \int d\Omega_k\ \tfrac{1}{2}\langle u_\alpha(\vec{k},t) u_\alpha(-\vec{k},t) \rangle \ ,
\end{align}
from which we define the energy spectrum
\begin{equation}
 E(k,t) = k^2 \int d\Omega_k\ \tfrac{1}{2}\langle u_\alpha(\vec{k},t) u_\alpha(-\vec{k},t) \rangle \ .
\end{equation}
For an isotropic field in the continuum, the angular integral is performed to give
\begin{align}
 \label{eq:iso_Ek}
 E(k,t) &= 4\pi k^2 \tfrac{1}{2}\langle u_\alpha(\vec{k},t) u_\alpha(-\vec{k},t) \rangle \ .
\end{align}
We see that the energy spectrum is intimately linked with the correlation tensor. In fact, by considering the Fourier transform of the isotropic correlation tensor, we can show \cite{Batchelor:1953-book,MoninYaglom:vol2,davidson:2004-book}
\begin{equation}
 \label{eq:second_corr}
 C(r) = \tfrac{1}{2} C_{\alpha\alpha}(\vec{r}) = \int dk\ E(k,t)\ \frac{\sin kr}{kr} \ , \qquad E(k,t) = \frac{2}{\pi} \int dr\ C(r)\ kr\sin kr \ .
\end{equation}
In a similar way, the third-order correlation can be connected to the transfer spectrum, which we will meet shortly. This will be discussed further in chapters \ref{chp:forced} and \ref{chp:fKHE}.

\subsubsection{The Lin equation}\label{subsec:lin_equation}
Starting from the Navier-Stokes equation in the form of equation (\ref{eq:forced_nse}), we can form an equation governing how the energy spectrum changes in time. Put simply, we form the Navier-Stokes equation for the modes $\vec{k}$ and $-\vec{k}$, multiplied (from the left) by $u_\alpha$ for the opposite mode, then averaged. So, we have two equations:
\begin{align}
 \left\langle u_\alpha(-\vec{k},t) \left(\partial_t + \nu_0 k^2\right) u_\alpha(\vec{k},t) \right\rangle &= M_{\alpha\beta\gamma}(\vec{k}) \int d^3j\ \left\langle u_\alpha(-\vec{k},t) u_\beta(\vec{j},t) u_\gamma(\vec{k}-\vec{j},t) \right\rangle \nonumber \\
 &\qquad+ \left\langle u_\alpha(-\vec{k},t) f_\alpha(\vec{k},t) \right\rangle \\
 \left\langle u_\alpha(\vec{k},t) \left(\partial_t + \nu_0 k^2\right) u_\alpha(-\vec{k},t) \right\rangle &= M_{\alpha\beta\gamma}(-\vec{k}) \int d^3j\ \left\langle u_\alpha(\vec{k},t) u_\beta(\vec{j},t) u_\gamma(-\vec{k}-\vec{j},t) \right\rangle \nonumber \\
 &\qquad+ \left\langle u_\alpha(\vec{k},t) f_\alpha(-\vec{k},t) \right\rangle \ .
\end{align}
Adding the two equation together and dropping the time argument, one obtains
\begin{align}
 \left( \frac{\partial}{\partial t} + 2\nu_0 k^2 \right) \langle u_\alpha(-\vec{k}) u_\alpha(\vec{k}) \rangle &= M_{\alpha\beta\gamma}(\vec{k}) \int d^3j\ \langle u_\alpha(-\vec{k}) u_\beta(\vec{j}) u_\gamma(\vec{k}-\vec{j}) \rangle \nonumber \\
 &\qquad+ M_{\alpha\beta\gamma}(-\vec{k}) \int d^3j\ \langle u_\alpha(\vec{k}) u_\beta(\vec{j}) u_\gamma(-\vec{k}-\vec{j}) \rangle \nonumber \\
 &\qquad+ \langle f_\alpha(\vec{k}) u_\alpha(-\vec{k})\rangle + \langle f_\alpha(-\vec{k}) u_\alpha(\vec{k})\rangle \ .
\end{align}
Since $u_\alpha(-\vec{k}) = u^*_\alpha(\vec{k})$, the left hand side of this equation is real. In fact, since $M_{\alpha\beta\gamma}(-\vec{k}) = M^*_{\alpha\beta\gamma}(\vec{k})$, the two equations are related by conjugation, so we may write
\begin{align}
 \label{eq:nearly_Lin}
 \left( \frac{\partial}{\partial t} + 2\nu_0 k^2 \right) C_{\alpha\alpha}(\vec{k}) = 2\overline{T}(\vec{k}) + 2 \realpart\left[ \langle f_\alpha(\vec{k}) u_\alpha(-\vec{k}) \rangle \right] \ ,
\end{align}
with the transfer density (where the overline does not represent a mean value)
\begin{equation}
 \overline{T}(\vec{k}) = \realpart\ M_{\alpha\beta\gamma}(\vec{k}) \int d^3j\ \langle u_\alpha(-\vec{k}) u_\beta(\vec{j}) u_\gamma(\vec{k}-\vec{j})\rangle \ .
\end{equation}
Note that $\realpart$ indicates that we are interested in the real part of an expression, rather than $Re$ which is a generic Reynolds number.

The final term on the RHS involving the force correlation is found using the relationship of Novikov \cite{novikov:1965-functionals} for a functional of the velocity $R[\vec{u}]$,
\begin{align}
 \langle f_\alpha(\vec{k},t) R[\vec{u}] \rangle &= \frac{1}{2} \int dt' \int d^3k'\ \langle f_\alpha(\vec{k},t) f_\beta(\vec{k'},t') \rangle \left\langle \frac{\delta R[\vec{u}]}{\delta u_\beta(\vec{k'},t')} \right\rangle \nonumber \\
 \langle f_\alpha(\vec{k},t) u_\alpha(-\vec{k},t) \rangle &= \frac{1}{2} \int dt' \int d^3k'\ \langle f_\alpha(\vec{k},t) f_\beta(\vec{k'},t') \rangle\ \left\langle \frac{\delta u_\alpha(-\vec{k},t)}{\delta u_\beta(\vec{k'},t')} \right\rangle \ .
\end{align}
If the system is isotropic and the forcing is assumed to obey
\begin{align}
 \label{eq:forcing_density}
 \langle f_\alpha(\vec{k},t) f_\beta(\vec{k'},t') \rangle = 2 P_{\alpha\beta}(\vec{k}) F(k) \delta(\vec{k}+\vec{k'}) \delta(t-t') \ ,
\end{align}
then
\begin{align}
 \langle f_\alpha(\vec{k},t) u_\alpha(-\vec{k},t) \rangle &= 2 F(k) \ ,
\end{align}
since the $P_{\alpha\beta}(\vec{k}) P_{\alpha\beta}(\vec{k}) = P_{\alpha\alpha}(\vec{k}) = 2$. Note that this is a real quantity. The averaged equation becomes
\begin{align}
 \left( \frac{\partial}{\partial t} + 2\nu_0 k^2 \right) C(k) = \overline{T}(\vec{k}) + 2F(k) \ .
\end{align}
Under the assumption of isotropy, we trivially multiply through by $4\pi k^2$ to obtain
\begin{align}
 \left( \frac{\partial}{\partial t} + 2\nu_0 k^2 \right) 4\pi k^2 C(k) &= 4\pi k^2 \overline{T}(\vec{k}) + 8\pi k^2 F(k) \ , \qquad\textrm{or} \nonumber \\
  \label{eq:Lin}
  \frac{\partial}{\partial t} E(k) + D(k) &= T(k) + W(k) \ .
\end{align}
This is known as the \emph{Lin equation}. We have defined:
\begin{itemize}
 \item The energy spectrum (again)
 \begin{align}
  E(k,t) &= 4\pi k^2 C(k) = 2\pi k^2 C_{\alpha\alpha}(\vec{k}) = 2\pi k^2 \langle u_\alpha(-\vec{k},t) u_\alpha(\vec{k},t) \rangle \ ;
 \end{align}
 
 \item The dissipation spectrum
 \begin{align}
  D(k,t) &= 2\nu_0 k^2 E(k,t) \ ;
 \end{align}

 \item The work spectrum
 \begin{align}
  W(k,t) &= 4\pi k^2 \langle u_\alpha(-\vec{k},t) f_\alpha(\vec{k},t) \rangle = 8\pi k^2 F(k) \ ;\ \textrm{and}
 \end{align}

 \item The transfer spectrum
 \begin{align}
  T(k,t) &= 4\pi k^2 \overline{T}(\vec{k}) = 4\pi k^2 \realpart\left[ M_{\alpha\beta\gamma}(\vec{k}) \int d^3j\ \langle u_\alpha(-\vec{k},t) u_\beta(\vec{j},t) u_\gamma(\vec{k}-\vec{j},t)\rangle \right] \ .
 \end{align}

\end{itemize}

The Lin equation expresses the balance of energy in mode $k$. In words it reads:
``change in energy of wavenumber $k$ = - energy dissipated + energy transferred to mode $k$ + energy input''.

\subsubsection{The transfer spectrum and conservation of energy}
The transfer spectrum quantifies the amount of energy transferred into mode $k$ due to its non-linear coupling to all other modes. Just as the integral over the energy spectrum measures the total kinetic energy, it can be shown that the transfer spectrum does no work on the system; it simply moves energy around. As such,
\begin{align}
 \label{eq:Tk_vanish}
 \int_0^\infty dk\ T(k,t) &= \realpart \int d^3k \int d^3j\ M_{\alpha\beta\gamma}(\vec{k}) \langle u_\alpha(-\vec{k}) u_\beta(\vec{j}) u_\gamma(\vec{k}-\vec{j})\rangle = 0 \ .
\end{align}
This is because the transfer spectrum is antisymmetric under the interchange of $\vec{k}$ and $\vec{j}$, as has been shown elsewhere, for example \cite{thesis:ajyoung,thesis:msalewski,McComb:1990-book}. We can integrate the energy balance equation to obtain
\begin{equation}
 \frac{\partial E(t)}{\partial t} = \varepsilon_W - \varepsilon \ ,
\end{equation}
since the integral over the dissipation and work spectra give the total energy dissipation rate and total energy input rate, respectively,
\begin{equation}
 \varepsilon = \int dk\ 2\nu_0 k^2 E(k,t) \qquad\qquad\textrm{and}\qquad\qquad \varepsilon_W = \int dk\ W(k,t) \ .
\end{equation}
Thus, for an unforced system ($\varepsilon_W = 0$) the change in energy is due to the dissipation rate, $\varepsilon$, and energy is lost; whereas, in the forced case the system adjusts itself until the dissipation rate matches the input rate and stationary ($\partial_t E(t) = 0$) turbulence is achieved. This does not mean that the system is not changing at all, just that the statistical properties are stationary.

\subsubsection{The energy cascade}\label{subsec:cascade}
The transfer spectrum is responsible for the redistribution of energy by modal coupling. It tries to create an equipartition of the total energy between all the modes of the system. However, the factor of $k^2$ present in the dissipation spectrum causes the high modes (small length-scales) to lose energy quicker than lower modes. Dissipation is very much a high-wavenumber effect. This is symmetry breaking \cite{McComb:2012,mccomb:2004-book} and creates a cascade of energy from low $k$ modes to high, or large length-scales to small. This is a very important mechanism and has been attributed to Richardson \cite{Richardson:1963}. It can be physically interpreted as eddies of large length-scale decaying into ever smaller eddies, creating ever smaller length-scales, but we note that the cascade is very much a spectral process and as such there is no `cascade' in real space \cite{McComb:2011-bulletin}, only an interpretation of its effects.

Energy transfer is clearly a very important process, as it generates finer and finer structures at smaller and smaller length-scales, until the viscosity dominates and the energy is lost. To measure the flux of energy flowing through a particular wavenumber, we can consider the transport power spectrum,
\begin{equation}
 \Pi(k,t) = \int_k^\infty dj\ T(j,t) = - \int_0^k dj\ T(j,t) \ .
\end{equation}
This measures the amount of energy flowing from wavenumbers less than $k$ to those greater than $k$. The transfer spectrum can be seen to be the (negative of the) derivative of the transport power spectrum. The maximum value that this spectrum attains is defined as the inertial flux, $\varepsilon_T$.

If the input of energy is occurring at very low modes and dissipation at high modes, one could ask what is happening with the intermediate modes. Since they are not receiving energy directly from the forcing, they are only excited by non-linear interaction; and if viscosity is small enough that dissipation is negligible, then they must be simply transferring all the energy that is passed to them. The high dissipation wavenumbers can only dissipate energy that has been passed down to them, which implies that, if the production and dissipation scales are well separated, the intermediate wavenumbers must transfer
\begin{equation}
 \Pi(k,t) = \varepsilon_T = \varepsilon \ .
\end{equation}
Indeed, for the turbulence to be statistically steady, the input rate must match the dissipation rate also. Thus dissipation seems to be a passive process dictated by the rate of inertial transfer. This is the inertial (sub)range of scales: a region where the effects of dissipation are not felt and there is a local equilibrium between scales. It is associated with self-similarity and scaling behaviour, as we shall see presently.

Before we do, we note that this is obviously not the case during free decay, when there is no input of energy. Since it requires a finite amount of time for energy to filter down the cascade by non-linear interaction to the dissipative scales, we expect the dissipation rate to always be greater than the transfer rate. This is because we are essentially dissipating what had been the transfer rate at a previous time, but the system has lost energy in the interim.

\section{The contribution of Kolmogorov}\label{sec:kol}
In a series of papers in 1941, Kolmogorov \cite{Kolmogorov:1991p1251,Kolmogorov:1941b,Kolmogorov:1991p138} introduced two hypotheses which produced some of the most influential results available in isotropic turbulence, commonly referred to as K41. The first of these hypotheses extends the ideas of the previous section by considering there to be a large separation between the large scale motion and dissipation. The two scales are essentially decoupled and do not directly influence one another, thus becoming statistically independent. The large scales instead simply advect the small scales through the volume. The dynamics of the dissipative scales operates on significantly shorter timescales \cite{TennekesLumley:1972} than the large scale motion and as such remain in statistical equilibrium. Any anisotropy introduced by the large scales is quickly removed.

\subsection{The universal equilibrium range}\label{subsec:kol_range}

Since the small scales are not directly influenced by the large scales, it seems natural to consider the small scale behaviour as universal in the sense that they are independent of how the turbulence is generated and/or sustained. This independence of the small dissipation scales from the large scale motion implies that their size should only depend on viscosity and the rate of energy transferred to them (or lost by them). By dimensional analysis, the only combination of these two parameters is
\begin{equation}
 \eta = \left( \frac{\nu_0^3}{\varepsilon} \right)^{1/4} \ ,
\end{equation}
which is known as the \emph{Kolmogorov microscale}. Since dimensional analysis cannot provide an absolute relationship (there could be an unknown constant), this is used to estimate the size of the dissipation scales and as such can be considered to be a lower bound on the size of small scales generated by the turbulence. The reciprocal of $\eta$ is usually defined as the dissipation wavenumber,
\begin{equation}
 \label{eq:k_d}
 k_d = \frac{1}{\eta} \ ,
\end{equation}
and any spectral method should ensure that all wavenumbers up to this are included in the analysis. In practice, it may be necessary to include even larger wavenumbers to capture all the dynamics of the system \cite{McComb:2012,McComb:2001p212}. We can also show \cite{TennekesLumley:1972,davidson:2004-book} that the Kolmogorov and integral scales are related by
\begin{equation}
 \eta \sim L\ R_L^{-3/4} \ .
\end{equation}
This highlights that, by increasing the Reynolds number whilst maintaining $L$ constant, the scale at which dissipation takes over is reduced.

In a similar manner, we may construct a characteristic velocity and timescale for the dissipative scales, with the combinations
\begin{equation}
 v = (\nu_0 \varepsilon)^{1/4} \ , \qquad\qquad \tau_d = \left( \frac{\nu_0}{\varepsilon}\right)^{1/2}\ .
\end{equation}
If we now construct a Reynolds number using these characteristic scales, we see that
\begin{equation}
 Re = \frac{\eta v}{\nu_0} = 1 \ ,
\end{equation}
and we are very much concerned with the scales for which viscous forces are important.
We can go further and deduce a form for the energy spectrum within the universal equilibrium range based on only $\varepsilon$ and $\nu_0$,
\begin{equation}
 \label{eq:Espec_UER}
 E_{U}(k) = \varepsilon^{1/4} \nu_0^{5/4} f(k\eta) \ ,
\end{equation}
where $f(k\eta)$ is a dimensionless function with the dimensionless argument $k\eta$ \cite{McComb1995:RepProgPhys}.

\subsection{Energy spectrum in the inertial subrange}\label{subsec:kol_energy}
Perhaps Kolmogorov's most famous result was a consequence of his second hypothesis \cite{Kolmogorov:1991p138}. As discussed above, as the separation between the large and dissipative scales increases, there exists an intermediate range of scales $\eta \ll r \ll L$ for which dissipation is negligible and only inertial transfer plays a role --- the inertial subrange. If these scales are taken to also be universal, then we can take the form of the universal equilibrium energy spectrum even further since it must now be independent of viscosity.

The dimensionless function $f(k\eta)$ is taken to be of power-law form,
\begin{equation}
 f(k\eta) = \alpha (k\eta)^\beta \ ,
\end{equation}
with $\alpha,\beta$ constants, so that we find (inserting the expression for $\eta$)
\begin{equation}
 E_K(k) = \alpha \nu_0^{(5+3\beta)/4} \varepsilon^{(1-\beta)/4} k^\beta \ .
\end{equation}
This can only be independent of viscosity if $\beta = -5/3$, and we have reached the Kolmogorov energy spectrum for the inertial range,
\begin{equation}
 \label{eq:K41_Ek}
 E_K(k) = \alpha \varepsilon^{2/3} k^{-5/3} \ .
\end{equation}
The constant $\alpha$ cannot be determined from this dimensional analysis and must be found by comparison to experimental data (see section \ref{subsec:kol_const}). This result has stood the test of time and is supported by a significant amount of evidence (see, for example, figure 2.4 in McComb \cite{McComb:1990-book} or figure 5.17 in Davidson \cite{davidson:2004-book}). The scales express self-similarity and scale invariance due to the power-law form of the energy spectrum.

After publishing this original form for the energy spectrum, the use of the (stationary) globally averaged energy dissipation rate rather than a locally averaged dissipation rate was questioned by Landau \cite{Landau:1944,LandauLifshitz:1959}. Since the local dissipation rate would fluctuate, intermittency would need to be accounted for. Kolmogorov later revised his hypotheses \cite{Kolmogorov:1962} with log-normal intermittency corrections to the energy spectrum in the inertial range. There has been a great deal of interest in deriving intermittent corrections to the Kolmogorov exponent, for example the beta and multi-fractal models \cite{frisch:1995-book,Boffetta:2008p1365}. More information on intermittency can be found in \cite{Vassilicos:2001} and a discussion on Kolmogorov's 1941 and 1962 theories in \cite{McComb:2011-bulletin}. It should be noted that Kraichnan \cite{Kraichnan:1967p1360} argued that ``the strongest intermittency is expected in the dissipation range where intrinsic Reynolds number are very small'' but, despite this; ``It does not follow that intermittency increases with decrease of scale size in the inertial range, in violation of Kolmogorov's hypotheses.''

Small deviations from the Kolmogorov exponent have been measured in a number of experiments and numerical investigations, suggesting the need for intermittency corrections. However, it should be stressed that the effects of finite Reynolds number cannot be ruled out as their origin \cite{McComb:2011-bulletin}. This will be discussed further in section \ref{sec:intermit_K41}.

\subsection{Structure functions}\label{subsec:kol_sf}
Kolmogorov \cite{Kolmogorov:1991p138} also studied the behaviour of the (longitudinal) structure functions, defined as
\begin{equation}
 \label{eq:sf_def}
 S_n(r) = \left\langle (\delta u_L)^n \right\rangle \ , \qquad\qquad \delta u_L = \Big( \vec{u}(\vec{x}+\vec{r}) - \vec{u}(\vec{x}) \Big) \cdot \frac{\vec{r}}{r} \ ,
\end{equation}
for stationary isotropic turbulence, where $\delta u_L$ is known as the longitudinal velocity increment. For the second- and third-order structure functions, we have
\begin{align}
 S_2(r) &= \langle u_L^2(\vec{x}+\vec{r})\rangle -2 \langle u_L(\vec{x}+\vec{r}) u_L(\vec{x}) \rangle + \langle u_L^2(\vec{x})\rangle \nonumber \\
 \label{eq:sf2_corr}
 &= 2u^2 - 2C_{LL}(r) \\
 S_3(r) &= \langle u_L^3(\vec{x}+\vec{r})\rangle -3 \langle u^2_L(\vec{x}+\vec{r}) u_L(\vec{x}) \rangle + 3 \langle u_L(\vec{x}+\vec{r}) u^2_L(\vec{x}) \rangle - \langle u_L^3(\vec{x})\rangle \nonumber \\
 \label{eq:sf3_corr}
 &= 6 C_{LL,L}(r) \ .
\end{align}
Kolmogorov showed how they satisfy a form of the K\'arm\'an-Howarth equation \cite{Karman:1938p153},
\begin{equation}
 \left( \frac{\partial}{\partial r} + \frac{4}{r} \right) \left( 6\nu_0 \frac{\partial S_2(r)}{\partial r} - S_3(r) \right) = 4\varepsilon \ ,
\end{equation}
which, upon integrating with respect to $r$, yields
\begin{equation}
 \label{eq:S3_with_diss}
 S_3(r) = - \frac{4}{5} \varepsilon r + 6\nu_0 \frac{\partial S_2(r)}{\partial r} \ .
\end{equation}
If the Reynolds number is taken to be very large such that $\nu_0 \to 0$ (or scales $r \gg \eta$ are considered), the viscous term can be neglected and we have an analytic form for the third-order structure function,
\begin{equation}
 S_3(r) = -\frac{4}{5} \varepsilon r \ ,
\end{equation}
which is referred to as the 4/5-law. By considering the skewness of the probability distribution, he also showed how the second-order moment should have the form
\begin{equation}
 S_2(r) = C \varepsilon^{2/3} r^{2/3} \ .
\end{equation}
This is entirely equivalent to the $k^{-5/3}$ energy spectrum, since the two are related by a Fourier transform. Despite Kolmogorov only presenting forms for the second- and third-order structure functions, the form has been generalised to an expression for $S_n(r)$ as
\begin{equation}
 \label{eq:sf_gen_form}
 S_n(r) = C_n (\varepsilon r)^{n/3} \ ,
\end{equation}
with $C_3 = -4/5$. It was not until much later that higher order structure functions could be measured \cite{Atta:1970p1207}. This is because they require very accurate measurement of the velocity increment and are highly sensitive to rare events (the tails of the PDF). The required degree of isotropy also necessitates a very large ensemble, and increases with $n$ \cite{Fukayama:1999p904}. Measurement of the scaling exponent of $S_n(r)$ for higher orders have found deviations from the K41 values of $n/3$ and have been associated with the need for intermittency corrections. It should be noted that the existence of the higher orders depends on the details of the tails of the PDF \cite{frisch:1995-book}. Further discussion about intermittency and the scaling exponents of the structure functions can be found in section \ref{sec:intermit_K41}.

\section{The statistical closure problem}\label{sec:intro_stat}
We now briefly outline the main problem faced when constructing a statistical theory of turbulence. Once again, the non-linear term is the cause of our frustration! Consider writing the Navier-Stokes equation given in equation \eqref{eq:NSE} in the schematic from,
\begin{equation}
 \mathscr{L}_0 u = \mathscr{M}uu + f \ ,
\end{equation}
where the operator $\mathscr{L}_0 = \partial_t + \nu_0 k^2$ and $\mathscr{M}$ represents the non-linear convolution. To illustrate our point, we take $f = 0$, but this is not necessary.

The goal of any statistical theory is to compute the correlation $\langle uu \rangle$, either in Fourier space (as is the case here) or configuration space. We therefore try to find an equation describing the evolution of $\langle uu \rangle$ by multiplying by $u$ and averaging,
\begin{equation}
 \mathscr{L}_0 \langle uu \rangle = \mathscr{M} \langle uuu \rangle \ .
\end{equation}
We see that a solution for $\langle uu \rangle$ requires the knowledge of the third-order moment, $\langle uuu \rangle$. We can try to form an equation for the order $n$ moment in the same way, obtaining
\begin{equation}
 \mathscr{L}_0 \langle u^n \rangle = \mathscr{M} \langle u^{n+1} \rangle \ ,
\end{equation}
and we see that we \emph{always} have one more unknown than we do equations. This is the \emph{closure problem} of statistical physics. Without the introduction of an approximation or physical assumption to close the set of equations at some order, we cannot proceed. Note that this highlights the non-Gaussian nature of the probability distribution, since the odd-order moments of a Gaussian-distributed random variable vanish and we would not be faced with the closure problem. Approaches to closure will be discussed in chapter \ref{chp:statistical}.

\section{Thesis overview}

This thesis is organised as follows: Chapters 2 -- 6 relate to numerical work which has been performed, while chapters 7 and 8 focus on analytic work.

Chapter 2 introduces the concept of direct numerical simulation of the Navier-Stokes equations by pseudospectral methods. Along with the basic equations and methodology, we discuss several computational challenges which need to be overcome. It is hoped that this review will aid the development of future DNS codes and provide a solid background.

A series of numerical experiments designed to validate the \dns\ code that has been produced as part of this project are presented in chapter 3. Comparison to available results from the literature allow us to conclude that the code is capable of reproducing established results and performing as expected.

Numerical investigation of turbulence undergoing free decay is considered in chapter 4. We introduce criteria for determining an evolved time based on dynamic properties of the velocity field and compare results for measurements made at these different times. We discuss the decay of evolved velocity fields generated using forced simulations, enabling the `cascade timescale' to be measured. Using a range of evolved time criteria, we show how $u^3/L$ is a better surrogate for inertial flux than the dissipation rate.

Chapter 5 presents DNS data for a number of investigations of stationary turbulence, some of which may also be compared to the literature. These include structure functions and their scaling exponents, calculated using extended self-similarity. We then show calculation of the structure functions from the energy and transfer spectra.

Chapter 6 focuses on an analytic treatment of the K\'arm\'an-Howarth equation for forced turbulence, which we then investigate using results from numerical simulation. The \emph{dissipation anomaly} is discussed and we present data for forced and decaying turbulence. Using a range of criteria for the evolved time, we show how the behaviour of the dimensionless dissipation coefficient is sensitive to measurement time. A model for the local energy balance (as expressed by the KHE) is obtained and fitted to DNS data. The model predicts $\Ceps = \Ceps(\infty) + C_L/R_L$ and this is shown to be in excellent agreement with data.

Moving to the purely analytic portion of this thesis, chapter 7 introduces the concept of the Renormalization Group and its application to turbulence. We then present a detailed analysis of a disagreement in the methodology used for an examination of the infra-red properties of stirred hydrodynamics by Forster, Nelson and Stephens, along with its resolution.

Chapter 8 provides an overview of statistical closures and renormalized perturbation theories in turbulence, before presenting work in progress in the development of a new statistical theory. The current status as well as future planned development and application is discussed.

Finally, while each chapter summarises its conclusions, we present our findings in chapter 9.

\chapter{Direct numerical simulation of isotropic turbulence}

\section{Numerical simulation}
Numerical simulation has become a common and useful tool for the study of turbulence and turbulent flow. It is used extensively for various problems, ranging from weather prediction and vehicle/building design, to the study of magnetohydrodynamics. One of their main advantages is the control they allow over the flow, making experimentally difficult or idealised flows more accessible. Moreover, specific initial conditions are \emph{exactly} reproducible (within the limits of the numerical precision used). There have been a number of positive comparisons between results obtained through numerical simulation and accepted experimental results. Results from simulations can be used to help validate models and theories, or help with interpretation (or even discovery) of new phenomena.

\subsection{Direct numerical simulation}\label{subsec:DNS}
Turbulence is a complex problem with a large range of interacting scales separated by several orders of magnitude, all of which need to be considered in order to capture the relevant physics. The Kolmogorov length scale, $\eta = \left(\nu_0^3/\varepsilon\right)^{1/4}$, is commonly quoted as the smallest length-scale which needs to be resolved to accurately reproduce the correct behaviour. Simulation of all these scales explicitly, from the large, energy-containing scales down to $\eta$, requires a large amount of computing power but does not involve any further assumptions about the behaviour of the fluid. This is direct numerical simulation (DNS) which, due to very good correlation between DNS and experimental data, has also become known as `numerical experiment'. There are, of course, many techniques to reduce this computational workload. Large-eddy simulation (LES) \cite{Lesieur2005:LES} attempts to model the effects of the smallest, so-called \textit{subgrid}, scales on the larger scales which are directly simulated. This relies on model equations for the underlying physics, and as such is open to some discrepancy. The Reynolds-averaged Navier-Stokes (RANS) \ref{subsec:Reynolds_equation} equations rely on a model for the Reynolds stresses and are used to find time-averaged statistics. Since both of these simulate fewer scales, they require considerably fewer grid points (for instance, LES results are an order of magnitude quicker to obtain). However, they cannot match the accuracy of full DNS, and it is this method which is adopted here.

Essentially, DNS attempts to numerically solve a partial differential equation, and as such we are interested in calculating derivatives. In configuration space, this can be done using an Eulerian or Lagrangian prescription. A Lagrangian mesh is attached to the material under consideration, and moves and deforms with it; whereas an Eulerian framework is attached to a spatial domain and all materials move and deform within it. Both have advantages: for example, a Lagrangian mesh can be more computationally efficient as it does not require a grid outwith the desired region and allows the history of field values at a particular point in the material to be tracked. On the other hand, they are difficult to apply to cases with extreme deformations, whereas an Eulerian mesh does not care how the material deforms \cite{Liu_Liu:2003-book}.

Grid-based methods include (Eulerian) finite difference (FDM) \cite{Hirsch:2007-book,Tannehill_Anderson_Pletcher:1997-book}, finite volume (FVM) \cite{Tannehill_Anderson_Pletcher:1997-book,Hirsch:1990-book} and (Lagrangian) finite element (FEM) \cite{Strang_Fix:1973-book,Zienkiewicz_Taylor:2000-book} methods. These use local information to estimate derivatives at grid points or volume integrals. Finite difference methods discretise the domain as a (regular) grid of points and solve the equations at these sites using finite difference operators. The finite volume method, which is more common in computational fluid dynamics,
converts divergence terms in volume integrals into surface integrals, which are then evaluated as fluxes. Since the flux leaving one volume (through a particular surface) is identical to that entering a neighbouring volume, these methods are conservative. Finite element methods use mesh discretisation of the domain into elements and a numerically stable ordinary differential equation (ODE) approximation to the PDE under study. Both FEM and FVM can be easily formulated to allow for unstructured meshes, allowing effort to be concentrated on the area surrounding a complicated object (for example, the flow over an aerofoil). However, with FEM the mesh will adapt with material deformation. This can be achieved for Eulerian meshes by mesh re-zoning or multi-meshing, but not with the same efficiency or accuracy. It is also possible to combine Eulerian and Lagrangian meshes. See \cite{Liu_Liu:2003-book} for further discussion.

There are also mesh-free methods (see \cite{Liu_Liu:2003-book} for more information), such as: Space-Time Meshfree Collocation Method \cite{Netuzhylov_Zilian:2009-mesh_free}, which use sampling from the Halton point\footnote{The Halton sequence generates quasi-random numbers that cover the space more evenly for small sample sizes than traditional pseudorandom number generators.} set to generate low-error approximations with significantly fewer evaluation points; Smoothed-Particle Hydrodynamics \cite{Liu_Liu:2003-book}, which has been used successfully for a variety of problems from as early as 1977; Diffuse Element Method and Method of Finite Spheres. These will not be discussed further here.

An alternative approach are spectral methods, which have been a standard analytical tool since the mid-nineteenth century and are based on the expansion of the solution in a set of global, orthogonal polynomials $\{\phi_n\}$. They are non-dissipative and, if constructed carefully, may offer exponential convergence to a highly accurate solution. Numerical spectral methods, originally applied to partial differential equations by meteorologists, have become an extremely powerful tool in large-scale numerical simulation.

Unlike finite difference schemes, spectral approximation use global information (the expansion functions are defined on the whole domain) about the function to estimate its derivatives. In many cases, spectral methods do not suffer from phase errors often associated with finite difference techniques \cite{Hussaini_Zang:1987-spectral_methods}. Spectral methods have proven to be viable alternatives to traditional methods for many other applications, including magnetohydrodynamics, compressible flows and boundary layers. However, they require a regular lattice, so if a complex geometry is needed then FEM or FVM may be more appropriate.

A comparison of statistical results obtained by finite difference and spectral methods was done by Rai and Moin \cite{Rai_Moin:1991-fdiff_comp}. They found that results obtained by spectral methods were closer to experiment than finite difference. However, it should be noted that it required about three-times as many grid points. A typical comparison of the error obtained when approximating a derivative by finite difference and spectral approaches is given in \cite{Hussaini_Zang:1987-spectral_methods}.

Spectral methods do have their disadvantages, however. For example, non-linearities introduce difficulties since simple multiplications become convolutions in the spectral-representation. An effective procedure to overcome this problem was found in the 1970s, pioneered by Orszag \cite{Orszag:1969-num_methods, Orszag:1969-trans_method, Orszag:1971-DNS_full} and Patterson \& Orszag \cite{Patterson_Orszag:1971-Alias_removal,Orszag_Patterson:1972-num_sim_3d}, whereby the calculation of convolutions could efficiently be avoided by shifting back and forth between configuration- and spectral-representations. This technique is called \emph{pseudospectral}, and combined with the development of fast transformation algorithms (available for both the Chebyshev and Fourier expansions below) allowed spectral methods to be reduced from $\ord{N^2}$ (for the convolution) to $\ord{N \log N}$ and become competitive methods of evaluation.

\subsection{Spectral Methods}\label{subsec:2:spectral_methods}

\subsubsection{Galerkin approximation}
Using a set of orthogonal basis functions, the exact solution $u(x,t)$ may be expanded as an infinite series
\begin{equation}
 u(x,t) = \sum\limits_{n} a_n(t) \phi_n(x)\ ,
\end{equation}
where the basis $\{\phi_n\}$ is assumed to be time- and linearly independent. The fundamental unknowns are now the expansion coefficients, $a_n(t)$, and their classical form --- or \emph{spectrum} --- can be obtained using the orthogonality of the basis functions (with respect to a certain weight) and an inner product,
\begin{equation}
 \label{eq:theory:chebyshev_spectrum}
 a_n(t) = c_n \int dx\ u(x,t) \phi_n(x) w(x)\ .
\end{equation}

The Galerkin approximation is constructed through the truncated series \cite{Gottlieb_Orszag:1977}
\begin{equation}
 u_N(x,t) = \sum\limits_{n=0}^{N} a_n(t) \phi_n(x)\ .
\end{equation}
The choice of basis functions should reflect the properties of the domain of interest. For a bound Cartesian spatial domain normalised to $[-1,1]$ which is non-periodic, an appropriate class of expansion functions are the Jacobi polynomials, such as the Legendre or Chebyshev polynomials (of the first kind), as used by the \emph{channelflow} code (section \ref{subsec:currently_avail}).

For a bound, periodic domain $D = \left\{x : x \in [0,L]\right\}$, the exact solution, $u(x,t)$, may be expressed by a complete Fourier series as
\begin{equation}
 u(x,t) = \sum_{n=-\infty}^{\infty} a_n(t) e^{i\frac{2\pi n}{L}x}\ ,
\end{equation}
which employs the periodic nature of the complex exponential on the same interval. The integer $n$ is known as the wavenumber. This may be approximated by the truncated series
\begin{equation}
 u_N(x,t) = \sum_{n=-\frac{N}{2}}^{\frac{N}{2}-1} a_n(t) e^{i\frac{2\pi n}{L} x} = \sum_{k} a_k(t) e^{ikx}\ ,
\end{equation}
where $N$ is taken to be even (usually a power of $2$) and we introduce the \emph{conjugate momentum}, $k$
\begin{equation}
k = \frac{2\pi n}{L} \ ,\qquad n \in \left\{-\frac{N}{2}, \ldots, \frac{N}{2}-1\right\}\ .
\end{equation}
Due to periodicity, the choice of including $\pm N/2$ is arbitrary; we here choose to include $n = -N/2$ (although this mode is set to zero anyway). The spectrum of the above expansion is then
\begin{equation}
 \label{eq:theory:inverse_transform}
 \tilde{u}(k,t) \equiv a_k(t) = \frac{1}{L} \int_0^{L} dx\ u(x,t) e^{-ikx}\ ,
\end{equation}
although throughout this work we drop the tilde and instead write $u(k,t)$ for the Fourier coefficients, as it is clear from the arguments whether we are dealing with the configuration space solution or its Fourier counterpart. It is common to take $L = 2\pi$ so that the conjugate momenta are equivalent to their corresponding integer wavenumber.

\subsection{Collocation methods}\label{sec:theory:collocation}
Collocation methods are an approximation based in configuration space in which the spatial continuum is represented discretely by the values at $N$ special collocation points, $x_j \in D$. The optimal choice for these points are often the extrema of $\phi_N(x)$, reducing the effects of Gibbs' phenomenon and resulting in an extremely accurate approximation. For the Chebyshev expansion above, the collocation points in $[-1,1]$ are the extrema of $T_N(\textrm{cos\ }\theta) = \textrm{cos}(N\theta)$, and are thus
\begin{equation}
 x_j = \textrm{cos}\left(\frac{\pi j}{N}\right)\ , \qquad j \in \{0,\ldots,N\}\ .
\end{equation}
Chebyshev collocation techniques have been used in boundary layer and channel flow problems with non-periodic flows.

For the Fourier series, the collocation points in $[0,L]$ are the extrema of $\re{e^{i\pi Nx/L}}$, which are
\begin{equation}
 x_j = \frac{L j}{N} = aj\ , \qquad j \in \{0,\ldots,N-1\}\ ,
\end{equation}
where $a = L/N$ is the lattice spacing.

At the collocation points, we ensure that our truncated approximation is equal to the full solution, $u_N(x_j,t) = u(x_j,t)$, so that
\begin{equation}
 u(x_j,t) = \sum_{k} u(k,t) e^{ikx_j}\ ,
\end{equation}
and the Fourier coefficients are found by
\begin{equation}
 \label{eq:theory:inverse_transform_discrete}
 u(k,t) = \frac{1}{N} \sum_{j=0}^{N-1} u(x_j,t) e^{-ikx_j} = \frac{1}{L} a\sum_{x} u(x,t) e^{-ikx}\ .
\end{equation}

\subsection{Pseudospectral methods}\label{sec:theory:pseudo}
Pseudospectral methods (section \ref{subsec:DNS}) use a mixture of evaluation in configuration- and spectral-space, with fast transformation algorithms to move between the two. Whilst in configuration-space, collocation methods are used, as in we deal with the values of the function at the collocation points; whereas spectral methods provide better approximations of derivatives since they are based on global basis functions. Most Fourier-based pseudospectral methods are algebraically equivalent to collocation methods \cite{chqz:2006-book}.

\section{The basics of writing a pseudospectral DNS}
While the \emph{techniques} associated with creating a pseudospectral code for solving partial differential equations (such as the incompressible Navier-Stokes equation) are well documented in the literature, a comprehensive description of what is actually involved is hard to come by. Here, we attempt to present a detailed description specifically for the Navier-Stokes equation, along with discussion of the various choices made.

\subsection{Tackling the non-linear term}
We start our description with a discussion of the equations we intend to solve. The Navier-Stokes equations in configuration space, which govern the behaviour of an incompressible Newtonian fluid, are written
\begin{equation}
 \partial_t \vec{u} - \nu_0 \nabla^2 \vec{u} + \vec{\nabla}p = -(\vec{u}\cdot\vec{\nabla})\vec{u}\ , \qquad\qquad \vec{\nabla}\cdot\vec{u} = 0 \ ,
\end{equation}
where the density has been taken to be unity. There are several different ways to evaluate the non-linear term on the RHS of the equation. These are:
\begin{align}
 \textrm{\textit{the convection form}}\qquad &(\vec{u}\cdot\vec{\nabla})\vec{u}\ ; \\
 \textrm{\textit{the divergence form}}\qquad &\vec{\nabla}\cdot(\vec{u}\otimes\vec{u}) \ ; \\
 \textrm{\textit{the skew-symmetric form}}\qquad &\tfrac{1}{2}\left[(\vec{u}\cdot\vec{\nabla})\vec{u} + \vec{\nabla}\cdot(\vec{u}\otimes\vec{u}) \right] \ ; \quad\text{and}\\
 \textrm{\textit{the rotational form}}\qquad &(\vec{\nabla} \times \vec{u})\times\vec{u} + \tfrac{1}{2}\vec{\nabla}u^2 \ .
\end{align}
The symbol $\otimes$ represents an outer- or tensor-product. These expressions are identical, provided we are considering an incompressible fluid such that $\vec{\nabla}\cdot\vec{u} = 0$. In practice, the rotational form is the cheapest to compute via pseudospectral methods. However, we must be aware of aliasing errors introduced to high wavenumbers due to the discrete transform. The skew-symmetric form does not suffer from such errors, but is significantly more computationally expensive. Since it is the average of the convection and divergence forms, one can alternate between them with a result that is as stable as the skew-symmetric form and similar to the rotational form in cost. Zang \cite{Zang:1991-forms} recommends either the skew-symmetric (or alternating) form without dealiasing or the rotational form with dealiasing.

We focus on the rotational form of the non-linear term, and so the equation of interest is
\begin{align}
 \left(\partial_t - \nu_0 \nabla^2\right) \vec{u}(\vec{x},t) &= \vec{u}(\vec{x},t)\times(\vec{\nabla} \times \vec{u}(\vec{x},t)) - \vec{\nabla}\left[p(\vec{x},t) + \tfrac{1}{2}u^2(\vec{x},t)\right] \nonumber \\
 &= \vec{u}(\vec{x},t)\times\vec{\omega}(\vec{x},t) - \vec{\nabla}\left[p(\vec{x},t) + \tfrac{1}{2}u^2(\vec{x},t)\right] \\
 &= \vec{W}(\vec{x},t) - \vec{\nabla}\left[p(\vec{x},t) + \tfrac{1}{2}u^2(\vec{x},t)\right] \ ,
\end{align}
where the vorticity has been defined $\vec{\omega}(\vec{x},t) = \vec{\nabla} \times \vec{u}(\vec{x},t)$ and non-linear term
\begin{equation}
 \vec{W}(\vec{x},t) = \vec{u}(\vec{x},t)\times\vec{\omega}(\vec{x},t)
\end{equation}
will be used.

Transforming to Fourier space, we have
\begin{equation}
 \left(\partial_t + \nu_0 k^2\right) \vec{u}(\vec{k},t) = \vec{W}(\vec{k},t) - i\vec{k} \mathscr{F}\left[p(\vec{x},t) + \tfrac{1}{2}u^2(\vec{x},t)\right] \ ,
\end{equation}
where we use $\mathscr{F}[\cdots]$ to represent the transform of those terms which we do not wish to deal with explicitly, for reasons which we discuss presently. The continuity condition has become $\vec{k}\cdot\vec{u}(\vec{k},t) = 0$ and so by taking a scalar product of the equation above with the wavevector $\vec{k}$ it reduces to
\begin{equation}
 0 = \vec{k}\cdot \vec{W}(k,t) - ik^2 \mathscr{F}\left[p(\vec{x},t) + \tfrac{1}{2}u^2(\vec{x},t)\right] \ ,
\end{equation}
which is rearranged to give us an expression for the terms we do not wish to deal with explicitly
\begin{equation}
 i\mathscr{F}\left[p(\vec{x},t) + \tfrac{1}{2}u^2(\vec{x},t)\right] = \frac{1}{k^2}\vec{k}\cdot\vec{W}(\vec{k},t) \ .
\end{equation}
Inserting this into the Navier-Stokes equation in Fourier space above, we have
\begin{equation}
 \left(\partial_t + \nu_0 k^2\right) \vec{u}(\vec{k},t) = \vec{W}(\vec{k},t) - \vec{k} \frac{1}{k^2}\vec{k}\cdot\vec{W}(\vec{k},t) \ ,
\end{equation}
or, written in component notation,
\begin{align}
 \left(\partial_t + \nu_0 k^2\right) u_\alpha(\vec{k},t) &= W_\alpha(\vec{k},t) -  \frac{k_\alpha k_\beta}{k^2} W_\beta(\vec{k},t) \nonumber \\
 &= \left[\delta_{\alpha\beta} - \frac{k_\alpha k_\beta}{k^2}\right] W_\beta(\vec{k},t) \nonumber \\
 \label{eq:non-linear}
 &= P_{\alpha\beta}(\vec{k}) W_\beta(\vec{k},t) \ ,
\end{align}
which defines the projection operator $P_{\alpha\beta}(\vec{k})$ in Fourier space.

In summary, the continuity condition has allowed us to eliminate the pressure field in terms of the non-linearity, and has led to the introduction of the projection operator. This ensures that the velocity field remains solenoidal (or divergenceless). Since the Navier-Stokes equations in Fourier space are more commonly seen in an alternative form, we briefly outline their equivalence for completeness below.

In configuration space, the non-linear term is defined using the completely anti-symmetric Levi-Civita symbol as
\begin{align}
 W_\alpha(\vec{x},t) &= \epsilon_{\alpha\beta\gamma} u_\beta(\vec{x},t) \omega_\gamma(\vec{x},t) \nonumber \\
 &= \epsilon_{\alpha\beta\gamma} \int d^3j\ e^{i\vec{j}\cdot\vec{x}} u_\beta(\vec{j},t) \int d^3q\ e^{i\vec{q}\cdot\vec{x}} \omega_\gamma(\vec{q},t) \ .
\end{align}
Taking the inverse Fourier transform back to $k$-space, we find
\begin{align}
 W_\alpha(\vec{k},t)
 &= \epsilon_{\alpha\beta\gamma} \int d^3j\ \int d^3q\ u_\beta(\vec{j},t)\ \omega_\gamma(\vec{q},t)\ \left( \left(\frac{1}{2\pi}\right)^3 \int d^3x\ e^{i(\vec{j}+\vec{q}-\vec{k})\cdot\vec{x}}\right) \nonumber \\
 &= \epsilon_{\alpha\beta\gamma} \int d^3j\ \int d^3q\ u_\beta(\vec{j},t)\ \omega_\gamma(\vec{q},t)\ \delta(\vec{j}+\vec{q}-\vec{k}) \ ,
\end{align}
or, in the discrete form,
\begin{equation}
 W_\alpha(\vec{k},t) = \epsilon_{\alpha\beta\gamma} \sum_{\vec{j}} u_\beta(\vec{j},t)\ \omega_\gamma(\vec{k}-\vec{j},t) \ .
\end{equation}
Using the definition of the vorticity in Fourier space, $\omega_\gamma(\vec{\kappa},t) = \epsilon_{\gamma\mu\nu} i\kappa_\mu u_\nu(\vec{\kappa},t)$ and the identities
\begin{equation}
 \epsilon_{\alpha\beta\gamma} \epsilon_{\gamma\mu\nu} = \delta_{\alpha\mu} \delta_{\beta\nu} - \delta_{\alpha\nu} \delta_{\beta\mu} \ , \qquad\qquad j_\beta u_\beta(\vec{j},t) = 0 \ ,
\end{equation}
we rewrite the non-linear term in Fourier space as
\begin{align}
 W_\alpha(\vec{k},t) &= \epsilon_{\alpha\beta\gamma} \epsilon_{\gamma\mu\nu} \sum_{\vec{j}} u_\beta(\vec{j},t)\ i(k_\mu - j_\mu)\ u_\nu(\vec{k}-\vec{j},t) \nonumber \\
 &= \left( \delta_{\alpha\mu} \delta_{\beta\nu} - \delta_{\alpha\nu} \delta_{\beta\mu} \right) \sum_{\vec{j}} u_\beta(\vec{j},t)\ i(k_\mu - j_\mu)\ u_\nu(\vec{k}-\vec{j},t) \\
 &= i\sum_{\vec{j}} \Big[ (k_\alpha - j_\alpha)\ u_\beta(\vec{j},t)\ u_\beta(\vec{k}-\vec{j},t) - (k_\beta - \cancel{j_\beta})\ u_\beta(\vec{j},t)\ u_\alpha(\vec{k}-\vec{j},t) \Big] \nonumber \ .
\end{align}
Next, we act on the non-linear term with the projection operator and, using the fact that $P_{\alpha\beta}(\vec{k})k_\alpha = P_{\alpha\beta}(\vec{k})k_\beta = 0$, write
\begin{align}
 P_{\alpha\beta}(\vec{k}) W_\beta(\vec{k},t) &= \frac{1}{2i} P_{\alpha\beta}(\vec{k}) \sum_{\vec{j}} \Big[ 2j_\beta\ u_\gamma(\vec{j},t)\ u_\gamma(\vec{k}-\vec{j},t) + 2k_\gamma\ u_\gamma(\vec{j},t)\ u_\beta(\vec{k}-\vec{j},t) \Big] \ .
\end{align}
We then perform a change of variables of $\vec{j} \to \vec{k}-\vec{j}$ on one of the two copies of each term on the RHS, since this doesn't affect the overall sum:
\begin{align}
 P_{\alpha\beta}(\vec{k}) W_\beta(\vec{k},t) &= \frac{1}{2i} P_{\alpha\beta}(\vec{k}) \sum_{\vec{j}} \Big[ j_\beta\ u_\gamma(\vec{j},t)\ u_\gamma(\vec{k}-\vec{j},t) + k_\gamma\ u_\gamma(\vec{j},t)\ u_\beta(\vec{k}-\vec{j},t) \nonumber \\
 &\qquad + (\cancel{k_\beta} - j_\beta)\ u_\gamma(\vec{k}-\vec{j},t)\ u_\gamma(\vec{j},t) + k_\gamma\ u_\gamma(\vec{k}-\vec{j},t)\ u_\beta(\vec{j},t) \Big] \nonumber \\
 &= \frac{1}{2i} k_\gamma\ P_{\alpha\beta}(\vec{k}) \sum_{\vec{j}} \Big[ u_\gamma(\vec{j},t)\ u_\beta(\vec{k}-\vec{j},t) + u_\gamma(\vec{k}-\vec{j},t)\ u_\beta(\vec{j},t) \Big] \nonumber \\
 &= \frac{1}{2i} \Big[ k_\beta P_{\alpha\gamma}(\vec{k}) + k_\gamma\ P_{\alpha\beta}(\vec{k}) \Big] \sum_{\vec{j}} u_\beta(\vec{j},t) u_\gamma(\vec{k}-\vec{j},t) \nonumber \\
 \label{eq:non-lin-term}
 &= M_{\alpha\beta\gamma}(\vec{k}) \sum_{\vec{j}} u_\beta(\vec{j},t) u_\gamma(\vec{k}-\vec{j},t) \ ,
\end{align}
which defines the vertex operator $M_{\alpha\beta\gamma}(\vec{k})$. This is the form of the non-linear term in the Navier-Stokes equations in Fourier space most commonly encountered and, as we have shown, it is equivalent to the form in equation (\ref{eq:non-linear}). This will be important when we come to look at the transfer spectrum later on.

\subsection{Introduction to \dns}
A pseudospectral DNS code has been developed in the course of this project, evaluating the rotational form of the non-linear term discussed above. This code, referred to as \dns, has been benchmarked using results from a previous code and another, freely-available code (section \ref{sec:benchmark}), as well as results in the literature, to ensure that it performs as expected. This section will attempt to clarify as clearly as possible what the code actually does and how, before we move on to some more general calculation of statistics and parameters. This is followed by discussion of several important topics which need to be considered, along with justification for the choices made in \dns.

The goal is to exploit the use of Fast Fourier Transform (FFT) algorithms to switch between configuration and Fourier space to efficiently perform certain operations. The general structure of the algorithm is thus outlined as:
\begin{enumerate}
 \item Calculate $\vec{u}(\vec{k},t)$ and $\vec{\omega}(\vec{k},t) = i\vec{k}\times \vec{u}(\vec{k},t)$ in Fourier space.
 \item Transform to configuration space using a FFT: $\vec{u}(\vec{x},t)$ and $\vec{\omega}(\vec{x},t)$.
 \item Calculate the non-linear interaction: $\vec{W}(\vec{x},t) = \vec{u}(\vec{x},t) \times \vec{\omega}(\vec{x},t)$.
 \item Transform back to Fourier space to find $\vec{W}(\vec{k},t)$, then act with the projection operator on it.
 \item Integrate the equation of motion forwards in time using some time-stepping procedure (see section \ref{subsec:time_advance}) to find $\vec{u}(\vec{k},t+\delta t)$.
\end{enumerate}

In configuration space, we create a cubic lattice of size $N\times N\times N$ which stores, at each point, three real numbers corresponding to the velocity field $\vec{u}(\vec{x},t)$. (We will restrict our attention to cubic domains, but this is not necessary --- one can perform a simulation on a $N_x \times N_y \times N_z$ lattice with trivial extensions to the discussion here.) Obviously, there are numerous ways to store this data in memory, depending on which index one requires to be contiguous in memory (occupy adjacent memory addresses). For the FFT library we will focus on (called FFTW), it will help to have each component scalar field contiguous, so that the data is stored, for example, as {\ttfamily field(i,x,y,z)}, where {\ttfamily i} $= 1,2,3$ varies most slowly and labels the vector index. Many codes also calculate and store the pressure field, and this can be included by allowing a fourth value for {\ttfamily i}.

When we take the Fourier transform of the configuration-space field, we are required to store $N^3$ complex numbers. However, since the configuration-space data is purely real, we have Hermitian symmetry, $\vec{u}(-\vec{k},t) = \vec{u}^*(\vec{k},t)$, with $*$ signifying standard complex conjugation. This is an important symmetry, as it allows us to recreate half our volume from the other half! So instead of $N^3$ complex numbers we need only store $N \times N \times (N/2 + 1)$. Thus, if we create the field to be $N \times N \times (N+2)$ real numbers, we can perform \emph{in place} transformations and avoid having to store both the real and spectral fields simultaneously. Note that this does require \emph{padding} to be added to the real-space array, and element {\ttfamily (x,y,z)} should be accessed as {\ttfamily [(N+2)*(x*N+y)+z]} rather than {\ttfamily [N*(x*N+y)+z]} for the unpadded array. With this data layout, the FFTW library can compute the real-to-complex FFT in place and it will store wavenumbers $n_{x,y} \in \{ -N/2, \cdots, N/2 - 1 \}$ (a total of $N$ elements) and $n_z \in \{0, \cdots, N/2\}$ (a total of $N/2 + 1$ elements); and the complex-to-real FFT vice versa.

\begin{figure}[tb!]
 \centering
 \includegraphics[width=0.9\textwidth]{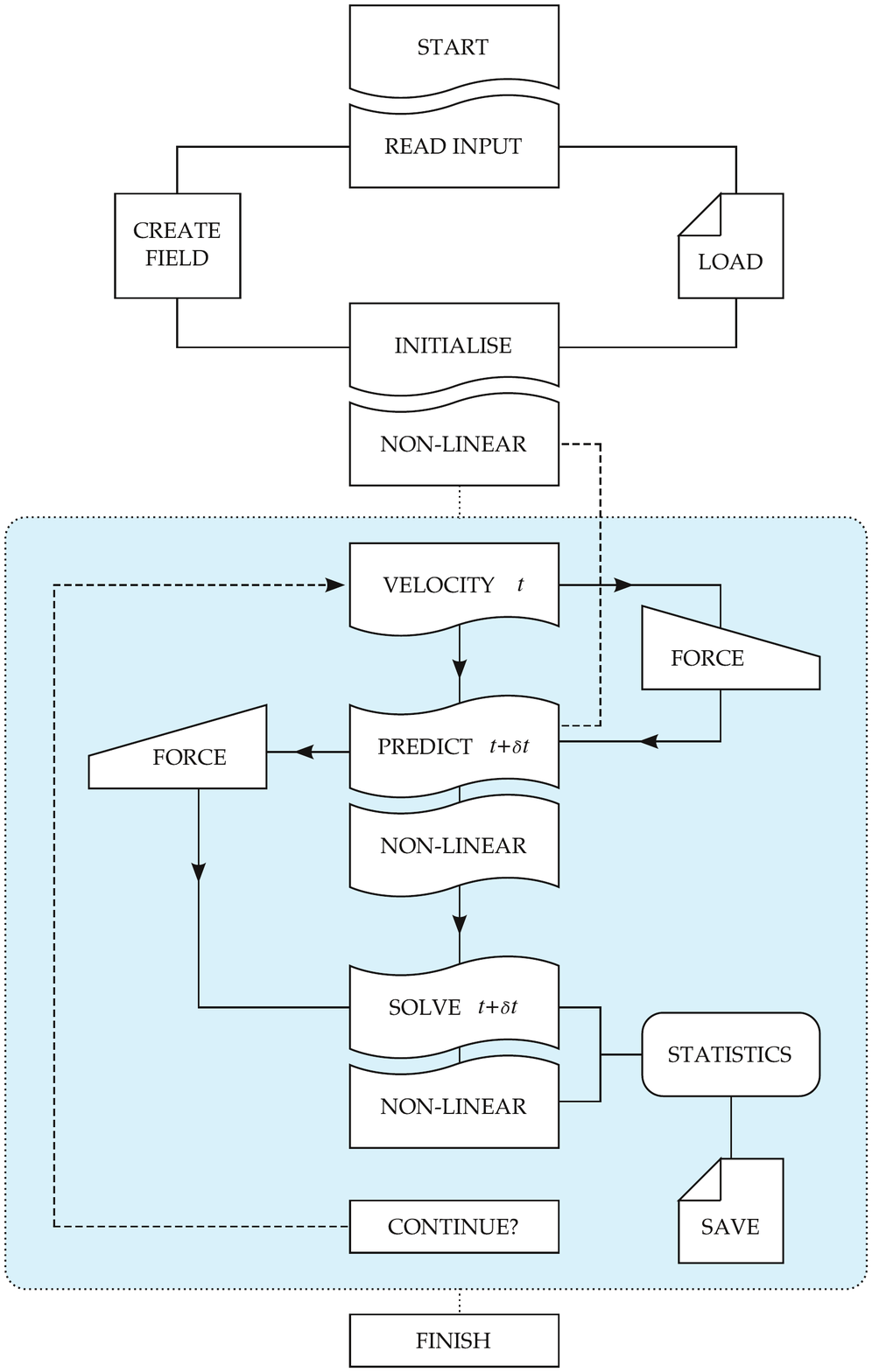}
 \caption{Schematic overview of the \dns\ code structure.}
 \label{fig:chp2:code}
\end{figure}

The order of the data in Fourier space calculated by FFTW is perhaps not what one would expect, as it is not the order which the wavenumbers are listed above. Instead, in each of the directions $x,y$ the wavenumbers are stored as:
\begin{equation}
 0, 1, \cdots, N/2-1, \pm N/2, -N/2 + 1, \cdots, -1 \ .
\end{equation}
We have written $\pm N/2$ as due to the periodicity they are the same value, and in our simulations this mode is set to zero. This may be surprising as the mode $n = 0$ is not in the centre of the spectral array, and as such needs to be considered when coding the data structure. The (complex) element {\ttfamily (i,j,k)} can be accessed using something like:
\begin{verbatim}
  if (i < 0) i+=N;  if (j < 0) j+=N;  return [(N/2+1)*(i*N+j)+k];
\end{verbatim}

\vspace{1em}
\noindent A schematic overview of the code structure can be seen in figure \ref{fig:chp2:code}, with the various components discussed in the sections that follow.

\clearpage

\subsection{Forcing}\label{subsec:forcing}
We now extend our analysis to include the study of statistically-steady turbulence. Since dissipation removes energy from the system, an unforced turbulent system will simply decay in time. Turbulence is non-sustaining, and the effect of the non-linear transfer is to move energy to smaller and smaller length scales, where dissipation is more efficient. Hence if we wish to maintain the system in a turbulent state we must input energy --- we must force the system.

The Navier-Stokes equation gains an extra term corresponding to this energy input, $\vec{f}(\vec{x},t)$, and we require that this term is solenoidal (divergenceless) to ensure the continuity condition is not violated by the forcing. In Fourier space, the Navier-Stokes equation becomes
\begin{equation}
 \label{eq:forced_nse}
 \left(\partial_t + \nu_0 k^2\right) u_\alpha(\vec{k},t) = P_{\alpha\beta}(\vec{k}) W_\beta(\vec{k},t) + f_\alpha(\vec{k},t) \ .
\end{equation}

The method of energy input depends on what one is interested in studying. For instance, a random force drawn from a Gaussian distribution could be used. This can be done to all length-scales, or only certain regions. This does, however, introduce another source of randomness into the system, and for studying the nature of turbulence itself it may not be appropriate. How could one distinguish between the random fluctuations due to turbulence and an artefact of the random forcing? A discussion of methods of random forcing is given in Alvelius \cite{Alvelius:1999p1379}, while Overholt and Pope \cite{Overholt:1998p557} present a deterministic scheme.

Another deterministic method, proposed by Machiels \cite{Machiels:1997-predictability}, is implemented here. It is technically a form of negative damping, and is applied to a band of wavenumbers such that
\begin{equation}
 f_\alpha(\vec{k},t) = \left\{ \begin{array}{ll}
                                 (\epsw/2E_f) u_\alpha(\vec{k},t) & \textrm{if } 0 < k < k_f \\
				 & \\
				 0 & \textrm{otherwise}
                                \end{array}
 \right. \ ,
\end{equation}
where $\epsw$ is the desired energy input rate, $k_f$ is the maximum forced wavenumber and $E_f$ is the total amount of energy contained in the forced band. The total amount of energy that is provided to the system (per unit time) is then found by integrating the work spectrum (see section \ref{subsec:lin_equation})
\begin{align}
 \int dk\ W(k,t) &= \int dk\ 4\pi k^2 \left\langle u_\alpha(-\vec{k},t) f_\alpha(\vec{k},t) \right\rangle \nonumber \\
 &= \frac{\epsw}{E_f} \int_0^{k_f} dk\ 4\pi k^2 \tfrac{1}{2} \left\langle u_\alpha(-\vec{k},t) u_\alpha(\vec{k},t) \right\rangle \nonumber \\
 &= \frac{\epsw}{E_f} \int_0^{k_f} dk\ E(k,t) \ .
\end{align}
This means that the system is receiving a constant rate of energy input, unlike the random forcing method, where $\epsw = \epsw(t)$ will also fluctuate in time. Other forms of deterministic forcing may be enforced, see the literature discussion in section \ref{subsec:DA_lit}.

\subsection{Time advancement}\label{subsec:time_advance}
Numerical integration of a differential equation is not a new topic, and many techniques have been developed over the years for doing just this. They range from the basic, single-step Euler methods to higher-order multi-step methods which can be much more stable. The classical integrator of choice is the fourth-order Runge-Kutta (RK4) method, although due to memory restrictions its application to this problem is complicated. (There are low-storage higher-order methods available -- see appendix D of \cite{chqz:2006-book} for more information.)

We first solve for the viscous term implicitly through the use of an integrating factor:
\begin{align}
 e^{\nu_0 k^2 t} \left(\partial_t + \nu_0 k^2\right) u_\alpha(\vec{k},t) &= e^{\nu_0 k^2 t} \left[ P_{\alpha\beta}(\vec{k}) W_\beta(\vec{k},t) + f_\alpha(\vec{k},t) \right] \\
 \partial_t \left[e^{\nu_0 k^2 t} u_\alpha(\vec{k},t) \right] &= e^{\nu_0 k^2 t} \left[ P_{\alpha\beta}(\vec{k}) W_\beta(\vec{k},t) + f_\alpha(\vec{k},t) \right] \\
 e^{\nu_0 k^2 t} u_\alpha(\vec{k},t) &= \int_{-\infty}^t ds\ e^{\nu_0 k^2 s} \left[ P_{\alpha\beta}(\vec{k}) W_\beta(\vec{k},s) + f_\alpha(\vec{k},s) \right] \ .
\end{align}
So far, this has been exact. We now introduce the time-step $\delta t$ such that our solution becomes
\begin{align}
 e^{\nu_0 k^2 (t+\delta t)} u_\alpha(\vec{k},t+\delta t) &= \int_{-\infty}^{t+\delta t} ds\ e^{\nu_0 k^2 s} \left[ P_{\alpha\beta}(\vec{k}) W_\beta(\vec{k},s) + f_\alpha(\vec{k},s) \right] \nonumber \\
 &= e^{\nu_0 k^2 t} u_\alpha(\vec{k},t) + \int_t^{t+\delta t} ds\ e^{\nu_0 k^2 s} \left[ P_{\alpha\beta}(\vec{k}) W_\beta(\vec{k},s) + f_\alpha(\vec{k},s) \right] \ ,
\end{align}
 but we are still left to discretise the integral on the RHS. This is done using Heun's method --- a second-order predictor-corrector algorithm. We take one step using the forward Euler method to the predictor, then refine our solution using a second evaluation and the predicted value.

The approximations to the integral
\begin{align}
I_\alpha &= \int_t^{t+\delta t} ds\ e^{\nu_0 k^2 s} \left[ P_{\alpha\beta}(\vec{k}) W_\beta(\vec{k},s) + f_\alpha(\vec{k},s) \right] \nonumber
\end{align}
are found at the lower- and upper-boundaries,
\begin{align}
 I_\alpha(t) &= \delta t\ e^{\nu_0 k^2 t} \left[ P_{\alpha\beta}(\vec{k}) W_\beta(\vec{k},t) + f_\alpha(\vec{k},t) \right]
\end{align}
and
\begin{align}
 I_\alpha(t+\delta t) &= \delta t\ e^{\nu_0 k^2 (t+\delta t)} \left[ P_{\alpha\beta}(\vec{k}) W_\beta(\vec{k},t+\delta t) + f_\alpha(\vec{k},t+\delta t) \right]\ ,
\end{align}
respectively. The latter requires the non-linear and forcing terms at $t+ \delta t$, so we use a predicted solution based on the lower limit approximation
\begin{equation}
 e^{\nu_0 k^2 (t+\delta t)} u^P_\alpha(\vec{k},t+\delta t) = e^{\nu_0 k^2 t} u_\alpha(\vec{k},t) + I_\alpha(t) \ ,
\end{equation}
such that
\begin{equation}
 \label{eq:predicted_velocity}
 u^P_\alpha(\vec{k},t+\delta t) = e^{-\nu_0 k^2 \delta t} \left[ u_\alpha(\vec{k},t) + \delta t\ \Big( P_{\alpha\beta}(\vec{k}) W_\beta(\vec{k},t) + f_\alpha(\vec{k},t) \Big) \right] \ ,
\end{equation}
to approximate them, and the upper limit is approximated as
\begin{align}
 I^P_\alpha(t+\delta t) = \delta t\ e^{\nu_0 k^2 (t+\delta t)} \left[ P_{\alpha\beta}(\vec{k}) W^P_\beta(\vec{k},t+\delta t) + f^P_\alpha(\vec{k},t+\delta t) \right]\ .
\end{align}
The final solution is taken to be the average of the upper- and lower- contributions,
\begin{equation}
 I_\alpha = \tfrac{1}{2} \left[I_\alpha(t) + I^P_\alpha(t+\delta t) \right] \ ,
\end{equation}
and the velocity field takes the value
\begin{align}
 \label{eq:corrected_velocity}
 u_\alpha(\vec{k},t+\delta t) &= e^{-\nu_0 k^2 \delta t} u_\alpha(\vec{k},t) \\
 &\quad+ \tfrac{1}{2}\delta t \Bigg[ e^{-\nu_0 k^2 \delta t} \Big( P_{\alpha\beta}(\vec{k}) W_\beta(\vec{k},t) + f_\alpha(\vec{k},t) \Big) \nonumber\\
 &\qquad+ \Big( P_{\alpha\beta}(\vec{k}) W^P_\beta(\vec{k},t+\delta t) + f^P_\alpha(\vec{k},t+\delta t) \Big) \Bigg] \nonumber \ .
\end{align}

In summary, we use the velocity field $\vec{u}(\vec{k},t)$ to evaluate the non-linear (and possibly forcing) term and use them to find the predicted solution $\vec{u}^P(\vec{k},t+\delta t)$ given by equation (\ref{eq:predicted_velocity}). Using this predicted solution, we re-evaluate the non-linear (and forcing) term and construct the corrected solution as given by equation (\ref{eq:corrected_velocity}). Thus each time-step requires two evaluations of the non-linear term. In terms of storage, we can write
\begin{equation}
 \vec{u}(\vec{k},t+\delta t) = \tfrac{1}{2} \left[ \vec{u}^P(\vec{k},t+\delta t) + \vec{u}^C(\vec{k},t+\delta t) \right] \ ,
\end{equation}
where
\begin{equation}
 u^C_\alpha(\vec{k},t+\delta t) = e^{-\nu_0 k^2 \delta t} u_\alpha(\vec{k},t) + \delta t \Big( P_{\alpha\beta}(\vec{k}) W^P_\beta(\vec{k},t+\delta t) + f^P_\alpha(\vec{k},t+\delta t) \Big) \ ,
\end{equation}
and we see that we need to keep two copies of the velocity field and one copy of each of the non-linear and forcing terms for the duration of the time-step.

In practice, it is convenient to pre-calculate the non-linear term at the \emph{end} of the time-step, as the process of doing so allows access to the velocity field in real space at the new time (for example, to write out the velocity field, vorticity field; calculate structure functions; etc.) and assists with the calculation of the transfer spectrum (see section \ref{subsubsec:transfer}).

\subsection{Initial field generation}\label{subsec:initial_field}
Initial conditions for the velocity field need to be carefully considered if one wishes to study a turbulent system. Since the initial condition will not itself be a solution of the Navier-Stokes equation, we need to pick something that is similar in form to a developed flow, otherwise we will spend a huge computational effort simply reaching a solution. To do this, we generate our initial field to be a random Gaussian field, but distributed according to a certain energy spectrum, $E(k,0)$, which we choose.

This can be done in configuration space or Fourier space, with the latter accomplished using a method proposed by Orszag \cite{Orszag:1969-num_methods}. In this case, depending on the seed given to the random number generator used to populate the field, the actual energy spectrum obtained will vary slightly from realisation to realisation. The down side of this method is exposed when the computation is spread over multiple processes which do not share the same memory. This is because generating the field in this manner requires non-local knowledge of the field --- data which resides with another process. This can be worked around using a complicated series of exchanges between nodes.

\dns\ instead generates its initial field in configuration space, filling all sites with a Gaussian random number (of mean 0, variance 1). This is then Fourier transformed to $k$-space where truncation and the projection operator are applied to ensure that the field is solenoidal. From this, the current energy spectrum, $E_c(k)$, is evaluated (see section \ref{subsubsec:energy_spectrum}). Each mode is then rescaled using
\begin{equation}
 u_\alpha(\vec{k},0) = u_\alpha(\vec{k},0) \cdot \sqrt{\frac{E(k,0)}{E_c(k)}} \ .
\end{equation}
This procedure exactly reproduces the desired initial energy spectrum, with no variation (although the actual values of the field are unique to each realisation). The \emph{hit3d} code (see section \ref{subsec:currently_avail}) also uses this method of field generation.

The initial spectra which have been considered here can be characterised by two forms:
\begin{itemize}
 \item Standard:
  \begin{equation}
   \label{eq:chp3:standard_spec}
   E(k,0) = C_1 k^{C_2} \exp (-C_3 k^{C_4}) \ ;
  \end{equation}
 \item von K\'arm\'an:
 \begin{equation}
  \label{eq:chp3:vKar_spec}
  E(k,0) = C_1 \left(k/C_2 \right)^{C_3} \left[1 + (k/C_2)^2 \right]^{-(3C_3 + 5)/6} \exp\left(-(k/C_4)^2\right) \ ,
 \end{equation}
\end{itemize}
with the constants $C_i$ chosen to set the distribution of energy among the modes. Particular values currently of interest are given in table \ref{tbl:init_spectra}. The von K\'arm\'an spectrum exhibits the Kolmogorov $k^{-5/3}$ scaling at large $k$ (scale set by $C_2$), damped by an exponential at larger $k$ (scale set by $C_4$) and goes as $k^{C_3}$ at low $k$. The Kolmogorov spectrum is just a special case of the standard spectrum. The difference between initial conditions which behave as $k^2$ and $k^4$ at low wavenumber is thought to be of interest \cite{Goto:2009p144}. This is because a low $k$ expansion of the energy spectrum goes as $E(k) = Ak^2 + Bk^4 + \ord{k^6}$, where we cannot have both $A \neq 0$ and $B \neq 0$ \cite{McComb:2012}.

\begin{table}
\begin{center}
\begin{tabular}{llllll}
ID & Spectrum & $C_1$ & $C_2$ & $C_3$ & $C_4$ \\
\toprule
S5 & Standard 5 & 0.001702 & 4 & 0.08 & 2 \\ 
\hline
S7 & Standard 7 & 0.08 & 2 & 0.082352309 & 2 \\
\hline
S8 & Standard 8 & 0.031913 & 2 & 0.08 & 2 \\
\hline
K41 & Kolmogorov & 1 & -5/3 & 0 & 0 \\
\midrule
vKA & von K\'arm\'an A & 0.05 & 2 & 2 & 10

\end{tabular}
\caption{Current values used for defining the initial energy spectrum.}
\label{tbl:init_spectra}
\end{center}
\end{table}

\newpage

\section{Aliasing errors}\label{sec:aliasing}
\begin{figure}[tb!]
 \centering
 \includegraphics[width=0.7\textwidth]{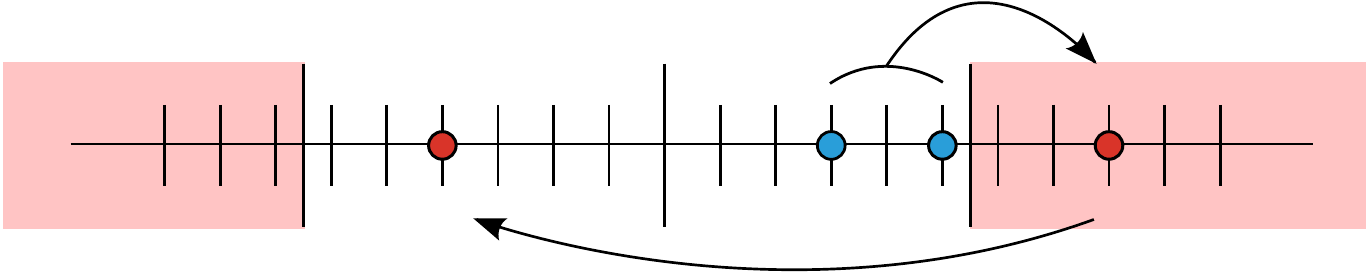}
 \caption{Illustration of the erroneous mode coupling due to aliasing. The shaded region is outwith the resolution of the simulation, but the inability of the transform to distinguish between $k$ and its aliases $k\pm 2\kcut$ leads to a spurious contribution to a mode within the simulation.}
 \label{fig:alias_modes}
\end{figure}

In pseudospectral DNS methods, the use of discrete, finite Fourier transforms to evaluate the non-linear term introduces aliasing errors. These are caused by two modes coupling to create a contribution to an unresolved mode. Due to the periodicity of the discrete Fourier transform, this results in a spurious contribution to a resolved mode. This is illustrated in figure \ref{fig:alias_modes}. For simplicity, we discuss the origin of these erroneous couplings in $1$-dimension, then extend the result for our use in $3$-dimensions.

The non-linear term which is evaluated is given by
\begin{equation}
 W(k) = \sum\limits_{j+l=k} u(j) \omega(l)\ ,
\end{equation}
where the momenta in the sum are restricted to be below the maximum resolved wavenumber, $\lvert j\rvert , \lvert l\rvert \leq \kcut$. This is achieved by first forming the product in configuration space and using a Fast Fourier Transform (FFT) algorithm to find its Fourier representation.

\begin{figure}[tb]
 \centering
 \includegraphics[width=0.7\textwidth]{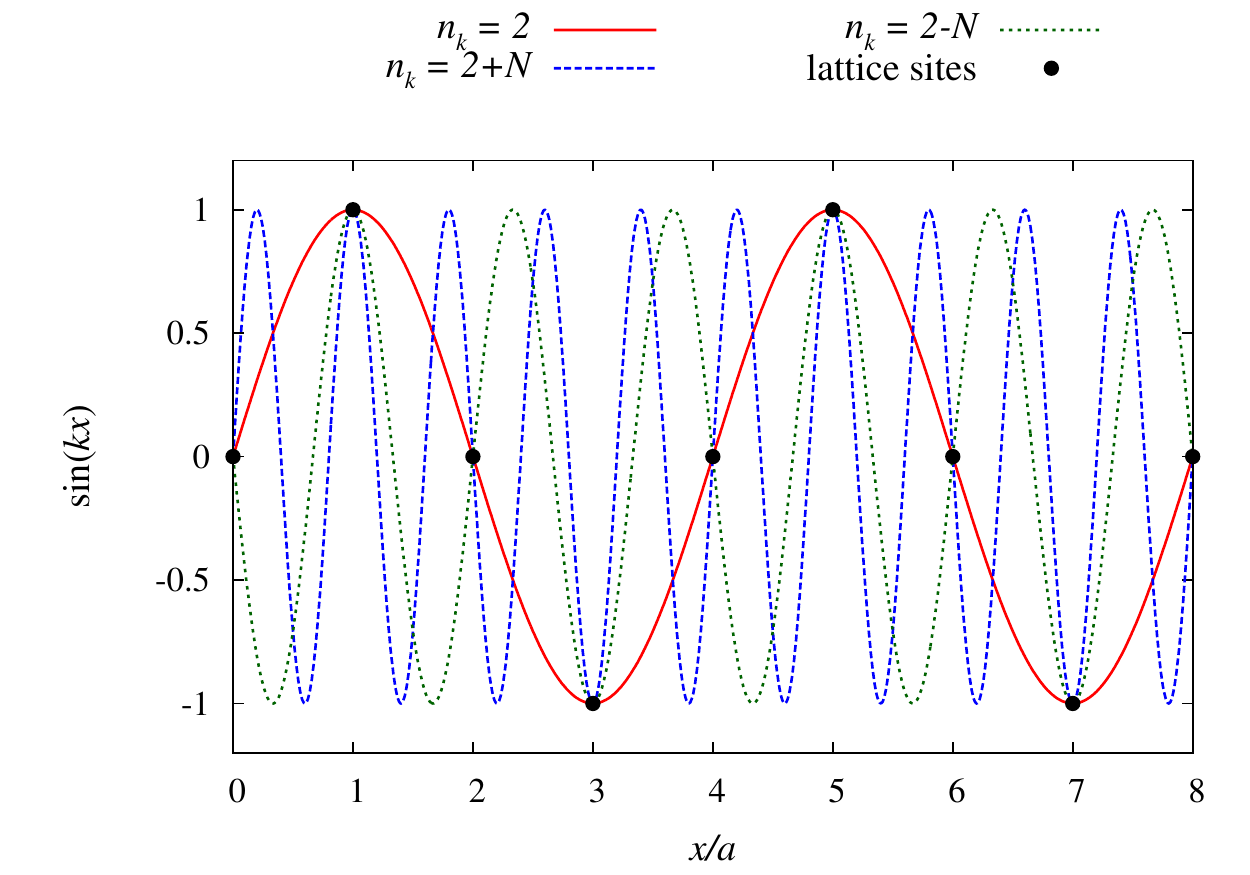}
 \caption{Example showing why modes $k$ and its aliases $k \pm 2\kcut$ are indistinguishable on the lattice of size $N=8$. Since only the values on the lattice sites are known, we have no information about which mode is actually present.}
 \label{fig:alias_sin}
\end{figure}

The velocity and vorticity fields ($u(j)$ and $\omega(l)$, respectively) are first transformed to configuration space by
\begin{equation}
 u(x) = \sum\limits_{j} u(j) e^{i j\cdot x} \ ,\qquad \omega(x) = \sum\limits_{l} \omega(l) e^{i l\cdot x} \ ,
\end{equation}
and the non-linear product formed as
\begin{equation}
 W(x) = u(x) \omega(x) = \sum\limits_{j,l} u(j) \omega(l) e^{i (j+l)\cdot x} \ .
\end{equation}
If the expression above is then transformed back to Fourier space, what has been calculated is
\begin{align}
 \hat{W}(k) &= \tfrac{1}{N} \sum\limits_{x} W(x) e^{-ik\cdot x} \\
 &= \tfrac{1}{N} \sum\limits_{x} \sum\limits_{j,l} u(j) \omega(l) e^{i (j+l-k)\cdot x} \\
 \label{eq:chp3:nonlin_alias1}
 &= \sum\limits_{j,l} u(j) \omega(l) \left[\tfrac{1}{N}\sum\limits_{m_x = 0}^{N-1} e^{i \frac{2\pi}{N}(n_j+n_l-n_k)\cdot m_x}\right] \ ,
\end{align}
where the last line uses the fact that, for our discrete, finite system, positions $x$ are confined to a lattice of spacing $a = L/N$, such that $x = m_x a$ with $m_x \in \{0,\ldots,N-1\}$, and momenta are quantised as $k = 2\pi n_k/L$ with $n_k \in \{-N/2,\ldots,N/2 - 1\}$. Due to the periodicity of the exponential on $[0,2\pi]$, we may add/subtract any integer multiple of $2\pi$ from the exponent; specifically,
\begin{equation}
 e^{i \frac{2\pi}{N}(n_j+n_l-n_k)\cdot m_x} = e^{i \frac{2\pi}{N}(n_j+n_l-(n_k\pm bN))\cdot m_x} \ ,\qquad b \in \mathbb{N} \ ,
\end{equation}
and the value is not changed. Since we can do this for any $b \in \mathbb{N}$, instead of just $\delta_{n_j+n_l,n_k}$ we actually find
\begin{align}
 \label{eq:delta_mod}
 \tfrac{1}{N}\sum\limits_{m_x = 0}^{N-1} e^{i \frac{2\pi}{N}(n_j+n_l-n_k)\cdot m_x} 
 &= \sum_{b\in\mathbb{N}} \delta_{n_j+n_l,n_k\pm bN} \ .
\end{align}
Alternatively, this is written
\begin{equation}
 e^{i \frac{2\pi}{N}(n_j+n_l-(n_k\pm bN))\cdot m_x} = e^{i (j+l-(k\pm 2b\kcut))\cdot x} \ ,\qquad b \in \mathbb{N} \ ,
\end{equation}
where $\kcut$ is the momentum cutoff due to the finite size of the lattice (it is the largest momentum the lattice can support, $\kcut = \pi/a$) and
\begin{equation}
 \label{eq:all_aliases}
 \tfrac{1}{N}\sum\limits_{x} e^{i (j+l-(k\pm 2b\kcut))\cdot x} = \sum_{b \in \mathbb{N}}\delta_{j+l,k\pm 2b\kcut} \ .
\end{equation}
This equation highlights all possible aliases to $k$, as the transform cannot distinguish between any $k\pm 2b\kcut$ (or wavenumbers $n_k \pm bN$). This is shown in figure \ref{fig:alias_sin}.

However, we are only interested in $\lvert k\rvert \leq \kcut$, as this is all that our lattice can resolve. Since the momenta satisfy $\lvert j\rvert , \lvert l\rvert \leq \kcut$, the extrema of the sum $j+l$ are $j+l = \pm 2\kcut$, and as such we only need to consider $b = 1$ in equation (\ref{eq:all_aliases}). This is because $j$ and $l$ cannot couple to any $k\pm 2b\kcut$ with $b > 1$ and still satisfy $\lvert k\rvert \leq \kcut$. For example, with $b=2$ we have, at the upper limit, $j+l = 2\kcut = k \pm 4\kcut$. Clearly, the solutions are $k = -2\kcut, 6\kcut$, which are not supported by the lattice. Figure \ref{fig:alias_modes} illustrates modes $j$ and $l$ coupling to $k + 2\kcut$ and being aliased as mode $k$.

Continuing from equation \eqref{eq:chp3:nonlin_alias1}, the transformed product, $\hat{W}(k)$, is therefore given by
\begin{equation}
 \hat{W}(k) = \sum\limits_{j+l=k} u(j) \omega(l) + \sum\limits_{j+l=k\pm 2\kcut} u(j) \omega(l) \ ,
\end{equation}
where the first term is the desired result, $W(k)$, and the second is an erroneous contribution to the mode $k$ caused by the aliases $k\pm 2\kcut$.

In $3$-dimensions, aliasing can occur in more than one dimension at the same time, and so we instead have 7 aliasing terms:
\begin{align}
 \label{eq:3d_aliased}
 \hat{W}_{\alpha}(\vec{k}) &= \epsilon_{\alpha\beta\gamma}\sum\limits_{\vec{j}+\vec{l}=\vec{k}} u_\beta(\vec{j}) \omega_\gamma(\vec{l}) \nonumber \\
&+ \epsilon_{\alpha\beta\gamma}\left[\sum\limits_{\vec{j}+\vec{l}=\vec{k}\pm2\vec{\kcut}_1} + \sum\limits_{\vec{j}+\vec{l}=\vec{k}\pm2\vec{\kcut}_2} + \sum\limits_{\vec{j}+\vec{l}=\vec{k}\pm2\vec{\kcut}_3} \right]\ u_\beta(\vec{j}) \omega_\gamma(\vec{l}) \nonumber \\
&+ \epsilon_{\alpha\beta\gamma}\left[\sum\limits_{\vec{j}+\vec{l}=\vec{k}\pm2\vec{\kcut}_{1}\pm2\vec{\kcut}_{2}} + \sum\limits_{\vec{j}+\vec{l}=\vec{k}\pm2\vec{\kcut}_{1}\pm2\vec{\kcut}_{3}} + \sum\limits_{\vec{j}+\vec{l}=\vec{k}\pm2\vec{\kcut}_{2}\pm2\vec{\kcut}_{3}} \right]\ u_\beta(\vec{j}) \omega_\gamma(\vec{l}) \nonumber \\
&+ \epsilon_{\alpha\beta\gamma}\sum\limits_{\vec{j}+\vec{l}=\vec{k}\pm2\vec{\kcut}_{1}\pm2\vec{\kcut}_{2}\pm2\vec{\kcut}_{3}} u_\beta(\vec{j}) \omega_\gamma(\vec{l}) \ ,
\end{align}
where $\vec{\kcut}_i = \kcut_i \vec{e}_i$ corresponds to aliasing in the direction labelled by $i$. This allows for the cutoff to be different in each direction, but we will usually consider $\kcut_i = \kcut$ for all $i$. The first term in equation (\ref{eq:3d_aliased}) is our desired result; all others need to be removed somehow, as the aliasing causes energy to appear to be transferred to the higher wavenumbers quicker than it should and the system decays faster. We now turn our attention to removing these unphysical couplings.

\subsubsection{Truncation}\label{sec:truncation}
In the previous section we saw how modes coupling to contribute to an unresolved wavenumber can be mistaken by the discrete Fourier Transform as a contribution to one within the simulation. Figure \ref{fig:alias_modes} gave a graphical interpretation of this. But what if the shaded region \emph{was} kept within the simulation but set to zero?

This is the basis of dealiasing by truncation of the velocity field. Suppose that we set the field to zero for all wavenumbers above a new cutoff, $u_\alpha(\vec{k}) = 0$ whenever $\left\lvert k_i\right\rvert \geq K_i$ for any component $k_i$, where $K_i \leq \kcut_i$. The aliases to $k_i$ arise when
\begin{equation}
 p_i + q_i = k_i \pm 2\kcut_i \ .
\end{equation}
For the alias to be irrelevant, we require (at least one of) the following to be satisfied:
\begin{equation}
 \left\lvert p_i \right\rvert \geq K_i \ ,\qquad\qquad \left\lvert q_i \right\rvert \geq K_i \ ,\qquad\qquad \left\lvert k_i \right\rvert \geq K_i \ .
\end{equation}
The first two conditions ensure that the contribution due to the alias is zero, with the third placing the mode in the truncation region (thus we are not interested in its value, nor any aliasing errors it may suffer from).

\begin{figure}[tb]
 \centering
 \subfigure[One mode lies in the truncation band]{
  \label{sfig:trunc_one}
  \includegraphics[width=0.7\textwidth]{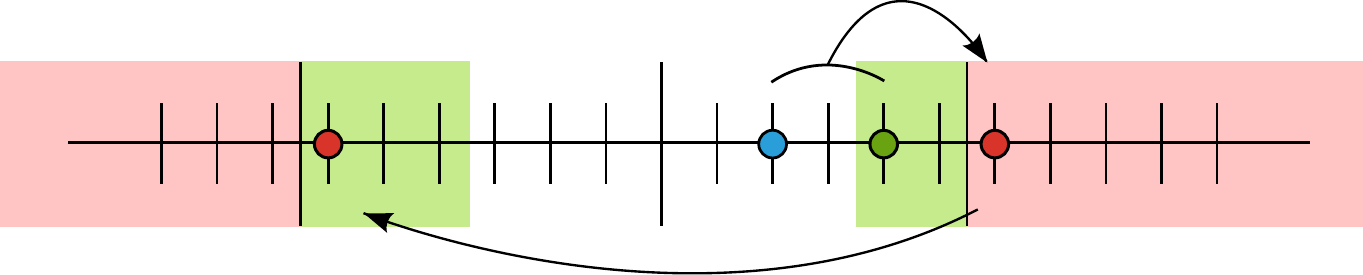}
 }
 \subfigure[Both modes lie in the truncation band]{
  \label{sfig:trunc_both}
  \includegraphics[width=0.7\textwidth]{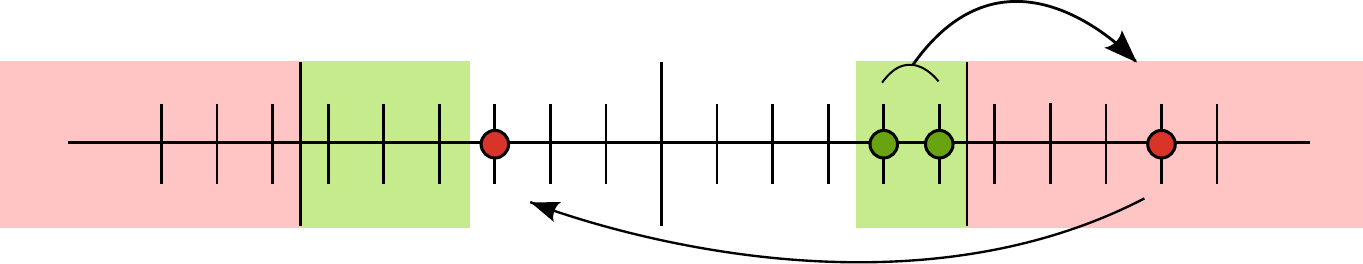}
 }
 \subfigure[Both modes lie on $\ktop = 4$]{
  \label{sfig:trunc_on_K}
  \includegraphics[width=0.7\textwidth]{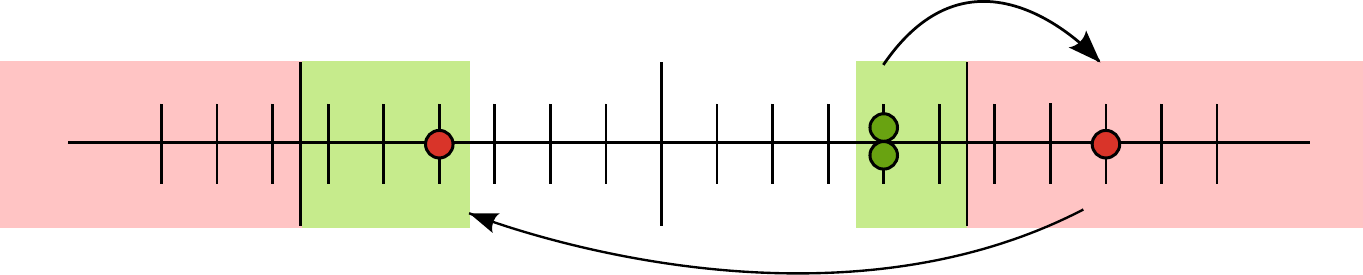}
 }

 \caption{Illustrations of mode couplings. In these figures, $2\kcut = 12$ so that $\ktop = 2\kcut/3 = 4$. The field is truncated for $k \geq 4$, shown in light green. As can be seen, truncating for $\ktop < 4$ also prevents contributions from aliases, but retains fewer modes.}
 \label{fig:alias_trunc}
\end{figure}

We now consider the condition for all aliases to lie outside region of interest: that is, we want $\left\lvert k_i \right\rvert \geq K_i$ for all $\left\lvert p_i \right\rvert, \left\lvert q_i \right\rvert < K_i$. If we focus on the alias with $k_i > 0$ to the correct mode $p_i + q_i$ (so that $p_i, q_i < 0$) then we want $p_i + q_i + 2\kcut_i = k_i \geq K_i$, and this gives 
\begin{equation}
 \label{eq:cut_cond}
 K_i \leqslant p_i + q_i + 2\kcut_i \ .
\end{equation}
We can find a minimum upper bound on this relation by considering the smallest possible value that the right hand side can take, namely when $p_i = q_i = -K_i$. If, instead, we focus on the alias with $k_i < 0$ (so that $p_i, q_i > 0$), then we need $p_i + q_i - 2\kcut_i = k_i \leqslant -K_i$ which rearranges to $K_i \leqslant -p_i - q_i + 2\kcut_i$. Both this and equation (\ref{eq:cut_cond}) can be expressed as
\begin{equation}
 K_i \leqslant -\left\lvert p_i\right\rvert - \left\lvert q_i\right\rvert + 2\kcut_i \ ,
\end{equation}
which gives a minimum upper bound when $\left\lvert p_i\right\rvert = \left\lvert q_i\right\rvert = K_i$ of
\begin{equation}
 \label{eq:cut_ineq}
 K_i \leqslant \frac{2}{3}\kcut_i \ ;
\end{equation}
provided this is satisfied, aliasing effects will be irrelevant for all $\left\lvert p_i\right\rvert, \left\lvert q_i\right\rvert < K_i$. See figure \ref{fig:alias_trunc} for a graphical clarification of this result. As $K_i$ is decreased, the number of modes we retain in the simulation becomes smaller. When we take $\kcut_i = \kcut$ for a cubic lattice, the least wasteful choice is, clearly,
\begin{equation}
 K_i = \ktop = \frac{2}{3}\kcut \ .
\end{equation}

We have arrived at the so-called $2/3$-rule. It states that if we define our velocity field on a (cubic) lattice with cutoff $\kcut$ but truncate the field at $\left\lvert k_i\right\rvert = \ktop$, then the result for the evaluation of the convolution sum will be alias-free for all $\left\lvert k_i\right\rvert < \ktop$. This is a fast, simple way to obtain de-aliased results, but the disadvantage is abundantly clear: we are investing a large amount of resources into simulating modes which are thrown away.

\begin{figure}[!tb]
 \centering
 \subfigure[Patterson and Orszag \cite{Patterson_Orszag:1971-Alias_removal}]{
  \label{sfig:trunc_orszag}
  \includegraphics[width=0.4\textwidth]{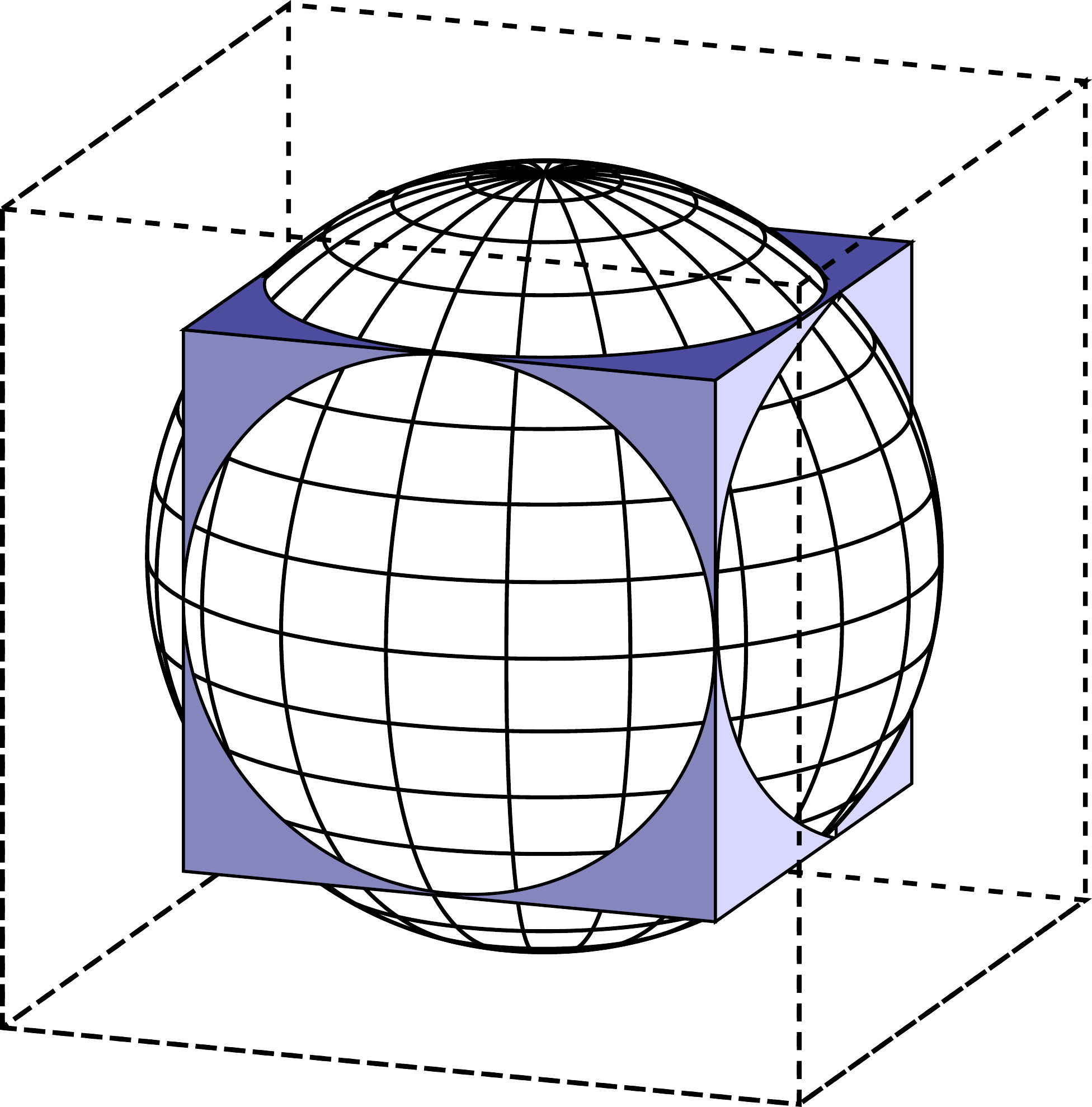}
 }
 \subfigure[Rogallo \cite{Rogallo:1981_num_exp}]{
  \label{sfig:trunc_rogallo}
  \includegraphics[width=0.4\textwidth]{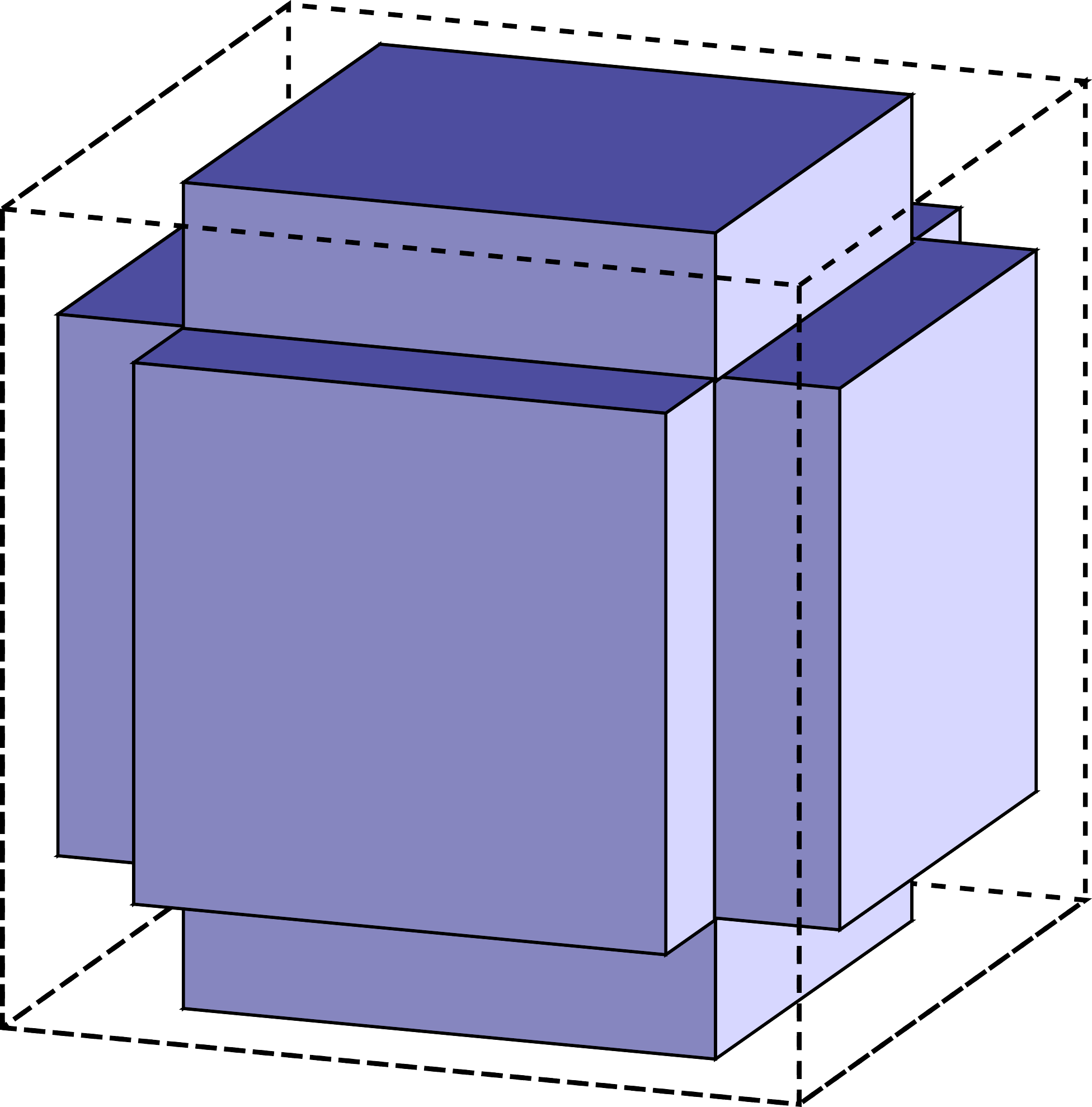}
 }
 \caption{Illustration of truncation methods. The solid cube in the centre of each figure represents the region completely free from aliasing errors, which can support a maximum momentum of $\kmax = 2\kcut/\sqrt{3} = 1.15\kcut$ and has a volume of retained modes $2.37\kcut^3$. Figure (a) can support up to $\kmax = 2\sqrt{2}\kcut/3 = 0.94\kcut$ with a volume of $3.51\kcut^3$. The later scheme by Orszag \cite{Orszag:1971-DNS_full} based on an 18-sided polygon has a volume $4.44\kcut^3$. Figure (b) can support a maximum momentum of $\kmax = \sqrt{17}\kcut/3 = 1.37\kcut$ with a volume of $3.56 \kcut^3$. The total volume of the simulation is $8\kcut^3$.}
 \label{fig:trunc_orszag_rog}
\end{figure}

In practice, this ideal case is not adhered to. Since any truncation will remove some aliasing errors, other standards have been developed. Patterson and Orszag \cite{Patterson_Orszag:1971-Alias_removal} showed how the spherical truncation, shown in figure \ref{sfig:trunc_orszag}, at $\kmax^2 = 2\ktop^2$ (which results in $\kmax = \frac{2\sqrt{2}}{3}\kcut$) eliminates aliasing errors in two or more directions (leaving only the possibility of aliasing in a single direction at once). Later, Orszag \cite{Orszag:1971-DNS_full} introduced a less severe truncation leading to the same result, in which truncation takes place outside an octodecahedron (18-sided polygon), although notes that for the simulation of isotropic turbulence the spherical truncation may be more natural. An additional scheme was used by Rogallo \cite{Rogallo:1981_num_exp}, in which the truncation is only made for modes having two or more components with $\left\lvert k_i\right\rvert \geq \ktop$. This results in the complete removal of aliasing in more than a single direction at once, but preserves more modes than the spherical truncation. This can then be supported by random grid shifts (see next section) to minimise the remaining aliasing error. This truncation is illustrated in figure \ref{sfig:trunc_rogallo}.

Note: $\ktop$ is the maximum wavenumber for any one direction, and so cubic truncation whenever $k_i > \ktop$ will still support a maximum wavenumber $\kmax = \ktop\sqrt{3}$. In this project, we employ an even more severe spherical truncation, truncating all modes with $k = \vmod{k} \geq \frac{2\kcut}{3}$. This completely removes aliasing errors and is more natural for the isotropic field under consideration. In this case, $\kmax = \ktop$.

\vspace{1em}

\noindent{\bfseries Technical aside:}
Technically, we do not require $\left\lvert p_i\right\rvert = \left\lvert q_i\right\rvert = K_i$ to produce an alias with $k_i \geq K_i$. Taking the number of lattice sites in the new and original regions to be $M_i$ and $N_i$, respectively, we have $K_i = \frac{\pi}{L_i}M_i$ and $\kcut_i = \frac{\pi}{L_i}N_i$. The required combination is in fact $\left\lvert p_i\right\rvert = \left\lvert q_i\right\rvert = \frac{\pi}{L_i}(M_i-1)$. In this case, our condition becomes $M_i \leqslant 2N_i - 2M_i + 2$ or $M_i \leqslant \frac{2}{3}(N_i + 1)$. Since the choice $M_i \leqslant 2N_i/3$ satisfies the previous condition, this is often used instead. Putting back in factors, this is identical to $K_i \leqslant \frac{2}{3}\kcut_i$.

\subsubsection{Grid shifting}\label{sec:grid_shift}
It is the opinion of the present writer that the tradition of presenting this analysis for the $1$-dimensional case is not satisfactory, since its extension to higher dimensions is \emph{not} trivial. Instead, we here start in $d$-dimensions before explicitly calculating the 2-dimensional case.

As a first step, we drop vector indices for the fields such that our desired result
\begin{equation}
 W_{\alpha}(\vec{k}) = \epsilon_{\alpha\beta\gamma}\sum\limits_{\vec{j}+\vec{l}=\vec{k}} u_\beta(\vec{j}) \omega_\gamma(\vec{l}) \qquad\textrm{becomes}\qquad W(\vec{k}) = \sum\limits_{\vec{j}+\vec{l}=\vec{k}} u(\vec{j}) \omega(\vec{l}) \ ,
\end{equation}
as these \emph{can} be trivially restored at the end of the calculation.

Consider a phase shift applied to the velocity field (before the Fourier Transform back to configuration space is found), then
\begin{equation}
 \tilde{u}(\vec{k}) = e^{i\vec{k}\cdot\vec{\Gamma}} u(\vec{k}) \qquad\qquad\textrm{and}\qquad\qquad \tilde{\omega}(\vec{k}) = e^{i\vec{k}\cdot\vec{\Gamma}} \omega(\vec{k}) \ .
\end{equation}
The (discrete) transform is then found to be
\begin{equation}
 u^{\vec{\Gamma}}(\vec{x}) = \sum\limits_{\vec{k}} \tilde{u}(\vec{k}) e^{i\vec{k}\cdot\vec{x}} \ ,
\end{equation}
and similarly for the vorticity, $\omega(\vec{k})$. This can also be written
\begin{equation}
 \label{eq:shifted_FT}
 u^{\vec{\Gamma}}(\vec{x}) = \sum\limits_{\vec{k}} u(\vec{k}) e^{i\vec{k}\cdot(\vec{x}+\vec{\Gamma})} = u(\vec{x}+\vec{\Gamma}) \ ,
\end{equation}
and we see that the phase shift has had the same effect as shifting the grid by $\vec{\Gamma}$. We can then construct the desired product on this shifted grid,
\begin{equation}
 W^{\vec{\Gamma}}(\vec{x}) = u^{\vec{\Gamma}}(\vec{x}) \omega^{\vec{\Gamma}}(\vec{x}) \ ,
\end{equation}
and return to Fourier space, to find
\begin{align}
 \tilde{W}^{\vec{\Gamma}}(\vec{k}) &= \left(\frac{1}{N}\right)^d \sum\limits_{\vec{x}} W^{\vec{\Gamma}}(\vec{x}) e^{-i\vec{k}\cdot\vec{x}} \nonumber \\
 &= \left(\frac{1}{N}\right)^d \sum\limits_{\vec{x}} \sum\limits_{\vec{p}} \sum\limits_{\vec{q}} u(\vec{p})  \omega(\vec{q})\ e^{i(\vec{p}+\vec{q})\cdot\vec{\Gamma}}\ e^{i(\vec{p}+\vec{q}-\vec{k})\cdot\vec{x}} \\
 W^{\vec{\Gamma}}(\vec{k}) &= \sum\limits_{\vec{p}} \sum\limits_{\vec{q}} u(\vec{p}) \omega(\vec{q}) e^{i(\vec{p}+\vec{q}-\vec{k})\cdot\vec{\Gamma}} \prod\limits_{j=1}^{d} \left(\frac{1}{N} \sum\limits_{x_j}  e^{i(p_j+q_j-k_j)\cdot x_j}\right) \ ,
\end{align}
where in the last line we phase shift back to our original lattice\footnote{Alternatively, this can be seen by noting that equation \eqref{eq:shifted_FT} is just a shifted Fourier transform, such that
\begin{equation}
 W^{\vec{\Gamma}}(\vec{k}) = \left(\frac{1}{N}\right)^d \sum\limits_{\vec{x}} W(\vec{x}+\vec{\Gamma}) e^{-i\vec{k}\cdot(\vec{x}+\vec{\Gamma})}
  = \left(\frac{1}{N}\right)^d \sum\limits_{\vec{x}} \sum\limits_{\vec{p}} \sum\limits_{\vec{q}} u(\vec{p}) \omega(\vec{q})\ e^{i(\vec{p}+\vec{q}-\vec{k})\cdot(\vec{x}+\vec{\Gamma})} \ .
\end{equation}
This highlights the equivalence to evaluating $W^{\vec{\Gamma}}(\vec{k})$ on different grids.}, using $e^{-i\vec{k}\cdot\vec{\Gamma}} \tilde{W}^{\vec{\Gamma}}(\vec{k})$.

Notice that, so far, this is valid in an arbitrary number of dimensions, $d$. Concentrating on the expression in the brackets, we can use the result of equation (\ref{eq:all_aliases}) to write
\begin{equation}
 \frac{1}{N} \sum\limits_{x_j}  e^{i(p_j+q_j-k_j)\cdot x_j} = \sum\limits_{b \in \mathbb{N}} \delta_{p_j+q_j,k_j\pm 2b\kcut_j} \ ,
\end{equation}
where $\kcut_j$ is the cutoff in dimension labelled by $j$ and $b = 0$ is our desired result. In section \ref{sec:aliasing}, we argued that, due the the finite extent of the lattice, only aliases with $b = 1$ can be resolved, and as such
\begin{equation}
 \frac{1}{N} \sum\limits_{x_j}  e^{i(p_j+q_j-k_j)\cdot x_j} = \delta_{p_j+q_j,k_j} + \delta_{p_j+q_j,k_j+2\kcut_j} + \delta_{p_j+q_j,k_j-2\kcut_j} \ .
\end{equation}
From this equation, we see that every dimension is aliased independently, and that in each dimension we get two erroneous contributions, corresponding to $k_j\pm2b\kcut_j$.

The total effect of this on the calculation,
\begin{equation}
 W^{\vec{\Gamma}}(\vec{k}) = \sum\limits_{\vec{p}} \sum\limits_{\vec{q}} u(\vec{p}) \omega(\vec{q}) e^{i(\vec{p}+\vec{q}-\vec{k})\cdot\vec{\Gamma}} \prod\limits_{j=1}^{d} \left[ \delta_{p_j+q_j,k_j} + \delta_{p_j+q_j,k_j \pm 2\kcut_j}
 \right] \ ,
\end{equation}
is more easily seen if we now select $d = 2$ as an example. In this case, the product over $\delta$-functions gives us
\begin{align}
 \prod\limits_{j=1}^{2} \Big[ \delta_{p_j+q_j,k_j} + \delta_{p_j+q_j,k_j \pm 2\kcut_j} 
 \Big]
 = \delta_{\vec{p}+\vec{q},\vec{k}} + \delta_{\vec{p}+\vec{q},\vec{k}\pm2\vec{\kcut}_1} + \delta_{\vec{p}+\vec{q},\vec{k}\pm2\vec{\kcut}_2} + \delta_{\vec{p}+\vec{q},\vec{k}\pm2\vec{\kcut}_1\pm2\vec{\kcut}_2}
 \ ,
\end{align}
which may be used to remove one of the sums, leaving
\begin{align}
 W^{\vec{\Gamma}}(\vec{k}) &= \Bigg[ \hat{\sum\limits_{0}} + e^{\pm i 2\vec{\kcut}_1\cdot\vec{\Gamma}} \hat{\sum\limits_{2\vec{\kcut}_1}} + \ e^{\pm i 2\vec{\kcut}_2\cdot\vec{\Gamma}} \hat{\sum\limits_{2\vec{\kcut}_2}} + \ e^{\pm i 2\vec{\kcut}_1\cdot\vec{\Gamma}} e^{\pm i 2\vec{\kcut}_2\cdot\vec{\Gamma}} \hat{\sum\limits_{2\vec{\kcut}_1 \pm 2\vec{\kcut}_2}} \Bigg] u(\vec{p}) \omega(\vec{q}) \ ,
\end{align}
where $\vec{\kcut}_i = \kcut_i \vec{e}_i$ as defined above in equation (\ref{eq:3d_aliased}) and we introduce the temporary notation
\begin{equation}
 \hat{\sum\limits_{\vec{X}}} = \sum\limits_{\vec{p}+\vec{q}=\vec{k}\pm\vec{X}} \ .
\end{equation}
This shows that the terms split into ($d+1$) categories: no aliasing; aliasing in one direction only; aliasing in two directions simultaneously; ... ; aliasing in $d$ directions simultaneously.

Using the fact that the cutoff $\kcut_j = \pi N_j/L_j = \pi/a_j$, where $a_j$ is the lattice spacing,
\begin{equation}
 W^{\vec{\Gamma}}(\vec{k}) = \left[\hat{\sum\limits_{0}}
+ \ e^{\pm 2\pi i\frac{\Gamma_1}{a_1}} \hat{\sum\limits_{2\vec{\kcut}_1}}
+ \ e^{\pm 2\pi i\frac{\Gamma_2}{a_2}} \hat{\sum\limits_{2\vec{\kcut}_2}}
+ \ e^{\pm 2\pi i \left(\frac{\Gamma_1}{a_1}+\frac{\Gamma_2}{a_2}\right)} \hat{\sum\limits_{2\vec{\kcut}_1 \pm 2\vec{\kcut}_2}} \right] u(\vec{p}) \omega(\vec{q}) \ .
\end{equation}
As can be seen, our desired result (the first sum in the brackets) does not have any residual phase associated with it, whereas all other terms do. So if we compute this using several different, specifically chosen $\vec{\Gamma}$, it would be possible to cancel these additional terms.

Let's consider the simplest case: Examining the first phase in the above equation, the simplest choice allowing the contribution to be cancelled comes when the phase is $\pm 1$, corresponding to $\Gamma_1 = 0, a_1/2$. If we sum the results of using these values for $\Gamma_1$, we have
\begin{equation}
 W^{(0,\Gamma_2)}(\vec{k}) + W^{(\frac{a_1}{2},\Gamma_2)}(\vec{k}) = \left[2\sum\limits_{\vec{p}+\vec{q}=\vec{k}}
+ \ 2e^{\pm 2\pi i\frac{\Gamma_2}{a_2}} \sum\limits_{\vec{p}+\vec{q}=\vec{k}\pm2\vec{\kcut}_2} \right] u(\vec{p}) \omega(\vec{q}) \ .
\end{equation}

A similar analysis holds for the second phase, giving $\Gamma_2 = 0, a_2/2$. Combining all combinations, we see that
\begin{align}
 W^{(0,0)}(\vec{k}) + W^{(\frac{a_1}{2},0)}(\vec{k}) &= \left[2\sum\limits_{\vec{p}+\vec{q}=\vec{k}}
+ \ 2 \sum\limits_{\vec{p}+\vec{q}=\vec{k}\pm2\vec{\kcut}_2} \right] u(\vec{p}) \omega(\vec{q}) \nonumber \ , \\
 W^{(0,\frac{a_2}{2})}(\vec{k}) + W^{(\frac{a_1}{2},\frac{a_2}{2})}(\vec{k}) &= \left[2\sum\limits_{\vec{p}+\vec{q}=\vec{k}}
- \ 2 \sum\limits_{\vec{p}+\vec{q}=\vec{k}\pm2\vec{\kcut}_2} \right] u(\vec{p}) \omega(\vec{q}) \ ,
\end{align}
thus
\begin{equation}
 \sum\limits_{\vec{p}+\vec{q}=\vec{k}}
 u(\vec{p}) \omega(\vec{q}) = \tfrac{1}{4} \left[ W^{(0,0)}(\vec{k}) + W^{(\frac{a_1}{2},0)}(\vec{k}) + W^{(0,\frac{a_2}{2})}(\vec{k}) + W^{(\frac{a_1}{2},\frac{a_2}{2})}(\vec{k}) \right] \ .
\end{equation}
The desired, alias-free result has been found by evaluating the term on four shifted grids and summing the results. More generally, we can write
\begin{equation}
 W(\vec{k}) = \sum\limits_{\vec{p}+\vec{q}=\vec{k}}
 u(\vec{p}) \omega(\vec{q}) = \frac{1}{2^d}\sum\limits_{\vec{\Gamma} \in S} W^{\vec{\Gamma}}(\vec{k}) \ ,
\end{equation}
where in 2-dimensions we saw the set $S = \left\{(0,0), (0,\frac{a_2}{2}),(\frac{a_1}{2},0), (\frac{a_1}{2},\frac{a_2}{2})\right\}$. In $d$-dimensions, this set consists of $2^d$ $d$-dimensional vectors, comprising all possible combinations of $(\Gamma_1, \ldots, \Gamma_d)$ --- all possible unique shifted grids.

The current choice $\Gamma_i = 0, a_i/2$ corresponds to no shift and a shift of half a lattice spacing, respectively. However, this is not the only possible choice which removes all aliasing errors. Orszag \cite{Orszag:1971-DNS_full} also describes the use of the shifts $\Gamma_i = \pm a_i/4$, as it turns out that the only necessary condition is that they be separated by half a grid spacing: $\left\lvert\Gamma_i^{(1)} - \Gamma_i^{(2)}\right\rvert = a_i/2$.

In 3-dimensions, we have the set
\begin{align*}
 S = \bigg\{& \left(0,0,0\right), \left(0,0,\frac{a_3}{2}\right), \left(0,\frac{a_2}{2},0\right), \left(0,\frac{a_2}{2},\frac{a_3}{2}\right), \\
 &\left(\frac{a_1}{2},0,0\right), \left(\frac{a_1}{2},0,\frac{a_3}{2}\right),\left(\frac{a_1}{2},\frac{a_2}{2},0\right), \left(\frac{a_1}{2},\frac{a_2}{2},\frac{a_3}{2}\right) \bigg\}\ .
\end{align*}
This means that for aliasing errors to be completely removed we must evaluate the convolution on eight grids and combine the results. This involves a significant additional computational cost.
Grid shifting does allow us to retain more modes with the same use of memory, but one cannot overlook the additional time requirements.

As a final step, we restore the vector indices, resulting in
\begin{equation}
 W_\alpha(\vec{k}) = \epsilon_{\alpha\beta\gamma}\sum\limits_{\vec{p}+\vec{q}=\vec{k}}
 u_\beta(\vec{p}) \omega_\gamma(\vec{q}) = \frac{1}{2^d}\sum\limits_{\vec{\Gamma} \in S} W_\alpha^{\vec{\Gamma}}(\vec{k}) \ .
\end{equation}

\subsubsection{Random grid shifting}

Rogallo \cite{Rogallo:1981_num_exp} suggested a procedure that would reduce (single-direction) aliasing errors without the huge additional computation work of repeated evaluation of the convolution. In this method,
we simply apply a random shift to the grid before evaluation, and then back again after. This will have spurious aliasing errors with a certain phase shift. At the next evaluation (which can be within the same time-step, depending on the time-integration algorithm employed) the opposite shift is applied, with the hope that the aliasing errors generated at one evaluation will be nearly cancelled at the next. The important step is that multiple evaluations of the non-linear term on different grids are not performed at the same time-steps, saving computation but preventing the errors from being cancelled exactly.

Note that this method does not help with the higher-order aliases, by which we mean aliasing in more than one direction simultaneously. As such, truncation such as those depicted in figure \ref{fig:trunc_orszag_rog}, which remove 2- and 3-directional aliases, should be used in conjunction with this method.

To start, we generate a random shift vector
\begin{equation}
 \vec{\Gamma}^{(1)} = a \vec{R} = \frac{\vec{R} L}{N} \ ,
\end{equation}
where $R_\alpha$ are random numbers on the interval $[0, 1)$. The phase $\exp i\vec{k}\cdot\vec{\Gamma}^{(1)}$ is applied to all modes, so after calculating the non-linear term (and phase-shifting back to our original grid) we find that the (single-direction) alias terms are multiplied by phases like $\exp \pm i2\vec{\kcut}_i\cdot\vec{\Gamma}^{(1)}$. At the next evaluation, we shift by a second vector $\vec{\Gamma}^{(2)}$. If we choose the shift so that the resulting phase on the aliasing errors is given by
\begin{equation}
 e^{\pm i2\vec{\kcut}_i\cdot\vec{\Gamma}^{(2)}} = - e^{\pm i2\vec{\kcut}_i\cdot\vec{\Gamma}^{(1)}} \ ,
\end{equation}
where once again $\vec{\kcut}_i = \kcut \vec{e}_i$, then, assuming that the change in the non-linear term at each time-step is small, the aliasing errors cancel one another, without the need to perform multiple evaluations at each step. To do this, we note that
\begin{equation}
 - e^{\pm i2\vec{\kcut}_i\cdot\vec{\Gamma}^{(1)}} = e^{\pm i2\vec{\kcut}_i\cdot\vec{\Gamma}^{(1)}} e^{\pm i\pi} \ ,
\end{equation}
so that
\begin{align}
 2\vec{\kcut}_i\cdot\vec{\Gamma}^{(2)} &= 2\vec{\kcut}_i\cdot\vec{\Gamma}^{(1)} \pm \pi \\
 &= 2\kcut a R_i \pm \kcut a \ .
\end{align}
From this equation, it is trivial to see that the required shift must be
\begin{equation}
 \Gamma^{(2)}_\alpha = a \left[ R_\alpha \pm \frac{1}{2} \right] \ .
\end{equation}
Again, providing the shifts are separated by half a lattice spacing, the phases will be equal and opposite.

Schematically, we represent the exact result of the non-linear term by $W$, the evaluation with the first phase shift by $W^{(1)}$, the second evaluation by $W^{(1,2)}$, and the aliasing errors with a hat, such that
\begin{align}
 W^{(1)} &= W + \theta \hat{W} \\
 W^{(1,2)} &= W^{(1)} - \theta \hat{W}^{(1)}
\end{align}
and $\theta$ is the resultant phase shift on the aliasing errors. By assuming that the non-linear term does not vary significantly between the evaluations (for example, using a very small time-step) so that the aliasing errors are similar $\hat{W}^{(1)} \simeq \hat{W}$, we have
\begin{align}
 W^{(1,2)} &\simeq W + \theta \left(\hat{W} - \hat{W}\right) \nonumber \\
 &\simeq W \ ,
\end{align}
and the aliasing errors will \emph{nearly} cancel each other, leaving just the result we require.

This may be used alongside partial truncation allowing more modes to be retained in the simulation, and is a popular method to efficiently \emph{reduce} aliasing errors.

\newpage
\section{Some currently available DNS codes}\label{subsec:currently_avail}
In addition to the \dns\ code written here, there are a number of freely available DNS codes available online. These include (but are not limited to):
\begin{description}
 \item [hit3d]{A pseudospectral code based on Fourier-decomposition for the simulation of homogeneous, isotropic turbulence in a periodic box. This parallel code, developed by the fluids group at Stanford, is written in FORTRAN 90 and is available under the GPL from \url{http://code.google.com/p/hit3d/}.}
\item [channelflow]{A serial pseudospectral code written in C++ which uses a mixture of Fourier- and Chebyshev-decomposition to simulate flow in a non-periodic (in one dimension) channel. Available at \url{http://www.channelflow.org/dokuwiki/doku.php/start}.}
\item [OpenFOAM]{An open-source CFD package based on finite volume methods (allowing unstructured grids). This freely available parallel code is written in C++ and can be found at \url{http://www.openfoam.com/}.}
\end{description}
The reader should visit internet addresses listed above for further information.

\newpage

\section{Calculating statistics}\label{sec:calc_stats}
Here we describe the calculation of the main quantities of interest for the simulation. But first we discuss a technique called shell-averaging which will be used when finding spectra.

\subsection{Spectra}
\subsubsection{Shell-averaging}\label{subsubsec:shell_average}
Consider a quantity $A(\vec{k})$ which can be measured from the system at all $\vec{k}$. This function depends on the wavevector, so if we instead wish to to study the one-dimensional quantity $A(\kappa)$ we would simply integrate (sum) over the angular directions, such that
\begin{equation}
 \label{eq:simple_ang_sum}
 A(\kappa) = \sum_{\vec{k} : \lvert\vec{k}\rvert = \kappa} A(\vec{k}) \ ,
\end{equation}
where we note that $\kappa$ is not necessarily an integer.

Points with $\lvert\vec{k}\rvert = \kappa$ lie on the surface of a sphere of radius $\kappa$ in the Fourier-space volume 
and as such, for our 3-dimensional lattice, $\kappa = 0,1,\sqrt{2},\sqrt{3}, \cdots, \kmax$. Due to our lattice being a Cartesian framework, the majority of points do not sit on the surface of any sphere with integer radius; for example, the point $(1,1,1)$ has $k = \sqrt{3}$, which sits between $\kappa = 1$ and $\kappa = 2$.

\begin{figure}[tb]
 \centering\includegraphics[width=0.7\textwidth]{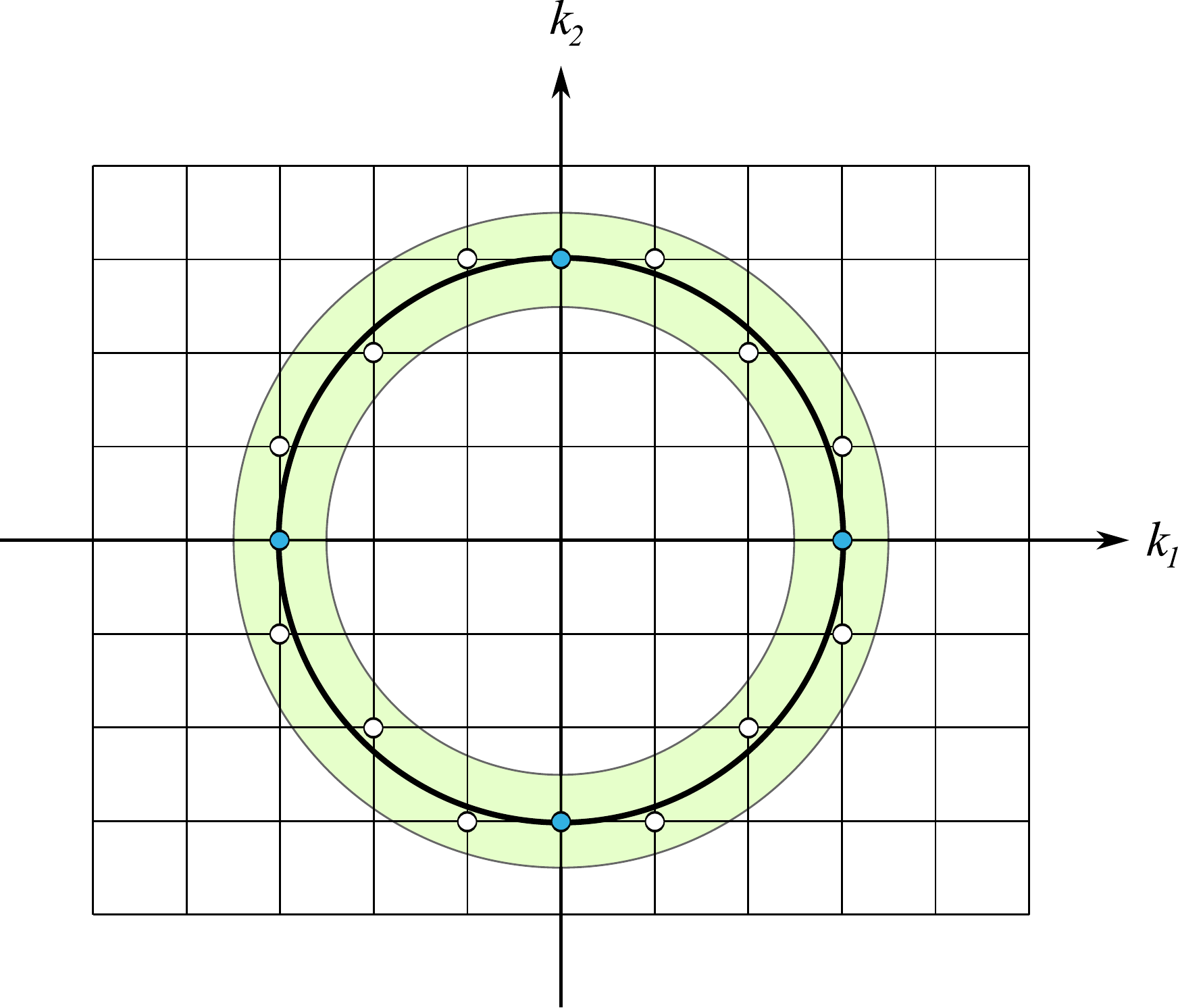}
 \caption{The $\Delta k = 1$ shell average in 2-dimensions. Filled points have $\vmod{k} = 3$ and would be used alone to calculate an un-averaged quantity. Whereas, the open points satisfy $2.5 \leq k < 3.5$ and would also contribute to a shell averaged quantity.}
 \label{fig:shell_average}
\end{figure}

Instead of considering contributions due only to points on the surface of these spheres with integer radius, \emph{shell-averaging} considers the contributions from all points which lie in a shell of thickness $\Delta k$, such that wavenumbers $n-\tfrac{1}{2}\Delta k \leq \kappa < n+\tfrac{1}{2}\Delta k$ all contribute to integer wavenumber $n$. For non-overlapping shells that fill the entire space, we take $\Delta k = 1$. This case is shown in figure \ref{fig:shell_average}. The effect of this procedure is to smooth the statistics by including a larger number of points. The average is then expressed for integer wavenumber as
\begin{equation}
 A(n) = \frac{1}{\Delta k} \sum_{n-\tfrac{1}{2}\Delta k \leq \kappa < n+\tfrac{1}{2}\Delta k} \sum_{\vec{k} : \lvert\vec{k}\rvert = \kappa} A(\vec{k}) \ ,
\end{equation}
where $\Delta k$ is essentially the number of (unit) shells being included in the average. This is expressed more succinctly as
\begin{equation}
 \label{eq:shell_av}
 A(n) = \frac{1}{\Delta k} \sum_{\vec{k} \in {\mathbb S}_n} A(\vec{k}) \ ,
\end{equation}
where the set ${\mathbb S}_n = \{ \vec{k} : n-\tfrac{1}{2}\Delta k \leq \lvert\vec{k}\rvert < n+\tfrac{1}{2}\Delta k \}$ contains all the points in the shell. \\

\noindent\textbf{An aside on isotropy}\\
Since the system is isotropic, one can better approximate the continuum by acknowledging that the density of states for the surface of radius $n$ is $4\pi n^2$, thus
\begin{align}
 A(n) &= 4\pi n^2 \langle A(\vec{n}) \rangle \nonumber \\
 &= \frac{4\pi n^2}{P_n} \sum_{\vec{k} \in {\mathbb S}_n} A(\vec{k}) \ ,
\end{align}
where $P_n$ is the number of points in the set ${\mathbb S}_n$.

Whilst technically being more accurate, this improvement is not used here, since the conservation of energy satisfied by the discrete spectra (the energy balance equation) is violated by this technique. For example, the isotropically-averaged transfer spectrum does not integrate to zero, see section \ref{sec:bench_TS}.
If this is required, it is easily performed in post-processing, since the code does save the number of points in each shell.

This also highlights how an isotropic quantity is evaluated using shell averaging,
\begin{equation}
 \label{eq:shell_av_iso}
 4\pi k^2 \langle A(\vec{k}) \rangle = \frac{1}{\Delta k} \sum_{\vec{k} \in {\mathbb S}_n} A(\vec{k}) \ .
\end{equation}

\subsubsection{Energy spectrum}\label{subsubsec:energy_spectrum}
The energy spectrum for the isotropic continuum was defined in equation \eqref{eq:iso_Ek} to be
\begin{equation*}
 E(k,t) = 4\pi k^2\ \tfrac{1}{2}\langle u_\alpha(\vec{k},t) u_\alpha(-\vec{k},t) \rangle \ ,
\end{equation*}
which is evaluated by shell averaging using the relationship in equation \eqref{eq:shell_av_iso} as simply
\begin{equation}
 E(k,t) = \frac{1}{\Delta k} \sum_{\vec{k} \in {\mathbb S}_k} \tfrac{1}{2} u_\alpha(\vec{k},t) u_\alpha(-\vec{k},t) \ ,
\end{equation}
As such the total energy of the system is simply given by
\begin{align}
 E(t) &= \Delta k\sum_k E(k,t) \nonumber \\
 &= \Delta k\sum_k \frac{1}{\Delta k} \sum_{\vec{k} \in {\mathbb S}_k} \tfrac{1}{2} u_\alpha(\vec{k},t) u_\alpha(-\vec{k},t) \nonumber \\
 &= \sum_{\vec{k}} \tfrac{1}{2} u_\alpha(\vec{k},t) u_\alpha(-\vec{k},t) \ ,
\end{align}
which is just the addition of the energy contained in each wavevector, as expected.

\subsubsection{Transfer spectrum}\label{subsubsec:transfer}
We wish to simplify the evaluation of $T(k,t)$ as far as possible. In the time-advancement of the Navier-Stokes equation, we are required to compute $W_\alpha(\vec{k},t)$ at each time-step. In the continuum, the transfer spectrum has the form
\begin{align}
 T(k,t) &= 4\pi k^2 M_{\alpha\beta\gamma}(\vec{k}) \int d^3j\ \left\langle u_\alpha(-\vec{k},t) u_\beta(\vec{j},t) u_\gamma(\vec{k}-\vec{j},t) \right\rangle \nonumber \\
 &= 4\pi k^2 \left\langle u_\alpha(-\vec{k},t) W_\alpha(\vec{k},t) \right\rangle \ ,
\end{align}
which shows how the transfer spectrum can be calculated from the non-linear term, and hence why pre-calculating $W_\alpha(\vec{k},t)$ at the end of the time-step allows us to find the transfer spectrum at that time. Using shell averaging, this is approximated for the discrete lattice as
\begin{align}
 T(k,t) = \frac{1}{\Delta k} \sum_{\vec{k} \in {\mathbb S}_k} u_\alpha(-\vec{k},t) W_\alpha(\vec{k},t) \ .
\end{align}

\subsection{Post-processing}\label{subsec:stats_PP}
There are numerous parameters that can be calculated either during the simulation or from the spectra after the simulation, provided they are saved to disk regularly (this saves computation time). These include:
\begin{description}
 \item [Total energy]{This was mentioned at the end of section \ref{subsubsec:energy_spectrum} and is found by integrating the energy spectrum over all $k$
   \begin{equation}
    E(t) = \int dk\ E(k,t) \qquad\textrm{or}\qquad \Delta k \sum_n E(n,t) \ .
   \end{equation}
}
 \item [Root-mean-square (rms) velocity]{A characteristic velocity scale is found from the total energy, since the total energy is proportional to the velocity squared
 \begin{equation}
  E(t) = \tfrac{1}{2}\langle u^2(\vec{x},t)\rangle = \tfrac{1}{2} \left[ \langle u_x^2(\vec{x},t)\rangle + \langle u_y^2(\vec{x},t)\rangle + \langle u_z^2(\vec{x},t)\rangle \right] \ .
\end{equation}
 By assuming isotropy, we must have $\langle u_x^2(\vec{x},t)\rangle = \langle u_y^2(\vec{x},t)\rangle = \langle u_z^2(\vec{x},t)\rangle = u^2$ so that $E(t) = \tfrac{3}{2}u^2$, or
\begin{equation}
 u(t) = \sqrt{\tfrac{2}{3} E(t)} \ .
\end{equation}
}
 \item [Dissipation spectrum]{As shown in section \ref{subsec:lin_equation}, the dissipation spectrum has the form
 \begin{equation}
 D(k,t) = 2\nu_0 k^2 E(k,t)\ ,
\end{equation}
so is readily found from the energy spectrum. It quantifies the amount of energy being lost by dissipation from the different scales of turbulence motion.
}
 \item [Dissipation rate]{The dissipation rate quantifies the total amount of energy being lost (per unit time), so is simply the integral of the dissipation spectrum
 \begin{equation}
 \varepsilon(t) = \int dk\ D(k,t) \qquad\textrm{or}\qquad \Delta k \sum_n D(n,t) \ .
\end{equation}
}
 \item [Transport power spectrum]{The transfer spectrum (see section \ref{subsubsec:transfer}) shows the amount of energy entering (or leaving) each wavenumber $k$ due to non-linear interactions with all other modes. The transport power spectrum, $\Pi(k,t)$, instead shows the rate at which energy is being transferred \emph{through} mode $k$ from modes $j < k$ to $j > k$. It is found as
 \begin{align}
  \Pi(k,t) &= \int_k^\infty dj\ T(j,t) \qquad\textrm{or}\qquad \Delta k \sum_{j=k}^{\kmax} T(j,t) \ .
 \end{align}
}
 \item [Maximum inertial flux]{The transport power spectrum allows us to find the maximum flux, $\pimax$, as it is simply the maximum value of the spectrum. When an inertial range develops, the power spectrum should flatten and there will be a plateau at $\pimax$ as these modes are \emph{scale invariant} and simply pass the energy along. For stationary turbulence, we should also find that $\pimax = \varepsilon_W = \varepsilon$ once an inertial range has formed, since the system can only dissipate at high wavenumbers the energy that is passed along through the intermediate ones.}

 \item [Integral scale]{This gives a characteristic length-scale of the system based on large-scale structures. It was initially introduced with model fits to the correlation function $f(r) \sim e^{-r/L}$, see section \ref{subsec:lengthscales}. It is defined in Fourier space as
 \begin{equation}
 L(t) = \frac{3\pi}{4E(t)} \int dk\ \frac{E(k)}{k} \qquad\textrm{or}\qquad \frac{3\pi}{4E(t)} \Delta k \sum_n \frac{E(n)}{n} \frac{L}{2\pi} \ .
\end{equation}
Note that $E(n)$ has dimension of length $\times$ energy.
}
 \item [Taylor micro-scale]{Another characteristic length-scale, this time characterising the small-scale structures of the system. See section \ref{subsec:lengthscales} for more information. It is found as
 \begin{equation}
 \lambda(t) = \left(\frac{10 \nu_0 E(t)}{\varepsilon(t)}\right)^{1/2} = \left(\frac{15 \nu_0 u^2(t)}{\varepsilon(t)}\right)^{1/2} \ .
\end{equation}
}
 \item [Reynolds numbers]{An important dimensionless quantity for classifying a turbulent flow is the Reynolds number (see section \ref{subsec:1:reynolds_number}). It is defined as
 \begin{equation}
 Re = \frac{Ul}{\nu_0} \ ,
\end{equation}
where $U$ and $l$ are some characteristic (possibly time-dependent) velocity- and length-scales and $\nu_0$ is the kinematic viscosity. For example, we have the integral Reynolds number $R_L = uL/\nu_0$ and the Taylor-Reynolds number $R_\lambda = u\lambda/\nu_0$. }
 \item [Kolmogorov scale]{The Kolmogorov length-scale gives the approximate scale at which viscous effects become important and is given by
 \begin{equation}
  \eta(t) = \left(\frac{\nu_0^3}{\varepsilon(t)}\right)^{1/4} \ .
\end{equation}
In a simulation, it is vital that all scales down to the Kolmogorov scale are resolved. In Fourier space, we require that modes up to $k_d = 1/\eta$ be included. In reality, this should not be taken as a guarantee of a fully resolved simulation, as $\eta k_d > 1$ is often required.
}
 \item [Longitudinal velocity derivative skewness]{Also referred to as simply the skewness, the longitudinal velocity derivative skewness is one of the most sensitive parameters in quantifying turbulence. In real space, it is defined as
\begin{equation}
 S(t) = \frac{\left\langle (\partial_{1} u_1(\vec{x},t))^3\right\rangle}{\left\langle (\partial_{1} u_1(\vec{x},t))^2 \right\rangle^{3/2}} \ ,
\end{equation}
where $\partial_1 = \partial/\partial x_1$, or in Fourier space as
\begin{equation}
 S(t) = \frac{2}{35} \left( \frac{\lambda(t)}{u(t)} \right)^3 \int dk\ k^2 T(k,t) \ .
\end{equation}
It should be noted that pseudospectral methods have access to both of these methods, and there is often a discrepancy between what should be equivalent results. 
}
 \item[Structure functions]{Structure functions are found in configuration space by considering the correlations of the difference between two points. The $n^{th}$-order longitudinal structure function was introduced in equation \eqref{eq:sf_def} and can be defined as
\begin{equation}
 S_n(r) = \left\langle \Big[ \delta \vec{u}(\vec{r}) \cdot \hat{\vec{r}}\Big]^n \right\rangle = \left\langle \Big[\big(\vec{u}(\vec{x}+\vec{r},t)-\vec{u}(\vec{x},t)\big)\cdot \hat{\vec{r}}\Big]^n \right\rangle \ .
\end{equation}
}

 \item[Dissipative wavenumber]{In section \ref{subsec:kol_range} we introduced the dissipation wavenumber $k_d$ as the reciprocal of the Kolmogorov microscale. To quantify how well resolved a computation is, we consider the lowest wavenumber $k_{\textrm{diss}}$ such that
 \begin{equation}
  \int_0^{k_{\textrm{diss}}} dk\ 2\nu_0 k^2\ E(k,t) \geq 0.995\varepsilon \ .
 \end{equation}
 That is, the wavenumber up to which 99.5\% of the dissipation is accounted for. This should satisfy $k_{\textrm{diss}} < \kmax$ for the simulation to be well resolved.}
\end{description}

\newpage

\section{Parallel computation}
Parallelisation is not necessarily performed to increase performance: Instead, it could be simply impossible to store the data required in the memory of a single machine. This is indeed the case here. The memory required to store a 3-dimensional vector field on a lattice of size $N^3$ is $N^3 \cdot 24$ bytes, where 24 bytes corresponds to the memory needed for each grid site using double precision. For $N = 1024$, this equates to 24GB. One typically needs (at least) three vector fields to be stored for the duration of the simulation, which takes us up to 72GB; something that is not currently available on a single machine. The computing facilities available can support 2GB per process, so this requires at least 36 processes to run. In practice, more memory is needed for storing other variables and spectra.

Due to the convolution in Fourier space required to calculate the non-linear interactions, which requires access to all wave-vectors, it is not obvious that the Navier-Stokes equation can be evaluated in parallel. However, as noted above, the convolution sum becomes a local product in configuration space, which does not require knowledge of other grid points to be computed. As there exist a number of highly-optimised routines for performing Fast Fourier Transforms of parallel data, this problem can be efficiently written to run on a number of processes at once, spreading the workload.

The \dns\ code was extended using the \emph{OpenMPI} standard to run on (the surprisingly aptly named) \eddie, a large Linux-based cluster at the \emph{Edinburgh Compute and Data Facility}. Parallel jobs can be submitted to a queue and run on a large number of nodes. Since each process does not have access to all of the data, only the chunk that it is working on, several things need to be borne in mind when adapting serial routines to take advantage of this parallel capability. These are discussed below.

\subsection{Data decomposition}
A three-dimensional Fourier transform can be performed as 3 independent one-dimensional transforms, one after another. The FFT routine needs to have access to all the data in one dimension in order to do the transform. Our first task is, therefore, to decide how the data should be split between the processes, as this will determine the number of processes that can be used to study a certain size lattice. It should be noted that the lattice size $N$ should be divisible by the number of processes, $N_p$, such that $N/N_p \in \mathbb{N}^*$.

\begin{figure}[tb!]
 \centering
 \subfigure[1-dimensional]{
  \label{sfig:parallel_decomp_slabs}
  \includegraphics[width=0.3\textwidth]{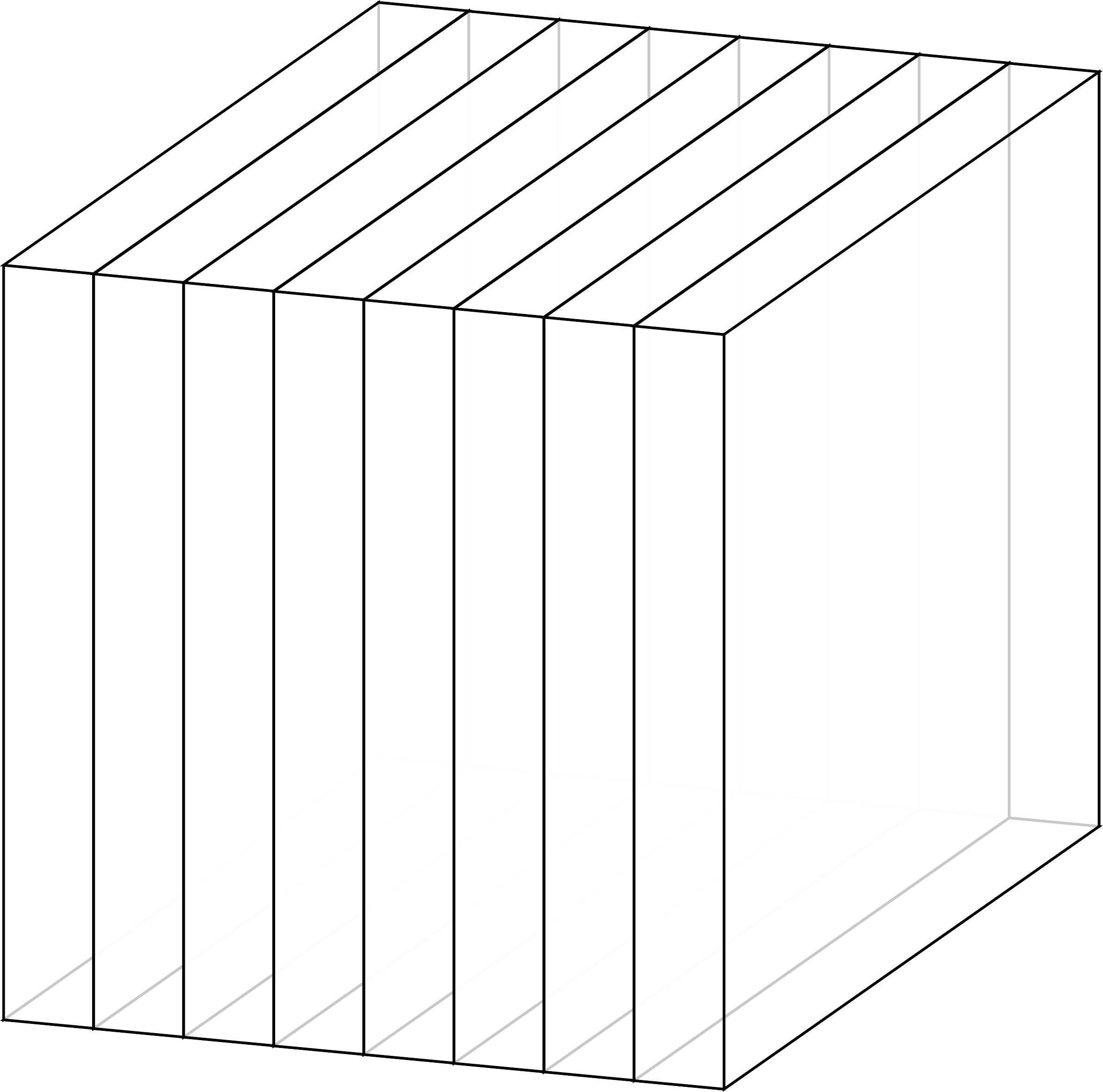}}
  \hspace{0.75in}
 \subfigure[2-dimensional]{
  \label{sfig:parallel_decomp_pencils}
  \includegraphics[width=0.3\textwidth]{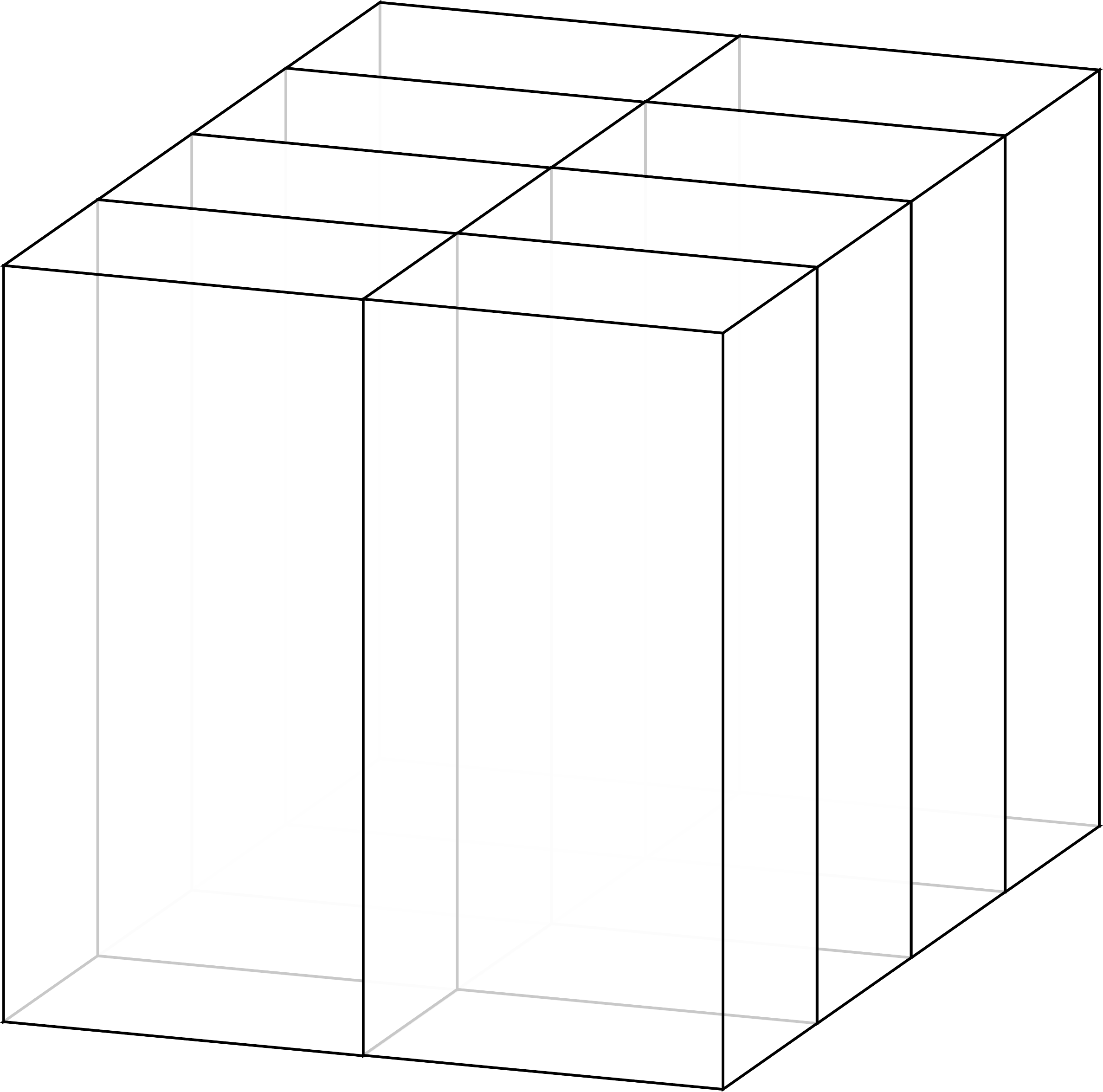}}
 \caption{Parallel decomposition of the domain onto 8 processes.}
 \label{fig:parallel_decomp}
\end{figure}

\begin{description}
 \item[1-dimensional decomposition]{As shown in figure \ref{sfig:parallel_decomp_slabs}, this decomposition splits the data in one direction amongst the processes, resulting in a series of \emph{slabs}. Each process thus stores all of the data for two directions locally, and is therefore capable of performing two of the one-dimensional transforms. To obtain the third direction in local memory, a transposition of the data is required. The data is then in the `wrong' order, and must be transposed back. Since we only split one direction, for a lattice of size $N^3$ each process stores an $N/N_p \times N \times N$ slab, with the maximum number of processes that can be used being $N_p = N$.}
 \item[2-dimensional decomposition]{Shown in figure \ref{sfig:parallel_decomp_pencils}, this decomposition splits the data in two directions, so each process stores a \emph{pencil} of the domain with one direction entirely stored locally. In order to perform the three transforms, we must use two transpositions: for example, if the data is stored with the $z$-direction contiguous, we perform one transposition so that the $y$-direction is local, then another so that $x$ is stored locally. A third transposition takes us back to having the $z$-direction stored locally. Each process locally stores an $N/N_{p_1} \times N/N_{p_2} \times N$ block, with a maximum number of processes $N_p = N^2$.}
\end{description}

The benefit of pencil decomposition is that one can use more processors, at most $N^2$ compared to $N$ for slab decomposition. The drawback is the extra data transpositions. This step requires communication between all the processes, as they each swap their data, and is a very costly procedure. Provided that the lattice is large enough that one side can be split over the number of available processes, slab decomposition prevents this extra workload.

\begin{figure}[tb!]
 \centering
 \subfigure[Configuration space]{
  \includegraphics[width=0.375\textwidth]{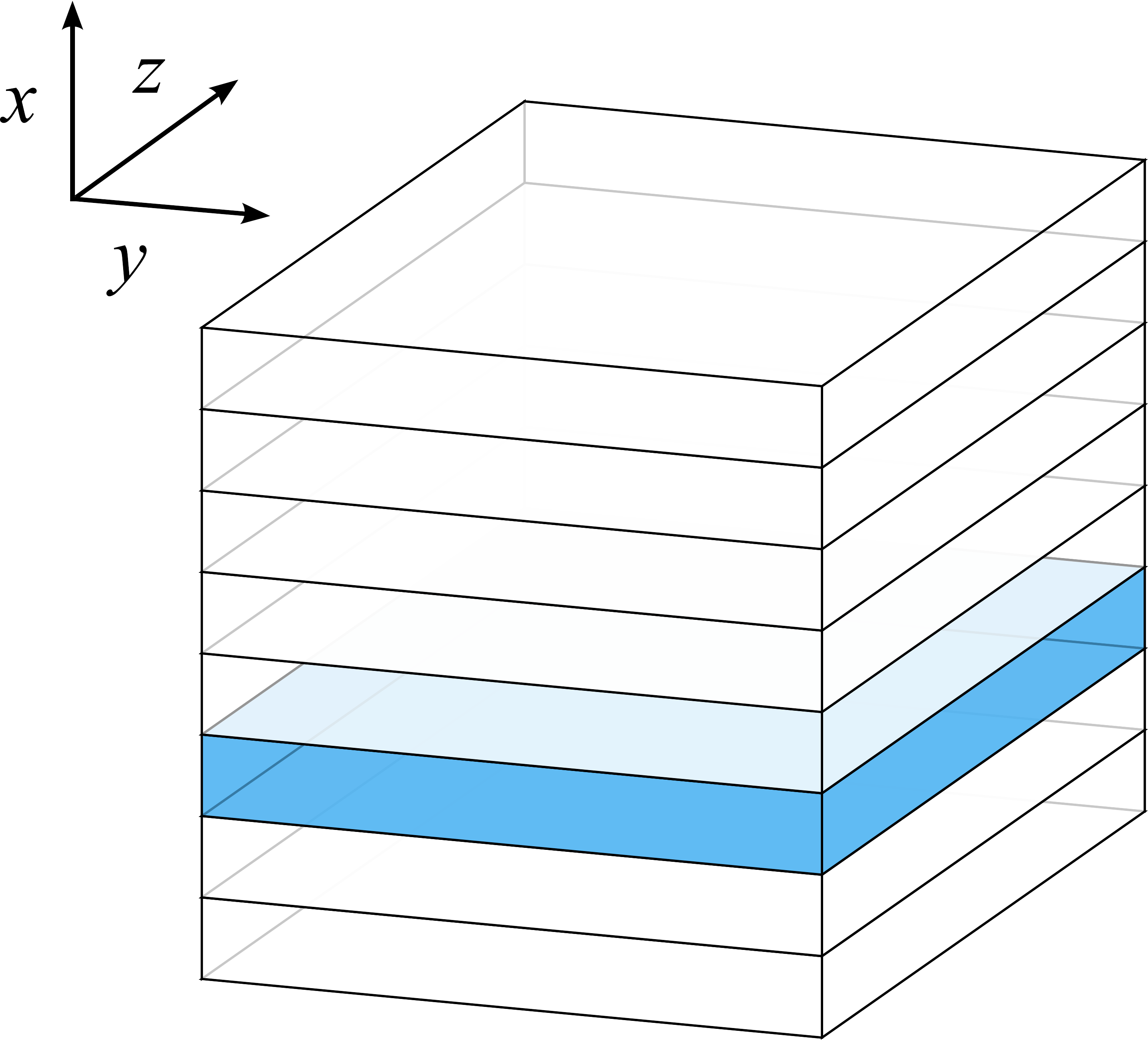}}
  \hspace{0.5in}
 \subfigure[Fourier space]{
  \includegraphics[width=0.375\textwidth]{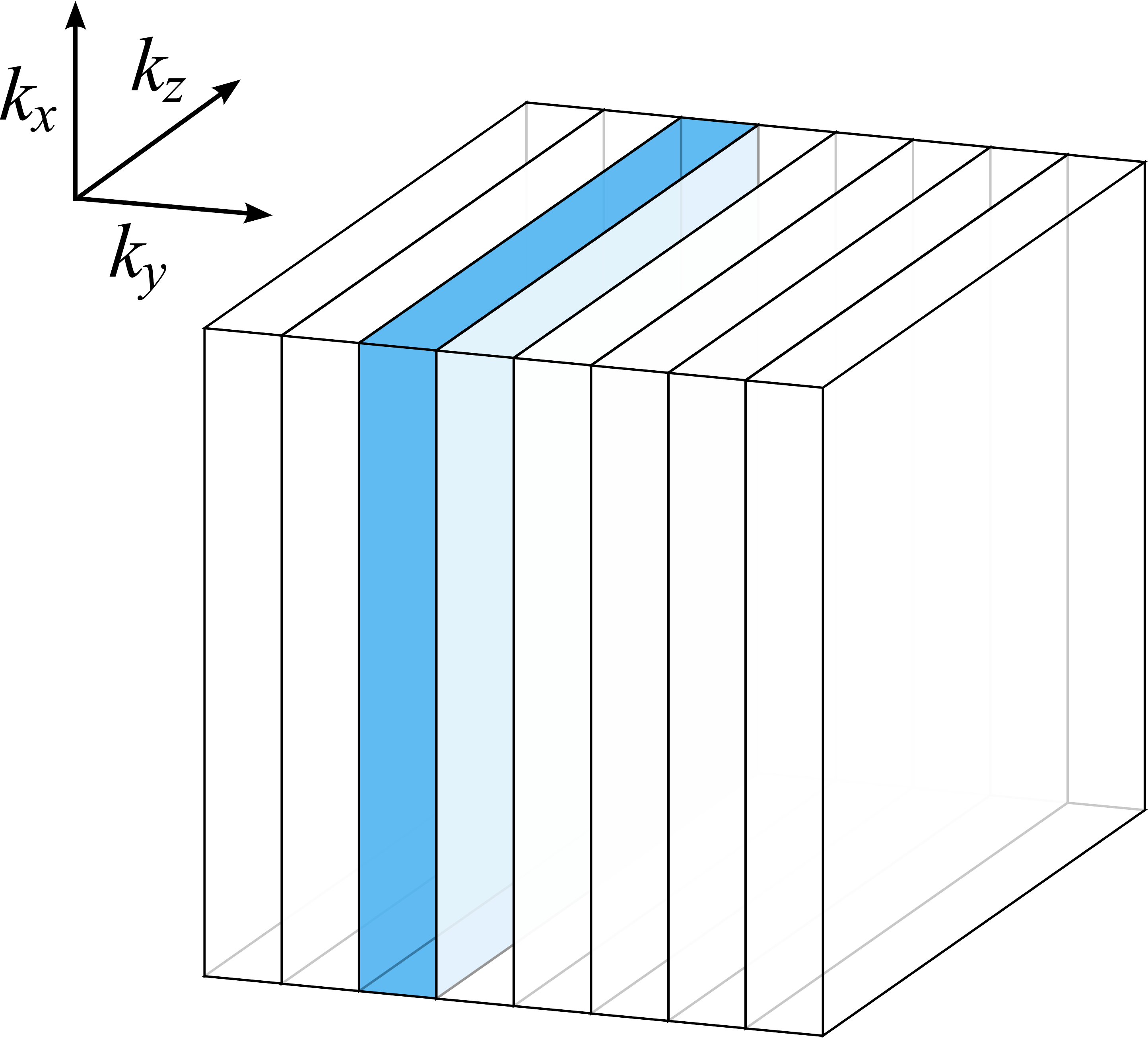}}
 \caption{Data layout in configuration and Fourier space, highlighted for one process. In real space, the data is split in the $x$-direction over the number of processes. The FFT to $k$-space must transpose the data, so it is then stored with the split in the $k_y$-direction, preventing an additional transposition step.}
 \label{fig:parallel_transpose}
\end{figure}

Since we are looking at $N \sim \ord{1000}$ and $N_p \sim \ord{100}$, \dns\ uses one-dimensional slab decomposition of the $x$-direction, but in fact goes one step further: The final transposition to restore the data order is ignored, and compensated for in the code. In configuration space, process $p = 0, \cdots, N_p-1$ locally stores a $N/N_p \times N \times (N+2)$ real block of the domain,
\begin{equation}
 p\frac{N}{N_p} \leq m_x^{(p)} < (p+1)\frac{N}{N_p} \ ,\qquad m_y^{(p)}, m_z^{(p)} = 0, \cdots, N-1 \ .
\end{equation}
After performing one transposition (one is unavoidable) the data stored locally is the complex block $N \times N/N_p \times (N/2+1)$, with the data  split in the $k_y$-direction,
\begin{equation}
 \qquad n_{k_x}^{(p)} = -\frac{N}{2} \cdots, \frac{N}{2}-1\ , \qquad n_{\textrm{min}}^{(p)} \leq n_{k_y}^{(p)} < n_{\textrm{min}}^{(p)} + \frac{N}{N_p} \ ,\qquad n_{k_z}^{(p)} = 0, \cdots, \frac{N}{2} \ ,
\end{equation}
where the minimum wavenumber
\begin{equation}
 n_{\textrm{min}}^{(p)} = \left\{ \begin{array}{ll}
                                 p N/N_p & \textrm{for}\quad 0 \leq p < N_p/2 \\
				 & \\
				 p N/N_p - N & \textrm{for}\quad N_p/2 \leq p < N_p
                                \end{array}
			\right.\ .
\end{equation}
This is illustrated in figure \ref{fig:parallel_transpose}.

Most other operations are local in Fourier- or configuration space. However, due to the distribution of the data, several minor modifications need to be made to the code, including the calculation of spectra. Each process calculates its contribution to, say, $E(k,t)$ for all values of $k$ from the modes it stores. The contributions are then collected together and summed. The forcing procedure also needs attention, since each process stores a list of the modes it has which need to receive energy input, and the method employed here requires knowledge of the energy contained in all modes which lie in the forcing band, see section \ref{subsec:forcing}.

\subsection{Checkpointing}
While \eddie\ is an extremely useful resource, it has an upper limit on the amount of time a single job can be submitted to run for of 48 hours. This may sound like a long time, but the large lattices considered in this project need considerably longer than this to reach steady state and obtain statistics. For this reason, checkpointing was implemented to the code where, at a user defined interval, the entire velocity field is saved to disk. For $N = 1024$, this requires 56GB storage per realisation. Routines for loading from a checkpoint were also written. Each process saves and loads from its own checkpoint files, but input data for the simulation is read only by the master process and shared with all others. This prevents different processes from accidentally running with different parameters, in the case that a checkpoint file is modified. The program also includes a number of commandline options to override information in the checkpoint file, so that, for example, a decaying simulation can easily be run from and evolved, stationary field.

The first generation of this code simply saved the field, but was later updated to export in the \texttt{VTK Rectilinear file} format, as this can be directly opened by a number of freely-available visualisation programs, such as \emph{Paraview}\footnote{Paraview is available from \url{http://www.paraview.org/}}.

\section{Code improvements}
While performing well in our validation experiments (chapter \ref{sec:benchmark}), the code still offers numerous areas for development and improvement.

\begin{itemize}
 \item \textbf{Field interpolation:} Initial fields for large lattice sizes can be created from evolved smaller lattice simulations using interpolation. The lattice sites of the higher resolution simulation which lie between those of the coarser grid are approximated in some way from the values at the known sites. By using this interpolated configuration as the initial condition, it is hoped that convergence to fully developed turbulence is improved compared to a random initial field.
 \begin{figure}[htb]
  \centering
  \vspace{1em}
  \includegraphics[width=0.7\textwidth]{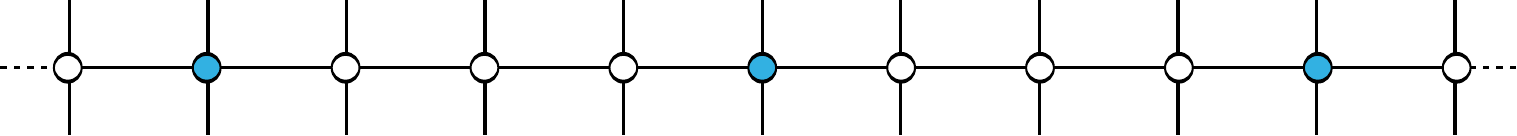}
  \caption{Interpolation of a lower resolution field, shown here for grid ratio $N_{\textrm{new}}/N_{\textrm{old}} = 4$. Filled sites correspond to both lattices, open sites must be interpolated.}
  \label{fig:interp}
 \end{figure}

 \item \textbf{PDF sampling:} Instead of being drawn from a Gaussian probability distribution, the initial field could be filled with random numbers satisfying a different distribution. Indeed, the probability distribution for $\vec{u}$ could be measured for an evolved field and used to generate a new realisation. Once again, convergence could be improved compared to a Gaussian initial field.

 \item \textbf{Time integration:} The current code uses a two-step predictor-corrector algorithm to evaluate the time integration. It would be interesting to implement a higher-order numerical integration scheme, such as a fourth-order \emph{Runge-Kutta} (RK4), as this would reduce the error at each step and allow for larger step sizes to be taken, possibly compensating for the additional computational load. Another improvement would be to introduce an adaptive step-size, where the error is monitored and the step-size altered accordingly. This would allow for large steps to be taken when they can, and the step-size reduced when necessary.

 \item \textbf{Forcing scheme:} The effect of different forcing schemes on the statistics of stationary turbulence could be investigated by implementing alternative forcing methods. These include the additional schemes mentioned in section \ref{subsec:forcing}.

 \item \textbf{Checkpointing/restart of passive scalar:} The advection of a passive scalar was implemented in section \ref{sec:scalar}. This would allow for simulation of a scalar to be restarted and longer times investigated.

 \item \textbf{Statistics for passive scalar:} Statistics for the scalar field, such as `energy' and transfer spectra and the quantities derived from them, could be included allowing for quantitative investigation of the properties of the field to be made.

 \item \textbf{Active scalar:} In the advection of a passive scalar, the scalar quantity has no effect on the underlying turbulent velocity field. The equations can be modified so that, instead, the scalar does directly influence the field, leading to very different behaviour.

 \item \textbf{Lagrangian tracers:} These passive particles can be added to the velocity field and used to trace the movement of particles in the fluid. Indeed, they can be made inertial and their size can be modified. Buoyancy can also be controlled, although we cannot impose gravity without choosing a direction and breaking isotropy.

 \item \textbf{Magnetohydrodynamics:} The velocity field can be coupled to a magnetic field and used to study the equations of MHD. This is non-trivial code development, but would be of interest for many reasons such as the simulation of plasmas or even large-scale magnetic fields in the universe.
\end{itemize}

\chapter{Verification of the \dns\ code}\label{sec:benchmark}

In an attempt to verify that the code is behaving as expected, a number of benchmarking simulations have been run for comparison to previous results. These are detailed below.

We start by considering the stability of the time integration along with the energy conservation of the non-linear term. A comparison of results for decaying turbulence to those obtained by Quinn \cite{thesis:apquinn} is then presented for a selection of Reynolds numbers. To further show that the code is behaving as expected, identical initial conditions were run using our DNS and the freely-available \emph{hit3d} code for both decaying and forced turbulence. The Taylor-Green vortex is then considered along with a fit to the energy spectrum. After this, we consider the isotropy of the system and various time-averaged quantities for stationary turbulence and draw comparison to the literature. Finally, advection of a passive scalar was implemented and a simple test performed. The chapter ends with a summary of our findings.

\section{Time-step and energy conservation}\label{sec:bench_TS}

\begin{figure}[tb!]
 \subfigure[Stability of total energy with time-step]{
  \centering
  \includegraphics[width=0.475\textwidth,trim=3px 0 3px 0,clip]{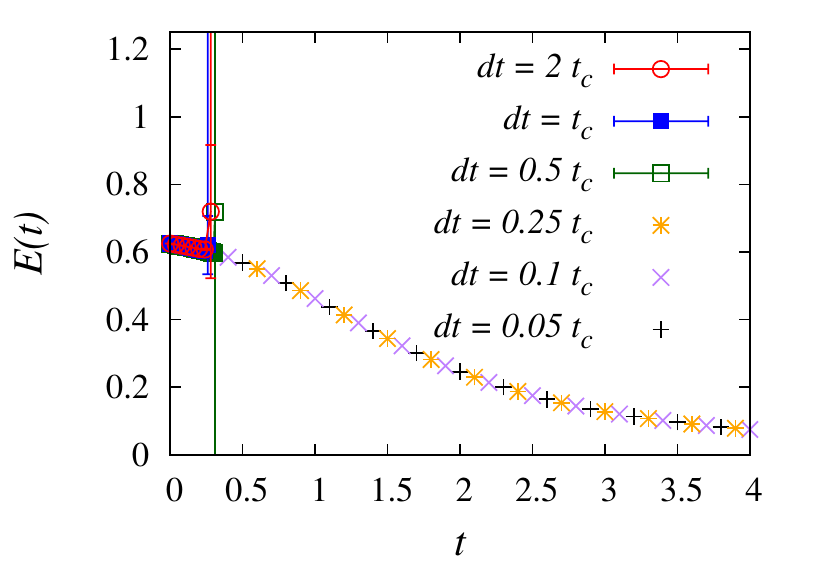}
  \label{fig:time_step}
 }
 \subfigure[Energy conserving nature of $T(k,t)$]{
  \centering
  \includegraphics[width=0.475\textwidth,trim=3px 0 3px 0,clip]{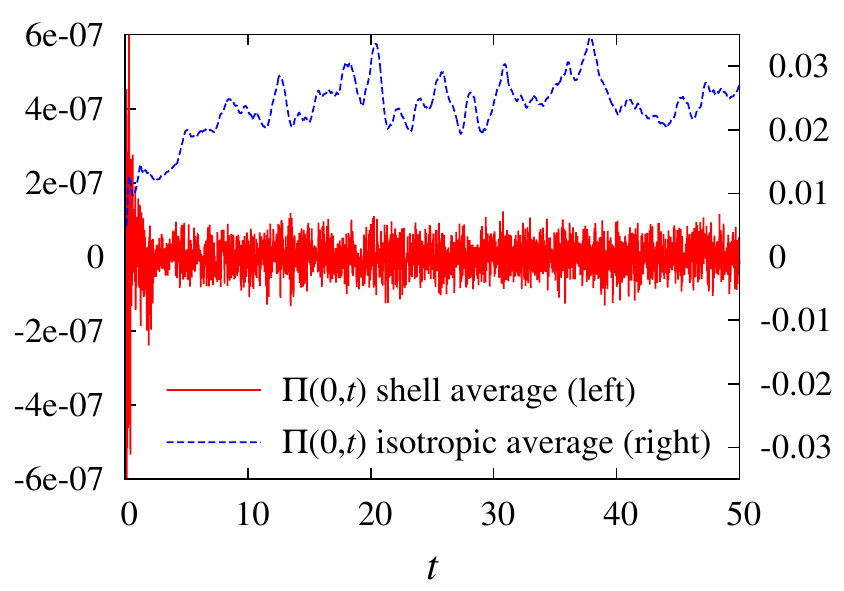}
  \label{fig:pi0}
 }
 \caption{Basic numerical verification of integration time-step and energy conservation.}
\end{figure}

The stability of any numerical integrator is sensitive to the size of the step taken in each iteration. Too large, and the results may be spurious or, worse, look fine but be incorrect. In turbulence, there are numerous time-scales one could choose as a way of determining what constitutes too large a step. However, essentially the most important ones are the convection and dissipation scales. These are defined as
\begin{equation}
 t_c = \frac{1}{u\ \kmax} \qquad\qquad\textrm{and}\qquad\qquad t_\nu =  \frac{1}{\nu_0 \kmax^2}\ ,
\end{equation}
respectively, where $\kmax$ is the highest wavenumber simulated on the lattice.
As a criterion, we therefore ensure that the time-step we take be smaller than both of these scales, $dt < \min(t_c,t_\nu)$. Variation of the total energy using different time-steps is shown in figure \ref{fig:time_step} for an ensemble of decaying simulations with $\nu_0 = 0.002$ on a $256^3$ lattice, such that $t_c < t_\nu$. We use $u(0)$ to define $t_c$, and find that provided $dt < 0.25 t_c(0)$ the integration remains stable. We will typically use $dt \leq 0.1 t_c$.

It is also important that we check that the non-linear term is conserving energy. To do this, we look at the integral over the transfer spectrum, since it was shown in equation \eqref{eq:Tk_vanish} that this must vanish. This is the same as looking at
\vspace{-0.75em}
\begin{equation}
 \Pi(0,t) = \int_0^\infty dk\ T(k,t) = \Delta k \sum_{k = k_{\textrm{min}}}^{\kmax} T(k,t) \ .
\end{equation}
This has been done for all simulations, and the time series for the $R_\lambda \sim 280$ simulation is shown in figure \ref{fig:pi0}. The figure also shows how the isotropic shell average mentioned in section \ref{subsubsec:shell_average} does not preserve this condition. Time averaging over the (stationary part of the) time series --- see section \ref{sec:time_av} --- we find
\begin{equation}
 \left\langle \Pi(0) \right\rangle_{\textrm{shell}} = -5.7 \times 10^{-9} \qquad\qquad\textrm{and}\qquad\qquad \left\langle \Pi(0) \right\rangle_{\textrm{iso}} = 0.026 \ .
\end{equation}
For the shell average, this is basically zero, but the isotropic average is clearly non-zero.

\section{Decaying turbulence}
Whilst the thesis of A. P. Quinn \cite{thesis:apquinn} is benchmarking a numerical computation of the LET theory of turbulence, it provides DNS results which we can compare against. Simulations were run for the same conditions (lattice size, viscosity) and are discussed below. The comparisons in this section start with initial spectrum S5, see equation \eqref{eq:chp3:standard_spec} and table \ref{tbl:init_spectra}. It should be noted that no de-aliasing is performed (by neither us nor Quinn) and only simple isotropic truncation is used. This method of truncation sets to zero any wavenumber that does not fit within the sphere of radius $\kcut$.

An additional important point which should be borne in mind is the method of initial velocity field generation. The previous generation code constructs its initial field following the method of Orszag \cite{Orszag:1969p250}, which introduces slight variation among the initial energy spectrum of each realisation. More information can be found in section \ref{subsec:initial_field}. As such, there are no error bars plotted for our initial energy spectra.

Error bars plotted in this section are purely statistical, calculated as the standard deviation from the mean of the ensemble.

\subsection{$R_\lambda(0) \simeq 3$ decaying turbulence}

\begin{figure}[tbp!]
 \centering
 \subfigure[Scaled energy and dissipation rate]{\label{sfig:64_0.1_energy_diss}
  \includegraphics[width=0.485\textwidth,trim=2px 0 10px 0,clip]{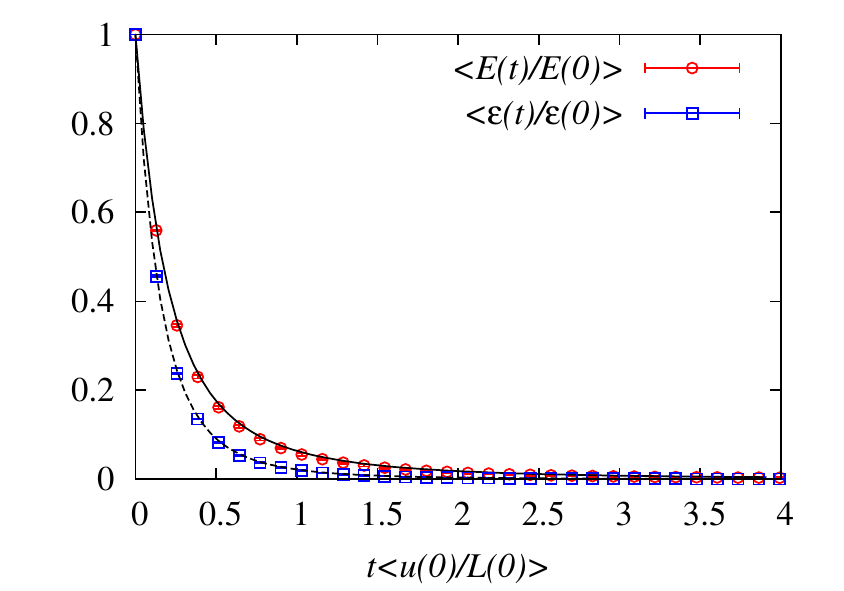}}
 \subfigure[Scaled integral and Taylor length-scales]{\label{sfig:64_0.1_length}
  \includegraphics[width=0.485\textwidth,trim=2px 0 10px 0,clip]{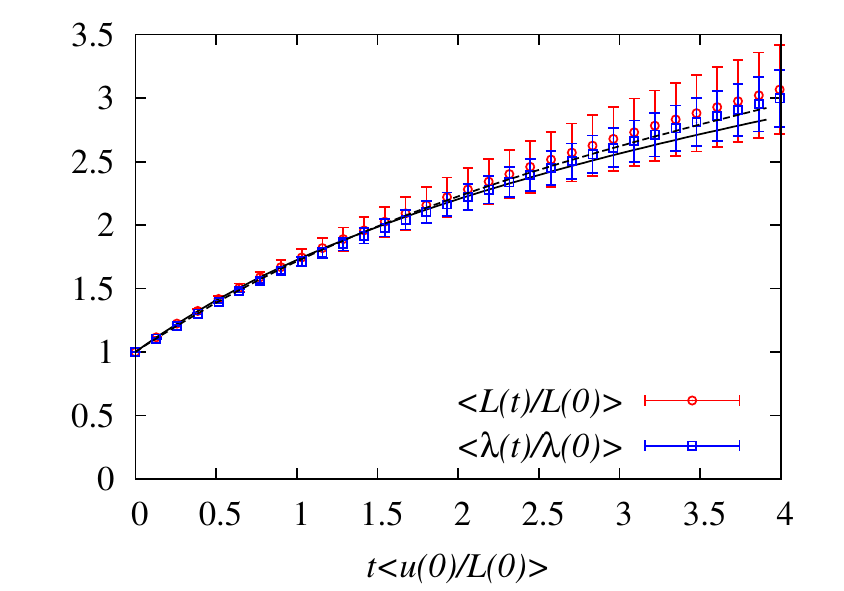}}
 \subfigure[Reynolds numbers]{\label{sfig:64_0.1_reynolds}
  \includegraphics[width=0.485\textwidth,trim=2px 0 10px 0,clip]{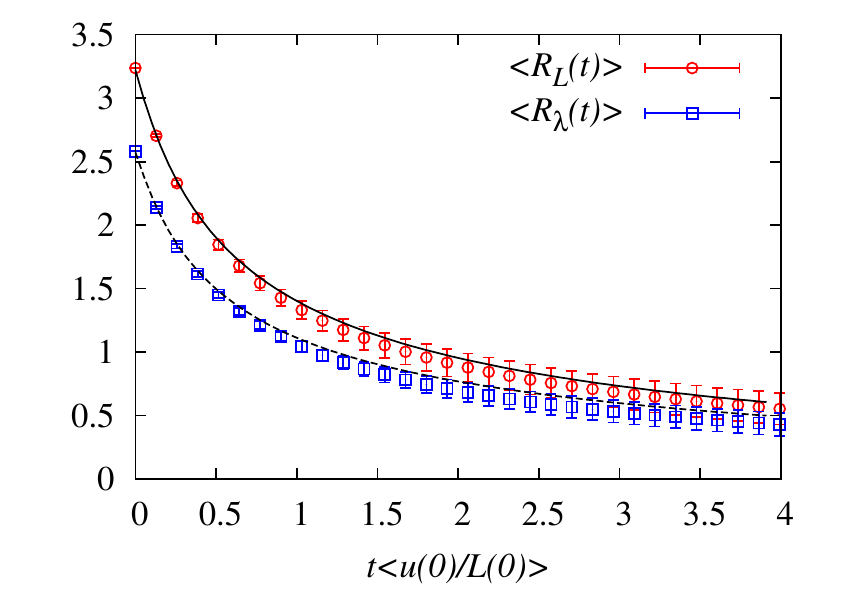}}
 \subfigure[Velocity derivative skewness]{\label{sfig:64_0.1_skewness}
  \includegraphics[width=0.485\textwidth,trim=2px 0 10px 0,clip]{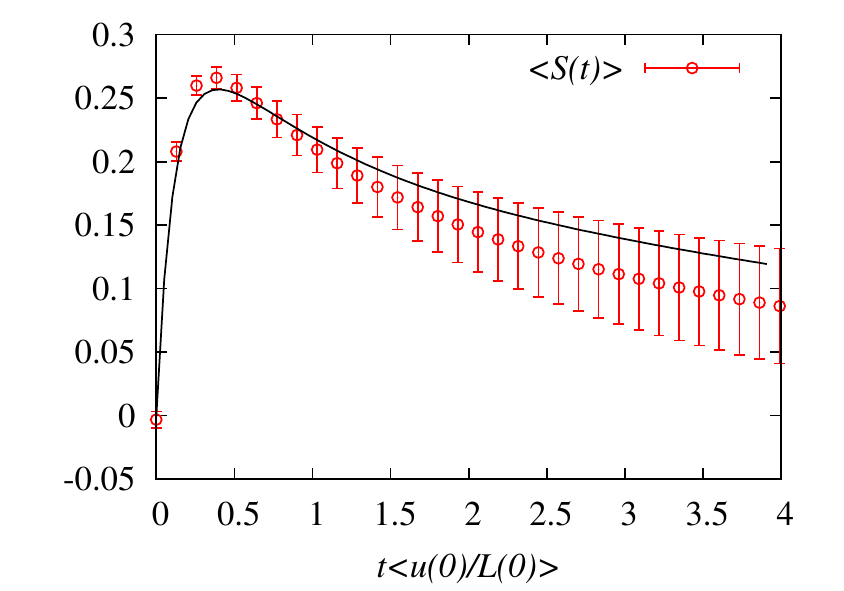}}
 \subfigure[Scaled dissipation spectra]{\label{sfig:64_0.1_D}
  \includegraphics[width=0.485\textwidth,trim=2px 0 10px 0,clip]{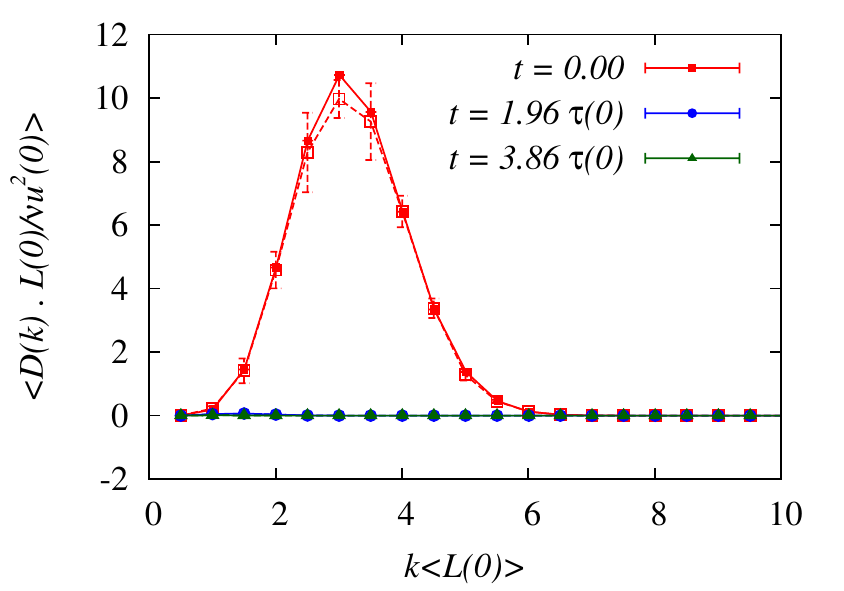}}
 \subfigure[Scaled transfer spectra]{\label{sfig:64_0.1_T}
  \includegraphics[width=0.485\textwidth,trim=2px 0 10px 0,clip]{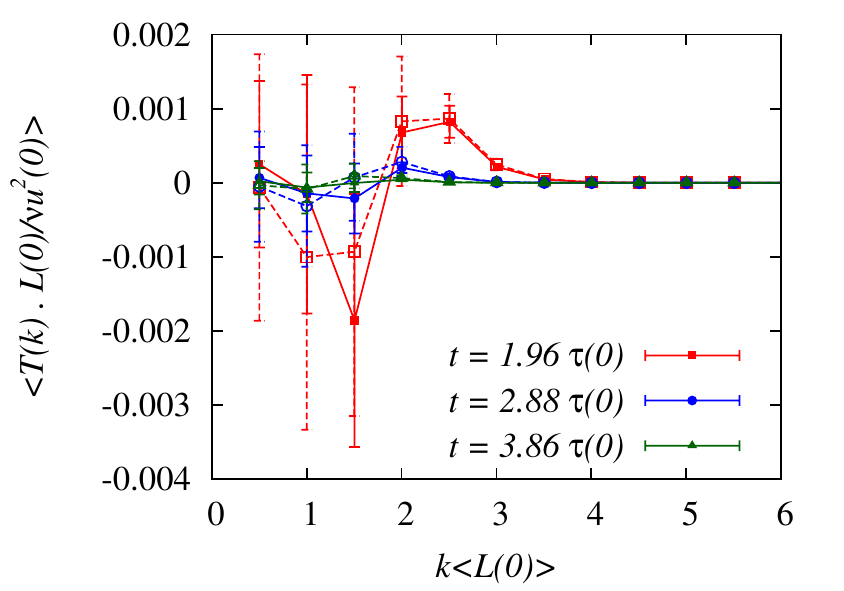}}
 \caption{Parameters and spectra for $R_\lambda(0) \simeq 3$. For parts (a)--(d): {\Large\color{red}$\circ$} and {\small\color{blue}$\square$} mark our DNS data. For comparison, we provide (------) Quinn's result for quantities plotted with {\Large\color{red}$\circ$}; (--~--~--) Quinn's result for quantities plotted with {\small\color{blue}$\square$}. For parts (e) and (f): (-----{\small$\blacksquare$}-----) Filled points, solid line show our DNS data; (-- --{\small$\square$}-- --) open points, dashed line show Quinn's results for the same times.}
 \label{fig:64_0.1_compare}
\end{figure}

This comparison used an ensemble of 10 realisations of a $64^3$ velocity field with $\nu_0 =  0.1$, giving an initial Taylor-Reynolds number $R_\lambda(0) = 2.55$. See section 6.5 of Quinn's thesis. All error bars represent three standard deviations, with the exception of figure \ref{sfig:64_0.1_skewness} where only one standard deviation is plotted, following Quinn.

Figures \ref{sfig:64_0.1_energy_diss}--\subref{sfig:64_0.1_skewness} show the time decay of total energy, dissipation rate, integral and Taylor length-scales, Reynolds numbers and skewness as functions of (scaled) time. Plotted with this are the data obtained by Quinn for comparison, with good agreement. As can be seen, the total energy and dissipation rate simply decay from unity. The error bounds are very tight, also a feature of Quinn. The integral and Taylor length-scales both vary from unity at $t = 0$ to just below 3 at $t = 4 \tau(0)$, with the error on the integral scale being slightly larger than the Taylor microscale, in agreement with Quinn. The Reynolds number based on the integral scale, $R_L$, drops from $\sim 3.25$ at $t = 0$ to just above 0.5 at $t = 4\tau(0)$, while $R_\lambda$ (based on the Taylor scale) drops from $\sim 2.5$ to just below 0.5 at the same times. The velocity derivative skewness as calculated in Fourier space peaks around $t \sim 0.4\tau(0)$ at a value of $\sim 0.26$, before falling to $\sim 0.1$ at $t = 4\tau(0)$. Our simulation appears to reduce its skewness slightly quicker, though remains within one standard deviation. The cause of this discrepancy is unknown.

Turning our attention to the scaled spectra in figures \ref{sfig:64_0.1_D} and \subref{sfig:64_0.1_T}, we once again see very good agreement. We choose to compare dissipation and transfer spectra, since the dissipation spectrum is simply a scaling of the energy spectrum. The dissipation spectrum shows a peak of $\sim 11$ at around $k\langle L(0)\rangle \simeq 3$, with Quinn a little lower at around 10. This could be accounted for by the difference in initial condition generation, with Quinn not exactly recreating the desired spectrum. However, the value of 11 actually sits just outside the error bars plotted by Quinn. The transfer spectra show several small differences to Quinn's, despite our effort to plot similar times. At $t = 1.96 \tau(0)$, Quinn has the transfer spectrum varying from about -0.001 to just below 0.001, whereas we have variation from about -0.002 to just below 0.001. However, the extremely large error bars in both plots do not rule out agreement.

\subsection{$R_\lambda(0) \simeq 26$ decaying turbulence}

This comparison used an ensemble of 10 realisations of a $64^3$ velocity field with $\nu_0 =  0.01$, giving an initial Taylor-Reynolds number $R_\lambda(0) = 25.54$. See section 6.6 of Quinn's thesis. All error bars represent three standard deviations.

\begin{figure}[tbp!]
 \centering
 \subfigure[Scaled energy and dissipation rate.]{\label{sfig:64_0.01_energy_diss}
  \includegraphics[width=0.475\textwidth,trim=2px 0 10px 0,clip]{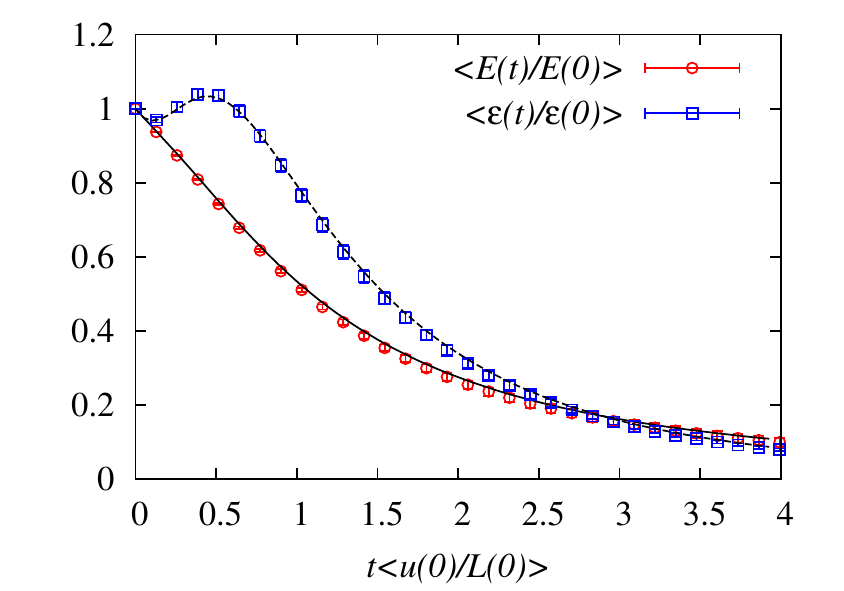}}
 \subfigure[Scaled integral and Taylor length-scales.]{\label{sfig:64_0.01_length}
  \includegraphics[width=0.475\textwidth,trim=2px 0 10px 0,clip]{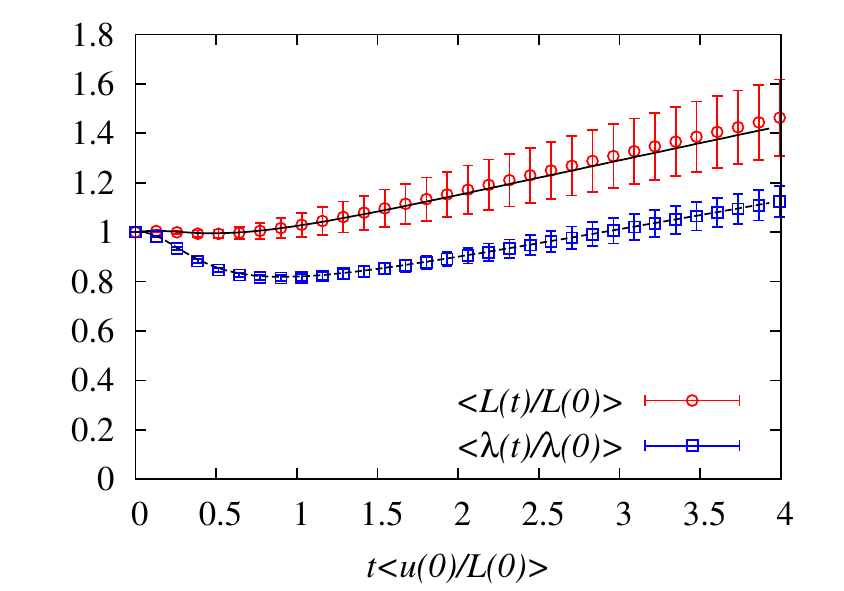}}
 \subfigure[Reynolds numbers.]{\label{sfig:64_0.01_reynolds}
  \includegraphics[width=0.475\textwidth,trim=2px 0 10px 0,clip]{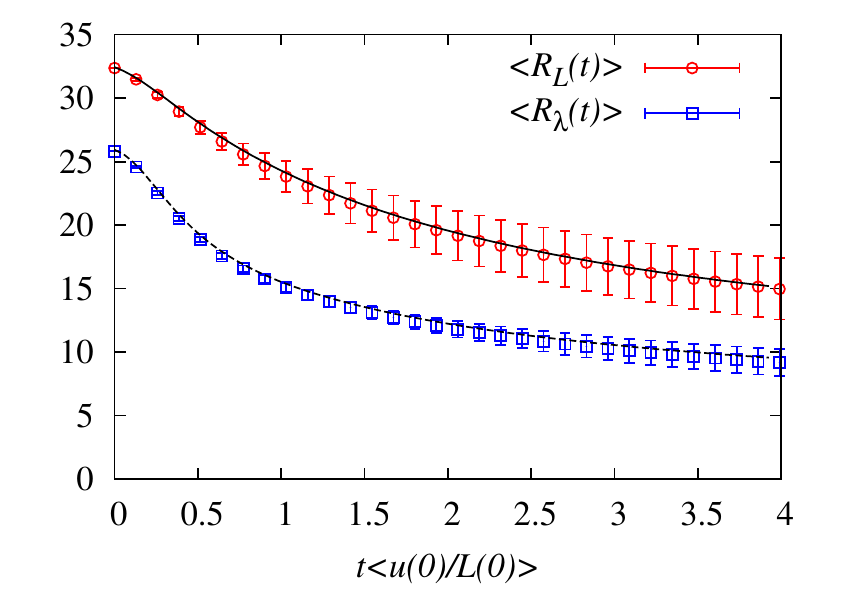}}
 \subfigure[Velocity derivative skewness.]{\label{sfig:64_0.01_skewness}
  \includegraphics[width=0.475\textwidth,trim=2px 0 10px 0,clip]{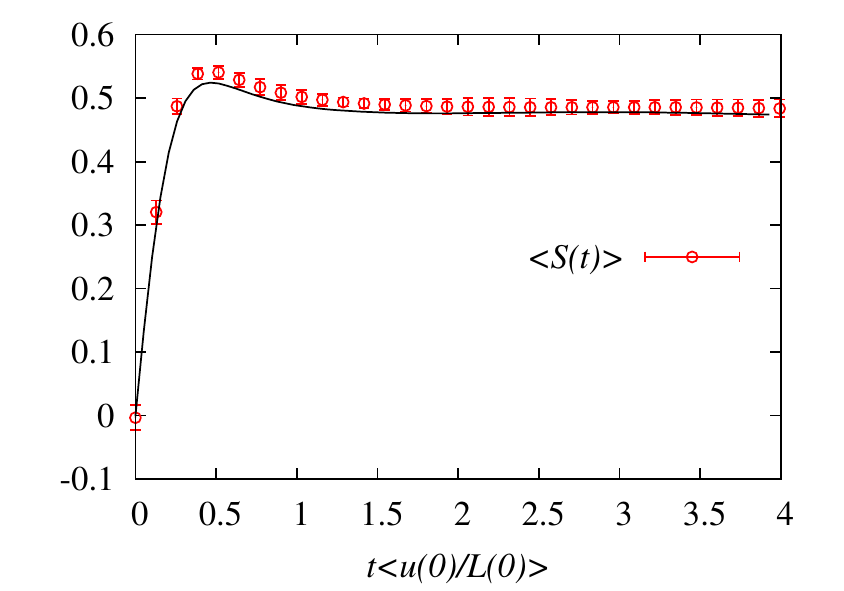}}
 \subfigure[Scaled dissipation spectra.]{\label{sfig:64_0.01_D}
  \includegraphics[width=0.475\textwidth,trim=2px 0 10px 0,clip]{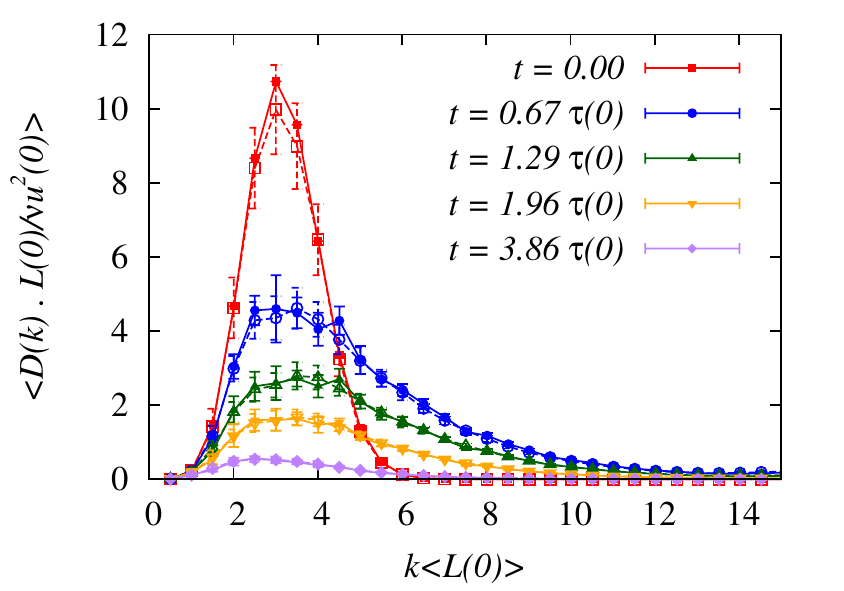}}
 \subfigure[Scaled transfer spectra.]{\label{sfig:64_0.01_T}
  \includegraphics[width=0.475\textwidth,trim=2px 0 10px 0,clip]{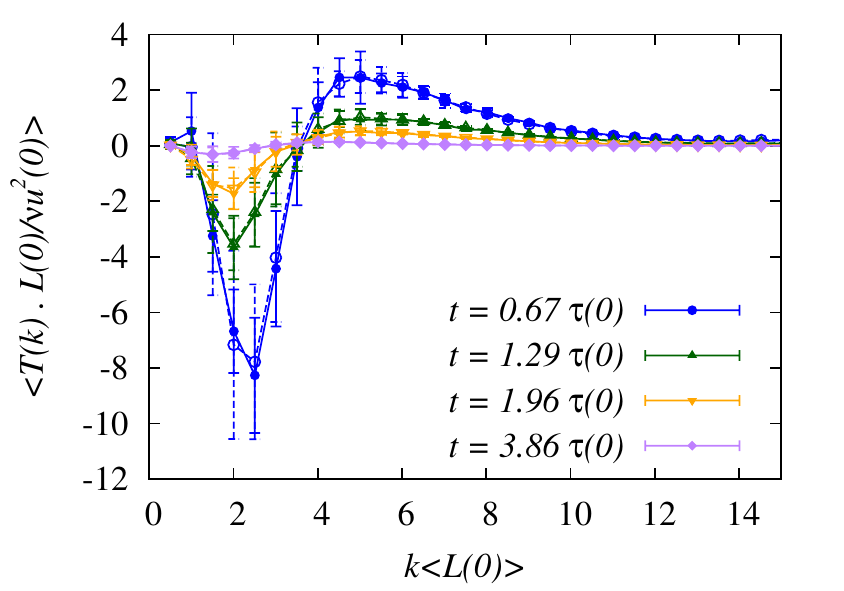}}
 \caption{Parameters and spectra for $R_\lambda(0) \simeq 26$. For parts (a)--(d): {\Large\color{red}$\circ$} and {\small\color{blue}$\square$} mark our DNS data. For comparison, we provide (------) Quinn's result for quantities plotted with {\Large\color{red}$\circ$}; (--~--~--) Quinn's result for quantities plotted with {\small\color{blue}$\square$}. For parts (e) and (f): (-----{\small$\blacksquare$}-----) Filled points, solid line show our DNS data; (-- --{\small$\square$}-- --) open points, dashed line show Quinn's results for the same times.}
 \label{fig:64_0.01_compare}
\end{figure}

Figures \ref{sfig:64_0.01_energy_diss}--\subref{sfig:64_0.01_skewness} show the time-variation of various parameters, along with the data obtained by Quinn for comparison. Once again, the agreement is seen to be very good. As with the lower Reynolds number comparison, total energy can be seen to simply decay, this time to around 0.1 at $t = 4\tau(0)$. Whereas, the dissipation rate initially decreases before increasing and peaking around $t = 0.5 \tau(0)$ at a value just above 1, before decaying to just below 0.1 at $t = 4\tau(0)$. Both the integral and Taylor length-scales initially decrease before increasing to around 1.4 and 1.1, respectively. Once again, error bounds are tighter for the Taylor microscale. The integral Reynolds number is seen to decrease from about 32 down to $\sim 15$, while $R_\lambda$ drops from $\sim 25$ to just below 10 in the same time. Error bars are tighter for the Taylor-Reynolds number, in agreement Quinn. Instead of increasing then decaying, velocity derivative skewness is seen to peak around 0.55 just before $t = 0.5 \tau(0)$, then develops a plateau at a value just below 0.5. In contrast, Quinn peaks around the same time but at a slightly lower value of 0.525, before settling slightly lower, with the latter scraping the lower error bound on our result. 
The spectra in figures \ref{sfig:64_0.01_D} and \subref{sfig:64_0.01_T} show excellent agreement for all times.

\vspace{1em}
\noindent Despite the difference in initial energy spectrum and a slightly lower Reynolds number, these results also compare favourably to those published in Herring, Riley, Patterson and Kraichnan \cite{Herring:1973p1105}, demonstrating familiar features to those seen above for the decay of energy and dissipation rate, as well as the dissipation and transfer spectra.

\subsection{$R_\lambda(0) \simeq 95$ decaying turbulence}

This comparison used an ensemble of 10 realisations of a $128^3$ velocity field with $\nu_0 =  0.0027$, giving an initial Taylor-Reynolds number $R_\lambda(0) = 94.59$. See section 6.7 of Quinn's thesis.

\begin{figure}[tbp!]
 \centering
 \subfigure[Scaled energy and dissipation rate]{\label{sfig:128_0.0027_energy_diss}
  \includegraphics[width=0.475\textwidth,trim=2px 0 10px 0,clip]{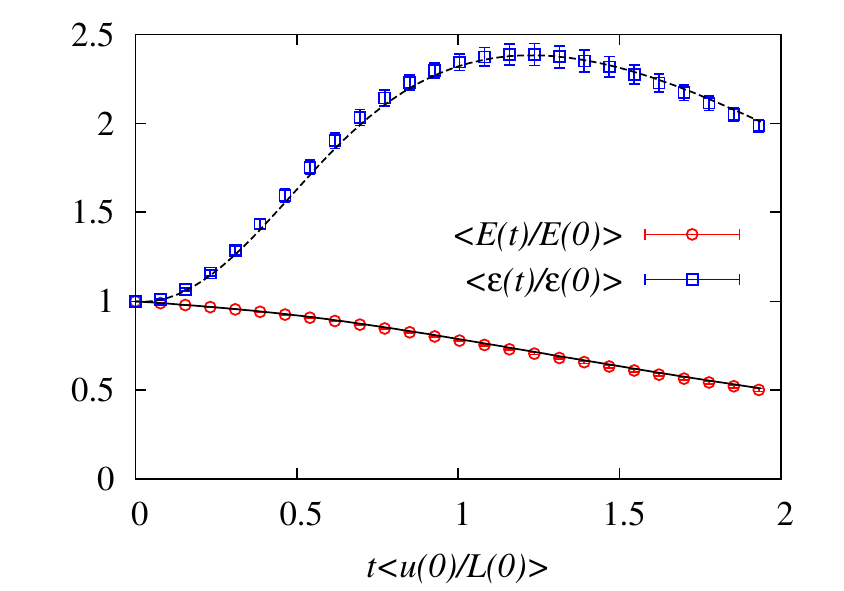}}
 \subfigure[Scaled integral and Taylor length-scales]{\label{sfig:128_0.0027_length}
  \includegraphics[width=0.475\textwidth,trim=2px 0 10px 0,clip]{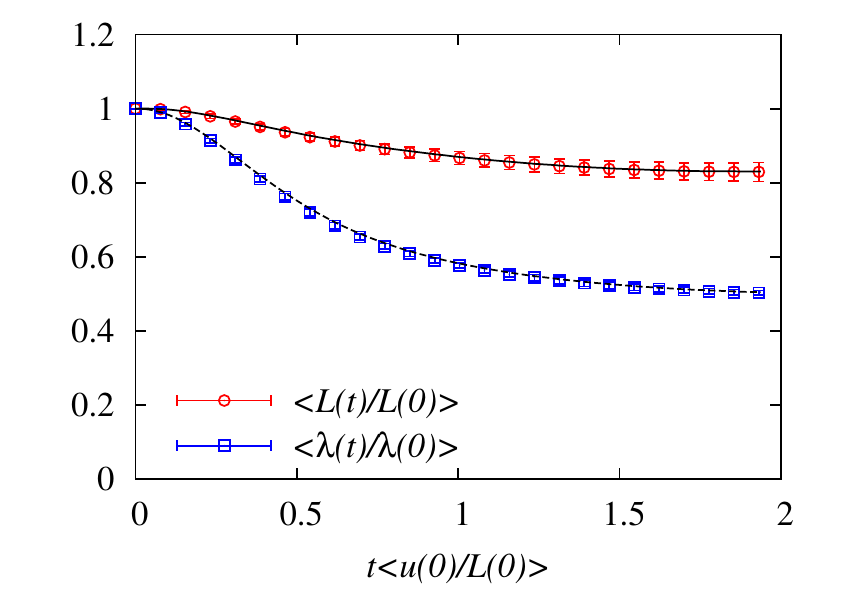}}
 \subfigure[Reynolds numbers]{\label{sfig:128_0.0027_reynolds}
  \includegraphics[width=0.475\textwidth,trim=2px 0 10px 0,clip]{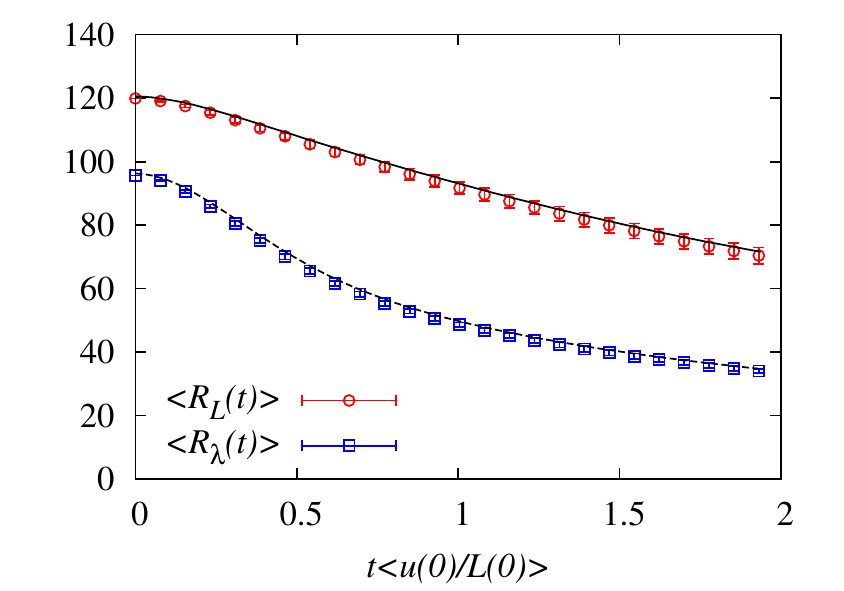}}
 \subfigure[Velocity derivative skewness]{\label{sfig:128_0.0027_skewness}
  \includegraphics[width=0.475\textwidth,trim=2px 0 10px 0,clip]{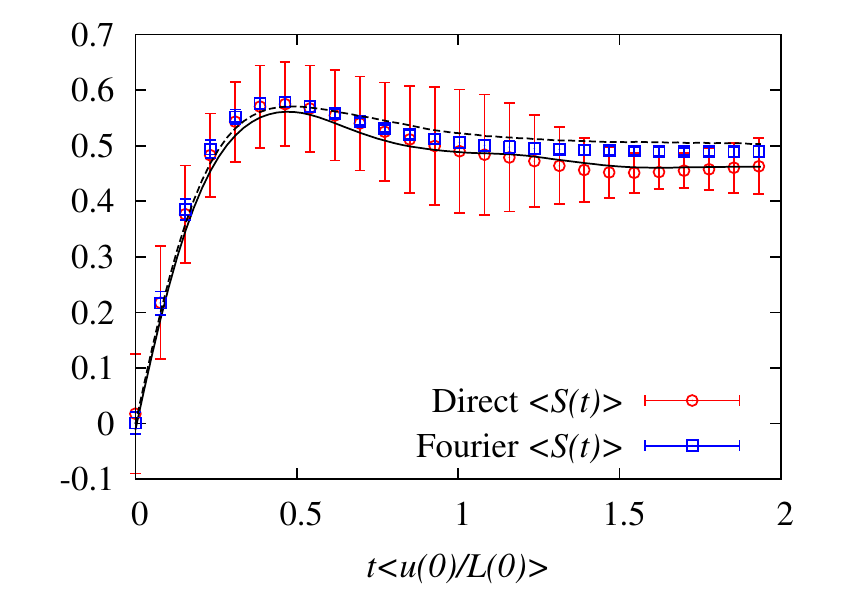}}
 \subfigure[Scaled dissipation spectra]{\label{sfig:128_0.0027_D}
  \includegraphics[width=0.475\textwidth,trim=2px 0 10px 0,clip]{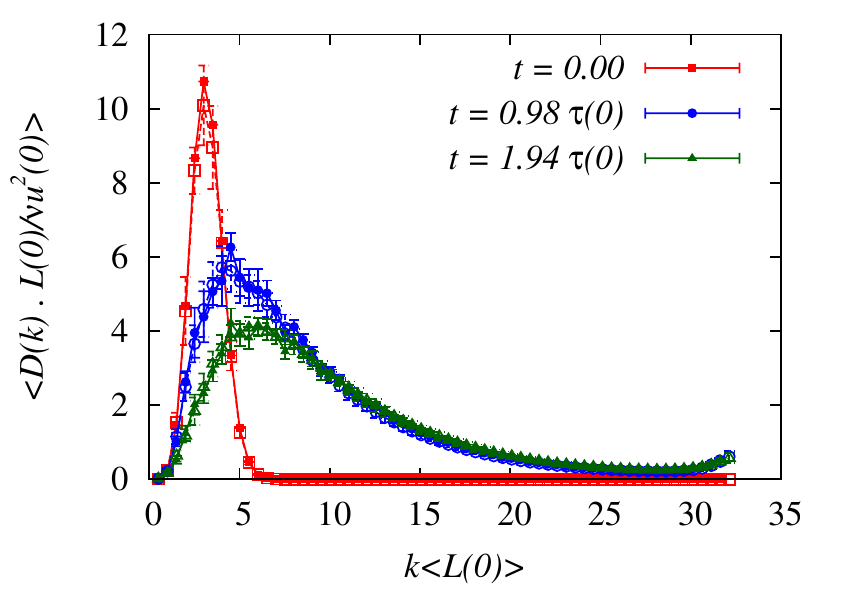}}
 \subfigure[Scaled transfer spectra]{\label{sfig:128_0.0027_T}
  \includegraphics[width=0.475\textwidth,trim=2px 0 10px 0,clip]{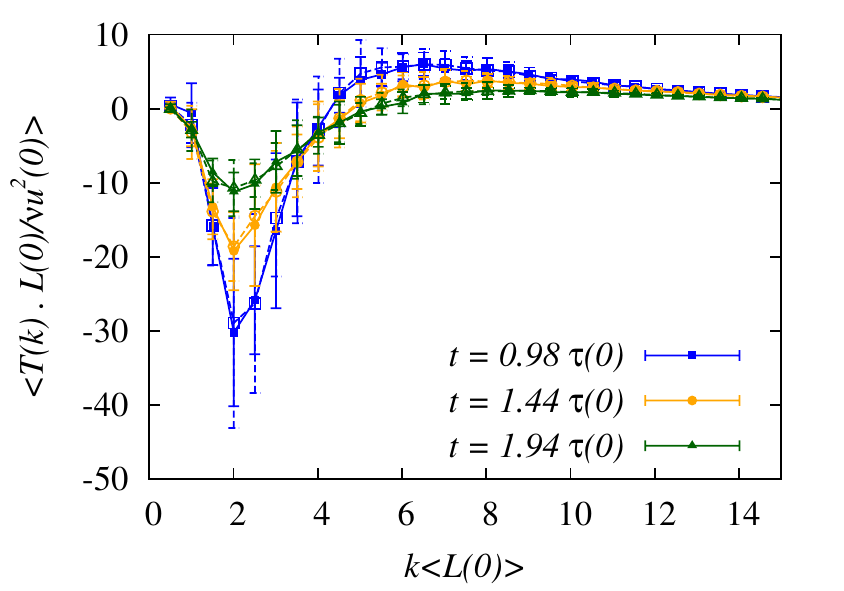}}
 \caption{Parameters and spectra for $R_\lambda(0) \simeq 95$. For parts (a)--(d): {\Large\color{red}$\circ$} and {\small\color{blue}$\square$} mark our DNS data. For comparison, we provide (------) Quinn's result for quantities plotted with {\Large\color{red}$\circ$}; (--~--~--) Quinn's result for quantities plotted with {\small\color{blue}$\square$}. For parts (e) and (f): (-----{\small$\blacksquare$}-----) Filled points, solid line show our DNS data; (-- --{\small$\square$}-- --) open points, dashed line show Quinn's results for the same times.}
 \label{fig:128_0.0027_compare}
\end{figure}

Figures \ref{sfig:128_0.0027_energy_diss}--\subref{sfig:128_0.0027_skewness} show the time variation of the same parameters as the previous two comparisons, this time only plotted up to $t = 2\tau(0)$ to match Quinn. The total energy decays to 0.5, in this time, while the dissipation rate peaks just below 2.5 at about $t = 1.25 \tau(0)$ before dropping to just below 2 at $t = 2\tau(0)$. The integral scale decreases to just above 0.8 and the Taylor microscale decreases quicker to 0.5. The integral Reynolds number $R_L$ decays from about 120 down to $\sim 65$ and the Taylor Reynolds number $R_\lambda$ from $\sim 95$ to $\sim 35$. The skewness once again peaks just before $t = 0.5 \tau(0)$ at a value this time just above 0.55, before reaching a plateau value of just below 0.5. This is in good agreement with Quinn, although again our result sits just above at the boundary of error. The figure also shows the skewness measured directly from the real-space velocity field, for comparison to figure 6.39. The agreement is once again very reassuring, with the real-space calculation peaking around the same time and value, before levelling off slightly lower at $\sim 0.45$ and with larger error bars.

The spectra in figures \ref{sfig:128_0.0027_D} and \ref{sfig:128_0.0027_T} show remarkable resemblance to Quinn for all times. The dissipation spectra in figure \ref{sfig:128_0.0027_D} not only peak at the same values as those presented in figure 6.32 of Quinn, but displays the same upturn after $k\langle L(0)\rangle = 30$ for the later two times. This kink is likely an artefact of the simulation being under-resolved, which was then investigated by Quinn. It could also be a result of aliasing errors, since this increases the transfer of energy into the higher modes. The transfer spectra are extremely well matched for all times.

\subsection{$R_\lambda(0) \simeq 129$ decaying turbulence}

\begin{figure}[tbp!]
 \centering
 \subfigure[Scaled energy and dissipation rate]{\label{sfig:256_0.002_energy_diss}
  \includegraphics[width=0.485\textwidth,trim=2px 0 10px 0,clip]{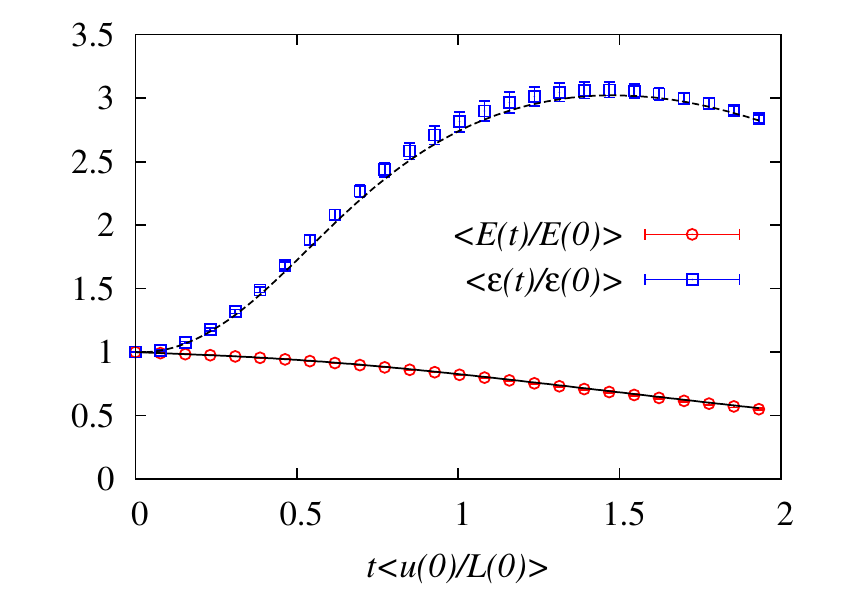}}
 \subfigure[Scaled integral and Taylor length-scales]{\label{sfig:256_0.002_length}
  \includegraphics[width=0.485\textwidth,trim=2px 0 10px 0,clip]{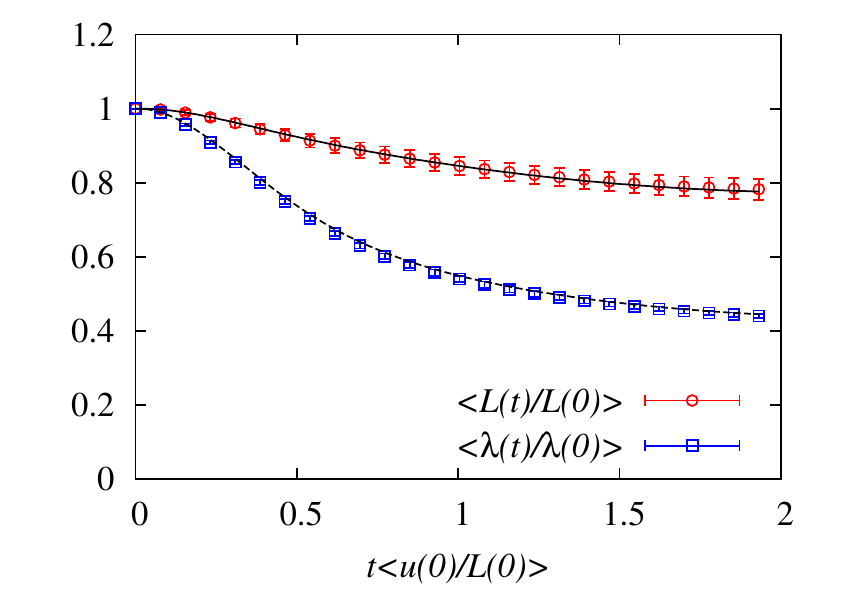}}
 \subfigure[Reynolds numbers]{\label{sfig:256_0.002_reynolds}
  \includegraphics[width=0.485\textwidth,trim=2px 0 10px 0,clip]{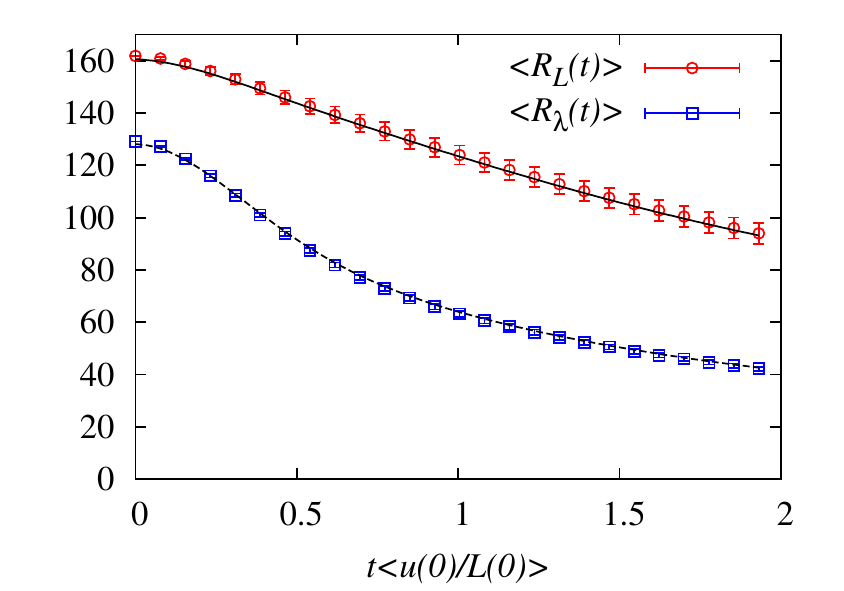}}
 \subfigure[Velocity derivative skewness]{\label{sfig:256_0.002_skewness}
  \includegraphics[width=0.485\textwidth,trim=2px 0 10px 0,clip]{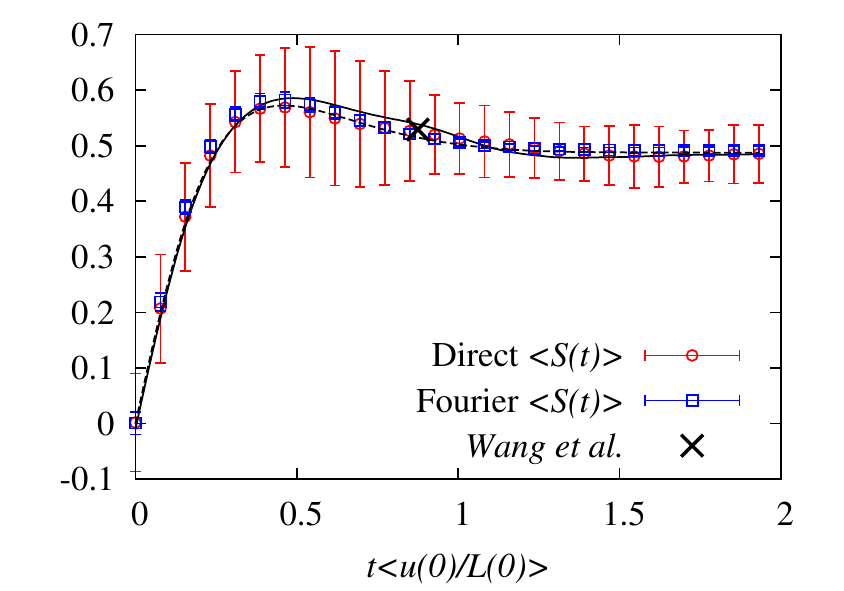}}
 \subfigure[Scaled dissipation spectra]{\label{sfig:256_0.002_D}
  \includegraphics[width=0.485\textwidth,trim=2px 0 10px 0,clip]{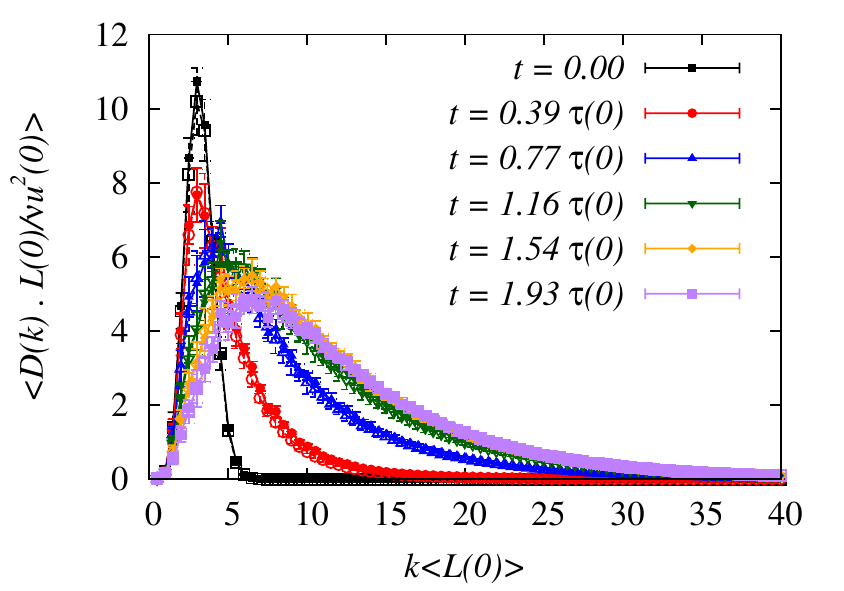}}
 \subfigure[Scaled transfer spectra]{\label{sfig:256_0.002_T}
  \includegraphics[width=0.485\textwidth,trim=2px 0 10px 0,clip]{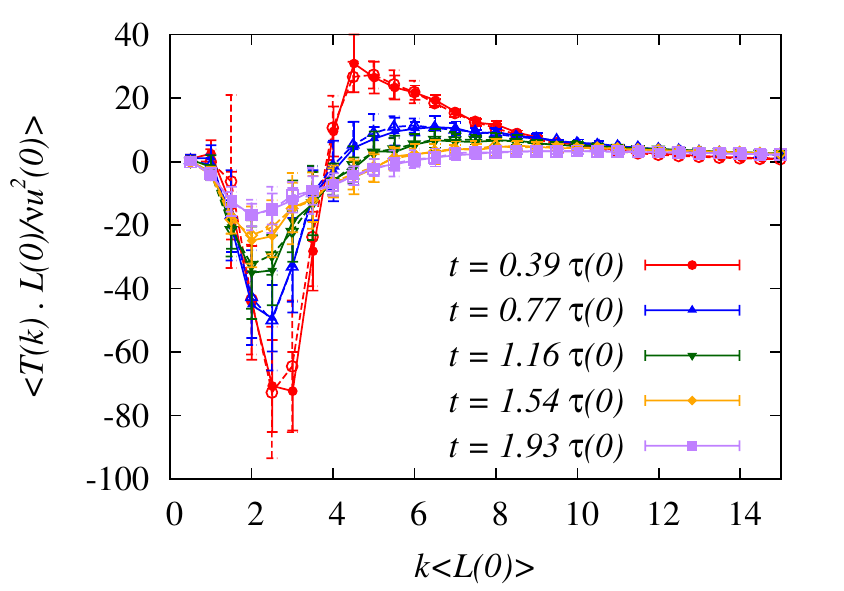}}
 \caption{Parameters and spectra for $R_\lambda(0) \simeq 129$. For parts (a)--(d): {\Large\color{red}$\circ$} and {\small\color{blue}$\square$} mark our DNS data. For comparison, we provide (------) Quinn's result for quantities plotted with {\Large\color{red}$\circ$}; (--~--~--) Quinn's result for quantities plotted with {\small\color{blue}$\square$}. For parts (e) and (f): (-----{\small$\blacksquare$}-----) Filled points, solid line show our DNS data; (-- --{\small$\square$}-- --) open points, dashed line show Quinn's results for the same times.}
 \label{fig:256_0.002_compare}
\end{figure}

This comparison used an ensemble of 10 realisations of a $256^3$ velocity field with $\nu_0 =  0.002$, giving an initial Taylor-Reynolds number $R_\lambda(0) = 127.70$. See section 6.8 of Quinn's thesis.

Figure \ref{sfig:256_0.002_energy_diss} shows the total energy decaying to 0.5 while the dissipation rate peaks around $t = 1.4\tau(0)$ at about 3.2, both in agreement with Quinn. The peak in the dissipation rate is a little on the high side, but well within the quoted error. The values of $\varepsilon$ at $t = 2\tau(0)$, however, do match well. Figure \ref{sfig:256_0.002_length} shows both the integral Taylor scales decreasing to just below 0.8 and just above 0.4, respectively. The Reynolds numbers shown in figure \ref{sfig:256_0.002_reynolds} can be seen to decrease from 160 to $\sim 90$ for $R_L$ and $\sim 130$ to 40 for $R_\lambda$. These are in excellent agreement with Quinn. Figure \ref{sfig:256_0.002_skewness} plots both the Fourier- and real-space calculations of the velocity derivative skewness with good agreement, and shows the two calculations converging as Reynolds number is increased. Also plotted is a value for the skewness found by Wang, Chen, Brasseur and Wyngaard \cite{Wang:1996p1041} during a decaying simulation. $R_\lambda = 68.1$ was used to find the time when the simulations agreed, $t\langle u(0)/L(0)\rangle = 0.875$, and the value for $S$ plotted at this time. All the spectra in figures \ref{sfig:256_0.002_D} and \subref{sfig:256_0.002_T} show excellent agreement.

\section{Comparison with \emph{hit3d}}
\emph{hit3d} is a freely-available pseudospectral DNS code, see section \ref{subsec:currently_avail}. A small modification was made to this code to correct a missing numerical factor of $0.5$ when computing the forcing to be applied. As such, the actual input rate was double that specified in the input file. Communication with the code developers revealed that their use had only relied on the system being stationary, not the actual value of the dissipation rate. Since we are interested in this quantity, the code was corrected accordingly.

\subsection{Decaying turbulence}
This comparison was run from a normalised Kolmogorov initial spectrum (see section \ref{subsec:initial_field}). They were run using the same time-step $dt = 0.001$ and statistics were collected at an interval of $0.1$. Viscosity used was 0.005 on a $128^3$ lattice and both simulations used $2/3$-rule for full de-aliasing. Quantities plotted are shell-averaged.

\begin{figure}[b!]
 \centering
 \subfigure[Fluctuation of total energy]{ \includegraphics[width=0.475\textwidth,trim=2px 0 10px 0,clip]{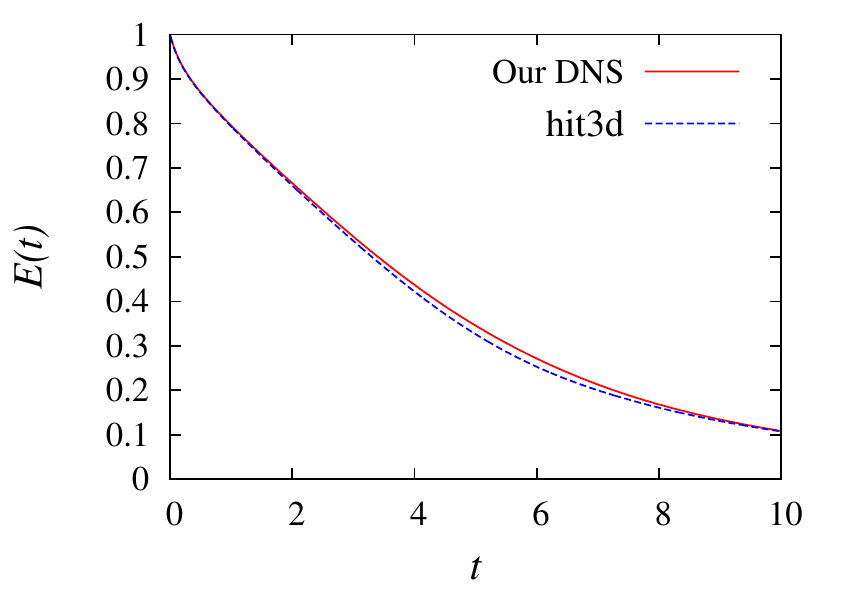} }
 \subfigure[Fluctuation of dissipation rate]{ \includegraphics[width=0.475\textwidth,trim=2px 0 10px 0,clip]{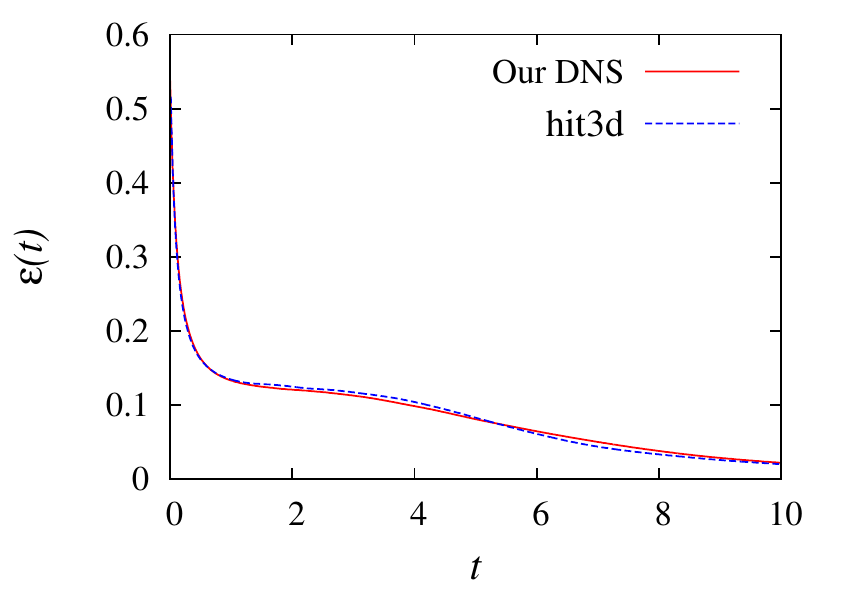} }
 \subfigure[Fluctuation of Taylor-Reynolds number]{ \includegraphics[width=0.475\textwidth,trim=2px 0 10px 0,clip]{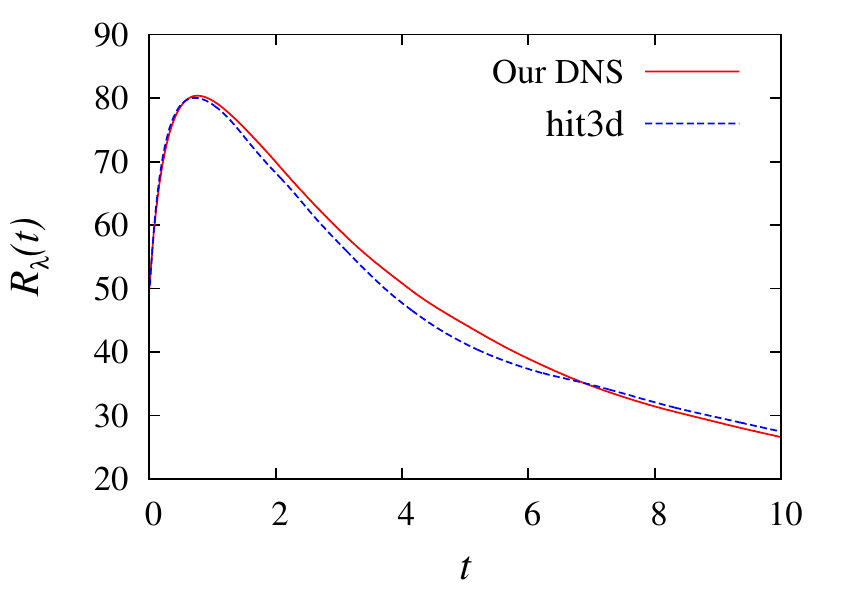} }
\subfigure[Fluctuation of velocity derivative skewness]{ \includegraphics[width=0.475\textwidth,trim=2px 0 10px 0,clip]{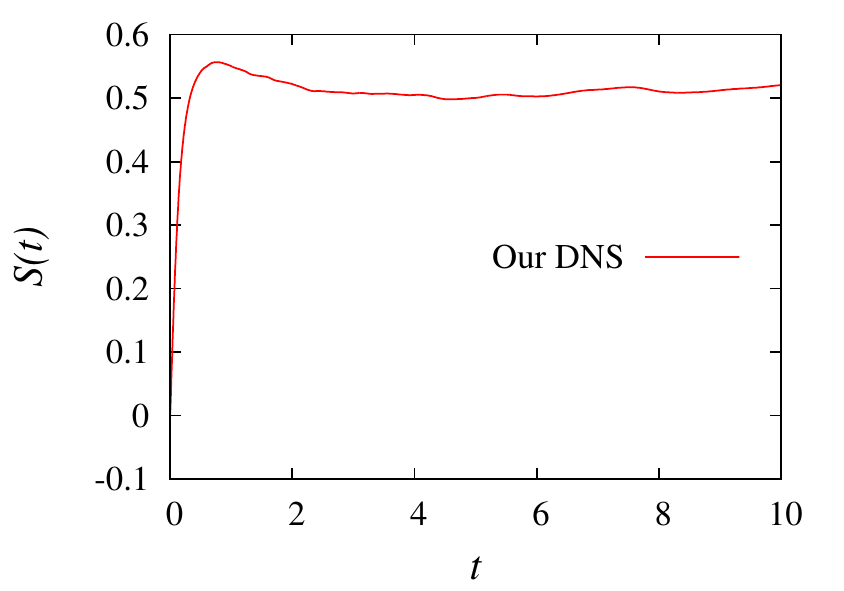} }
 \caption{Comparison with \emph{hit3d} for decaying simulations.}
 \label{fig:hit3d_decay_comp}
\end{figure}

In each figure, the agreement between the two codes is very good, despite only being presented for a single realisation. Velocity derivative skewness is only plotted for our code since it is not accessible from \emph{hit3d}.

\subsection{Forced turbulence}
This comparison was run from a normalised Kolmogorov initial spectrum using a forcing rate $\epsw = 0.1$ with this energy being inserted into the lowest \emph{two} shells (see section \ref{subsec:forcing} for details). They were run using the same time-step $dt = 0.001$ and statistics were collected at an interval of $0.1$. Viscosity used was 0.005 on a $128^3$ lattice and both simulations used $2/3$-rule for full de-aliasing. The results from both codes follow similar paths and eventually fluctuate around equal steady state values. The agreement between the two codes is reassuring.

Since the turbulence is stationary, if we consider only the period after the initial transient to our steady state we can obtain a value for the mean quantities. By sampling the data every large eddy turnover time ($\sim 2$) in the period $t \in [20,45]$, we find, for our DNS data: 
\begin{align}
 E &= 0.500 \pm 0.014 \ , \qquad \varepsilon &= 0.0968 \pm 0.0050\ \qquad\text{and}\qquad R_\lambda &= 58.8 \pm 2.3 \ . 
\end{align} 
These should be compared to the results obtained by \emph{hit3d}:
\begin{align}
 E &= 0.512 \pm 0.014 \ ,\qquad \varepsilon &= 0.0966 \pm 0.0070 \qquad\text{and}\qquad R_\lambda &= 60.3 \pm 2.8 \ .
\end{align}
The agreement is excellent, particularly for the dissipation rate. All quantities agree within one standard deviation. 

\begin{figure}[tb!]
 \centering
 \subfigure[Fluctuation of total energy]{ \includegraphics[width=0.475\textwidth,trim=2px 0 10px 0,clip]{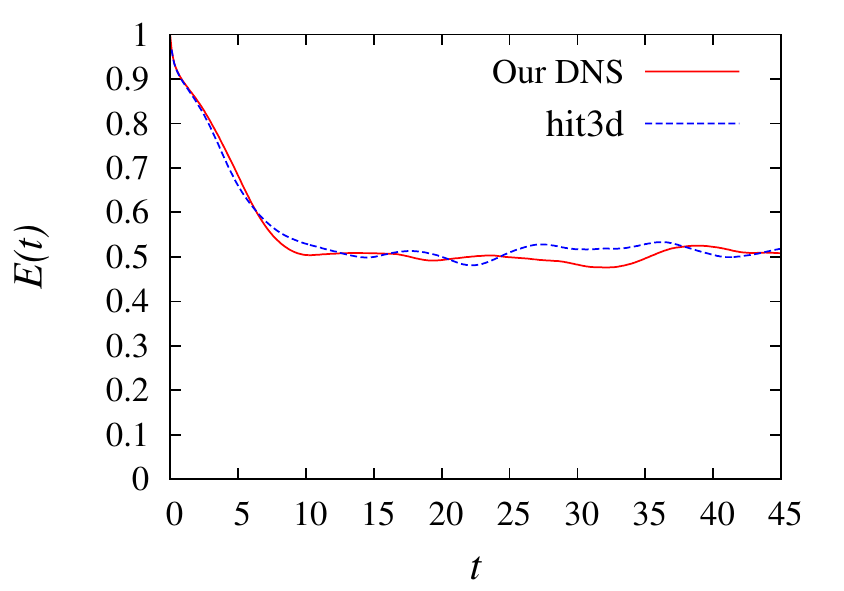} }
 \subfigure[Fluctuation of dissipation rate in time]{ \includegraphics[width=0.475\textwidth,trim=2px 0 10px 0,clip]{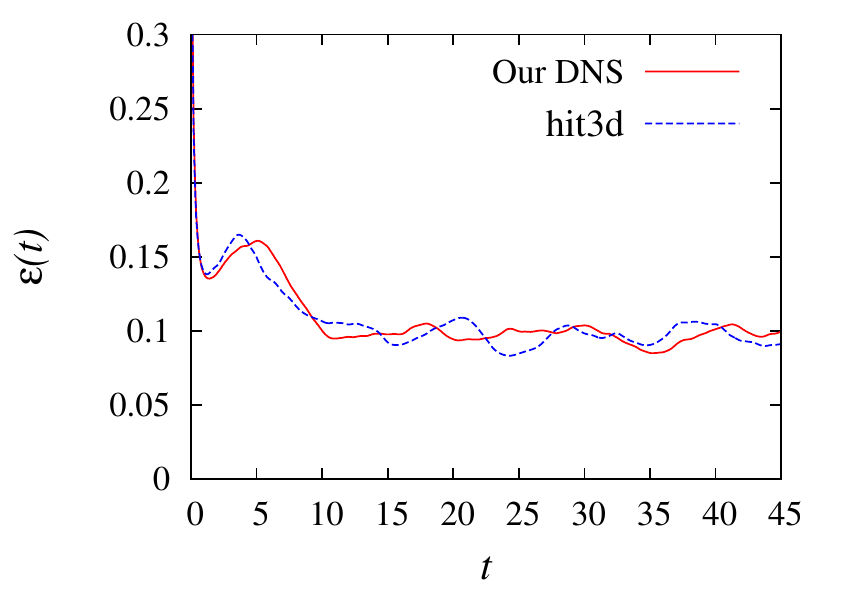} }
 \subfigure[Fluctuation of Taylor-Reynolds number]{ \includegraphics[width=0.475\textwidth,trim=2px 0 10px 0,clip]{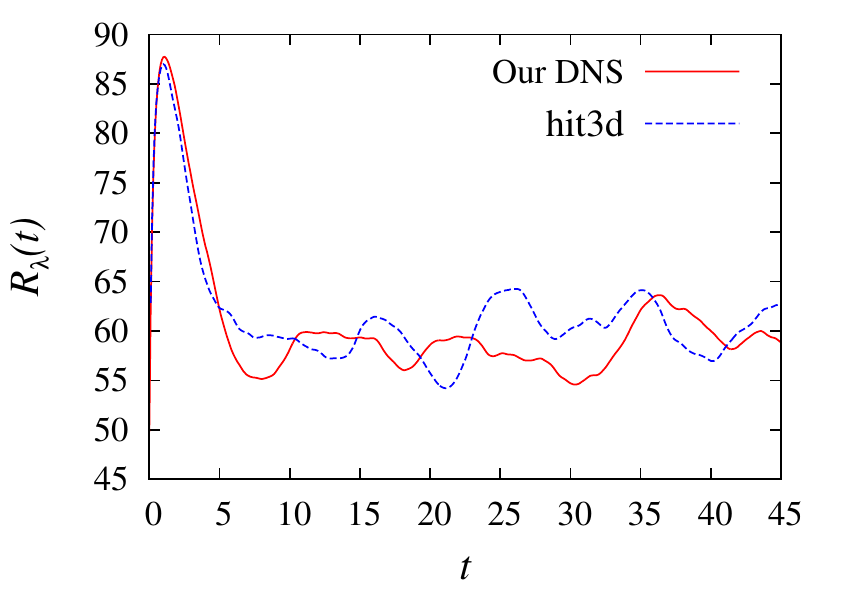} }
\subfigure[Fluctuation of velocity derivative skewness]{ \includegraphics[width=0.475\textwidth,trim=2px 0 10px 0,clip]{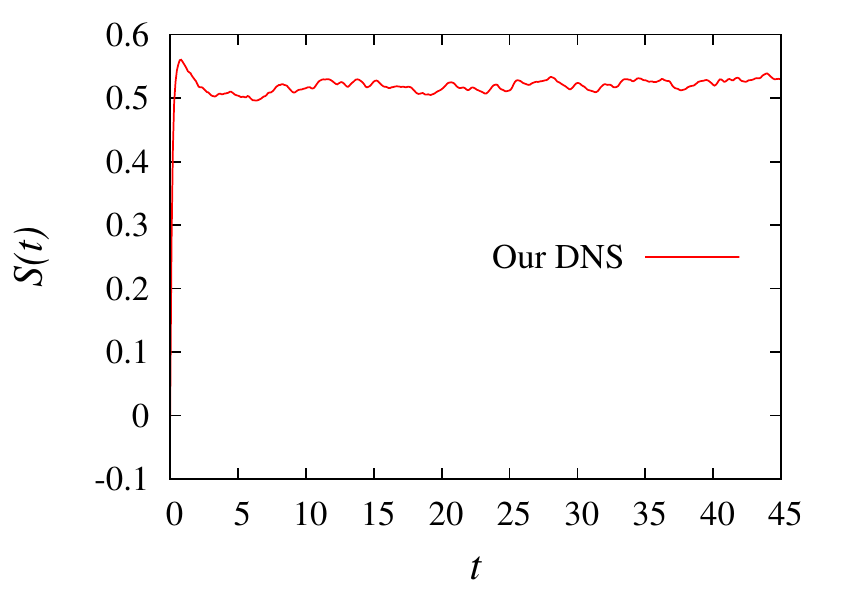} }
 \caption{Comparison with \emph{hit3d} for forced simulations.}
 \label{fig:hit3d_comp}
\end{figure}

\clearpage

\section{Taylor-Green vortex}

Taylor and Green, for whom this vortex is name, studied the evolution of a velocity field starting from an initial condition of the form \cite{Taylor:1937p499}
\begin{align}
 u_x(\vec{x},0) &= A \cos (ax) \sin(by) \sin(cz) \nonumber \\
 u_y(\vec{x},0) &= B \sin (ax) \cos(by) \sin(cz) \\
 u_z(\vec{x},0) &= C \sin (ax) \sin(by) \sin(cz) \nonumber \ ,
\end{align}
and later specialised on the case $a = b = c$, $A = -B$, $C = 0$. The same problem was later studied in depth by Brachet, Meiron, Orszag, Nickel, Morf and Frisch \cite{Brachet:1983p411}, who further restricted their attention to $a = b = c = A = -B = 1$ with $C = 0$. They also applied a shift $x_i \to x_i - \pi/2$ such that $\cos(x_i) \to \sin(x_i)$ and $\sin(x_i) \to -\cos(x_i)$. We follow suit, and the initial conditions under consideration are now
\begin{align}
 u_x(\vec{x},0) &= \sin(x) \cos(y) \cos(z) \nonumber \\
 u_y(\vec{x},0) &= -\cos(x) \sin(y) \cos(z) \\
 u_z(\vec{x},0) &= 0 \nonumber \ .
\end{align}
Results in this section should be compared to section 5.1 of the thesis by Young \cite{thesis:ajyoung} and \cite{Brachet:1983p411}. The flow is inviscid, $\nu_0 = 0$. Here we used a $256^3$ lattice, as Brachet \etal\ performed simulations of grids ranging from $32^3$ to $256^3$ and we compare to the highest resolution runs. Whereas, the results in Young are from a $128^3$ lattice. We have used full de-aliasing by isotropic velocity field truncation for $k \geq N/3 = 85$.

\begin{figure}[tbf!]
 \centering
 \subfigure[$t = 0$]{ \includegraphics[width=0.45\textwidth,trim=60px 0 60px 0, clip]{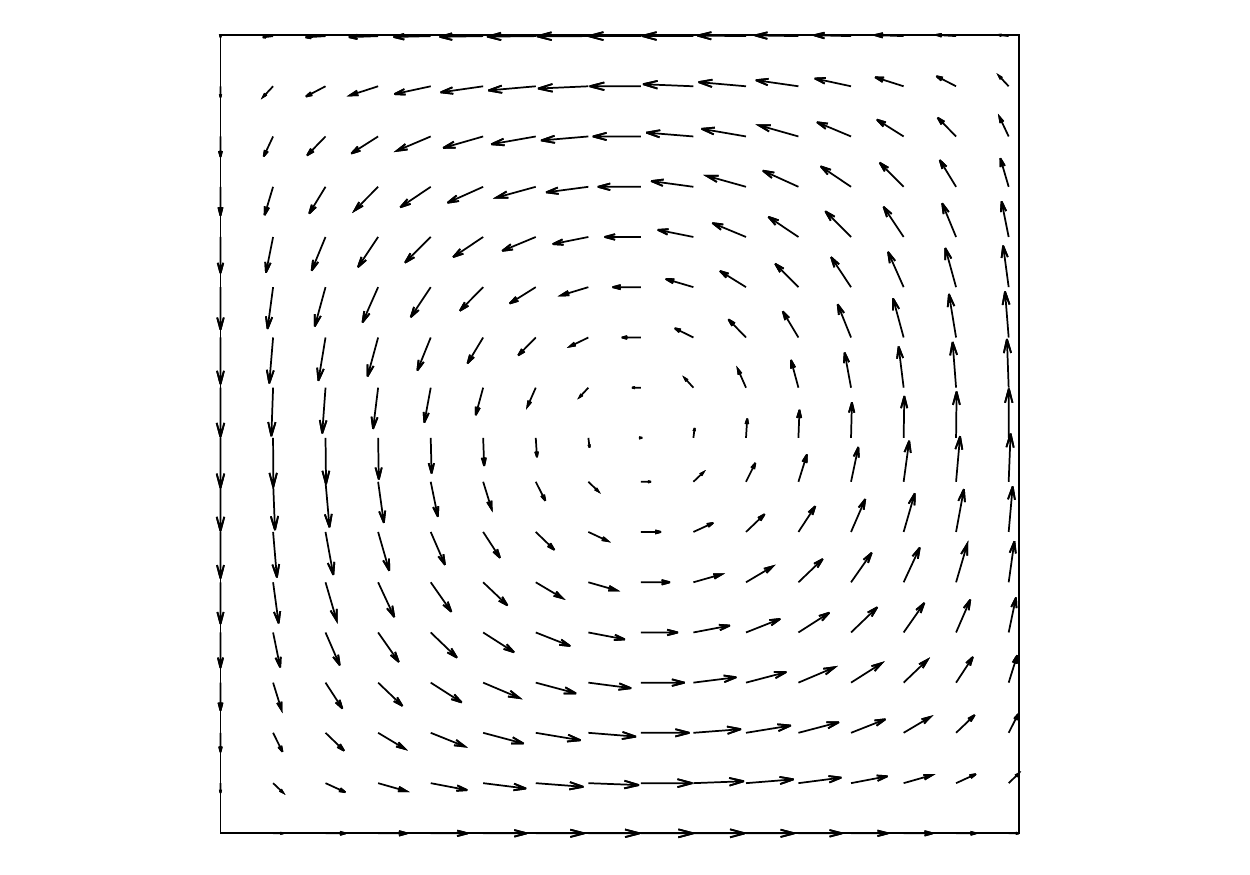} }
 \subfigure[$t = 1$]{ \includegraphics[width=0.45\textwidth,trim=60px 0 60px 0, clip]{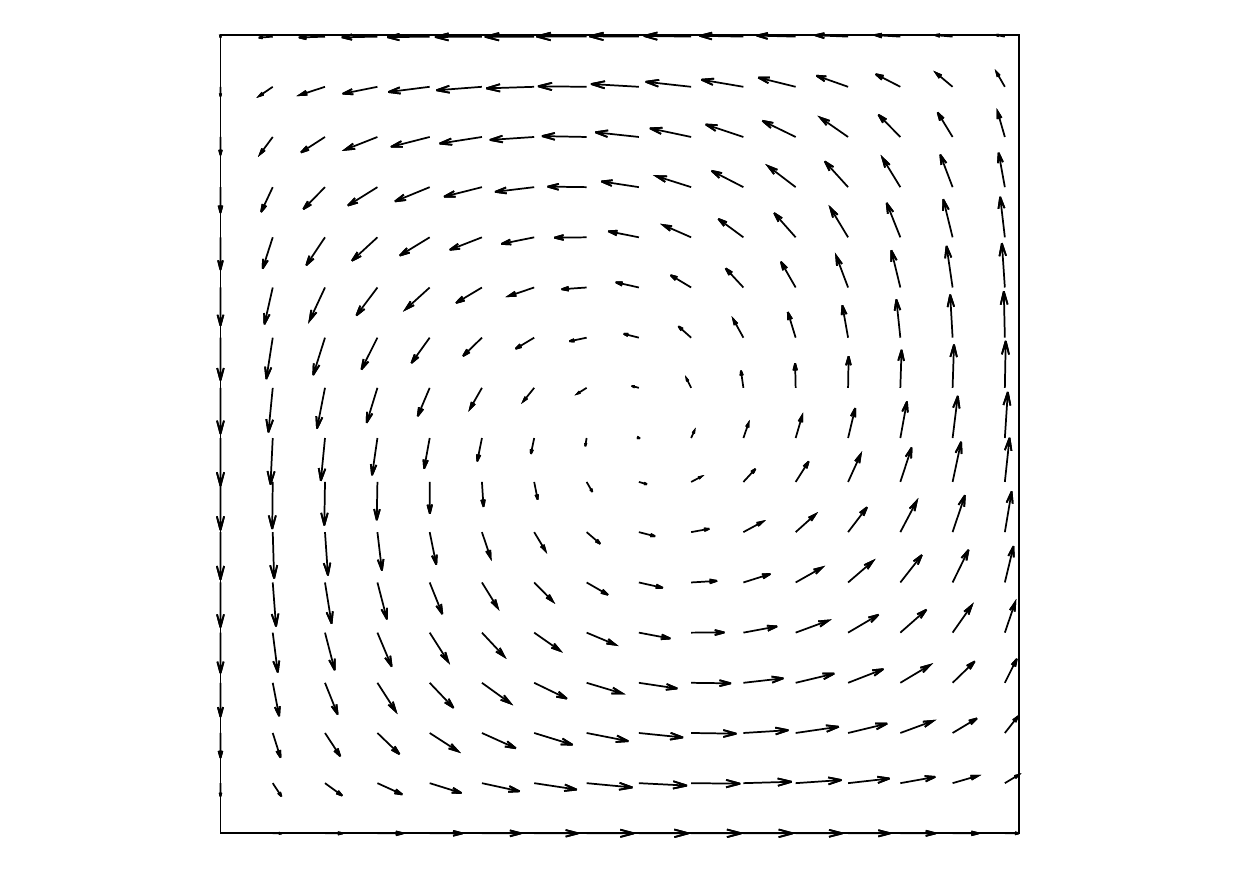} }
 \subfigure[$t = 2$]{ \includegraphics[width=0.45\textwidth,trim=60px 0 60px 0, clip]{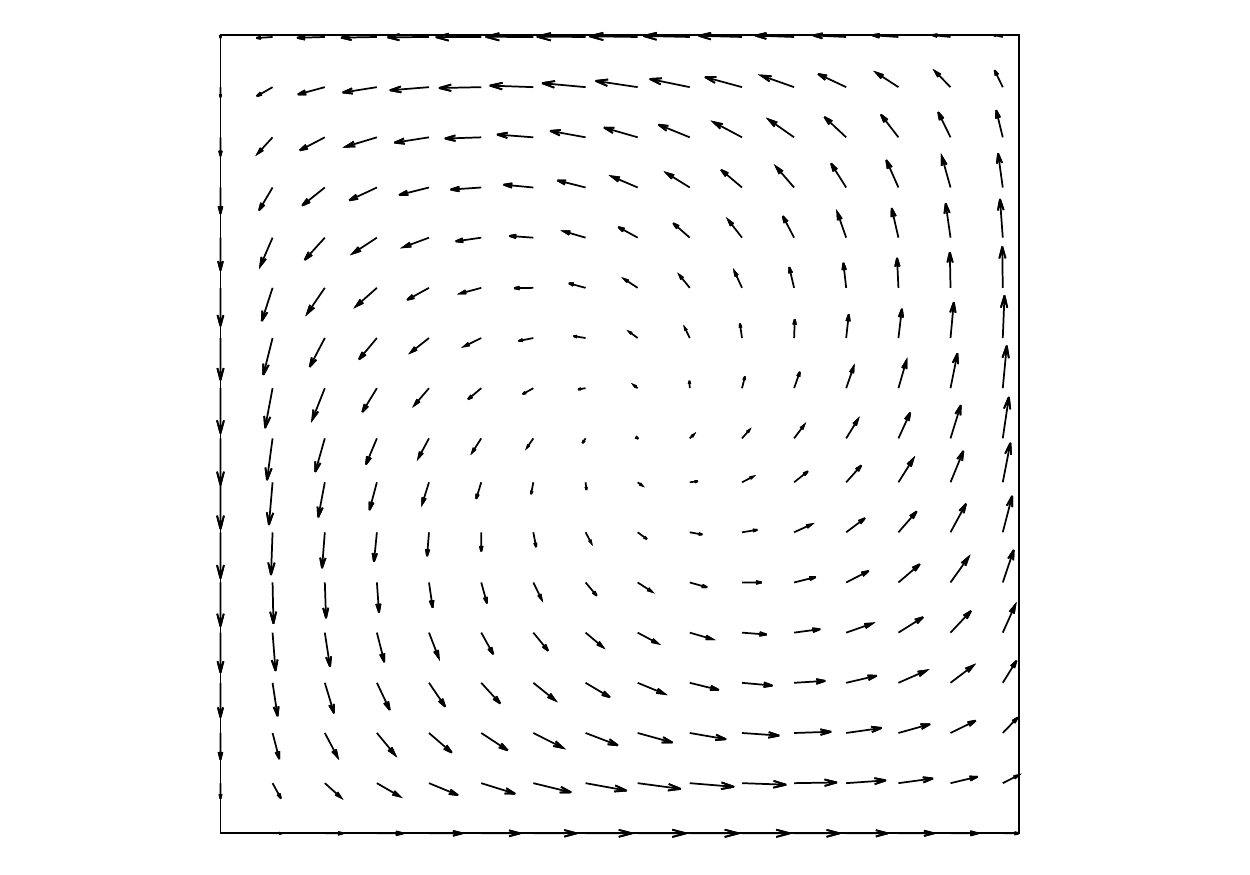} }
\subfigure[$t = 3$]{ \includegraphics[width=0.45\textwidth,trim=60px 0 60px 0, clip]{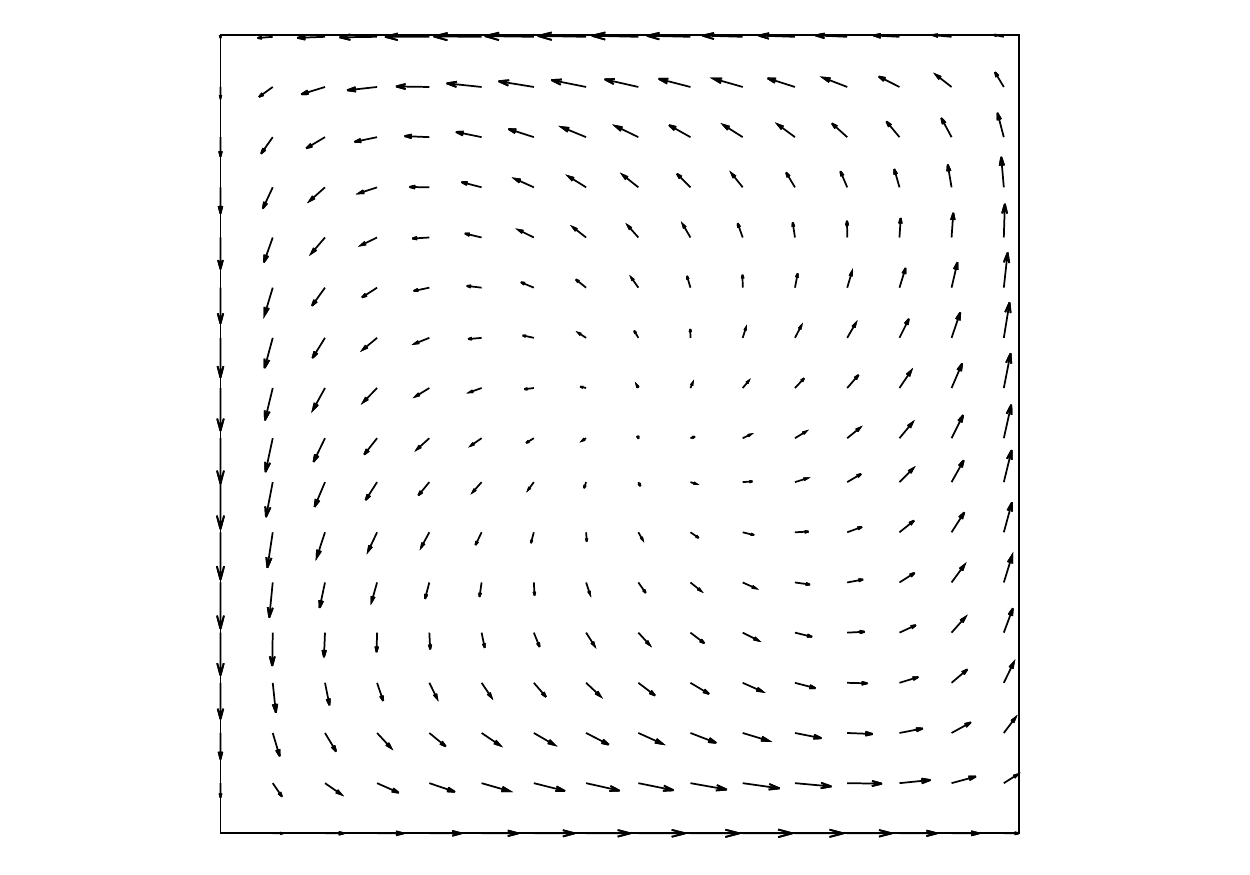} }
 \caption{Taylor-Green velocity field in the $z = 0$ plane.
 }
 \label{fig:tg_slice}
\end{figure}

\begin{figure}[tbf!]
 \centering
 \subfigure[$t = 0$]{ \includegraphics[width=0.45\textwidth,trim=27px 20px 27px 20px, clip]{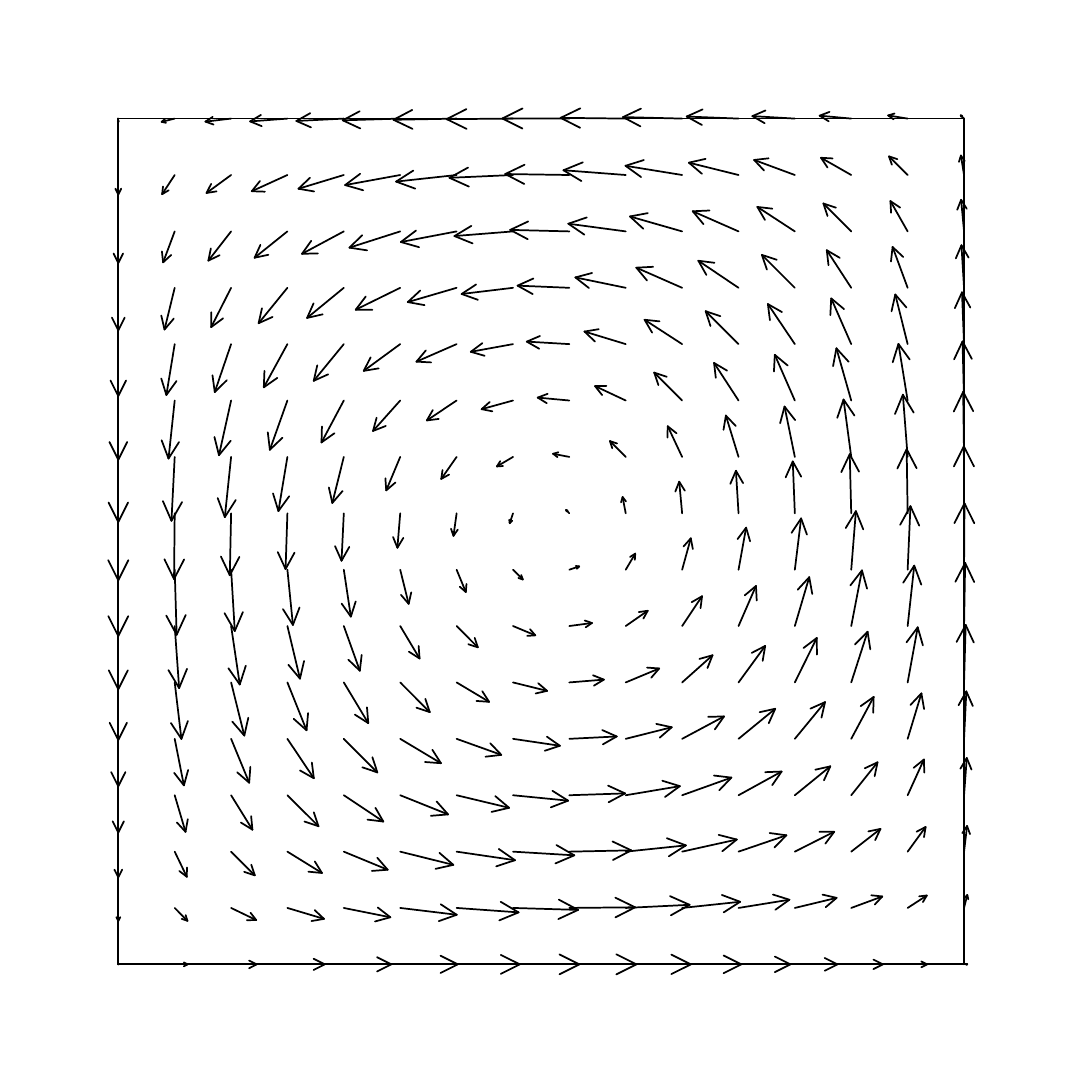} }
 \subfigure[$t = 1$]{ \includegraphics[width=0.45\textwidth,trim=27px 20px 27px 20px, clip]{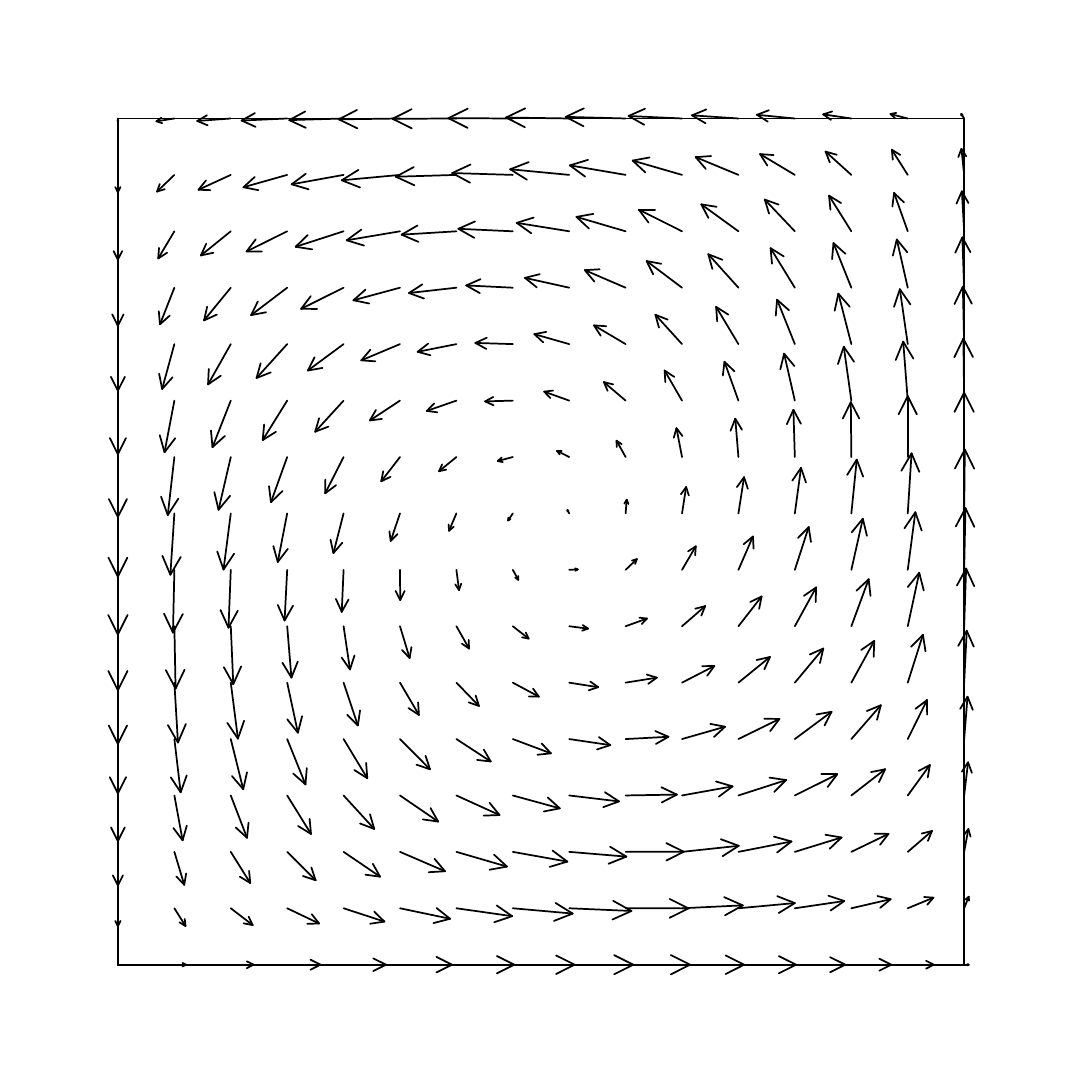} }
 \subfigure[$t = 2$]{ \includegraphics[width=0.45\textwidth,trim=27px 20px 27px 20px, clip]{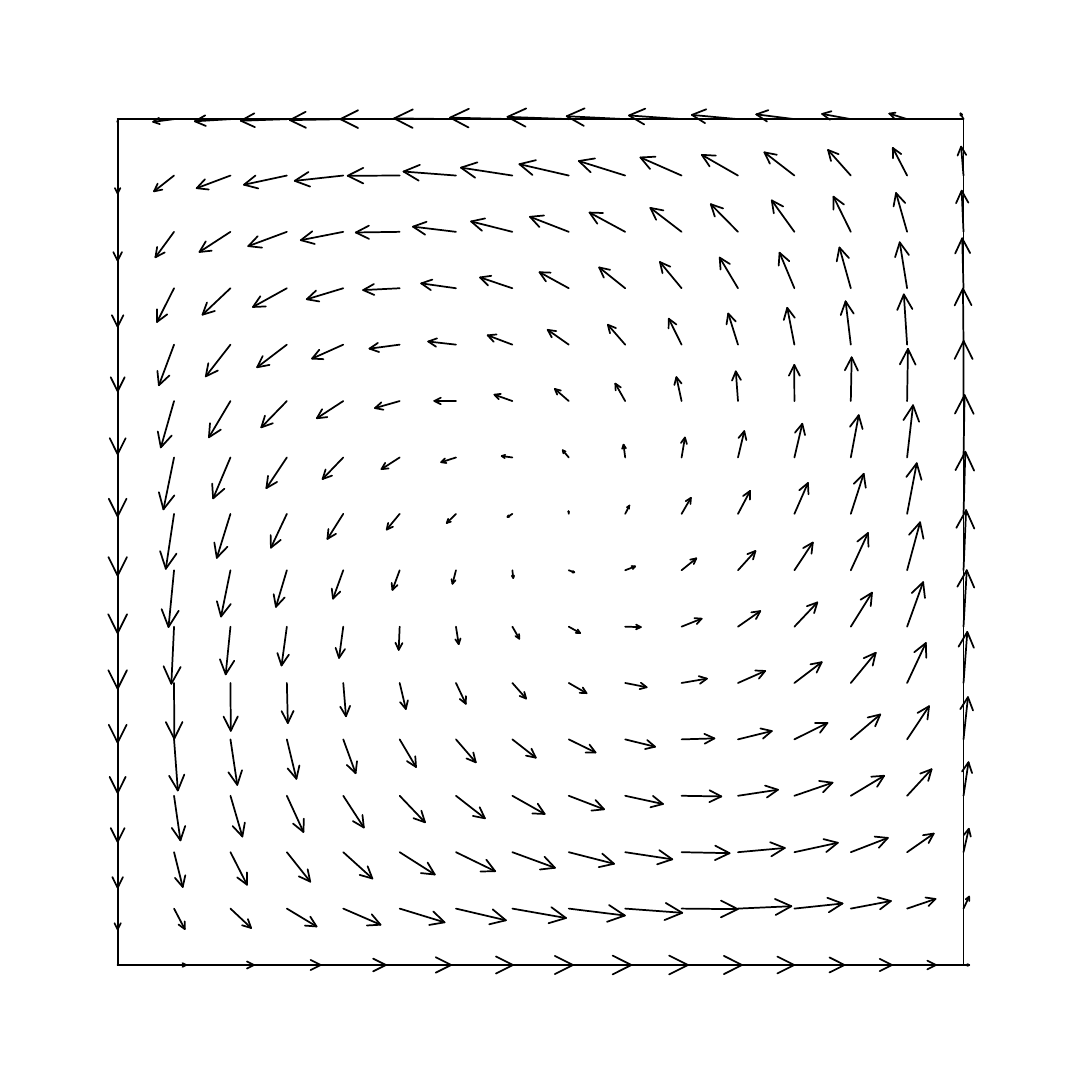} }
 \subfigure[$t = 3$]{ \includegraphics[width=0.45\textwidth,trim=27px 20px 27px 20px, clip]{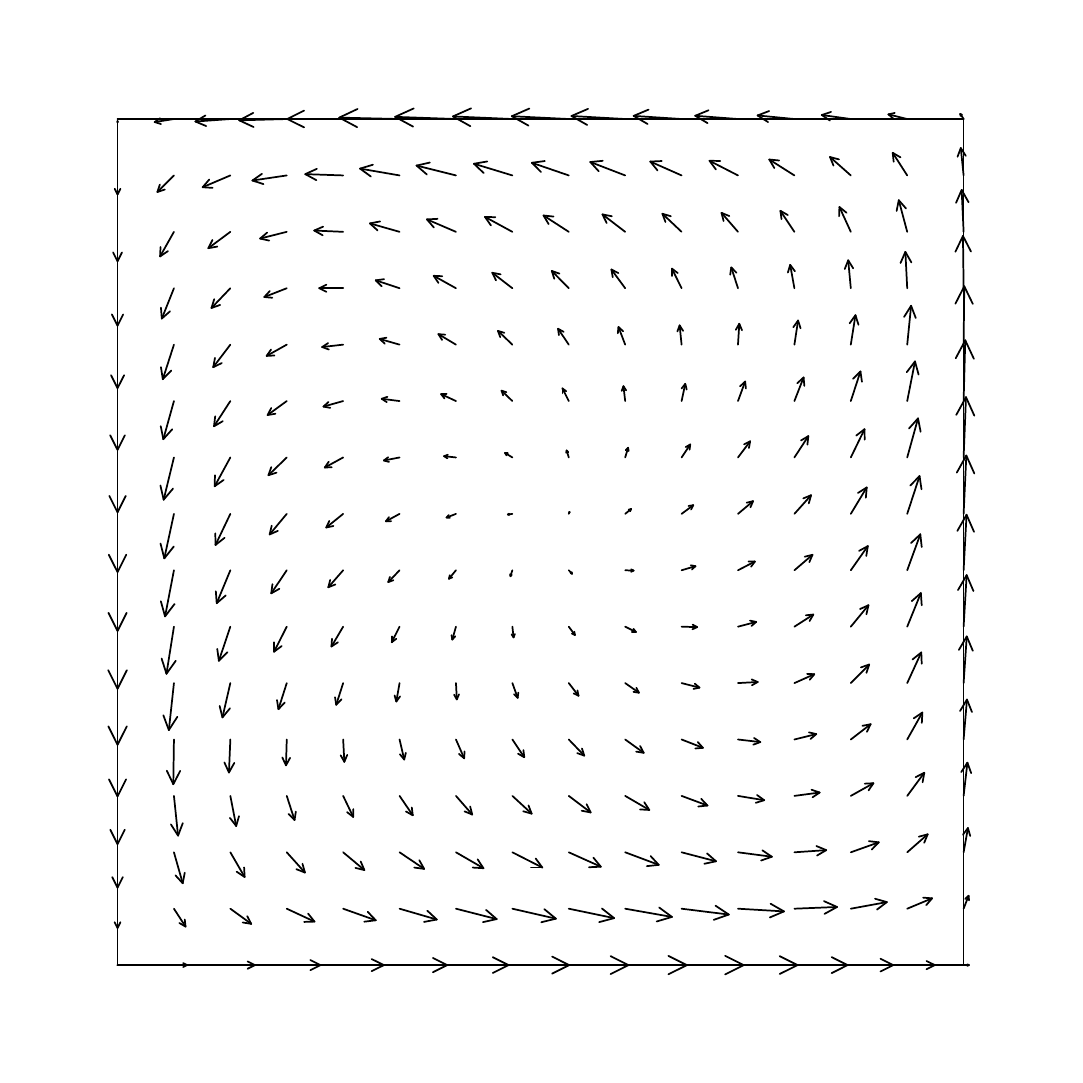} }
 \caption{Taylor-Green velocity field in the $z = 0$ plane reproduced from Young \cite{thesis:ajyoung}.}
 \label{fig:tg_slice_Young}
\end{figure}

\begin{figure}[tbf!]
 \centering
 \subfigure[$t = 0$]{ \includegraphics[width=0.45\textwidth,trim=60px 0 60px 0, clip]{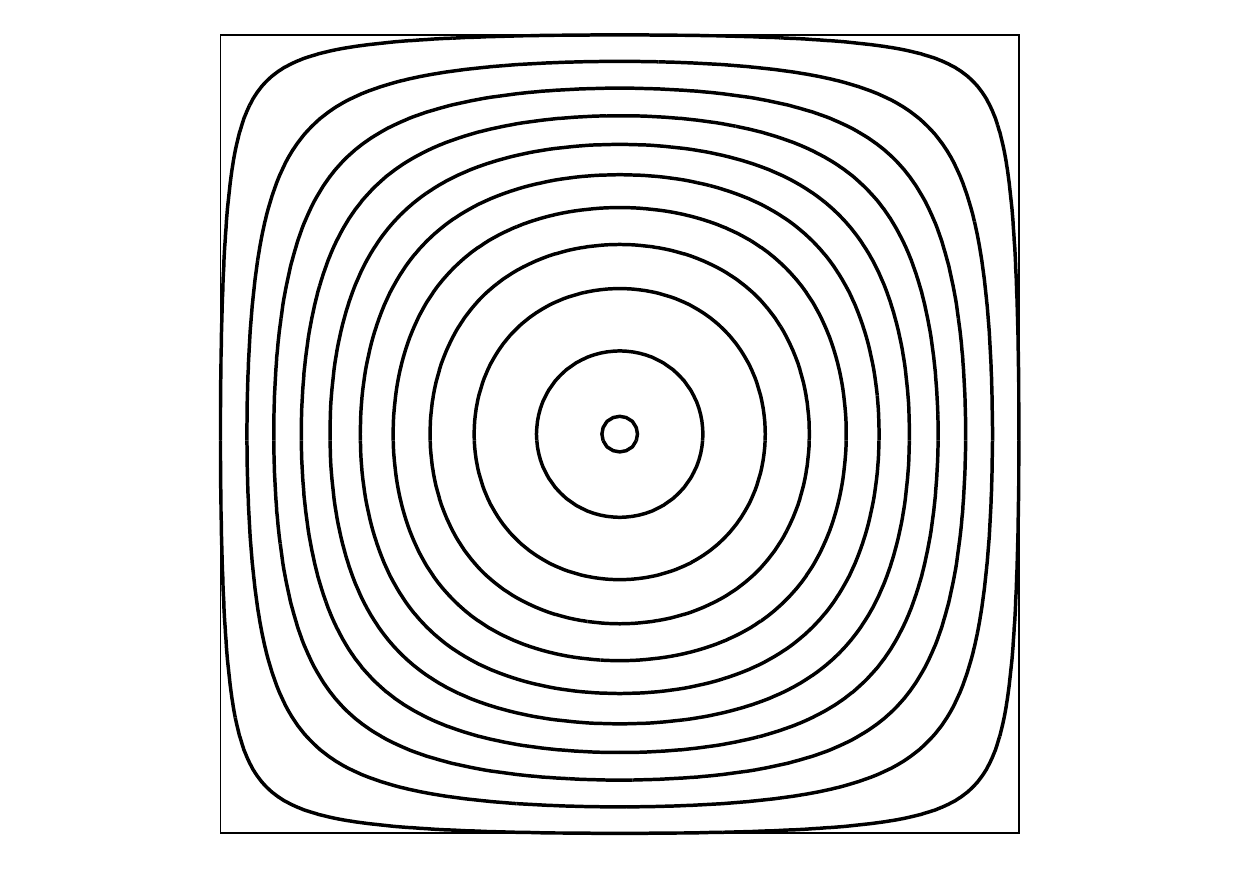} }
 \subfigure[$t = 1$]{ \includegraphics[width=0.45\textwidth,trim=60px 0 60px 0, clip]{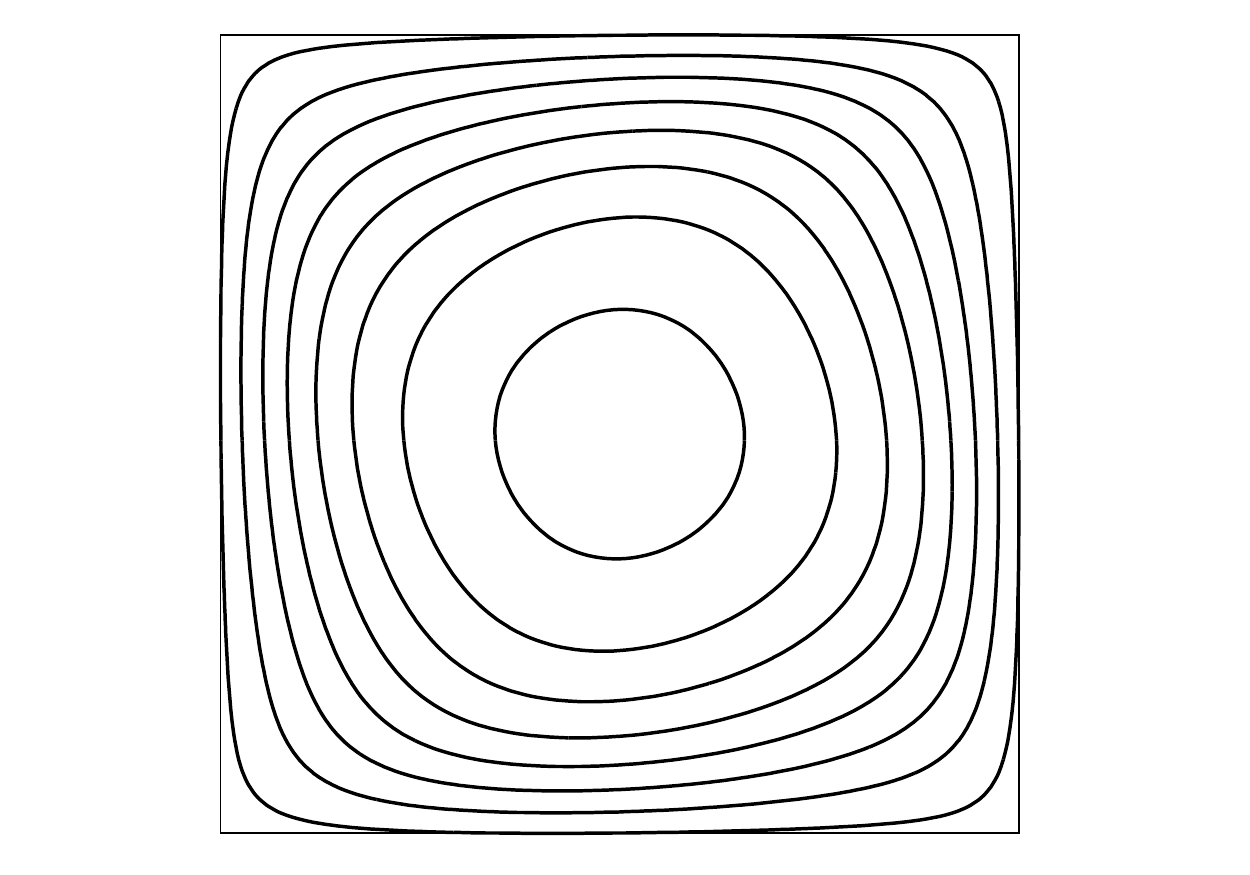} }
 \subfigure[$t = 2$]{ \includegraphics[width=0.45\textwidth,trim=60px 0 60px 0, clip]{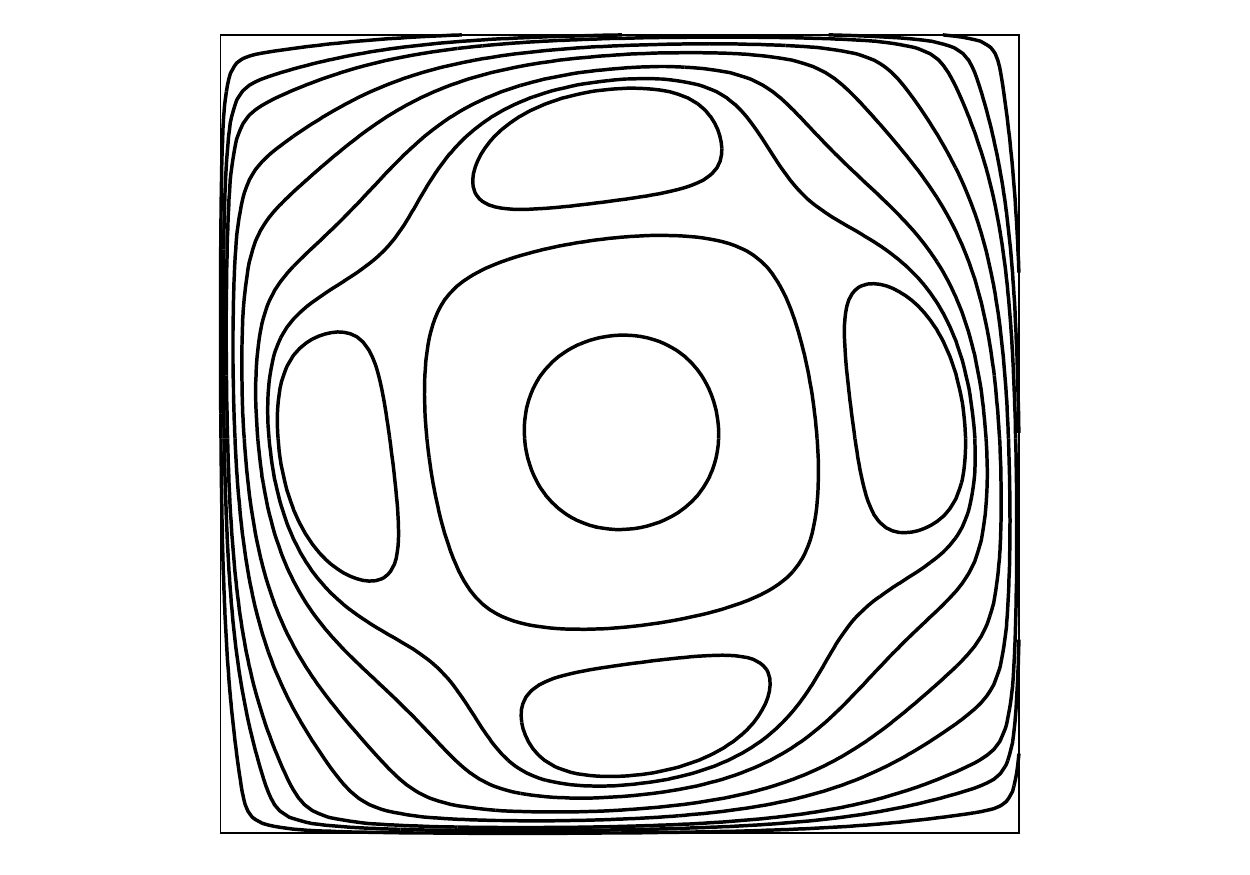} }
\subfigure[$t = 3$]{ \includegraphics[width=0.45\textwidth,trim=60px 0 60px 0, clip]{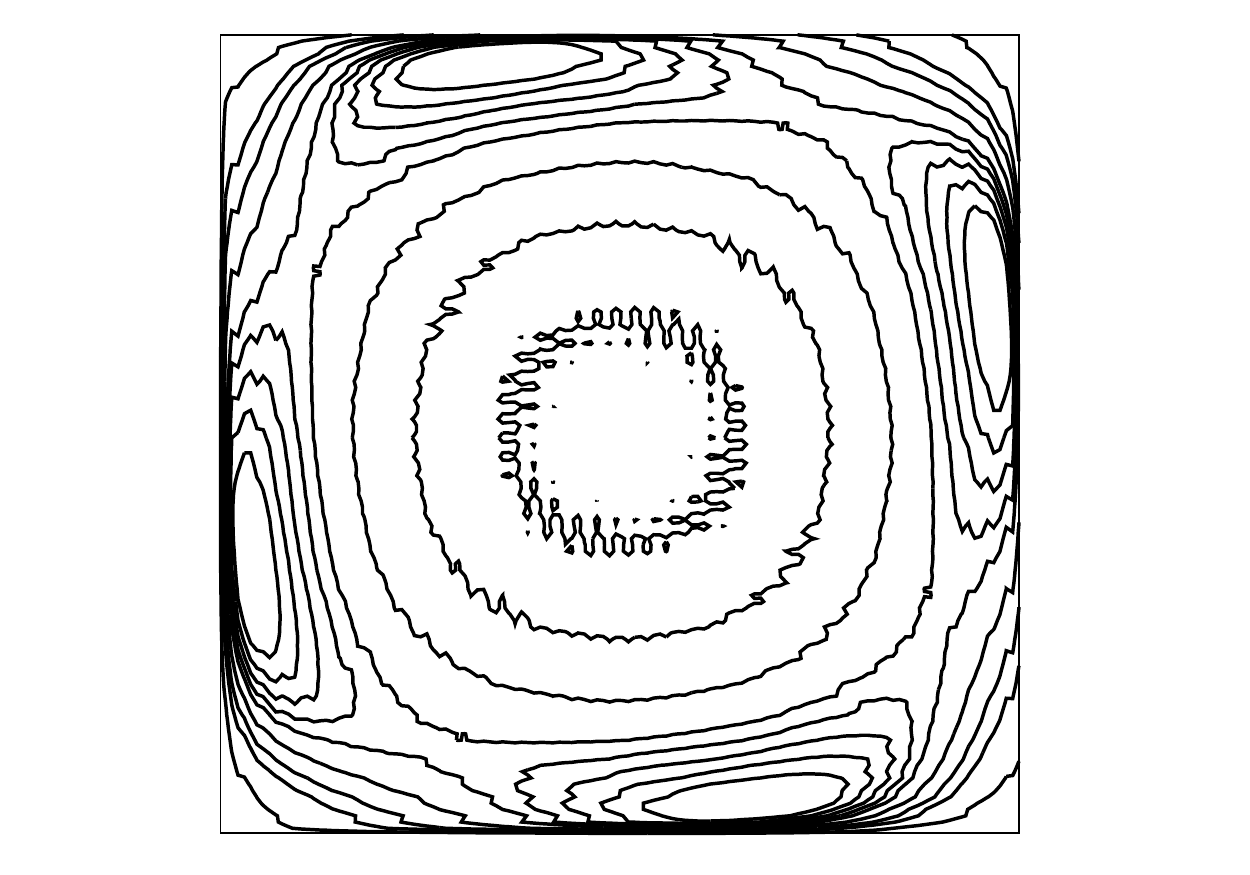} }
 \caption{Isovorticity contours in the $z = 0$ plane.
 }
 \label{fig:tg_contours}
\end{figure}

\begin{figure}[tbf!]
 \centering
 \subfigure[$t = 0$]{ \includegraphics[width=0.45\textwidth,trim=25px 20px 25px 20px, clip]{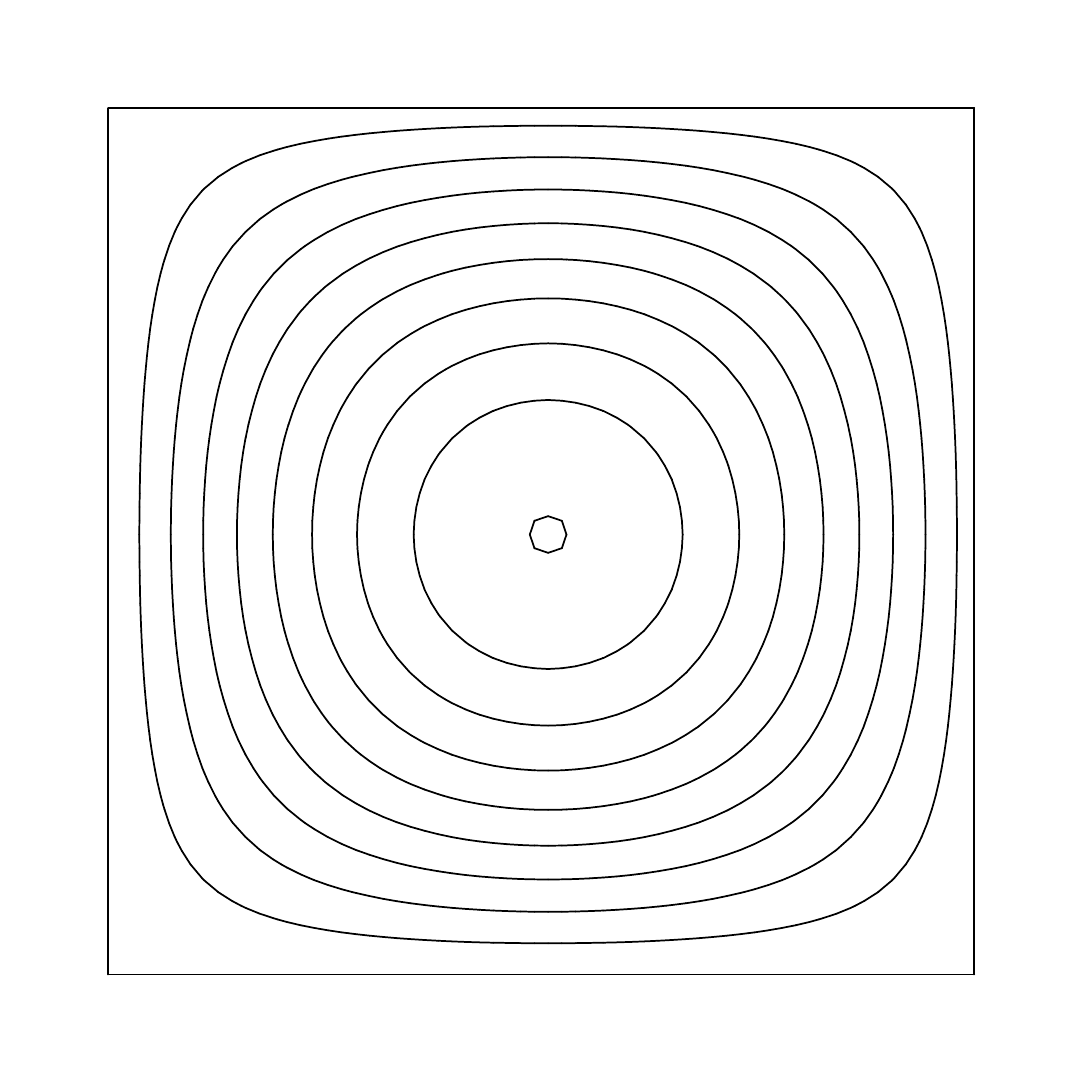} }
 \subfigure[$t = 1$]{ \includegraphics[width=0.45\textwidth,trim=25px 20px 25px 20px, clip]{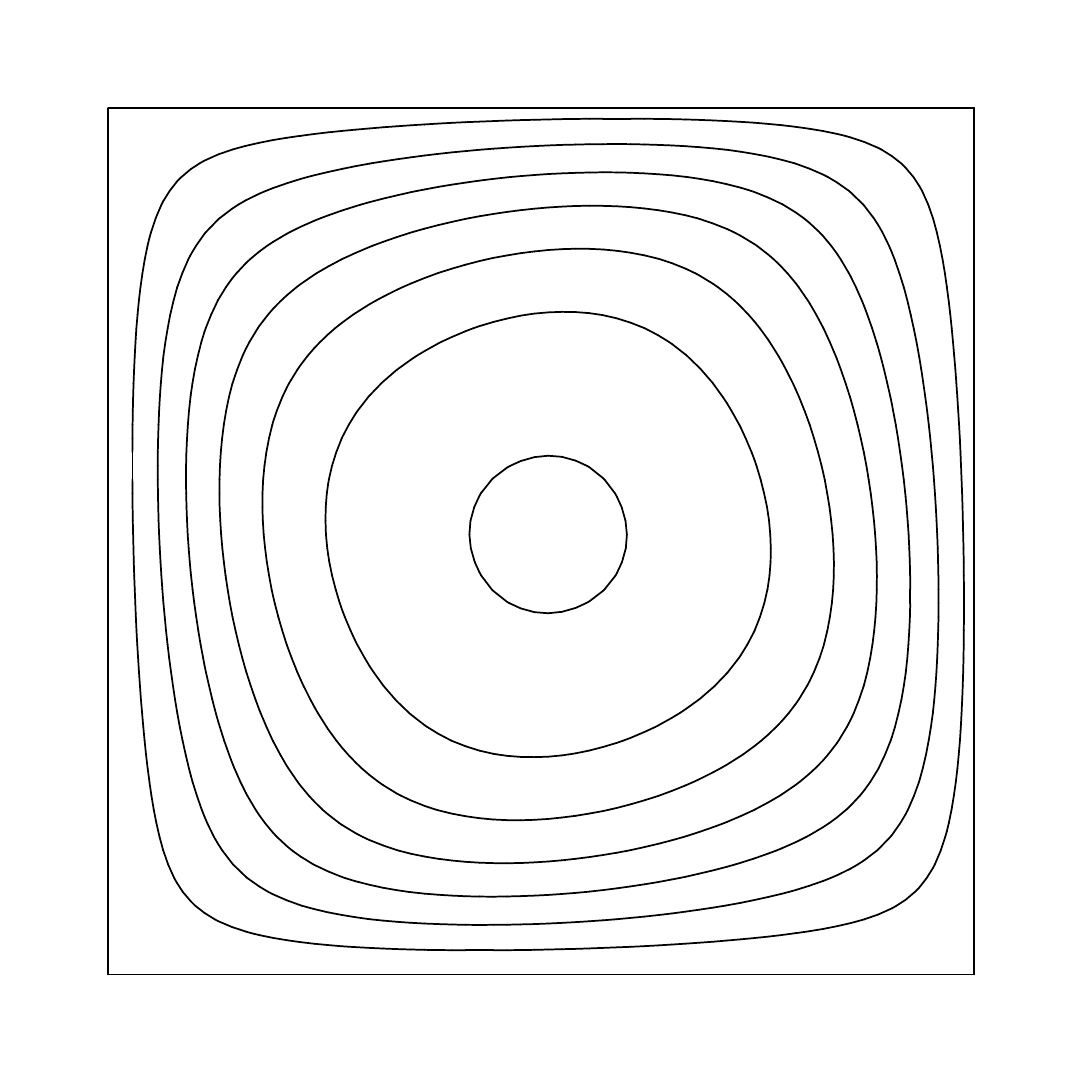} }
 \subfigure[$t = 2$]{ \includegraphics[width=0.45\textwidth,trim=25px 20px 25px 20px, clip]{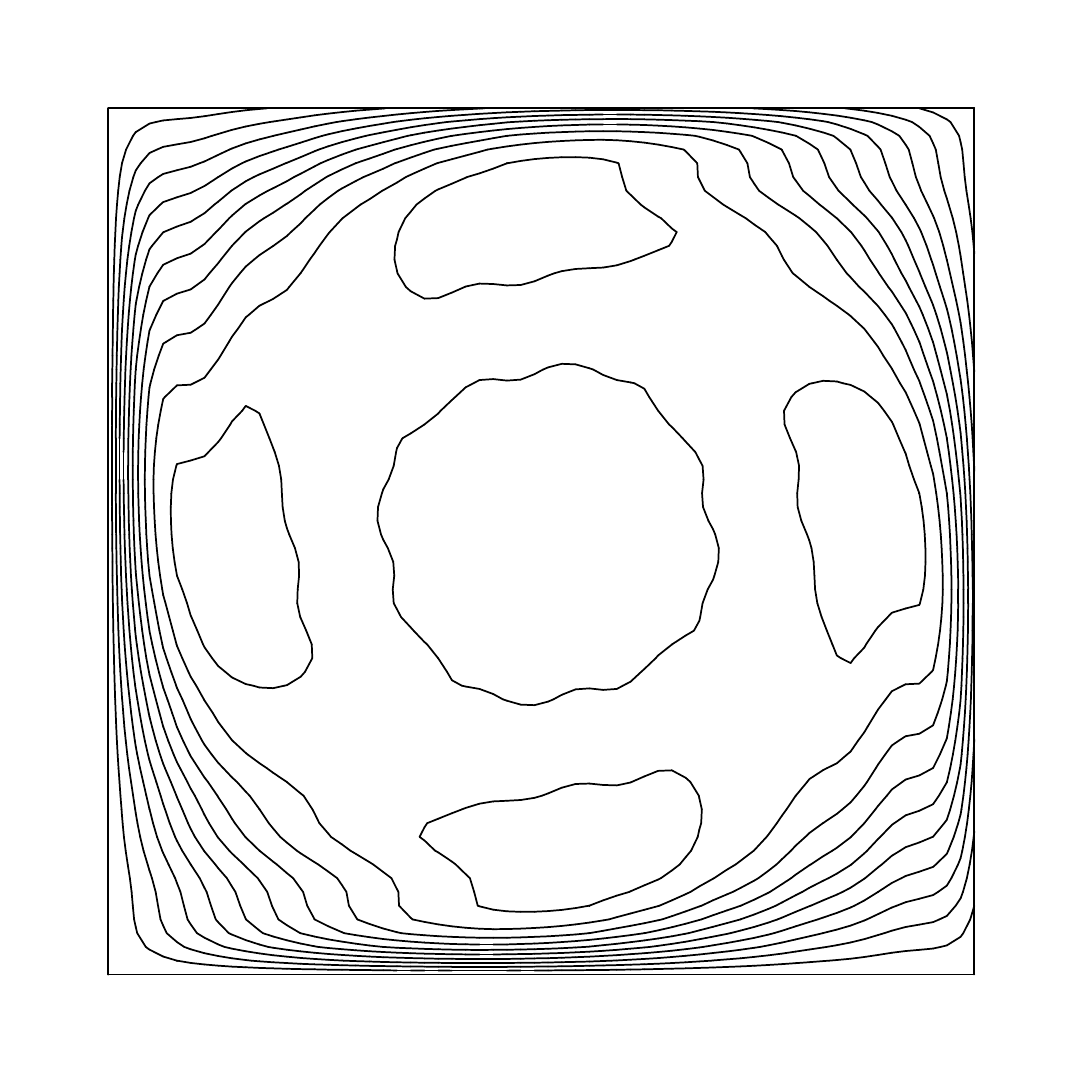} }
 \subfigure[$t = 3$]{ \includegraphics[width=0.45\textwidth,trim=25px 20px 25px 20px, clip]{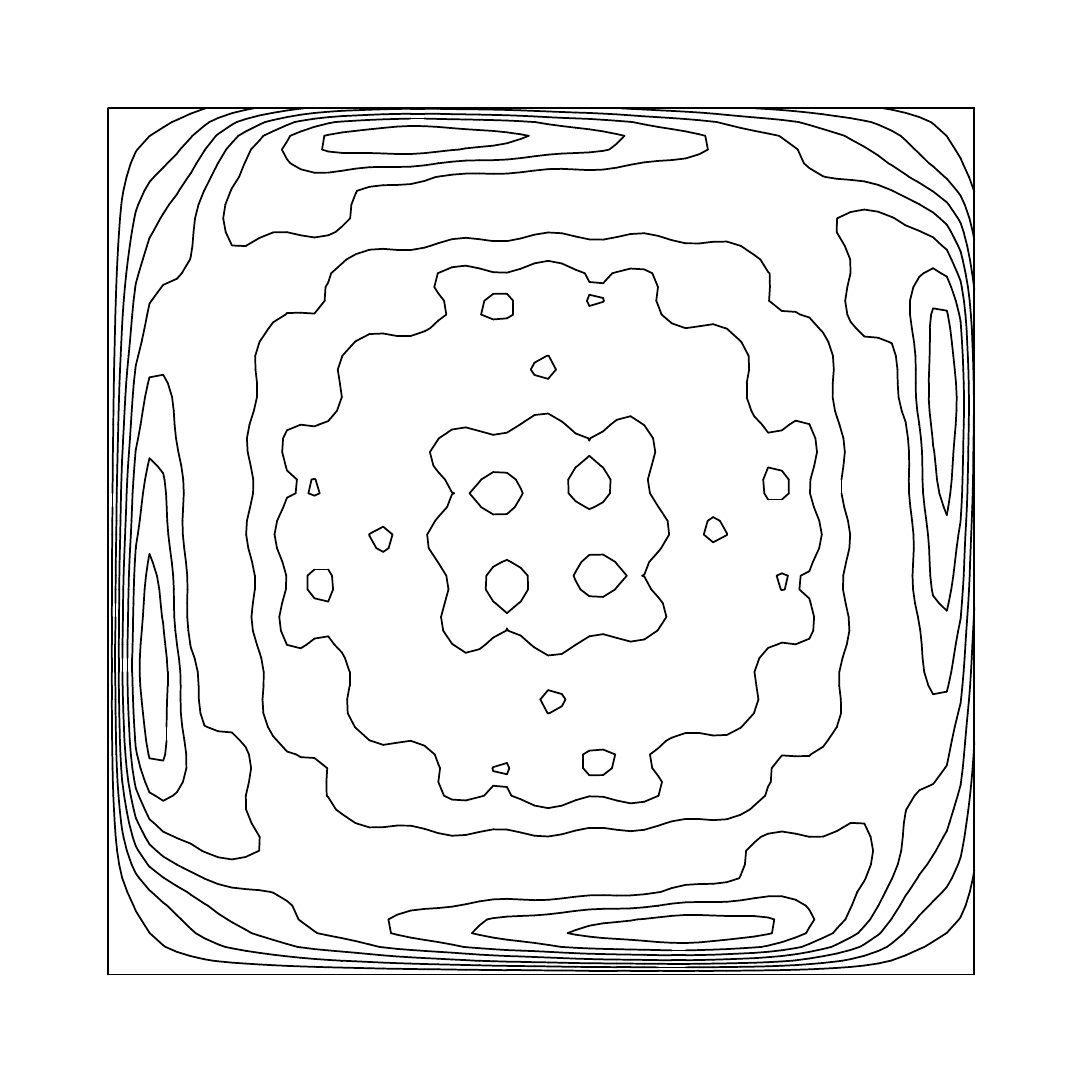} }
 \caption{Isovorticity contours in the $z = 0$ plane reproduced from Young \cite{thesis:ajyoung}.}
 \label{fig:tg_contours_Young}
\end{figure}

\begin{figure}[!p]
 \centering
 \includegraphics[width=0.9\textwidth,trim=130px 445px 145px 70px, clip]{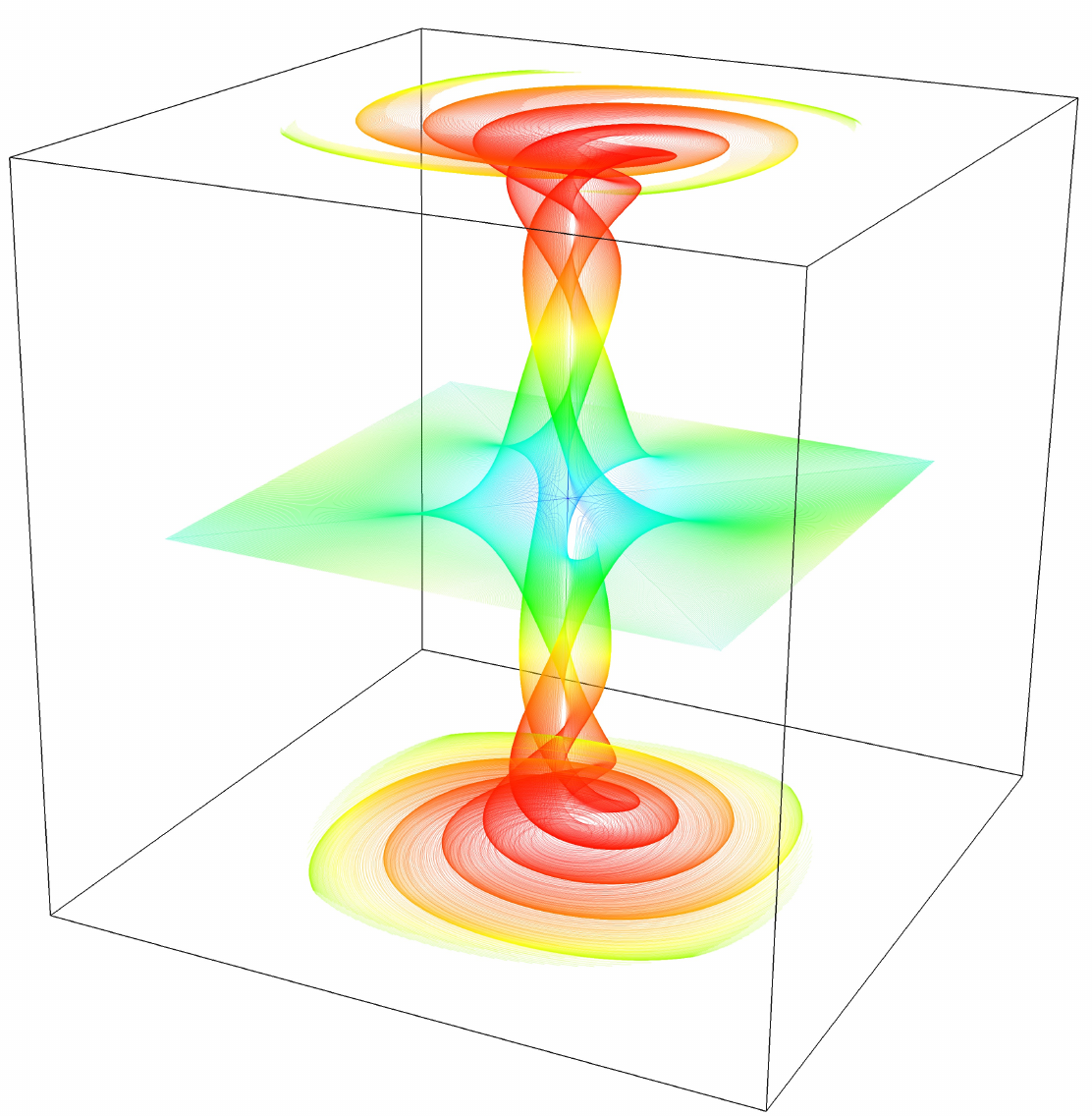}
 \caption{Visualisation of the Taylor-Green vortex at $t = 1.0$. Streamlines and local velocity vectors have been plotted, coloured by the magnitude of vorticity, $\vmod{\omega}$.}
 \label{fig:tg_vortex}
\end{figure}

The flow generated by the initial conditions above possesses certain properties due to symmetry. In particular, there is no flow of mass or momentum through any plane $x,y$ or $z = n\pi$, for $n \in \mathbb{Z}$. The full system of size $L = 2\pi$ will therefore contain 8 isolated sub-domains. Despite simulating the full system, it is common to consider the contents of one of these sub-domains, as illustrated in figures \ref{fig:tg_slice} -- \ref{fig:tg_vortex}. 

The velocity field plots in figure \ref{fig:tg_slice} are almost indistinguishable from those in Young (figures 5.2 -- 5.5), reproduced in figure \ref{fig:tg_slice_Young}. The contour plots for various values of $\omega = \left\lvert\vec{\omega}\right\rvert$ presented in figure \ref{fig:tg_contours} show good agreement (see figures 5.9 -- 5.9, reproduced in figure \ref{fig:tg_contours_Young}). It should be highlighted that our results are from a $256^3$ lattice and are therefore smoother than those of Young, which are from a $128^3$ lattice. Attention should therefore be directed to figure 1(b) of Brachet \textit{et al.} \cite{Brachet:1983p411} for comparison.

\clearpage

\begin{figure}[tbf!]
 \centering
 \subfigure[Shell averaged with $\Delta k = 1$]{
  \label{sfig:tg_energy_spec:dk1}
  \includegraphics[width=0.85\textwidth]{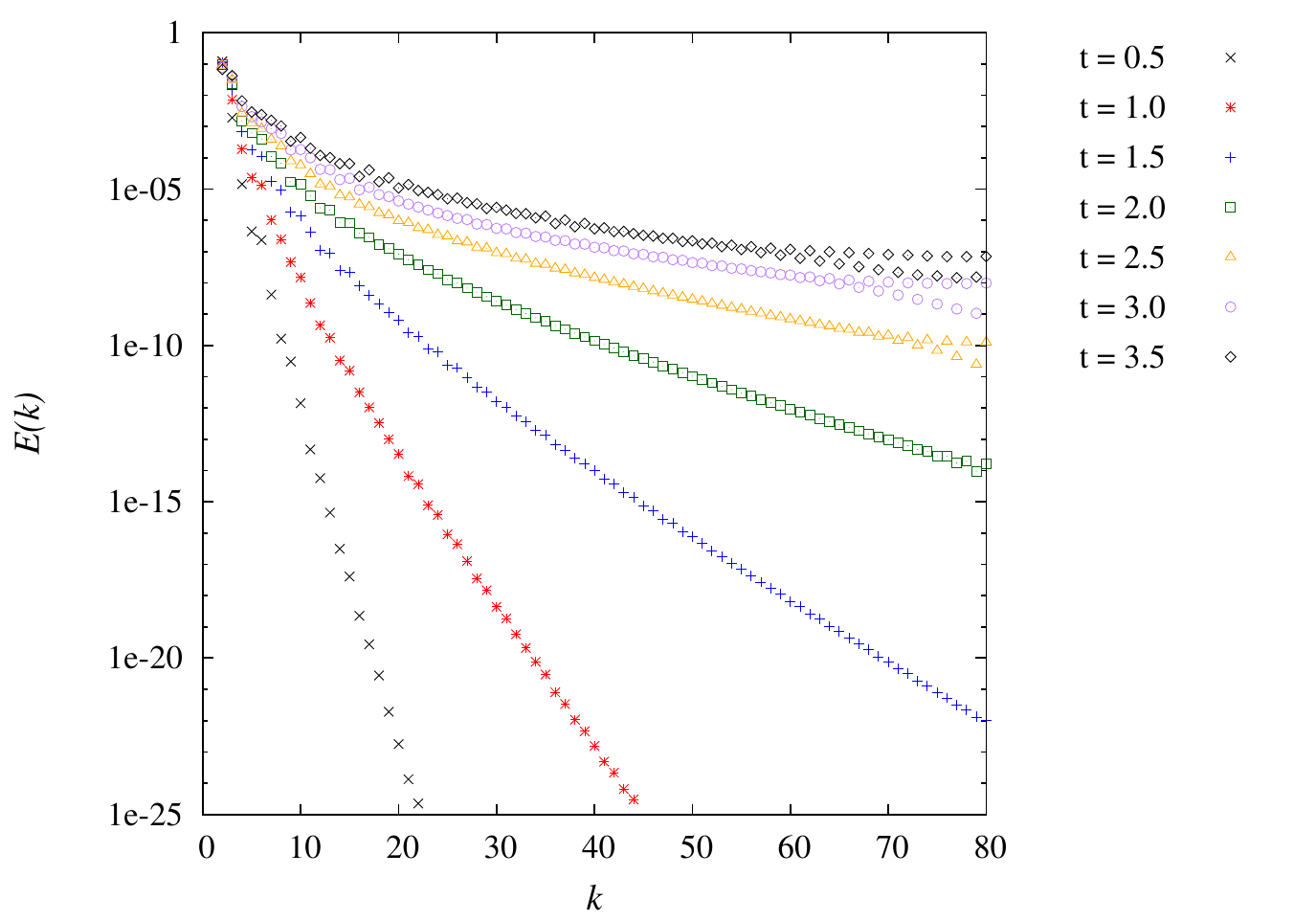}
 }\vspace{1em}
 \subfigure[Shell averaged with $\Delta k = 2$]{
  \label{sfig:tg_energy_spec:dk2}
  \includegraphics[width=0.85\textwidth]{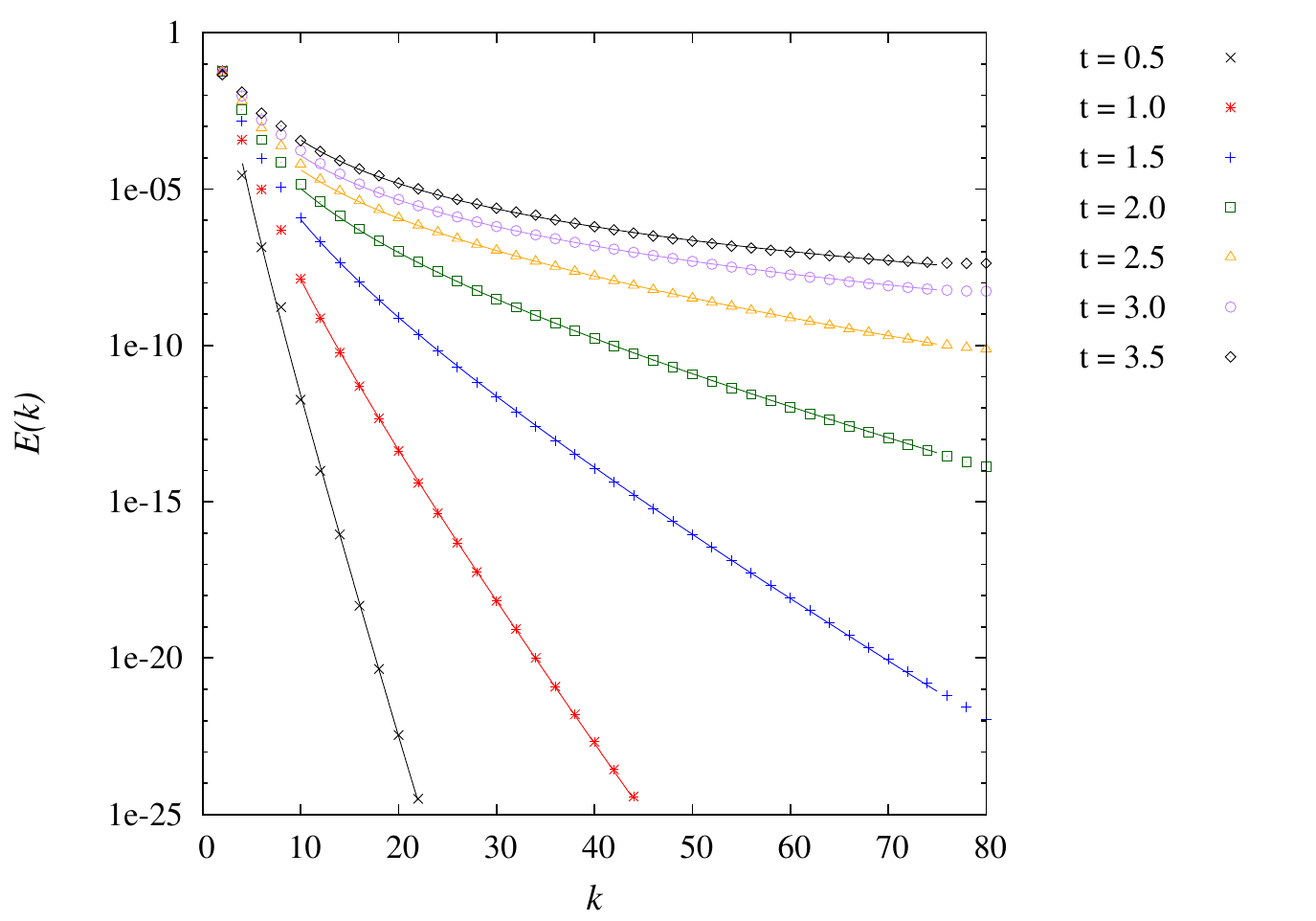}
 }
 \caption{Evolution of the energy spectrum for the Taylor-Green vortex. The solid lines in (b) show the fit of Brachet \etal\ \cite{Brachet:1983p411}.}
 \label{fig:tg_energy_spec}
\end{figure}

The time evolution of the energy spectrum allows us to make a quantitative comparison. Figure \ref{fig:tg_energy_spec} shows the energy spectrum at various times, shell-averaged with $\Delta k = 1,2$, and should be compared to figure 3 of Brachet \etal\ \cite{Brachet:1983p411} or (noting the log-scale) figures 5.10 and 5.11 of Young \cite{thesis:ajyoung}. Comparison of the energy spectra gives excellent agreement. We note the oscillatory behaviour for larger $k$ observed by Brachet \textit{et al.} for $t = 2.5$ and above was also present here. This was eliminated by increasing the width of the shell average to $\Delta k = 2$, as can be seen in figure \ref{sfig:tg_energy_spec:dk2}.

The energy spectrum was then assumed by Brachet \etal\ to take the form
\begin{equation}
 \label{eq:tgfit}
 E(k,t) = A(t) k^{-n(t)} e^{-2\delta(t) k} \ ,
\end{equation}
and a least-squares fit of $\log E(k,t)$ was used to find $A(t), n(t)$ and $\delta(t)$; values of the latter two may be found in \cite{Brachet:1983p411} and are listed in table \ref{tbl:tg_fit}. For comparison, we fit the form \eqref{eq:tgfit} to our data using the values of $n(t), \delta(t)$ for the $256^3$ computation. For fitting, we used the range $10 \leq k \leq 75$ as used in the original work, with $4 \leq k \leq 22$ used for $t=0.5$. The only exception was $t = 1.0$ when we used the range $10 \leq k \leq 44$, which was due to our data flattening off. This was also observed by Young \cite{thesis:ajyoung}. The parameter $A(t)$ was found by fitting
\begin{equation}
 \log E(k,t) = \log A(t) - n(t) \log k - 2\delta(t) k
\end{equation}
to the data, as this was far more accurate than fitting equation \eqref{eq:tgfit} directly. The obtained fits are plotted as solid lines in figure \ref{sfig:tg_energy_spec:dk2}, where we also only plot every other point so they can be seen clearly. Since $A(t)$ is just a scaling and does not affect the shape of the curves, once again we conclude that the agreement is excellent.

See also Brachet, Meneguzzi, Vincent, Politano and Sulem \cite{Brachet:1992p1170} for a more recent, larger-scale investigation of the Taylor-Green vortex.

\begin{table}[tb!]
\begin{center}
\begin{tabular}{l|lll}
$t$ & $A(t)$ & $n(t)$ & $\delta(t)$ \\ 
\hline
 0.5 & 221.4 & 4.31 & 1.107 \\ 
 1.0 & 15.04 & 5.14 & 0.451 \\ 
 1.5 & 3.592 & 4.86 & 0.192 \\ 
 2.0 & 1.643 & 4.50 & 0.08  \\ 
 2.5 & 1.242 & 4.18 & 0.034 \\ 
 3.0 & 5.122 & 4.59 & 0.005 \\ 
 3.5 & 10.91 & 4.56 & -0.002 \\
\end{tabular}
\vspace{1em}
\caption[Parameters for the fit to the energy spectrum of the Taylor-Green vortex proposed by Brachet \etal\ (1983)]{Parameters for the fit to the energy spectrum of the Taylor-Green vortex given in equation \eqref{eq:tgfit}.}
\label{tbl:tg_fit}
\end{center}
\end{table}

\clearpage

\section{Isotropy}
Since we are attempting to simulate isotropic turbulence, it is important to ensure that the velocity field does indeed satisfy this property. This is done using the method presented in Young \cite{thesis:ajyoung}.

\begin{figure}[tbf!]
 \centering
  \includegraphics[width=0.75\textwidth]{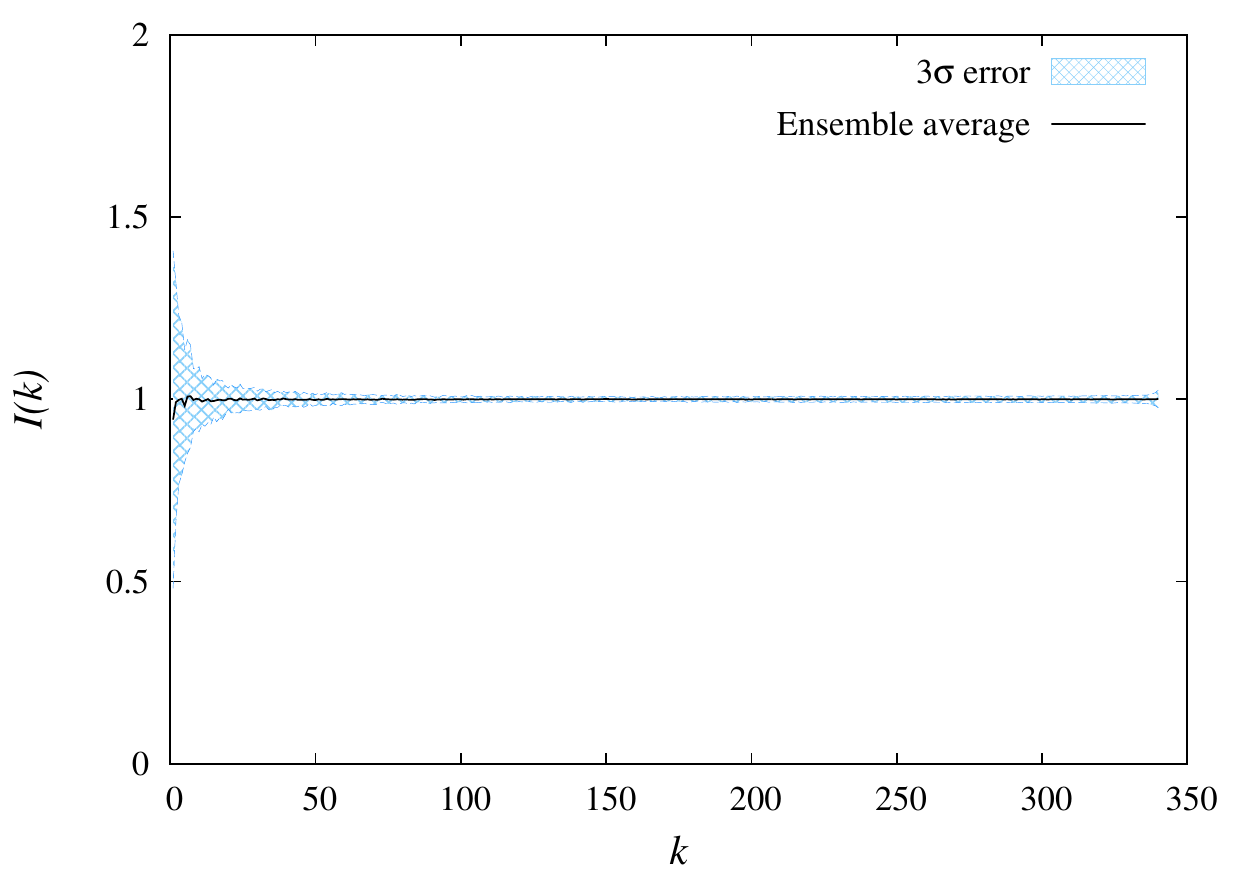}
 \caption{Ensemble averaged isotropy spectrum for an $N = 1024$ lattice.}
 \label{fig:isotropy_spec}
\end{figure}

A random unit vector $\vec{z}(\vec{k})$ which is not parallel to $\vec{k}$ (that is, it satisfies $\vec{z}(\vec{k}) \cdot\hat{\vec{k}} \neq 1$) is chosen for all wavevectors, and from it we define two mutually orthogonal unit vectors
\begin{align}
 \vec{e}_1(\vec{k}) = \frac{\vec{k}\times\vec{z}(\vec{k})}{\lvert \vec{k}\times\vec{z}(\vec{k}) \rvert} \ , \qquad\qquad \vec{e}_2(\vec{k}) = \frac{\vec{k}\times\vec{e}_1(\vec{k})}{\lvert \vec{k}\times\vec{e}_1(\vec{k}) \rvert} \ .
\end{align}
These are used to compute the average energy in these two directions,
\begin{equation}
 I_j(k,t) = \left\langle \lvert \vec{e}_j(\vec{k}) \cdot \vec{u}(\vec{k},t) \rvert^2 \right\rangle\ , \qquad j = 1,2 \ ,
\end{equation}
which should be the same for isotropic turbulence. A measure of the degree of isotropy is, therefore, the ratio
\begin{equation}
 I(k,t) = \sqrt{\frac{I_1(k,t)}{I_2(k,t)}} \ .
\end{equation}
As seen plotted in figure \ref{fig:isotropy_spec}, while individual realisations fluctuate, the ensemble average is close to 1 for all values of $k$. The increase in the deviation from unity as one moves towards low $k$ is due to the resolution of these shells, since they contain fewer points the statistics are not as good.

A representative value can be obtained by averaging over all of Fourier space. Values for a variety of simulation sizes can be found in table \ref{tbl:isotropy} and are very satisfactory, allowing us to conclude that there is not any significant deviation from isotropy in our simulations. The uncertainty on the scale of the mean, $\sigma / \langle I \rangle$, decreases as $N$ is increased, since the large $k$ modes are more isotropic than the low $k$ modes and we are including more of them in the simulation.

\begin{table}[tb!]
\begin{center}
\begin{tabular}{r||cccc}
$N$ & 128 & 256 & 512 & 1024 \\
\hline
 $\langle I \rangle \pm \sigma$ & $1.002 \pm 0.009$ & $1.005 \pm 0.008$ & $0.9979 \pm 0.0077$ & $1.0002 \pm 0.0034$
\end{tabular}
\vspace{1em}
\caption{Representative values for the total isotropy for various lattice sizes.}
\label{tbl:isotropy}
\end{center}
\end{table}

\section{Time-averaged statistics}\label{sec:time_av}
To help establish the reliability of the code, we look at several key turbulence parameters
and compare our results to those obtained by other authors.

So far, shell and ensemble averaging have been used to present time-varying statistics such as the fluctuation of total energy or the dissipation rate in forced and decaying simulations. These quantities were presented as time series. For stationary turbulence, once we reach steady state, rather than run multiple simulations an ensemble can be generated by looking at the field at various times. If the sample time between realisations is longer than the typical correlation time scales of the system, we can consider the times to be uncorrelated realisations of the flow. From this new ensemble, we can calculate a single mean value for various parameters of the stationary flow and their associated error.

First, we must discard the transient data collected while our system evolved from its initial condition into a stationary solution of the Navier-Stokes equation. Typically, this takes around 10 eddy turnover times. The remaining data is then sampled every $\Delta t$ and used to calculated a mean value. Here, $\Delta t = \tau = L/u$, the eddy turnover time (although it could be argued that $\Delta t = \tau/2$ is sufficient). For the simulations in this work, we collect data for at least 15$\tau$ after the transition period. Time averaged value for the parameter $A$ is then calculated as
\begin{equation}
 \overline{A} = \frac{1}{T} \sum_{t_i \in \mathbb{T}} A(t_i) \ ,
\end{equation}
where T is the number of realisations in our ensemble, $\mathbb{T}$. The overline indicates an average over time, if the system is ergodic then this becomes equivalent to an ensemble average, and we write $\overline{A} = \langle A \rangle$. An estimate of the error is given by the standard deviation,
\begin{equation}
 \sigma_A^2 = \langle A^2 \rangle - \langle A \rangle^2 \ ,
\end{equation}
although we occasionally refer to the \emph{standard error} on the mean, denoted $\hat{\sigma}$, by which we mean
\begin{equation}
 \hat{\sigma} = \frac{\sigma}{\sqrt{T}} \ .
\end{equation}

\subsection{Kolmogorov constant}\label{subsec:kol_const}
\begin{figure}[tbf!]
 \centering
  \includegraphics[width=0.8\textwidth]{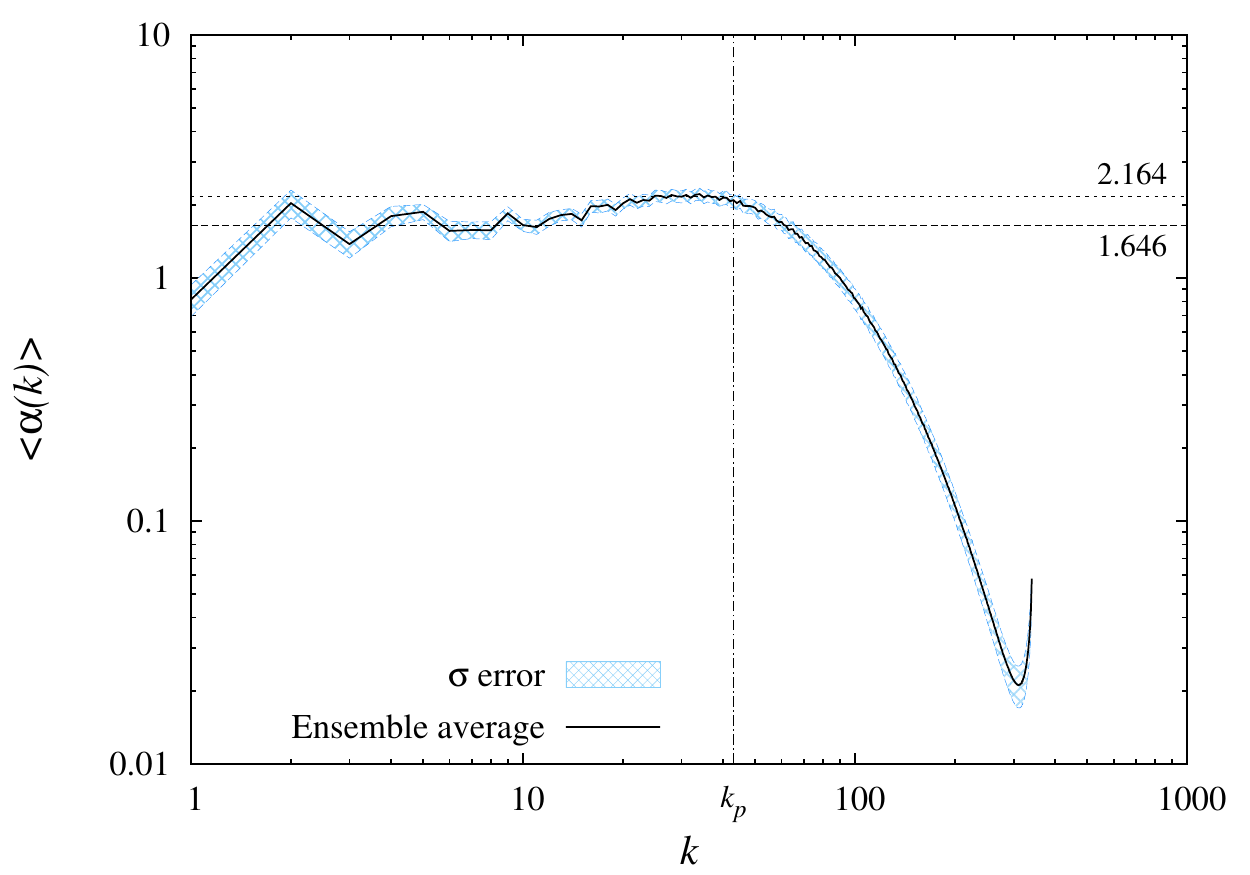}
 \caption{Time averaged compensated energy spectrum for an $R_\lambda \sim 280$ simulation. (-- -- --) Averaged value of $\alpha$. ($\cdots$) Anomalous plateau. (-- $\cdot$ --) $k_p \sim 43$.}
 \label{fig:kol_const_E}
\end{figure}

In Fourier space, the famous Kolmogorov spectrum for the inertial range, given in equation \eqref{eq:K41_Ek}, involves a constant, $\alpha$, known as the Kolmogorov constant. Rearranging the K41 energy spectrum in terms of a wavenumber dependent $\alpha(k)$ gives the \emph{compensated} energy spectrum,
\begin{equation}
 \alpha(k) = \varepsilon^{-2/3} k^{5/3} E(k) \ ,
\end{equation}
which shows the variation of this `constant' with wavenumber. Regions in which this spectrum is flat thus take the Kolmogorov form, $k^{-5/3}$, with $\alpha = \textrm{constant}$. Figure \ref{fig:kol_const_E} shows the compensated energy spectrum for an $N = 1024$ simulation. The spectrum has been time averaged with $\Delta t = \tau$, allowing us to plot an estimate of the error.

As noted by Yeung and Zhou \cite{Yeung:1997p1746}, there appears to be two plateaus: one at lower $k$ and one at medium $k$. In the paper, the authors highlight how the location of the inertial range has been misidentified in many numerical simulations, causing the value of $\alpha$ to be overestimated. They present arguments for the plateau at lower $k$ corresponding to the actual inertial range. This is based on the peak of the dissipation spectrum coinciding with the higher plateau, hence it cannot correspond to inertial behaviour. This is also observed in our data, with the peak of the dissipation spectrum at $k_p \sim 43$, indicated in figure \ref{fig:kol_const_E}. Ishihara, Gotoh and Kaneda \cite{Ishihara:2009p165} also provide a discussion of this misidentification.

\begin{figure}[tbf!]
 \centering
  \includegraphics[width=0.8\textwidth]{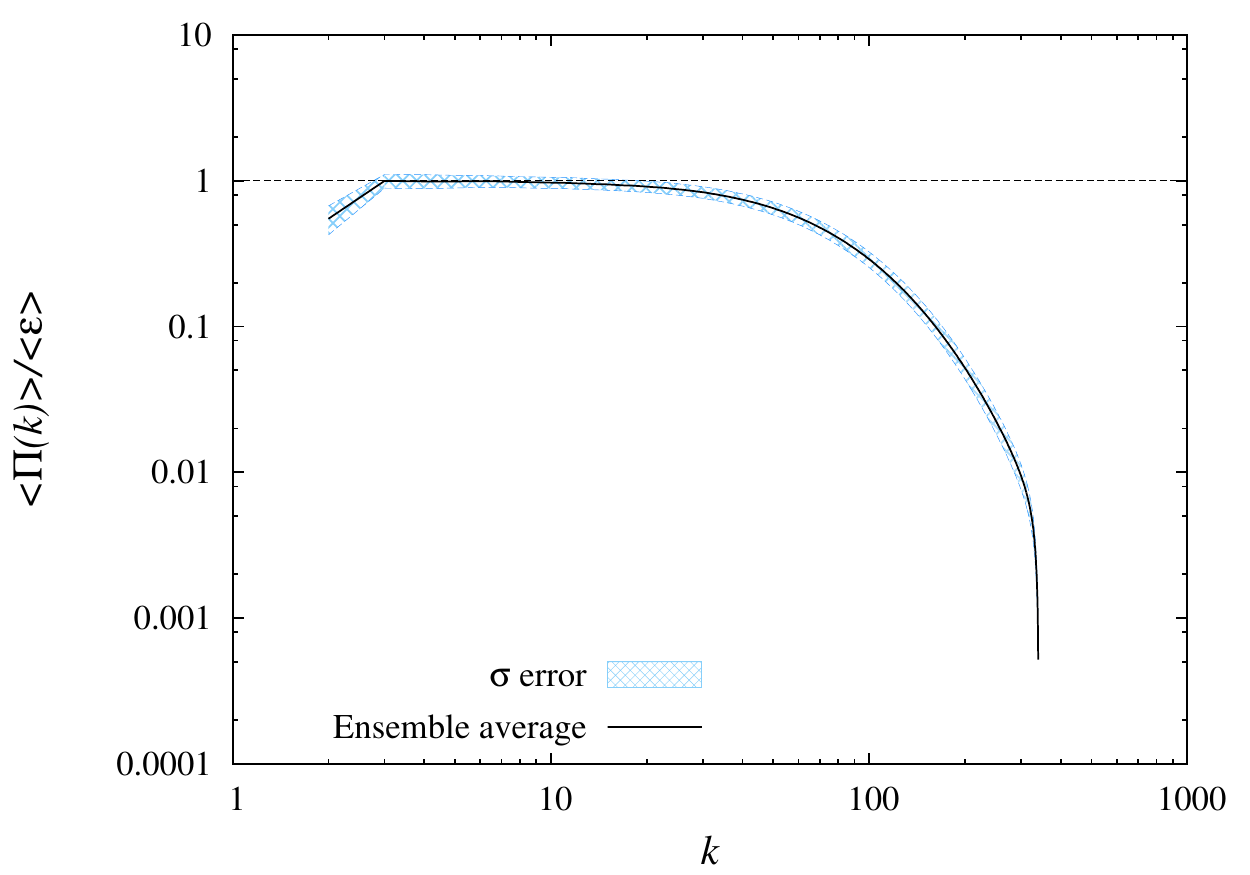}
 \caption{Time averaged scaled transport power spectrum for $R_\lambda \sim 280$.}
 \label{fig:kol_const_P}
\end{figure}

To find an estimate for the value of $\alpha$, we turn our attention to the scaled transport power spectrum. As mentioned in section \ref{subsec:cascade}, in the inertial subrange of wavenumbers, energy is transferred at the dissipation rate, such that the flux through a wavenumber satisfies
\begin{equation}
 \Pi(k,t) = \varepsilon(t) \ ,
\end{equation}
making $\langle \Pi(k) \rangle/\varepsilon$ a simple test for an inertial range. In figure \ref{fig:kol_const_P}, this can be seen to be unity for the range $3 \leq k \leq 7$, corresponding to the lower $k$ plateau in figure \ref{fig:kol_const_E}. To obtain a mean value for this plateau, we average over the range to find the value
\begin{equation}
 \alpha = 1.646 \pm 0.144 \ .
\end{equation}
This value is highlighted in figure \ref{fig:kol_const_E} by the dashed line, along with the value corresponding to the `anomalous' plateau of 2.164 (dotted line).

Ishihara \etal\ \cite{Ishihara:2009p165} found $\alpha = $ 1.5 -- 1.7 in their high-$R_\lambda$ simulations, placing our result within their range. In fact, studying the data found in Gotoh and Fukayama \cite{Gotoh:2001p653}, one finds the value for their most similar Reynolds number, $R_\lambda = 284$, to be 1.64, in excellent agreement with the above. They quote an average value of $\alpha = 1.65 \pm 0.05$, and our result agrees within one error unit. Yeung and Zhou \cite{Yeung:1997p1746} found a value of 1.62 for $R_\lambda = 240$. Note that the Kolmogorov constant can be measured from one- or three-dimensional energy spectra using the relation $\alpha = (55/18) C_1$, where $C_1$ is measured from one-dimensional spectra \cite{Yeung:1997p1746}. Comparison can then also be made to the experimental value obtained by Sreenivasan \cite{Sreenivasan:1995p154} of $C_1 = 0.53 \pm 0.055$ which gives $\alpha = 1.62 \pm 0.17$. Mydlarski and Warhaft \cite{Mydlarski:1996p178} found the experimental value $C_1 = 0.51$, giving $\alpha = 1.56$.
Further values for comparison obtained using DNS and LES can be found in \cite{Yeung:1997p1746,Gotoh:2001p653,Wang:1996p1041}. A discussion can also be found in Monin and Yaglom volume 2 \cite{MoninYaglom:vol2}.

\begin{figure}[bp!]
 \centering
  \subfigure[Our DNS data]{
   \label{sfig:kol_const_both}
   \includegraphics[width=0.53\textwidth]{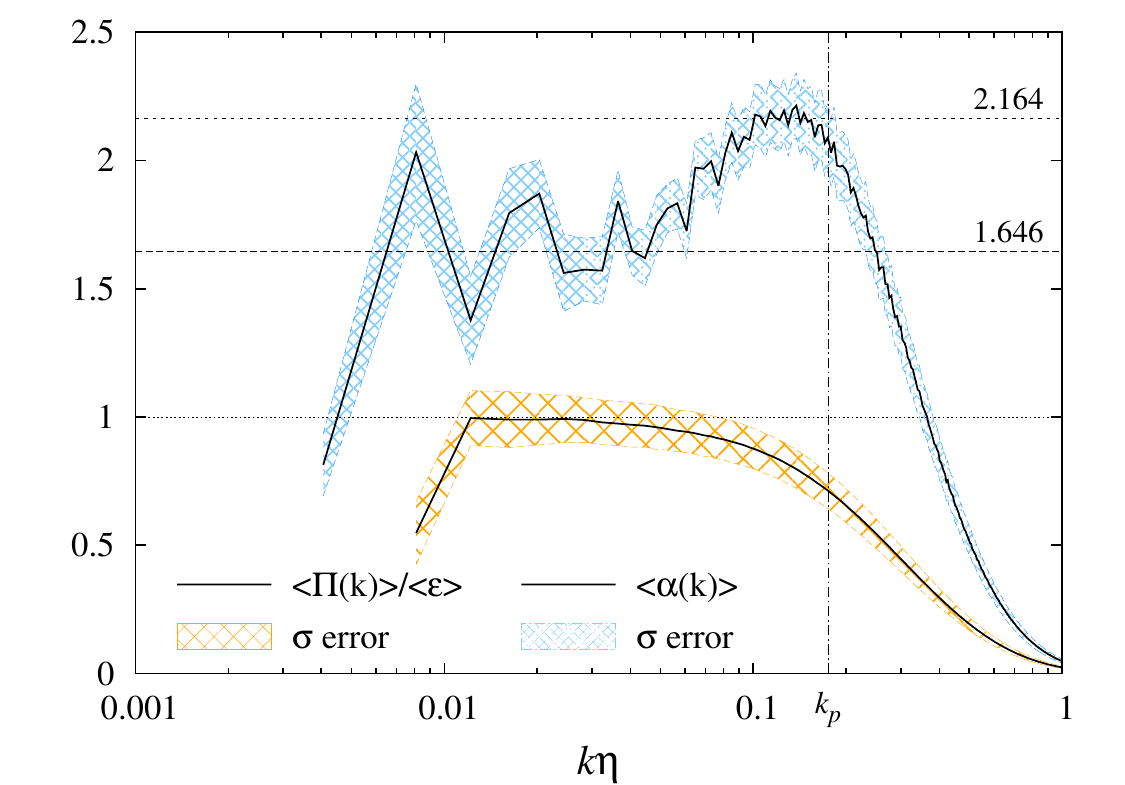}
  }
  \subfigure[Reproduced from Ishihara \etal\ \cite{Ishihara:2009p165}]{
   \label{sfig:kol_const_both_Ishihara}
   \vspace{-1em}\includegraphics[width=0.42\textwidth,trim=10px -8.5px 10px 0,clip]{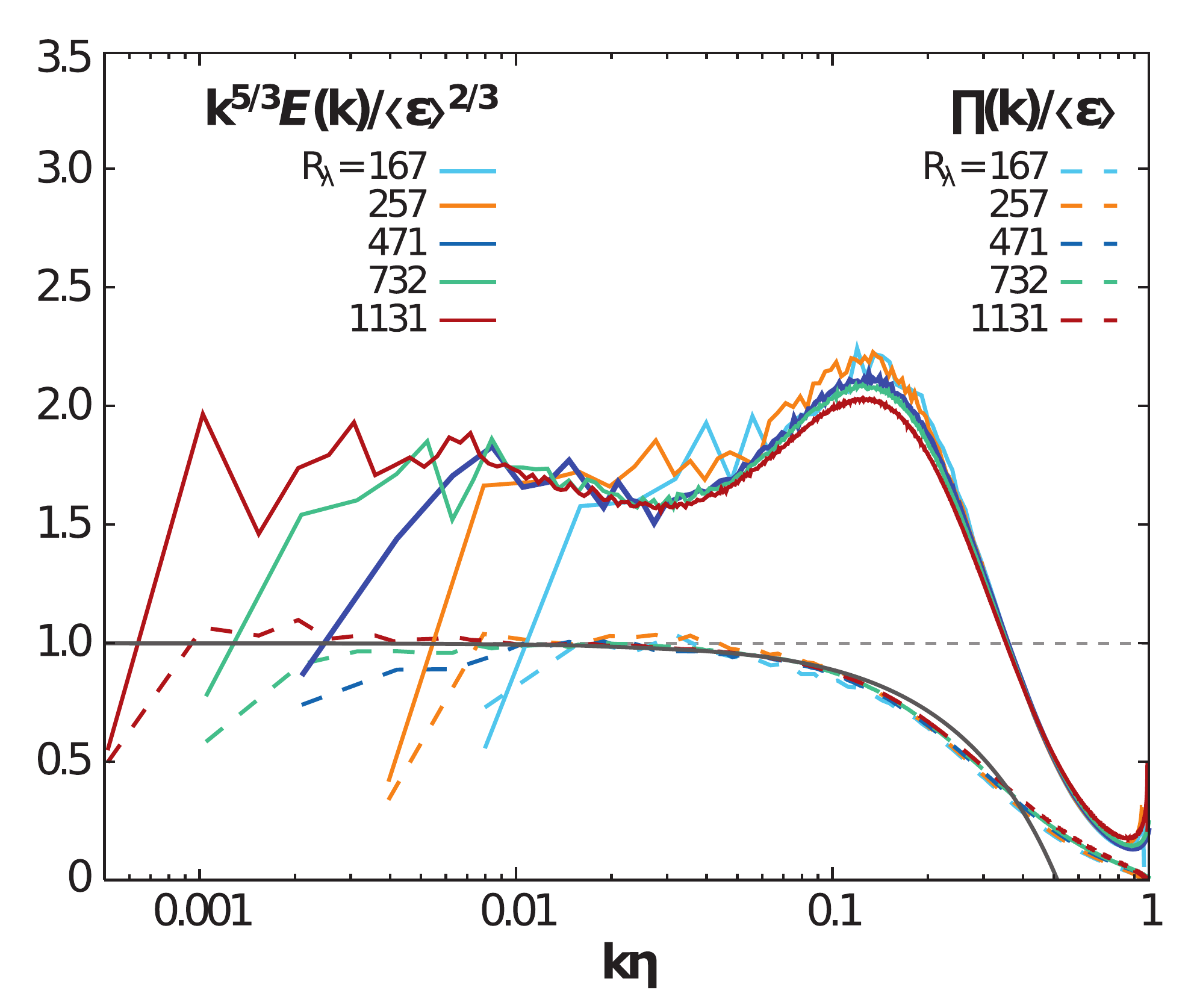}
  }
 \caption{Compensated energy and scaled transport power spectra for $R_\lambda = 276$.}
 \label{fig:kol_const_both}
\end{figure}

Finally, figures \ref{fig:kol_const_E} and \ref{fig:kol_const_P} are presented in figure \ref{sfig:kol_const_both} for direct comparison to figure 3(a) of Ishihara \etal\ \cite{Ishihara:2009p165}, reproduced as figure \ref{sfig:kol_const_both_Ishihara}.

\subsection{Longitudinal velocity derivative skewness}\label{subsec:skewness}
The skewness, as defined in section \ref{subsec:stats_PP}, is a very sensitive parameter. It is computed in both real and Fourier space to obtain values
\begin{equation}
 S_x = 0.551 \pm 0.015 \qquad\qquad \textrm{and} \qquad\qquad S_k = 0.557 \ ,
\end{equation}
respectively.
The Fourier-space result has been calculated using the time averaged transfer spectrum, and as such it is difficult to associate an error with it. However, agreement with the real-space result is excellent.

This can be compared to other stationary simulations such as Ishihara \etal\ \cite{Ishihara:2009p165}, who find $S \sim 0.5$, or Machiels \cite{Machiels:1997p148} who quotes a result of $S = 0.51$ for $R_\lambda \simeq 190$. Vincent and Meneguzzi \cite{Vincent:1991p325} found a value of $S = 0.5$ for $R_\lambda \sim 150$, which is the same as Kerr \cite{Kerr:1985p191} for $R_\lambda < 80$. Gotoh, Fukayama and Nakano \cite{Gotoh:2002p627} performed a series of simulations on $512^3$ and $1024^3$ grids. For $R_\lambda = 284$, the closest Reynolds number to that used here, they found $S = 0.531$. The average value of their $R_\lambda = 284$ and 381 runs gives $S = 0.553$. Jim\'enez, Wray, Saffman and Rogallo \cite{Jimenez:1993p634} found $S = 0.525$ for $R_\lambda = 168.1$. Wang, Chen, Brasseur and Wyngaard \cite{Wang:1996p1041} found a value of $S = 0.545$ for the largest forced run with $R_\lambda = 195$. Sreenivasan and Antonia \cite{Sreenivasan:1997p334} comment on skewness increasing monotonically with Reynolds number and present a collection of data from DNS and experiment to support this. This can also be observed in \cite{Ishihara:2007p612}.

\subsection{Dissipation-scaled energy spectrum}
\begin{figure}[btf!]
 \centering
  \includegraphics[width=0.7\textwidth]{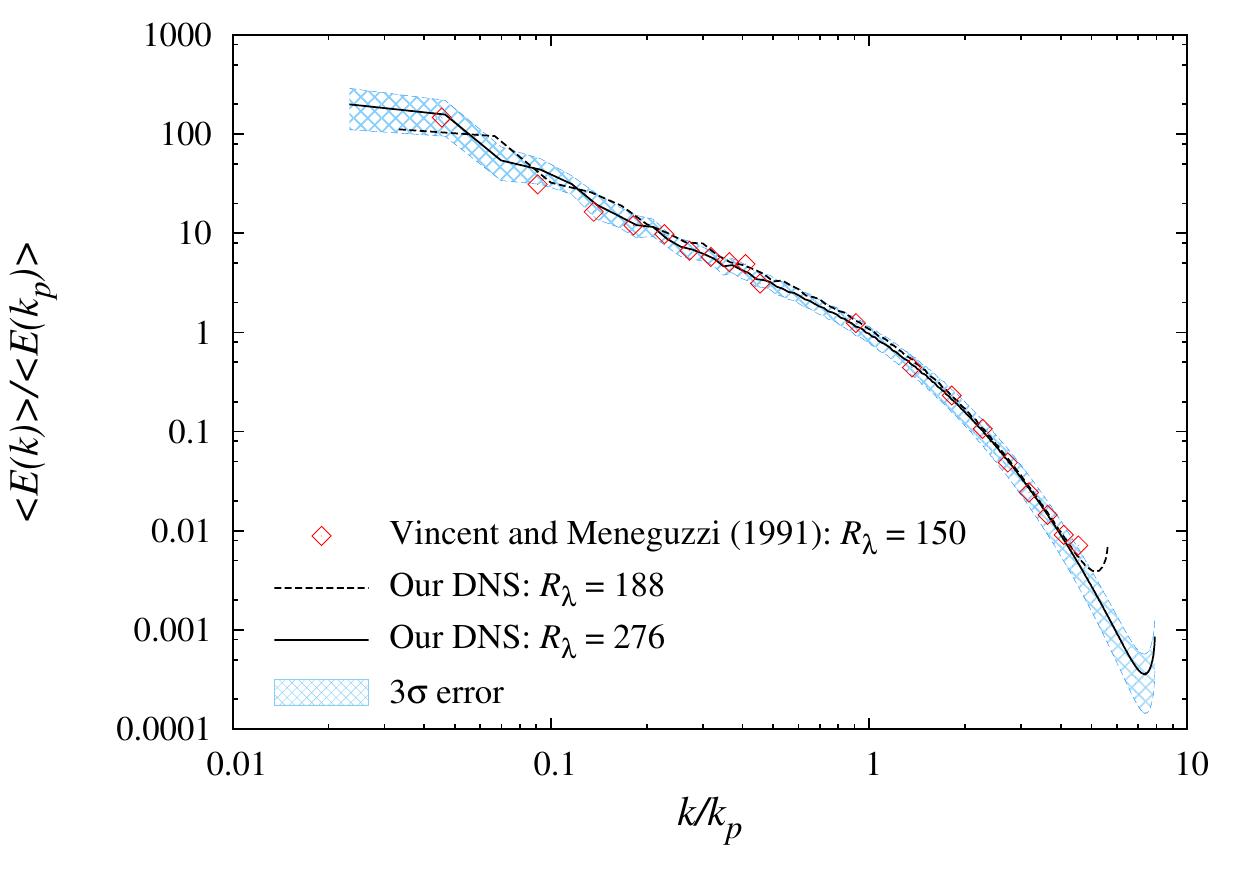}
 \caption{Comparison of dissipation-scaled energy spectra.}
 \label{fig:scaled_E}
\end{figure}

She, Chen, Doolen, Kraichnan and Orszag \cite{She:1993p506} found that the energy spectra from various Reynolds numbers collapse when scaled on the peak of the dissipation spectrum; that is, $k/k_p$ and $E(k)/E(k_p)$. The authors present the collapse of DNS for Reynolds numbers $R_\lambda \sim 70$ to 200, along with experimental data. In figure \ref{fig:scaled_E} we plot our own DNS results for two Reynolds numbers, along with data points from Vincent and Meneguzzi \cite{Vincent:1991p325} for $R_\lambda = 150$ for comparison. Note that the points have been extracted by hand from their figure. The data is seen to collapse well. The error shown is that for $R_\lambda = 276$.

\vfill

\section{Advection of a passive scalar}\label{sec:scalar}
As an additional test of the code reliability, the advection of a passive scalar was implemented. A passive scalar is an additional scalar quantity at each lattice site whose equation of motion depends on the underlying velocity, but does not itself influence the velocity field. It is therefore a good way to study the mixing nature of the turbulent field as it transports this quantity around.

The scalar is denoted $\theta(\vec{x},t)$ in real space and satisfies the equation of motion
\begin{equation}
 \frac{\partial \theta(\vec{x},t)}{\partial t} + \vec{u}(\vec{x},t)\cdot\vec{\nabla} \theta(\vec{x},t) = \kappa \nabla^2 \theta(\vec{x},t) + f_\theta(\vec{x},t) \ ,
\end{equation}
where $\kappa$ is the thermal diffusivity of the scalar and $f_\theta$ is a forcing term.
We can define a useful dimensionless parameter,
\begin{equation}
 \textrm{Pr} = \frac{\nu_0}{\kappa} \ ,
\end{equation}
known as the Prandtl number (or Schmidt number when $\theta$ is a concentration and $\kappa$ the mass diffusivity), which effectively compares whether convection by the velocity field ($\textrm{Pr} > 1$) or conduction ($\textrm{Pr} < 1$) is dominant, much like the Reynolds number compares inertial and viscous forces.

We deal with the equation of motion in Fourier space,
\begin{equation}
 \left(\frac{\partial}{\partial t} + \kappa k^2 \right) \theta(\vec{k},t) = -i k_\alpha \int d^3j\ u_\alpha(\vec{j},t) \theta(\vec{k}-\vec{j},t) \ .
\end{equation}
In the code, this is evaluated by noting that, if we define
\begin{equation}
 X_\alpha(\vec{k},t) = \int d^3j\ u_\alpha(\vec{j},t) \theta(\vec{k}-\vec{j},t) \ ,
\end{equation}
then in real space this convolution is simply
\begin{equation}
 X_\alpha(\vec{x},t) = u_\alpha(\vec{x},t) \theta(\vec{x},t) \ .
\end{equation}
The procedure is then similar to the time evolution of the velocity field detailed in section \ref{subsec:time_advance}: We Fourier transform to real space and find $X_\alpha(\vec{x},t)$, then transform back to obtain $X_\alpha(\vec{k},t)$. The predicted and corrected solutions are then
\begin{align}
 \theta^P(\vec{k},t+\delta t) &= e^{-\kappa k^2 \delta t} \bigg[ \theta(\vec{k},t) - \delta t \Big( ik_\alpha X_\alpha(\vec{k},t)) \bigg] \\
 \theta(\vec{k},t+\delta t) &= \frac{1}{2} \bigg[ e^{-\kappa k^2 \delta t} \theta(\vec{k},t) + \theta^P(\vec{k},t+\delta t) - \delta t \Big( ik_\alpha X^P_\alpha(\vec{k},t+\delta t)\Big) \bigg] \ .
\end{align}
We can reuse the memory allocated to store the non-linear term for the velocity field, so we need one extra scalar field to store $\vec{k}\cdot\vec{X}(\vec{k},t)$ and one to store our predicted solution, $\theta^P(\vec{k},t)$. Thus, each scalar field requires a total of three additional scalar fields to be stored. Our code is currently written to calculate a single scalar field.

We now focus on the test case we have studied. This was done for $f_\theta = 0$ and $\textrm{Pr} = 1$. The initial distribution for the scalar field was a `hot' slab ($\theta = 1$) placed in the centre of the domain surrounded by `cold' ($\theta = 0$). This is introduced to two evolved stationary flows with $R_\lambda \sim 100$ and $R_\lambda \sim 280$ (\texttt{f512f}, \texttt{f1024a}; see tables \ref{tbl:summary_sims}). As time progresses, the turbulent field spreads the `heat' among the system. Our initial condition violates isotropy and so breaks the symmetry of our system. The scalar is therefore transferred until we restore this symmetry with the scalar randomly distributed.

\begin{figure}[htb]
 \centering
 \subfigure[$t = 0$]{
  \includegraphics[width=0.31\textwidth,trim=130px 430px 130px 40px, clip]{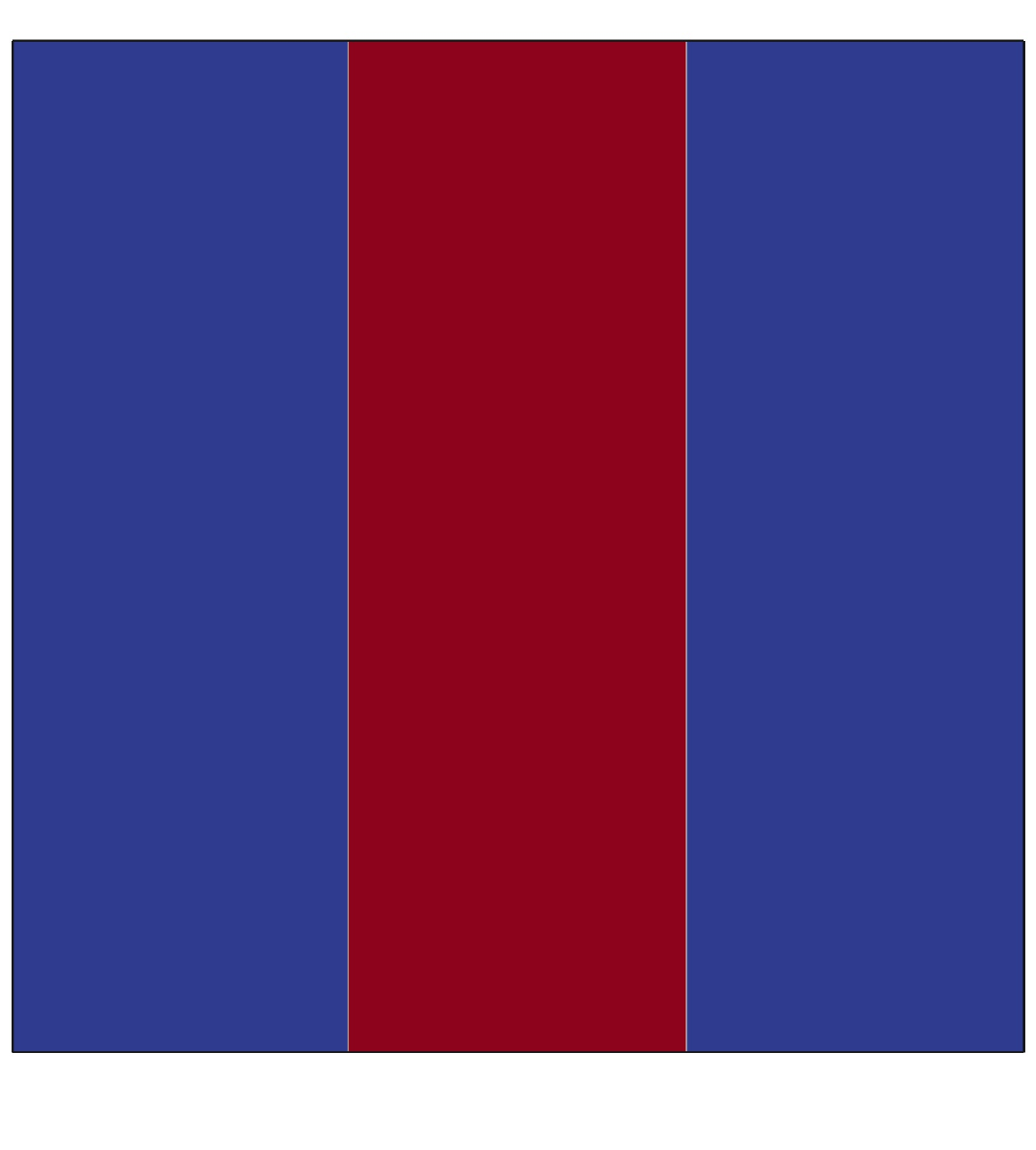}
 }
 \subfigure[$t = 1$]{
  \includegraphics[width=0.31\textwidth,trim=130px 430px 130px 40px, clip]{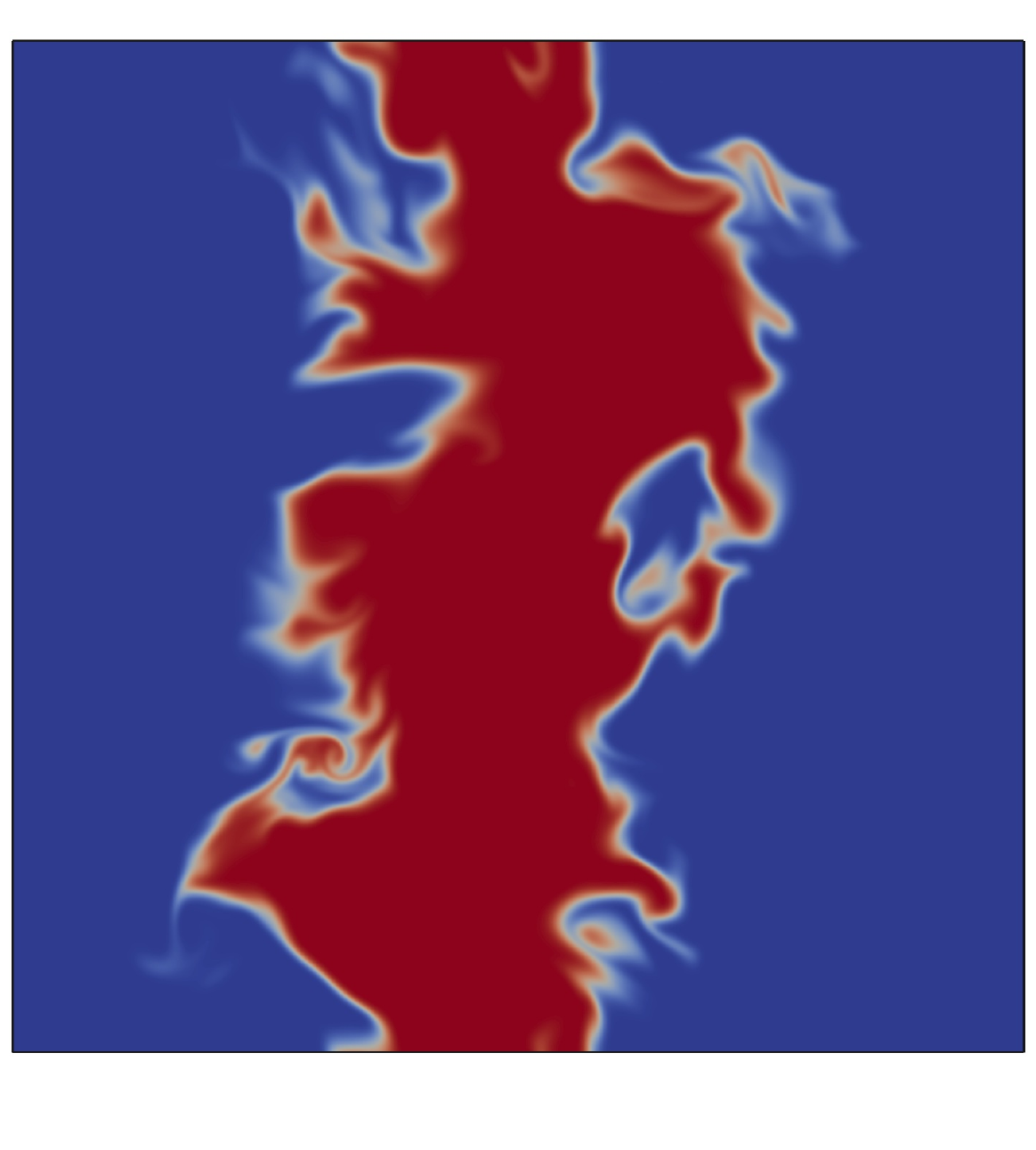}
 }
 \subfigure[$t = 2$]{
  \includegraphics[width=0.31\textwidth,trim=130px 430px 130px 40px, clip]{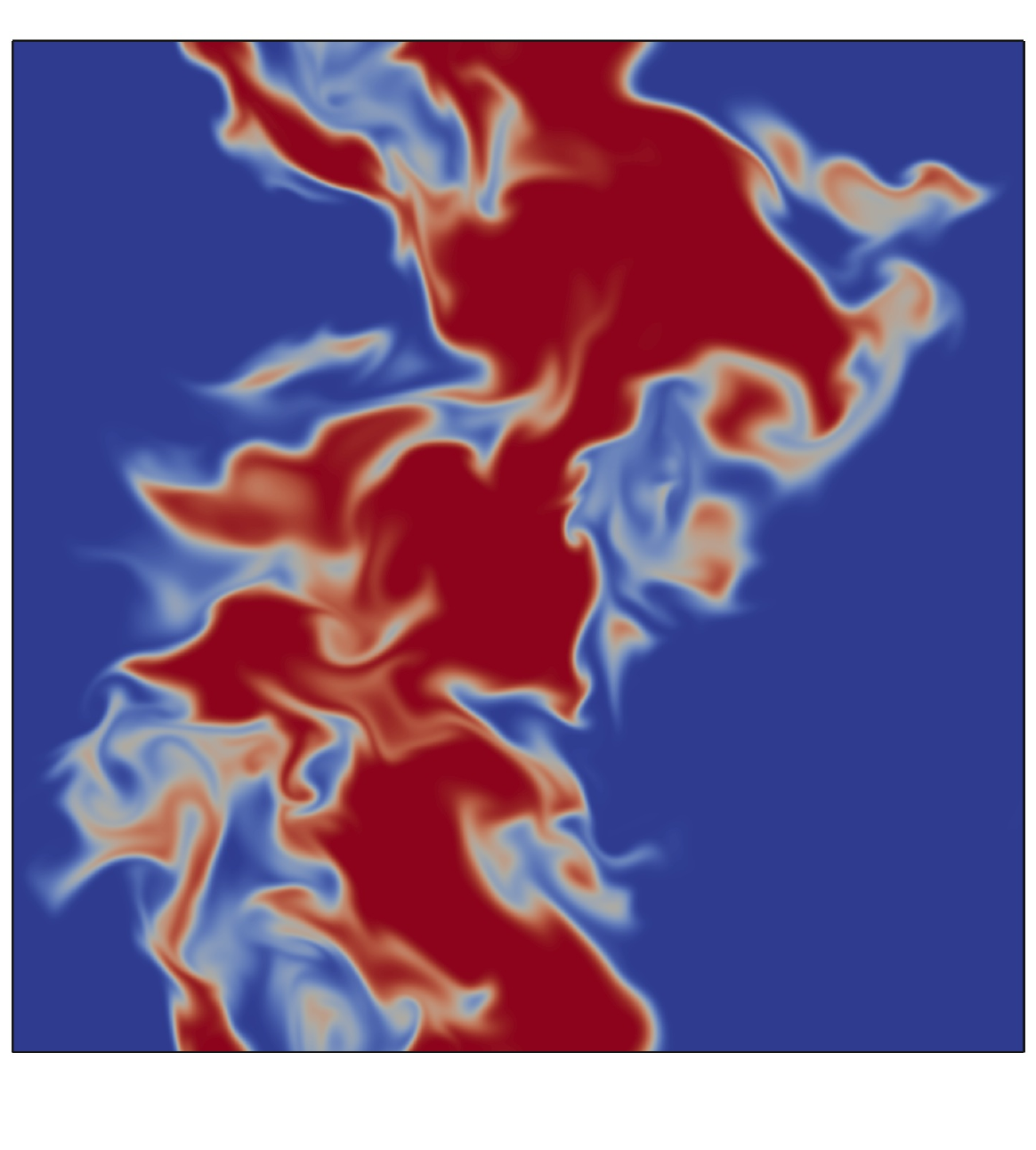}
 } \\
 \subfigure[$t = 3$]{
  \includegraphics[width=0.31\textwidth,trim=130px 430px 130px 40px, clip]{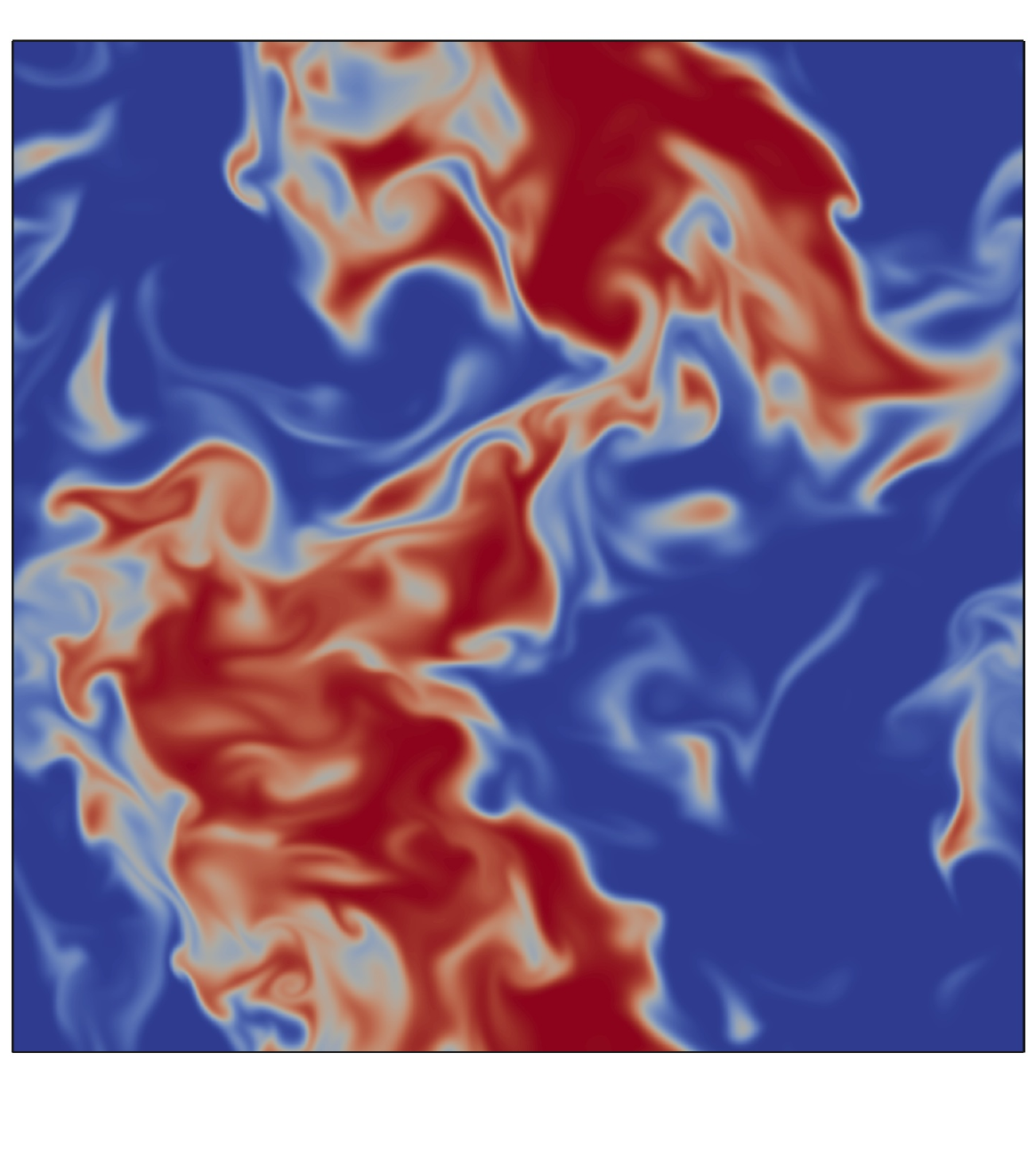}
 }
 \subfigure[$t = 4$]{
  \includegraphics[width=0.31\textwidth,trim=130px 430px 130px 40px, clip]{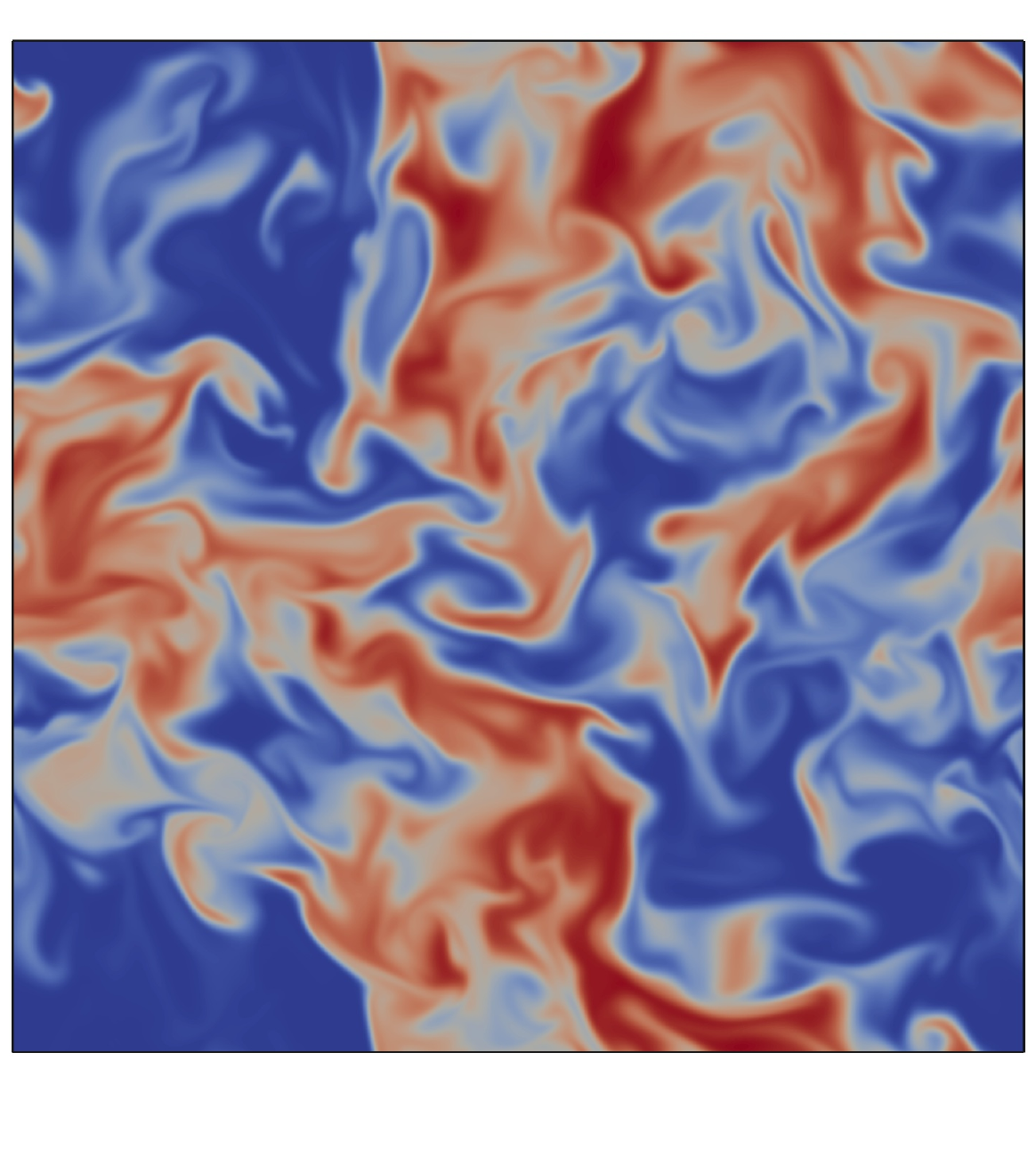}
 }
 \subfigure[$t = 5$]{
  \includegraphics[width=0.31\textwidth,trim=130px 430px 130px 40px, clip]{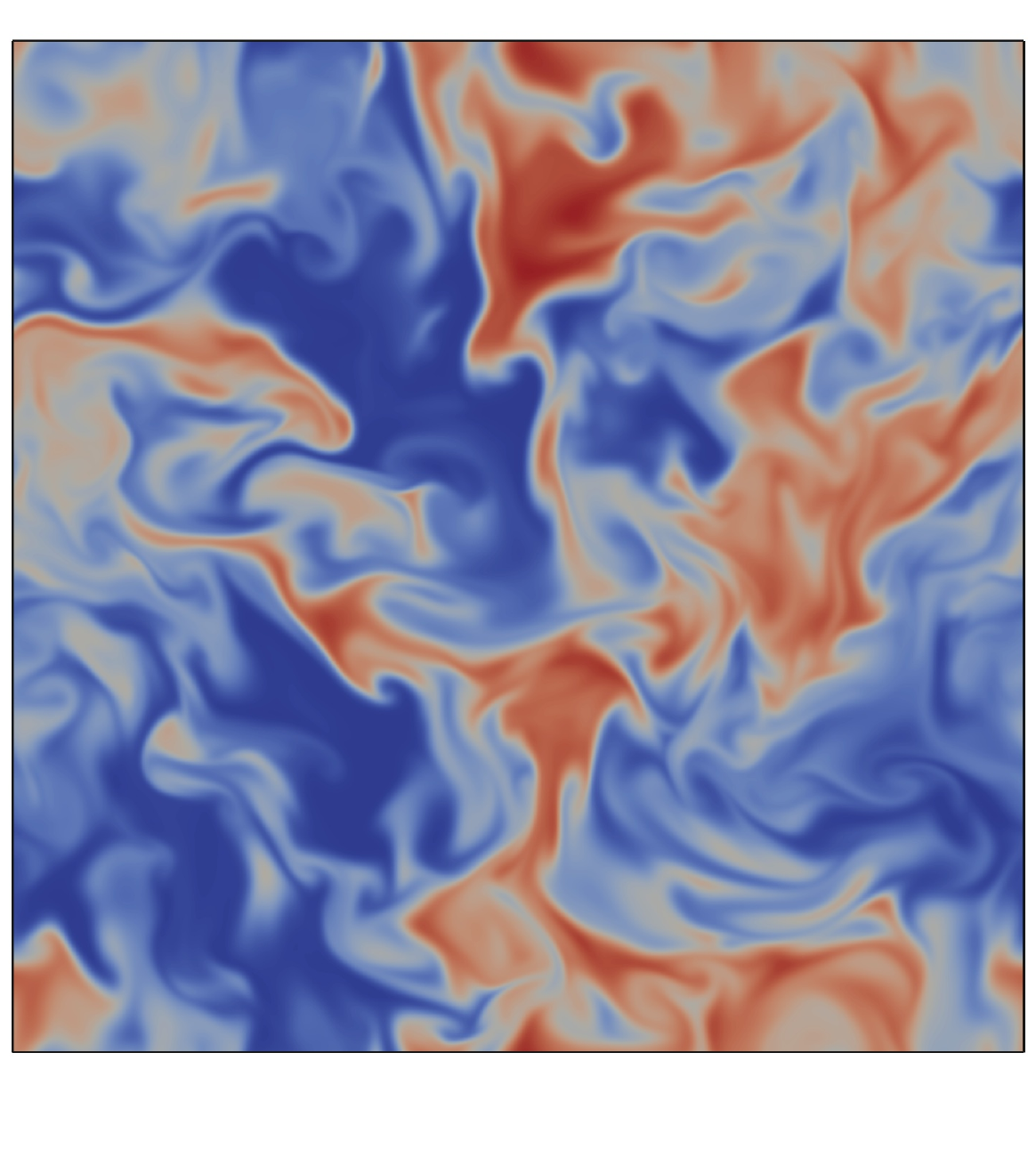}
 } \\
 \subfigure[$t = 6$]{
  \includegraphics[width=0.31\textwidth,trim=130px 430px 130px 40px, clip]{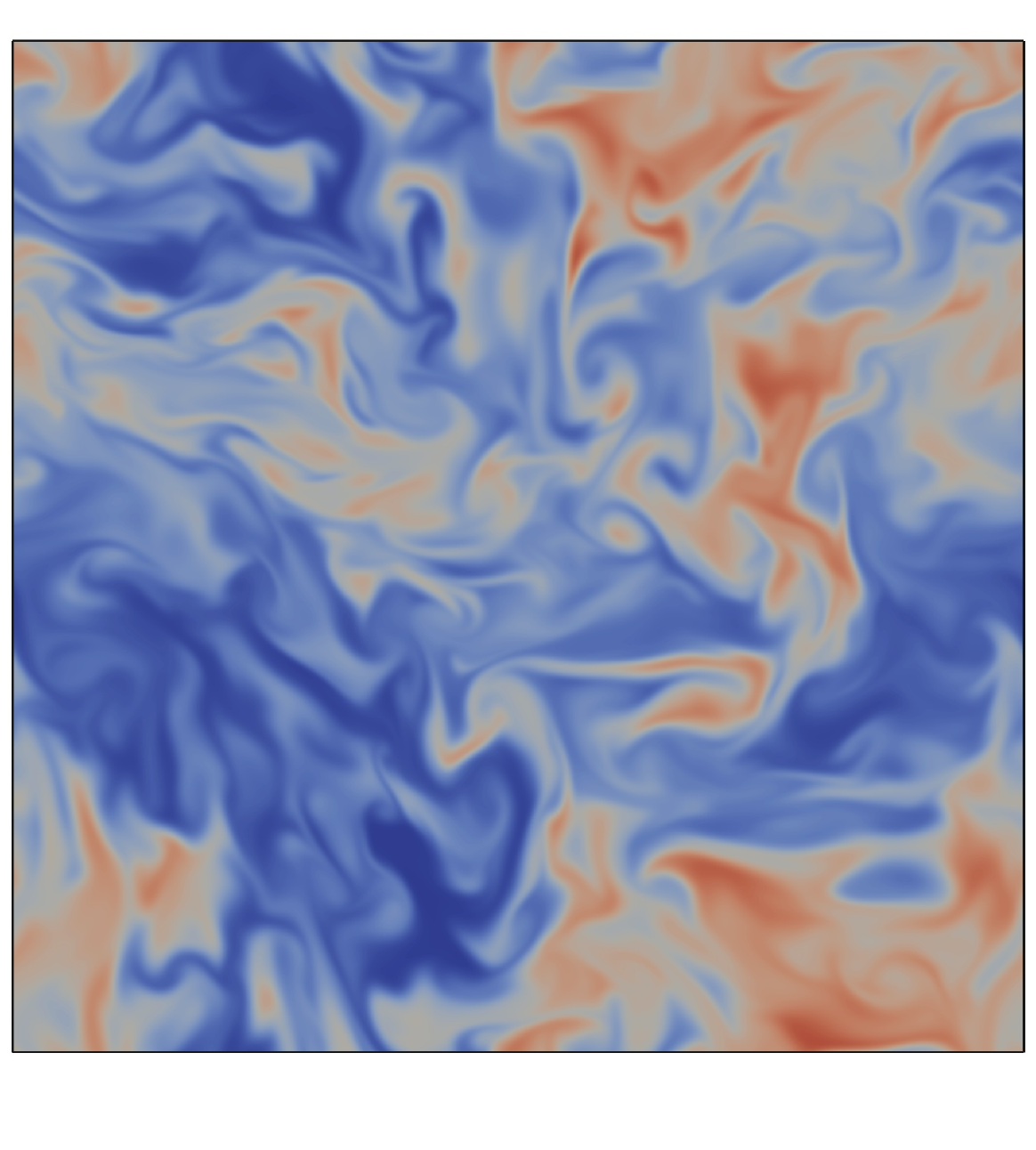}
 }
 \subfigure[$t = 7$]{
  \includegraphics[width=0.31\textwidth,trim=130px 430px 130px 40px, clip]{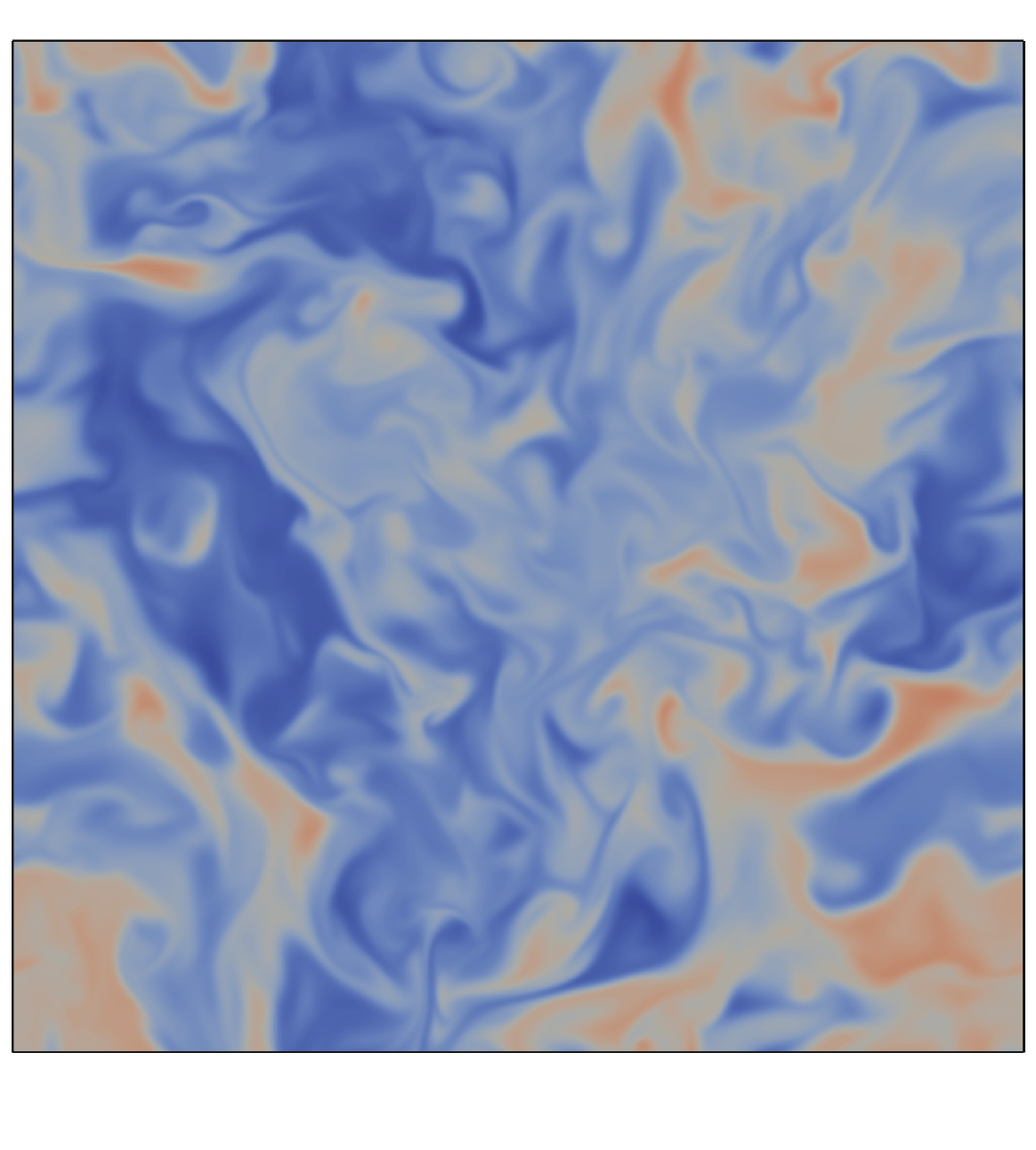}
 }
 \subfigure[$t = 8$]{
  \includegraphics[width=0.31\textwidth,trim=130px 430px 130px 40px, clip]{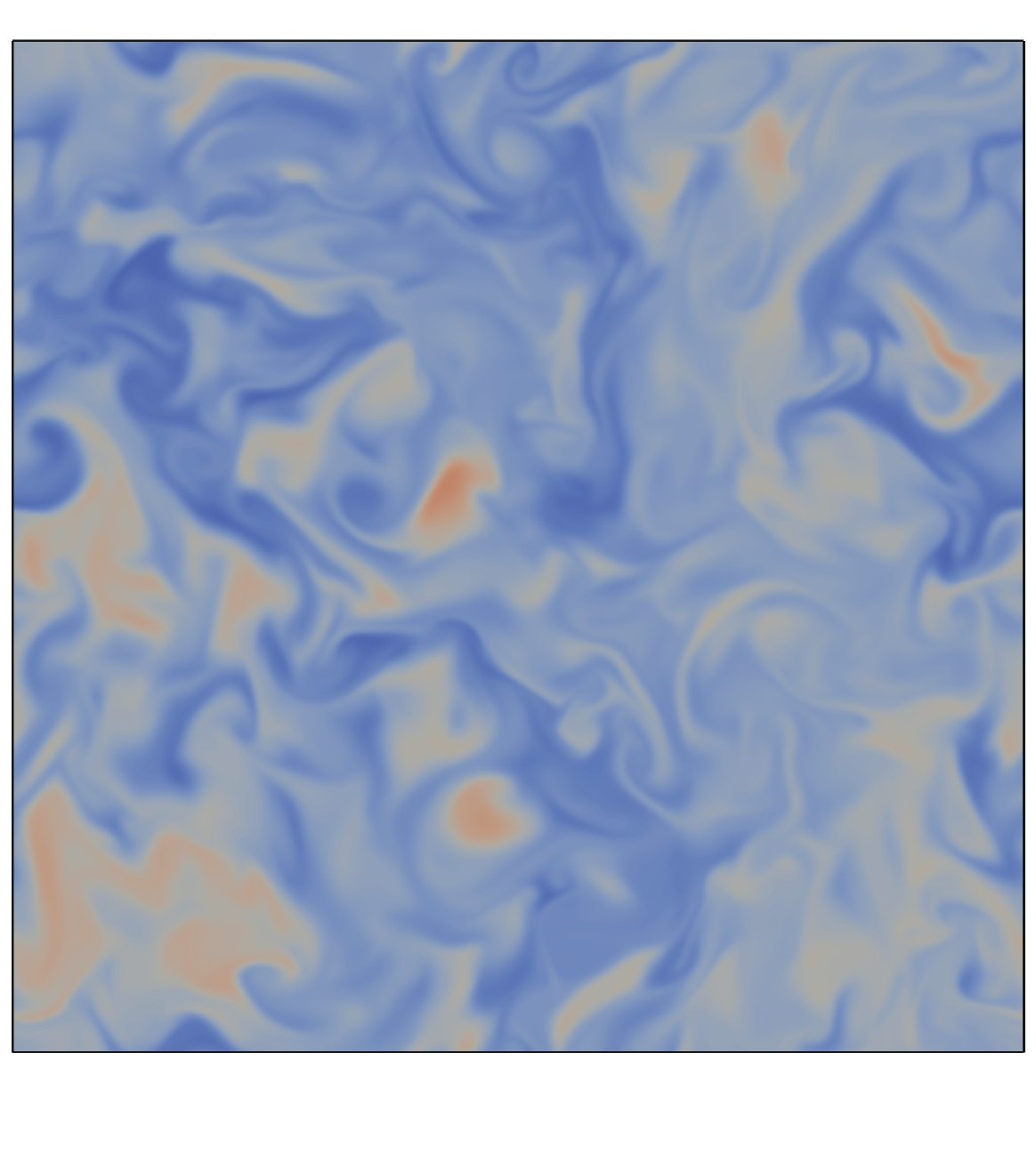}
 }
 \caption{Time slices ($z = 0$ plane) of the advection of a passive scalar field by an $R_\lambda \sim 100$ velocity field on a $512^3$ lattice. Coloured by $\theta$. Note that the colour `fades' due to the lack of forcing of the scalar, without which it decays.}
 \label{fig:ps512}
\end{figure}

\begin{figure}[htb]
 \centering
 \subfigure[$t = 0$]{
  \includegraphics[width=0.39\textwidth,trim=145px 475px 145px 50px, 
  clip]{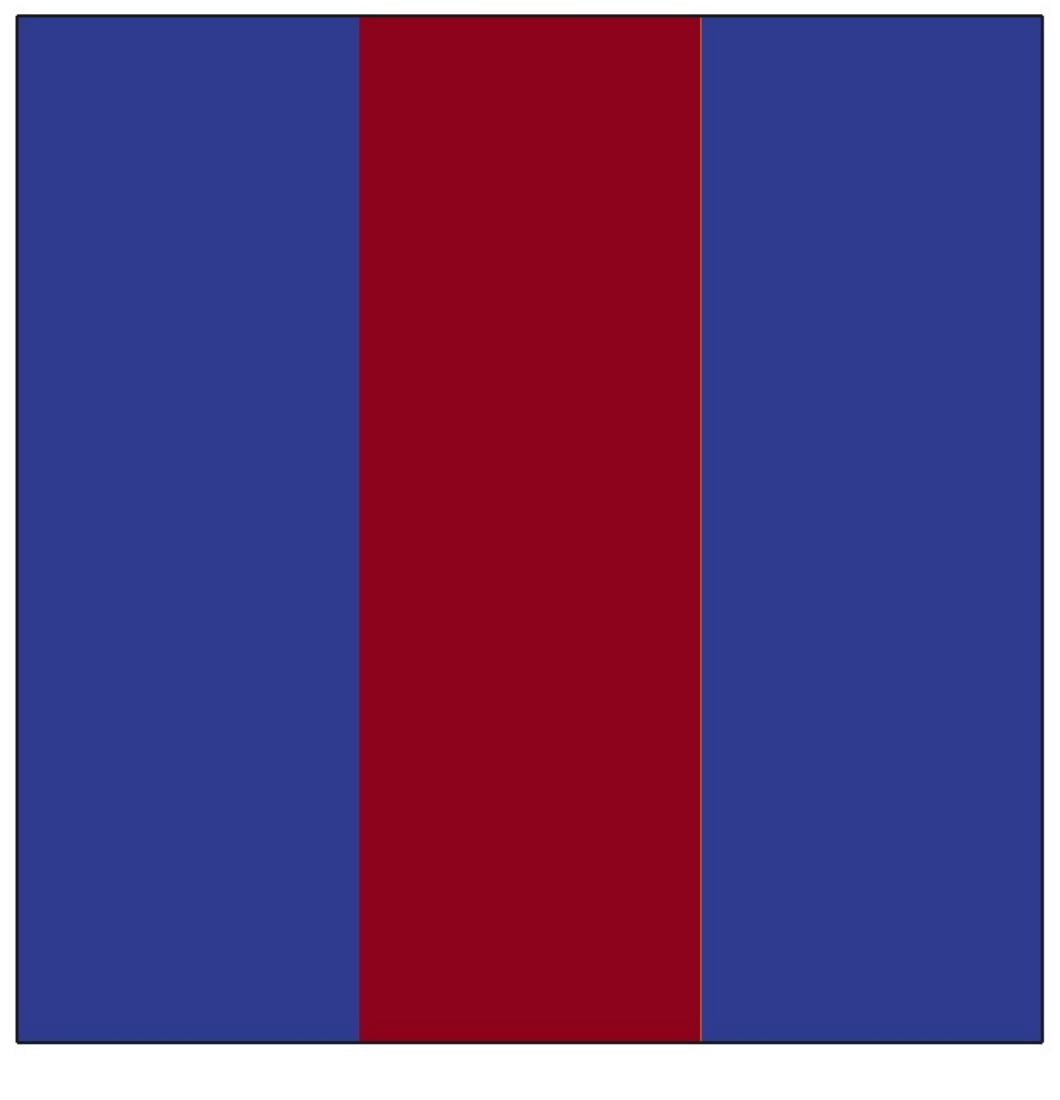}
 }\hspace{0.5in}
 \subfigure[$t = 0.25$]{
  \includegraphics[width=0.39\textwidth,trim=145px 475px 145px 50px, clip]{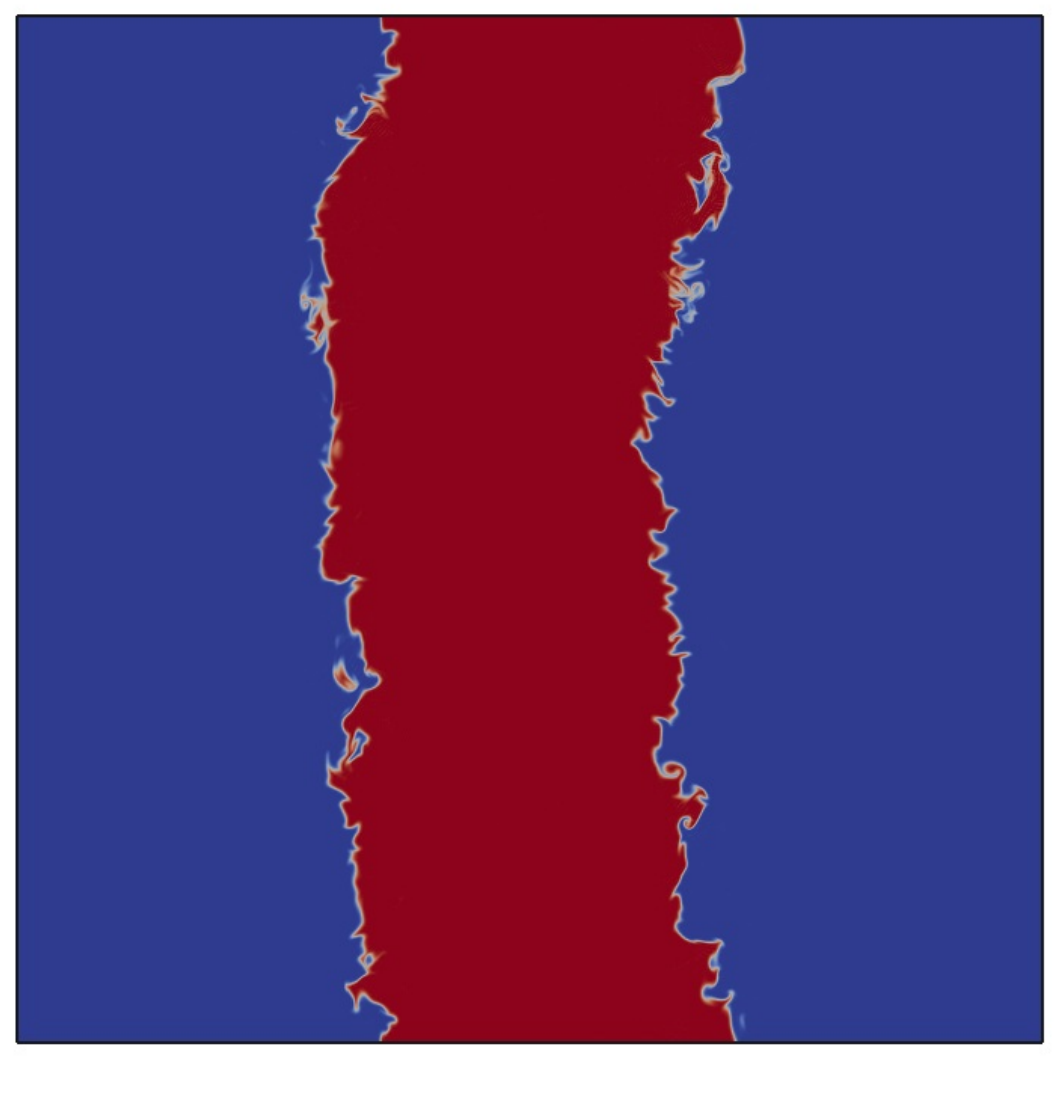}
 } \\
 \subfigure[$t = 0.5$]{
  \includegraphics[width=0.39\textwidth,trim=145px 475px 145px 50px, clip]{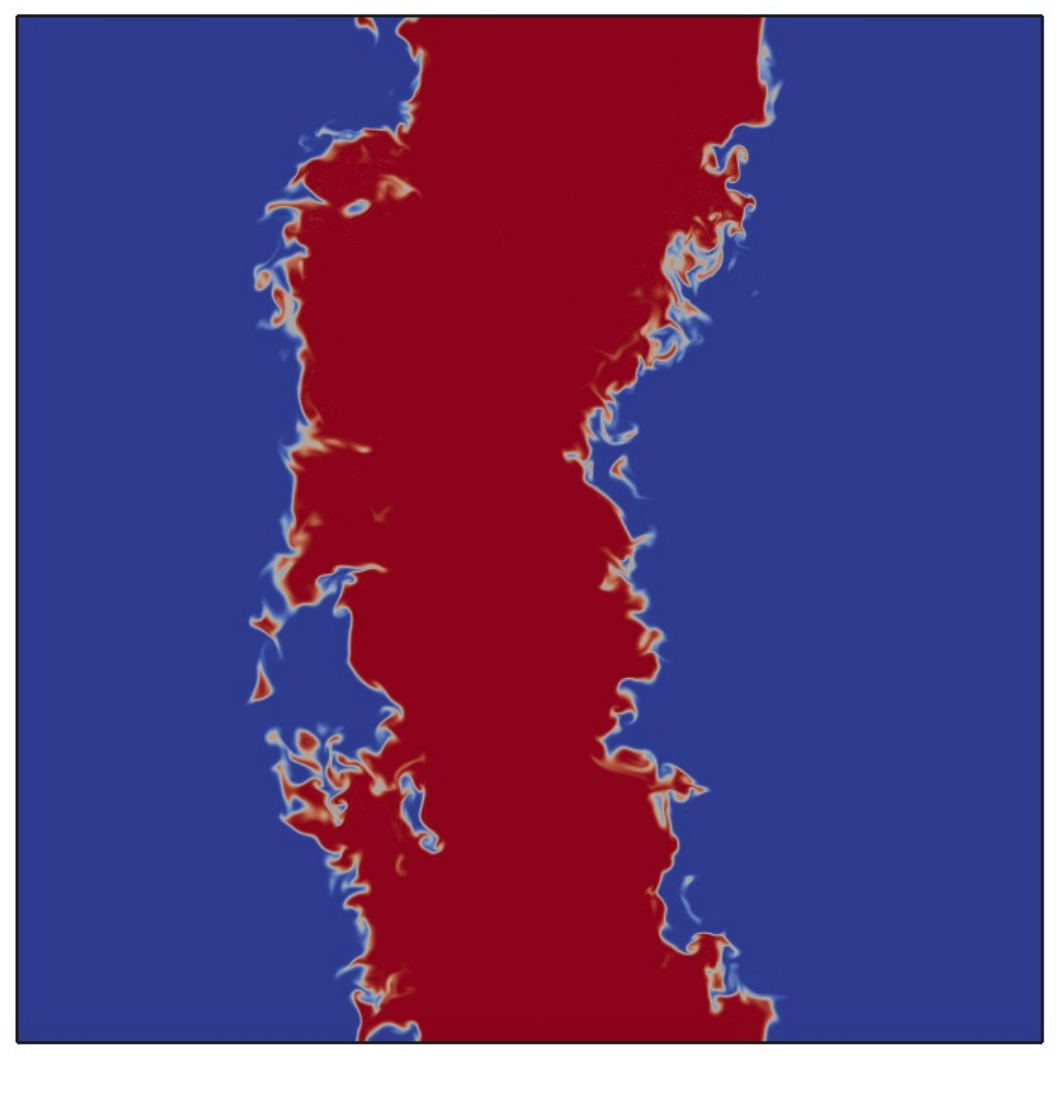}
 }\hspace{0.5in}
 \subfigure[$t = 0.75$]{
  \includegraphics[width=0.39\textwidth,trim=145px 475px 145px 50px, clip]{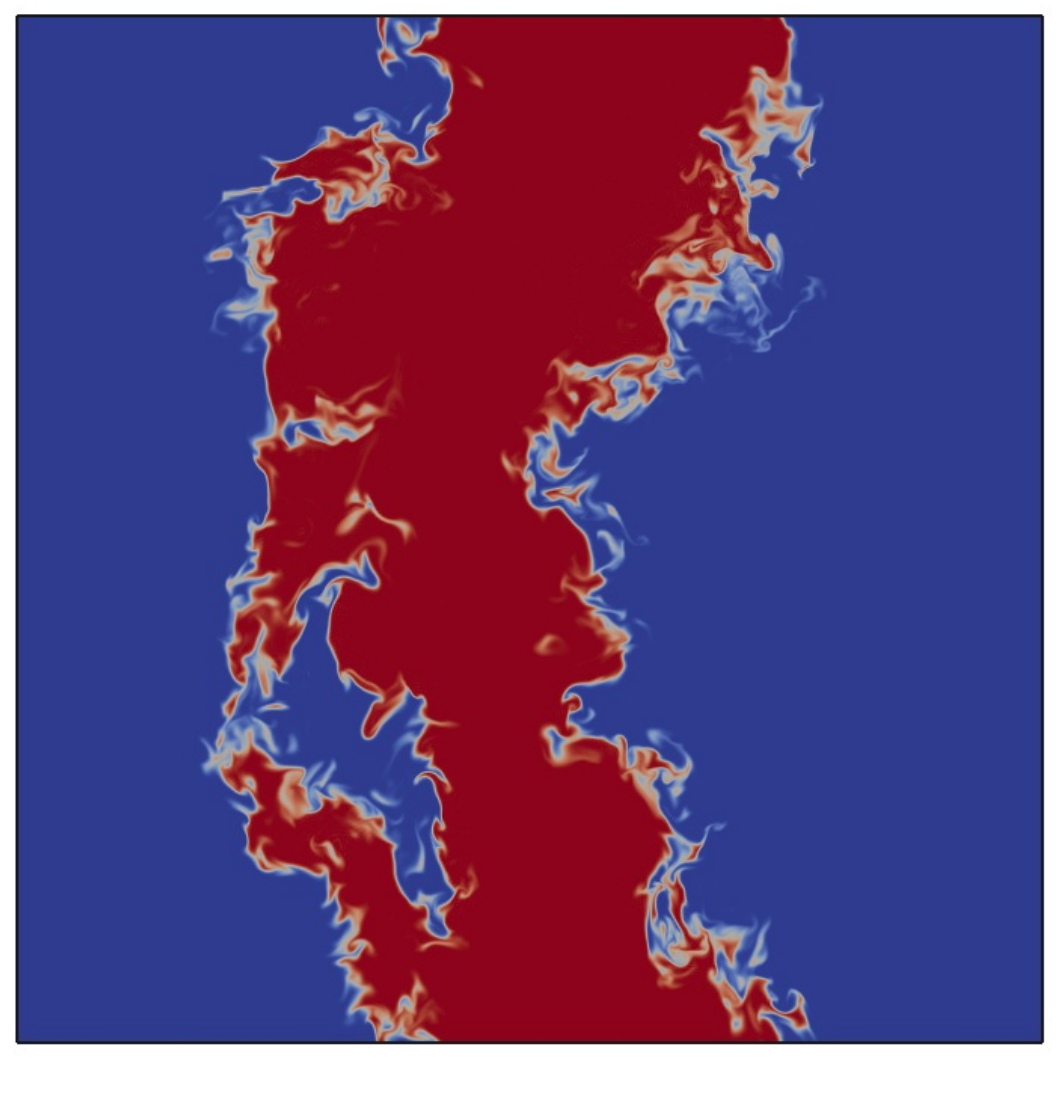}
 } \\
 \subfigure[$t = 1$]{
  \includegraphics[width=0.39\textwidth,trim=145px 475px 145px 50px, clip]{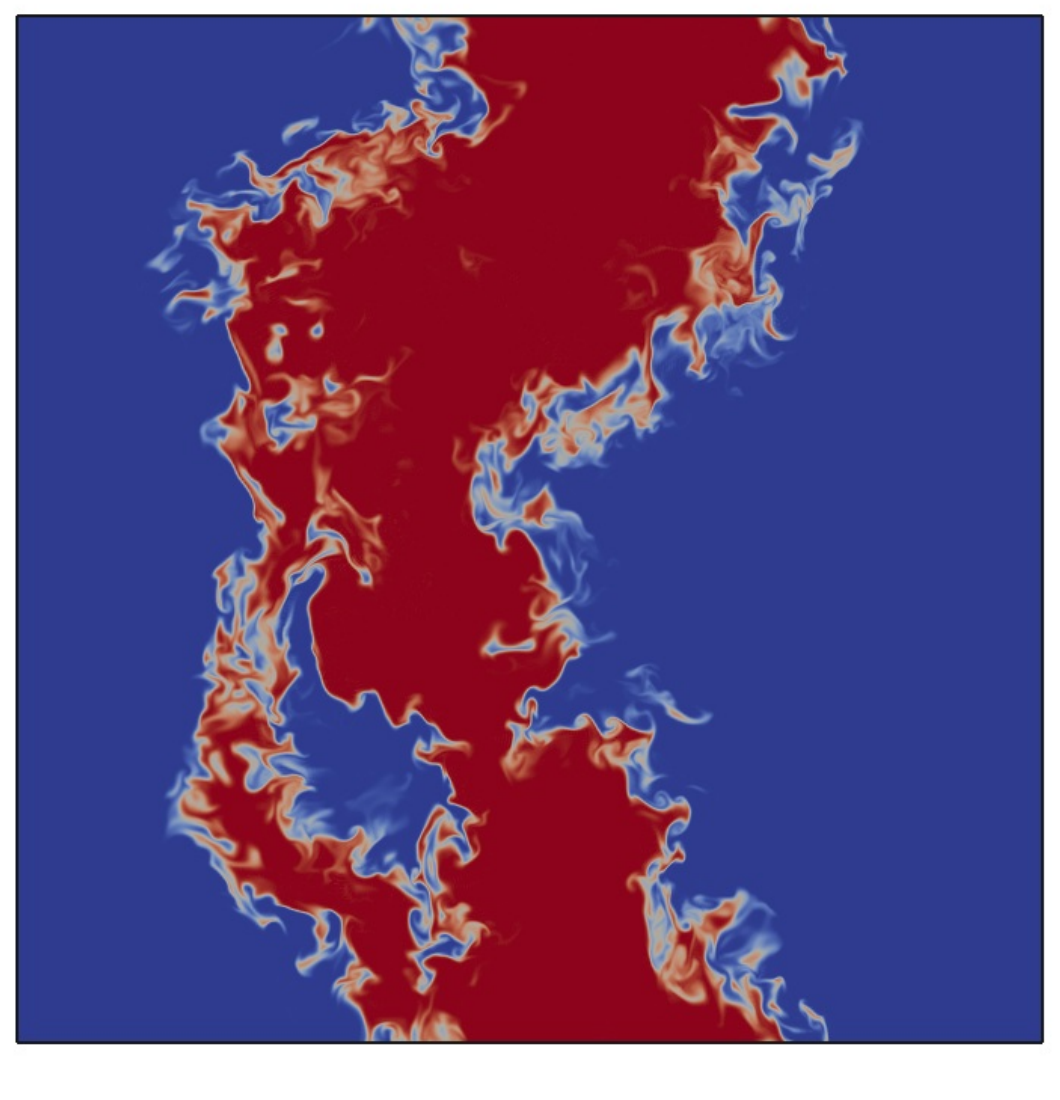}
 }\hspace{0.5in}
 \subfigure[$t = 1.81$]{
  \includegraphics[width=0.39\textwidth,trim=145px 475px 145px 50px, clip]{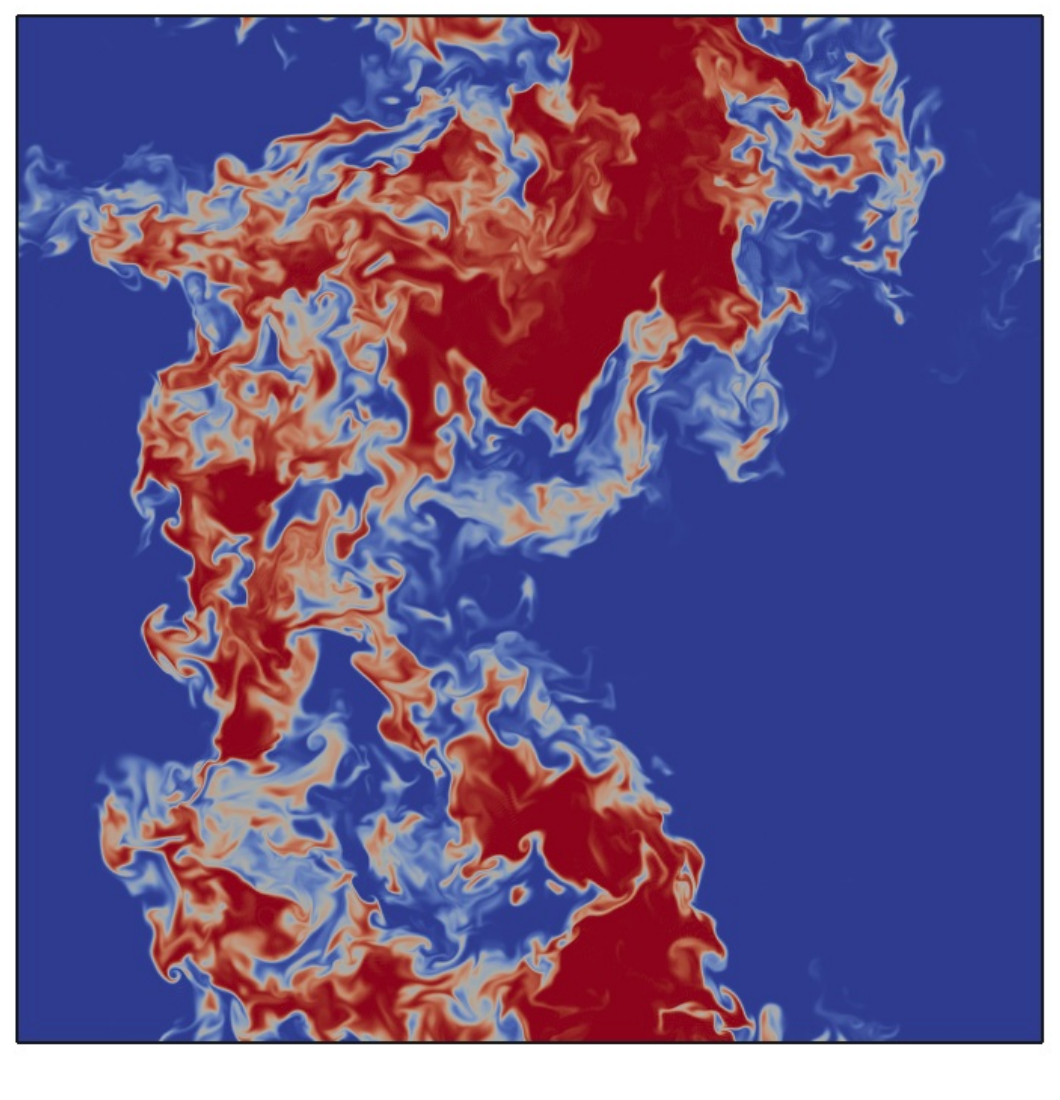}
 }
 \caption{Higher resolution time slices ($z = 0$ plane) of the advection of a passive scalar field by an $R_\lambda \sim 280$ velocity field on a $1024^3$ lattice. Coloured by $\theta$.}
 \label{fig:ps1024}
\end{figure}

The time development in the $z = 0$ plane is shown in figures \ref{fig:ps512} and \ref{fig:ps1024}. Due to time constraints on \eddie\ and because checkpointing of the scalar has not been implemented, this could only be run up to $t = 1.81s$ for the high resolution case. Nevertheless, the results clearly show the velocity field breaking up the hard boundaries and spreading the scalar across the domain, towards the isotropic configuration.

\section{Further validation tests}
\label{sec:further_tests}
There are other validation tests that are commonly used to check DNS data and which will be discussed later in more depth: coherent structures (section \ref{sec:structures}); structure functions (section \ref{sec:sf}); and the dimensionless dissipation coefficient (section \ref{sec:DA}).

\clearpage

\section{Concluding remarks}
We have performed several tests in order to ascertain the reliability of the code produced in this project, as detailed above in this chapter:

\begin{enumerate}
 \item We have established the magnitude of the time-step that is required to keep integration errors small and to prevent simulations from diverging, and that shell averaging preserves $\Pi(0,t) = 0$.

 \item In direct comparison to DNS results for decaying turbulence obtained by a previous-generation code and used by Quinn \cite{thesis:apquinn} to test a numerical computation of the  LET closure, we find excellent agreement. This is seen for all parameters for which there was data available, as well as energy (not shown), dissipation and transfer spectra. Comparison to the skewness data of Wang \etal\ \cite{Wang:1996p1041} for decaying turbulence also supports our results.

 \item We compared results obtained from our DNS code to those from another freely-available pseudospectral code, \textit{hit3d}. This was done for both decaying and forced runs under as similar conditions as possible. The qualitative agreement (and quantitative for the forced case) even for a single realisation provides us with confidence in the performance of our code.

 \item We studied the Taylor-Green vortex, which has become a standard test problem. This allowed us to demonstrate qualitative agreement between our results for the streamline visualisation, velocity field plots and isovorticity contours to those obtained by Young \cite{thesis:ajyoung} and/or Brachet \etal\ \cite{Brachet:1983p411}. A more quantitative comparison was possible using a fit to the shell averaged energy spectrum.

 \item A test of the degree of isotropy did not show any significant deviation for a range of lattice sizes.

 \item The time averaged Kolmogorov constant and skewness give good quantitative agreement with the literature, and the energy spectrum scaled using the peak of the dissipation spectrum, $k_p$, also shows good collapse of data.

 \item The advection of a passive scalar shows the expected behaviour of turbulent mixing. While not being a rigorous test, it helps support confidence in the code.

\end{enumerate}

We therefore conclude that the \dns\ code that has been produced is performing as expected and is capable of reproducing accepted numerical results. Confidence may be placed in new results obtained from it.

\chapter{Numerical investigation of decaying isotropic turbulence}

The \dns\ code has been used to run a number of simulations of homogeneous, isotropic turbulence for both decaying and forced systems. This chapter focuses on the data collected from our decaying runs. As far as possible, the initial condition was kept as consistent as possible and variation was introduced by changing a single variable, the viscosity, to explore a range of Reynolds number. This was done by maintaining the initial energy spectrum. The decaying simulations are detailed in the following section and summarised in table \ref{tbl:summary_decay_sims}. Time evolution of parameters and spectra are shown in figures \ref{fig:decay_TS} and \ref{fig:decay_spectra}.

\vspace{1.5em}
\noindent The content of section \ref{sec:Taylor_surr} was published in McComb, Berera, Salewski and Yoffe \cite{McComb:2010p250}.

\newpage
\section{Summary of decaying simulations performed}
A series of simulations has been performed to study the properties of decaying turbulence. The system is initialised as a Gaussian random field with a desired energy spectrum $E(k,0)$, as described in section \ref{subsec:initial_field}. An ensemble is generated by using different seeds to the random number generator, which results in a set of unique velocity field configurations (but all with the same initial energy spectrum). The initial fields are then allowed to decay and their spectra sampled at regular intervals. Since the Gaussian initial condition does not describe fully developed turbulence, the velocity field has to be allowed to evolve before measurement will produce statistics characteristic of turbulence (rather than the initial conditions).

As well as shell averaging, at each measurement time the spectra are also ensemble averaged. These averaged spectra are then used to calculate statistics for the velocity field. The simulations use full dealiasing by truncation of the velocity field according to the $2/3$-rule, see section \ref{sec:truncation}.

Instead of starting the decaying simulations from a random Gaussian field, it is possible to use an evolved stationary field from a forced simulation. This was done here for our highest Reynolds number stationary run, \frun{f1024a}, and will be discussed in section \ref{sec:decay_from_forced}.

\subsection*{Statistics and spectra}
\begin{figure}[tbp]
 \begin{center}
  \includegraphics[width=\textwidth,trim=0 0 0 375px,clip]{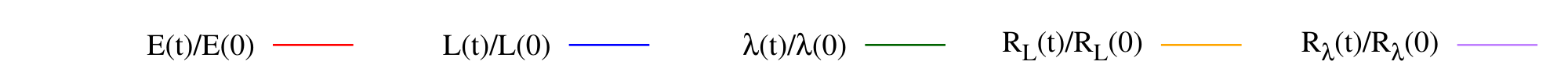}
  \subfigure[Run \drun{d128d}, $R_\lambda(0) = 2.58$]{
   \label{fig:decay_TS_d128d}
   \includegraphics[width=0.475\textwidth,trim=2px 0 10px 0,clip]{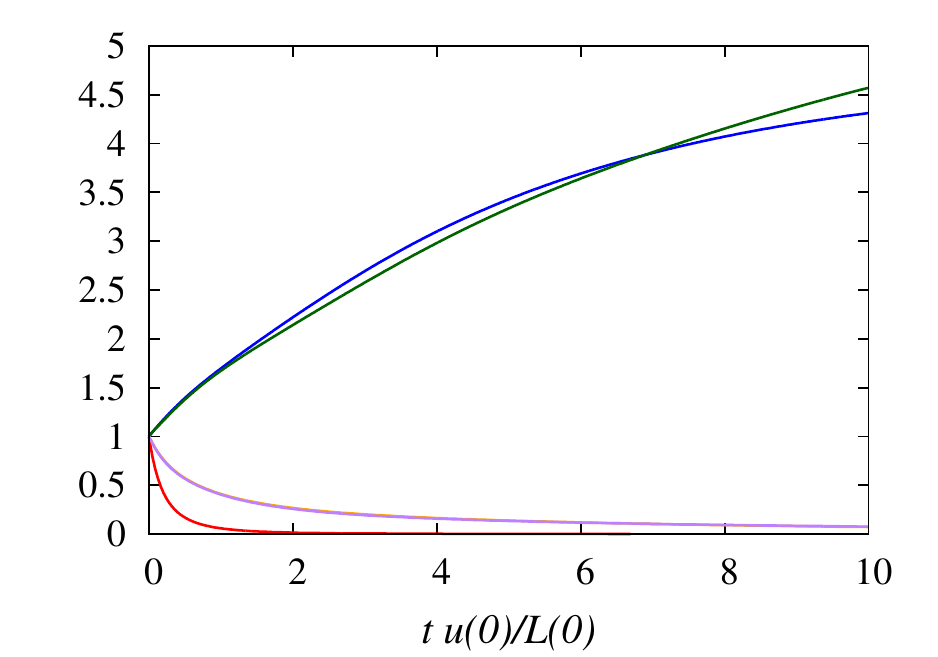}
  }
  \subfigure[Run \drun{d128a}, $R_\lambda(0) = 25.8$]{
   \includegraphics[width=0.475\textwidth,trim=2px 0 10px 0,clip]{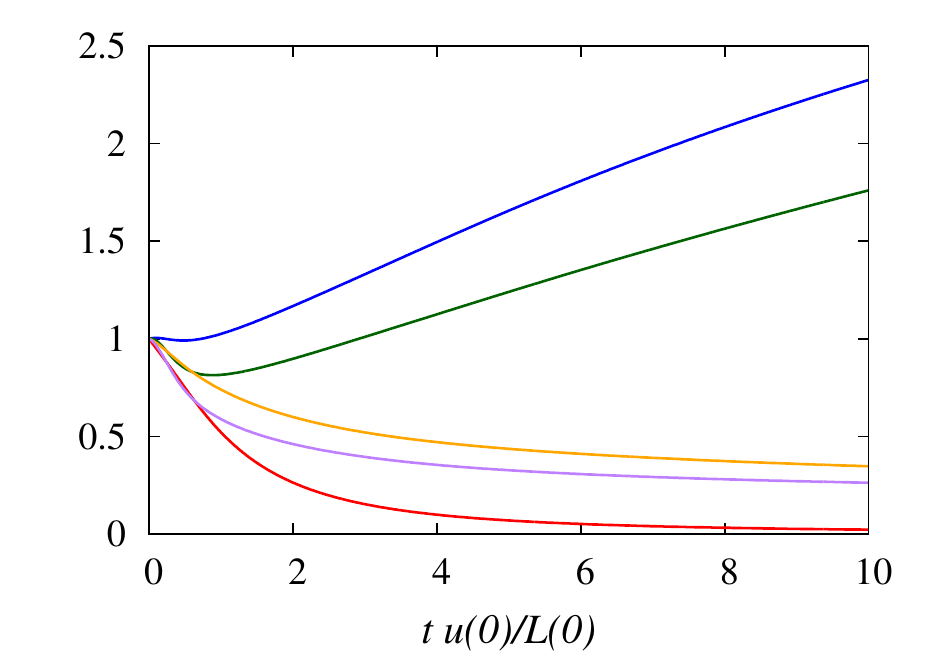}
  }
  \subfigure[Run \drun{d128c}, $R_\lambda(0) = 51.6$]{
   \includegraphics[width=0.475\textwidth,trim=2px 0 10px 0,clip]{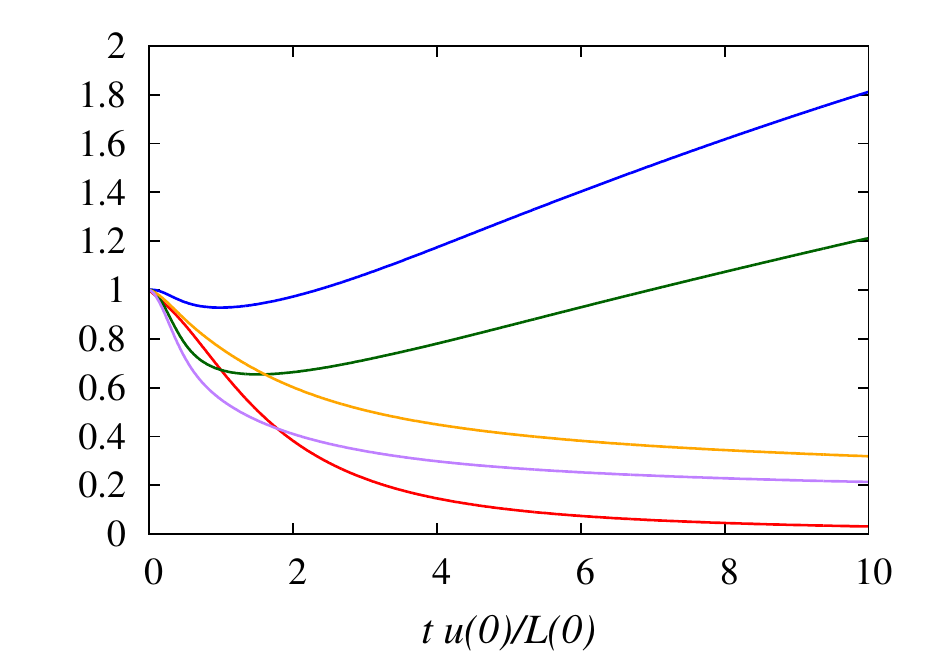}
  }
  \subfigure[Run \drun{d256a}, $R_\lambda(0) = 103.3$]{
   \includegraphics[width=0.475\textwidth,trim=2px 0 10px 0,clip]{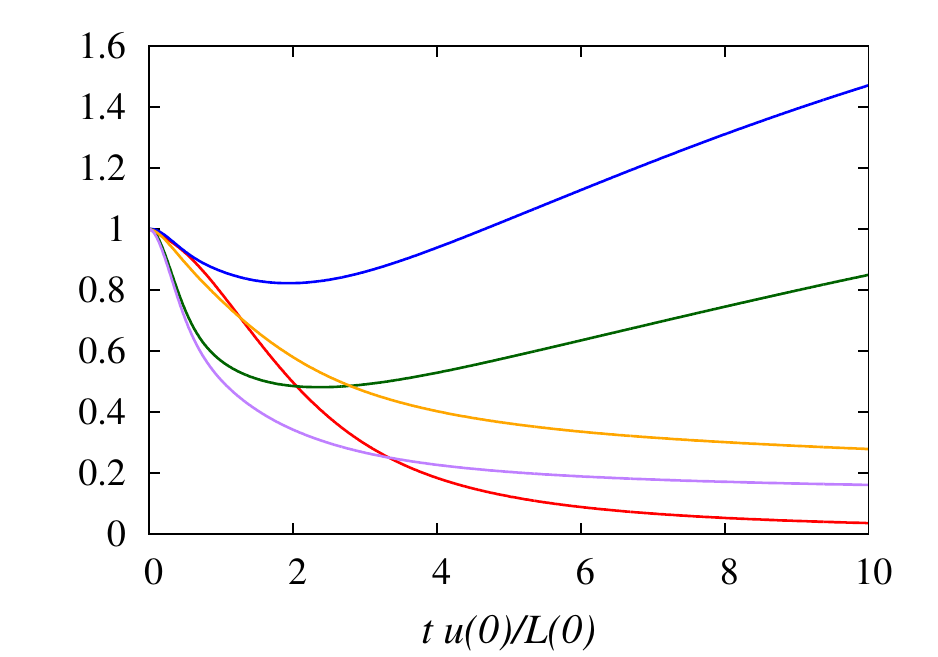}
  }
  \subfigure[Run \drun{d256b}, $R_\lambda(0) = 143.4$]{
   \includegraphics[width=0.475\textwidth,trim=2px 0 10px 0,clip]{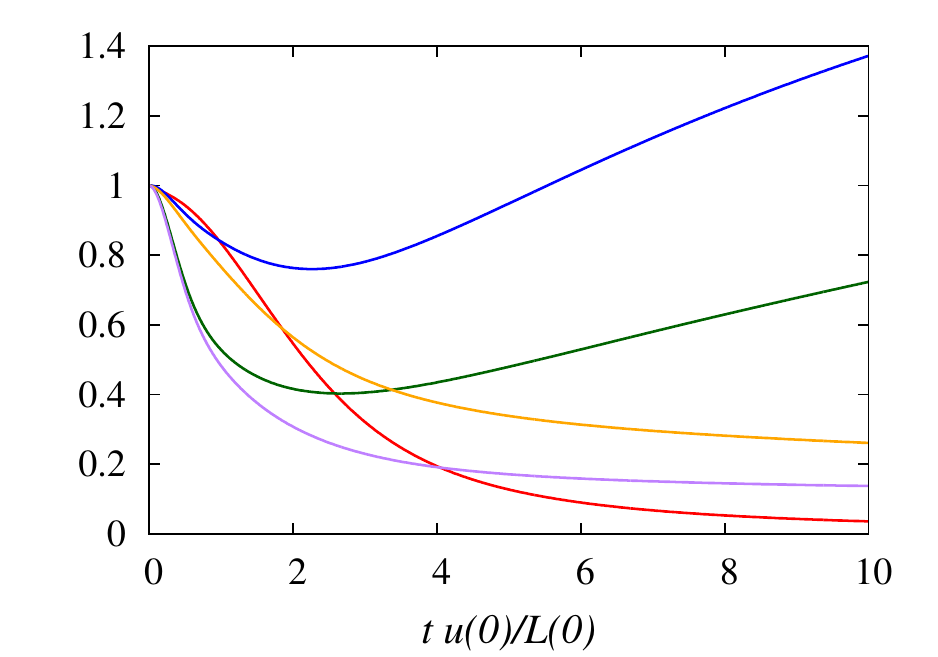}
  }
  \subfigure[Run \drun{d512a}, $R_\lambda(0) = 358.6$]{
   \includegraphics[width=0.475\textwidth,trim=2px 0 10px 0,clip]{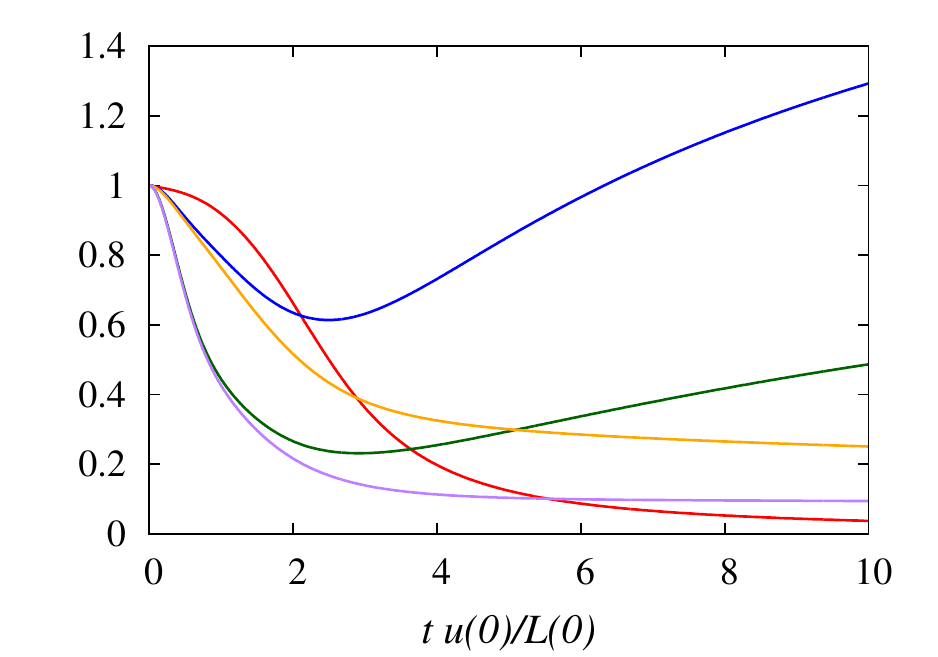}
  }
 \end{center}
 \caption[Time evolution of parameters for decaying turbulence, scaled by their initial value at $t = 0$.]{Time evolution of various parameters, scaled by their initial value at $t = 0$. The legend can be found at the top of the page and applies to all parts. Dissipation, maximum inertial flux and velocity derivative skewness can be found in figures \ref{fig:tevo_eps}, \ref{fig:tevo_pi} and \ref{fig:tevo_skew}, respectively, where they are discussed further.}
 \label{fig:decay_TS}
\end{figure}

A summary of the decaying simulations which have been performed can be found in table \ref{tbl:summary_decay_sims}.
\begin{table}[tbp!]
\begin{center}
\begin{tabular}{r|llll|ll|l}
ID & $N$ & $\nu_0$ & \# & $E(k,0)$ & $R_L(0)$ & $R_\lambda(0)$ & $t_{\textrm{max}}$ \\
\hline
\hline
\texttt{d128d}  & 128  & 0.1    & 10  & S5             & 3.24   & 2.58   & 10s \\
\texttt{d128e}  & 128  & 0.07    & 10  & S5            & 4.62   & 3.69   & 10s \\
\texttt{d128f}  & 128  & 0.05    & 10  & S5            & 6.47   & 5.16   & 10s \\
\texttt{d128g}  & 128  & 0.03    & 10  & S5            & 10.8   & 8.61   & 10s \\
\texttt{d128h}  & 128  & 0.02    & 10  & S5            & 16.2   & 12.9   & 10s \\
\texttt{d128a}  & 128  & 0.01    & 10  & S5            & 32.4   & 25.8   & 50s \\
\texttt{d128b}  & 128  & 0.007   & 10  & S5            & 46.2   & 36.9   & 50s \\
\texttt{d128c}  & 128  & 0.005   & 10  & S5            & 64.7   & 51.6   & 50s \\
\texttt{d256a}  & 256  & 0.0025  & 10  & S5            & 129.5  & 103.3  & 50s \\
\texttt{d256b}  & 256  & 0.0018  & 10  & S5            & 179.8  & 143.4  & 50s \\
\texttt{d512a}  & 512  & 0.00072 & 10  & S5            & 449.5  & 358.6  & 40.16s \\
\texttt{d1024}  & 1024 & 0.0002  & 5   & \frun{f1024b} & 3828.2 & 353.7  & 6.18s \\
\end{tabular}
\caption{Summary of the main decaying simulations that have been run and their parameters.}
\label{tbl:summary_decay_sims}
\end{center}
\end{table}
Figure \ref{fig:decay_TS} presents the time evolution of the total energy, length-scales and Reynolds numbers, scaled by their initial values at the beginning of the decay, $t = 0$. Evolution of the dissipation rate, maximum inertial flux and velocity derivative skewness can be found in figures \ref{fig:tevo_eps}, \ref{fig:tevo_pi} and \ref{fig:tevo_eps}, respectively, where they are discussed in more detail.

As time progresses, the total energy and Reynolds numbers are seen to decay towards zero, as expected. This decay becomes quicker as we move to lower Reynolds numbers, since dissipation has an increasingly dominant role. For initial Reynolds numbers $R_\lambda \gtrsim 100$, the Reynolds numbers can be initially seen to follow the decay of their associated length-scale. This is because the decay of energy is comparatively slow, and Reynolds number is proportional to the length-scale and $\sqrt{E(t)}$. The decay of the length-scales shows that the system is creating smaller characteristic scales. As the decay progresses, the length-scales experience a turning point and begin to increase. At this point, which is slightly later for $\lambda$ than $L$, the system no longer needs such small scales to dissipate energy.

Despite this increase in length-scales, the energy decays sufficiently quickly to ensure that the Reynolds numbers fall off. For lower Reynolds numbers, the length-scales exhibit very little, if any, decay before they start to increase. Run \drun{d128d} shown in figure \ref{fig:decay_TS_d128d} displays no decay of length-scales and can be considered as an example of viscous decay rather than developing turbulence.

\begin{figure}[tbp]
 \begin{center}
  \subfigure[Run \drun{d128a}, $R_\lambda(0) = 25.8$]{
   \includegraphics[width=0.475\textwidth]{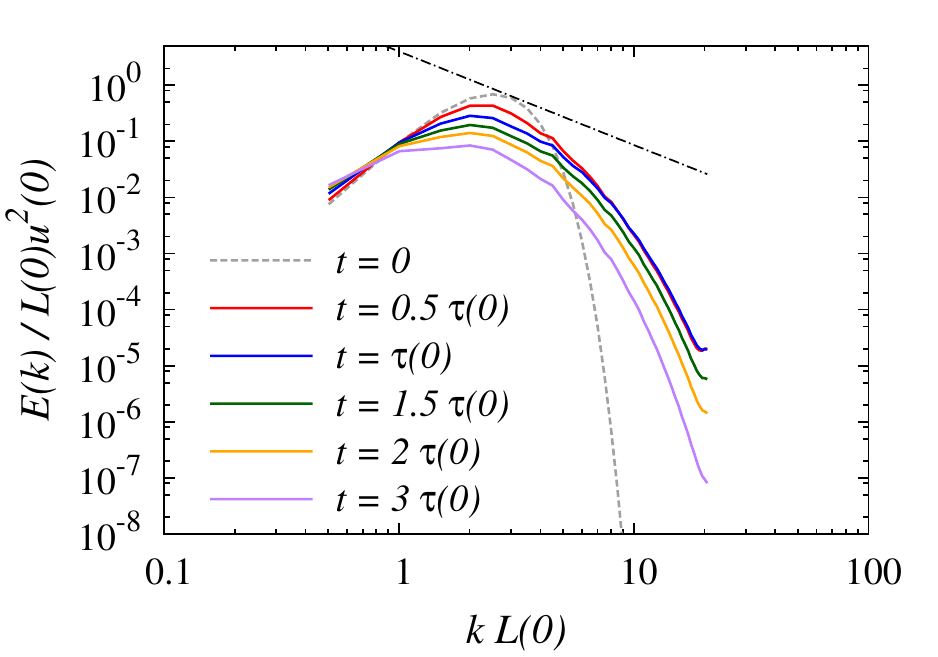}
   \includegraphics[width=0.475\textwidth]{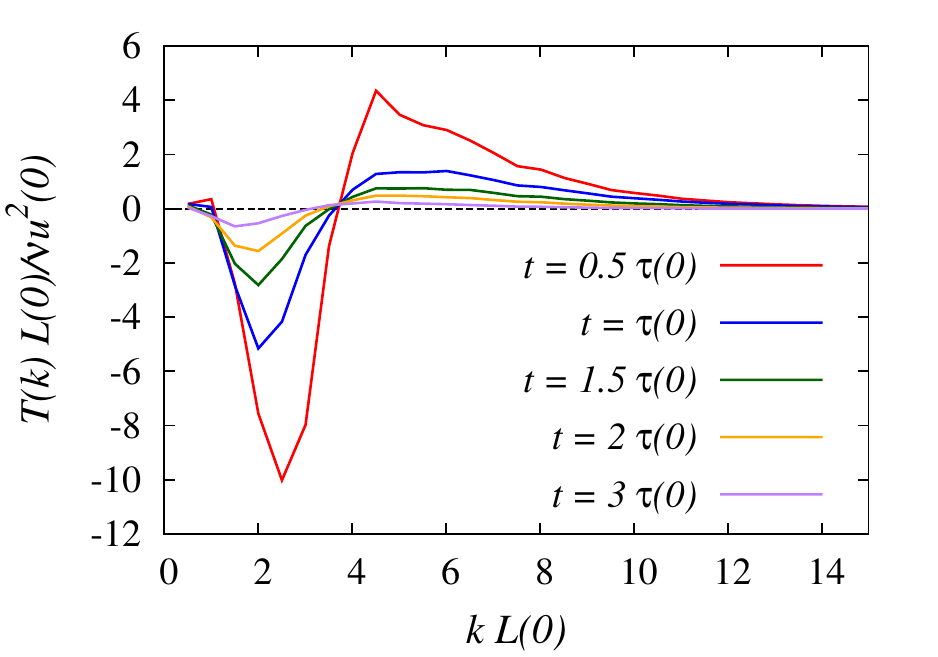}
  }
  \subfigure[Run \drun{d256b}, $R_\lambda(0) = 143.4$]{
   \includegraphics[width=0.475\textwidth]{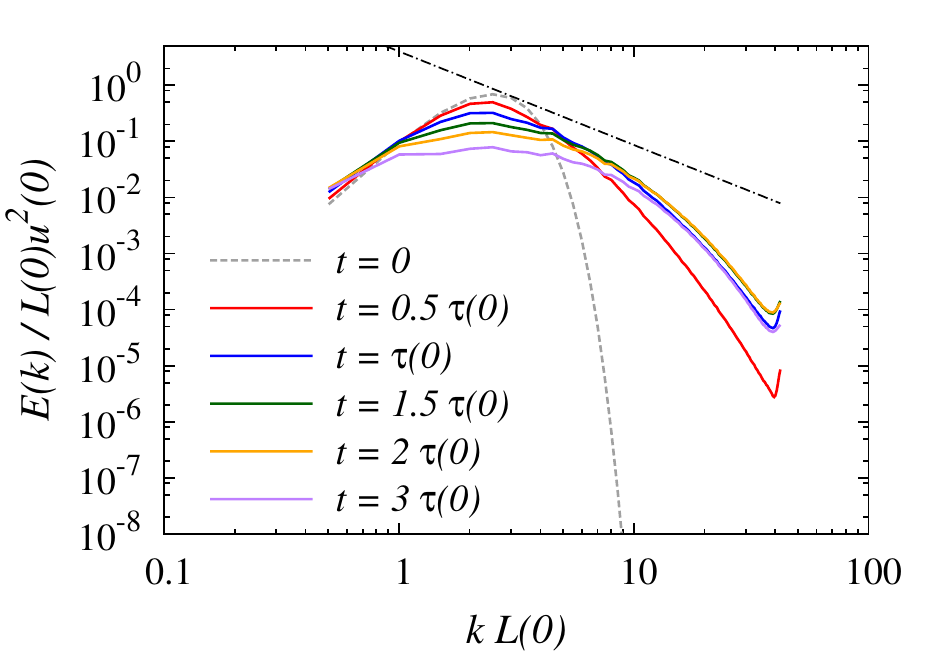}
   \includegraphics[width=0.475\textwidth]{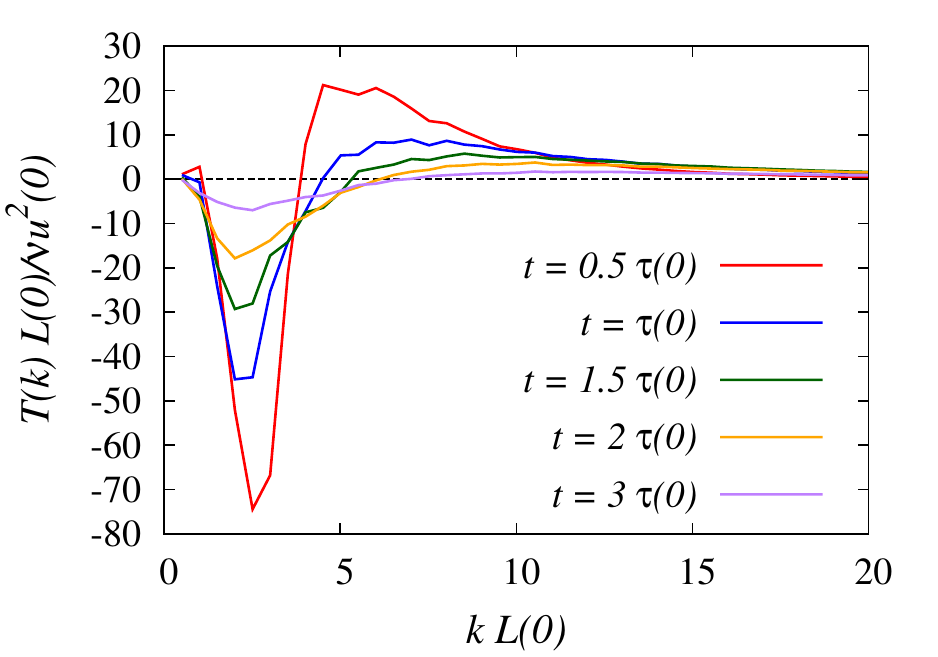}
  }
  \subfigure[Run \drun{d512a}, $R_\lambda(0) = 348.6$]{
   \includegraphics[width=0.475\textwidth]{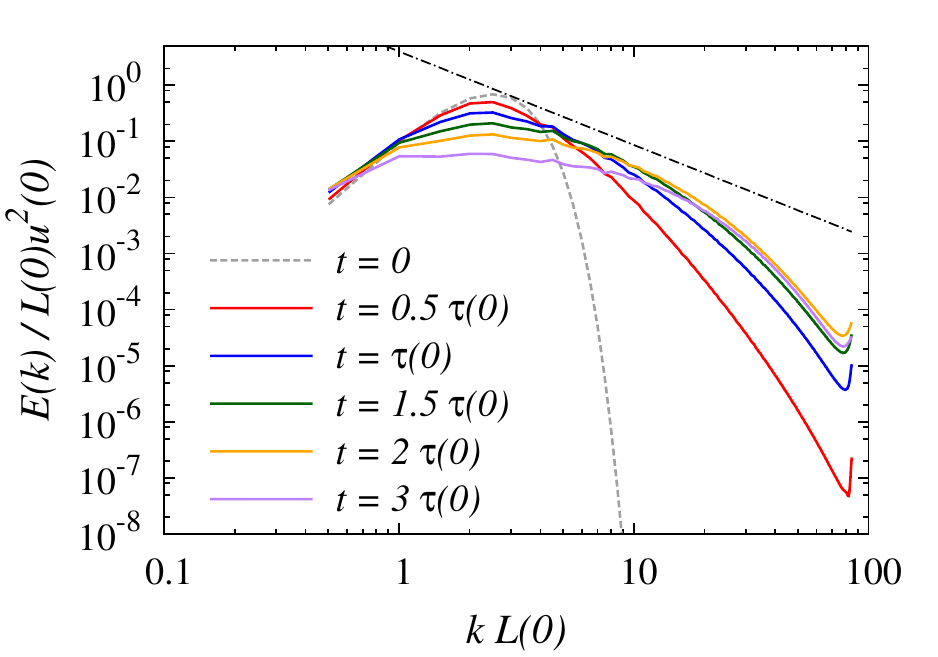}
   \includegraphics[width=0.475\textwidth]{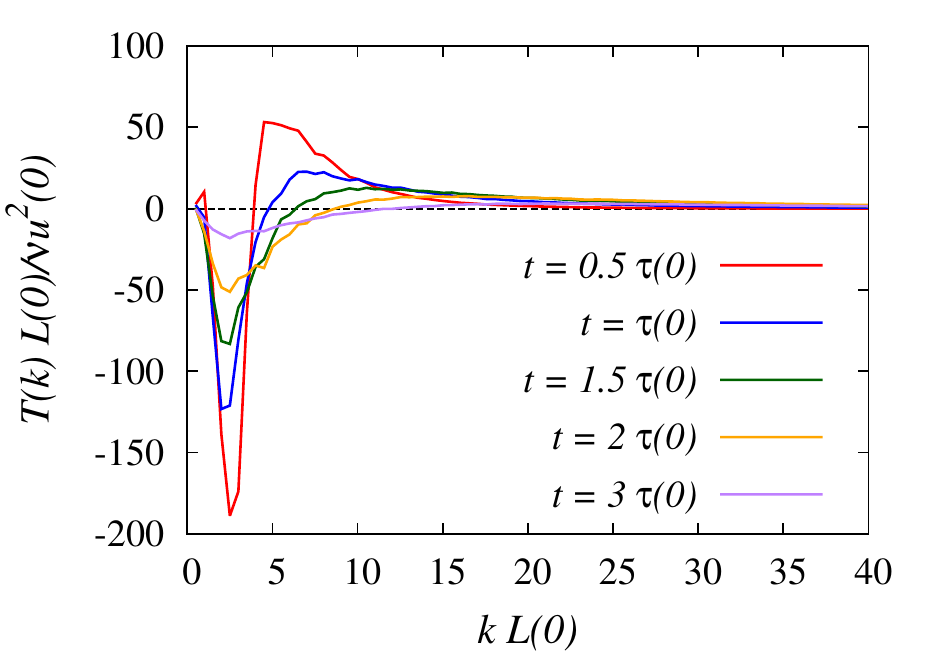}
  }
 \end{center}
 \caption{Energy and transfer spectra shown at a selection of times during the decay process for a selection of runs. (-- $\cdot$ --) shows Kolmogorov $k^{-5/3}$ behaviour for the energy spectra (left), while (-- -- --) shows zero for the transfer spectra (right).}
 \label{fig:decay_spectra}
\end{figure}

The (scaled) energy and transfer spectra for runs \drun{d128a},\drun{d256b} and \drun{d512a} are shown in figure \ref{fig:decay_spectra} for various times. The initial large eddy turnover time $\tau(0) = L(0)/u(0)$ has been used to scale the time. As time progresses, we see the shape of the energy spectrum change from $E(k,0)$ (dashed line) where the energy is entirely located in the low wavenumbers. The tail of the spectrum lifts as energy is transferred to higher modes by non-linear interactions. Eventually, dissipation wins and the spectrum decays in a self-similar fashion.

The (scaled) transfer spectra presented in figure \ref{fig:decay_spectra} (right column) is initially zero since the field is Gaussian. As time progresses it takes its characteristic shape, since the inertial transfer is removing energy from low modes (hence negative value) and depositing it in high modes. Due to the decaying nature of the system, the transfer spectrum reaches a maximum and starts to decay. The decay is not self-similar, with the zero crossing wavenumber $k_*(t)$ (where $T(k_*,t) = 0$) increasing in time.

\clearpage

\section{Determination of an evolved time}\label{sec:tevo}
When we run a pseudospectral numerical simulation, we usually start from some Gaussian random field with a prescribed energy spectrum. This initial condition is not characteristic of a fully developed solution of the Navier-Stokes equation, but an artificial initialisation which we hope resembles a suitable solution in some way. As we step forward in time and the non-linear term generates non-linear couplings, our velocity field eventually describes fully developed turbulence, and continues to do so for the rest of the simulation. The initial field configuration can have a significant impact on the amount of time it takes to reach this developed regime, which is one reason why it is important to consider an ensemble when talking about decaying turbulence.

For stationary turbulence, when presented with an extended time series for some fluctuating quantity, locating the steady state once it has been reached is relatively simple. One can also play it safe and associate a longer time to the transient, it does not matter since the statistics are stationary. This is not the case for decaying simulations as we do not develop a steady state, nor can we play it safe since the system is decaying and we risk losing the interesting information.

We start by considering the decay of turbulence generated by a grid placed in a wind tunnel, since this is highly relevant for experimental data. The fact that this is not completely incompressible should be borne in mind. Batchelor \cite{Batchelor:1953-book} defined the initial period of decay as the region where the total energy behaves as $E(t) \propto t^{-1}$. Note that this was originally presented for the spatial coordinate measured in the streamwise direction from the grid, as used in grid-generated turbulence as a measurement of time using $x = \overline{U}t$. This is in contrast to the final period of decay, when the Reynolds number becomes sufficiently low that the non-linear term can be ignored. In this case, one finds that the energy spectrum simply decays exponentially. Batchelor showed how this implied $E(t) \propto t^{-5/2}$, although this depends on the shape of the energy spectrum at the start of the decay \cite{thesis:msalewski}.

For a more recent discussion we turn to Davidson \cite{davidson:2004-book}, who describes four stages of evolution for decaying grid-generated turbulence: (1) transition to turbulence after the fluid passes the grid; (2) developed turbulence, where all length-scales $L$ to $\eta$ are excited; (3) small scale depletion; and finally (4) exponential decay. The second stage corresponds to the initial period of the decay described by Batchelor and used in most early work.

\subsection{Power-law decay}
This is the traditional method of evolved time identification, for example \cite{Wang:1996p1041,Bos:2007p128}. One identifies the time from which the total energy experiences power-law decay of the form
\begin{equation}
 E(t) \propto t^{-n} \ ,
\end{equation}
where $n$ is found to be in the range 1--1.7 \cite{Yakhot:2004p1376}. The decaying simulations of Wang, Chen, Brasseur and Wyngaard \cite{Wang:1996p1041} found $n = 1.47$ for $R_\lambda = 20.9$, while a higher exponent of $n = 1.81$ was found for their (lower resolution) $R_\lambda = 68.1$ and 132 simulations.

\begin{figure}[tbp]
 \begin{center}
  \includegraphics[width=0.7\textwidth]{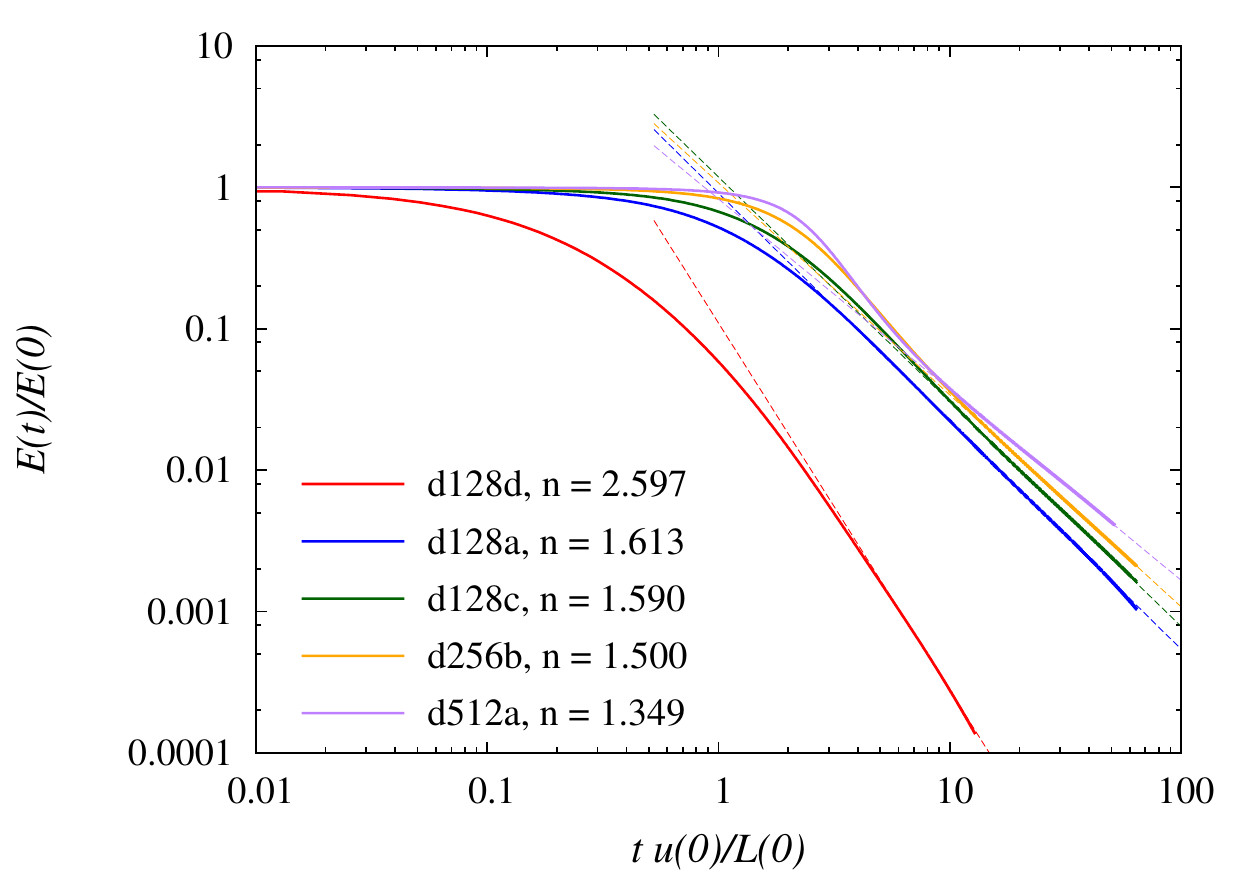}
 \end{center}
 \caption{Identification of power-law decay of the total energy.}
 \label{fig:tevo_powerlaw}
\end{figure}

Figure \ref{fig:tevo_powerlaw} shows the time evolution of the (scaled) total energy on a log-log plot. This shows that power-law decay of the total energy is observed from about 10 initial large eddy turnover times. The decay exponents are given in the figure legend, and with the exception of run \drun{d128d} are all within the range 1--1.7, decreasing as the Reynolds number is increased. Note that run \drun{d128d} has a comparatively `high' viscosity and low (initial) Reynolds number of just $R_\lambda = 2.6$. It is conceivable that this run involves only viscous decay and is not `turbulent' at all.

For decaying turbulence, the rate of change of the total energy is given by the dissipation rate, $dE/dt = -\varepsilon$, in which case the dissipation rate also exhibits power-law decay of the form
\begin{equation}
 \varepsilon \propto t^{-n-1} \ .
\end{equation}
This offers an additional method of identifying the region of power-law decay.

While finding the decay period and its associated exponent presents a challenge, there is not necessarily any difficulty in using this power-law decay of the total energy to define an evolved state. Instead, we focus on criteria based on the internal dynamics of the system through some measurable quantity. In this way, we attempt to define a time at/after which the turbulence can be considered to be evolved.

We note that, if the total energy exhibits power-law behaviour $E(t) \propto t^{-n}$, then for isotropic turbulence it follows that with \emph{any} exponent $n$ the Taylor microscale \emph{must} behave as $\lambda \propto t^{1/2}$. This condition is helpful when identifying power-law behaviour, since if it is not satisfied then the total energy is \emph{not} decaying as a power law.

\subsection{Peak dissipation rate}\label{subsec:tevo_epsPi}
\begin{figure}[tbp]
 \begin{center}
  \includegraphics[width=0.7\textwidth]{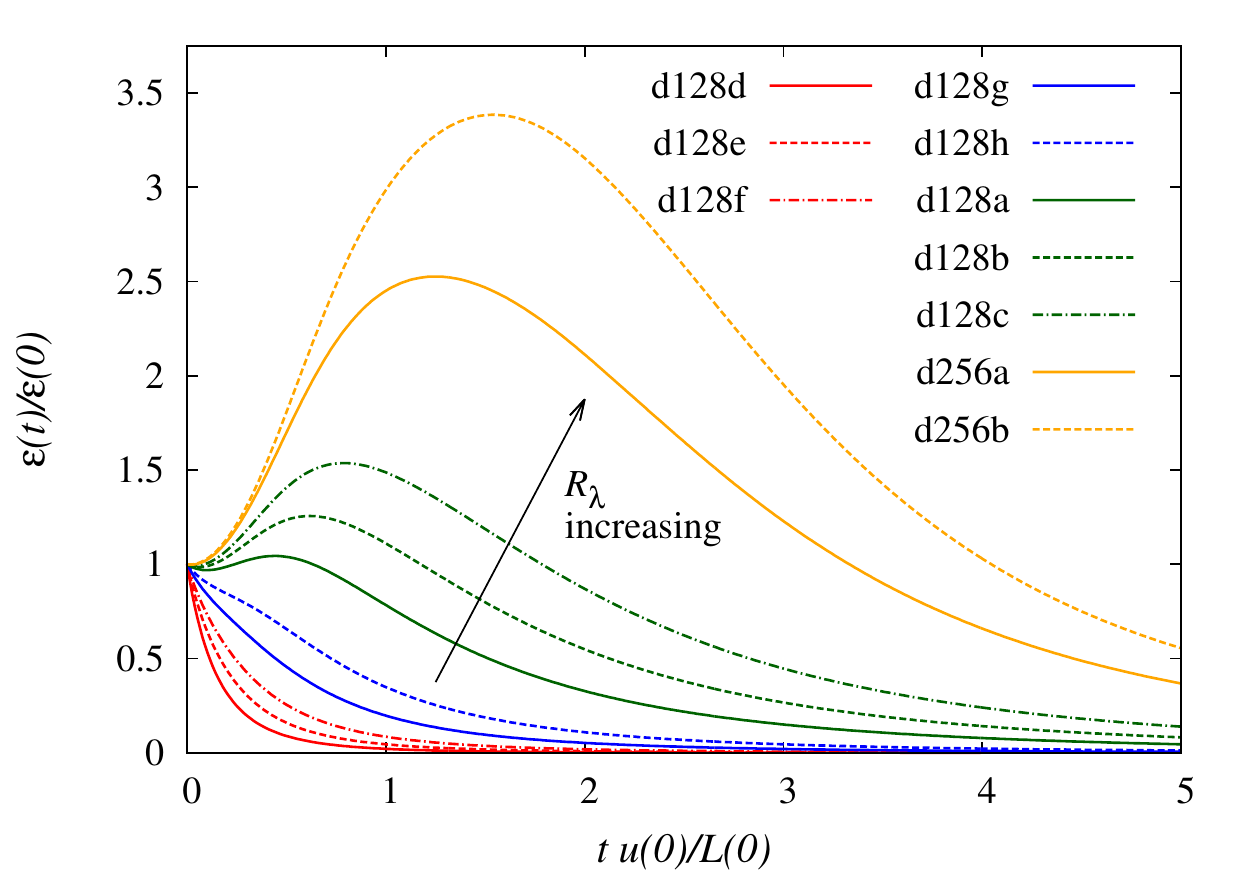}
 \end{center}
 \caption{Time evolution of the dissipation rate for a range of decaying simulations.}
 \label{fig:tevo_eps}
\end{figure}

Since turbulence is characterised as a highly dissipative phenomenon, the measurement of the dissipation rate may offer an indication of a developed state. The time evolution of the dissipation rate is plotted in figure \ref{fig:tevo_eps} for a selection of the decaying simulations performed in this investigation. For simulations with an initial Reynolds number $R_\lambda \gtrsim 25$, the dissipation rate initially increases; the non-linear term is busy transferring energy to smaller and smaller scales where it is dissipated more effectively. This happens until the energy has reached well into the dissipation range and the peak at $t = t_\varepsilon$ could therefore be thought of as indicating the position of maximum turbulent intensity. Since the Kolmogorov scale is based on $\varepsilon$, the peak signifies the excitation of the smallest length-scales. After this, the dissipation rate decays. 

Since a peak value lends itself to easy identification this is an attractive option. This criterion for determining the evolved time was used by Fukayama, Oyamada, Nakano, Gotoh and Yamamoto \cite{Fukayama:1999p904} for their decaying simulations.

\begin{figure}[tbp]
 \begin{center}
  \includegraphics[width=0.7\textwidth]{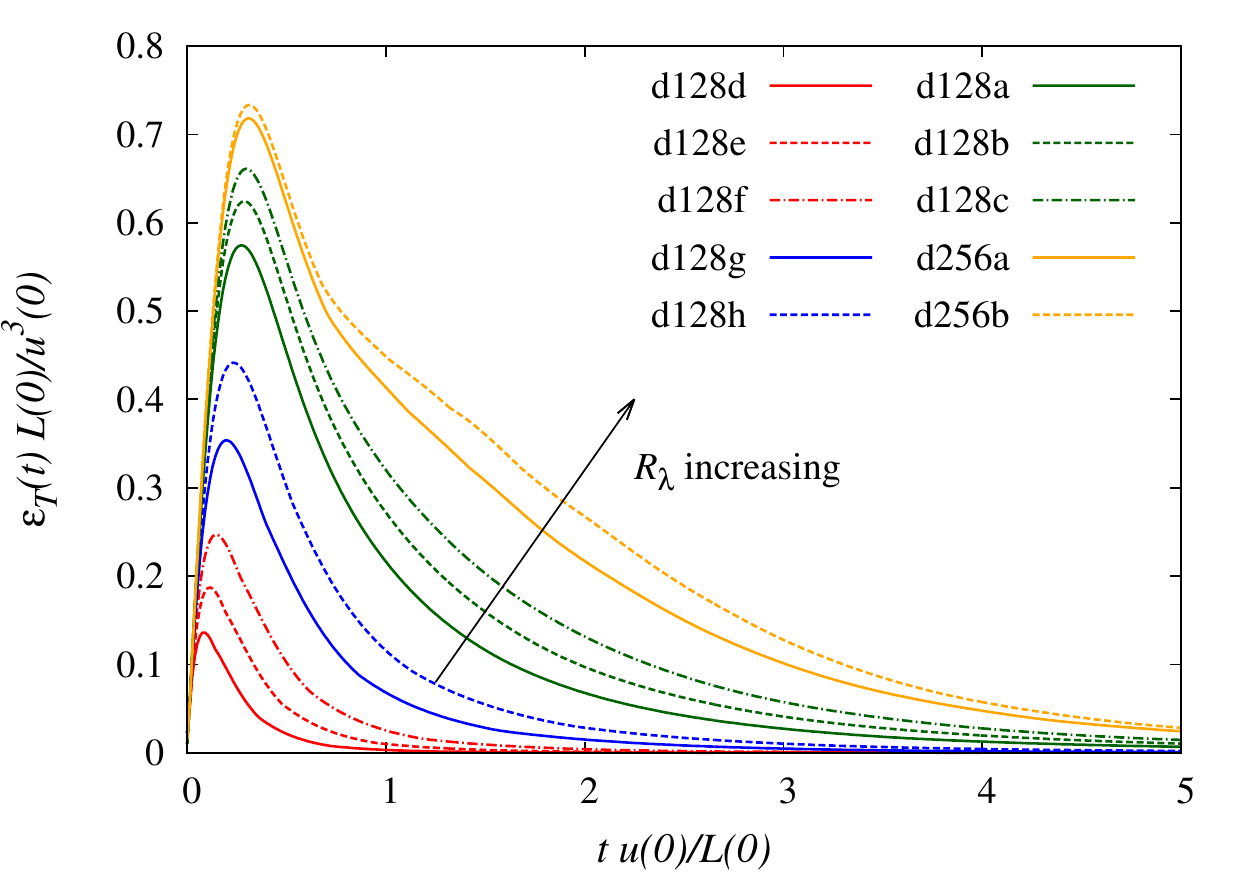}
 \end{center}
 \caption{Time evolution of the maximum inertial flux for a range of simulations.}
 \label{fig:tevo_pi}
\end{figure}

When the Reynolds number is below a certain threshold, dissipation dominates from the outset and the dissipation rate does not develop this peak. Instead, we consider the maximum inertial flux, $\varepsilon_T = \max \Pi(k,t)$, shown in figure \ref{fig:tevo_pi}. This quantity develops a peak at $t = t_\Pi$ and then decays away for all Reynolds numbers. Since the peak implies that the non-linear term is working the hardest, it could be considered as indicating the time at which we have established a fully-developed solution.

\begin{figure}[tbp]
 \begin{center}
  \subfigure[Coincidence of time $t_\Pi$ and inflection of $\varepsilon$]{
   \label{sfig:tevo_eps-pi}
   \includegraphics[width=0.54\textwidth,trim=3px 0 3px 0,clip]{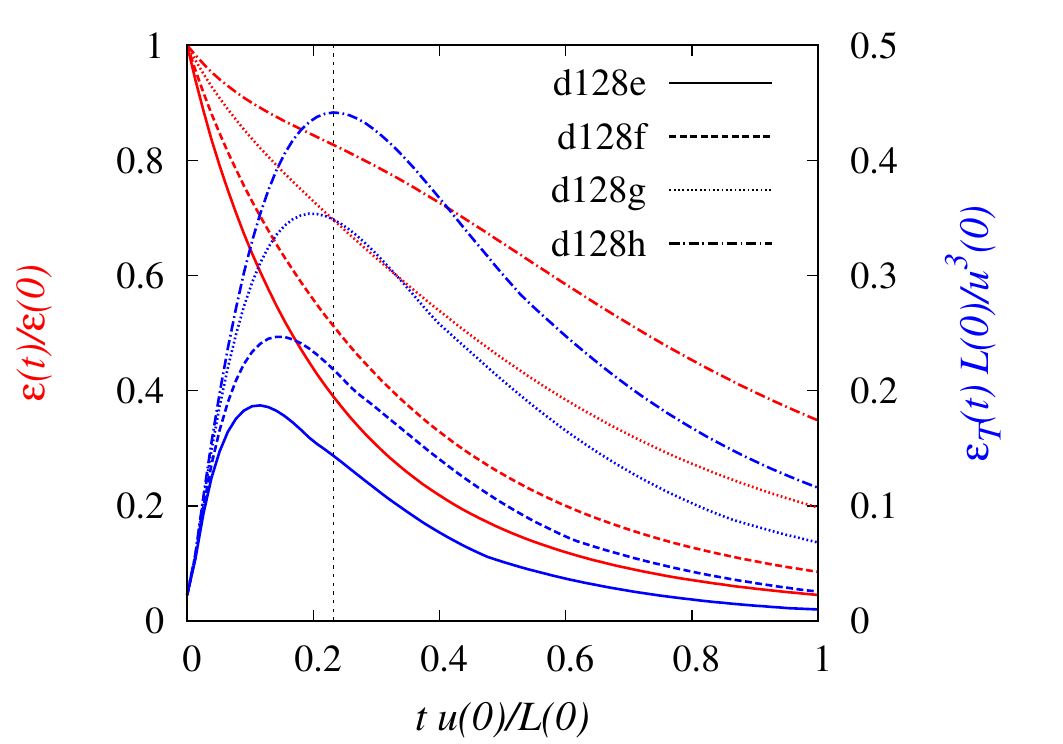}
  }
  \subfigure[Continuity of measured (dimensionless) dissipation rate]{
   \label{sfig:tevo_eps-pi_eps}
   \includegraphics[width=0.41\textwidth,trim=3px -15px 10px 0,clip]{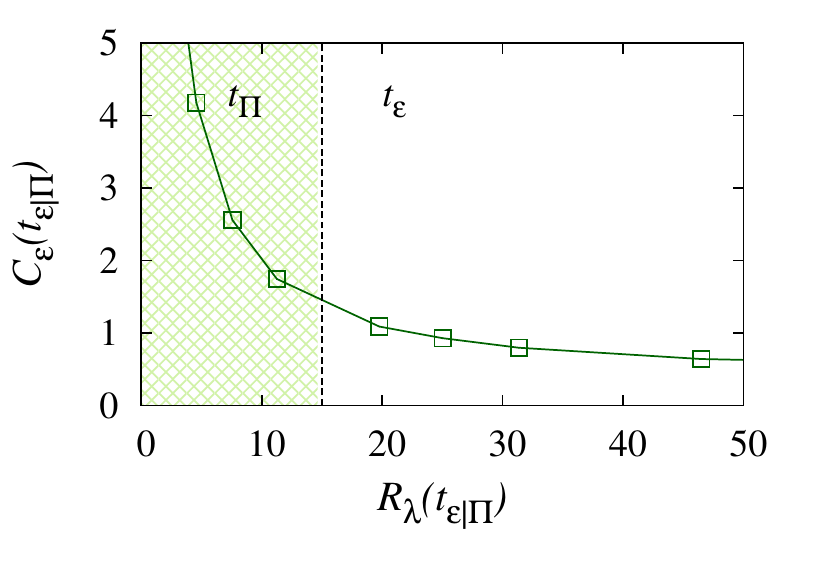}
  }
 \end{center}
 \caption{The use of $t_\Pi$ for low Reynolds number in the development of $t_{\varepsilon\vert\Pi}$.}
 \label{fig:tevo_eps-pi}
\end{figure}

The peak at $t_\Pi$ is seen to remain early in the time evolution of the system. This perhaps indicates that it alone is not representative of an evolved solution. However, when we consider the dissipation rate for the Reynolds numbers which did not develop a peak, we notice that there is an inflection point. In this case, the peak of the inertial flux corresponds nicely to the inflection point in the dissipation rate, as seen in figure \ref{sfig:tevo_eps-pi}. The vertical dotted line indicates the peak in $\varepsilon_T$ for run \drun{d128h}.

We therefore define a composite evolved time
\begin{equation}
 \label{eq:t_eps_Pi}
 t_{\varepsilon\vert\Pi} = \left\{ \begin{array}{ll}
  t_\varepsilon & \textrm{if peak }\varepsilon\textrm{ exists} \\
  t_\Pi & \textrm{otherwise}
 \end{array} \right. \ ,
 \end{equation}
which uses the time associated with the peak in the dissipation rate if it exists or, failing that, the peak in the inertial flux. This provides us with a continuous evolved time for all Reynolds numbers. Figure \ref{sfig:tevo_eps-pi_eps} shows that we do not encounter any discontinuity in the measured value of the dimensionless dissipation coefficient $\Ceps = \varepsilon L/u^3$ (introduced in section \ref{sec:Taylor_surr}) at $t_{\varepsilon\vert\Pi}$ as we go from one regime to the other.

We briefly mention the self-similar decay of the dissipation spectrum for completeness. As the decay starts, the inertial transfer of energy from low modes to high modes causes the dissipation spectrum to spread, as was seen, for example, in figure \ref{sfig:256_0.002_D}. At some point, the non-linear term is unable to transfer energy faster than it is dissipated and the spectrum has spread as far as it can. The highest wavenumber has been excited and the decay then proceeds in a self-similar manner, that is $D(k,t_e) > D(k,t>t_e)$ for all wavenumbers, $k$. This is not investigated further here. For more information, see Salewski \cite{thesis:msalewski}. It was shown that this time occurred later than peak in the dissipation rate but before the onset of power-law decay. It was also present for all Reynolds numbers investigated, unlike $t_\varepsilon$.

\subsection{Peak skewness}
Developed turbulence has a non-Gaussian probability distribution, as measured by the (negative value of) the velocity derivative skewness. What's more, the value of the skewness is not arbitrary but has been shown to have a value around 0.5 for stationary turbulence, see section \ref{subsec:skewness}. For our decaying simulations, it must start at zero (since our initial condition is Gaussian) but become non-zero as turbulence develops.

\begin{figure}[tbp]
 \begin{center}
  \includegraphics[width=0.7\textwidth]{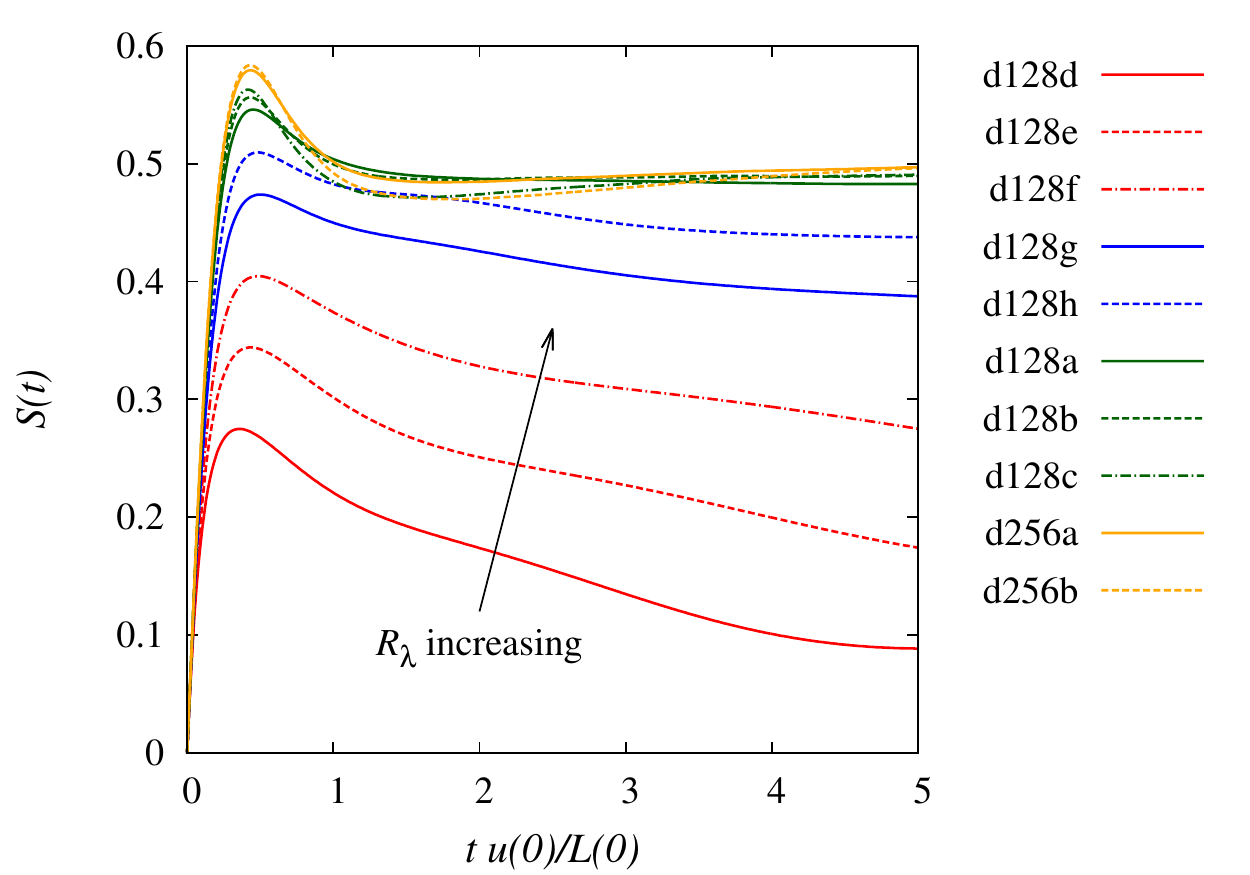}
 \end{center}
 \caption{Time evolution of the velocity derivative skewness. All Reynolds numbers display a peak, with higher $Re$ later developing a plateau around 0.5.}
 \label{fig:tevo_skew}
\end{figure}

The time variation of the skewness for a range of simulations is presented in figure \ref{fig:tevo_skew}. For large enough Reynolds numbers, a plateau appears to develop around 0.5, perhaps indicating that the probability distribution has reached its developed form. However, this is not the case for all Reynolds numbers. In fact, it is the same set of Reynolds numbers which presented us with a problem when looking for a peak in the dissipation rate. Despite this, there is a distinct peak in the skewness present for all Reynolds numbers, denoted $t_S$. Like the peak in the maximum inertial flux, this occurs very early in the evolution of the system and is seen to hardly increase as $Re$ increases.

\subsection*{A comment on the evolved times}
The use of power-law decay of the energy was seen to occur significantly later in the evolution of the system than $t_\varepsilon$, $t_\Pi$ or $t_S$ determined from the dynamics. This requires the simulation to run longer and, since it is decaying as it does so, limits us to lower Reynolds numbers from the same run. On the other hand, the earlier times may actually be located in the transition period and hence should not be used.

\section{Decay from a stationary field}\label{sec:decay_from_forced}
Besides starting from an artificial initial condition and waiting until a fully-developed solution has been reached, one can consider starting from a stationary field obtained during a forced simulation. In this way, the system is not decaying while we wait to develop an evolved solution. However, the field is a solution of the forced Navier-Stokes equation. As such, once the forcing is `switched off' there will still be a transient period as the system adapts to the lack of energy input.

To explore this setup, we created run \drun{d1024a} which involves an ensemble of five initial fields. The realisations are sampled at an interval of one large eddy turnover from our highest Reynolds number stationary simulation, run \texttt{f1024b} (see table \ref{tbl:summary_sims}). Each initial realisation is then run with the input rate set to zero. Due to time and computational constraints, an ensemble of five was all that could be generated. This is because it requires approximately 55GB to store each initial realisation and 48 hours to generate one from the next. The decay was run for 96 hours per realisation, allowing 6.38s of data to be collected. This corresponds to just over three steady-state turnover times. In total, this required over 28 days of run time (not including time spent copying data or waiting in job scheduling queues for 132 infiniband nodes to become available).

Since the decay proceeds from a stationary simulation, the initial values of the parameters at $t = 0$ are simply the steady state values from the forced run. It should be noted that a large eddy turnover based on these values, $\tau(0) = L/u = 1.94s$, is significantly longer than that for the Gaussian initial condition, $\tau(0) = 0.777s$ (S5).

\begin{figure}[tbp]
 \begin{center}
  \subfigure[Evolution of length-scales, Reynolds numbers and skewness]{
   \label{sfig:fdecay_params}
   \includegraphics[width=0.4775\textwidth,trim=5px 0 10px 0,clip]{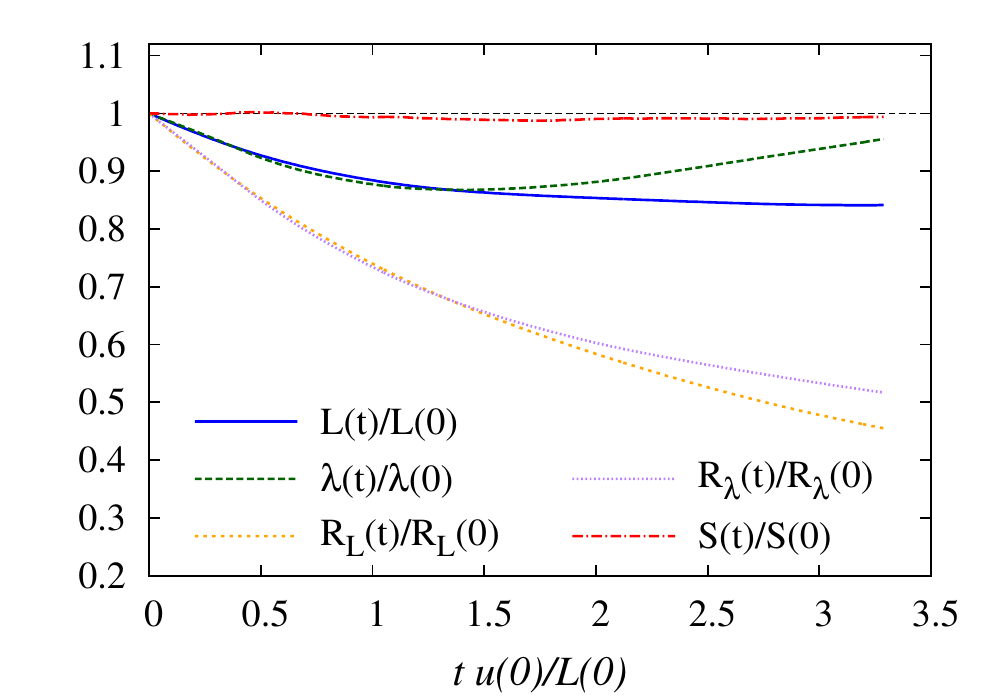}
  }
  \subfigure[Variation of energy, dissipation and inertial flux]{
   \label{sfig:fdecay_rates}
   \includegraphics[width=0.4775\textwidth,trim=5px 0 10px 0,clip]{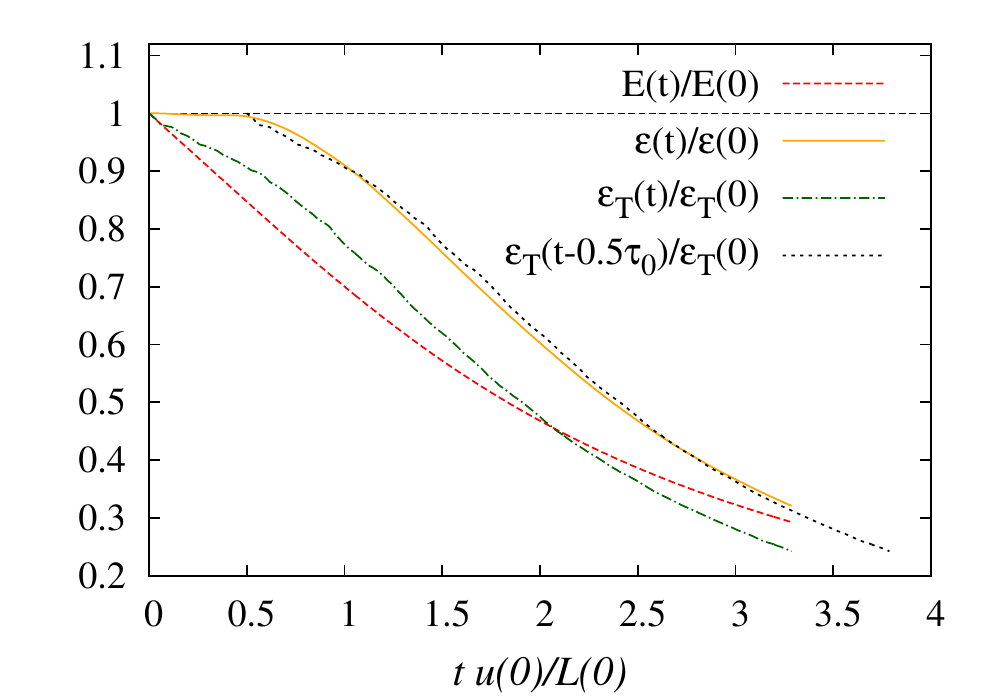}
  }
 \end{center}
 \caption{Time variation of key turbulence parameters for a decaying simulation started from a stationary evolved field at time $t = 0$.}
 \label{fig:fdecay}
\end{figure}

The time evolution of various parameters from the start of the decay (at $t = 0$) is shown in figure \ref{fig:fdecay}. Figure \ref{sfig:fdecay_params} shows the variation of length-scales and Reynolds numbers, along with the velocity derivative skewness. As expected, the skewness of the stationary field had already obtained its steady state value of around 0.55. Once the forcing is removed, there is little change in the skewness although it does appear to adopt a slightly lower value.

Turning our attention to figure \ref{sfig:fdecay_rates}, we notice straightaway that there is a period of about $0.5 \tau(0)$ after the start of the decay during which the dissipation rate remains constant at its steady state value from the forced simulation. This is very interesting and essentially measures the time it takes for energy to pass through the cascade, since during this time the high wavenumbers are not aware of the change that has occurred at low wavenumbers. The total energy and maximum inertial flux are both seen to start their decay from $t = 0$. The dissipation rate appears to mimic the maximum inertial transfer, with its curve shifted to the right by about $0.5\tau(0)$. This is observed for the full time range available here. To highlight this, we have also plotted the $\varepsilon_T$ curve shifted by exactly $0.5\tau(0)$, as shown by the black dotted line.

As an estimate for the time associated with the peak dissipation, we take the time where the dissipation rate starts to decay. This occurs around $0.41\tau(0)$. Due to its temporal location, we have left the steady state and have $\varepsilon_T(t) < \varepsilon(t)$. But the measured value of the dissipation rate will still closely resemble that of the forced system. This could be an indication that measurements at $t_\varepsilon$ in a decaying simulation are similar to the steady state of a stationary one. 

\section{Dependence of statistics on the choice of $t_e$}\label{subsec:tevo_dep}
Since the dynamical evolved times $t_{\varepsilon\vert\Pi}$ and $t_S$ occur early in the decay compared to the onset of power-law decay, it is interesting to compare how the choice of evolved time affects measurements of the system. We start with a consideration of the energy cascade.

In forced turbulence, energy enters the large scales through whatever forcing scheme is put in place at a rate $\varepsilon_W$. Once a steady state is reached, we must have a balance of energy in and out such that $\varepsilon_W = \varepsilon$. That is, the turbulence organises itself by creating scales such that it can dissipate energy at the same rate as it receives it. At low Reynolds numbers, there may be some loss of energy directly from the large scales due to viscosity, so the amount of energy passing through the cascade, measured by $\varepsilon_T = \max \Pi(k)$, can be less than $\varepsilon$. As Reynolds number increases, the loss of energy from large scales becomes negligible and all the energy lost passes through the cascade such that we measure $\varepsilon_T = \varepsilon$.

The story for decaying turbulence is quite different. Let us first consider Reynolds number sufficiently high that there is very little direct dissipation from the large scales. In our transition from the Gaussian initial condition with energy concentrated in the low wavenumbers to a field characteristic of developed turbulence, the inertial transfer of energy can in fact dominate over dissipation. The system needs to move the energy to the high wavenumbers so that it can be removed more efficiently. Therefore, we can measure $\varepsilon_T > \varepsilon$. We stress that this is \emph{only in the transition period} of the evolution. Once the system has settled to a developed solution and turbulence has been fully established, transfer can no longer occur quicker than dissipation. Since it takes a finite amount of time for the energy to filter down through the cascade to smaller length-scales, the energy transferred at time $t$ will be dissipated at a later time, $t + \Delta t$. If the turbulence is decaying, then $\varepsilon(t + \Delta t) < \varepsilon(t)$ and as such we find
\begin{equation}
 \varepsilon_T(t) < \varepsilon(t) \ .
\end{equation}
We therefore see that $t_\varepsilon$, the time corresponding to the peak of the dissipation rate, is the border between the two cases $\varepsilon_T(t) > \varepsilon(t)$ and $\varepsilon_T(t) < \varepsilon(t)$. Therefore, at this time one could imagine measuring $\varepsilon_T(t_\varepsilon) = \varepsilon(t_\varepsilon)$.

If we now reduce the Reynolds number such that our large scales are directly influenced by dissipation, the picture is blurred slightly. For $t \geq t_\varepsilon$, where before we had $\varepsilon(t) \geq \varepsilon_T(t)$, we now have strictly $\varepsilon(t) > \varepsilon_T(t)$. Although we can still achieve $\varepsilon_T(t) > \varepsilon(t)$ in the transition period. The point is that the peak of the dissipation rate is no longer associated with equality of transfer and decay rates, but this is a finite Reynolds number effect.

\begin{figure}[tbp]
 \begin{center}
  \includegraphics[width=0.7\textwidth]{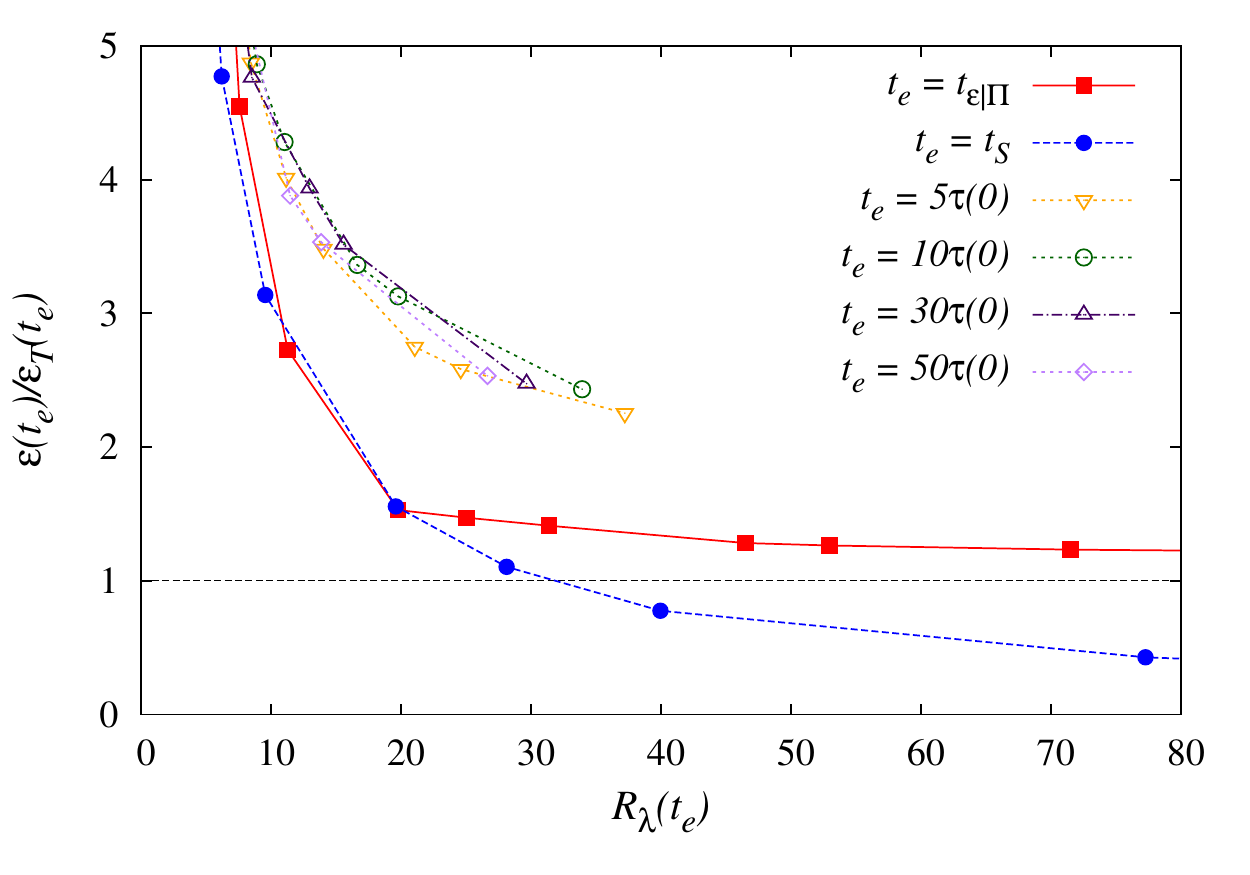}
 \end{center}
 \caption{The ratio $\varepsilon/\varepsilon_T$ measured using several criteria for the evolved time, $t_e$.}
 \label{fig:decay_eps-pi}
\end{figure}
Figure \ref{fig:decay_eps-pi} shows the variation of $\varepsilon(t_e)/\varepsilon_T(t_e)$ with Reynolds number, as measured using several criteria for the evolved time. The use of skewness is clearly a measurement made in the transition period as we have $\varepsilon_T(t_S) > \varepsilon(t_S)$. This initially behaved similarly to the measurement using $t_{\varepsilon\vert\Pi}$. However, the latter does not cross unity: We measure $\varepsilon(t_{\varepsilon\vert\Pi}) > \varepsilon_T(t_{\varepsilon\vert\Pi})$ for all Reynolds numbers. Unlike the curves measured late in the evolution, where power-law decay of the total energy is found, the $t_{\varepsilon\vert\Pi}$ curve could be asymptoting towards unity, which would be in agreement with the discussion above. The curves measured in the power-law period are in good agreement with one another. One could possibly use this ratio and the decay exponent to estimate the transient time for $\varepsilon_T(t)$ to pass through the cascade and be dissipated as $\varepsilon(t + \Delta t)$.

\subsection{The Taylor dissipation surrogate}\label{sec:Taylor_surr}
In 1935, Taylor \cite{Taylor:1935p308} introduced an expression for the dissipation rate which, for the case of isotropic turbulence of interest here, takes the form
\begin{equation}
 \varepsilon = \Ceps(Re) \frac{u^3}{L} \ .
\end{equation}
$\Ceps(Re)$ is referred to as Taylor's dissipation coefficient or the \emph{dimensionless dissipation rate}. The expression was put forward on dimensional grounds, and Batchelor \cite{Batchelor:1953-book} presented data to suggest that this Reynolds number dependent coefficient became a constant as $Re$ was increased. Later, Sreenivasan \cite{Sreenivasan:1984p139,Sreenivasan:1998p898} provided a compilation of experimental and numerical data which showed $\Ceps$ becoming constant for $R_\lambda \gtrsim 50$. This is known as the dissipation anomaly and a discussion is deferred to section \ref{sec:DA}.

\begin{figure}[tbp]
 \begin{center}
  \subfigure[$t_e = t_{\varepsilon\vert\Pi}$]{
   \label{sfig:decay_eps-pi-surr_eP}
   \includegraphics[width=0.475\textwidth,trim=3px 0 10px 0,clip]{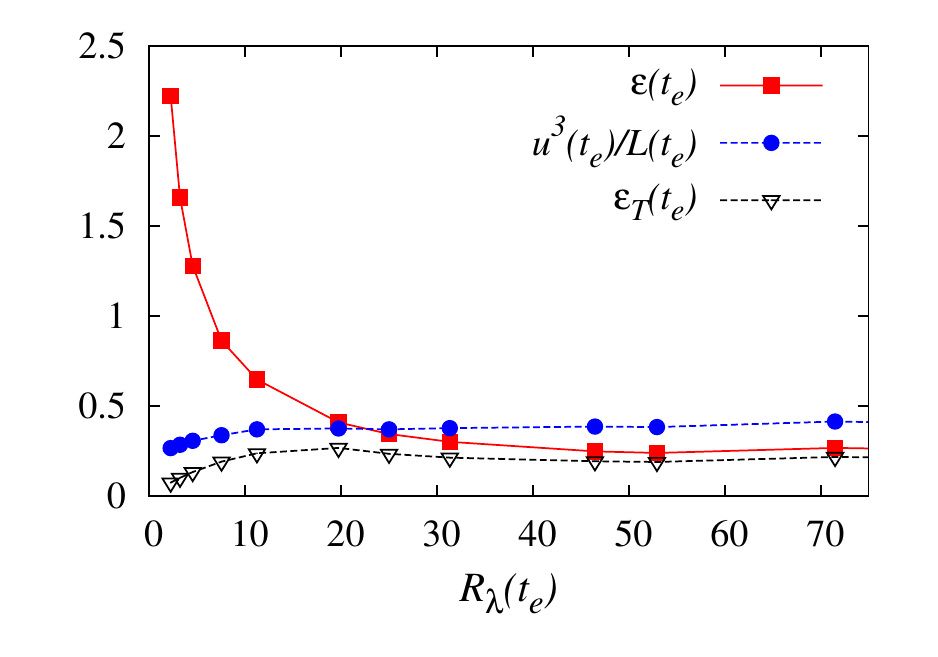}
  }
  \subfigure[$t_e = t_S$]{
   \label{sfig:decay_eps-pi-surr_skew}
   \includegraphics[width=0.475\textwidth,trim=3px 0 10px 0,clip]{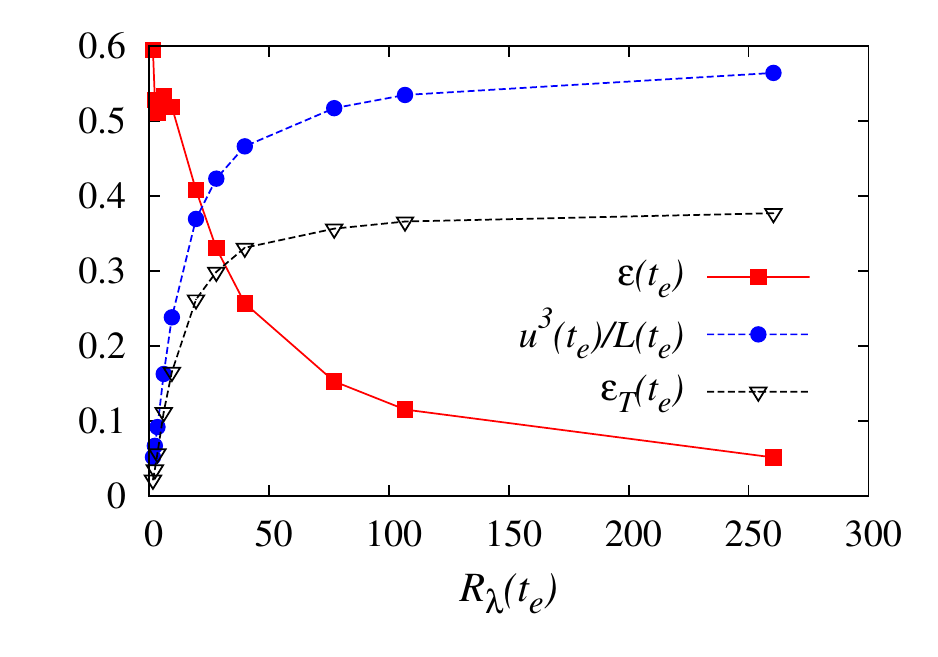}
  }
  \subfigure[$t_e = 5\tau(0)$]{
   \label{sfig:decay_eps-pi-surr_t5}
   \includegraphics[width=0.475\textwidth,trim=3px 0 10px 0,clip]{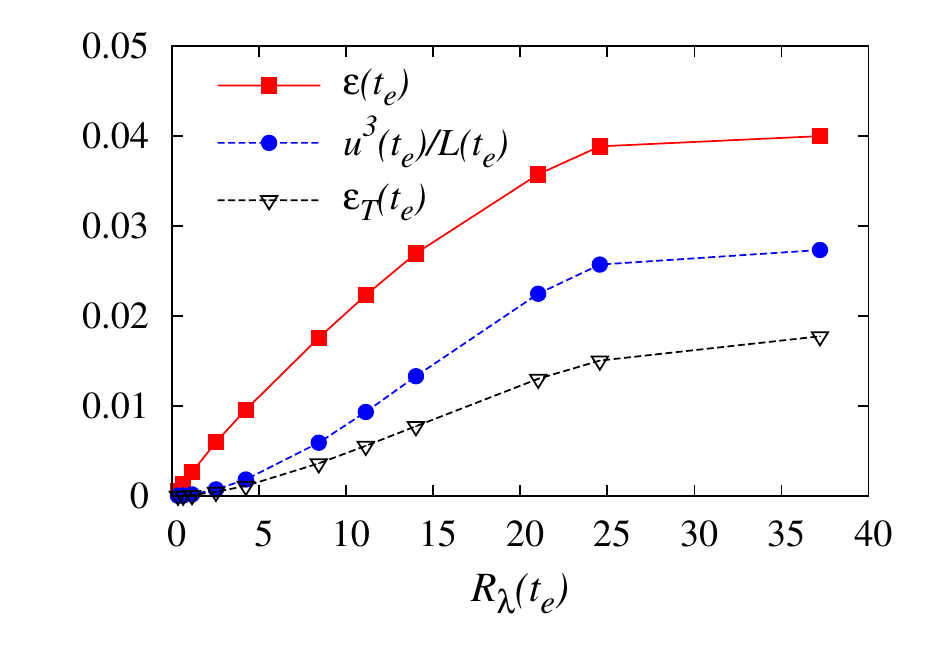}
  }
  \subfigure[$t_e = 10\tau(0)$]{
   \label{sfig:decay_eps-pi-surr_t10}
   \includegraphics[width=0.475\textwidth,trim=3px 0 10px 0,clip]{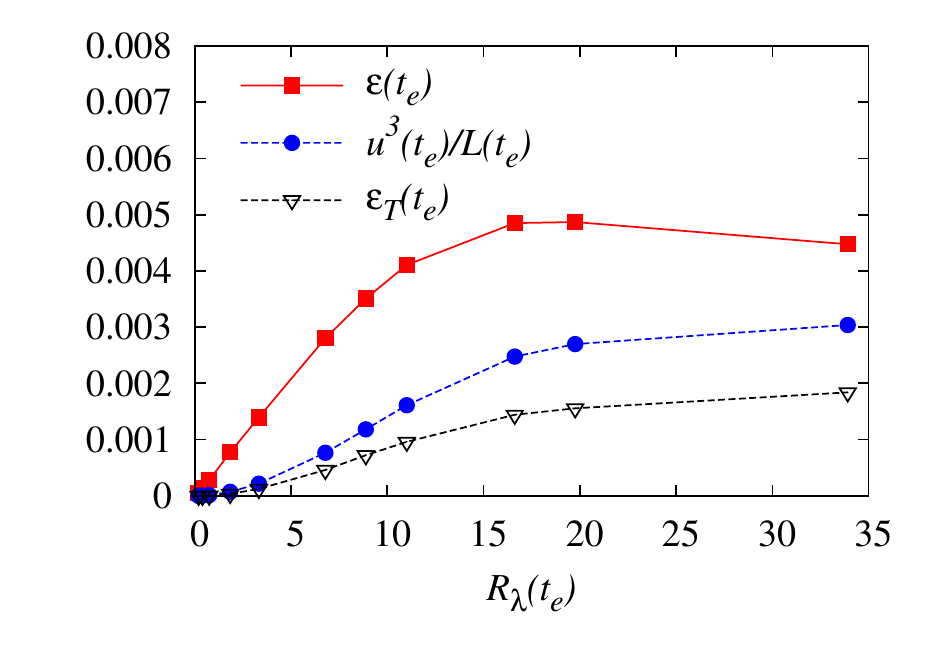}
  }
  \subfigure[$t_e = 30\tau(0)$]{
   \label{sfig:decay_eps-pi-surr_t30}
   \includegraphics[width=0.475\textwidth,trim=3px 0 10px 0,clip]{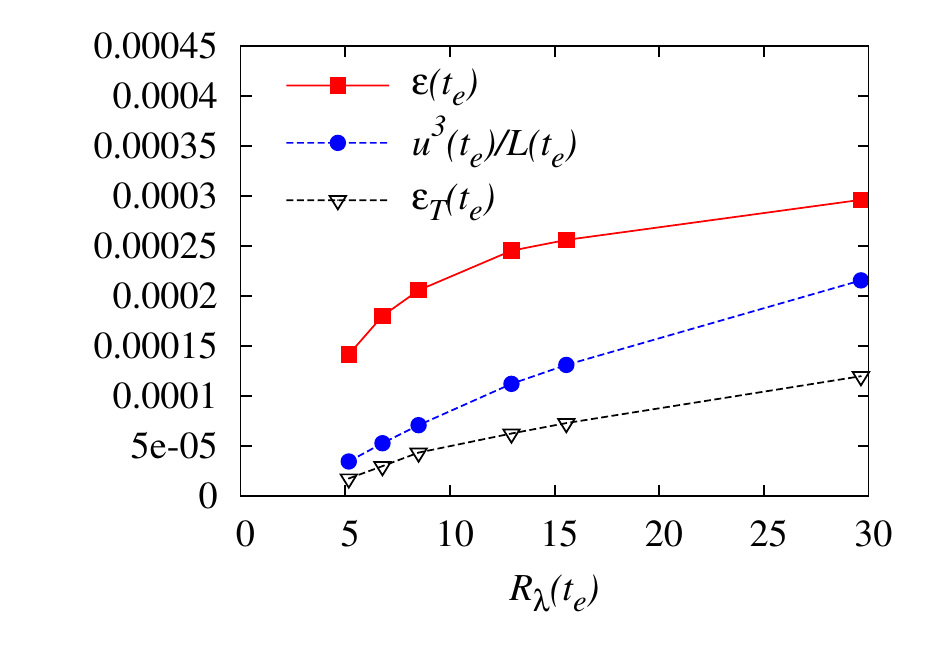}
  }
  \subfigure[$t_e = 50\tau(0)$]{
   \label{sfig:decay_eps-pi-surr_t50}
   \includegraphics[width=0.475\textwidth,trim=3px 0 10px 0,clip]{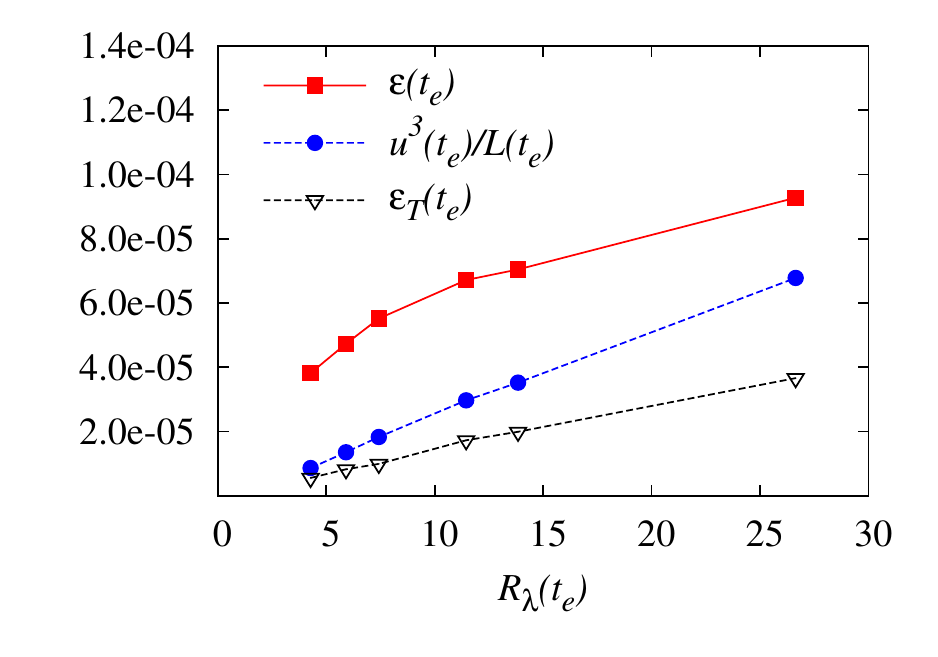}
  }
 \end{center}
 \caption{Comparison of the Taylor surrogate to dissipation and maximum inertial flux using different evolved time criteria.}
 \label{fig:decay_eps-pi-surr}
\end{figure}
Instead, we choose to focus on the behaviour of the dissipation surrogate, $u^3/L$. McComb, Berera, Salewski and Yoffe \cite{McComb:2010p250} showed how $u^3/L$ is a better surrogate for the maximum inertial flux than for the dissipation rate. This was presented using the $t_{\varepsilon\vert\Pi}$ criterion (see equation \eqref{eq:t_eps_Pi}) for a range of initial spectra with low wavenumber behaviour going as $k^2$ and $k^4$. This is reassuring, since in decaying turbulence it is thought that the initial condition can have a direct impact on the turbulence which stems from it. Figure \ref{sfig:decay_eps-pi-surr_eP} shows this behaviour for the initial spectrum used in these decaying simulations. As can be seen, at low Reynolds numbers the dissipation increases whereas the inertial flux and dissipation surrogate turn down towards zero.

Figure \ref{fig:decay_eps-pi-surr} shows the variation of the dissipation surrogate along with the dissipation and inertial transfer rates for a variety of evolved time criteria. We see a striking difference in the behaviour of the dissipation rate between the early measurement times and those associated with power-law decay. The peak in the skewness used in figure \ref{sfig:decay_eps-pi-surr_skew} also shows the dissipation rate behaving very differently to the surrogate and inertial flux. Whereas, measurement made once power-law decay has been established shows all three quantities going to zero as $R_\lambda \to 0$, figures \ref{sfig:decay_eps-pi-surr_t5}--\ref{sfig:decay_eps-pi-surr_t50}. Presented with only the power-law decay data, one could conclude that $u^3/L$ is a surrogate for dissipation, but we see that $u^3/L$ follows the maximum inertial flux for \emph{all} times which have been sampled here. This is compelling evidence that $u^3/L$ is connected to inertial flux rather than dissipation. This is a view supported by Tennekes and Lumley \cite{TennekesLumley:1972} who always use $u^3/L$ when discussing inertial transfer.

\section{Discussion}
The free-decay of isotropic, homogeneous turbulence from both Gaussian initial conditions and an evolved velocity field has been investigated using direct numerical simulation. Using the DNS data obtained, we have verified that the time-evolution of the energy and transfer spectra, as well as key statistical quantities derived from them, are as expected and in agreement with the literature.

A key concept which we have built upon is the determination of an evolved time for decaying turbulence. This is an issue for simulations starting from both Gaussian initial conditions and evolved velocity fields, but also for the collection of experimental data. Based on dynamical quantities measured from the velocity field, we outlined criteria for defining evolved times based on the peak dissipation rate (or maximum inertial flux) and peak skewness. These occur significantly earlier in the decay than power-law decay of the total energy. It was suggested that using $t_\varepsilon$ might compare more favourably to stationary results then being characteristic of decay.

The effect the evolved measurement time on statistical quantities was 
studied. Behaviour of the ratio $\varepsilon/\varepsilon_T$ was shown to 
diverge into three possible classes as (evolved) Reynolds number increases: 
$\varepsilon/\varepsilon_T < 1$ found using the peak skewness, an early 
measurement time; $\varepsilon/\varepsilon_T \to 1$ using the peak in the 
dissipation rate, an intermediate time; or $\varepsilon/\varepsilon_T > 1$ after power-law energy decay is observed. The latter was studied using four, progressively later, measurement times. The change of behaviour of the dissipation rate was also investigated, where it was found that $u^3/L$ was a better surrogate for the maximum inertial flux at all measurement times. This is in agreement with McComb, Berera, Salewski and Yoffe \cite{McComb:2010p250}, where is was shown using a variety of initial spectra for the hybrid dissipation-inertial flux evolved time. It would be useful to provide the analysis performed here for a variety of initial conditions.

For the case when decay was created by switching off the energy injection mechanism for an evolved, stationary velocity field, it was found that there is a finite time of about $0.5\tau(0)$, where $\tau(0)$ is the initial eddy turnover time, before the dissipation rate reacts to loss of energy from the system. This was interpreted as quantifying the time it took energy to pass through the energy cascade. In contrast, the maximum inertial flux was seen to begin its decay immediately, as did the measurement of the total energy. It would be interesting to perform this decay from a stationary field for a range of initial Reynolds numbers in an attempt to study the behaviour of the finite `cascade time'. The use stationary fields maintained by different forcing mechanisms and the effect on the cascade time would also be of interest.

\chapter{Numerical investigation of stationary isotropic turbulence}\label{chp:forced}

This chapter focuses on the data obtained from a series of forced turbulence simulations. As with the decaying case, the initial condition used the same energy spectrum to keep it as consistent as possible, with variation of Reynolds number introduced by changing only the viscosity. The input rate of energy into the large-scales was also maintained constant. The stationary simulations are detailed in the following section and summarised in tables \ref{tbl:summary_sims} and \ref{tbl:summary_stats}. The data generated by these simulations has been used to conduct several numerical experiments on the properties of the velocity field, and these are discussed in the proceeding sections of this chapter. As mentioned in section \ref{sec:further_tests}, some of the results presented here can also be considered as further evidence that the code is performing as expected, and comparison to the literature will be drawn where appropriate.

\newpage

\section{Summary of stationary simulations}\label{sec:forced_summary}
At present, the largest DNS of stationary homogeneous, isotropic turbulence for an incompressible fluid was carried out on the Earth Simulator in Japan \cite{Yokokawa:2006p169,Kaneda:2006p183}. They achieved double-precision floating point simulations on lattices up to $2048^3$, with a Taylor-Reynolds number of $R_\lambda = 732$. By using single-precision for the time integral and double-precision for the calculation of the non-linear term, this was pushed up to $4096^3$ with $R_\lambda = 1131$ \cite{Ishihara:2007p612,Ishihara:2009p165}. While we cannot compete with these numbers, we have obtained results on lattices of up to $1024^3$ (double-precision) with a steady state Reynolds number of $R_\lambda = 335$. This required 128 processes on \eddie\ using the \emph{infiniband} network, and required approximately 32 hours per large eddy turnover time, $\tau = L/u$.

The time evolution of forced isotropic turbulence has been simulated for a variety of Reynolds numbers. The system is initialised as a Gaussian random field using the method detailed in section \ref{subsec:initial_field} with initial spectrum S5 ($k^4$ low $k$ behaviour) and allowed to reach a steady state solution of the Navier-Stokes equations. Once this initial transient period has passed, the velocity field is sampled every large eddy turnover time, $\tau = L/u$, to create an ensemble. As well as shell averaging, the energy and transfer spectra are also averaged over this ensemble and used to calculate the statistics of the velocity field, as detailed in section \ref{sec:calc_stats}. This is beneficial for quantities derived from the transfer spectrum.

\subsection{Statistics and spectra}

Table \ref{tbl:summary_sims} provides a summary of the main stationary simulations which have been run, while \ref{tbl:summary_stats} summarises the mean values of the most common statistical quantities. Figure \ref{fig:forced_time_series} shows the evolution of these key parameters as the simulation progresses from its Gaussian initial condition to steady state. The quantities have been scaled by their time-averaged mean value (note that this is different for each run). As can be seen, after $t \sim 10 \langle L \rangle/\langle u \rangle$ most simulations have settled to their steady state solution. The figures also highlight how stationarity is a statistical concept --- fluctuations around the mean are expected and present in the system but they should vanish in an average. A single measurement need not necessarily represent a stationary system.

\begin{sidewaystable}[tbp!]
\begin{center}
\begin{tabular}{r|llll|ll|lll|llll}
ID & $N$ & $\nu_0$ & \# & $E(k,0)$ & $R_L$ & $R_\lambda$ & $\kmax$ & $\eta\ (\times 10^{-3})$ & $\kmax\eta$ & $t_{\textrm{max}}$ & $dt$ & $t_c$ & $t_\nu$ \\
\hline
\hline
\texttt{f64d}   & 64   & 0.09    & --- & S5 & 10.6   & 8.70  & 20  & 306  & 6.12 & 50    & 0.001 & 0.113 & 0.028 \\
\texttt{f64c}   & 64   & 0.07    & --- & S5 & 12.8   & 9.91  & 20  & 255  & 5.10 & 50    & 0.001 & 0.114 & 0.036 \\
\texttt{f64a}   & 64   & 0.05    & --- & S5 & 19.0   & 13.9  & 20  & 196  & 3.91 & 50    & 0.005 & 0.103 & 0.050 \\
\texttt{f64b}   & 64   & 0.02    & --- & S5 & 39.5   & 24.7  & 20  & 96.5 & 1.93 & 50    & 0.005 & 0.096 & 0.125 \\
\texttt{f128a}  & 128  & 0.01    & 101 & S5 & 82.7   & 42.5  & 41  & 57.1 & 2.34 & 149.9 & 0.001 & 0.042 & 0.059 \\
\texttt{f128b}  & 128  & 0.009   & --- & S5 & 88.2   & 44.0  & 41  & 52.5 & 2.15 & 50    & 0.001 & 0.042 & 0.066 \\
\texttt{f128c}  & 128  & 0.008   & --- & S5 & 101.4  & 48.0  & 41  & 48.0 & 1.96 & 50    & 0.001 & 0.042 & 0.074 \\
\texttt{f128d}  & 128  & 0.007   & --- & S5 & 105.7  & 49.6  & 41  & 43.3 & 1.77 & 50    & 0.001 & 0.042 & 0.085 \\
\texttt{f128e}  & 128  & 0.005   & 101 & S5 & 158.6  & 64.2  & 41  & 33.5 & 1.37 & 149.9 & 0.001 & 0.040 & 0.119 \\
\texttt{f256a}  & 256  & 0.0025  & --- & S5 & 284.6  & 89.3  & 84  & 20.0 & 1.67 & 50    & 0.001 & 0.020 & 0.057 \\
\texttt{f256b}  & 256  & 0.002   & 101 & S5 & 360.1  & 101.3 & 84  & 16.9 & 1.41 & 120   & 0.001 & 0.020 & 0.071 \\
\texttt{f256c}  & 256  & 0.0018  & --- & S5 & 432.6  & 113.3 & 84  & 15.5 & 1.31 & 50    & 0.001 & 0.019 & 0.079 \\
\texttt{f512a}  & 512  & 0.00072 & 15  & S5 & 1026   & 176.9 & 169 & 7.78 & 1.31 & 50    & 0.001 & 0.0095& 0.049 \\
\texttt{f512b}  & 512  & 0.0005  & --- & S5 & 1373   & 203.7 & 169 & 5.96 & 1.01 & 50    & 0.001 & 0.0097& 0.070 \\
\texttt{f512c}  & 512  & 0.01    & --- & S5 & 81.5   & 41.8  & 169 & 56.6 & 9.57 & 50    & 0.001 & 0.0102& 0.004 \\
\texttt{f512d}  & 512  & 0.005   & --- & S5 & 146.5  & 60.8  & 169 & 33.6 & 5.68 & 50    & 0.001 & 0.0100& 0.007 \\
\texttt{f512e}  & 512  & 0.0025  & --- & S5 & 287.8  & 89.4  & 169 & 19.8 & 3.35 & 50    & 0.001 & 0.0098& 0.014 \\
\texttt{f512f}  & 512  & 0.0018  & --- & S5 & 436.3  & 113.0 & 169 & 15.7 & 2.65 & 50    & 0.001 & 0.0095& 0.019 \\
\texttt{f512g}  & 512  & 0.001   & --- & S5 & 785.2  & 153.4 & 169 & 10.1 & 1.70 & 50    & 0.001 & 0.0095& 0.035 \\
\texttt{f1024a} & 1024 & 0.0003  & --- & S5 & 2415   & 276.2 & 340 & 4.05 & 1.38 & 50    & 0.0004& 0.0047& 0.029 \\
\texttt{f1024b} & 1024 & 0.0002  & --- & S5 & 3535   & 335.2 & 340 & 2.97 & 1.01 & 50    & 0.0004& 0.0047& 0.043
\end{tabular}
\caption{Summary of the forced simulations that have been run and their parameters.}
\label{tbl:summary_sims}
\end{center}
\end{sidewaystable}

\begin{sidewaystable}[tbp!]
\begin{center}
\begin{tabular}{r|llllllllllc}
ID & $R_L$ & $R_\lambda$ & $u$ & $L$ & $\lambda$ & $\varepsilon$ & $\varepsilon_T$ & $S$ & $k_{\textrm{diss}}$ & $k_d$ & $\Pi(0)\ (\times 10^{-9})$ \\
\hline
\hline
\texttt{f64d}   & 10.6   & 8.70  & 0.441 & 2.163 & 1.777 & 0.083 & 0.026 & 0.566 & 5   & 3   & 62.9  \\
\texttt{f64c}   & 12.8   & 9.91  & 0.440 & 2.041 & 1.578 & 0.081 & 0.031 & 0.615 & 6   & 4   & 5.20  \\
\texttt{f64a}   & 19.0   & 13.9  & 0.485 & 1.956 & 1.435 & 0.086 & 0.037 & 0.583 & 7   & 5   & -42.9 \\
\texttt{f64b}   & 39.5   & 24.7  & 0.523 & 1.512 & 0.943 & 0.092 & 0.060 & 0.554 & 13  & 10  & -64.0 \\
\texttt{f128a}  & 82.7   & 42.5  & 0.581 & 1.442 & 0.733 & 0.094 & 0.079 & 0.540 & 22  & 18  & -73.9 \\
\texttt{f128b}  & 88.2   & 44.0  & 0.578 & 1.374 & 0.686 & 0.096 & 0.083 & 0.533 & 24  & 19  & 60.1  \\
\texttt{f128c}  & 101.4  & 48.0  & 0.586 & 1.383 & 0.655 & 0.096 & 0.084 & 0.535 & 26  & 21  & -46.5 \\
\texttt{f128d}  & 105.7  & 49.6  & 0.579 & 1.279 & 0.600 & 0.098 & 0.088 & 0.531 & 29  & 23  & 20.7  \\
\texttt{f128e}  & 158.6  & 64.2  & 0.607 & 1.307 & 0.529 & 0.099 & 0.092 & 0.529 & 38  & 30  & -5.10 \\
\texttt{f256a}  & 284.6  & 89.3  & 0.600 & 1.185 & 0.372 & 0.098 & 0.095 & 0.522 & 64  & 50  & -58.1 \\
\texttt{f256b}  & 360.1  & 101.3 & 0.607 & 1.187 & 0.334 & 0.099 & 0.096 & 0.521 & 76  & 59  & 37.5  \\
\texttt{f256c}  & 432.6  & 113.3 & 0.626 & 1.243 & 0.326 & 0.100 & 0.099 & 0.525 & 80  & 65  & -69.7 \\
\texttt{f512a}  & 1026   & 176.9 & 0.626 & 1.181 & 0.204 & 0.102 & 0.100 & 0.537 & 162 & 129 & 11.8  \\
\texttt{f512b}  & 1373   & 203.7 & 0.608 & 1.129 & 0.167 & 0.099 & 0.098 & 0.518 & 168 & 168 & -30.9 \\
\texttt{f512c}  & 81.5   & 41.8  & 0.581 & 1.403 & 0.720 & 0.097 & 0.082 & 0.535 & 22  & 18  & -26.6 \\
\texttt{f512d}  & 146.5  & 60.8  & 0.589 & 1.243 & 0.516 & 0.098 & 0.093 & 0.525 & 38  & 30  & 38.0  \\
\texttt{f512e}  & 287.8  & 89.4  & 0.605 & 1.189 & 0.369 & 0.101 & 0.096 & 0.525 & 65  & 51  & -75.7 \\
\texttt{f512f}  & 436.3  & 113.0 & 0.620 & 1.267 & 0.328 & 0.096 & 0.096 & 0.535 & 83  & 64  & 22.1  \\
\texttt{f512g}  & 785.2  & 153.4 & 0.626 & 1.255 & 0.245 & 0.098 & 0.095 & 0.541 & 132 & 99  & 70.9  \\
\texttt{f1024a} & 2415   & 276.2 & 0.626 & 1.158 & 0.132 & 0.100 & 0.100 & 0.557 & 323 & 247 & -4.40 \\
\texttt{f1024b} & 3535   & 335.2 & 0.626 & 1.130 & 0.107 & 0.102 & 0.102 & 0.541 & 337 & 337 & -34.3 \\ 
\end{tabular}
\caption{Summary of the mean statistics for our forced simulations.}
\label{tbl:summary_stats}
\end{center}
\end{sidewaystable}

\begin{figure}[tb]
 \begin{center}
  \subfigure[Fluctuation of energy and dissipation rate]{
   \includegraphics[width=0.76\textwidth,trim=3px 35px 15px 40px,clip]{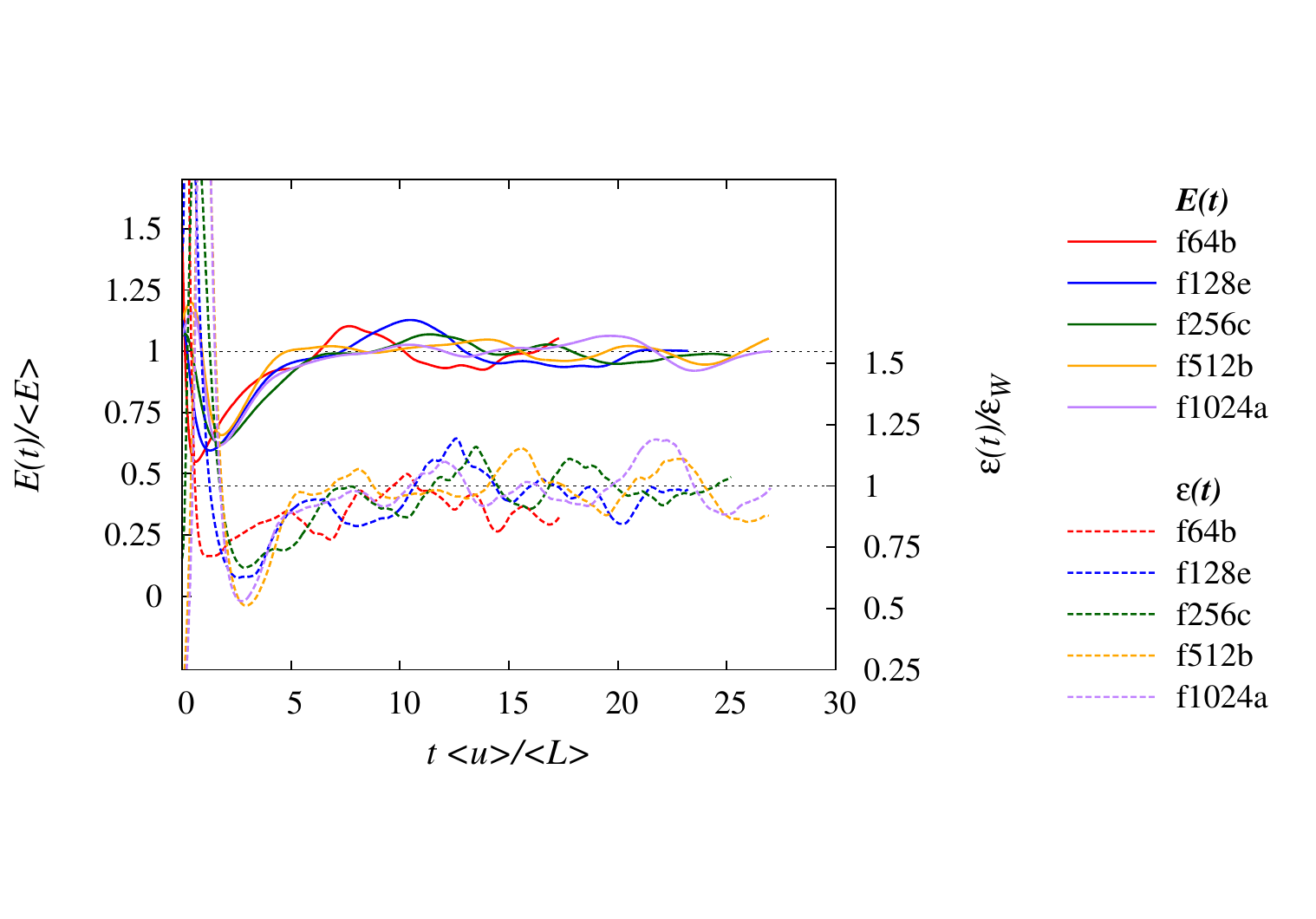}
  }
  \subfigure[Fluctuation of length-scales]{
   \includegraphics[width=0.76\textwidth,trim=3px 35px 15px 40px,clip]{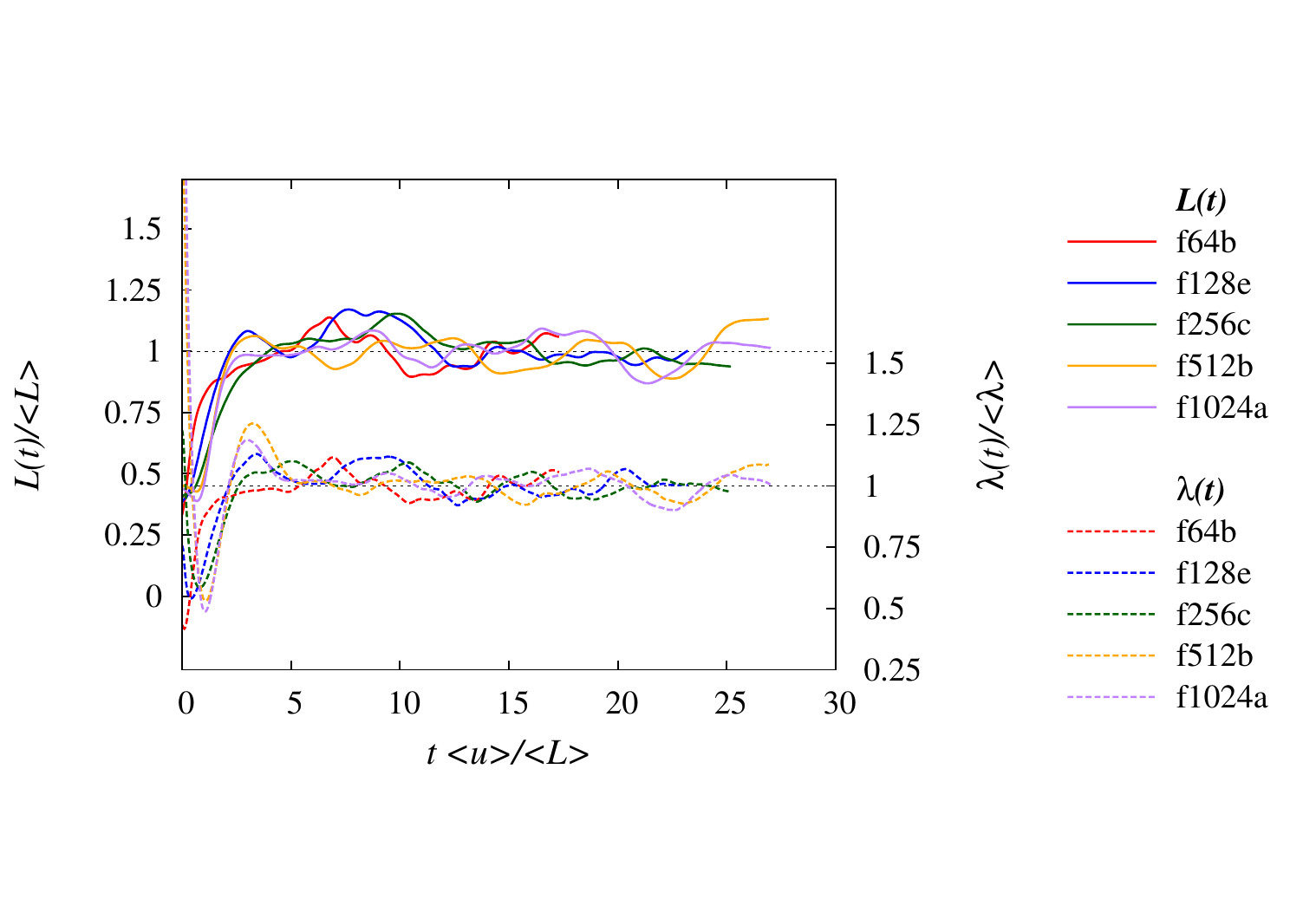}
  }
  \subfigure[Fluctuation of Reynolds numbers]{
   \includegraphics[width=0.76\textwidth,trim=3px 35px 15px 40px,clip]{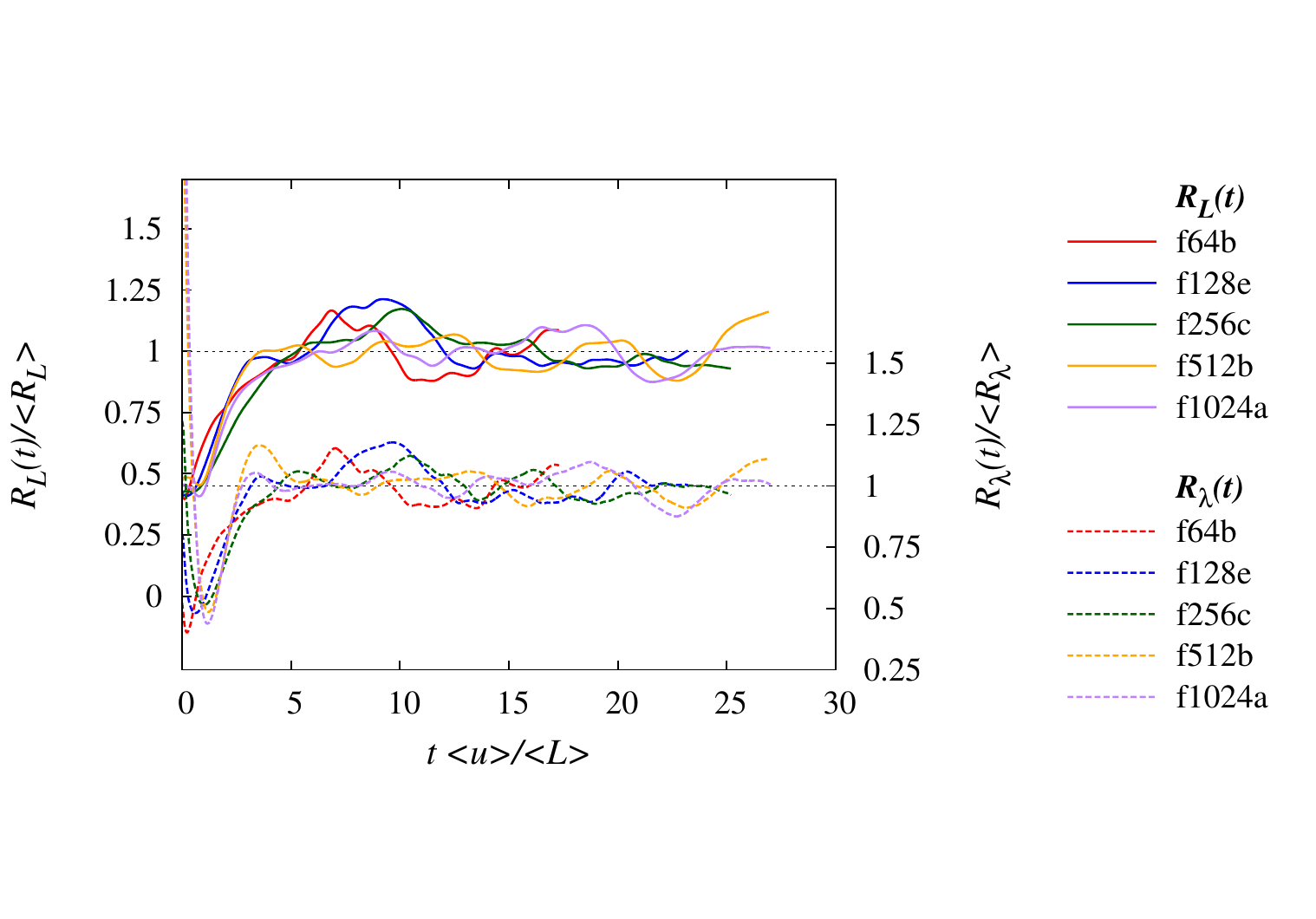}
  }
 \end{center}
 \captcont*{Continued overleaf$\ldots$}
\end{figure}

\begin{figure}[tb]
 \begin{center}
  \subfigure[Fluctuation of transfer spectrum measurements]{
   \label{sfig:forced_time_series_S-Pi0}
   \includegraphics[width=0.76\textwidth,trim=3px 35px 13px 40px,clip]{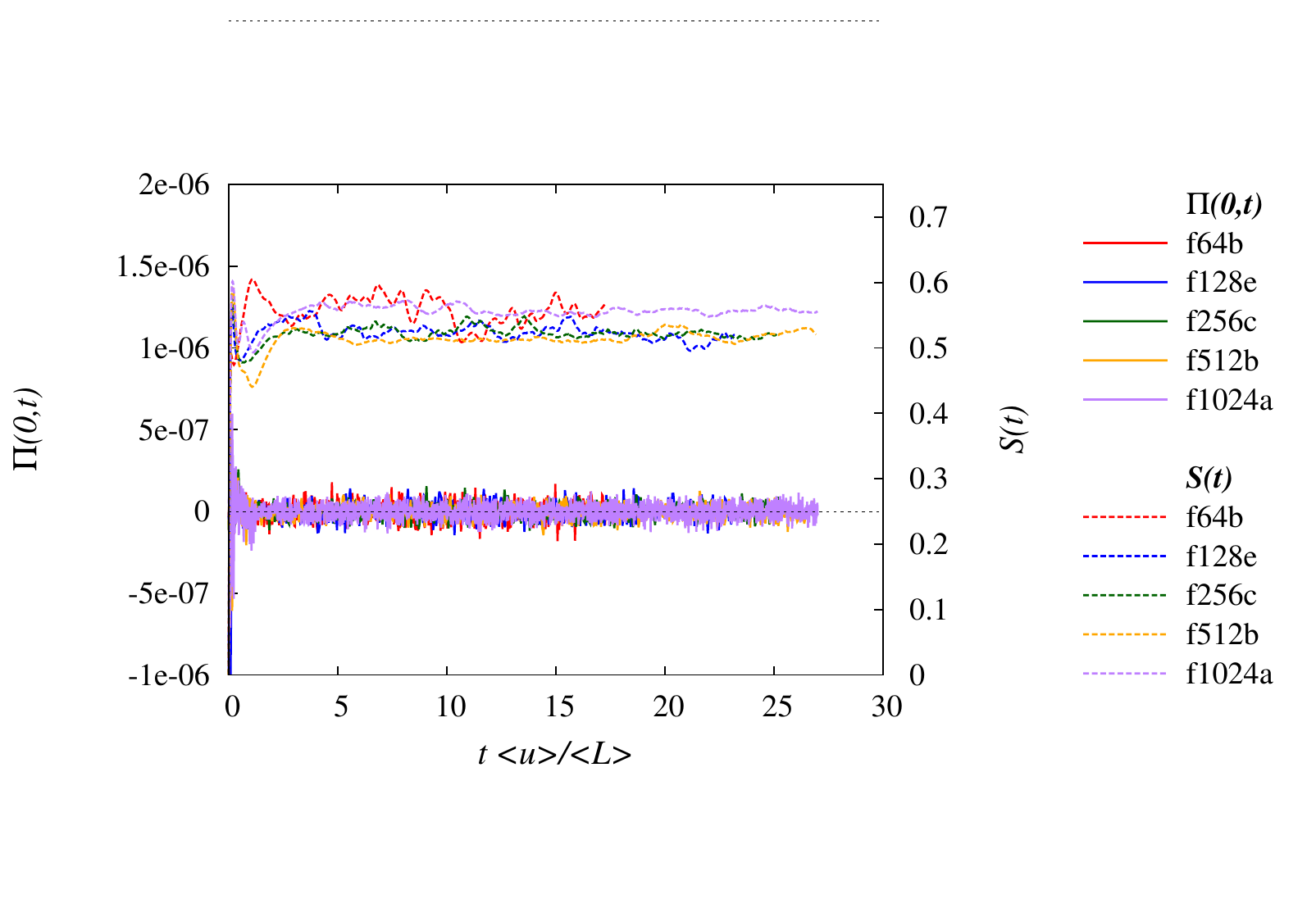}
  }
 \end{center}
 \caption{Time variation of key parameters for forced turbulence. Parts (a)--(c) scaled by their steady state mean value.}
 \label{fig:forced_time_series}
\end{figure}

The integral over the transfer spectrum, $\Pi(0,t)$, is shown in figure \ref{sfig:forced_time_series_S-Pi0} and can be seen to fluctuate around zero. The time-averaged values, shown in table \ref{tbl:summary_stats}, show $\Pi(0)$ to be consistently of order $10^{-8}$ or smaller, indicating that the non-linear term is conserving energy. 

\begin{figure}[tb]
 \begin{center}
  \includegraphics[width=0.65\textwidth,trim=0 90px 0 80px,clip]{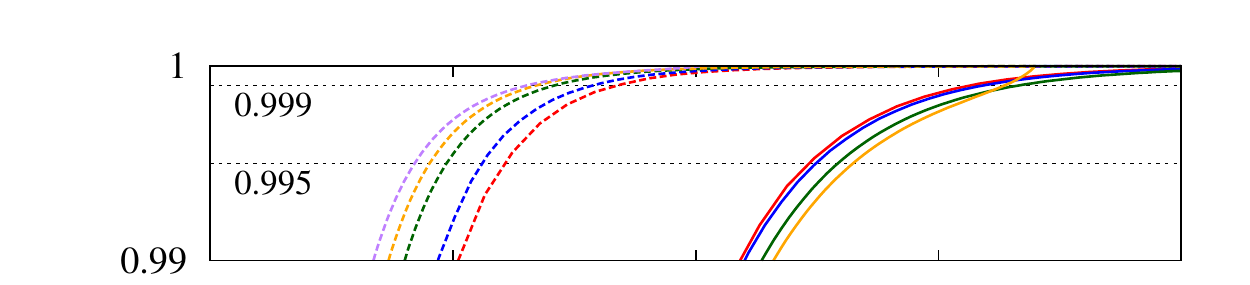}
  \includegraphics[width=0.65\textwidth,trim=0 0 0 10px,clip]{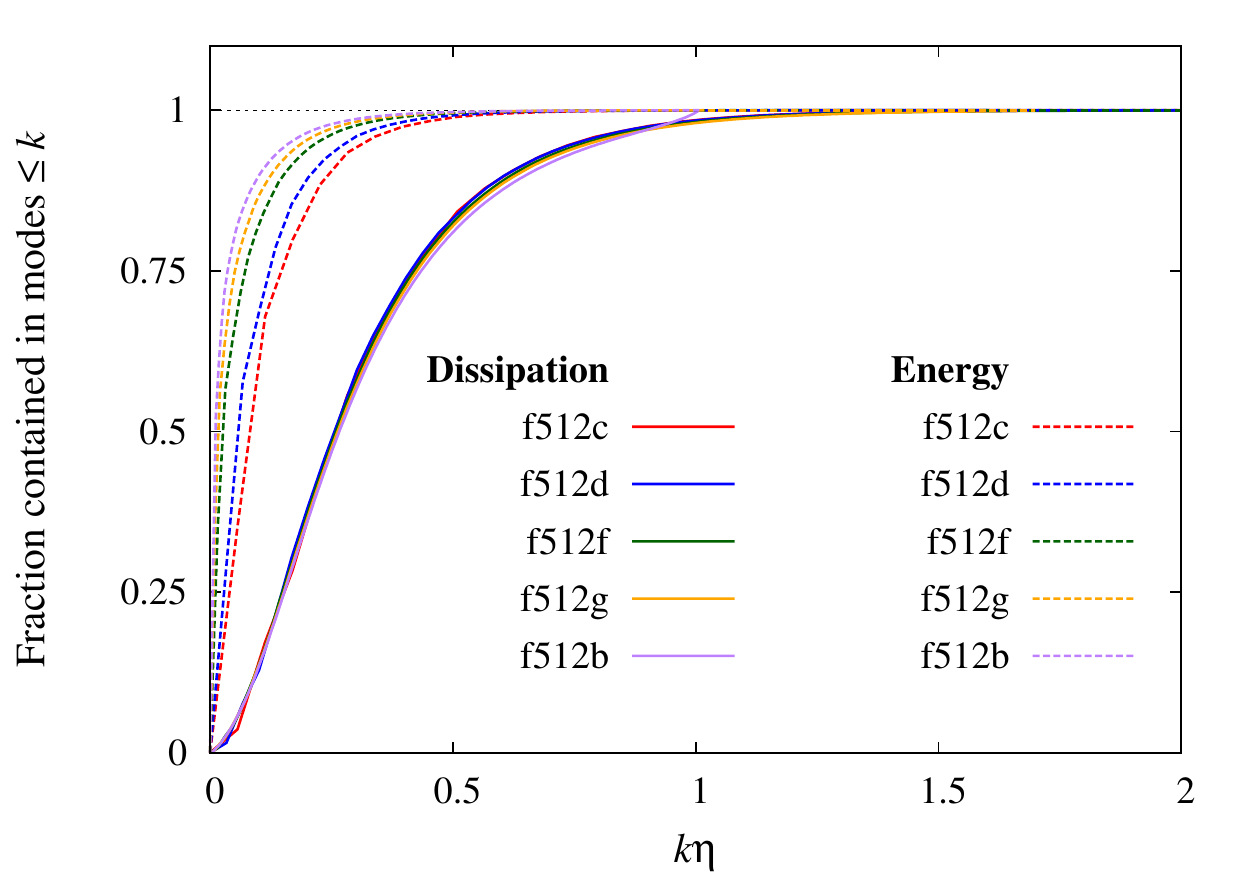}
 \end{center}
 \caption{Location of energy and dissipation rate in highly resolved simulations. To retain 99.5\% of dissipative dynamics, one must satisfy $\kmax\eta > 1.25$ while 99.9\% requires $\kmax\eta > 1.7$.}
 \label{fig:forced_resolved}
\end{figure}

A general rule for DNS is that one must satisfy $\kmax \eta > 1$, with $\kmax \eta = 1$ known as partially resolved. It has been suggested that one is actually required to satisfy $\kmax \eta > 1.5$ to capture the relevant dynamics. Therefore, a series of `highly resolved' runs was performed, by which we mean that $\kmax \eta > 1.5$, see runs \frun{f512c-g}. This allowed us to explore the distribution of energy and dissipation without artefacts caused due to the system being under-resolved. Figure \ref{fig:forced_resolved} shows our results. We plot the total energy (and dissipation rate) accounted for up to mode $k$ and normalise by the total, thus
\begin{equation}
 \frac{1}{\langle E \rangle} \int_{k_{\textrm{min}}}^k dk\ E(k) \ \quad\qquad\textrm{and}\quad\qquad \frac{1}{\langle \varepsilon \rangle} \int_{k_{\textrm{min}}}^k dk\ 2\nu_0 k^2\ E(k) \ .
\end{equation}
We also plot the partially resolved run \frun{f512b} for comparison, which can be seen to kick up unphysically as it reaches $k\eta = 1$. This also occurs for run \frun{f512g} as it reaches its cutoff $\kmax\eta = 1.7$. The energy really is contained in much lower wavenumbers (larger length-scales) than the dissipative loss. By $k\eta \sim 0.5$ we have already accounted for virtually all the energy, but only around 75\% of the dissipation rate. The additional graphic in figure \ref{fig:forced_resolved} shows a close up of the final percentile. This highlights two key points: First, if we want to include 99.5\% of dissipative dynamics, we must use $\kmax\eta \simeq 1.25$. Whereas, to include 99.9\% requires $\kmax\eta \simeq 1.7$. Second, as Reynolds number is increased, energy is contained in progressively lower $k\eta$ while dissipation is pushed to higher $k\eta$.

\begin{figure}[tb]
 \begin{center}
  \subfigure[Energy spectra scaled with $\eta$]{
   \label{sfig:forced_spec_Ekol}
   \includegraphics[width=0.475\textwidth,trim=0 0 10px 0,clip]{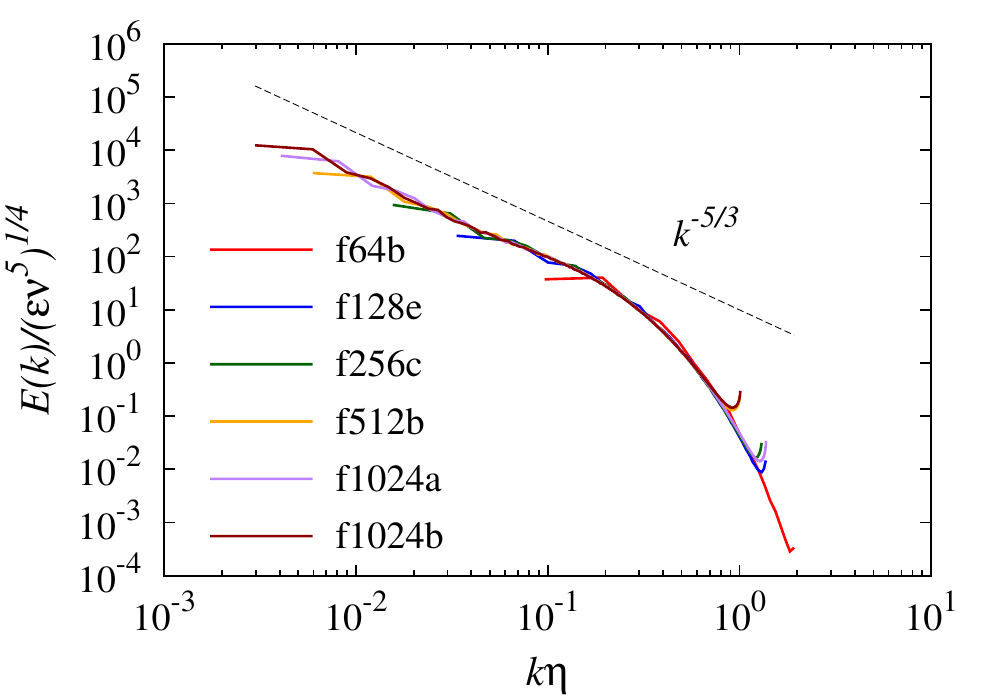}
  }
  \subfigure[Energy spectra scaled with the integral scale]{
   \label{sfig:forced_spec_EL}
   \includegraphics[width=0.475\textwidth,trim=0 0 10px 0,clip]{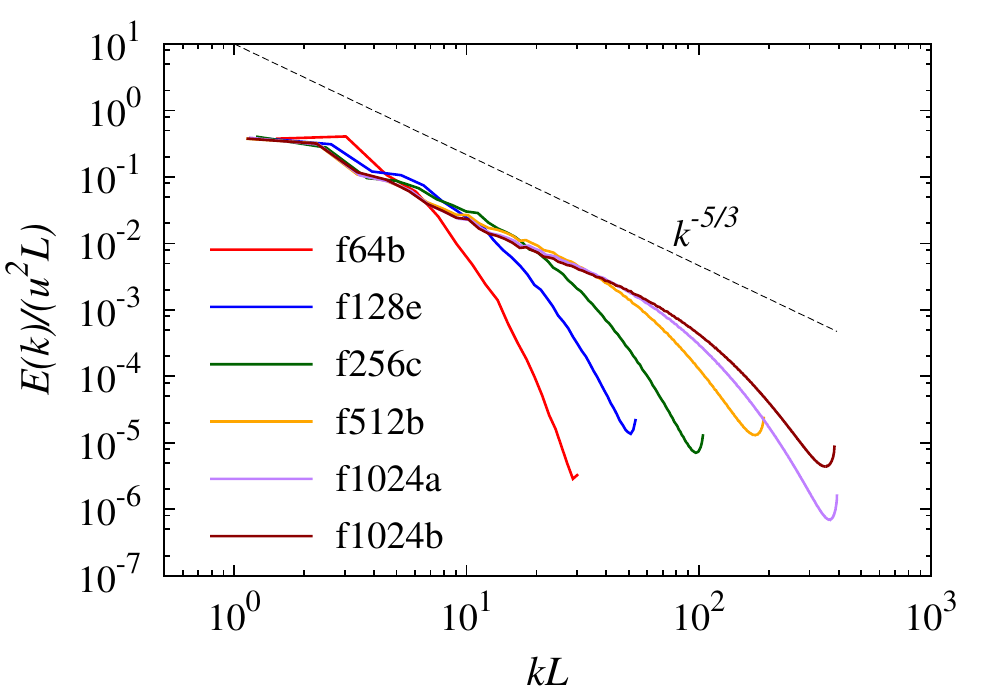}
  }
  \subfigure[The compensated energy spectrum]{
   \label{sfig:forced_spec_Ecomp}
   \includegraphics[width=0.475\textwidth,trim=0 0 10px 0,clip]{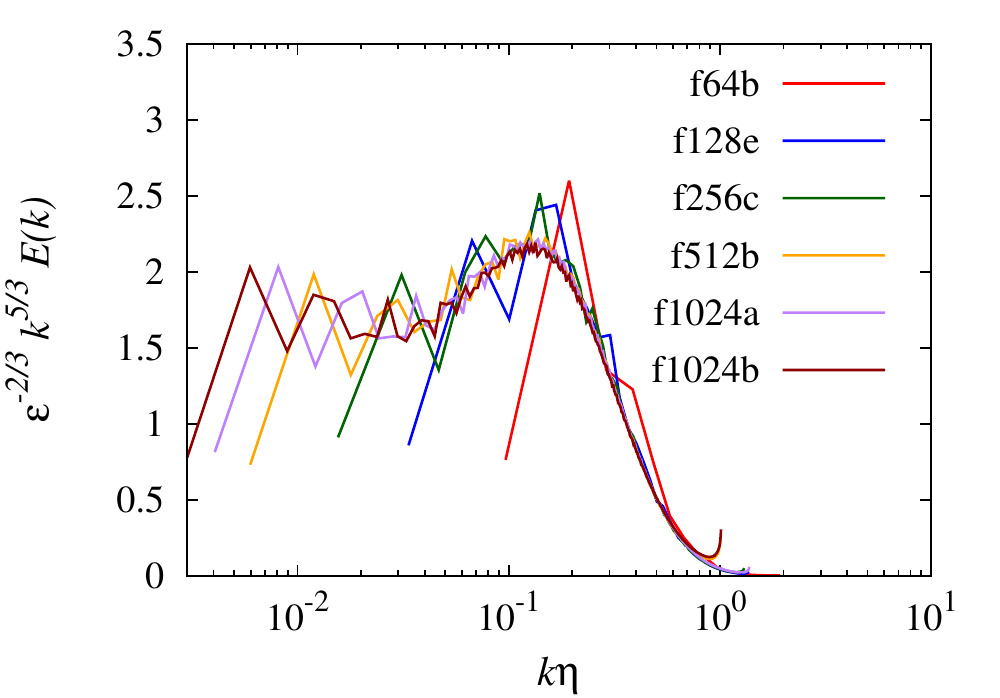}
  }
  \subfigure[Transport power spectrum]{
   \label{sfig:forced_spec_P}
   \includegraphics[width=0.475\textwidth,trim=0 0 10px 0,clip]{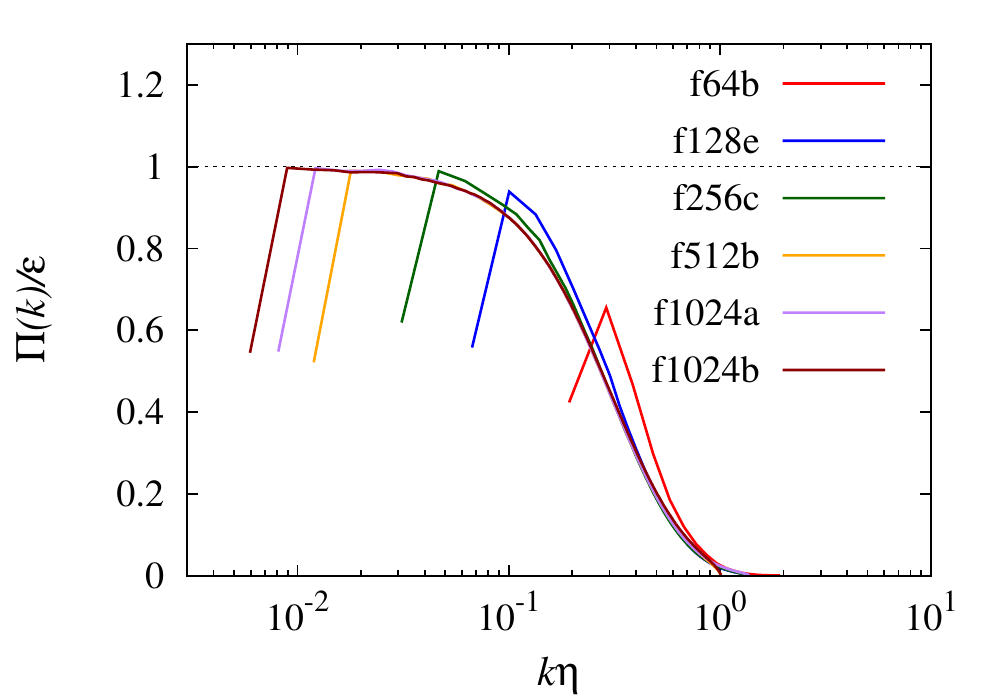}
  }
  \subfigure[Transfer spectra scaled with $\eta$]{
   \label{sfig:forced_spec_T}
   \includegraphics[width=0.475\textwidth,trim=0 0 10px 0,clip]{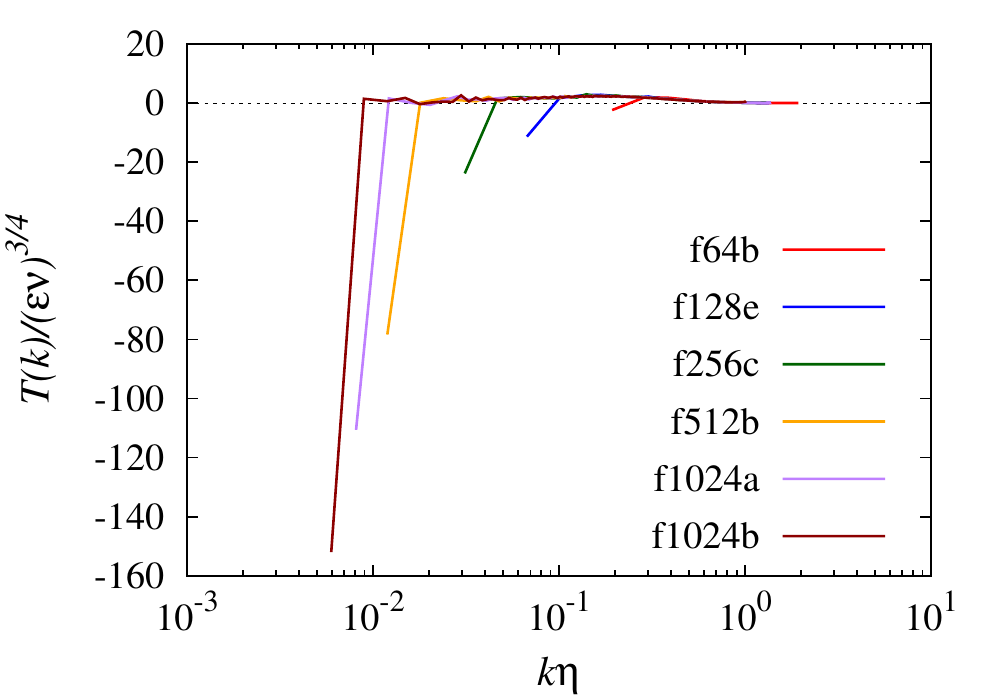}
  }
  \subfigure[Transfer spectra scaled with the integral scale]{
   \label{sfig:forced_spec_Tu3}
   \includegraphics[width=0.475\textwidth,trim=0 0 10px 0,clip]{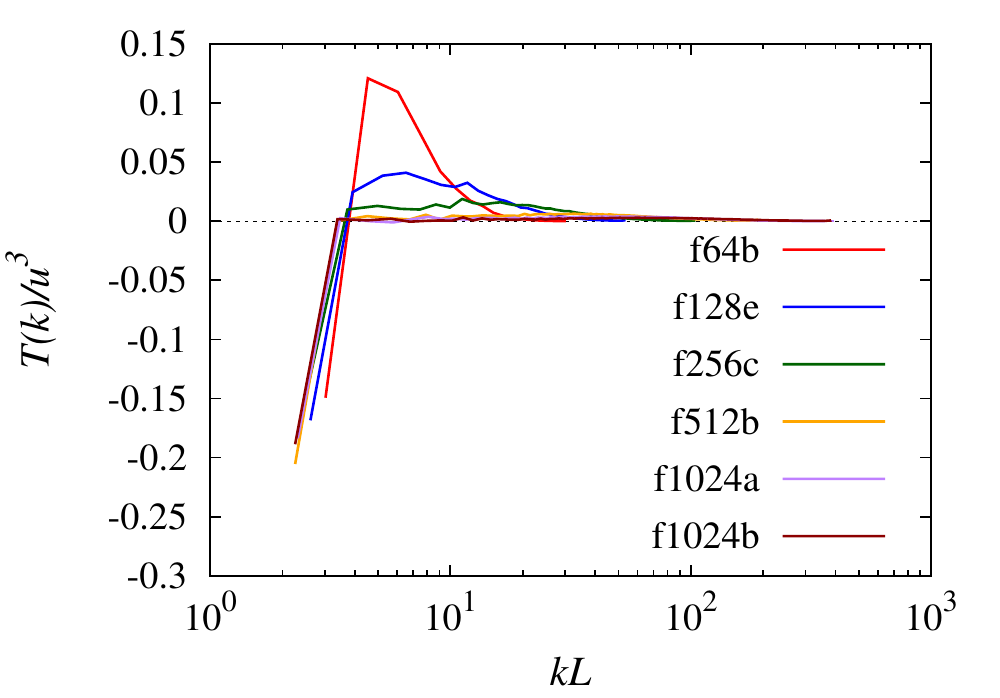}
  }
 \end{center}
 \caption{Energy, transfer and transport power spectra for a variety of Reynolds numbers.}
 \label{fig:forced_spec}
\end{figure}

The energy spectra taken for a selection of runs are presented in figure \ref{fig:forced_spec}. Figure \ref{sfig:forced_spec_Ekol} is scaled using the Kolmogorov length-scale and the appropriate combination of dissipation range variables $\varepsilon$ and $\nu_0$, as seen in equation \eqref{eq:Espec_UER}. The collapse of all runs is very good. The slope of the data can be seen to be slightly shallower than K41 for a period, hinting at $-5/3 + \mu$ with $\mu > 0$. This is not in agreement with Kaneda, Ishihara and Yokokawa, Itakura and Uno \cite{Kaneda:2003p134} who found $\mu \simeq -0.1$ by considering the compensated energy spectrum for the high Reynolds number simulations performed on the Earth Simulator. This correction could be Reynolds number dependent and vanish as $Re \to \infty$, making it a finite Reynolds number effect. An analysis of the Reynolds number variation of this exponent would help determine whether K41 is an asymptotic theory or not. Unfortunately, the data obtained here, presented in figure \ref{sfig:forced_spec_Ecomp}, did not offer a large enough range to measure this exponent properly. The compensated energy spectra should be compared to those obtained by Ishihara, Gotoh and Kaneda \cite{Ishihara:2009p165}, which were presented in figure \ref{sfig:kol_const_both_Ishihara}. Figure \ref{sfig:forced_spec_EL} shows the energy spectrum scaled using the integral scale, for comparison. The slope here also looks to be shallower than $k^{-5/3}$.

The scaled transfer spectra are shown in figures \ref{sfig:forced_spec_T} and \ref{sfig:forced_spec_Tu3}.

\clearpage

\subsection{The Kolmogorov prefactor}
Figure \ref{sfig:forced_spec_Ecomp} shows the compensated energy spectrum, as seen in section \ref{subsec:kol_const}. Note the pronounced curl up of the tail for the partially resolved run \frun{f512b}. This is also the case for \frun{f1024b} (not plotted). The figure shows how a plateau could be identified for a low $k\eta$ range for runs with $R_\lambda > 113$. This plateau can be used to find a value for the Kolmogorov constant, $\alpha$, and can be seen to lie around 1.6 -- 1.7. The values measured for the four runs for which a plateau could be found are given in table \ref{tbl:kol_const}, using a simple average and an error-weighted fit. The transport power spectra shown in figure \ref{sfig:forced_spec_P} were used to define the fit region. Note that the peak associated with $k\eta \sim 0.1-0.2$ is not associated with an inertial range (section \ref{subsec:kol_const}). For runs with lower Reynolds number, a plateau cannot be identified.

\begin{table}[tb!]
 \begin{center}
  \begin{tabular}{r|cccc}
   ID & $R_\lambda$ & $\alpha$ & $\alpha'$ \\
   \hline\hline
   \frun{f512a}  & 176.9 & $1.663 \pm 0.218$ & $1.632 \pm 0.172$ \\
   \frun{f512b}  & 203.7 & $1.625 \pm 0.166$ & $1.621 \pm 0.165$ \\
   \frun{f1024a} & 276.2 & $1.636 \pm 0.177$ & $1.646 \pm 0.144$ \\
   \frun{f1024b} & 335.2 & $1.643 \pm 0.136$ & $1.625 \pm 0.119$
  \end{tabular}
 \end{center}
 \caption{Measured values of the Kolmogorov constant. Found by identifying the range of wavenumbers where $\Pi(k) \simeq \varepsilon$ and averaging over those points. The value $\alpha$ is obtained by a simple average over the range, whereas $\alpha'$ is calculated using an error-weighted fit on the range.}
 \label{tbl:kol_const}
\end{table}

\subsection{Reynolds number dependence of statistics}\label{subsec:forced_Rdep}
\begin{figure}[tb]
 \begin{center}
  \subfigure[Onset of an inertial subrange]{
   \label{sfig:forced_Rvar_eps-pi}
   \includegraphics[width=0.59\textwidth]{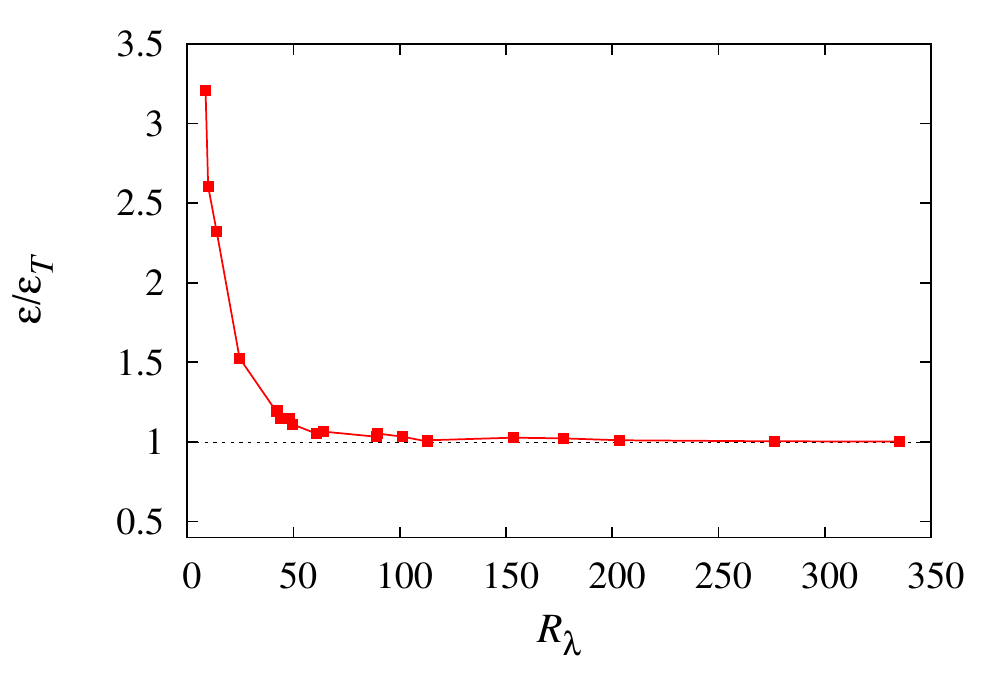}
  }
  \subfigure[Variation of a selection of parameters]{
   \label{sfig:forced_Rvar_params}
   \includegraphics[width=0.59\textwidth]{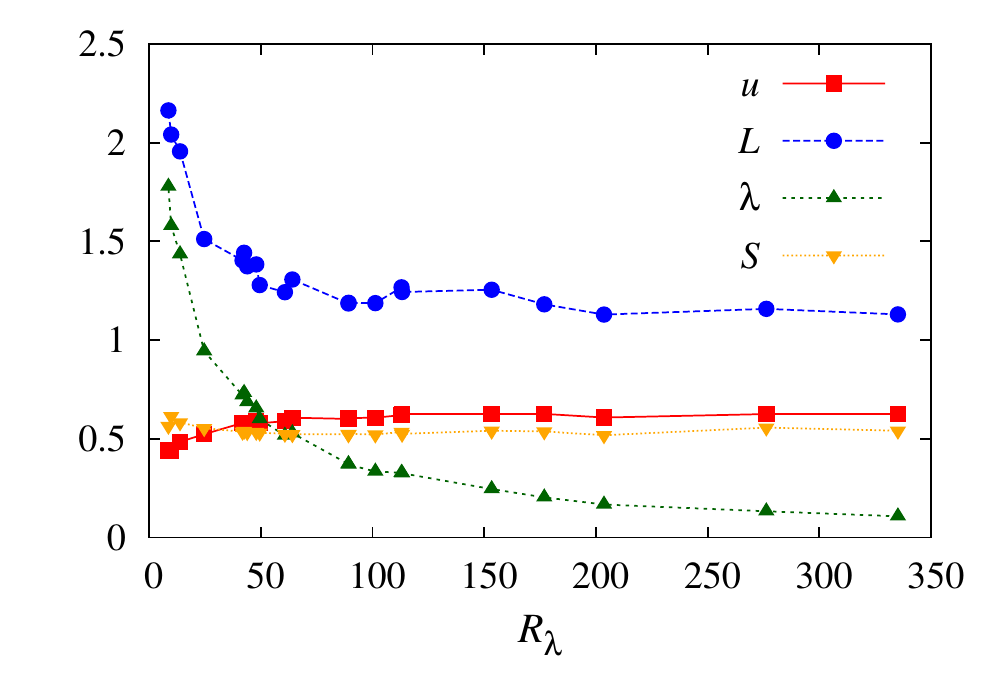}
  }
  \subfigure[The Taylor surrogate, $u^3/L$]{
   \label{sfig:forced_Rvar_eps-pi-surr}
   \includegraphics[width=0.59\textwidth]{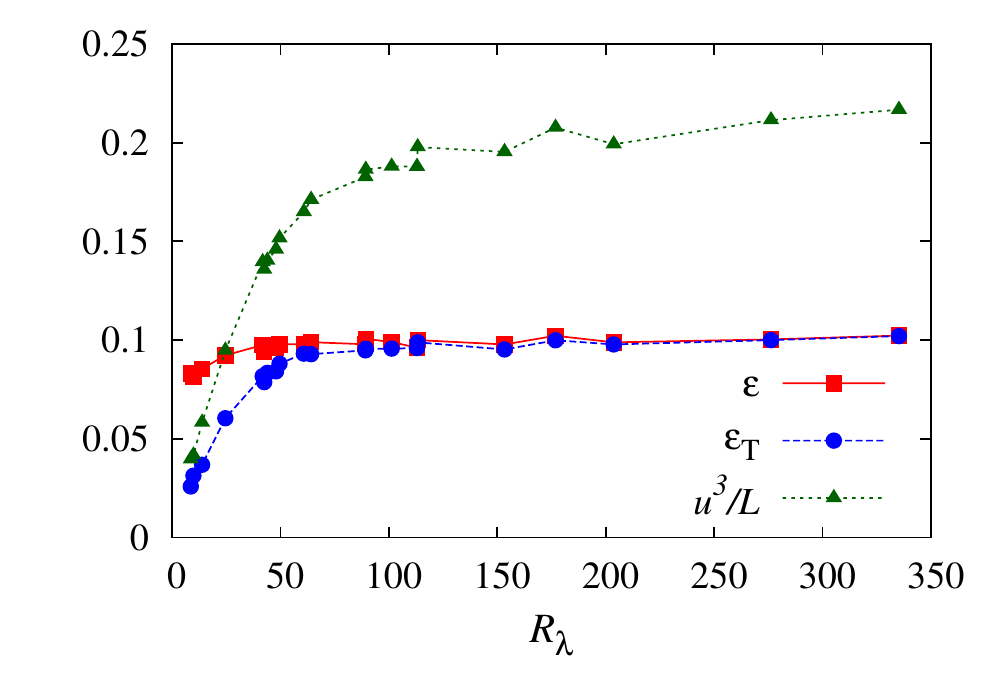}
  }
 \end{center}
 \caption{Reynolds number variation of key parameters for forced turbulence.}
 \label{fig:forced_Rvar}
\end{figure}

We now look at how the values of some important parameters vary with increasing Reynolds number. As mentioned in section \ref{subsec:kol_const}, an indication of the presence of a inertial subrange in a stationary system is a range of wavenumbers for which the transport power, or flux of energy through that wavenumber, is equal to the dissipation rate, $\Pi(k) = \varepsilon$. When this is the case, we find that the maximum transfer rate $\varepsilon_T = \max \Pi(k) = \varepsilon$. As such, a study of $\varepsilon_T/\varepsilon$ will give unity for stationary systems in which the integral and dissipation scales are sufficiently well separated that an inertial subrange can form. The variation of this quantity with Reynolds number is presented in figure \ref{sfig:forced_Rvar_eps-pi}. This should be compared to decaying turbulence, section \ref{subsec:tevo_dep}, where the maximum transport can never quite reach the dissipation rate. Note that, for Reynolds numbers $R_\lambda > 113$, we basically find $\varepsilon_T = \varepsilon$, perhaps indicating the presence of an inertial subrange.

Figure \ref{sfig:forced_Rvar_params} shows the Reynolds number variation of the steady state value of the rms velocity, integral and Taylor length-scales, and the velocity derivative skewness. We see the skewness remain more or less constant, just above 0.5, for the range of Reynolds numbers available. The length-scales are both seen to decrease as $Re$ increases. However, the integral scale looks like it may have reached a plateau, whereas the same cannot be said for the Taylor microscale. The rms velocity initially increases then appears to stay constant. We would expect the rms velocity to increase as the Reynolds number increases since there are more modes excited. This may still be the case, but as the majority of energy is located in low wavenumbers the increase is small.

The Taylor surrogate $u^3/L$, discussed in section \ref{sec:Taylor_surr}, is compared in figure \ref{sfig:forced_Rvar_eps-pi-surr} to the dissipation rate and inertial flux, $\varepsilon_T$. Once again we see that the surrogate is better matched to the behaviour of the inertial flux than the dissipation rate. This is in agreement with the findings of McComb \etal\ \cite{McComb:2010p250}.

\clearpage


\section{Visualisation of coherent structures}\label{sec:structures}

\subsection{Identification of coherent structures}
Over the years, there have been many attempts to define a vortex in such a way that they may be identified in flow visualisation, whether that be experimental data from real flows or DNS data as studied here. The na\"ive definition of a vortex as a region of high vorticity can be misleading since there is no particular value above which vorticity can be universally regarded as being high. In fact, even in the absence of vortices there can exist areas of high vorticity in parallel shear flows \cite{Haller:2005p618}. This creates a difficulty in finding unambiguous criteria which can isolate a unique vortex.

Jeong and Hussain \cite{Jeong:1995p1073} summarised and compared a selection of methods available. They highlighted that any criteria should be Galilean invariant, and found that previous indicators of a vortex, such as streamlines, isovorticity and minima in the local pressure are not suitable for use in unsteady flow. Haller \cite{Haller:2005p618} provides a comprehensive review of the definition of a vortex along with a variety of identification techniques.

Despite this, surfaces of isovorticity continue to be used for vortex identification and can produce good results in the case of isotropic turbulence. This section aims to compare the detection of coherent structures in visualisations of our DNS data for isotropic turbulence using isovorticity contours, magnitude of the strain rate and the $Q$-criterion.

\subsubsection{Isovorticity}
Surfaces of isovorticity connect regions which have the same magnitude of vorticity, $\vert\vec{\omega}\vert$. Since the core of a vortex is associated with high vorticity, with the value progressively dropping as we move away from the core, these surfaces form structures such as `worms' and `sheets'. See, for example, \cite{Jimenez:1993p634,Okamoto:2007p591}. Structures identified in the plane $z = 0$ using vorticity can be seen in figures \ref{sfig:structures512_vort} and \ref{sfig:structures1024_vort} for two different Reynolds numbers, as part of a comparison with other identification methods. As can be seen, the magnitude of vorticity shows a large amount of structure in the plane, and there are several regions of high vorticity that could be identified as being vortices. Three-dimensional structures can be seen in figures \ref{sfig:3dstructures256_vort} and \ref{sfig:3dstructures1024_vort} and show how the vorticity has organised itself into an entanglement of tubes or `worms', as observed by many other authors \cite{Ishihara:2009p165,Ishihara:2007p612,Jimenez:1993p634}. These should be compared to the Gaussian initial condition shown in figure \ref{sfig:256_Gaussian_vort}, which shows little in the way of organised structure.

\subsubsection{The $Q$-criterion}
The $Q$-criterion was originally proposed by Hunt, Wray and Moin \cite{Hunt:1988p1088} and is based on the invariants of the deformation tensor, $A$, whose elements are
\begin{equation}
 a_{ij} = \frac{\partial u_i}{\partial x_j} \ .
\end{equation}
The eigenvalues, $\lambda$, of this tensor are found by requiring that
\begin{equation}
 \det (A - \lambda \unitM) = 0 \ ,
\end{equation}
which in three-dimensions leads to the third-order characteristic equation
\begin{equation}
 \lambda^3 - P\lambda^2 + Q\lambda - R = 0 \ ,
\end{equation}
with the coefficients
\begin{align}
 P &= \tr(A) \\
 Q &= \tfrac{1}{2} \Big( \tr(A)^2 - \tr\big( A^2 \big) \Big) \\
 R &= \det A \ .
\end{align}
The coefficients are called the principle invariants of $A$, since the eigenvalues do not depend on the choice of basis vectors. We first note that
\begin{equation}
 \tr A = \frac{\partial u_i}{\partial x_i} = 0
\end{equation}
for an incompressible fluid, such as that considered here. Next, the deformation tensor can be decomposed into its symmetric and antisymmetric parts,
\begin{equation}
 S_{ij} = \tfrac{1}{2}\Big( a_{ij} + a_{ji} \Big) \qquad\qquad\textrm{and}\qquad\qquad \Omega_{ij} = \tfrac{1}{2}\Big( a_{ij} - a_{ji} \Big)\ ,
\end{equation}
which may be recognised as the strain and vorticity tensors, respectively. We can therefore evaluate the trace
\begin{align}
 \tr \big(A^2 \big) &= \tr\big( SS + S\Omega + \Omega S + \Omega\Omega \big) \nonumber \\
 &= \tr\big( SS \big) + \tr\big(S\Omega\big) + \tr\big(\Omega S\big) + \tr\big(\Omega\Omega \big) \nonumber \\
 &= \tr\big( SS^T \big) - \tr\big(S\Omega^T\big) + \tr\big(\Omega S^T\big) - \tr\big(\Omega\Omega^T \big) \ ,
\end{align}
where the last line used the symmetry of $S$ and $\Omega$. Since the trace has the properties $\tr(AB) = \tr(BA)$ and $\tr(A^T) = \tr(A)$, the two cross terms cancel to leave
\begin{equation}
 Q = \tfrac{1}{2} \Big( \Vert \Omega \Vert^2 - \Vert S \Vert^2 \Big) \ ,
\end{equation}
with the Euclidean matrix norm defined as $\Vert M \Vert^2 = \tr\big(MM^T\big)$. For the antisymmetric component, we have $\Vert\Omega\Vert^2 = \tfrac{1}{2} \vert\vec{\omega}\vert^2$, and the value of $Q$ is calculated as
\begin{equation}
 Q = \tfrac{1}{2} \Big( \tfrac{1}{2}\omega^2 - \Vert S \Vert^2 \Big) \ .
\end{equation}
$Q$ represents the local balance between shear strain rate and vorticity magnitude, and vanishes at a solid boundary (unlike $\lvert\vec{\omega}\rvert$) \cite{Jeong:1995p1073}. When $Q > 0$, the implication is that the vorticity tensor (quantifying that amount of rotation) is dominant over the strain rate tensor (which is related to dissipation) and there is a vortex. Figure \ref{sfig:structures512_Q} shows the $Q$-criterion for a two-dimensional slice through a $512^3$ evolved velocity field. As can be seen by comparison to \ref{sfig:structures512_vort} for the vorticity, the $Q$-criterion is more selective in what it considers to be coherent structures. Figures \ref{sfig:3dstructures256_Q} and \ref{sfig:3dstructures1024_Q} show the three-dimensional structures identified using the $Q$-criterion. By comparison to those obtained using vorticity, we once again see that this method is stricter with what it considers to be a vortex. Note also that the `sheet'-like structures obtained using vorticity are no longer present. Comparison should be made to the Gaussian initial condition plotted in figure \ref{sfig:256_Gaussian_Q}. See \cite{Jeong:1995p1073, Haller:2005p618} and the many references therein for more information.

\subsubsection{Rate-of-strain}
\begin{figure}[htb]
 \centering
 \subfigure[Vorticity]{
  \label{sfig:structures512_vort}
  \includegraphics[width=0.475\textwidth,trim=130px 460px 130px 30px, clip]{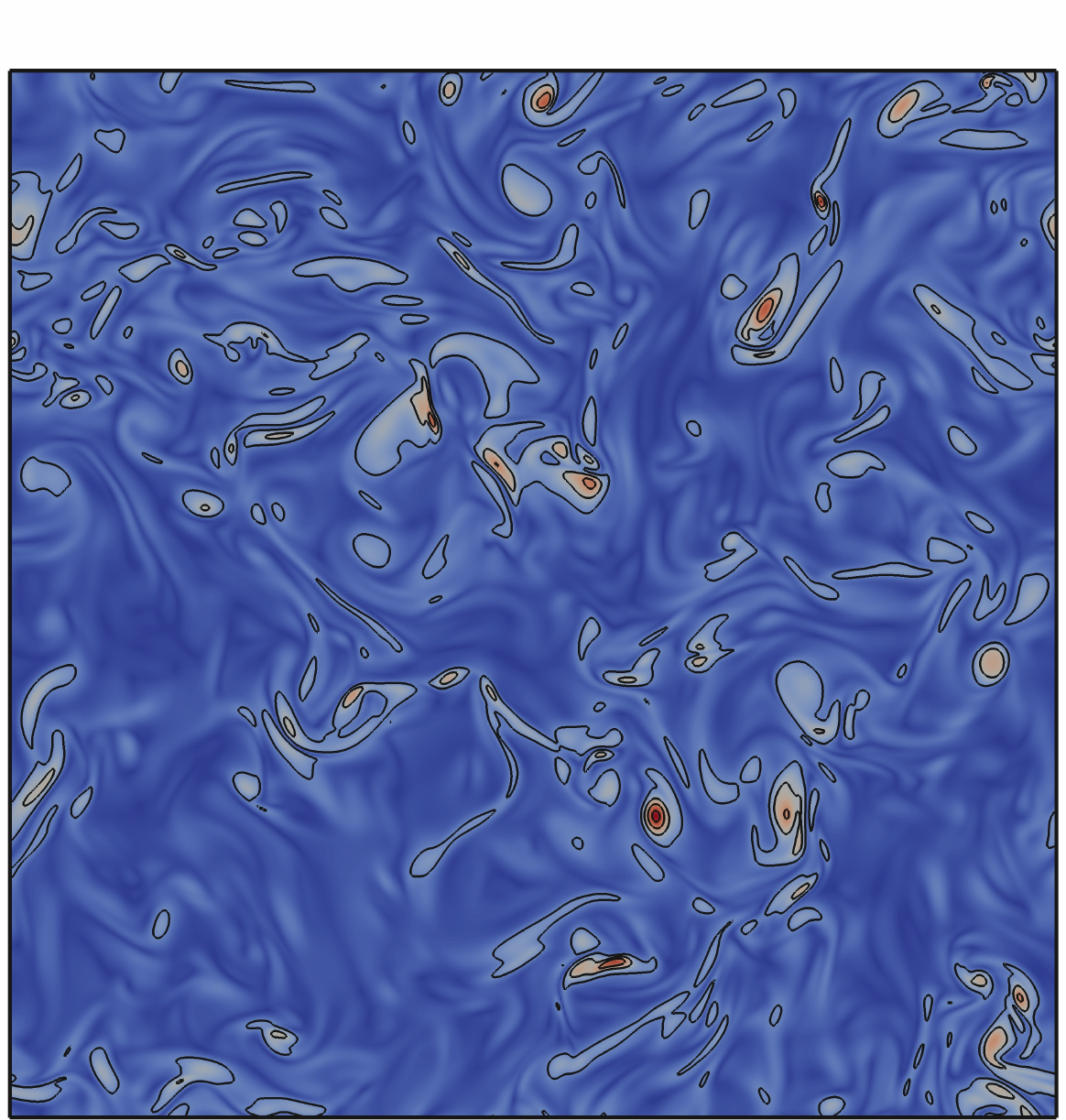}
 }\hfill
 \subfigure[Strain rate]{
  \label{sfig:structures512_strain}
  \includegraphics[width=0.475\textwidth,trim=130px 460px 130px 30px, clip]{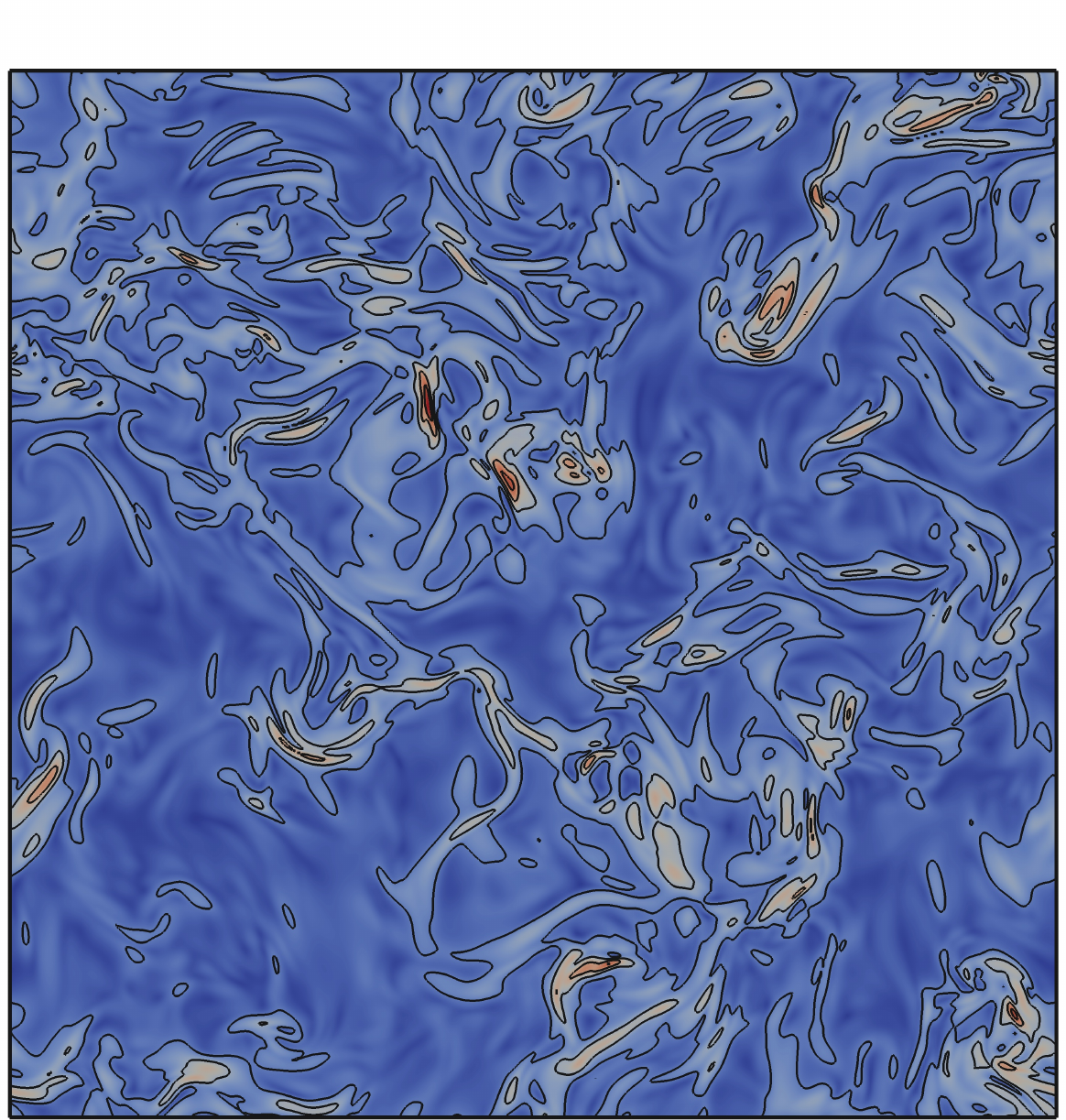}
 } \\
 \subfigure[$Q$-criterion]{
  \label{sfig:structures512_Q}
  \includegraphics[width=0.475\textwidth,trim=130px 460px 130px 30px, clip]{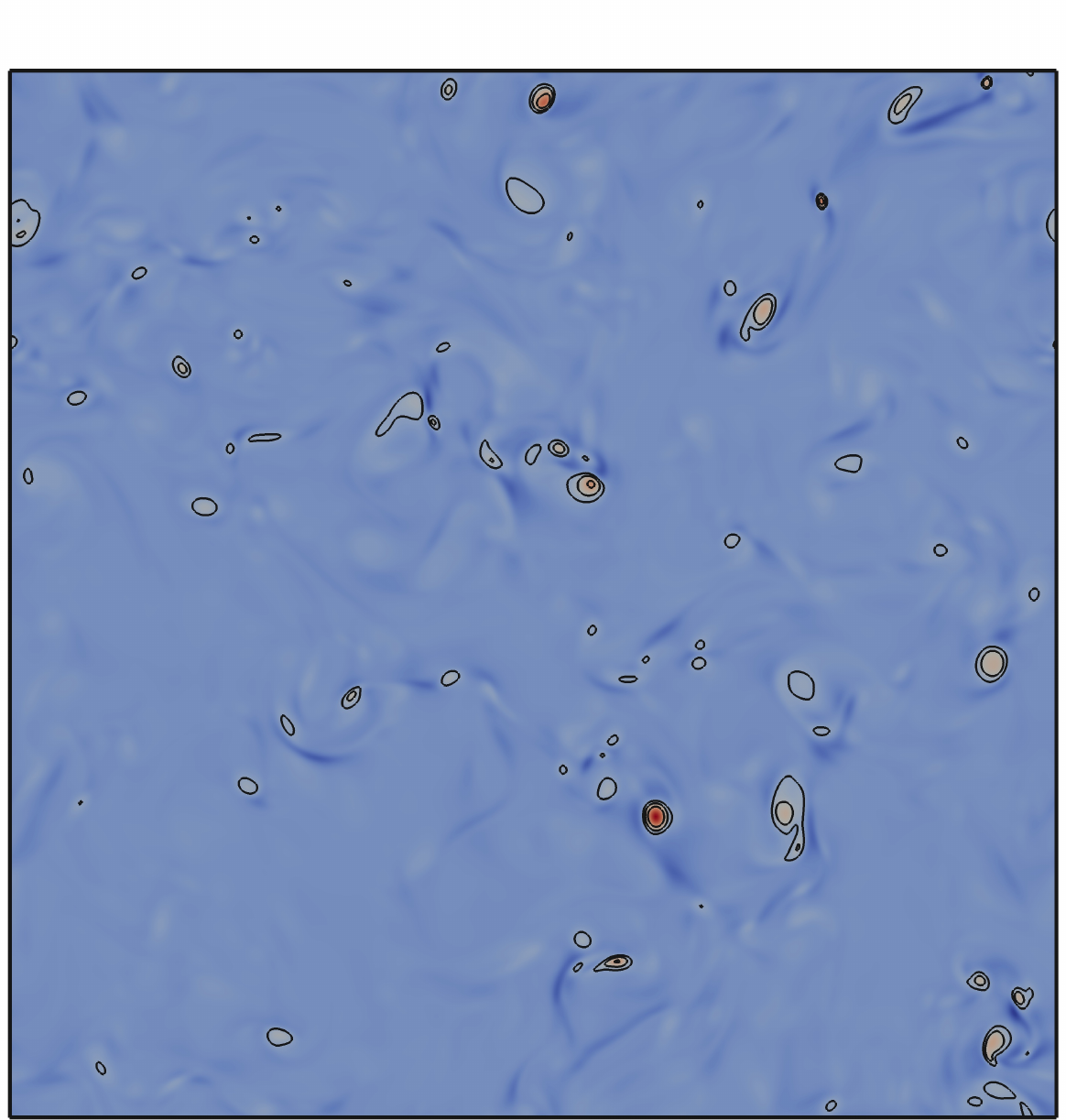}
 }
 \caption{Visualisation of the $z=0$ plane of an $R_\lambda \sim 115$ evolved velocity field from run \frun{f512f}, using: (a) vorticity, $\vmod{\omega}$; (b) magnitude of the strain rate tensor, $\Vert S \Vert$; and (c) $Q$-criterion. Contours for a range of values are also plotted. Note that the $Q$-criterion identifies far fewer structures. Contours for $Q$-criterion all have $Q \geq 0.1 Q_{\textrm{max}}$.}
 \label{fig:structures512}
\end{figure}

\begin{figure}[htb]
 \centering
 \subfigure[Velocity]{
  \label{sfig:structures1024_vel}
  \includegraphics[width=0.475\textwidth,trim=145px 475px 145px 50px, clip]{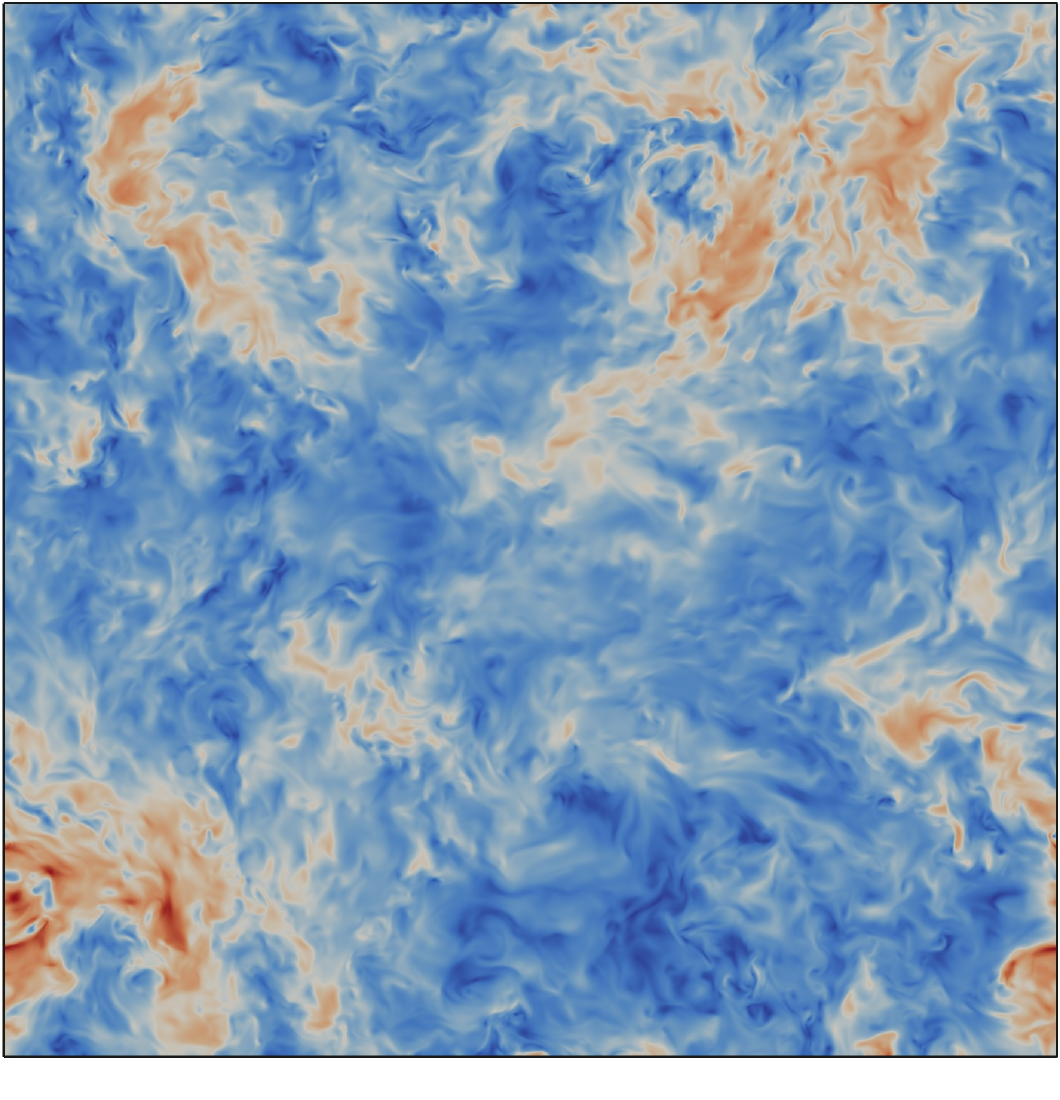}
 }\hfill
 \subfigure[Vorticity]{
  \label{sfig:structures1024_vort}
  \includegraphics[width=0.475\textwidth,trim=145px 475px 145px 50px, clip]{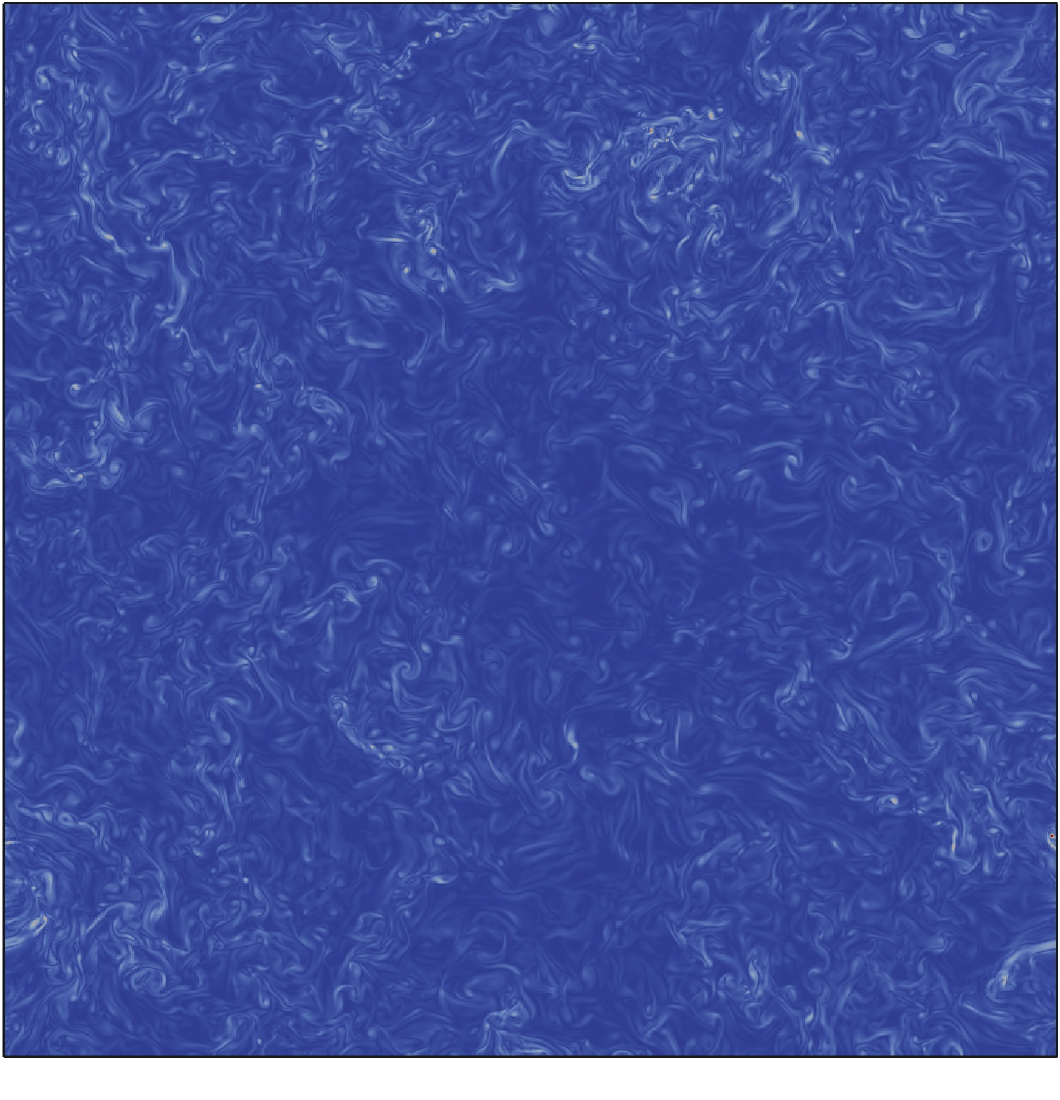}
 } \\
 \subfigure[Strain rate]{
  \label{sfig:structures1024_strain}
  \includegraphics[width=0.475\textwidth,trim=145px 475px 145px 50px, clip]{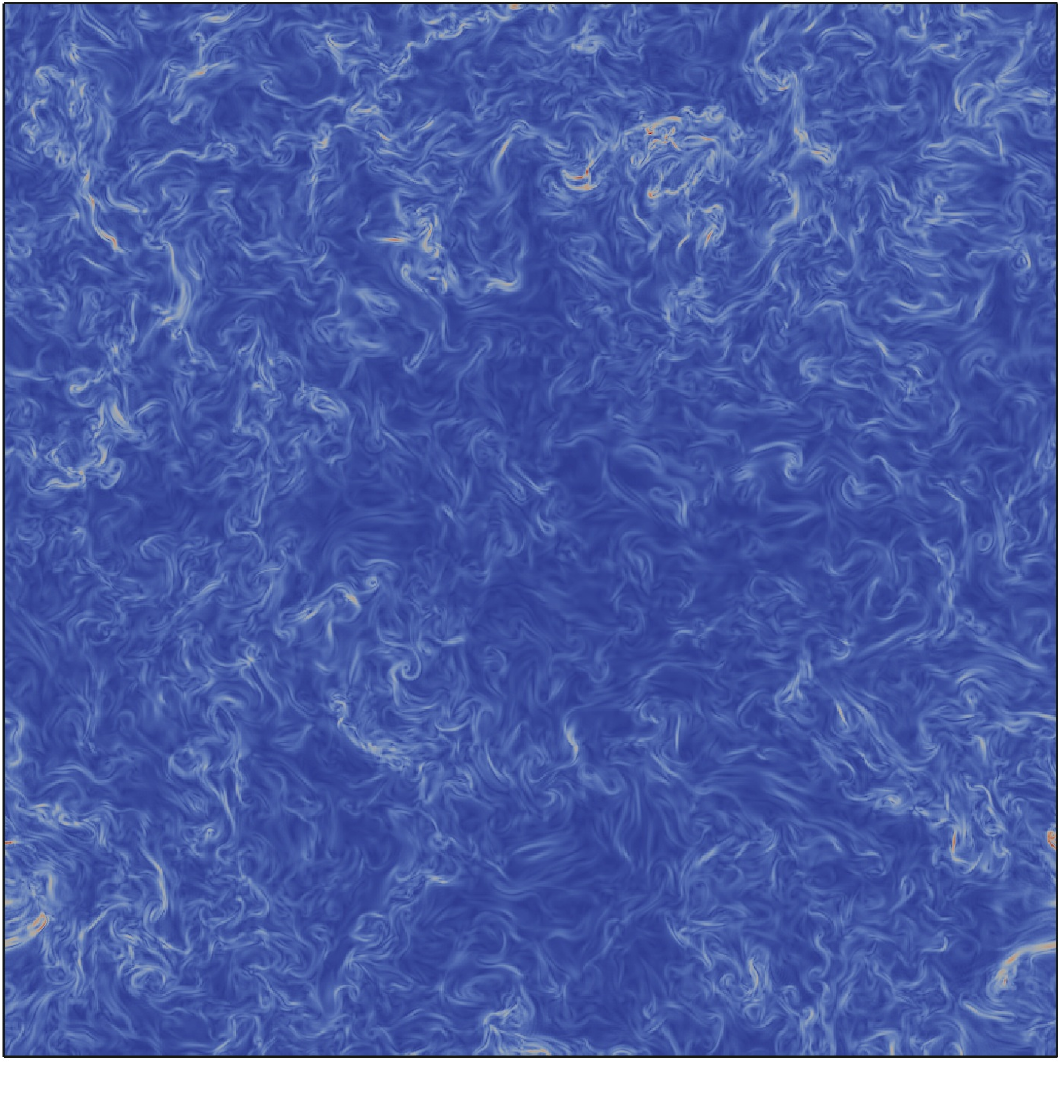}
 }
 \caption{A snapshot of (the $z=0$ plane of) the evolved velocity field from run \frun{f1024a}, coloured by: (a) $\vmod{u}$; (b) $\vmod{\omega}$; and (c) magnitude of the strain rate tensor, $\Vert S \Vert$. Contours not plotted due to the small size of the structures. Magnitude of velocity offers little in the way of identifying structures.}
  \label{fig:structures1024}
\end{figure}

\begin{figure}[htb]
 \centering
 \subfigure[Vorticity]{
  \label{sfig:3dstructures256_vort}
  \includegraphics[width=0.475\textwidth,trim=130px 430px 145px 70px, clip]{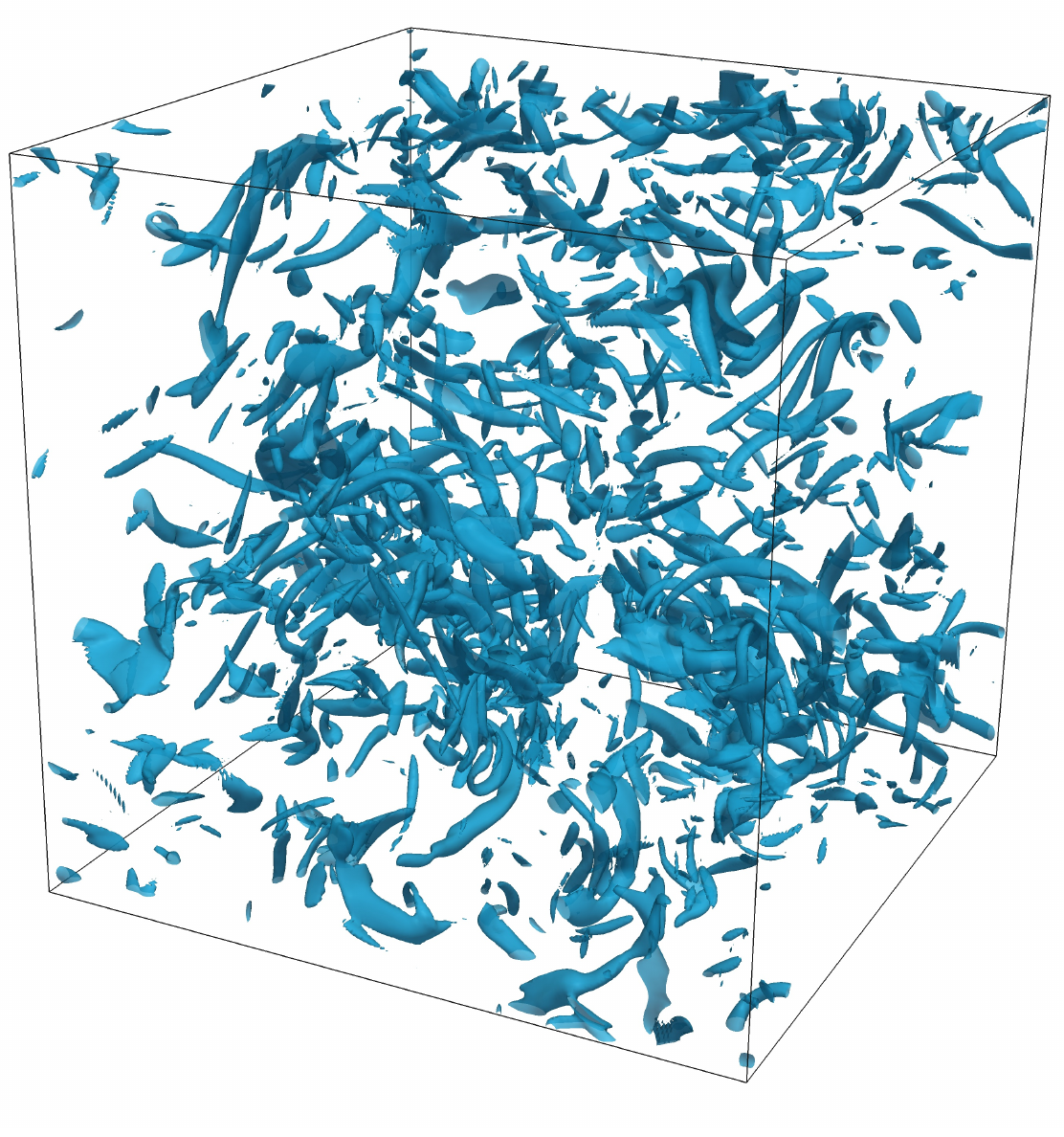}
 }
 \subfigure[Vorticity and rate-of-strain]{
  \label{sfig:3dstructures256_vort_strain}
  \includegraphics[width=0.475\textwidth,trim=130px 430px 145px 70px, clip]{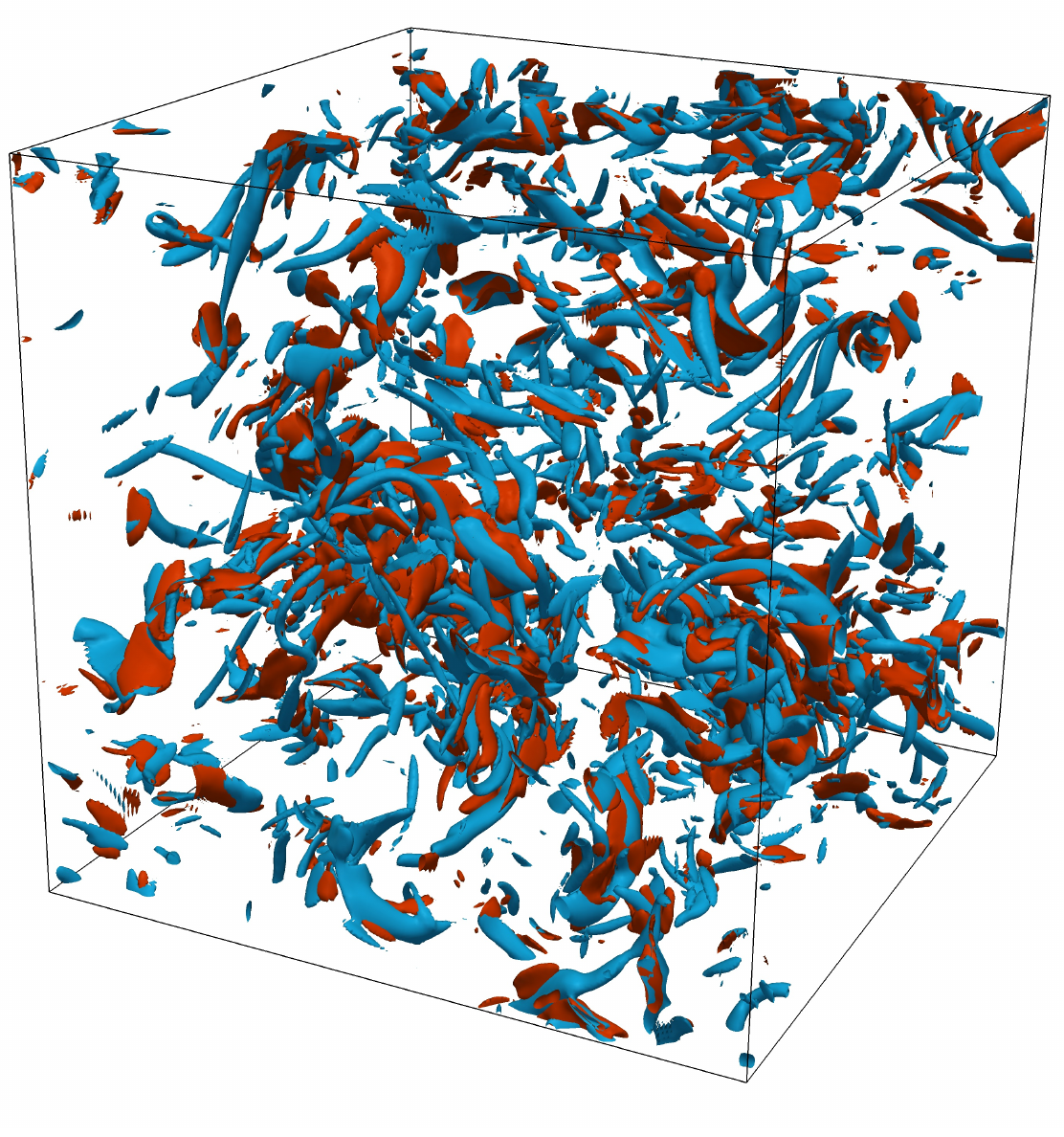}
 }
 \subfigure[Q-criterion]{
  \label{sfig:3dstructures256_Q}
  \includegraphics[width=0.64\textwidth,trim=130px 430px 145px 70px, clip]{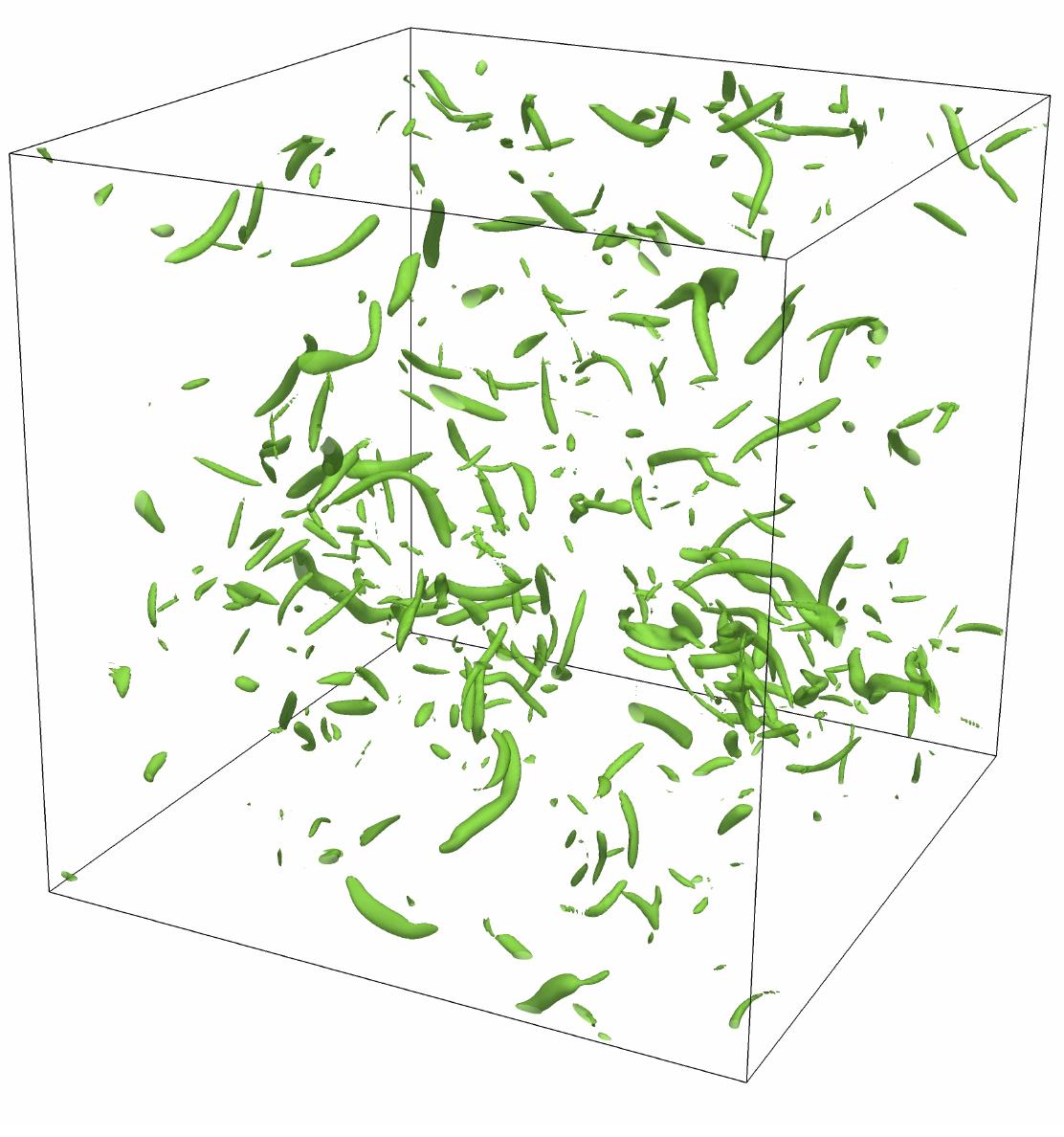}
 }
 \caption{Visualisation of turbulent structures in an $R_\lambda \sim 100$ evolved velocity field from run \frun{f256b}. Isosurfaces of (a) vorticity ($0.25 \omega_{\textrm{max}}$ plotted); (b) vorticity (blue) and strain rate ($0.4\Vert S\Vert_{\textrm{max}}$ plotted, red); and (c) Q-criterion ($0.1 Q_{\textrm{max}}$ plotted). Regions of high vorticity are seen to be correlated with areas of high strain. The $Q$-criterion can be seen to pick out fewer structures than just vorticity.}
 \label{fig:3dstructures256}
\end{figure}

\begin{figure}[htb]
 \centering
 \subfigure[Isovorticity]{
  \label{sfig:256_Gaussian_vort}
  \includegraphics[width=0.35\textwidth,trim=130px 430px 145px 70px, clip]{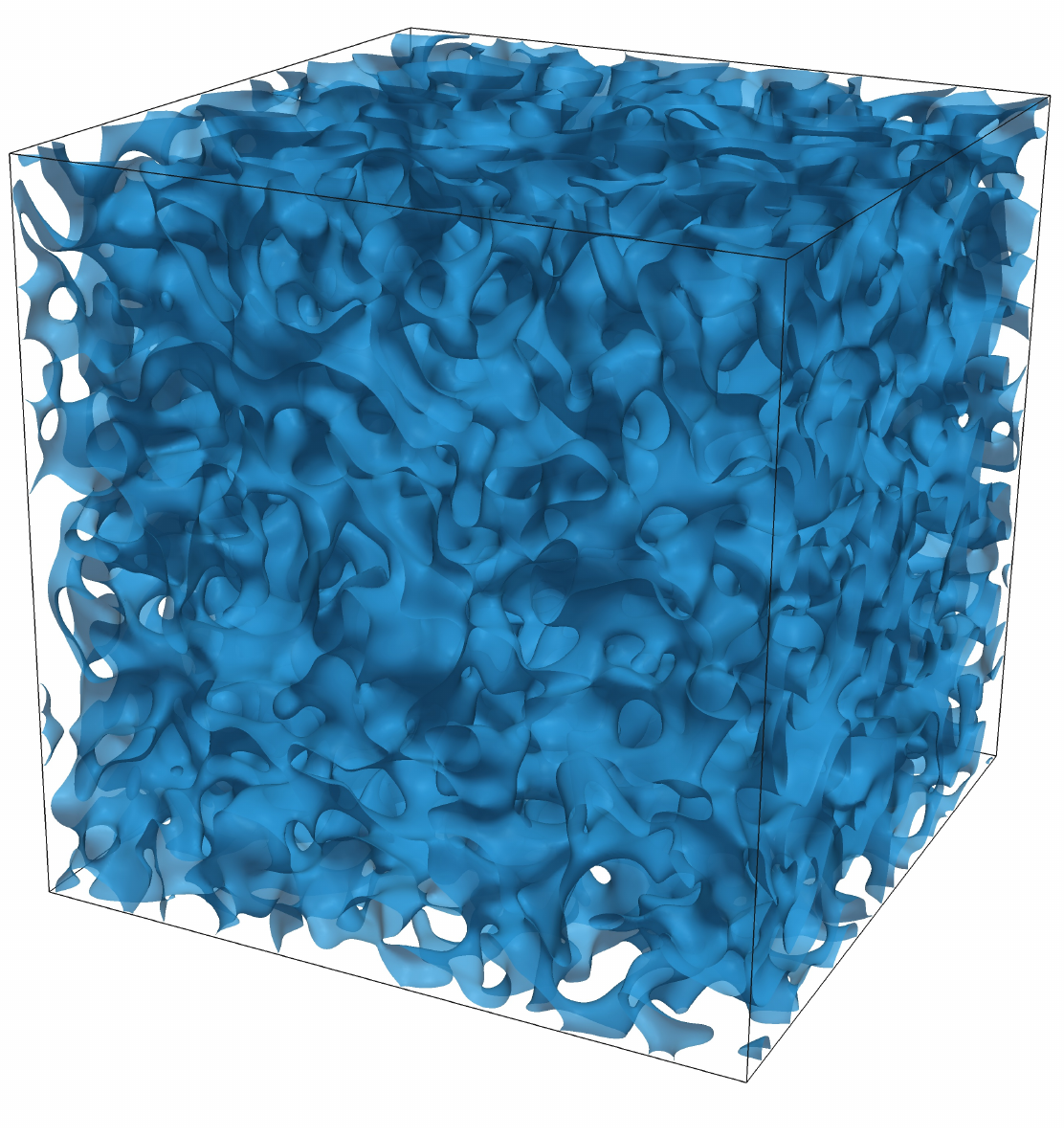}
 }\hspace{0.5in}
 \subfigure[Q-criterion]{
  \label{sfig:256_Gaussian_Q}
  \includegraphics[width=0.35\textwidth,trim=130px 430px 145px 70px, clip]{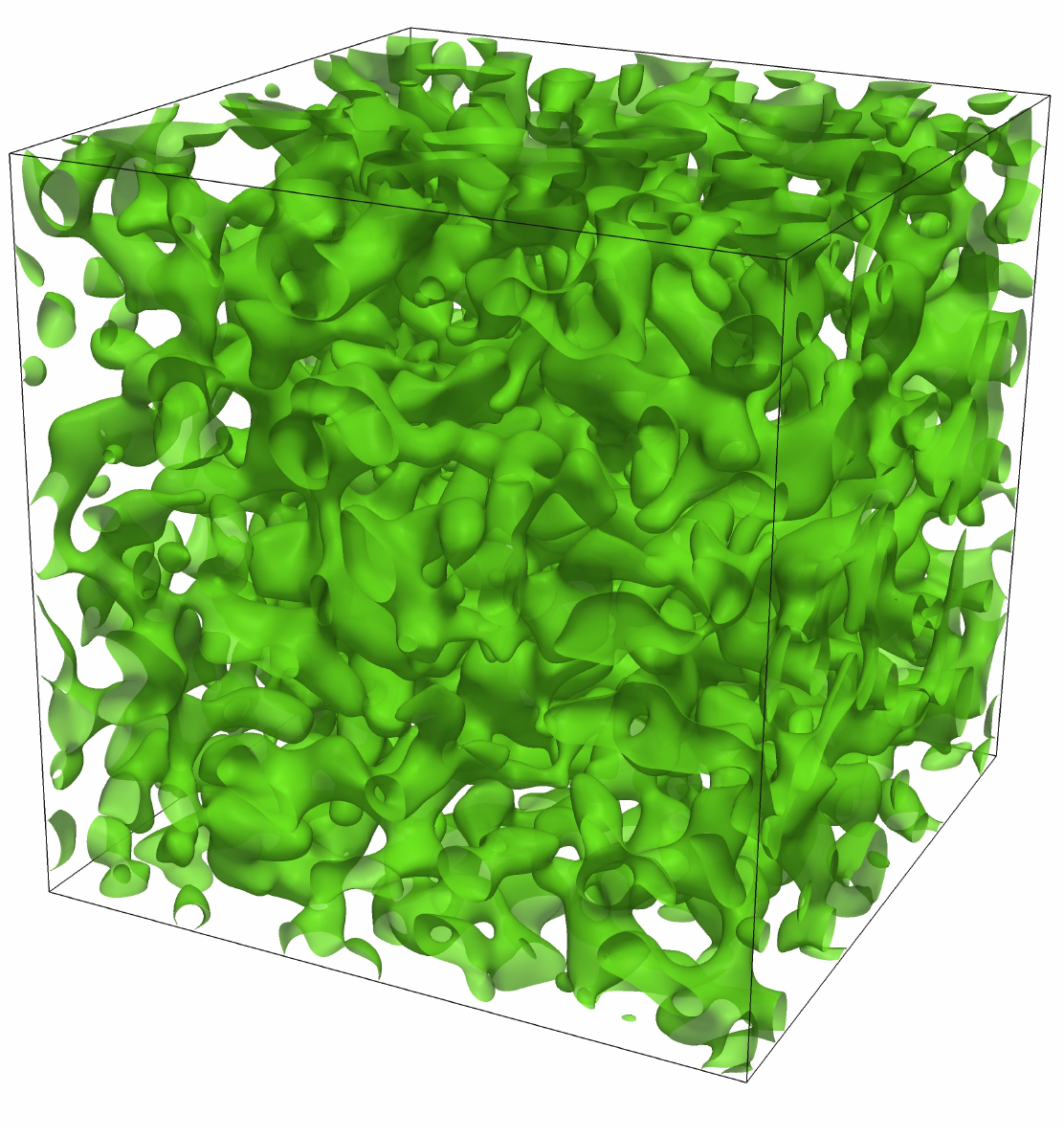}
 }
 \caption{Visualisation of structures in an $N = 256$ initial random Gaussian field. There is little evidence of coherent structure. The same surfaces have been plotted as figure \ref{fig:3dstructures256} above.}
 \label{fig:256_Gaussian}
\end{figure}

\begin{figure}[htb]
 \centering
 \subfigure[Vorticity]{
  \label{sfig:3dstructures1024_vort}
  \includegraphics[width=0.45\textwidth,trim=165px 465px 185px 60px, clip]{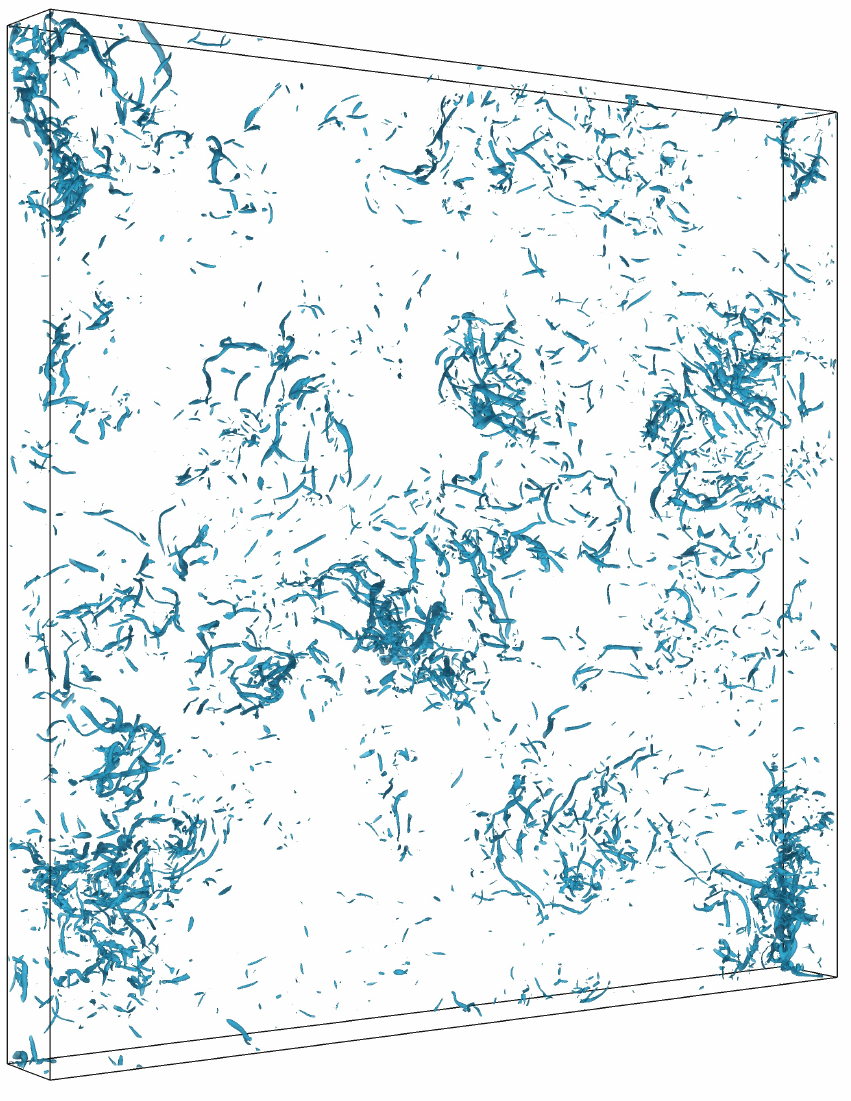}
 }\hfill
 \subfigure[Q-criterion]{
  \label{sfig:3dstructures1024_Q}
  \includegraphics[width=0.45\textwidth,trim=165px 465px 185px 60px, clip]{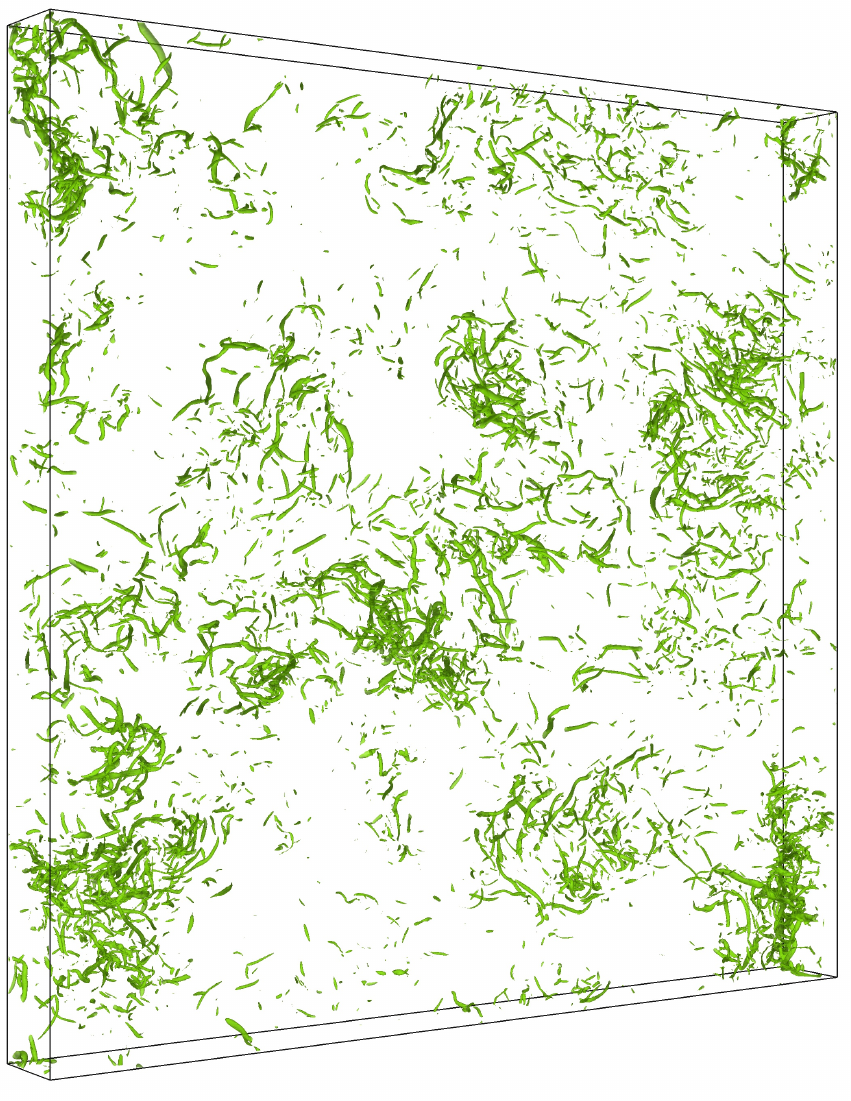}
 }
 \caption{Visualisation of turbulent structures in a $96\times 1024 \times 1024$ slice (due to memory constraints, the whole volume could not be rendered) of an $R_\lambda \sim 335$ evolved velocity field from run \frun{f1024b}. Isosurfaces of (a) vorticity ($0.25 \omega_{\textrm{max}}$ plotted) and (b) Q-criterion ($0.05 Q_{\textrm{max}}$ plotted).}
 \label{fig:3dstructures1024}
\end{figure}

The rate-of-strain tensor defined above can be connected to the dissipation rate, since
\begin{equation}
 \varepsilon = \frac{\nu_0}{2} \left\langle \left( \frac{\partial u_i}{\partial x_j} + \frac{\partial u_j}{\partial x_i} \right)^2 \right\rangle = 2\nu_0 \big\langle \Vert S \Vert^2 \big\rangle \ ,
\end{equation}
where the average is performed over space. This means that $2\nu_0 \Vert S \Vert^2$ gives a measure of the \emph{local} dissipation at point $\vec{x}$. Since $2\nu_0$ is just a scaling, the magnitude of the strain rate tensor indicates the strength of the dissipation and allows for the identification of dissipative structures. These are shown in figures \ref{sfig:structures512_strain} and \ref{sfig:structures1024_strain}: the former compares contours with the magnitude of vorticity and $Q$-criterion, discussed below, while the latter shows the structures for a higher Reynolds number on a larger lattice. Figure \ref{sfig:3dstructures256_vort_strain} shows the dissipative structures in three-dimensions, indicating that they are correlated with and attached to the regions of high vorticity, but that the two criteria are not indistinguishable.

\clearpage

\subsubsection{Other methods of identification}
The above criteria are not the only methods available for defining vortices. The article by Haller \cite{Haller:2005p618} discusses a much more complete list, and provides comparisons. We direct the reader to this paper for more information, as we will concern ourselves only with those discussed above. We do, however, mention two more as being of interest to the present author:
\begin{enumerate}
 \item The $\Delta$-criterion of Chong, Perry and Cantwell \cite{Chong:1990p1089} --- regions with
 \begin{equation}
  \Delta = \left( \frac{Q}{3} \right)^3 + \left( \frac{R}{2} \right)^2 > 0 \ .
 \end{equation}
 Note that this criterion is less restrictive than $Q > 0$.

 \item The $\lambda_2$-criterion of Jeong and Hussain \cite{Jeong:1995p1073}. If the eigenvalues of $S^2 + \Omega^2$ satisfy $\lambda_1 \geq \lambda_2 \geq \lambda_3$, then define as a vortex the regions where $\lambda_2 < 0$ (\textit{i.e.} there are two negative eigenvalues). This guarantees a local pressure minimum in a two-dimensional plane.

\end{enumerate}

\subsection{Persistence of structure under averaging}

Looking at the snapshots of the velocity field in the figures above, it can be seen that there are well-defined structures and a great deal of variation from point to point. The velocity field is said to be intermittent: there is a high degree of spatial variation. This intermittency becomes an issue when one considers the form for the structure functions, since they, and Kolmogorov's theory, are for an ensemble averaged system. The existence of fine-scale structure (the source of intermittency) in the ensemble is seen by many to be of direct importance. 

\begin{figure}[tbp!]
 \begin{center}
  \subfigure[$N = 1$]{
   \includegraphics[width=0.38\textwidth,trim=130px 460px 130px 30px, clip]{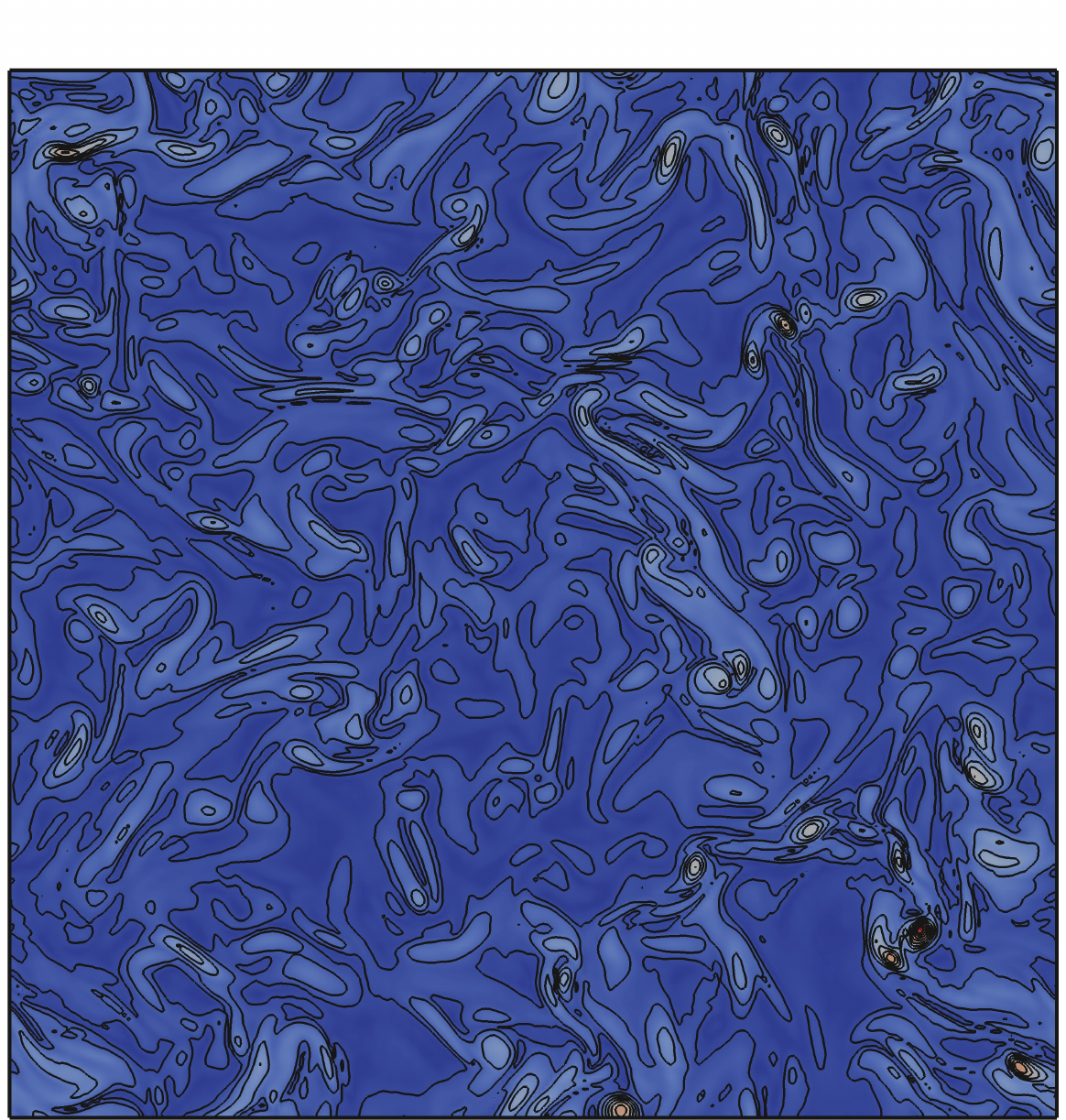}
  }
  \subfigure[$N = 2$]{
   \includegraphics[width=0.38\textwidth,trim=130px 460px 130px 30px, clip]{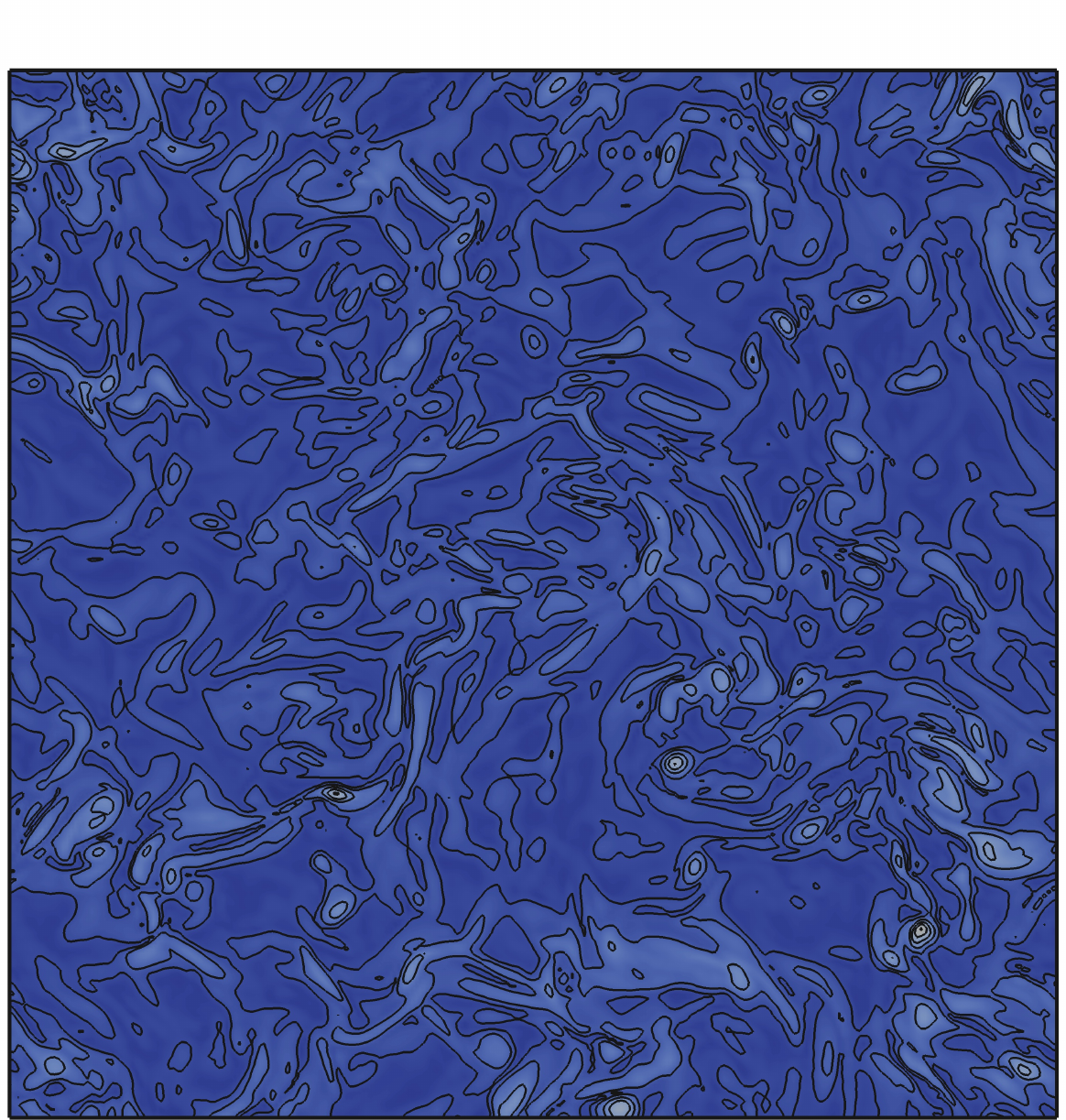}
  }
  \subfigure[$N = 5$]{
   \includegraphics[width=0.38\textwidth,trim=130px 460px 130px 30px, clip]{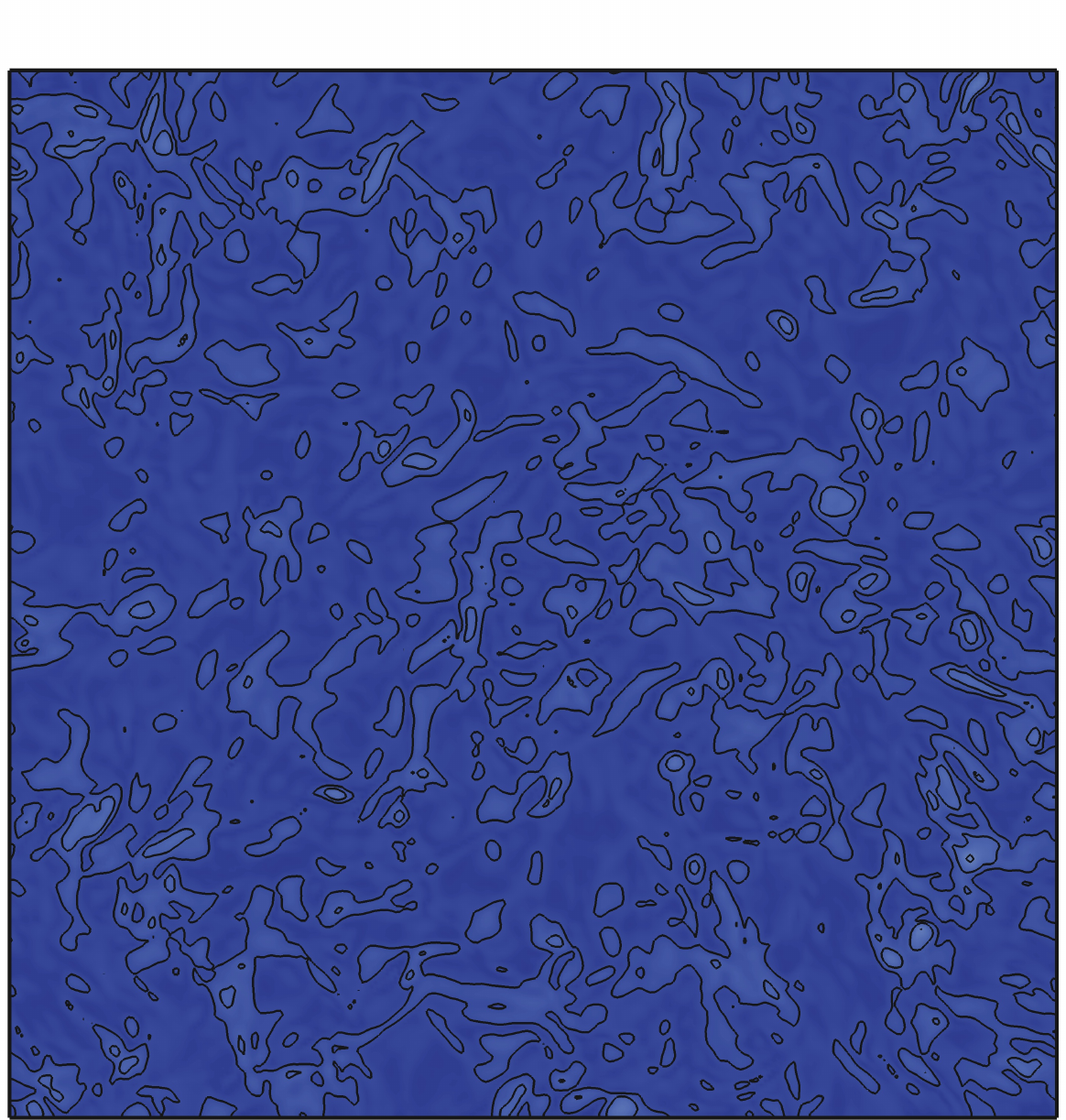}
  }
  \subfigure[$N = 10$]{
   \includegraphics[width=0.38\textwidth,trim=130px 460px 130px 30px, clip]{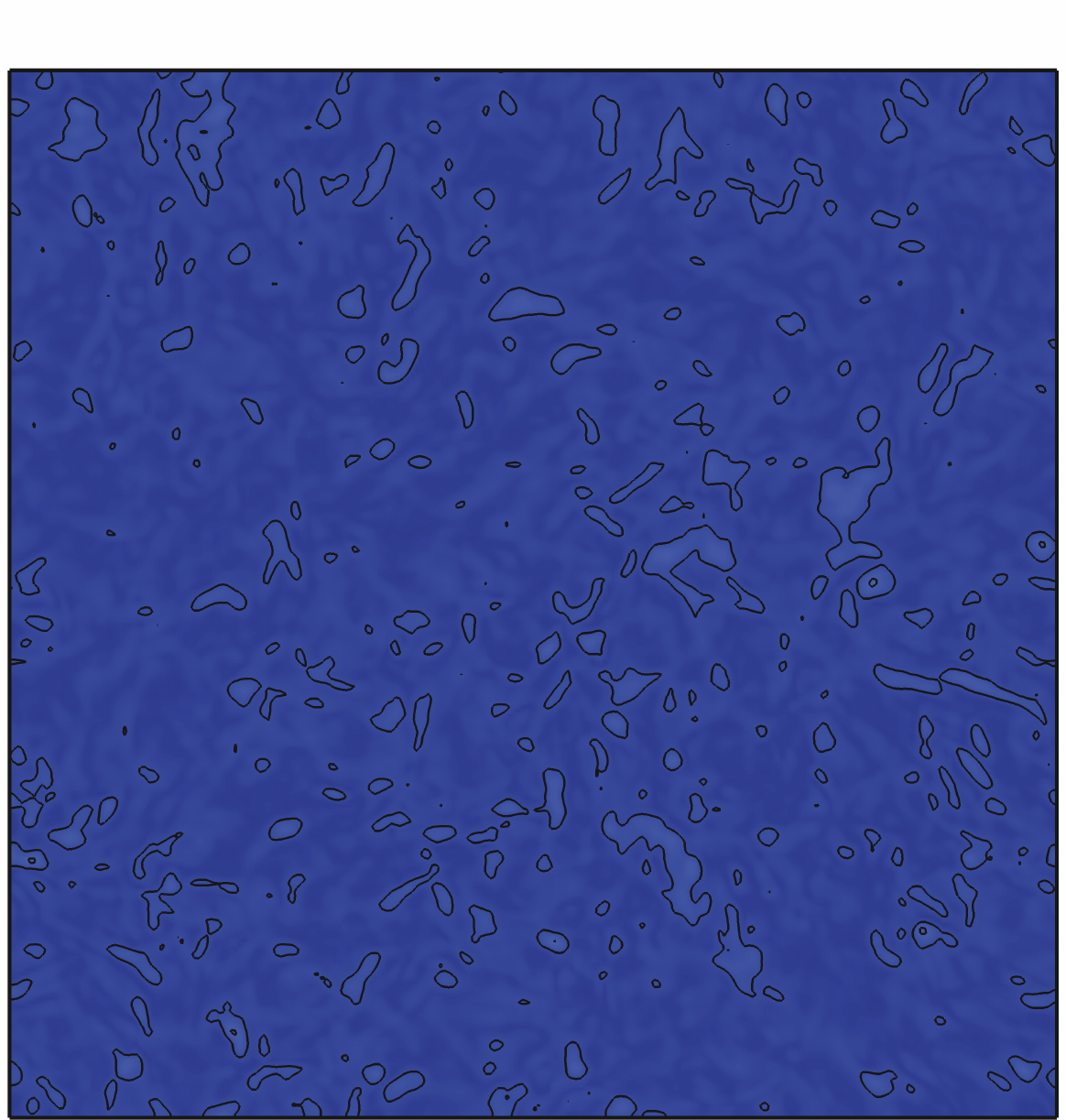}
  }
  \subfigure[$N = 25$]{
   \includegraphics[width=0.38\textwidth,trim=130px 460px 130px 30px, clip]{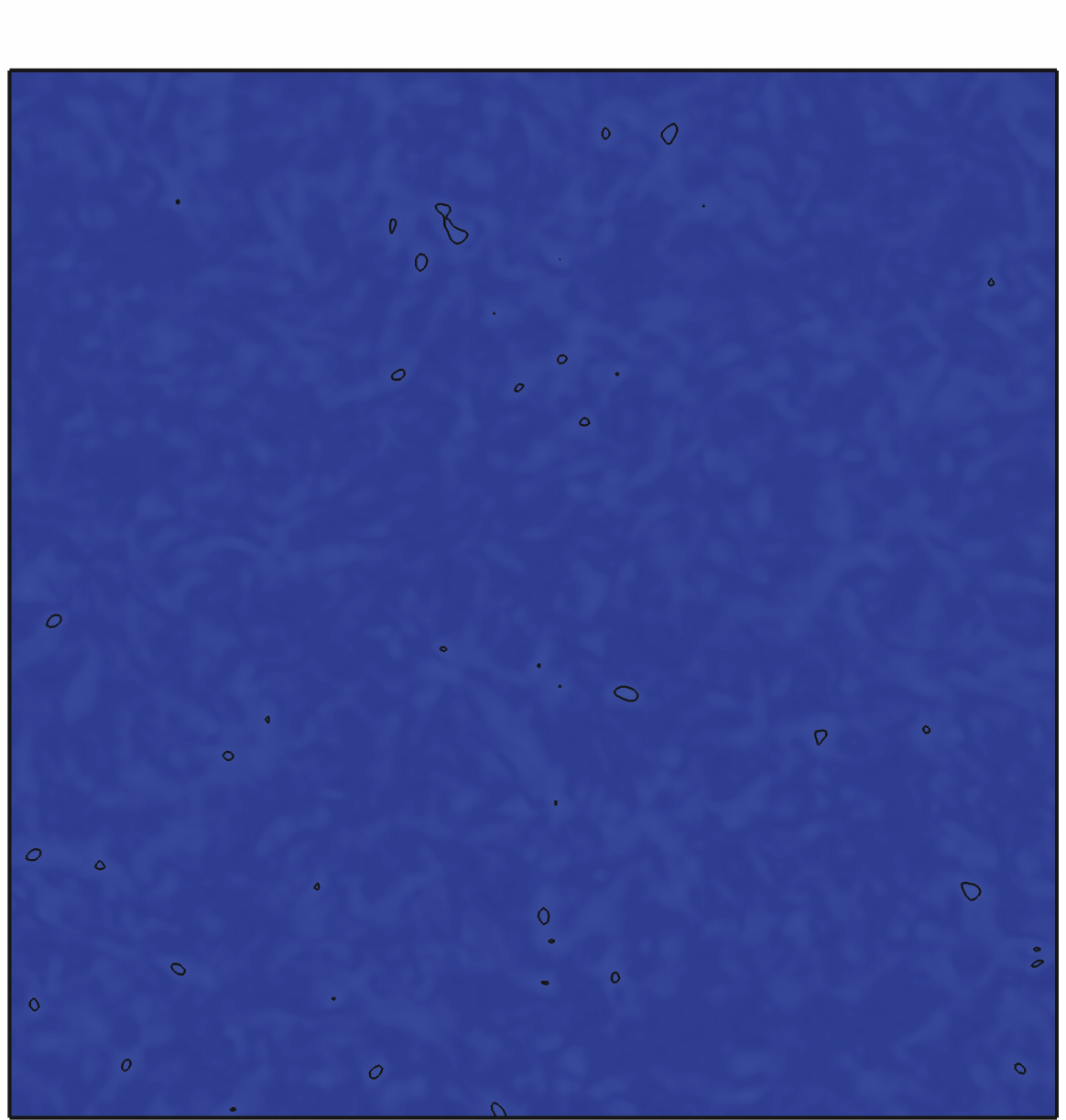}
  }
  \subfigure[$N = 46$]{
   \includegraphics[width=0.38\textwidth,trim=130px 460px 130px 30px, clip]{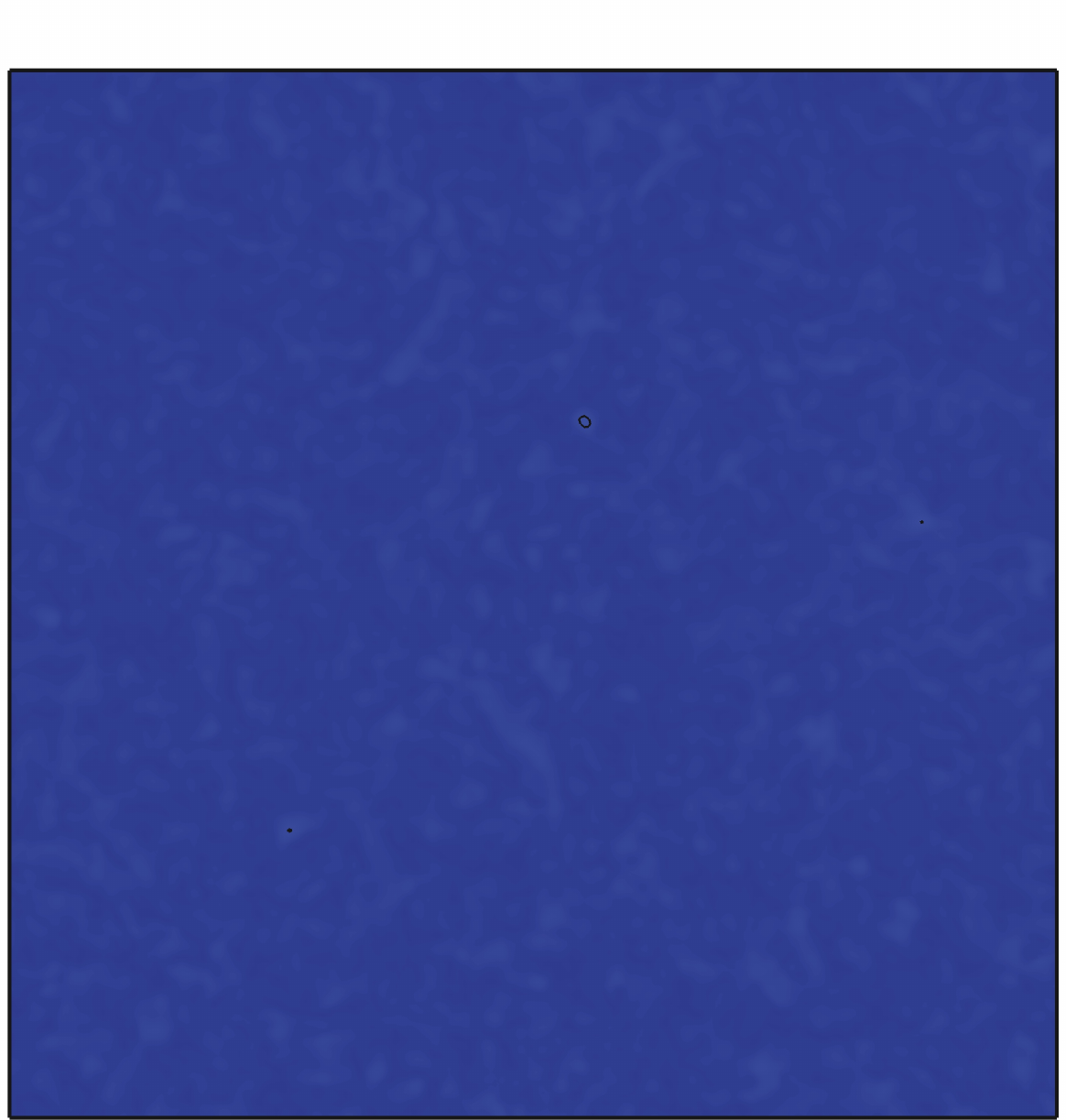}
  }
 \end{center}
 \caption{Contours plotted for 5\%, 10\%, 20\%, 30\%, 40\%, 50\%, 60\%, 70\%, 80\% and 90\% of $\omega_{\textrm{max}}$. $R_\lambda \sim 100$ on $256^3$ run \frun{f256b}.}
 \label{fig:str256_V}
\end{figure}

\begin{figure}[tbp!]
 \begin{center}
  \subfigure[$N = 1$]{
   \includegraphics[width=0.38\textwidth,trim=130px 460px 130px 30px, clip]{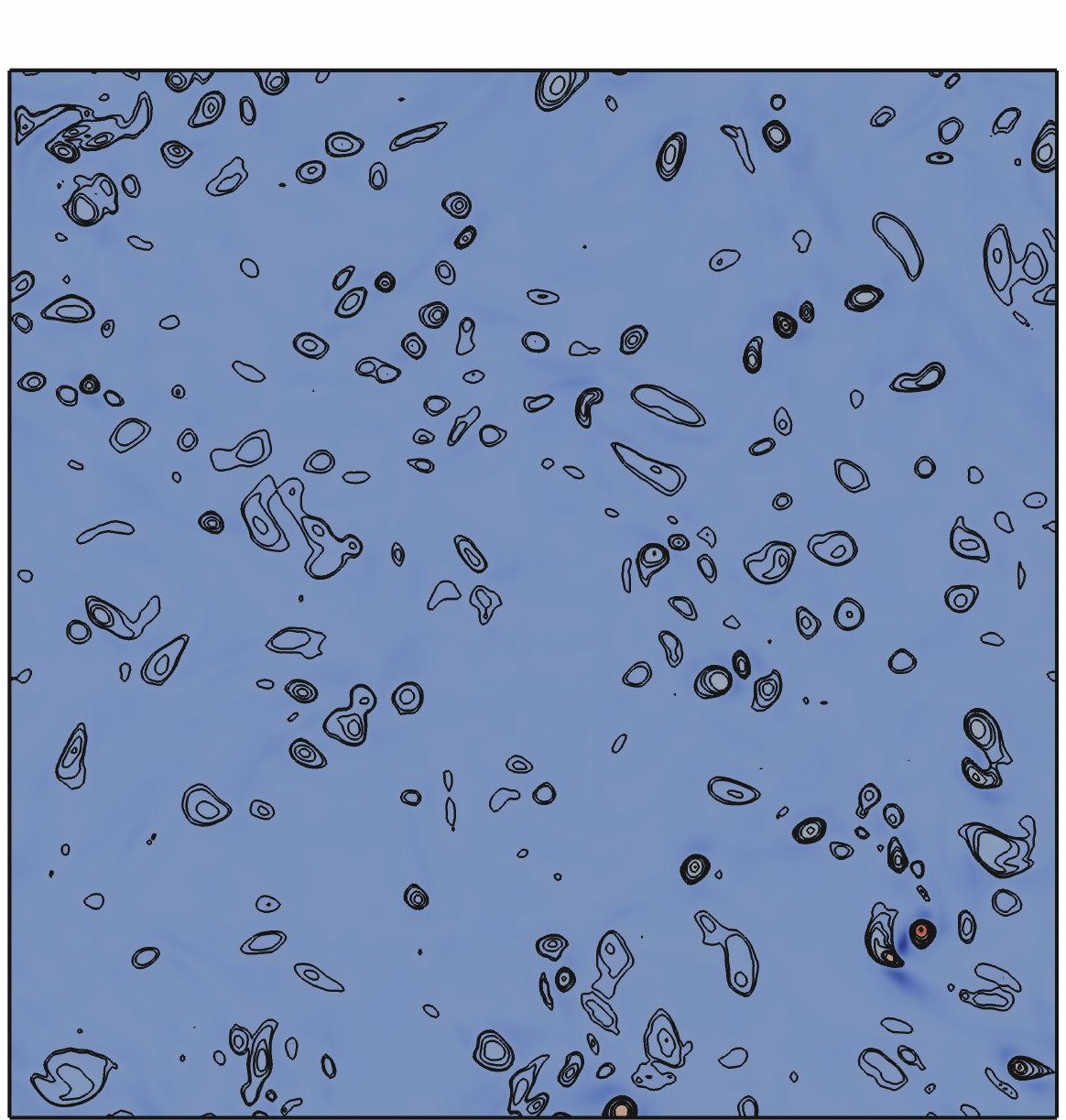}
  }
  \subfigure[$N = 2$]{
   \includegraphics[width=0.38\textwidth,trim=130px 460px 130px 30px, clip]{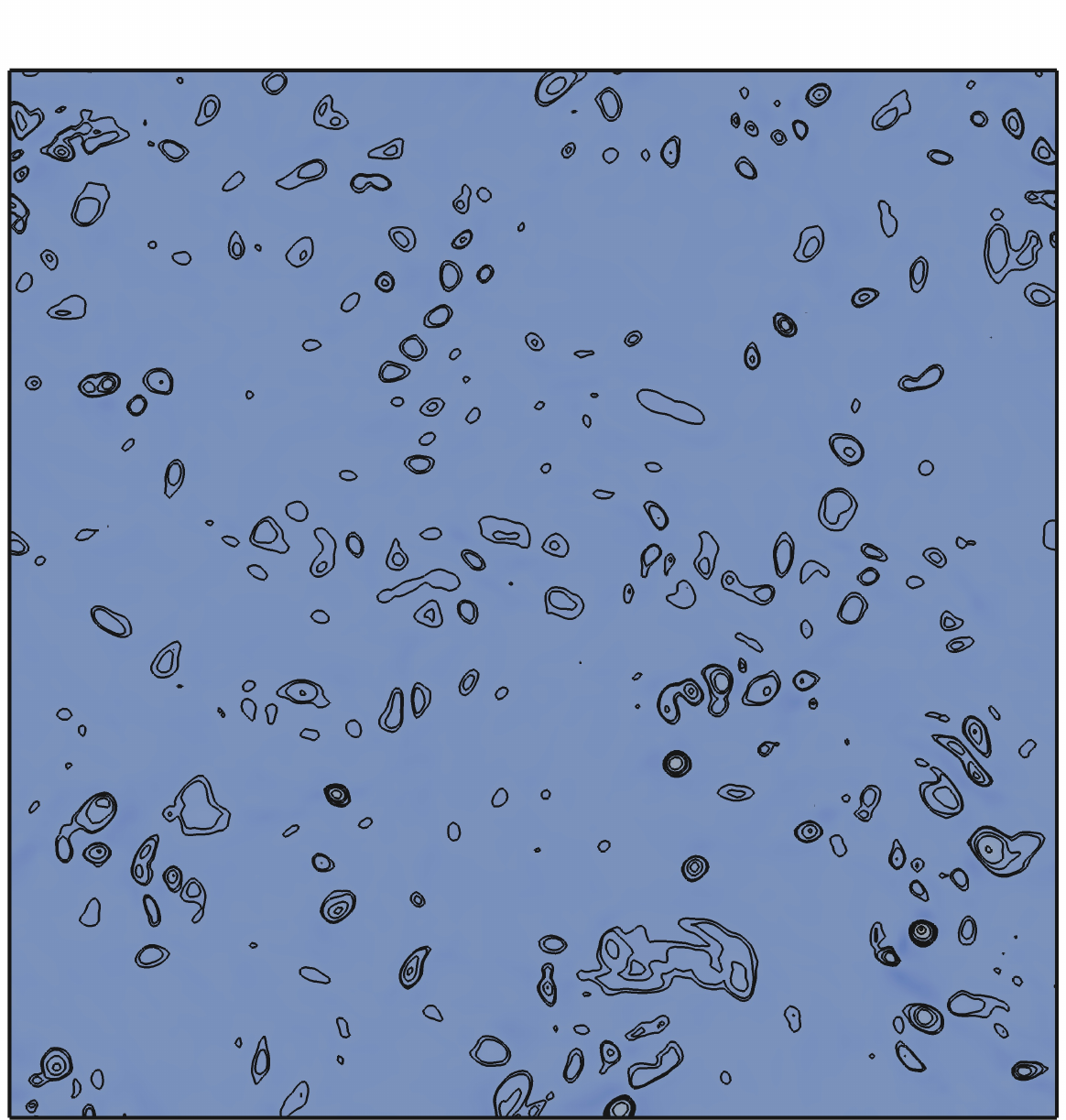}
  }
  \subfigure[$N = 5$]{
   \includegraphics[width=0.38\textwidth,trim=130px 460px 130px 30px, clip]{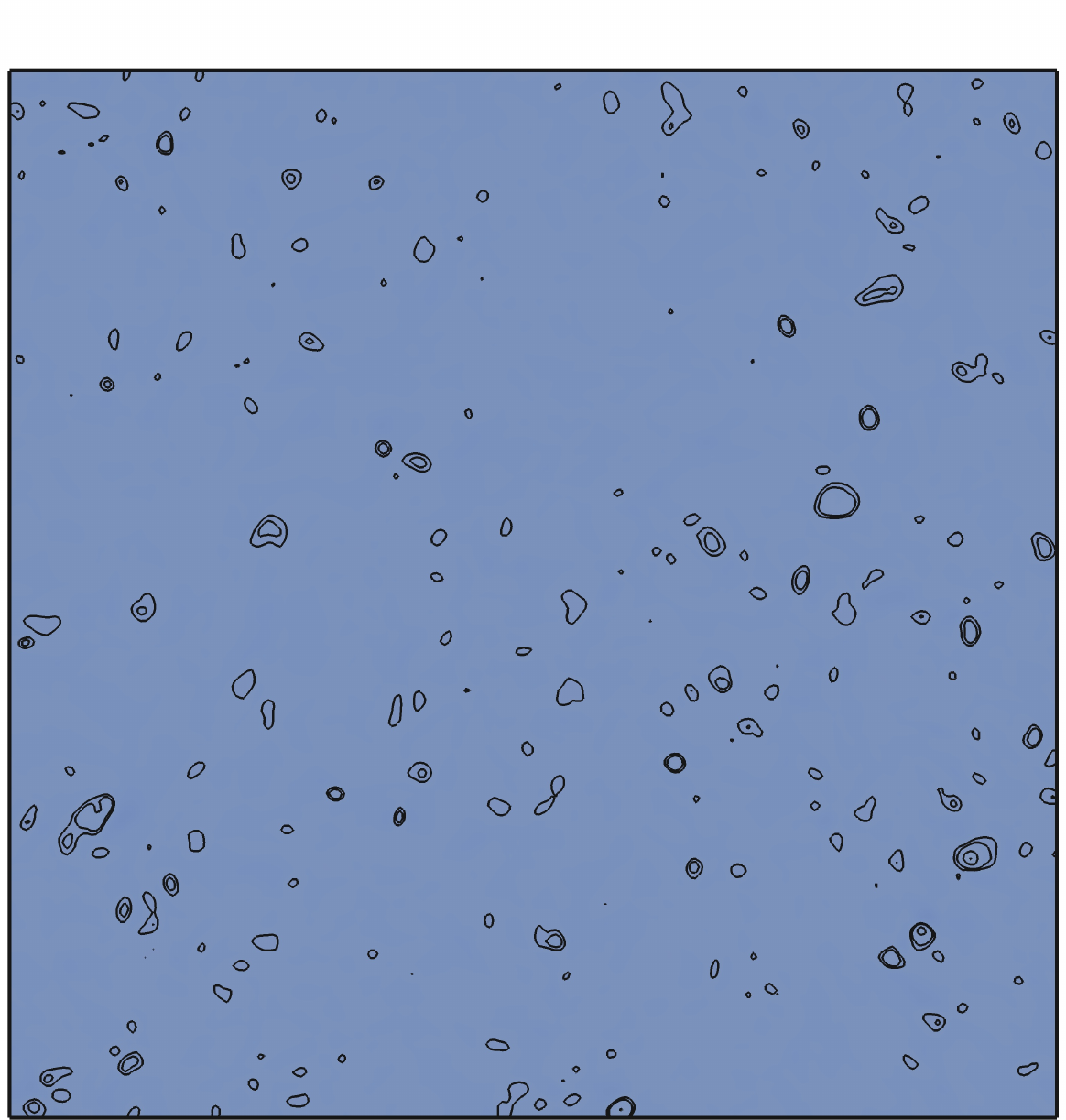}
  }
  \subfigure[$N = 10$]{
   \includegraphics[width=0.38\textwidth,trim=130px 460px 130px 30px, clip]{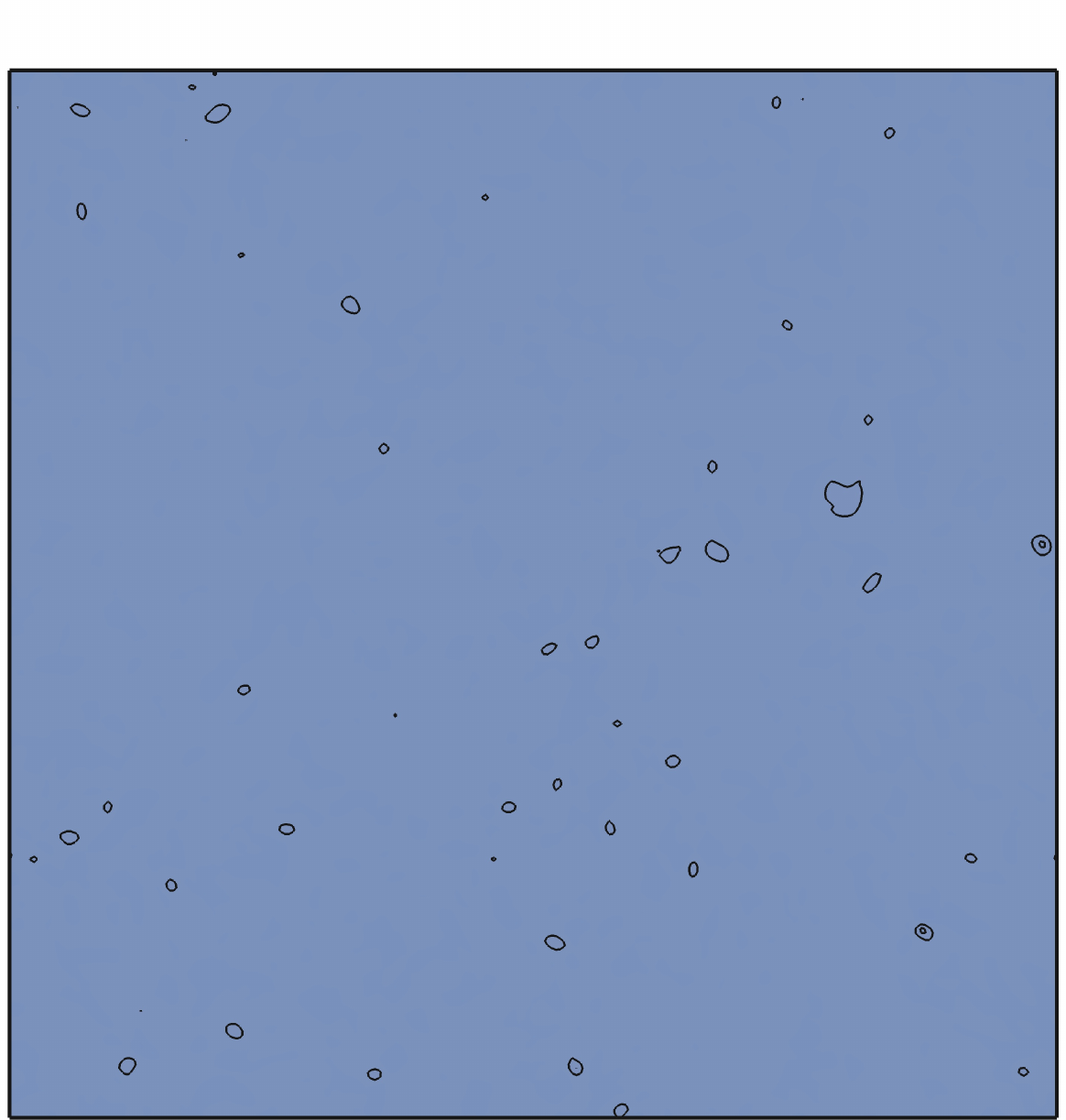}
  }
  \subfigure[$N = 25$]{
   \includegraphics[width=0.38\textwidth,trim=130px 460px 130px 30px, clip]{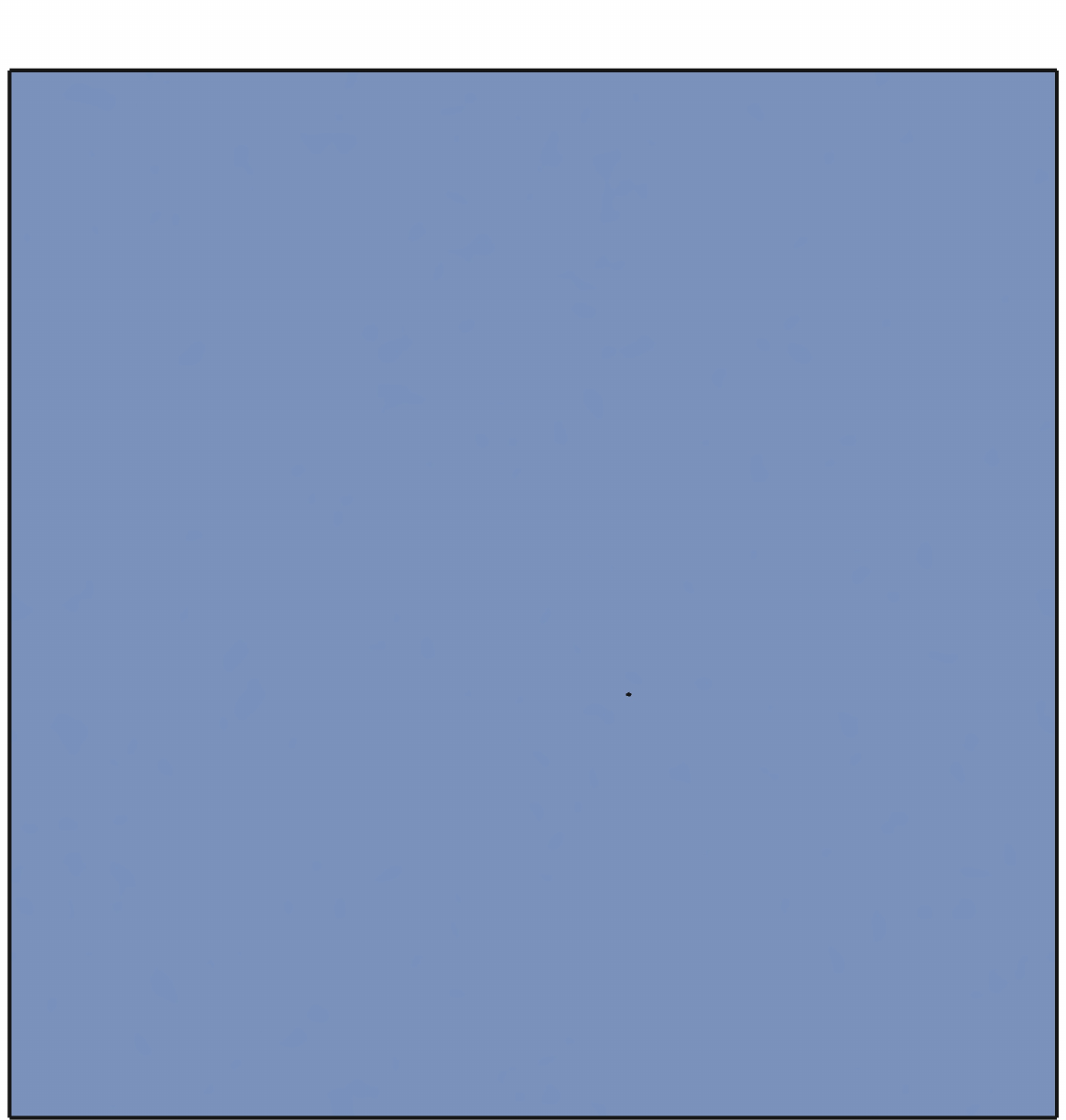}
  }
  \subfigure[$N = 46$]{
   \includegraphics[width=0.38\textwidth,trim=130px 460px 130px 30px, clip]{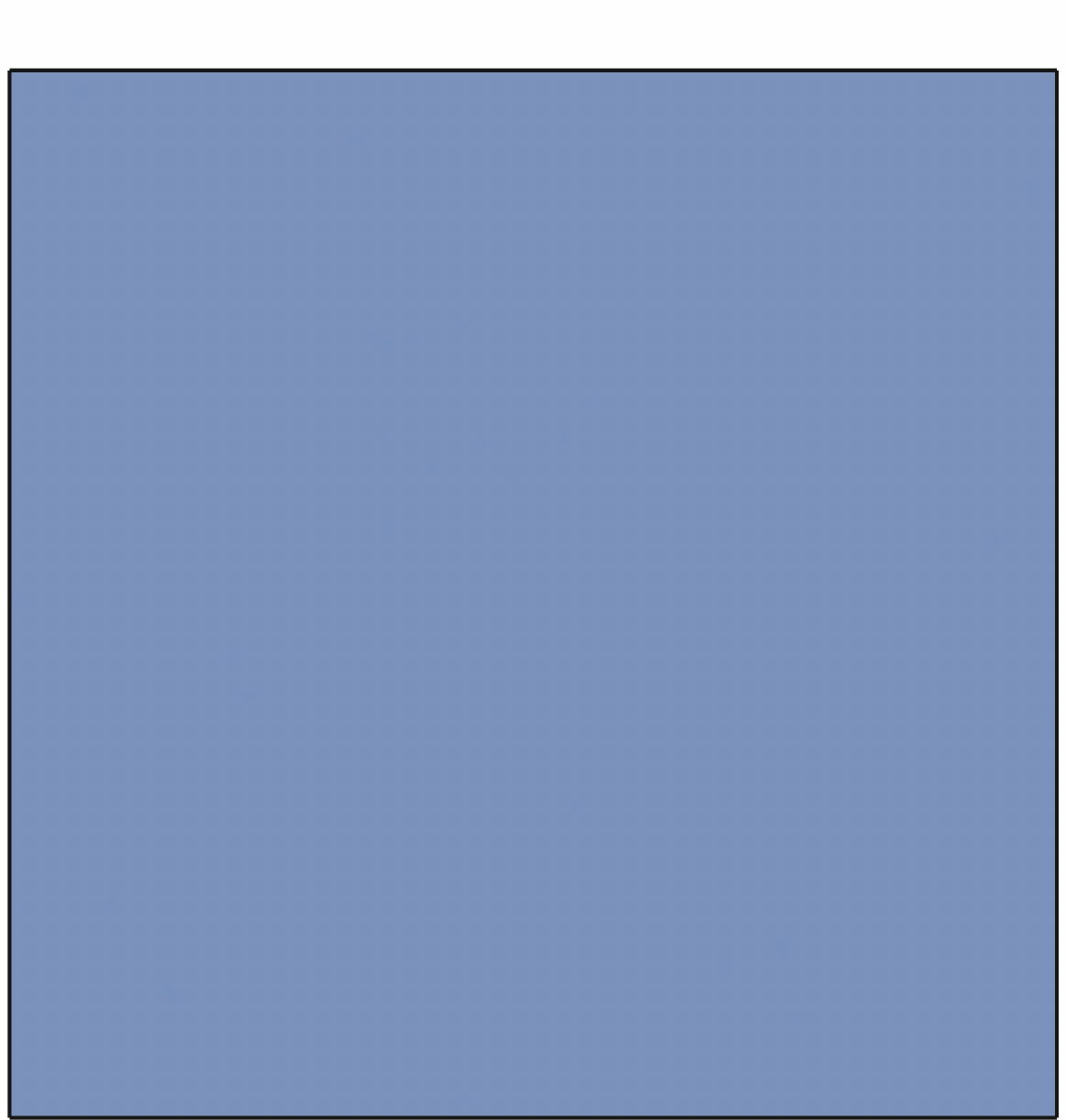}
  }
 \end{center}
 \caption{Contours plotted for 0.5\%, 1\%, 3\%, 5\%, 10\%, 20\%, 25\%, 50\%, 75\% and 90\% of $Q_{\textrm{max}}$. $R_\lambda \sim 100$ on $256^3$ run \frun{f256b}.}
 \label{fig:str256_Q}
\end{figure}

Due to the restriction of isotropy, it is impossible for coherent structure to exist in homogeneous, isotropic turbulence in anything other than an instantaneous sense \cite{McComb:2011-bulletin}. To test the amount residual coherent structures present in a finite ensemble, we ensemble average the velocity field. The set $\mathbb{S}_N$ contains $N$ realisations of the velocity field taken from a stationary simulation, sampled at an interval of one large eddy turnover time. The average is then found as
\begin{equation}
 \left\langle \vec{u}(\vec{x}) \right\rangle_N = \frac{1}{N} \sum_{n = 1}^N \vec{u}^{(n)}(\vec{x}) \ ,
\end{equation}
where $\vec{u}^{(n)}(\vec{x})$ is the $n$th member of $\mathbb{S}_N$. The resultant fields for a sample of $N$ are visualised in figures \ref{fig:str256_V} and \ref{fig:str256_Q} using vorticity and the $Q$-criterion, respectively. As can be seen in figure \ref{fig:str256_V}, the case $N = 1$ corresponding to a single realisation displays the expected mass of structures. When we add in another realisation ($N = 2$) we see definite reduction in the amount of structure present, and this becomes even more dramatic when we move to $N = 5$. By this point, we have no structures with higher vorticity than our second lowest contour. Proceeding to $N = 10$ and 25 the structures reduce further, until for $N = 46$ we see only two very small areas with vorticity as high as 5\% of $\omega_{\textrm{max}}$ obtained from the initial single realisation. A similar story is seen in figure \ref{fig:str256_Q} using the $Q$-criterion, with the difference that by $N = 46$ there are no structures present with even 0.5\% of $Q_{\textrm{max}}$ obtained in $N = 1$. Note that there appear to be more structures in figure \ref{fig:str256_Q} than \ref{sfig:structures512_Q} above due to the inclusion of significantly lower, less restrictive contour values.

From this, we conclude that as the ensemble size is increased, there remains less and less coherent structure in the velocity field. The effect of intermittency, clearly present in a single realisation, on statistical quantities should be investigated more thoroughly.

The constraint $\langle \vec{u} \rangle = 0$ does not itself imply that there is no coherent structure which could remain under averaging, since one could set up, for example, two counter-rotating vortices. However, these structures break isotropy, and it is this which prevents their presence under the ensemble averaging process. This test therefore assists in determining the degree to which isotropy is satisfied as the ensemble size is increased.

\subsection{The role of intermittency in K41 theory}\label{sec:intermit_K41}
When Kolmogorov derived the results for the second- and third-order structure functions, he did so based on several assumptions. First of all, the relevant equation is an expression of conservation of energy involving the \emph{statistical quantities} $S_2(r),S_3(r)$. This is the K\'arm\'an-Howarth equation \cite{Karman:1938p153}, which for stationary turbulence is taken to be
\begin{equation}
 \label{eq:KHE_kol}
 \varepsilon = -\frac{1}{4r^4} \frac{\partial}{\partial r} \left( r^4 \frac{\partial S_3(r)}{\partial r} \right) + \frac{3\nu_0}{2r^4} \frac{\partial}{\partial r} \left( r^4 \frac{\partial S_2(r)}{\partial r} \right) \ ,
\end{equation}
with caveat that this is only valid for scales unaffected (directly) by the action of forcing. See section \ref{sec:KHE} for further discussion of the applicability of this equation, particularly the origin of $\varepsilon$. For now, we note that by assuming the above we have already assumed an asymptotic form for the equation such that the `input' of energy is just inertial transfer, $\varepsilon_T = \varepsilon$.

As discussed in section \ref{sec:kol}, Kolmogorov showed how, as the viscosity is taken to zero (or we only consider scales for which dissipation is negligible, \textit{i.e.} the inertial subrange), then the form of the third- and second-order structure functions satisfy
\begin{equation}
 S_n(r) = C_n (\varepsilon r)^{n/3} \, \qquad n = 2,3 \ .
\end{equation}
The use of the intermittent local dissipation rate (instead of the average) or an additional characteristic length scale has led to a significant amount of work studying the discrepancy between measured exponents and the Kolmogorov predicted values, known as intermittency corrections. Since the relevant quantity in the inertial range of scales is actually the inertial flux, $\varepsilon_T$, which is equal to $\varepsilon$ (assuming that we indeed have sufficient separation of integral and dissipative scales for the formation of an inertial subrange), it seems unfounded to introduce the local dissipation rate.

We also mention that, in obtaining the above form of the K\'arm\'an-Howarth equation, we have introduced the second-order structure function by inserting
\begin{equation}
 C_{LL}(r) = u^2 - \tfrac{1}{2} S_2(r)
\end{equation}
into the term
\begin{equation}
 \frac{2\nu_0}{r^4} \frac{\partial}{\partial r} \left( r^4 \frac{\partial C_{LL}(r)}{\partial r} \right) \ .
\end{equation}
This has introduced a term
\begin{equation}
 \frac{2\nu_0}{r^4} \frac{\partial}{\partial r} \left( r^4 \frac{\partial u^2}{\partial r} \right) \ ,
\end{equation}
which is assumed to vanish since $u^2$ is a constant. However, if the details of intermittency must be kept, then one could consider including the \emph{local} rms velocity, $u^2(r)$, for which this term is not necessarily zero. This would not affect the form of $S_3(r)$ found by Kolmogorov, since it comes with a factor of $\nu_0$ which is taken to zero, but for finite Reynolds numbers it would need to be included.

Measurements of the scaling exponents of the structure functions do exhibit deviations from the Kolmogorov prediction and the deviation is seen to increase with order. The question is whether these really are due to intermittency or rather that the conditions required for K41 to hold are simply not satisfied at finite Reynolds number. In the latter case, K41 is an asymptotic theory and the correction to any exponent must vanish as $Re \to \infty$. Support for the latter case is becoming increasingly popular, once again see \cite{McComb:2011-bulletin} and the references therein.


\section{Structure functions and scaling exponents}\label{sec:sf}
\subsection{Computation of the structure functions}
Since the \dns\ code is pseudospectral, we have access to the velocity field in configuration space. This allows the structure functions to be evaluated for a range of orders. As noted by Fukayama, Oyamada, Nakano, Gotoh and Yamamoto \cite{Fukayama:1999p904}, the degree of isotropy and homogeneity of the ensemble is of critical importance for the calculation of the structure functions. As such, a large number of realisations is required to ensure that the probability distribution be sufficiently isotropic. For the case of stationary turbulence, this is done by taking a snapshot of the velocity field at some time interval, once the initial transient has died away. For decaying turbulence, this requires an ensemble of runs from varying initial configurations.

Fukayama \etal\ present their analysis of isotropy for both forced and free decay. They note that the conditions of isotropy are not satisfied for forced turbulence as well as they are for decaying turbulence. This lack of isotropy has a larger influence on odd-order structure functions than even ones and causes statistical convergence to be gradually lost as one examines higher order structure functions, since they become increasingly affected by the tails of the PDF. Anisotropy is more pronounced in the large scales, see figure \ref{fig:isotropy_spec} which measured the isotropy spectrum (low wavenumbers correspond to large length-scales).

To do this for forced turbulence, we therefore have to store and process a large number of realisations, which immediately places constraints on the size of the simulation this can reasonably be done for. For each site on our lattice, we calculate the longitudinal correlation of the velocity field with every other site. To improve isotropy, this is done in each of the three directions and averaged, but could be extended to use directions not parallel to the coordinate axes as well, as used in the angle-average by Taylor, Kurien and Eyink \cite{Taylor:2003p1581}. Thus, we are calculating
\begin{align}
 S_n(r) &= \frac{1}{3N^3} \sum_{\vec{x}} \bigg[ \Big( \big( u_x(\vec{x} + r\vec{e}_x) - u_x(\vec{x}) \big)^n + \big( u_y(\vec{x} + r\vec{e}_y) - u_y(\vec{x}) \big)^n \nonumber \\
 &\qquad\qquad\qquad+ \big( u_z(\vec{x} + r\vec{e}_z) - u_z(\vec{x}) \big)^n \Big) \bigg] \ .
\end{align}
This is obtained for each realisation, and then ensemble averaged to give the final result. 

We have performed a similar analysis to that of Fukayama \etal\ \cite{Fukayama:1999p904} for runs \frun{f128a}, \frun{f128e}, \frun{f256b} and \frun{f512a}, using an ensemble of 101 realisations for the first three and just 15 for the last. This was due to memory and time constraints, since it requires $\sim 8$ times more storage and computation for the $512^3$ lattice over the $256^3$, not to mention the additional run time to generate one sample from the next. The realisations were taken every half a large turnover time. This roughly coincides with the frequency of samples used by \cite{Fukayama:1999p904}, who took 126 realisations from $\sim 50$ large eddy turnovers for their $R_\lambda = 70$ simulation and 45 from 9 large eddy turnovers for their higher run with $R_\lambda = 125$. The forcing used here is negative damping of the lowest two wavenumber shells with the input rate maintained constant, whereas Fukayama \etal\ used Gaussian white noise introduced to the band $2 \lesssim k \lesssim 3$. Their simulations started from an initial Gaussian random field with energy spectrum $E(k,0) = c(k/k_0)^4 \exp\big[-2(k/k_0)^2\big]$ with $k_0 = 1$ or 3.

\begin{figure}[tb]
 \begin{center}
  \subfigure[Our DNS data]{
   \label{sfig:realspace_SF}
   \includegraphics[width=0.6\textwidth]{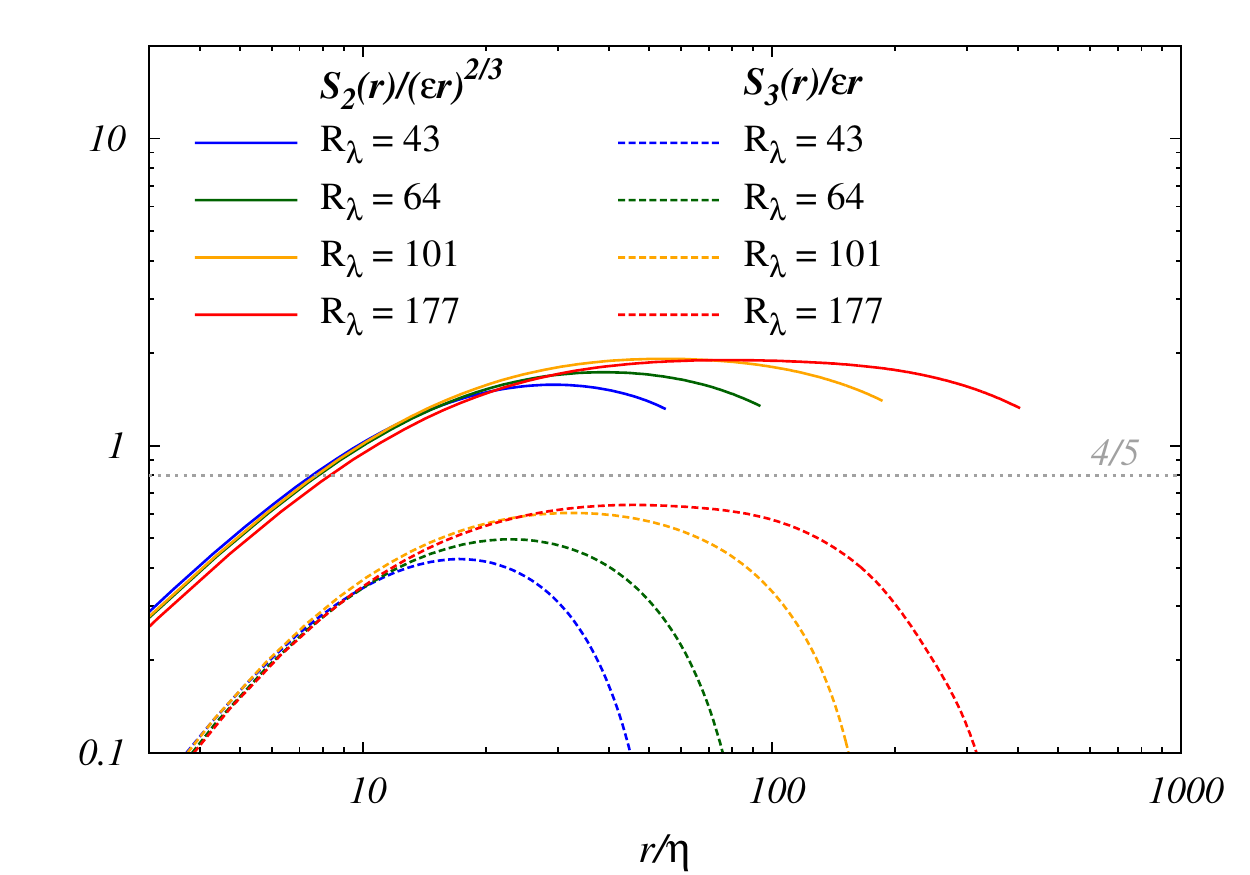}
  }
  \subfigure[Ishihara \etal\ \cite{Ishihara:2009p165}]{
   \label{sfig:realspace_SF_Ishihara}
   \includegraphics[width=0.352\textwidth,trim=10px -8.25em 12px 
   0,clip]{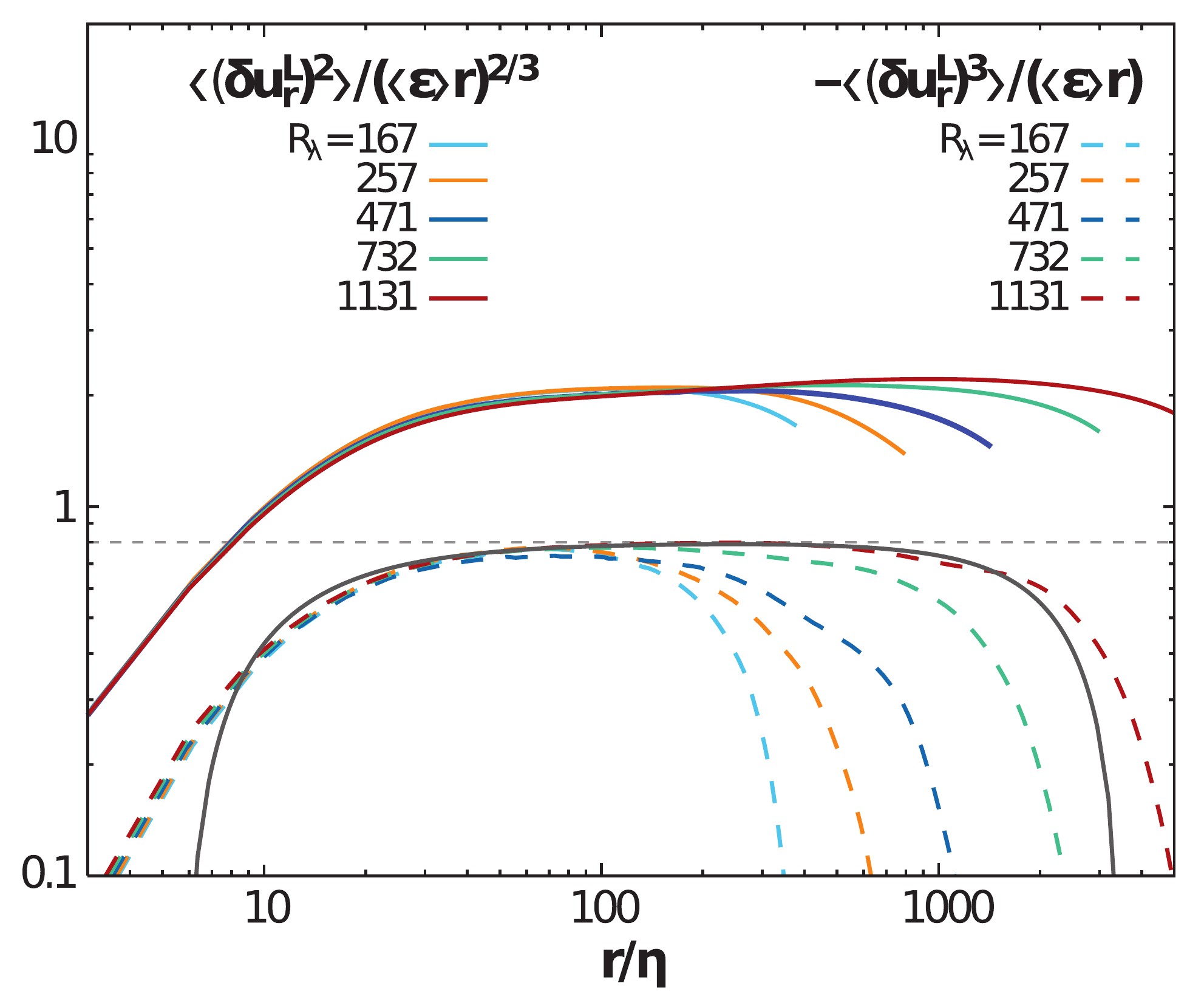}
  }
 \end{center}
 \caption{Second- and third-order (scaled) structure functions. These are calculated for runs \frun{f128a}, \frun{f128e}, \frun{f256b} and \frun{f512a}.}
 \label{fig:realspace_SF}
\end{figure}

Figure \ref{sfig:realspace_SF} presents the second- and third-order structure functions as calculated in configuration space from our ensembles of realisations. These may be compared to Ishihara, Gotoh and Kaneda \cite{Ishihara:2009p165}, where a curve is plotted for $R_\lambda = 167$. This is reproduced in figure \ref{sfig:realspace_SF_Ishihara}. We see that our data for $R_\lambda = 177$ is in good agreement with Ishihara \etal\ with $S_3(r)/\varepsilon r$ meeting the $r/\eta$ axis just below 4 and just above 300. The curve peaks at around $0.65$ for our data compared to just above $0.7$ for \cite{Ishihara:2009p165}.

\begin{figure}
 \begin{center}
  \includegraphics[width=0.75\textwidth]{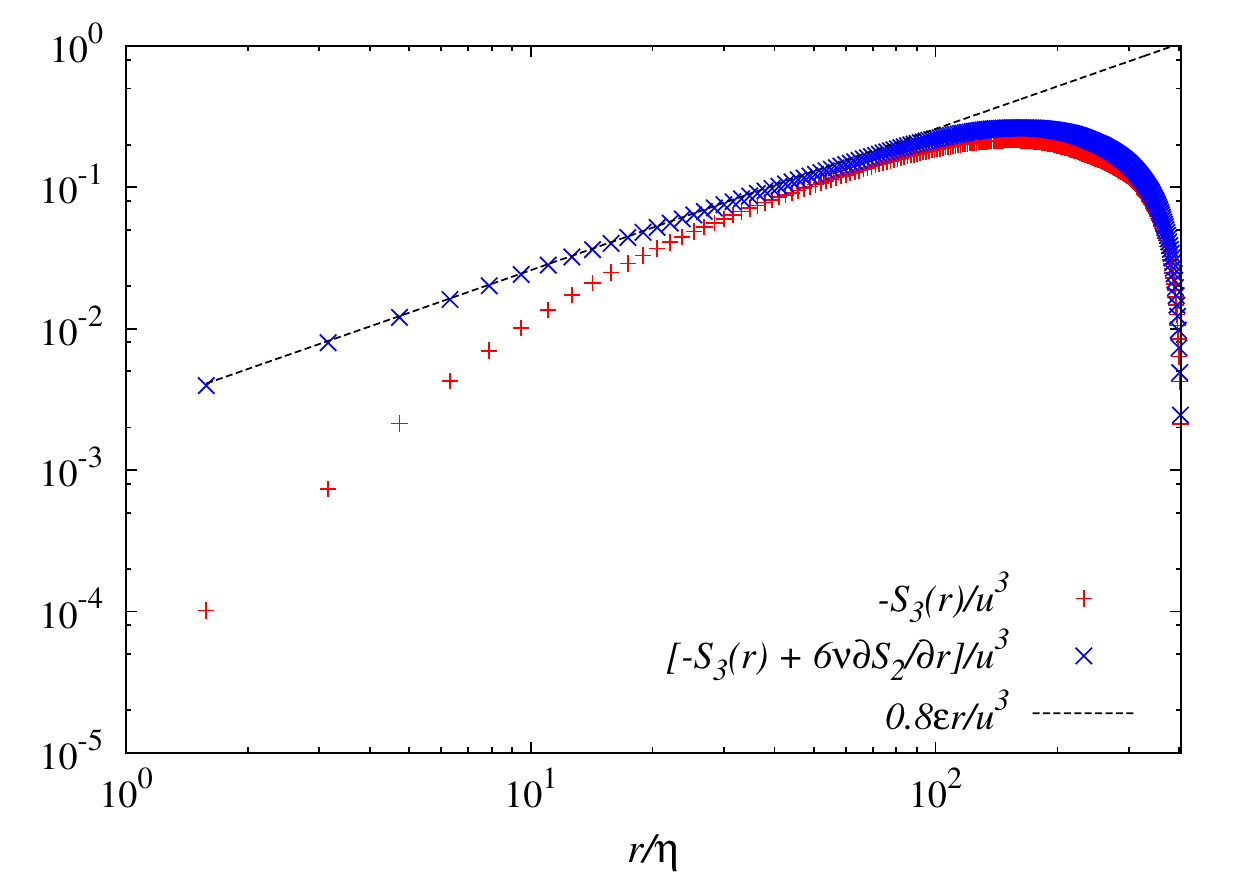}
 \end{center}
 \caption{The third-order structure function and its viscous correction for comparison to the K41 form, as evaluated for run \frun{f512a} with $R_\lambda = 176.9$.}
 \label{fig:SF_diss_correct}
\end{figure}

We also see that none of the simulations are seen to follow the K41 form for $S_3(r)$, which at these low Reynolds number is not surprising. Qian \cite{Qian:1997p903} found that a K41 scaling region could not be identified below $R_\lambda = 10,000$. This discrepancy can in part be accounted for by including the effects of dissipation. From equation \eqref{eq:KHE_kol} we can multiply through by $r^4$ and perform an integral over $r$ to obtain
\begin{equation}
 \label{eq:KHE_visc_correct}
 \frac{4}{5}\varepsilon r = -S_3(r) + 6\nu_0 \frac{\partial S_2(r)}{\partial r} \ .
\end{equation}
Measuring the second term on the right hand side should therefore correct for this disagreement. Figure \ref{fig:SF_diss_correct} shows the analysis for run \frun{f512a}. The third order structure function is seen never to meet the $4\varepsilon r/5$ line predicted by K41, whereas with the viscous correction agreement is seen up to around $r \sim 20\eta$ before the data and prediction once again diverge. This discrepancy at large scales is due to forcing, and will be discussed in section \ref{subsec:fKHE_SF}.

\subsection{Generalised structure functions}
The structure functions found from numerical simulation are assumed to behave as
\begin{equation}
 S_n(r) \sim r^{\zeta_n} \ ,
\end{equation}
which defines the scaling exponent $\zeta_n$. This can be calculated from DNS data and compared to the Kolmogorov result of $n/3$. If K41 is an asymptotic theory, we would expect to find $\zeta_n \to n/3$ as $Re \to \infty$.

\begin{figure}
 \begin{center}
  \includegraphics[width=0.75\textwidth]{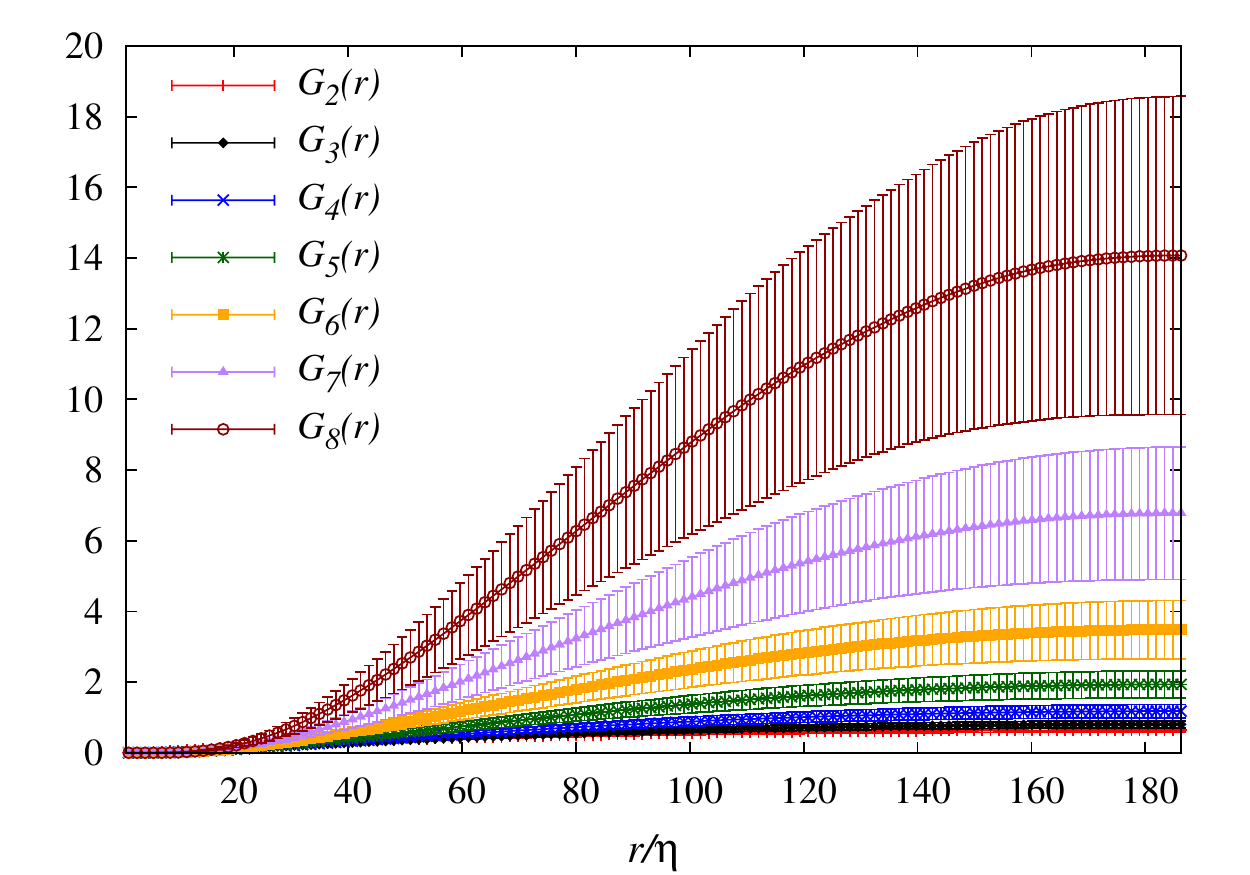}
 \end{center}
 \caption{Generalised structure functions of orders two to eight presented for run \frun{f128e} with $R_\lambda = 64.2$.}
 \label{fig:gen_SF}
\end{figure}

\begin{figure}[tb]
 \begin{center}
  \subfigure[]{
   \includegraphics[width=0.475\textwidth,trim=10px 0 10px 0,clip]{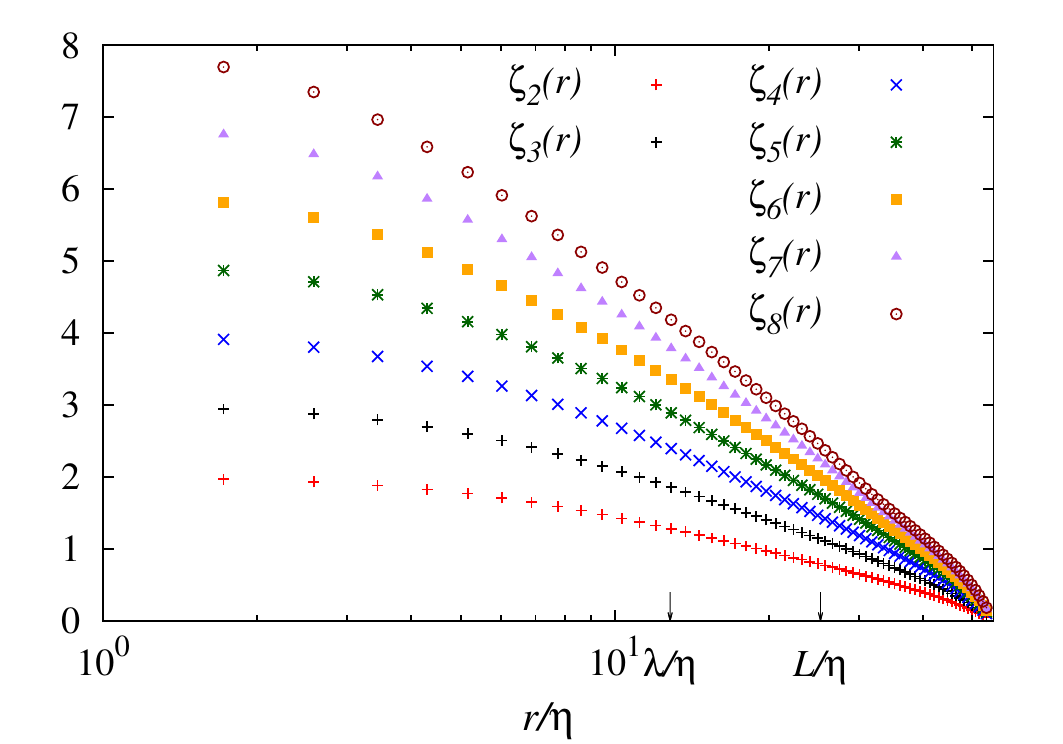}
  }
  \subfigure[]{
   \includegraphics[width=0.475\textwidth,trim=10px 0 10px 0,clip]{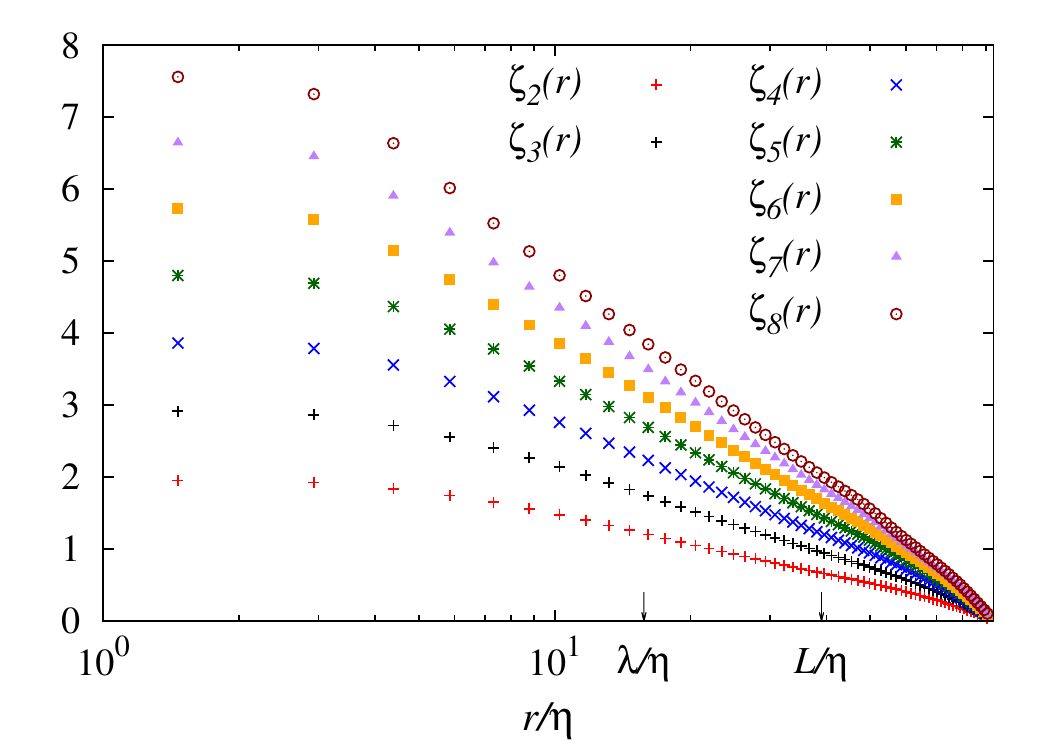}
  }
  \subfigure[]{
   \includegraphics[width=0.475\textwidth,trim=10px 0 10px 0,clip]{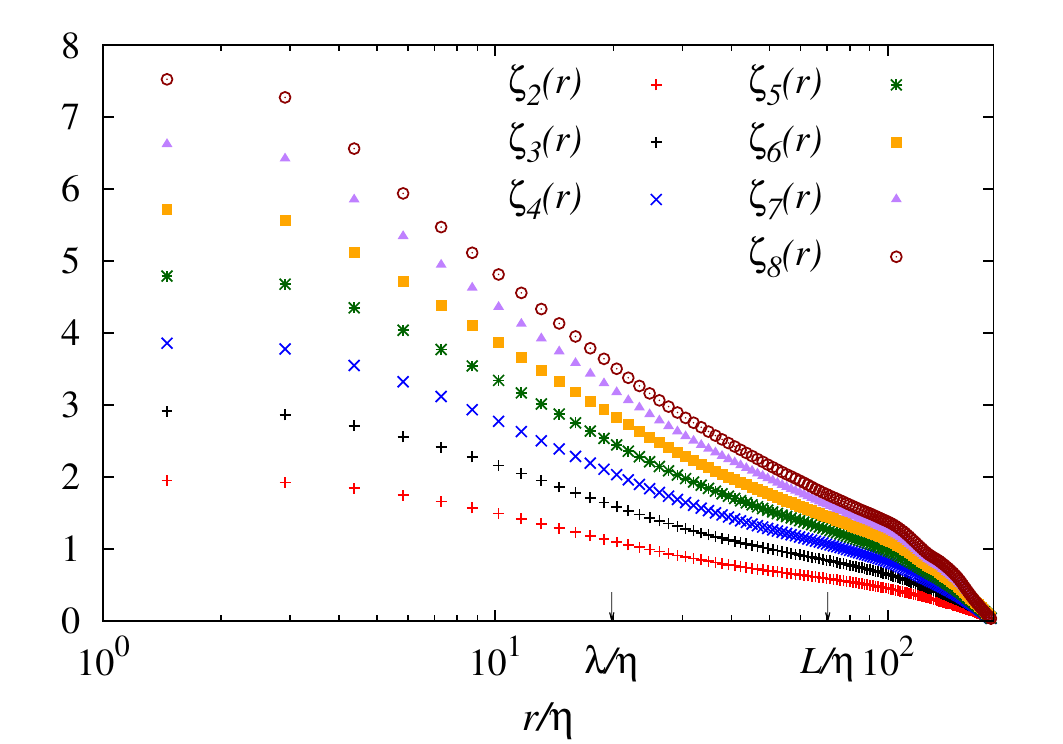}
  }
  \subfigure[]{
   \includegraphics[width=0.475\textwidth,trim=10px 0 10px 0,clip]{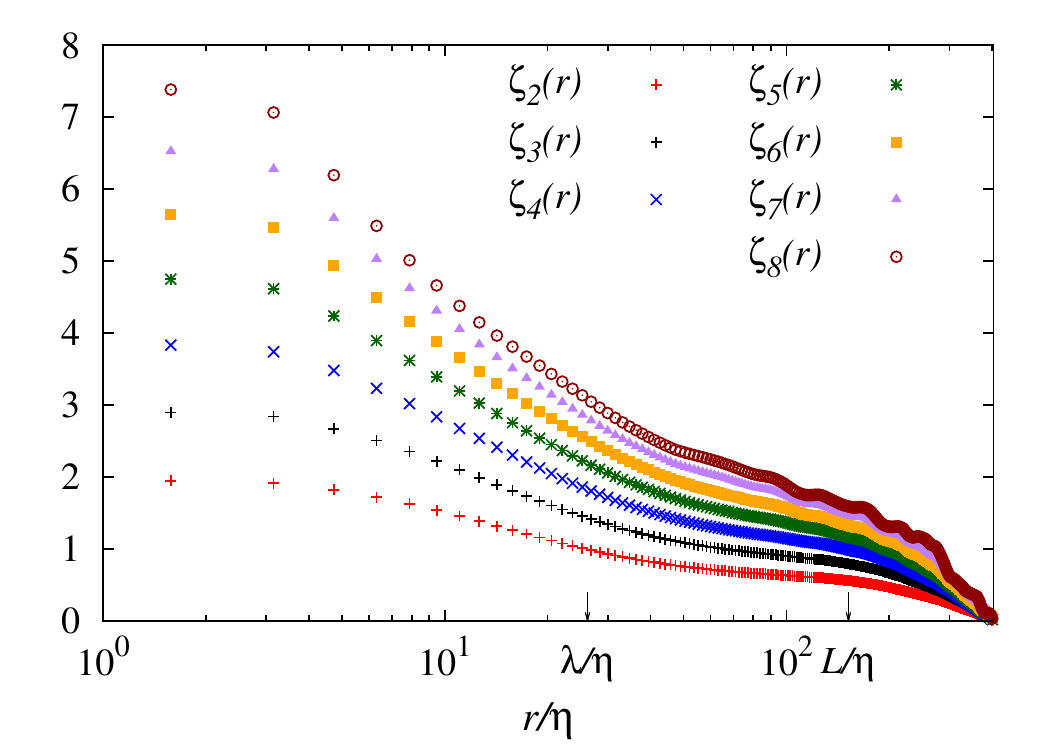}
  }
 \end{center}
 \caption{Local scaling exponents $\zeta_n(r)$ for run (a) \frun{f128a} with $R_\lambda = 42.5$; (b) \frun{f128e} with $R_\lambda = 64.2$; (c) \frun{f256b} with $R_\lambda = 101.3$; and (d) \frun{f512a} with $R_\lambda = 176.9$. Also indicated are the integral and Taylor length scales.}
 \label{fig:local_exp}
\end{figure}

Statistical convergence of even-order structure functions is significantly quicker than odd-order structure functions \cite{Benzi:1993:EPL,Fukayama:1999p904}. This is because the odd-orders involve a delicate balance of positive and negative values in the calculation of the average, making reliable evaluation more difficult. To overcome this, we introduce the \emph{generalised structure functions} \cite{Fukayama:1999p904,Sreenivasan:1993p1514,Stolovitzky:1993p1476,Benzi:1993p861}
\begin{equation}
 G_n(r) = \left\langle \lvert \delta u_L(r) \rvert^n \right\rangle \ ,
\end{equation}
such that the odd-orders can be obtained in a similar way to the even-orders. The generalised structure functions with $n \in \{2, \cdots, 8\}$ are shown in figure \ref{fig:gen_SF}. The error is seen to increase (even when scaled by the mean) as the order is increased.
We expect the scaling exponents will vary smoothly as $n$ is increased and, most importantly, that the scaling will be the same \cite{Benzi:1993:EPL,Benzi:1993p861,Benzi:1995p859}, that is
\begin{equation}
 G_n(r) \sim r^{\zeta'_n} \qquad\text{with}\qquad \zeta'_n = \zeta_n \ .
\end{equation}
The local scaling exponent can then by found by considering
\begin{equation}
 \zeta_n(r) = \zeta'_n(r) = \frac{d \log{G_n(r)}}{d \log{r}} \ ,
\end{equation}
and a region with a $\zeta_n(r) = \textrm{constant}$ plateau used to evaluate the inertial range exponent. This is plotted in figure \ref{fig:local_exp} for $n \in \{2, \cdots, 8\}$. Identifying a scaling region is, even on a semi-log plot, next to impossible, and gets progressively worse as the order of the generalised structure function is increased.

This difficulty was noted by Sreenivasan and Dhruva \cite{Sreenivasan:1998p1569} in experimental data at $R_\lambda = 19,500$ where they found a small scaling region from the local exponent of $S_3(r)$ of about half a decade --- considerably smaller than measured from the spectral density. They comment: ``what can be said of turbulence at much smaller $R_\lambda$? How can one say with any confidence there is scaling in turbulence, let alone determine the exponents with certainty?'' These are valid comments. If we cannot identify a scaling region at low Reynolds number, possibly due to the effects of forcing penetrating into the scaling region, then it is not clear how a measurement of the scaling exponent can be made.

Chen, Dhruva, Kurien, Sreenivasan and Taylor \cite{Chen:2005p911} comment that it is now believed that the structure functions exhibit `anomalous' scaling, in that the deviation from K41 increases non-linearly with the order of the structure function. This can be expressed more simply as $\zeta_{2n} \neq 2\zeta_n$. It should be borne in mind that the higher orders rely more heavily on the tails of the probability distribution. They present DNS data from a $1024^3$ simulation of isotropic turbulence and experimental data from atmospheric boundary layers. The anisotropy of the system is removed using a projection of the structure function in question against members of its $SO(3)$ group decomposition \cite{Arad:1998p1578,Arad:1999p1579}. This is essentially the same as performing an angle average, as developed by Taylor \etal\ \cite{Taylor:2003p1581}. This allowed Chen \etal\ to obtain plateaus for the local exponent of the structure functions for non-integer orders between -1 and 2 and measure the exponents directly. Note that their ensemble was 10 samples from one large eddy turnover time.

\subsection{Extended self-similarity}\label{subsec:ESS}
\begin{figure}[tb]
 \begin{center}
  \subfigure[]{
   \includegraphics[width=0.475\textwidth,trim=10px 0 10px 0,clip]{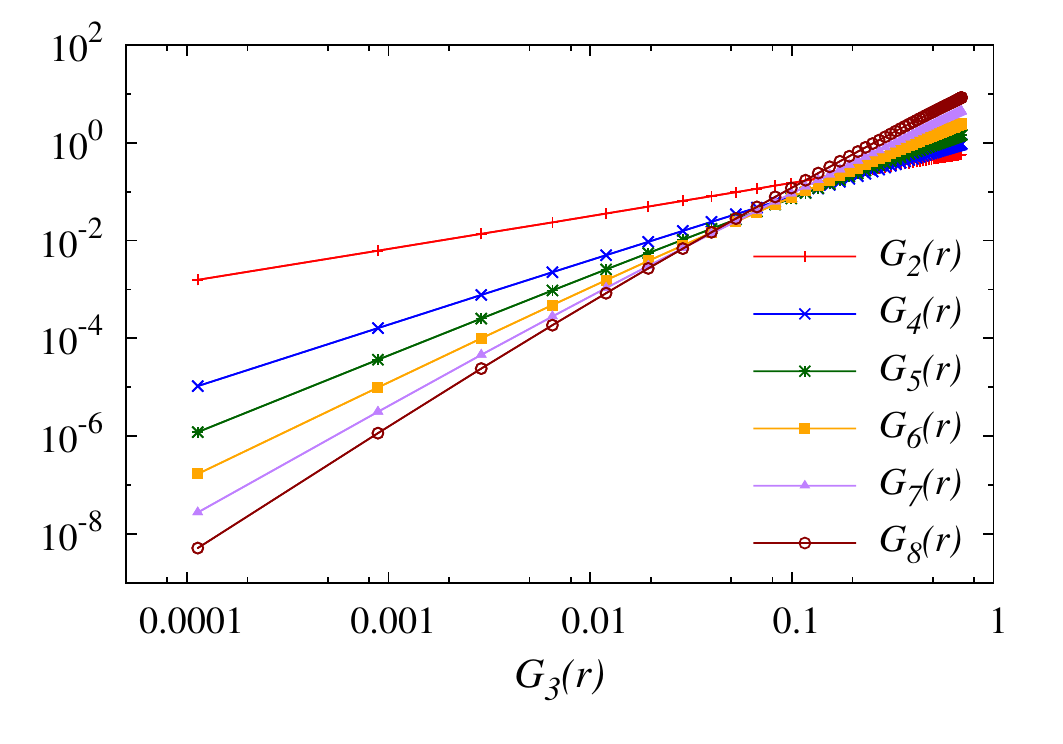}
  }
  \subfigure[]{
   \includegraphics[width=0.475\textwidth,trim=10px 0 10px 0,clip]{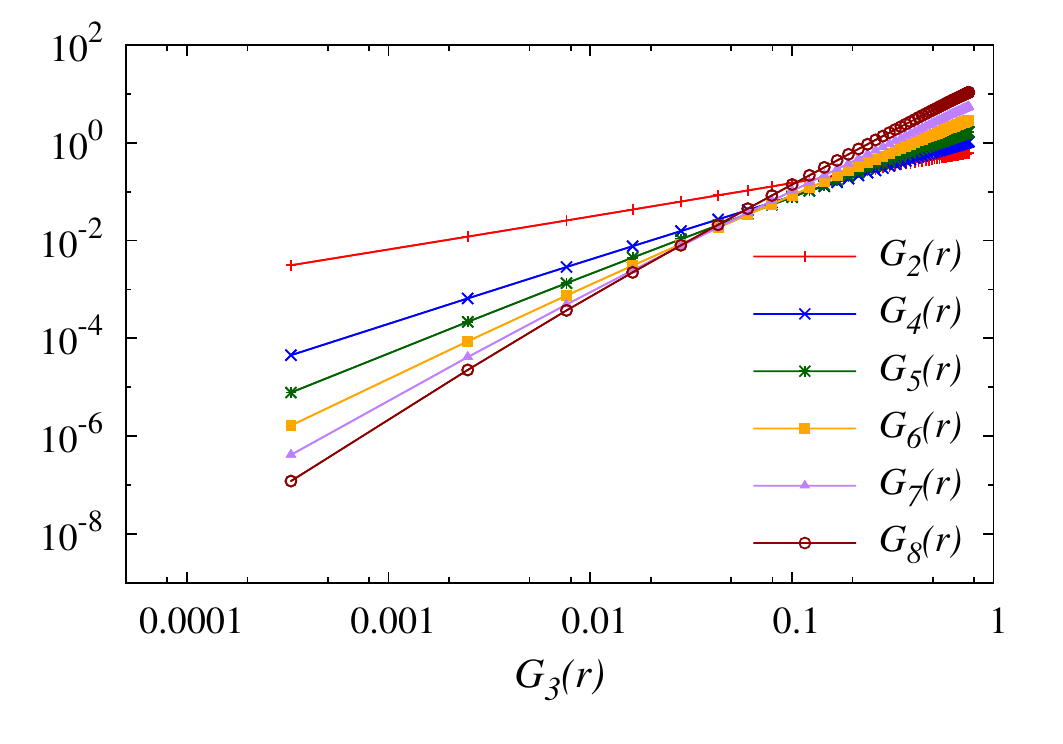}
  }
  \subfigure[]{
   \includegraphics[width=0.475\textwidth,trim=10px 0 10px 0,clip]{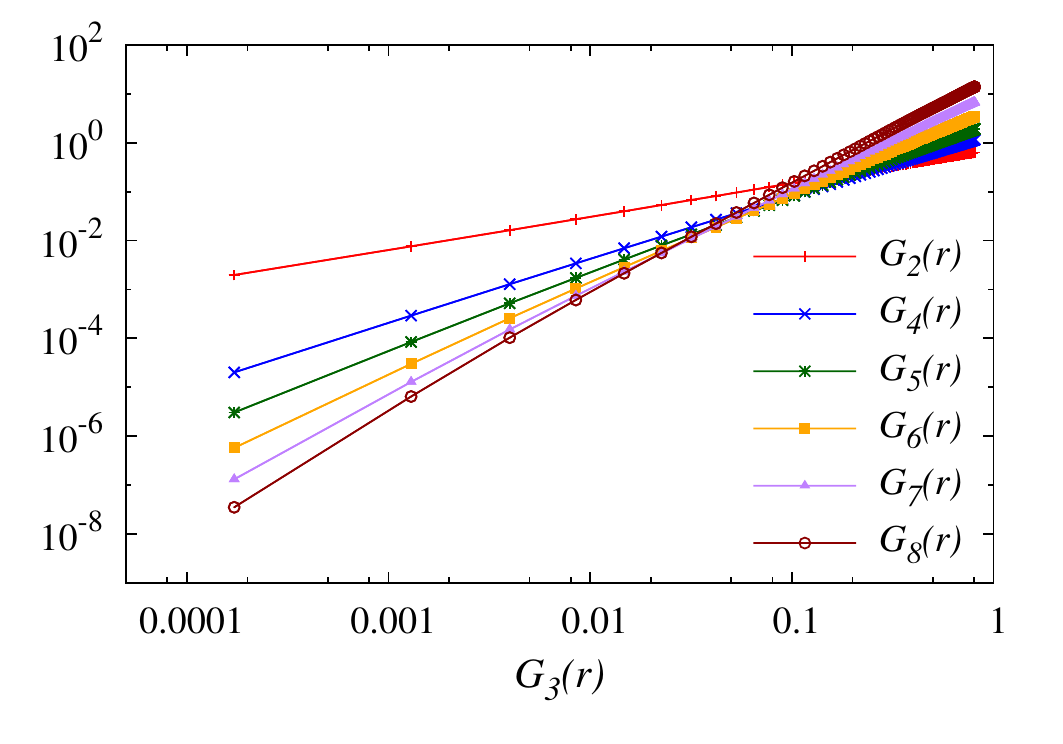}
  }
  \subfigure[]{
   \includegraphics[width=0.475\textwidth,trim=10px 0 10px 0,clip]{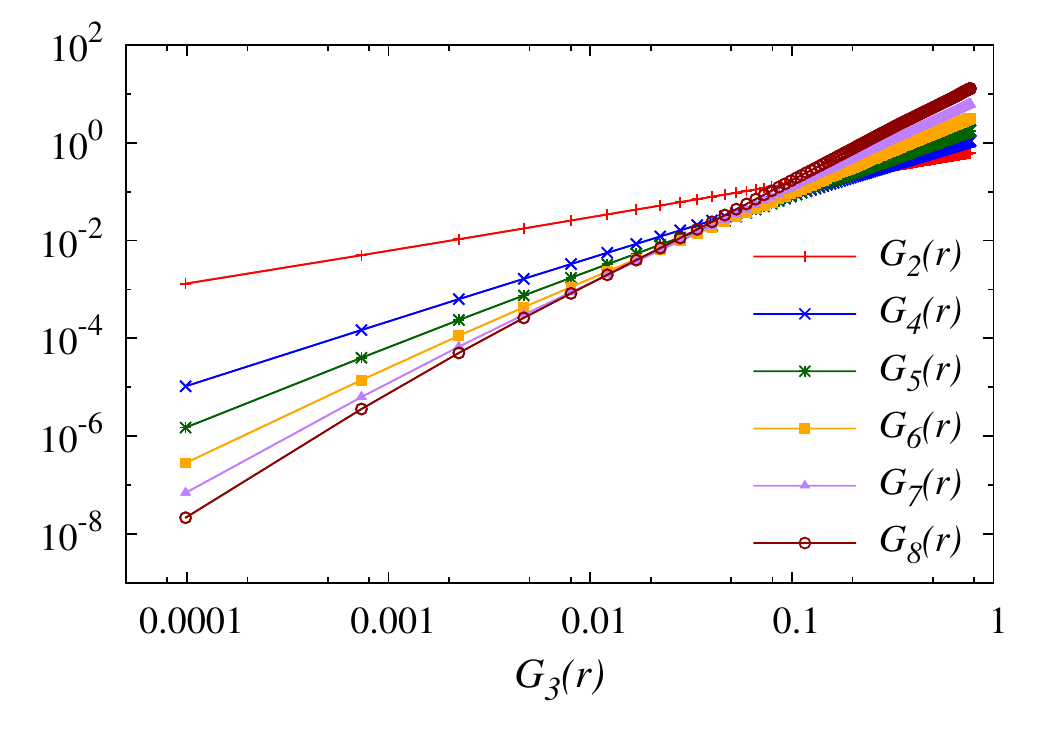}
  }
 \end{center}
 \caption{Generalised structure functions plotted against $G_3(r)$ to absorb any variation from K41. Plotted for run (a) \frun{f128a} with $R_\lambda = 42.5$; (b) \frun{f128e} with $R_\lambda = 64.2$; (c) \frun{f256b} with $R_\lambda = 101.3$; and (d) \frun{f512a} with $R_\lambda = 176.9$.}
 \label{fig:ESS_plots}
\end{figure}

To overcome this difficulty of identifying a plateau for the local scaling exponent $\zeta_n(r)$ for the generalised structure functions at low Reynolds numbers, we introduce the idea of extended self-similarity (ESS) \cite{Benzi:1993p861,Santangelo:1994p245}. This has been used to study the scaling exponents of the structure functions in a variety of experimental and numerical configurations \cite{Benzi:1995p859,Camussi:1996p1428,Camussi:1996p1458,Camussi:1997p1475,Gotoh:2002p627}. First, we assume that the Kolmogorov form for the third-order structure function is correct in the infinite Reynolds number limit, such that $\zeta_3 = 1$. We then speculate that any measured difference of $\zeta_3$ from unity can be used to compensate for the differences measured in the other scaling exponents. Essentially, we instead consider $S_n(r)$ or $G_n(r)$ to be a function of $\lvert S_3(r)\rvert$ or $G_3(r)$, respectively:
\begin{equation}
 S_n(r) \sim r^{\zeta_n} \sim \lvert S_3(r)\rvert^{\zeta^*_n} \qquad\text{or}\qquad G_n(r) \sim r^{\zeta'_n} \sim \big[ G_3(r) \big]^{\Sigma_n} \ ,
\end{equation}
with new scaling exponents $\zeta^*_n$ and $\Sigma_n$. Figure \ref{fig:ESS_plots} shows the generalised structure functions plotted instead against $G_3(r)$. Plotted on a log-log plot in this way, it can be seen that the lines appear very straight over an extended region, implying a constant exponent. If we assume that $\zeta'_3 = \zeta_3 = 1$, then we have
\begin{equation}
 \Sigma_n = \frac{\zeta'_n}{\zeta'_3} = \zeta'_n = \zeta_n \ ,
\end{equation}
and measurement of ESS exponent $\Sigma_n$ becomes a measurement of the actual exponent, $\zeta_n$.

\begin{figure}[tb]
 \begin{center}
  \subfigure[]{
   \includegraphics[width=0.475\textwidth,trim=10px 0 10px 0,clip]{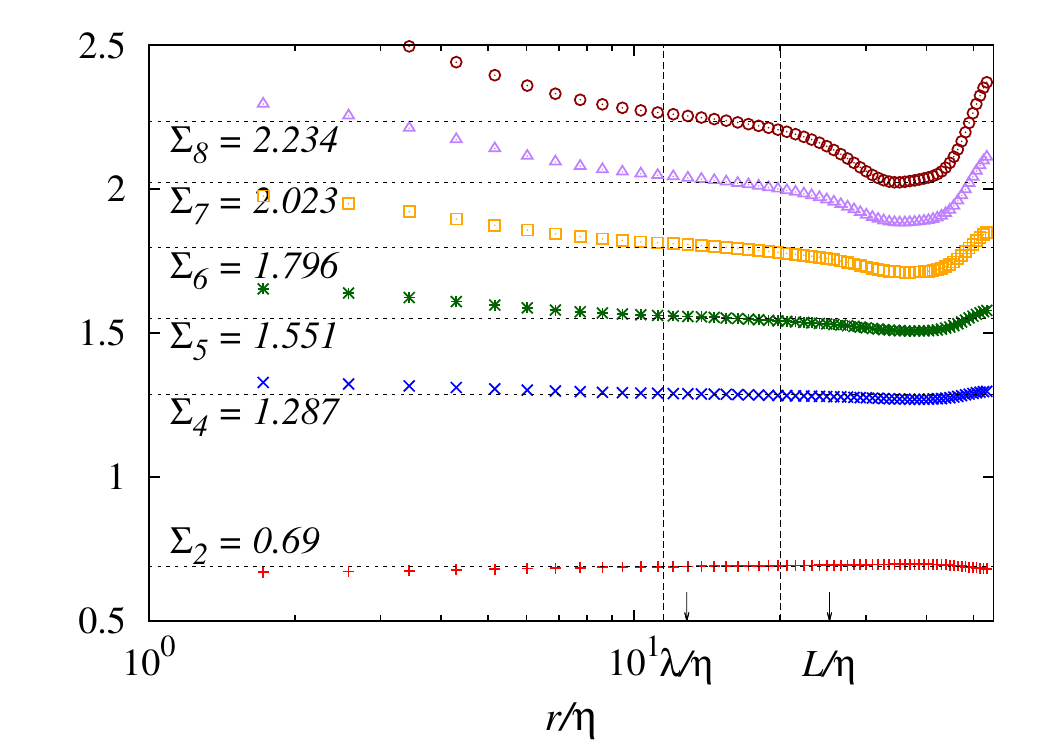}
  }
  \subfigure[]{
   \includegraphics[width=0.475\textwidth,trim=10px 0 10px 0,clip]{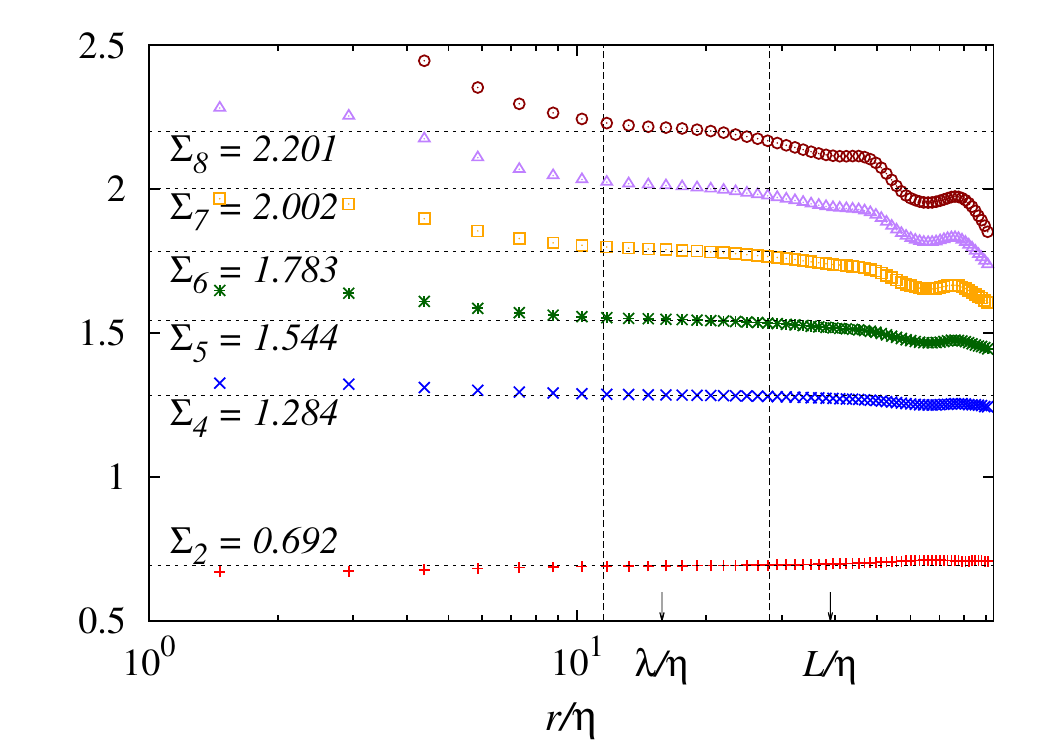}
  }
  \subfigure[]{
   \includegraphics[width=0.475\textwidth,trim=10px 0 10px 0,clip]{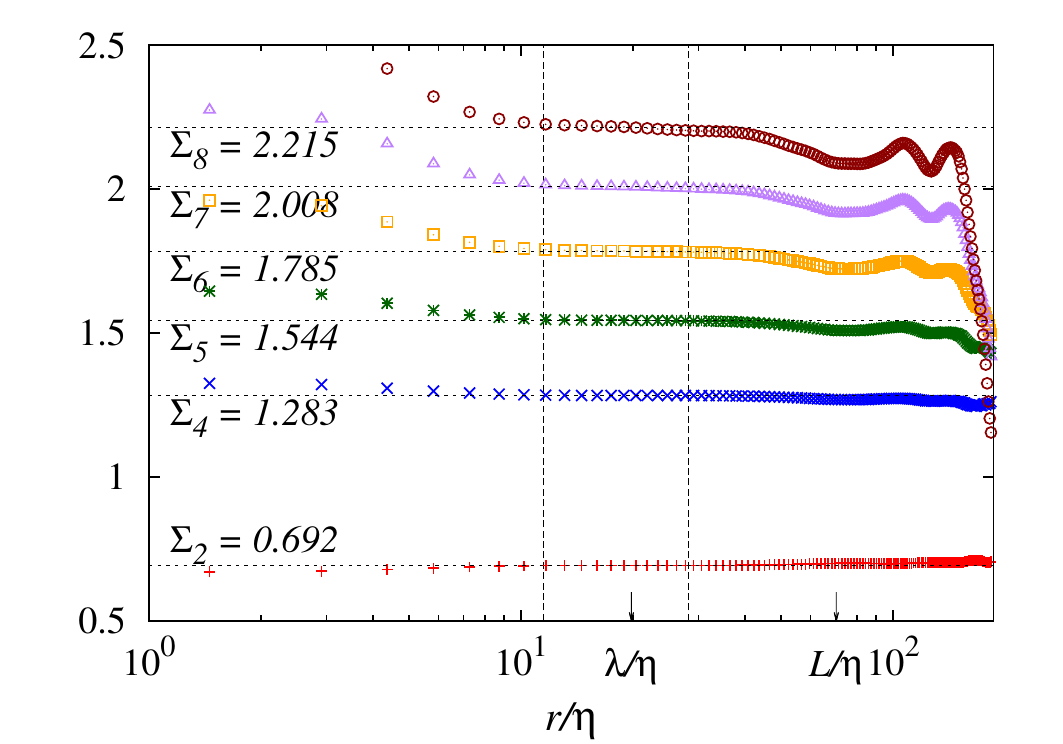}
  }
  \subfigure[]{
   \includegraphics[width=0.475\textwidth,trim=10px 0 10px 0,clip]{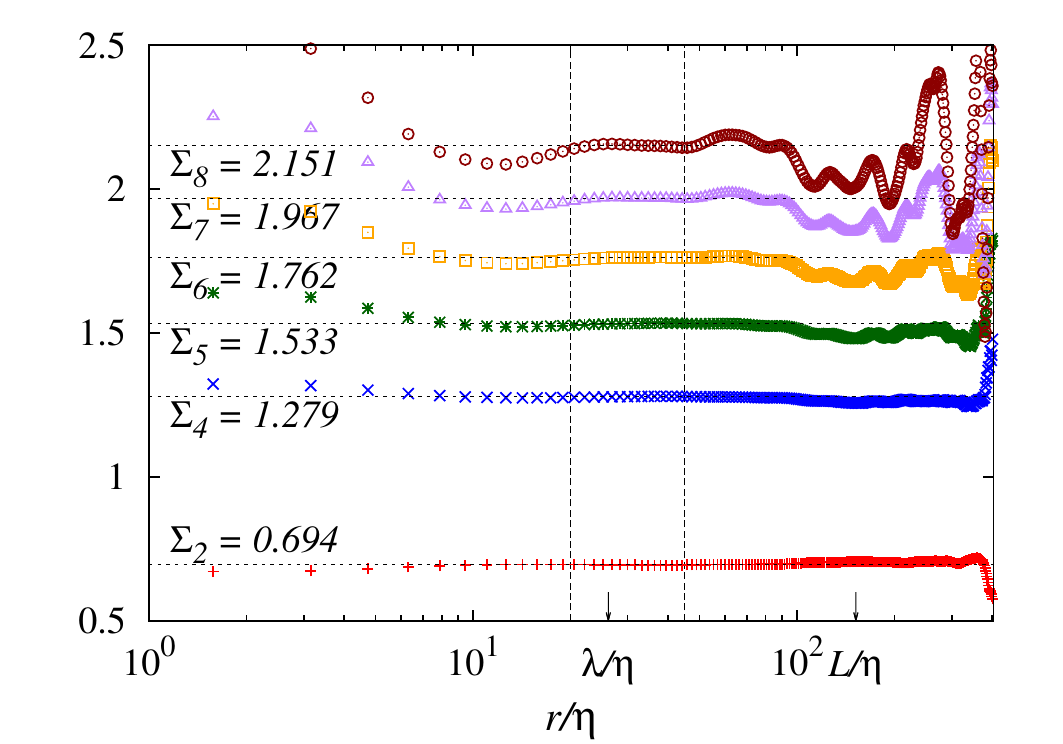}
  }
 \end{center}
 \caption{Local ESS scaling exponents $\Sigma_n(r)$ for run (a) \frun{f128a} with $R_\lambda = 42.5$; (b) \frun{f128e} with $R_\lambda = 64.2$; (c) \frun{f256b} with $R_\lambda = 101.3$; and (d) \frun{f512a} with $R_\lambda = 176.9$. Also indicated are the integral and Taylor length scales along with the fit range.}
 \label{fig:local_ESS_exp}
\end{figure}

The local ESS exponent, found as
\begin{equation}
 \Sigma_n(r) = \frac{d \log{G_n(r)}}{d \log{G_3(r)}} \ ,
\end{equation}
has been measured and is presented in figure \ref{fig:local_ESS_exp} for several orders. Notice that identifying a scaling region has become significantly easier, although the slower convergence of the higher-order generalised structure functions can still be seen. The figure also shows the range over which the values for the plateau have been calculated, and the location of the integral and Taylor microscales. Note how the Taylor microscale sits within the scaling region, whereas the Kolmogorov microscale and the integral scale sit to the left and right of the region, respectively.

\begin{table}[tb!]
 \begin{center}
  \begin{tabular}{l|llllll|ll|l}
   $R_\lambda$ & $\Sigma_2$ & $\Sigma_4$ & $\Sigma_5$ & $\Sigma_6$ & $\Sigma_7$ & $\Sigma_8$ & run ID & \# & Ref. \\
   \hline\hline
   K41& 0.667 & 1.333 & 1.667 & 2.000 & 2.333 & 2.667 & \multicolumn{2}{c|}{Theory} \\
   \hline
   42.5 & 0.690 & 1.287 & 1.551 & 1.796 & 2.023 & 2.234 & \frun{f128a} & 101 \\
   64.2 & 0.692 & 1.284 & 1.544 & 1.783 & 2.002 & 2.201 & \frun{f128e} & 101 \\
   101.3 & 0.692 & 1.283 & 1.544 & 1.785 & 2.008 & 2.215 & \frun{f256b} & 101 \\
   176.9 & 0.694 & 1.279 & 1.533 & 1.762 & 1.967 & 2.150 & \frun{f512a} & 15\\
   \hline
   70 & 0.690 & 1.288 & 1.555 & 1.804 & 2.037 & 2.254 & Run 4 & 126 & \multirow{2}{*}{\cite{Fukayama:1999p904}} \\
   125 & 0.692 & 1.284 & 1.546 & 1.788 & 2.011 & 2.217 & Run 5 & 45  \\
   \hline
   381 & 0.709 & 1.30 & 1.56 & 1.79 & 1.99 & 2.18 & & & \multirow{2}{*}{\cite{Gotoh:2002p627}} \\
   460 & 0.701 & 1.29 & 1.54 & 1.77 & 1.98 & 2.17 & & & \\
   \hline
    & 0.70 & 1.28 & 1.54 & 1.78 & 2.00 & 2.23 & \multicolumn{2}{c|}{Experiment} & \cite{Benzi:1995p859} \\
    & 0.696 & 1.279 & 1.538 & 1.778 & 2.001 & 2.211 & \multicolumn{2}{c|}{Theory} & \cite{She:1994p601}
  \end{tabular}
 \end{center}
 \caption[Measurement of the scaling exponents using extended self-similarity.]{Measurement of the scaling exponents from our DNS data using ESS. Presented with the K41 predicted values and results from Fukayama \etal\ \cite{Fukayama:1999p904}, Gotoh \etal\ \cite{Gotoh:2002p627} and Benzi \etal\ \cite{Benzi:1995p859}. The theoretical values predicted by She and L\'ev\^eque \cite{She:1994p601} are also provided.}
 \label{tbl:ESS_exponents}
\end{table}

Values of the ESS exponents computed from our DNS data are summarised in table \ref{tbl:ESS_exponents}. We also print those obtained by Fukayama \etal\ \cite{Fukayama:1999p904} (ESS), Gotoh \etal\ \cite{Gotoh:2002p627} (non-ESS) and Benzi, Ciliberto, Baudet and Chavarria \cite{Benzi:1995p859} (ESS) for comparison. The scaling exponents predicted by the theory of She and L\'ev\^eque \cite{She:1994p601} are also provided. Further values, and Reynolds number dependence, may be found in \cite{Belin96,Santangelo:1994p245}. As Reynolds number is increased, the measured values of the ESS exponent actually move away from the K41 result. This appears to contradict the assumption of K41 as an asymptotic theory. Note that the ESS exponent is not a constant for all $r$. Instead, the values appear to tend towards $n/3$ as $r \to 0$, from above for $n = 2$ and from below for $n \geq 3$. This is because the structure functions scale as $n$ in the limit $r \to 0$. Since the ESS exponent is $\zeta_n/\zeta_3$, we expect it to tend towards $n/3$. This is \emph{not} an indication that we are developing K41 scaling!

The use of generalised and standard structure functions with ESS is discussed in Grossmann, Lohse and Reeh \cite{Grossmann:1997p1566}. They show that without the use of generalised structure functions there is no ESS, and go on to study the scaling of the transverse structure functions. It is also unclear \emph{why} ESS improves the scaling, as mentioned in \cite{Sreenivasan:1998p1569,Meneveau:1996p1593}, where it is also noted that it is possible that the measured ESS exponents $\Sigma_n$ differ from those of the structure functions, $\zeta_n$, or even $\zeta^*_n$ measured using $\lvert S_3(r) \rvert$. We feel this is an important point.

\subsection{Comments on finite Reynolds number effects}
From the Reynolds number variation of the values of the ESS exponents presented in table \ref{tbl:ESS_exponents}, one could conclude that K41 is in fact not an asymptotic theory and that intermittency corrections are indeed needed. However, it should be borne in mind that the ensemble used must be large enough that the PDF be sufficiently isotropic for the generalised structure function of order $n$ to be reliably found. The failing of these values to approach K41 as Reynolds number is increased could be due to the ensemble size for the required degree of isotropy increasing and not being satisfied. Due to the memory and time required to store and process the larger lattice sizes needed for higher Reynolds number simulations, this becomes unfeasible. It is also not guaranteed that the ESS exponents are equivalent to scaling exponents of the structure functions.

Another possibility is finite Reynolds number effects once again, since we are still at relatively low Reynolds number. Indeed, figures \ref{fig:realspace_SF} and \ref{fig:SF_diss_correct} for the structure functions do not show any region where the K41 result for the third-order structure function is exactly obeyed. As discussed by George \cite{George:1992p673}, the scaling laws require that the low wavenumbers contribute nothing to dissipation and high wavenumbers contain no energy, neither of which are true at finite Reynolds numbers. That author notes that ``Kolmogorov's theory is at best an approximation for turbulence at finite Reynolds number.''

Qian \cite{Qian:1998p889,Qian:2000p854} developed a non-Gaussian model of the PDF for the velocity increment based on high-Reynolds number experimental data. This was used to study the standard and ESS scaling exponents for the structure functions. The model predicted anomalous scaling at finite Reynolds number but, unlike other models, found that as Reynolds number was increased the exponents approach their K41 values. This implies that deviation from K41 scaling is purely a finite Reynolds number effect.

\section{Exploiting the pseudospectral technique}\label{sec:exploit_PS}
The examination of correlation and structure functions has so far been restricted to real space, where a large ensemble of independent realisations of the velocity field is required to satisfy isotropy and obtain reliable results. As the Reynolds number is increased and we are forced to use larger lattice sizes to resolve the necessary scales, the size of a realisation increases as $N^3$. As such, finding statistics from them involves an increasing amount of computational effort. However, it is possible to access this information from our Fourier spectra.

Consider the two-point, single-time isotropic correlation tensor expressed in terms of its Fourier transform. This can be written as: 
\begin{align}
 C_{\alpha\beta}(\vec{r}) &= \langle u_\alpha(\vec{x},t) u_\beta(\vec{x}+\vec{r},t) \rangle \nonumber \\
 &= \int d^3k\ C_{\alpha\beta}(\vec{k})\ e^{i \vec{k}\cdot\vec{r}} \ .
\end{align}
By using a choice of suitable spherical polar coordinates where $\xi$ is the cosine of the angle between $\vec{k}$ and $\vec{r}$, we can perform the integral over the polar angle (to obtain a factor of $2\pi$) and the remaining azimuthal angular integral to find
\begin{align}
 C_{\alpha\beta}(\vec{r}) &= 2\pi\int dk\ k^2 \int_{-1}^1 d\xi\ P_{\alpha\beta}(\vec{k})\ C(k)\ e^{i kr\xi} \\
 &= 2\pi\int dk\ k^2\ P_{\alpha\beta}(\vec{k})\ C(k)\ \left[ \frac{e^{i kr}}{ikr} - \frac{e^{-i kr}}{ikr} \right] \\
 &= \int dk\ 4\pi k^2 C(k)\ P_{\alpha\beta}(\vec{k})\ \frac{\sin kr}{kr} \\
 &= \int dk\ E(k)\ P_{\alpha\beta}(\vec{k})\ \frac{\sin kr}{kr} \ .
\end{align}
Therefore, we can write the isotropic correlation function
\begin{align}
 \label{eq:R_Ek}
 C(r) = \tfrac{1}{2} C_{\alpha\alpha}(\vec{r}) &= \int dk\ E(k)\ \frac{\sin kr}{kr} \ .
\end{align}
This result was introduced in equation \eqref{eq:second_corr} and can be found discussed in \cite{Batchelor:1953-book,MoninYaglom:vol2,davidson:2004-book,Bos:2011p797} along with the (slightly more involved) derivation of the third-order result
\begin{align}
 \label{eq:CLLL_Tk}
 \frac{1}{2} \left( 3 + r\frac{\partial}{\partial r} \right) \left( \frac{\partial}{\partial r} + \frac{4}{r} \right) C_{LL,L}(r) = \int dk\ T(k)\ \frac{\sin kr}{kr} \ .
\end{align}
The (longitudinal) correlation functions were introduced in section \ref{subsec:long_trans_corr}. From these relationships, it is possible to find forms for the second- and third-order longitudinal correlation functions explicitly by performing integrals over $r$ analytically,
\begin{align}
 C_{LL}(r) &= \frac{2}{r^3} \int_0^r d\xi\ \xi^2\ C(\xi) \nonumber \\
 &= \frac{2}{r^3} \int dk\ \frac{E(k)}{k} \int_0^r d\xi\ \xi \sin{k\xi} \nonumber \\
 \label{eq:CLL_R}
 &= 2 \int dk\ E(k)\ \left[ \frac{\sin{kr} - kr\cos{kr}}{(kr)^3} \right]
\end{align}
and
\begin{align}
  C_{LL,L}(r) = 2r \int dk\ T(k)\ \left[ \frac{3\sin kr - 3kr\cos kr - (kr)^2\sin kr}{(kr)^5} \right] \ ,
\end{align}
respectively. Thus, given the shell-averaged spectra obtained in a pseudospectral simulation, one can in fact calculate \emph{averaged} real-space quantities. Note that we have the limits
\begin{align}
 \label{eq:spectral_limits}
 \lim_{r \to 0} \left( \frac{\sin{kr} - kr\cos{kr}}{(kr)^3} \right) &= \frac{1}{3} \ , \qquad\textrm{and} \\
 \label{eq:spectral_limits2}
 \lim_{r \to 0}\ r \left[ \frac{3\sin kr - 3kr\cos kr - (kr)^2\sin kr}{(kr)^5} \right] &= 0 \ ,
\end{align}
as these will be of use later.

It should be noted that this is also useful for the study of statistical closures based in $k$-space. One finds the evolution of the spectral density $\langle u(k) u(-k) \rangle$ and hence the energy spectrum. The above relationships allow one to consider real-space quantities using these Fourier-based methods.

\subsection{Structure functions}
The expressions above for the longitudinal correlation functions allow us to find spectral expressions for the structure functions of second- and third order. Using equations \eqref{eq:sf2_corr} and \eqref{eq:sf3_corr} for the structure functions in terms of the longitudinal correlation functions, we find (see, for example, Bos \etal\ \cite{Bos:2011p797})
\begin{align}
 S_2(r) &= 4 \int dk\ E(k)\ \left[ \frac{1}{3} - \frac{\sin{kr} - kr\cos{kr}}{(kr)^3} \right] \nonumber \\
 S_3(r) &= 12r \int dk\ T(k)\ \left[ \frac{3\sin kr - 3kr\cos kr - (kr)^2\sin kr}{(kr)^5} \right] \ .
\end{align}

These structure functions have been found in this manner from spectral DNS data for the forced simulations we have performed. Since it requires considerably less effort to process $N/3$ modes (since we are using the $2/3$ rule for velocity field truncation) of a realisation compared to its $N^3$ data points, this approach can be used very quickly on significantly larger lattices. The spectra can also be ensemble-averaged before being used to create the structure functions, as will be the case in the remainder of this section.

\begin{figure}[tbp!]
 \begin{center}
  \subfigure[Run \frun{f256b} with $R_\lambda = 101.3$]{
   \label{sfig:psr_sf_f256b}
   \includegraphics[width=0.75\textwidth]{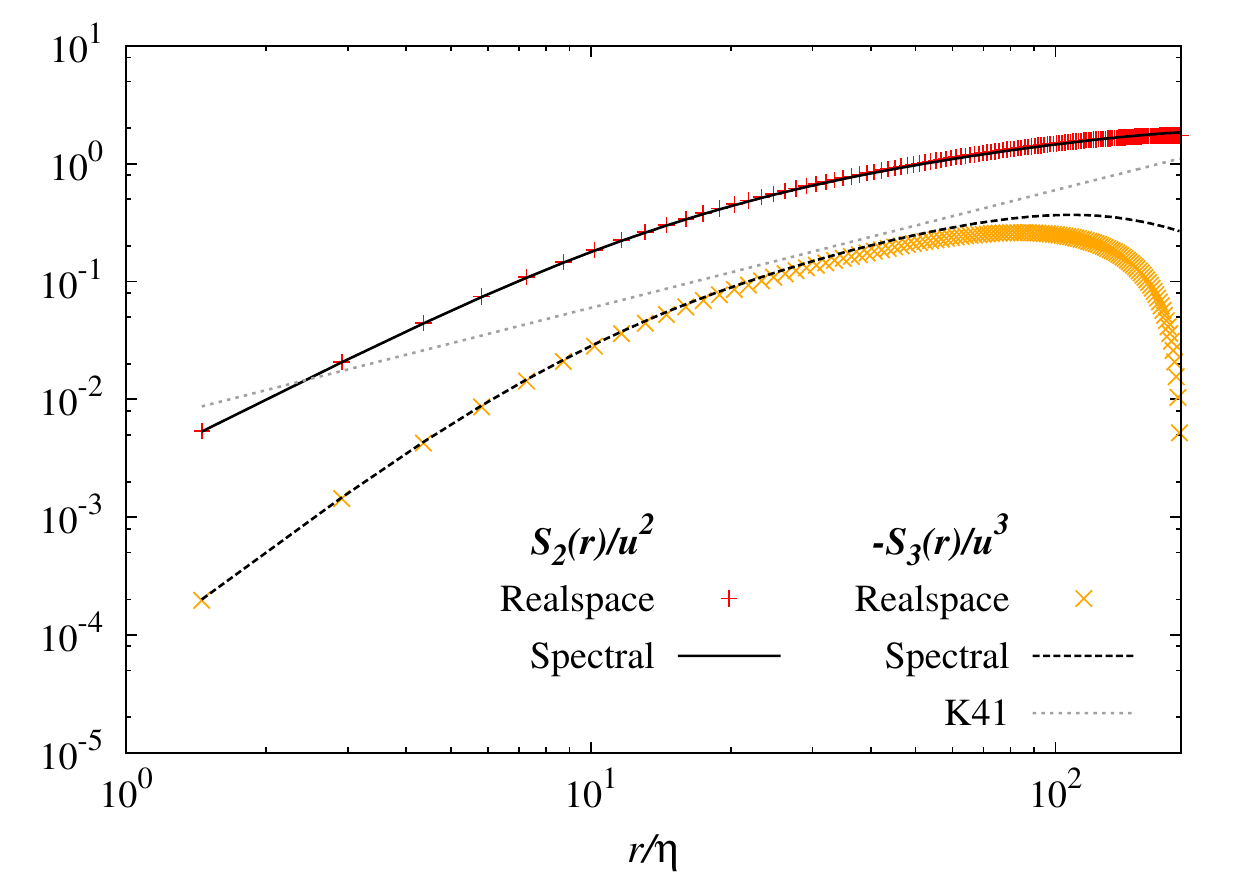}
  }
  \subfigure[Run \frun{f512a} with $R_\lambda = 176.9$]{
 \label{sfig:psr_sf_f512a}
   \includegraphics[width=0.75\textwidth]{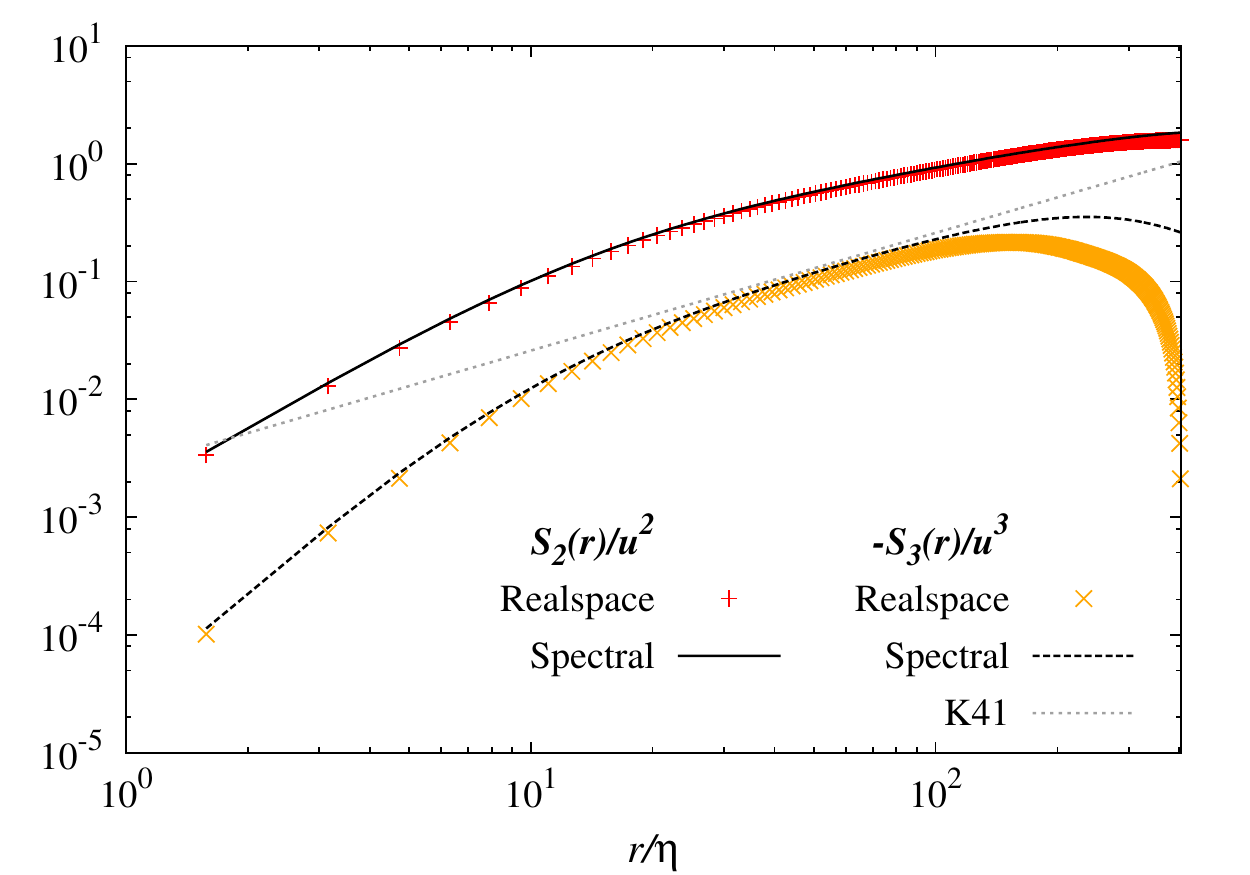}
  }
 \end{center}
 \caption{Comparison of structure functions calculated by traditional methods in real space with the spectral technique.}
 \label{fig:psr_sf}
\end{figure}

A comparison of the structure functions calculated from real-space correlations and using this spectral technique is given in figure \ref{fig:psr_sf} for runs \frun{f256b} and \frun{f512a} with a Reynolds number of $R_\lambda = 101.3$ and 176.9, respectively. Note that the agreement for $S_2(r)$ is very good for all $r$, whereas the two curves for $S_3(r)$ diverge as $r$ increases. This is most likely due to the periodicity of the structure functions: The real-space calculation is required to be an odd function of period $L = 2\pi$ and as such must go to zero at $r = \pi$. This is not necessarily the case for the transform of the shell- (and possibly ensemble) averaged spectra. On the other hand, $S_2(r)$ is an even function and can be seen to roll off slightly from the spectral form close to $r = \pi$, since it must have zero gradient at $r = 0$ and return to zero at $r = 2\pi$. Another possibility is the real-space ensemble requiring more realisations to ensure isotropy of the large scales than were available

\begin{figure}[tb]
 \begin{center}
  \includegraphics[width=0.75\textwidth]{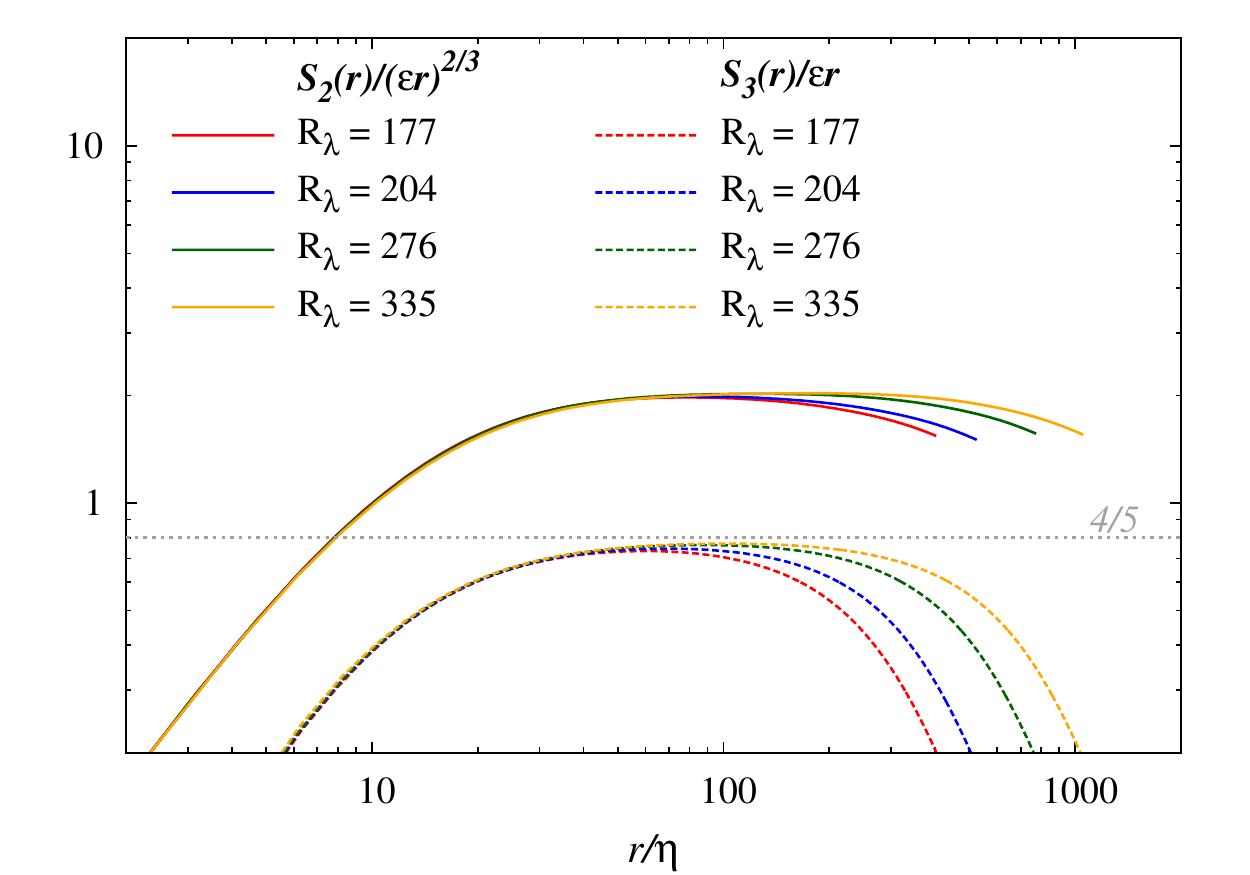}
 \end{center}
 \caption{Comparison of (scaled) second- and third-order structure functions calculated from energy and transfer spectra for our higher Reynolds number simulations, runs \frun{f512a,b} and \frun{f1024a,b}.}
 \label{fig:ps_sf_comp}
\end{figure}

Dimensionless structure functions calculated using the spectral method are presented in figure \ref{fig:ps_sf_comp}. This figure should be directly compared to the review by Ishihara, Gotoh and Kaneda \cite{Ishihara:2009p165} which presents dimensionless structure functions for the high-Reynolds number runs performed on the Earth Simulator. The agreement is very good and suggests that this method for calculating the structure functions is recovering the details of the real-space correlations.
\begin{figure}[tb]
 \begin{center}
  \includegraphics[width=0.75\textwidth]{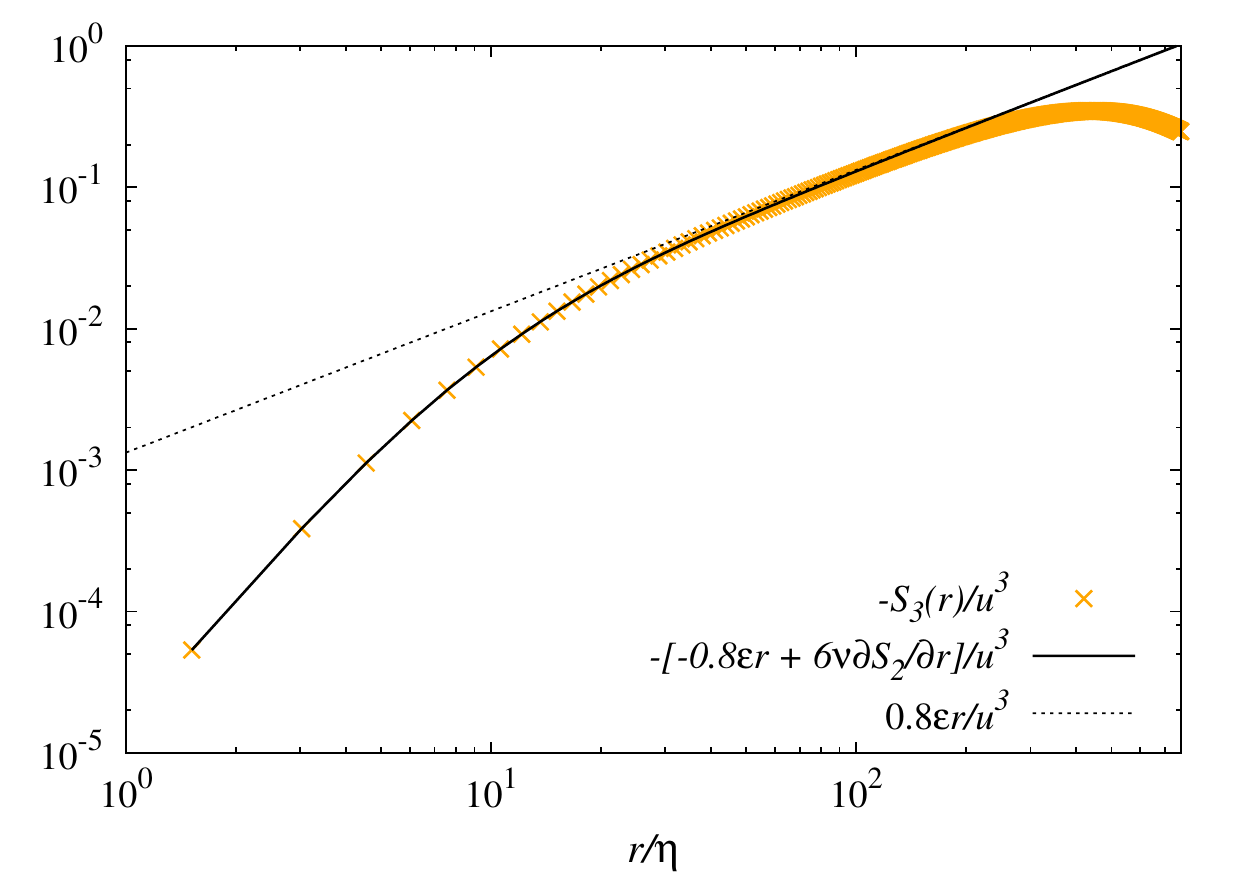}
 \end{center}
 \caption{Viscous correction to the K41 form for the third-order structure function, calculated from energy and transfer spectra for run \frun{f1024a} with $R_\lambda = 276.2$.}
 \label{ps_sf_1024a}
\end{figure}

We can also find the form of the viscous correction to the third-order structure function used by Fukayama \etal\ \cite{Fukayama:1999p904}, which was analysed in section \ref{sec:sf}. We rewrite equation \eqref{eq:KHE_visc_correct} as
\begin{equation}
 S_3(r) = -\frac{4}{5}\varepsilon r + 6\nu_0 \frac{\partial S_2(r)}{\partial r} \ ,
\end{equation}
where we evaluate using
\begin{align}
 \label{eq:visc_term_spectral}
 6\nu_0 \frac{\partial S_2(r)}{\partial r} &= 24\nu_0 \frac{\partial}{\partial r} \int dk\ E(k)\ \left[ \frac{1}{3} - \frac{\sin{kr} - kr\cos{kr}}{(kr)^3} \right] \nonumber \\
 &= 12r \int dk\ 2\nu_0 k^2\ E(k) \left[ \frac{3\sin{kr} - 3kr\cos{kr} - (kr)^2 \sin{kr}}{(kr)^5} \right] \ .
\end{align}
The derivative with respect to $r$ was performed analytically. Thus we can reproduce figure \ref{fig:SF_diss_correct} and show that the deviation of $S_3(r)$ from the Kolmogorov form at low $r$ is due to these length-scales being in the dissipation range, as shown in figure \ref{ps_sf_1024a}. The data is now expected to follow the solid curve, rather than the K41 dotted line. This is a much better match for scales up to $r \sim 20\eta$, but the larger scales are still not described.

\subsection{Scaling exponents}
Unfortunately, the ESS exponents cannot be obtained using this spectral technique as we only have access to $S_2(r)$ and $S_3(r)$. Expressions for the higher, even-order structure functions as a weighted integral over the spectra could potentially be derived but have not been done here. It is the opinion of the present writer that this would be an interesting study, since isotropy appears to be better satisfied by the ensemble-averaged spectra than the real-space correlation measurements for the same ensemble. Note that odd-order generalised structure functions present a problem, since the magnitude is taken before the average and it is unclear how this could be reproduced in Fourier space.

\begin{figure}[tbp!]
 \begin{center}
  \includegraphics[width=0.75\textwidth]{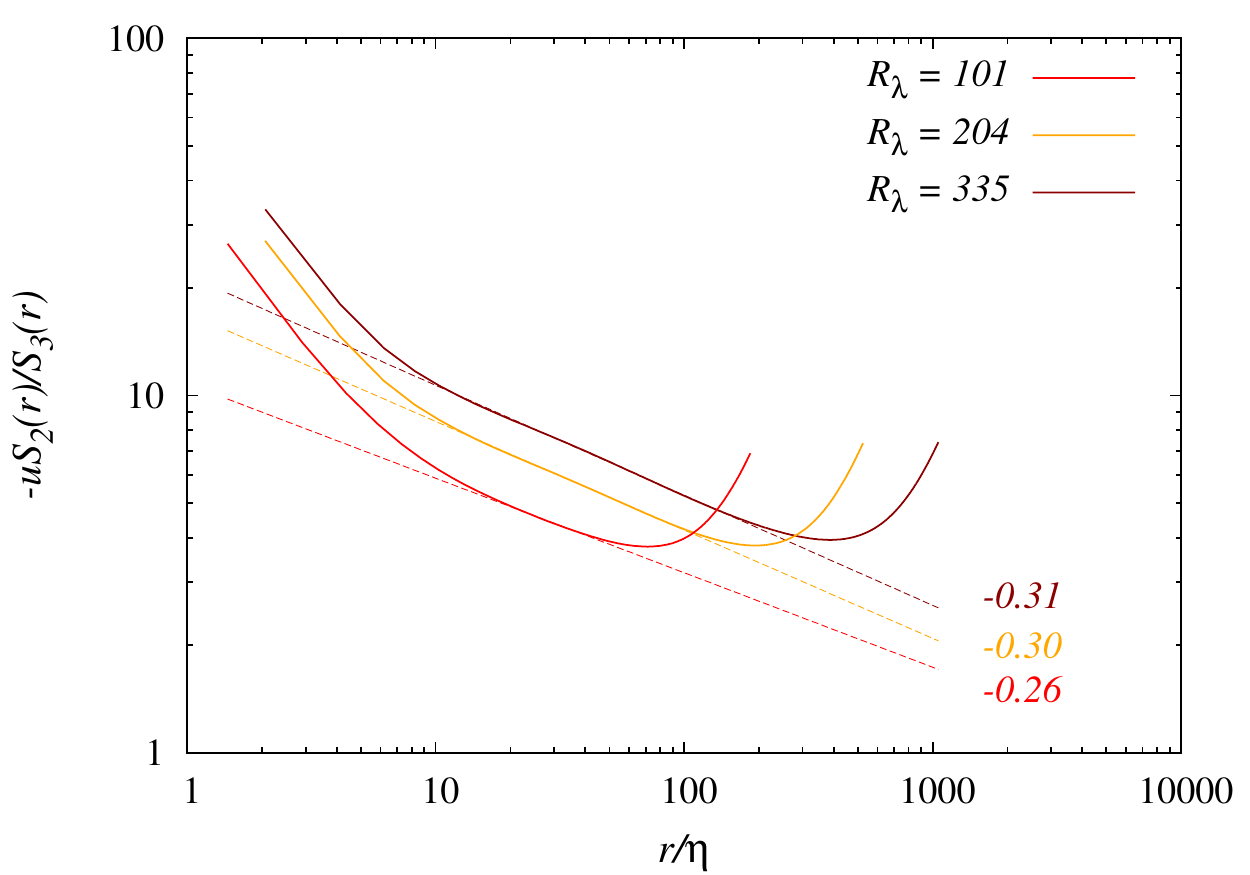}
 \end{center}
 \caption{Attempt to obtain the scaling exponent $\zeta_2$ using the structure functions obtained from energy and transfer spectra. The gradients marked are $\zeta_2 - 1$.}
 \label{fig:WDM_SS}
\end{figure}

The motivation for studying the generalised structure functions for ESS are clear and we would like to perform a similar analysis using our spectral technique.
We therefore focus on another presentation of data for which the cancellation of systematic `error' can be justified. Instead of plotting $G_n(r)$ against $G_3(r)$, we consider plotting $S_n(r) / \lvert S_3(r)\rvert$ against $r$. If the discrepancy between the measurements of the structure functions and their actual values (indicated by an overline) can be expressed as
\begin{equation}
 S_n(r) = \big( 1 + \epsilon(r) \big) \overline{S}_n(r) \ ,
\end{equation}
where $\epsilon(r)$ is a measure of the systematic error, then assuming that this error is of the same order for all $n$ and is small such that we may expand the denominator in a binomial expansion, we find
\begin{equation}
 \frac{S_n(r)}{S_3(r)} \simeq \frac{\overline{S}_n(r)}{\overline{S}_3(r)} \big( 1 - \epsilon(r) \big)\big( 1 + \epsilon(r) \big) \simeq \frac{\overline{S}_n(r)}{\overline{S}_3(r)} \big( 1 - \epsilon^2(r) \big) \ ,
\end{equation}
and the relative error has been reduced to order $\epsilon^2(r)$. This is shown in figure \ref{fig:WDM_SS}, where it can be seen that, while a scaling region can be identified, it is not as long as that generated using the ESS of the previous section. The exponent in this case becomes
\begin{equation}
 \frac{S_n(r)}{\lvert S_3(r) \rvert} \sim r^{\Xi_n} \ ,
\end{equation}
where $\Xi_n = \zeta_n - \zeta_3$ and we once again assume that $\zeta_3 = 1$. Values of $\zeta_2 - 1$ are printed in figure \ref{fig:WDM_SS}. It should be noted that in this case the the measured value of $\zeta_2$ is not increasing with Reynolds number but rather decreases towards 2/3.

\begin{figure}[tbp!]
 \begin{center}
  \includegraphics[width=0.75\textwidth]{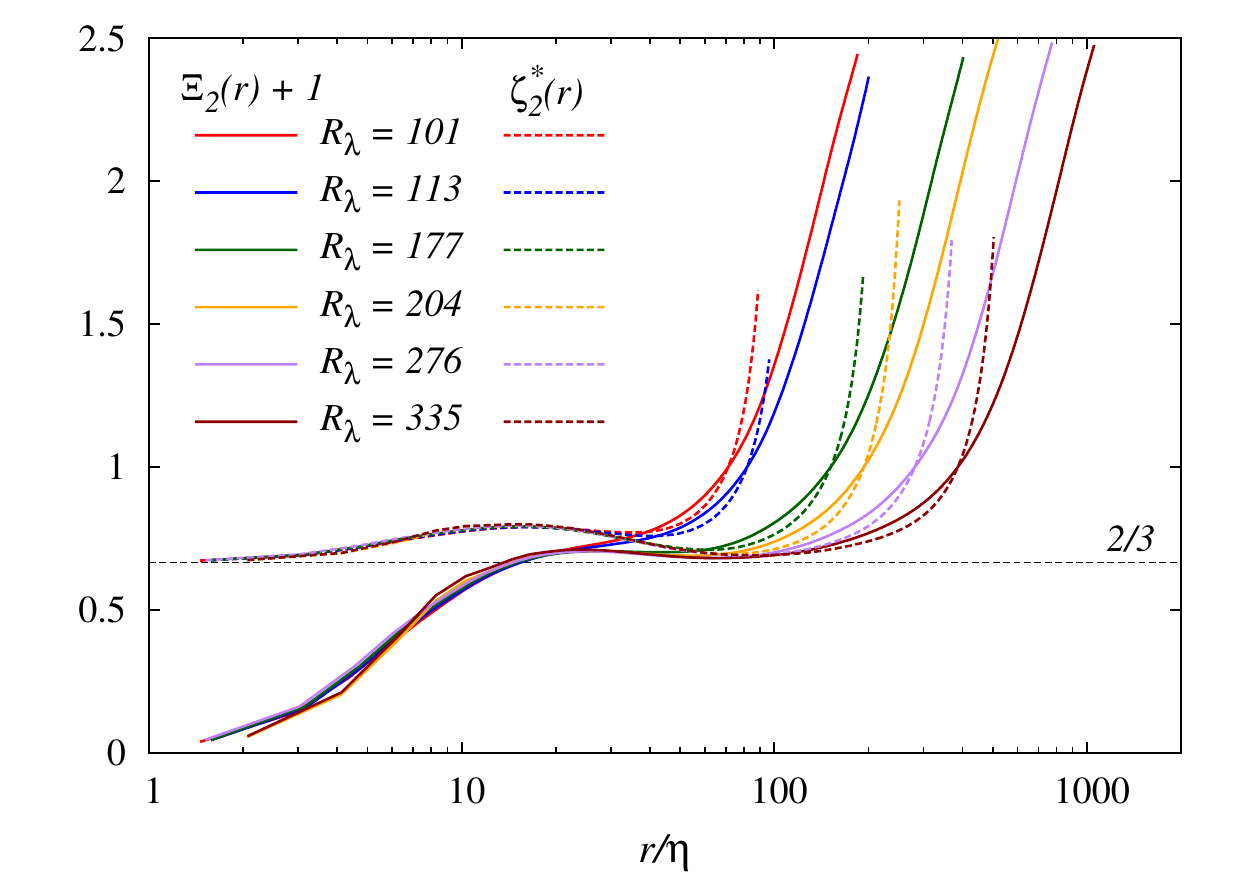}
 \end{center}
 \caption{Local scaling exponent $\Xi_2(r) = \zeta_2(r) - \zeta_3(r)$. We take $\zeta_3(r) = 1$ and plot $\zeta_2(r) = \Xi_2(r) + 1$. Also plotted (-- -- --) are the values of $\zeta^*_2(r) = \zeta_2(r)/\zeta_3(r)$. Horizontal dashed line marks the K41 prediction of $2/3$.}
 \label{fig:WDM_SS_exp}
\end{figure}

The local exponent, taking care to treat $S_3(r) < 0$ properly,
\begin{align}
 \Xi_n(r) = \frac{d \log \big( S_n / \lvert S_3 \rvert \big)}{d\log{r}} = \zeta_n(r) - \zeta_3(r)
\end{align}
can also be found and is plotted in figure \ref{fig:WDM_SS_exp}, along with $\zeta^*_2(r) = \zeta_2(r)/\zeta_3(r)$ for comparison (note that this is an ESS exponent but is calculated using $S_3(r)$ and so may be different from $\Sigma_2(r)$ calculated using $G_3(r)$). We see that there looks to be the development of a plateau, the value of which is moving towards 2/3 as Reynolds number is increased. Using the spectral technique, this is evaluated by noting
\begin{equation}
 \zeta_3(r) = \frac{12r}{S_3(r)} \int dk\ T(k)\ \left[ \frac{ \big(5(kr)^2 - 12\big) \sin{kr} - \big( (kr)^2 - 12 \big) kr\cos{kr}}{(kr)^5} \right] \ .
\end{equation}

We note that, since $S_n(r) \sim r^n$ as $r \to 0$ \cite{Sirovich94,Stolovitzky:1993p1476}, the exponent $\Xi_n(r) + 1 \to n - 2$, while $\Sigma_n(r) \to n/3$ (where this limit does not indicate K41). Due to the behaviour of $\Sigma_n(r)$ in this limit being the same as the K41 behaviour in the inertial range which we hope to isolate, we argue that $\Xi_n(r)$ is a more appropriate exponent with which to study inertial range scaling.

\section{Discussion}
Statistically stationary turbulence has been studied using direct numerical simulation, using a deterministic forcing scheme in which the rate of energy input is maintained constant. This sets our simulations apart from the literature, where most studies do not maintain $\varepsilon_W$ for each simulation, let alone a whole series of runs. This allowed a very systematic investigation with variation \emph{only} introduced through the (kinematic) viscosity. The time series for the evolution of statistical quantities were presented and clearly show the development of a steady-state. We note that, since stationarity is a statistical property, we expect fluctuations around a mean value and this is observed. The energy-conserving nature of the transfer spectrum is also well preserved for the duration of the runs. Scaled energy spectra show a good collapse onto the equilibrium range form, while the compensated energy spectra agree well with the literature. The Reynolds number dependence of the steady-state statistical quantities was also in agreement with $u^3/L$ being a better surrogate for the maximum inertial flux than the dissipation rate.

The condition placed on $\kmax\eta$, used to quantify how well `resolved' a simulation is, showed that the commonly reported $\kmax\eta = 1.5$ accounts for over 99.75\% of dissipation for the Reynolds numbers available. This makes it a suitable criterion, although we note that value of $\kmax\eta$ required to account for a fixed fraction of the total dissipation rate is Reynolds number dependent and increases with $Re$. In which case, as we move to larger and larger simulations, this condition should be carefully monitored.

Methods of vortex identification were discussed and used to visualise coherent structures in a realisation of the velocity field. They were then used to show that the remaining structure in an ensemble-average field decreases with an increase in the ensemble size, consistent with developing an isotropic ensemble.

The computation of structure functions in physical-space was performed for our lower-Reynolds number runs and results are consistent with the literature. The viscous correction to the third-order structure function in the K\'arm\'an-Howarth equation is shown to account for the deviation of $S_3(r)$ from the Kolmogorov form in the small scales, with disagreement maintained in the large scales. From the generalised structure functions, we show that the local scaling exponents do not offer a plateau with which we can calculate the scaling exponents. Instead, extended self-similarity is used, with which we obtain values for the ESS scaling exponents consistent with the literature.

The calculation of second- and third-order structure functions from the energy and transfer spectra was shown to give good agreement with their real-space computation, while requiring significantly less computational effort. While the relationship is well documented, we do not know of any direct comparison. We also note that this approach could also be used to study the structure functions for generated using spectral closures, which also does not appear to have been done.

Since the generalised structure functions are not available to a calculation from energy and transfer spectra, we propose an alternative to ESS and show that the scaling exponent for the second-order structure function instead approaches the K41 value of $2/3$. Expressions for higher-order structure functions in terms of the energy and transfer spectra could be developed, which would allow for a wider range or scaling exponents to be studied in this way. It would be interesting to see whether they also approach the K41 values of $n/3$, unlike those calculated using ESS. This could potentially exploit the fact that isotropy appears to be better satisfied in $k$-space than real space.

Since isotropy plays an important role in the calculation of structure functions and scaling exponents, we would like to implement `angle-averaging' to improve the isotropy of our physical-space computations. We expect this would improve the isotropy of the large scales, but note that it would involve an increase in computational work-load.

\chapter{Inertial transfer and dissipation of energy in isotropic turbulence}\label{chp:fKHE}

\section{The dissipation anomaly}\label{sec:DA}
In section \ref{sec:Taylor_surr} we introduced the surrogate expression for the dissipation,
\begin{equation}
 \varepsilon = \Ceps(Re) \frac{u^3}{L} \ ,
\end{equation}
put forward by Taylor \cite{Taylor:1935p308}. Originally, this was presented for some general characteristic length-scale, $l$. The quantity $u^3/l$ was considered to give a measure of the dissipation rate and was discussed by Batchelor \cite{Batchelor:1953-book} as having two interpretations: The first of these is the decay of energy $u^2$ in a time $l/u$. The second is the effect of an eddy viscosity acting on a shear to create a ``dissipation'' of energy `from the energy-containing eddies to smaller eddies'. The latter would instead be what we refer to as inertial flux, rather than dissipation. Batchelor also made use of the integral scale in his analysis. Tennekes and Lumley \cite{TennekesLumley:1972} provide a justification for the integral scale $L$ being the appropriate length-scale for a measure of the energy transfer. It was later shown by McComb, Berera, Salewski and Yoffe \cite{McComb:2010p250}, as well as sections \ref{sec:Taylor_surr} and \ref{subsec:forced_Rdep}, that $u^3/L$ is indeed a better surrogate for the inertial flux than the dissipation rate. Of course, as the Reynolds number increases and the viscous loss of energy from the large scales becomes negligible, dissipation becomes controlled by inertial flux and we expect $u^3/L$ to provide a good measure of both.

Recently, most work has focused on the Reynolds number dependent dimensionless dissipation rate,
\begin{equation}
 \Ceps(Re) = \frac{\varepsilon L}{u^3} \ .
\end{equation}
While the first evidence to suggest that $\Ceps$ became constant as Reynolds number is increased was presented by Batchelor \cite{Batchelor:1953-book}, it was the paper by Sreenivasan \cite{Sreenivasan:1984p139} which highlighted this behaviour. Experimental data from square-mesh grid-generated turbulence with $R_\lambda \in [10, 500]$ established that $\Ceps$ became a constant $\Ceps(\infty) \sim 1$ for $R_\lambda \gtrsim 50$. In an update, Sreenivasan \cite{Sreenivasan:1998p898} collected numerical data for both forced and freely-decaying turbulence. This showed a clear difference between forced and decaying turbulence and revealed that the asymptotic value could be sensitive to both initial conditions and the method of forcing.

\subsection{Comparison to the literature}\label{subsec:DA_lit}
Before presenting DNS data obtained from our simulations, we present a survey of established work with which we must compare. This will allow us to isolate key differences in the approaches used.

Jim\'enez, Wray, Saffman and Rogallo \cite{Jimenez:1993p634} investigated forced turbulence using the pseudospectral method with (incomplete) dealiasing by truncation and random phase shifts. An integrating factor was used to deal with the viscous term and time stepping was performed with a second-order variable-$dt$ Runge-Kutta algorithm. They obtained steady-state Reynolds numbers of $R_\lambda = 35, 61, 95$ and 170. Their $R_\lambda = 95$ run was maintained for $2\tau$ while $R_\lambda = 170$ ran for just $0.3\tau$, where $\tau = L/u$ is the large eddy turnover time. Average statistics were measured from 5-10 realisations. The forcing method involved a negative viscosity for modes with $k \leq 2.5$ which was modified to maintain $\kmax\eta$ constant. A value of $\Ceps(\infty) \simeq 0.7$ was found. They note that the upturn of the energy spectrum is caused by numerical artefacts resulting from the incorrect representation of the cascade mechanism at wavenumbers of the order of the resolution. 

Wang, Chen, Brasseur and Wyngaard \cite{Wang:1996p1041} used a pseudospectral code (developed by Chen and Shan \cite{Chen:1992p186}) to study both forced and decaying turbulence. Decaying simulations were either started from an initial condition which behaved as $k^2$ at low and $k^{-5/3}$ at high $k$, or from a developed stationary field. They considered the decay to be evolved when the total energy expressed power-law decay. The system was maintained stationary by holding the energy spectrum for the lowest two wavenumber shells with $k^{-5/3}$ and maintaining $E_f$, the energy in the forcing band. An asymptotic value of $\Ceps(\infty) \simeq 0.62$ was found for decay and $\Ceps(\infty) \simeq 0.42$ -- 0.49 for forced turbulence.

Yeung and Zhou \cite{Yeung:1997p1746} also used a pseudospectral code with incomplete alias removal by truncation and phase shifts. They used stochastic forcing to maintain $R_\lambda = 38, 90 140, 180$ and 240 over about four large eddy turnovers.

Cao, Chen and Doolen \cite{Cao:1999p899} used the same pseudospectral code and forcing scheme as \cite{Wang:1996p1041} above but with no dealiasing. The initial condition was $k^4 \exp(k/k_0)^2$ with $k_0 \sim 5$. Chen, Doolen, Kraichnan and She \cite{Chen:1993p491} also performed forced and decaying simulations, while Chen, Holm, Margolin and Zhang \cite{Chen:1999p419} presented a comparison of forced turbulence developed from an initial condition which behaved as $k^4$ for low $k$ and the Lagrangian $\alpha$-model.

Pearson, Krogstad and van de Water \cite{Pearson:2002p156} presented experimental results for $\Ceps$ for a number of shear flows, from which they found $\Ceps(\infty) \simeq 0.48$. Later, and rather against the trend, Pearson, Yousef, Haugen, Brandenburg and Krogstad \cite{Pearson:2004p147} performed a high-order finite difference study of a slightly compressible isothermal fluid. Stationarity was maintained by forcing the large scales with $\vec{f}(\vec{x},t) = \vec{f}_0 \cos [i\vec{k}\cdot\vec{x} + i\phi(t)]$ with $\phi(t) \in [-\pi,pi]$ and $1 \leq \lvert\vec{k}\rvert \leq 2$. Both $\vec{k}$ and $\phi(t)$ were chosen randomly at each time-step, ensuring that the forcing is $\delta$-correlated in time. They found a value of $\Ceps \simeq 0.5$.

Other experimental work of interest includes Burattini, Lavoie and Antonia \cite{Burattini:2005p142}, who studied the behaviour of $\Ceps$ for a variety of different flow types, and Mazellier and Vassilicos \cite{Mazellier:2008p141} where a variety of grid configurations (including fractal grids) were investigated in a wind tunnel. In the latter, they found that the behaviour of $\Ceps$ could be related to $\log{R_\lambda}$ and a flow- (and possibly weakly Reynolds number-) dependent constant, $C'_S$, which can be related to the average separation of stagnation points and characterises the large-scale structure. The variation in $C'_S$ causes the asymptotic value $\Ceps(\infty)$ as Reynolds number is increased to be significantly lower, of order $\sim 0.065$ and dependent on flow configuration. Thus $\Ceps(\infty)$ is not a universal constant.

Returning to DNS, Donzis, Sreenivasan and Yeung \cite{Donzis:2005p855} used stochastic forcing in a pseudospectral code with partial dealiasing to simulate Reynolds numbers up to $R_\lambda = 390$ on lattices ranging from $64^3$ to $1024^3$. They find a value of $\Ceps(\infty) \simeq 0.4$ and fit their data to the form $A[1+\sqrt{1 + (B/R_\lambda)^2}]$ with $A \simeq 0.2$ and $B \simeq 92$.

Pseudospectral DNS with Reynolds numbers up to $R_\lambda \sim 100$ were performed by Bos, Shao and Bertoglio \cite{Bos:2007p128}. The authors compared results with LES and the Eddy-Damped Quasi-Normal Markovian (EDQNM) closure with up to $R_\lambda \sim 2000$. A variety of initial conditions were studied based on exponential forms and the von K\'arm\'an spectrum (see section \ref{subsec:initial_field}). They found $\Ceps(\infty) \simeq 0.5$ for forced and $\Ceps(\infty) \simeq 1$ for decaying turbulence. No dependence was found on the initial condition \emph{once turbulence had developed}, but they did note that the von K\'arm\'an spectrum offered a shorter transient time to steady state. Forcing was implemented by maintaining the energy spectrum (either as von K\'arm\'an or K41) in the range $k \leq k_f$ with $k_f \in [2, 5]$. The LES and EDQNM data obtained shows an initial agreement between the $\Ceps$ data for decaying and forced turbulence before the curves split to their respective plateau.

Variation of initial condition was studied by Goto and Vassilicos \cite{Goto:2009p144} who considered
\begin{equation}
 E(k,0) = \left\{ \begin{array}{ll}
  Ck^q \exp \big[ -\tfrac{q}{2}(k/k_0)^2 \big] & k \leq k_0 \\
   & \\
  Ck^q \exp \big[ -\tfrac{q}{\alpha} (k/k_0)^\alpha + q/\alpha - q/2 \big] & k \geq k_0
 \end{array} \right. \ ,
\end{equation}
with $q = 2,4$; $\alpha = 1, 0.5$ and $k_0 = 5, 10$ and 15. The constant $C$ is chosen such that the total initial energy $E(0) = 1$. They found that different $q$ produced different curves for $\Ceps$ and different plateaus, highlighting a dependence on the initial large scales. Variation of $k_0$ did not influence the data severely. Stationarity was maintained by keeping the magnitude of $\vec{u}(\vec{k},t)$ fixed in the range $0 < k < k_f$ with $k_f = 2.4 k_0$. This is a huge forcing range compared to our and other numerical simulations, where forcing is constrained to the lowest few wavenumber shells. Note that, by keeping the magnitude of the velocity field constant they are maintaining the energy spectrum in the forcing band. The difference in $\Ceps(\infty)$ plateaus could therefore be associated with different large scale forcing rather than initial condition.

Finally, Kaneda, Ishihara, Yokokawa, Itakura and Uno \cite{Kaneda:2003p134} presented an analysis of $\Ceps$ data from the large Reynolds number simulations performed on the Earth Simulator. They used a pseudospectral code with complete dealiasing to study on lattices up to $4096^3$. Interpolation of the velocity field was used to generate the initial condition for larger runs, $256^3 \rightarrow 512^3 \rightarrow 1024^3 \rightarrow 2048^3 \rightarrow 4096^3$. They were able to obtain Reynolds numbers $R_\lambda = 94, 173, 268$ and 429 with $\kmax\eta = 2$. When this condition was relaxed to $\kmax\eta = 1$, this allowed $R_\lambda = 167, 257, 471, 732$ and 1201. Forcing was implemented by using a negative viscosity for the modes $k < 2.5$, modified to keep the total energy constant, as used by Vincent and Meneguzzi \cite{Vincent:1991p325} and Kerr \cite{Kerr:1985p191}. A value of $\Ceps(\infty) \simeq 0.4$ -- 0.5 was found.

\vspace{1em}
\noindent Turning to our results, we perform fully dealiased pseudospectral computation on lattices of size $64^3, 128^3, 256^3, 512^3$ and $1024^3$. Our initial condition is spectrum 5 with $k^4$ low $k$ behaviour and forcing is done by negative damping in the band $0 < k <2.5$ with fixed energy input rate \emph{for all runs}.

\begin{figure}[tbp!]
 \begin{center}
  \includegraphics[width=0.75\textwidth]{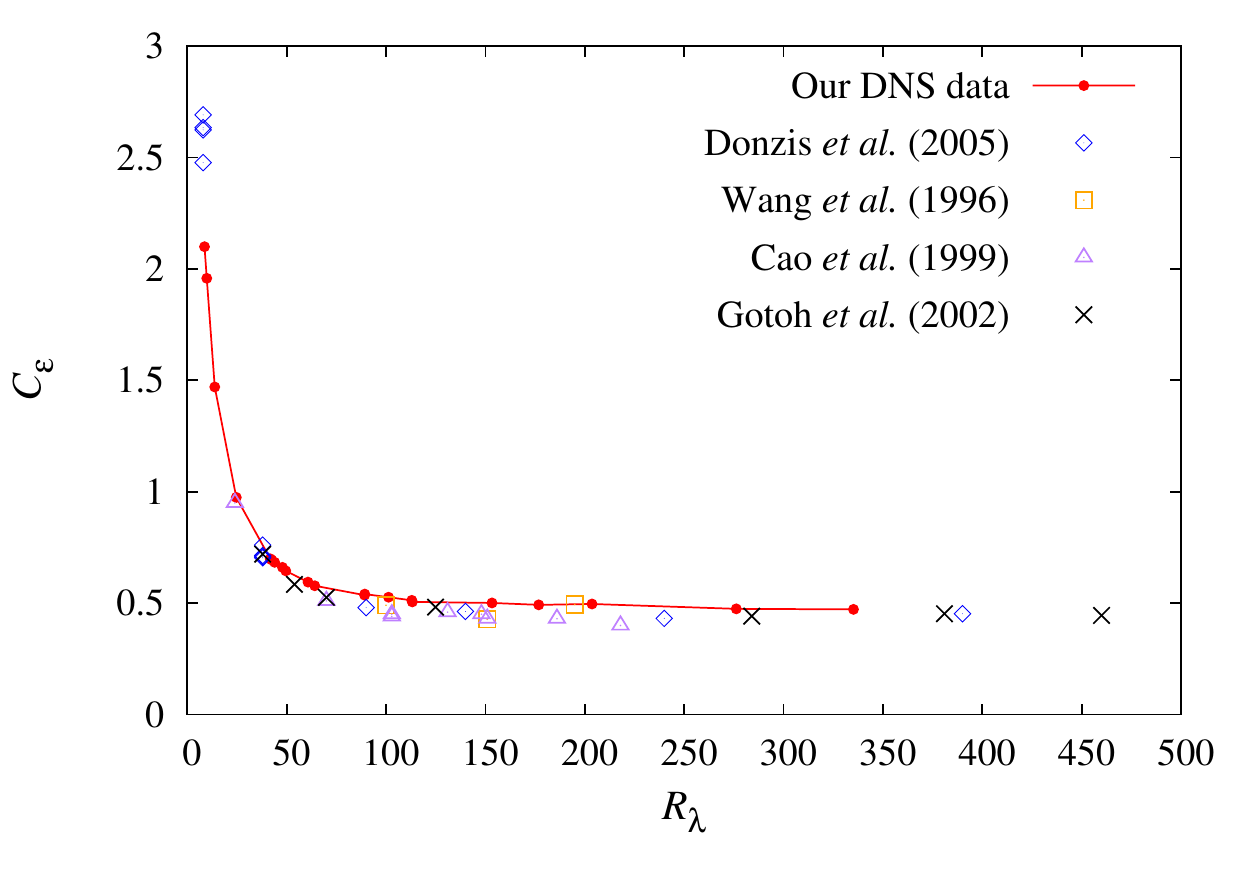}
 \end{center}
 \caption[DNS data for $\Ceps$ against $R_\lambda$ for forced turbulence, with a comparison to results from the literature.]{A comparison of our DNS data for the Reynolds number variation of the dimensionless dissipation rate in forced turbulence. Plotted also are the results obtained by Donzis \etal\ \cite{Donzis:2005p855}, Wang \etal\ \cite{Wang:1996p1041}, Cao \etal\ \cite{Cao:1999p899} and Gotoh \etal\ \cite{Gotoh:2002p627}.}
 \label{fig:DA_forced}
\end{figure}

We present our DNS data for stationary turbulence in figure \ref{fig:DA_forced}. Following the literature, $\Ceps$ is plotted against Reynolds number based on the Taylor microscale. For comparison, we also plot data obtained from a sample of numerical work. Our data agrees very well with the other numerical investigations, despite the variation in initial condition and forcing scheme used. A plateau of $\Ceps(\infty) \simeq 0.47$ is found.

\subsection{DNS results for free decay}
It would be nice to present a similar analysis of the Reynolds number behaviour of $\Ceps$ for decaying turbulence, but we are faced the problem of defining an evolved time, as discussed in section \ref{sec:tevo}. It was shown that selection of evolved time had a dramatic effect on the measurement of the dissipation rate, figure \ref{fig:decay_eps-pi-surr}.

\begin{figure}[tbp!]
 \begin{center}
  \hspace{0.25in}\includegraphics[width=0.8\textwidth,trim=85px 350px 15px 0px, clip]{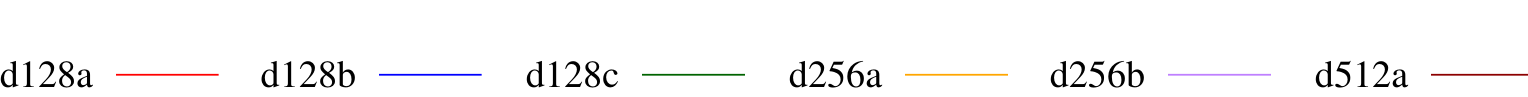}
  \subfigure[Time evolution of $\Ceps(t)$]{
   \label{sfig:Ceps_TS}
   \includegraphics[width=0.475\textwidth,trim=2px 0px 10px 2px, clip]{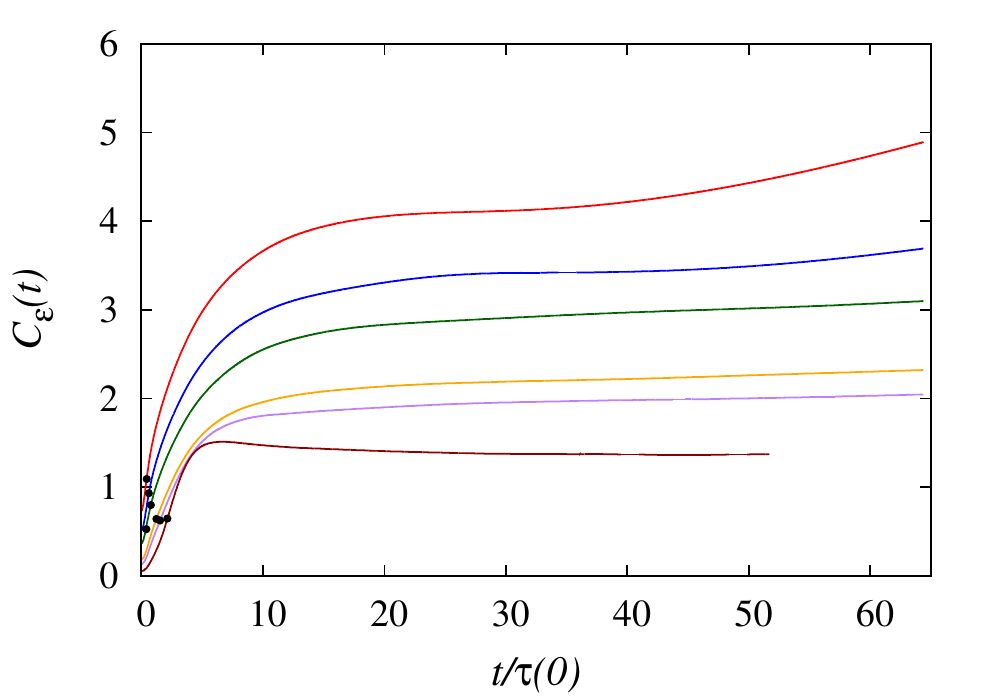}
  }
  \subfigure[Decay of $L/\lambda$]{
   \label{sfig:Ceps_TS_Ll}
   \includegraphics[width=0.475\textwidth,trim=0px 0px 10px 2px, clip]{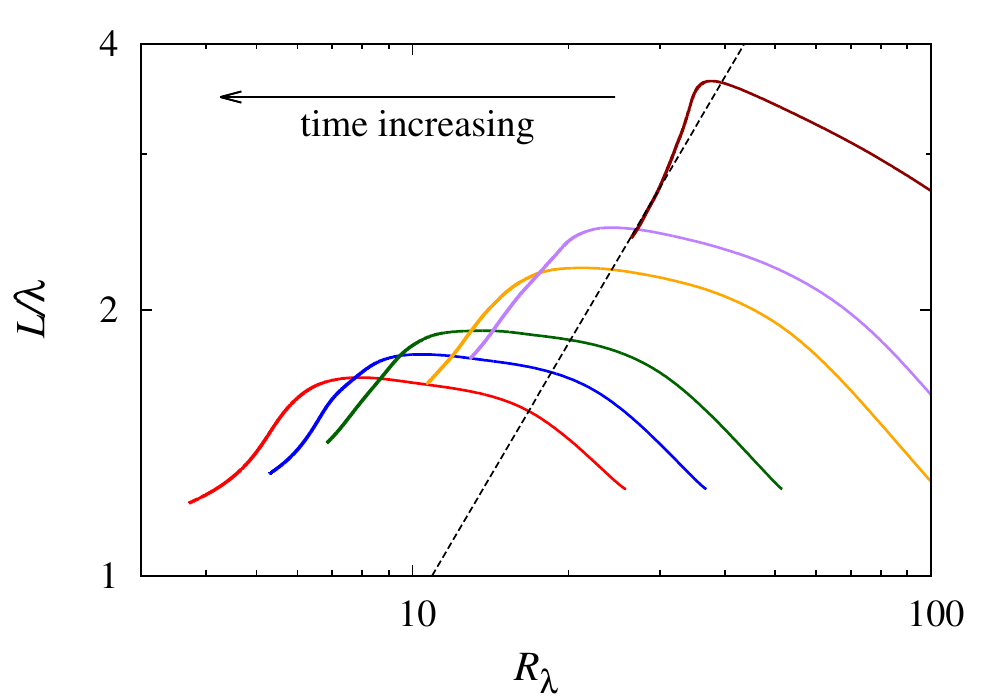}
  }
 \end{center}
 \caption{Time evolution of the dimensionless dissipation rate, $\Ceps(t) = \varepsilon(t) L(t)/u^3(t)$. The constant behaviour observed for a period of the decay in (a) implies $L/\lambda \sim R_\lambda$, which is supported by DNS data in (b) where the dashed line indicates $1/R_\lambda$.}
 \label{fig:Ceps_TS}
\end{figure}

We now look at the time evolution of $\Ceps(t) = \varepsilon(t) L(t) / u^3(t)$ and this is shown in figure \ref{fig:Ceps_TS} for a selection of decaying runs. The measurement time $t_{\varepsilon\vert\Pi}$ corresponding to the peak of the dissipation/transport spectrum are given as the solid points. We see that $t_{\varepsilon\vert\Pi}$ occurs very early in the decay while $\Ceps$ is still strongly time-dependent, but the quantity does seem to develop a plateau. It would seem promising that one should choose a time on this plateau, but there is a problem: Since the turbulence is decaying, while $\Ceps(t)$ is remaining constant the Reynolds number is still decaying. As such, we measure the same value for $\Ceps(Re)$ at various $Re$, and the curves are translated. This is not the case for the lower Reynolds numbers which do not develop a plateau. Since the low Reynolds numbers do not develop this plateau, it is possible that the curves will converge at sufficiently low Reynolds number. The upturn of the plateau at longer times is probably associated with the development of what Davidson refers to as periods of small scale depletion and exponential decay \cite{davidson:2004-book}.

The quantity $L/\lambda$ is essentially a Reynolds number, since it quantifies how small the small-scale motion, generated by the non-linear coupling, is compared to the large scales. We note that, from the definitions given in equation \eqref{eq:Clambda_forms}, the constant behaviour of $\Ceps(t)$ in time (and hence Reynolds number) observed for a period of the decay in figure \ref{sfig:Ceps_TS} implies that $L/\lambda \sim R_\lambda$ (or, equivalently, $(L/\lambda)^2 \sim R_L$ or $R_L \sim R_\lambda^2$) for the same period. This was found experimentally by Valente and Vassilicos \cite{Valente2012} for low $R_\lambda$ far downstream, and followed a region where $L/\lambda \sim \text{constant}$. Figure \ref{sfig:Ceps_TS_Ll} shows our DNS data for the $R_\lambda$ dependence of $L/\lambda$ and supports a linear relationship at low but not at higher $R_\lambda$, consistent with \cite{Valente2012}. Perhaps larger decaying numerical simulations would also support this $L/\lambda \sim \text{constant}$ region at larger $R_\lambda$.

\begin{figure}[tbp!]
 \begin{center}
   \includegraphics[width=0.75\textwidth]{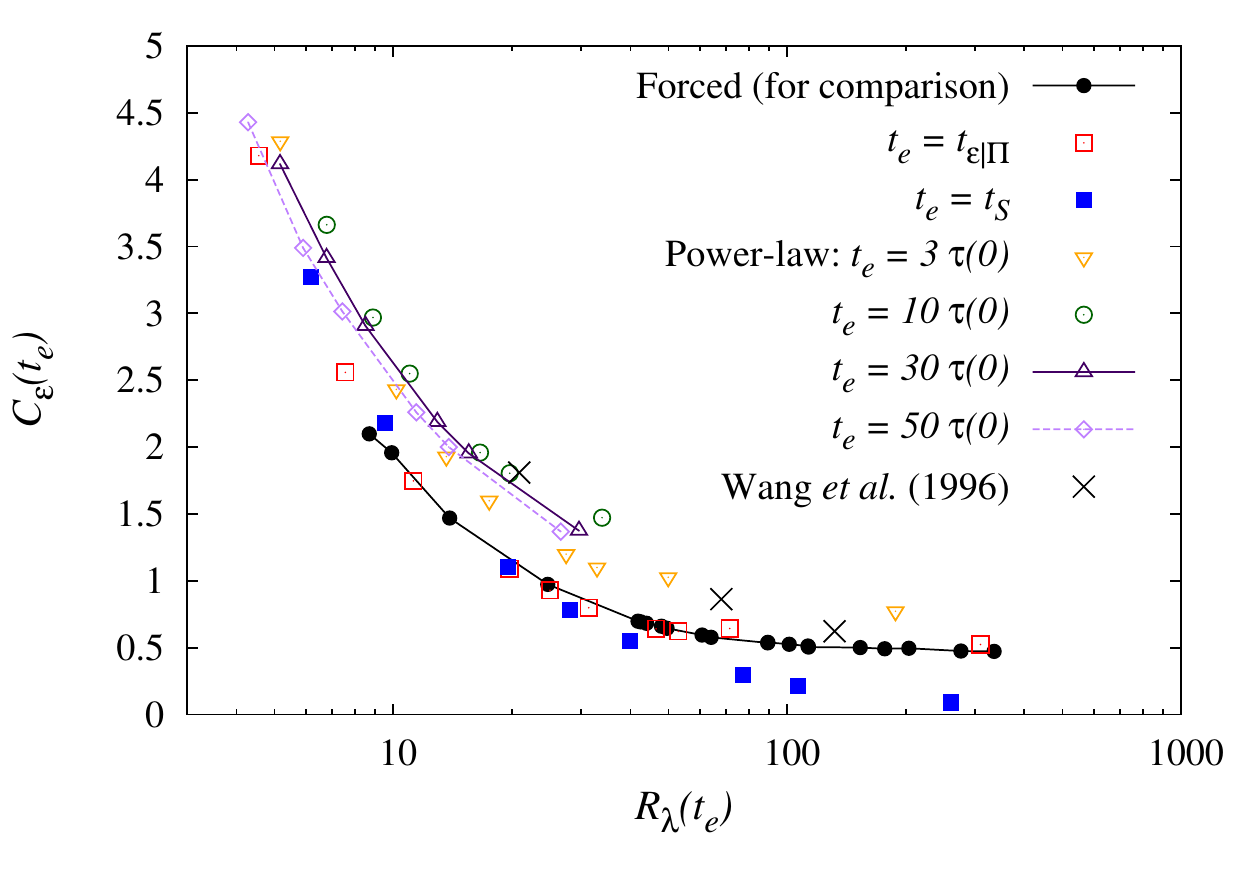}
 \end{center}
 \caption{The Reynolds number variation of the dimensionless dissipation rate for decaying turbulence. Plotted for various choices of the evolved time along with the forced case and decay data from Wang \etal\ \cite{Wang:1996p1041}, for comparison.}
 \label{fig:Ceps_decay}
\end{figure}

The values of $\Ceps(t_e)$ and the corresponding $R_\lambda(t_e)$ for a range of times during the decay are shown in figure \ref{fig:Ceps_decay}. The peak of the skewness, denoted $t_S$, is the earliest time, and for this measurement time the dimensionless dissipation coefficient does not develop a plateau but looks as though it decays to zero. The peak dissipation measurement time appears to match the forced case, consistent with the discussion of sections \ref{subsec:tevo_epsPi} and \ref{sec:decay_from_forced}. As we move to $t_e = 3\tau(0)$, the earliest time that can be connected to power-law decay of the total energy, we see very different behaviour. The curve follows a similar profile to the forced case only shifted up the $\Ceps$ axis. Moving to $t = 30\tau(0)$, which was seen in figure \ref{fig:Ceps_TS} to sit on the plateau of $\Ceps(t)$ for the simulations given, this shift has increased to about 0.5. At this time, if the curved continued in this manner we would expect to find a plateau around unity, in agreement with the literature.

We see that the choice of evolved time has a large effect on the conclusions we draw. Use $t_S$ and we go to zero; use $t_{\varepsilon\vert\Pi}$ and we measure the same asymptote as for forced; or look further into the decay and develop a progressively higher plateau. This latter case occurs until $\Ceps(t)$ becomes a constant, after which the curve moves down again. This is why the curves for $t = 30\tau(0)$ and $50\tau(0)$ are successively lower than that for $t = 10\tau(0)$, since $10\tau(0)$ roughly corresponds to the beginning of the $\Ceps(t)$ plateau. We conclude that the asymptote for decaying turbulence is in the range $0 \leq \Ceps(\infty) \lesssim 1.2$ depending on the choice of evolved time.

We also note an important distinction: for decaying turbulence, the behaviour of $\Ceps$ is largely due to variation of $\varepsilon$. Whereas, for forced turbulence as studied here with the dissipation rate kept constant, the decay and plateau of $\Ceps$ is entirely due to the behaviour of the surrogate, $u^3/L$; see figures \ref{fig:decay_eps-pi-surr} and \ref{sfig:forced_Rvar_eps-pi-surr}.

\subsection{Use of the Taylor microscale}
Ever since Batchelor \cite{Batchelor:1953-book} and Tennekes and Lumley \cite{TennekesLumley:1972} the surrogate has used the integral length-scale in its definition, yet results for $\Ceps$ are always plotted against $R_\lambda$. Indeed, Sreenivasan \cite{Sreenivasan:1984p139} compared the form $\Ceps = 15 R_\lambda^{-1} \sqrt{\pi/2}$ to the experimental data. It seems that the integral scale has been justified as the relevant length-scale but the Taylor microscale is always introduced.

In numerical simulation, the Taylor microscale is usually computed from the dissipation rate as
\begin{equation}
 \label{eq:Clambda_forms}
 \lambda^2 = \frac{15\nu_0 u^2}{\varepsilon} \qquad\quad \Longrightarrow \qquad\quad \Ceps = \frac{15\nu_0 L}{\lambda^2 u} = \frac{15}{R_L} \frac{L^2}{\lambda^2} = \frac{15}{R_\lambda} \frac{L}{\lambda} = 15\frac{R_L}{R_\lambda^2} \ .
\end{equation}
Of course, all of the forms given on the right fit the data perfectly, since they are just a rearrangement of our original definition. If one defined $\Ceps^\lambda$ using the Taylor microscale, one would find $\Ceps^\lambda = 15/R_\lambda$ in a similar way.

The problem is that $\lambda$ is not an independent variable due to its dependence on $\varepsilon$. On the other hand, the integral scale is not dependent on our measurement of $\varepsilon$. In section \ref{sec:model_DA}, in particular \ref{subsec:model_DA}, we develop a model based entirely on the integral scale. This is then shown to be in excellent agreement with DNS data.

\section{The K\'arm\'an-Howarth equation for forced turbulence}\label{sec:KHE}
The K\'arm\'an-Howarth equation (KHE) was introduced in 1938 by T. von K\'arm\'an and L. Howarth \cite{Karman:1938p153} and has become of central importance in the study of isotropic turbulence. In its original form, it presents a local (in $r$-space) balance of energy in the form of a dynamical equation for the evolution of the second-order longitudinal correlation function (see section \ref{subsec:long_trans_corr}) and is expressed in our notation as,
\begin{align}
 \frac{\partial C_{LL}}{\partial t} = \left( \frac{\partial}{\partial r} + \frac{4}{r} \right) C_{LL,L} + 2\nu_0 \left( \frac{\partial}{\partial r} + \frac{4}{r} \right) \frac{\partial C_{LL}}{\partial r} \ .
\end{align}
This can be compared to equation (51) of the original paper \cite{Karman:1938p153}. It is an exact relation for the statistical correlation functions derived from the Navier-Stokes equations \cite{MoninYaglom:vol2}.

By using the definition of the second- and third-order structure functions expressed in terms of the longitudinal correlation functions, this equation can be rewritten (but completely equivalently) as
\begin{align}
 \label{eq:KHE}
 \varepsilon_D = -\frac{3}{4}\frac{\partial S_2}{\partial t} -\frac{1}{4r^4} \frac{\partial}{\partial r} \Big( r^4 S_3 \Big) + \frac{3\nu_0}{2r^4}\frac{\partial}{\partial r} \left( r^4 \frac{\partial S_2}{\partial r} \right) \ ,
\end{align}
where the time rate of change of the total energy is expressed by
\begin{equation}
 \varepsilon_D \equiv - \frac{\partial E(t)}{\partial t} \ .
\end{equation}
Note that the longitudinal correlation functions, and hence the structure functions, are potentially functions of both separation $r$ and time $t$. In the case of decaying turbulence, we have
\begin{equation}
 \varepsilon_D = \varepsilon \ ;
\end{equation}
that is, the energy lost by the system has been dissipated. We therefore replace $\varepsilon_D$ with $\varepsilon$.

Let us now consider stationary turbulence. This requires the input of energy and, since we assume the steady state, the input must be equal to the dissipation rate. Hence, for stationary turbulence we write the KHE as
\begin{align}
 \label{eq:fKHE_orig}
 \varepsilon = -\frac{1}{4r^4} \frac{\partial}{\partial r} \Big( r^4 S_3 \Big) + \frac{3\nu_0}{2r^4}\frac{\partial}{\partial r} \left( r^4 \frac{\partial S_2}{\partial r} \right) \ ,
\end{align}
where the time derivative of the second-order structure function has been discarded since the system is statistically steady in time. The caveat is that we consider only scales which are not directly influenced by whatever forcing scheme we have used at the large scales. This equation has been used extensively in the literature, for example by Kolmogorov in his derivation of K41 \cite{Kolmogorov:1991p138}.

But for stationary turbulence, the system is \emph{not} losing energy and we have
\begin{equation}
 \varepsilon_D = 0 \ .
\end{equation}
As such, the origin of the dissipation rate is hidden. Also, what was regarded as a loss of energy on the LHS is now considered to be an input. Clearly, the relevant quantity at small scales is the amount of energy they receive through the non-linear cascade of energy to larger wavenumbers. This is the inertial transfer $\varepsilon_T$. If the Reynolds number is sufficiently high that the large and dissipative scales are well separated, then the transfer must become equal to the dissipation rate, $\varepsilon_T = \varepsilon$, since dissipation is a small-scale process.

\subsection{Derivation for forced turbulence}
The above discussion highlights a problem with the form of the KHE for stationary turbulence. The treatment of how the energy is injected into the systems seems rather vague. Attempts have been made to include a work term by measuring the correlation of the velocity field with the driving force \cite{frisch:1995-book,Fukayama:1999p904,Gotoh:2002p627}, however these approaches attempt to maintain $\varepsilon$ on the LHS and add an \emph{additional} term. Whereas, our discussion above implies that $\varepsilon$ \emph{is} the energy entering and passing through the inertial subrange of scales.

There are issues with defining the relevant work function in real space, so in an attempt to derive a KHE with a work term we start by considering the Lin equation, for which the work spectrum is well defined,
\begin{equation}
 \frac{\partial E(k,t)}{\partial t} = T(k,t) - 2\nu_0 k^2 E(k,t) + W(k,t) \ ,
\end{equation}
where $W(k,t) = 4\pi k^2 F(k,t)$ and $F(k,t)$ is the autocorrelation of the force. This was introduced in section \ref{subsec:lin_equation}.
In section \ref{sec:exploit_PS}, we saw how the real-space isotropic correlation functions could be related to the energy and transfer spectra in equations \eqref{eq:R_Ek} and \eqref{eq:CLLL_Tk}. As a starting point, we therefore consider multiplying the Lin equation by $\sin{kr}/kr$ and integrating over $k$,
\begin{align}
 \label{eq:trans_Lin}
 \int dk\ \frac{\partial E(k,t)}{\partial t}\ \frac{\sin kr}{kr} &= \int dk\ T(k,t)\ \frac{\sin kr}{kr} - \int dk\ 2\nu_0 k^2 E(k,t)\ \frac{\sin kr}{kr} \nonumber \\
 &\qquad+ \int dk\ W(k,t)\ \frac{\sin kr}{kr} \ .
\end{align}
It is a simple matter to show that
\begin{align}
 \label{eq:diss_trans}
 \left( \frac{\partial}{\partial r} + \frac{2}{r} \right) \frac{\partial}{\partial r} \int dk\ E(k,t)\ \frac{\sin kr}{kr} = - \int dk\ k^2\ E(k,t)\ \frac{\sin kr}{kr} \ ,
\end{align}
which will be useful when we transform the dissipation spectrum, and we define the work function as
\begin{align}
 \label{eq:work_trans}
 W(r,t) = \int dk\ W(k,t)\ \frac{\sin{kr}}{kr} \ .
\end{align}
Inserting equations \eqref{eq:R_Ek}, \eqref{eq:CLLL_Tk}, \eqref{eq:diss_trans} and \eqref{eq:work_trans} into the transformed Lin equation \eqref{eq:trans_Lin}, we find
\begin{align}
 \frac{\partial C(r,t)}{\partial t} &= \frac{1}{2} \left( 3 + r\frac{\partial}{\partial r} \right) \left( \frac{\partial}{\partial r} + \frac{4}{r} \right) C_{LL,L}(r) + 2\nu_0 \left( \frac{\partial}{\partial r} + \frac{2}{r} \right) C(r,t) + W(r,t) \ .
\end{align}

The isotropic correlation function can be expressed using equations \eqref{eq:R_iso_exp} and \eqref{eq:CNN_exp}
\begin{align}
 C(r) &= \tfrac{1}{2} \Big( C_{LL}(r) + 2 C_{NN}(r) \Big) \nonumber \\
 &= \frac{1}{2} \left( 3 + r \frac{\partial}{\partial r} \right) C_{LL}(r) \ ,
\end{align}
which allows us to write the transformed Lin equation in terms of the longitudinal correlation functions,
\begin{align}
 \frac{1}{2} \frac{\partial}{\partial t} \left(3 + r\frac{\partial}{\partial r}\right) C_{LL}(r) &= \frac{1}{2}\left(3 + r\frac{\partial}{\partial r}\right)\left(\frac{\partial}{\partial r} + \frac{4}{r}\right) C_{LL,L}(r) \\
 &\qquad+ \nu_0\left(\frac{\partial}{\partial r} + \frac{2}{r}\right) \frac{\partial}{\partial r} \left(3 + r\frac{\partial}{\partial r}\right) C_{LL}(r) + W(r) \nonumber \ .
\end{align}
Multiplying through by $2r^2$, we see that this is equivalent to
\begin{align}
 \frac{\partial}{\partial t} \frac{\partial}{\partial r}\Big[r^3 C_{LL}(r)\Big] &= \frac{\partial}{\partial r}\left[ r^3 \left(\frac{\partial}{\partial r} + \frac{4}{r} \right) C_{LL,L}(r)\right] \nonumber \\
 &\qquad+ 2\nu_0  \frac{\partial}{\partial r} \left[ r^2 \frac{\partial}{\partial r} \left(3 + r\frac{\partial}{\partial r}\right) C_{LL}(r) \right] + 2r^2 W(r) \ .
\end{align}
We now write the above equation for a new variable $r \to \xi$ and perform an integral over $\xi$ from 0 to $r$, after which we divide by $r^3$. This gives
\begin{align}
 \label{eq:fKHE_nearly}
 \frac{\partial}{\partial t} C_{LL}(r) &=  \left(\frac{\partial}{\partial r} + \frac{4}{r} \right) C_{LL,L}(r) + 2\nu_0 \frac{1}{r} \frac{\partial}{\partial r} \left(3 + r\frac{\partial}{\partial r}\right) C_{LL}(r)  + \frac{2}{r^3} \int^r_0 d\xi\ \xi^2 W(\xi) \ ,
\end{align}
since the terms associated with the lower limit vanish identically due to the presence of $r^3$ and $r^2$. This equation is starting to resemble the KHE, and we simplify the viscous term slightly to continue,
\begin{align}
  \frac{1}{r}\frac{\partial}{\partial r} \left( 3 C_{LL}(r) + r\frac{\partial C_{LL}(r)}{\partial r}\right) &= \frac{3}{r} \frac{\partial C_{LL}(r)}{\partial r} + \frac{1}{r}\frac{\partial C_{LL}(r)}{\partial r} + \frac{\partial^2 C_{LL}(r)}{\partial r^2} \nonumber \\
  &= \left( \frac{4}{r} + \frac{\partial}{\partial r} \right) \frac{\partial C_{LL}(r)}{\partial r} \ .
\end{align}
Equation \eqref{eq:fKHE_nearly}, with the simplification above, can now be written in terms of the structure functions as
\begin{equation}
 \label{eq:fKHE}
 \varepsilon_D = -\frac{3}{4}\frac{\partial S_2(r)}{\partial t} -\frac{1}{4r^4} \frac{\partial}{\partial r} \Big( r^4 S_3(r) \Big) + \frac{3\nu_0}{2r^4}\frac{\partial}{\partial r} \left( r^4 \frac{\partial S_2(r)}{\partial r} \right) - I(r) \ ,
\end{equation}
where the input term is given by
\begin{equation}
 \label{eq:input_spectral}
 I(r) \equiv \frac{3}{r^3} \int_0^r d\xi\ \xi^2 W(\xi) \ .
\end{equation}
By comparing to equation \eqref{eq:CLL_R}, one could identify the work term $I(r)$ with the longitudinal correlation of the velocity and force, but this is not pursued here.

Comparison to the KHE given in equation \eqref{eq:KHE} shows that the form of equation \eqref{eq:fKHE} is identical except from the explicit forcing term present on the RHS. This also shows the equivalence of the K\'arm\'an-Howarth and Lin equations, with one being local in configuration space and the other in $k$-space. We now have two cases:
\begin{enumerate}
 \item \textbf{Decaying turbulence}\newline
 There is no input of energy so that the work spectrum is zero, $W(k,t) = 0$ for all $k$. As such, the work function also vanishes, $W(r,t) = 0$ for all $r$. The time derivative of the total energy is simply the dissipation rate, $\varepsilon_D = \varepsilon$, and we recover equation \eqref{eq:KHE} for decaying turbulence.

 \item \textbf{Stationary turbulence}\newline
 There is no time dependence in the problem and as such all time derivatives are zero. The forced KHE is instead written
 \begin{equation}
  \label{eq:fKHE_stationary}
  I(r) = -\frac{1}{4r^4} \frac{\partial}{\partial r} \Big( r^4 S_3(r) \Big) + \frac{3\nu_0}{2r^4}\frac{\partial}{\partial r} \left( r^4 \frac{\partial S_2(r)}{\partial r} \right) \ ,
 \end{equation}
 which is now valid for all scales. This confirms that, in the original formulation of the forced KHE equation, $\varepsilon$ is indeed representing the energy entering and passing through the inertial subrange, $\varepsilon_T$, rather than dissipation. We expect that, as the separation of integral and dissipative scales increases, the range of length-scales unaffected by forcing will have $I(r) = \varepsilon_T = \varepsilon$ and equation \eqref{eq:fKHE_orig} will be valid.

\end{enumerate}

We end this section with a discussion of the input term derived above. In order to make our points, we consider the application of the spectral method described in the previous chapter. Since the work function is defined as equation \eqref{eq:work_trans}, we can insert this into equation \eqref{eq:input_spectral} and perform the spatial integral analytically to find
\begin{equation}
 I(r) = 3 \int dk\ W(k)\ \left[ \frac{\sin{kr} - kr\cos{kr}}{(kr)^3} \right] \ .
\end{equation}
First, in the limit $r \to 0$ we apply equation \eqref{eq:spectral_limits} to find that
\begin{equation}
 I(0) = \int dk\ W(k) = \varepsilon_W \ ,
\end{equation}
which is the input rate by definition. Hence for stationary turbulence, we therefore see that the limiting case is indeed $I(r) \to \varepsilon$ as $r \to 0$, as was expected in point 2 above.

Second, the form of the input term is \emph{qualitatively} similar to that introduced by Gotoh, Fukayama and Nakano \cite{Gotoh:2002p627}, but the interpretation is very different. In their paper, they define the input term through the equation (in our notation)
\begin{align}
 \label{eq:KHE_Gotoh02}
 \frac{4}{3} \varepsilon r &= - \frac{1}{3} \left( 4 + r\frac{\partial}{\partial r} \right) S_3(r) + 2\nu_0 \left( 4 + r\frac{\partial}{\partial r} \right) \frac{\partial S_2(r)}{\partial r} + I_{GFN}(r) \ ,
\end{align}
where we use $I_{GFN}(r)$ to distinguish their function from ours. (Note that they actually use $W(r)$ in the paper, but this could be confused with what we call the work function, the weighted integral of the work spectrum.) With this is mind, they define
\begin{align}
 I_{GFN}(r) &= 4r \int dk\ W(k) \left[ \frac{1}{3} - \frac{\sin{kr} - kr\cos{kr}}{(kr)^3} \right] \\
      &= \frac{4r}{3} \Big( \varepsilon_W - I(r) \Big) \ .
\end{align}
Since the turbulence is stationary, $\varepsilon_W = \varepsilon$ and this term has been engineered to simply cancel the dissipation rate present on the LHS. Indeed, the result is effectively the same as for our analysis. However, we argue that the presence of the dissipation rate on the LHS is erroneous for the case of stationary turbulence since it originated as $\varepsilon_D = 0$ and does not need to be cancelled in this manner. Instead, the input term $I(r)$ arises quite naturally from a consideration of the Lin equation. As we saw above, as $r$ becomes small, $I(r) \to \varepsilon_W$ and \emph{this} is the origin of $\varepsilon$ on the LHS of the original forced KHE, equation \eqref{eq:fKHE_orig}.

Using the fact that the forcing is confined to small wavenumbers, $I_{GFN}(r)$ is approximated as
\begin{equation}
 \label{eq:Gotoh_approx}
 I_{GFN}(r) \simeq \frac{2}{15} \varepsilon_W K^2 r^3 \ , \qquad\text{where}\qquad K^2 = \frac{\int dk\ k^2 W(k)}{\int dk\ W(k)} \ ,
\end{equation}
and used in figure 13 of Gotoh \etal\ \cite{Gotoh:2002p627} to account for the difference between data for the third-order structure function and K41 form at the large scales. This correction is seen to overestimate the difference at the larger scales.

We may also compare to Sirovich, Smith and Yakhot \cite{Sirovich94}, who essentially express
\begin{equation}
 X(r) = -\frac{4}{5}\varepsilon r + \frac{6}{r^4}\int_0^r d\xi\ \xi^4 \big\langle \delta u_L(\xi) \delta f_L(\xi) \big\rangle \ ,
\end{equation}
where, in a similar manner to the longitudinal velocity increment $\delta u_L(r)$ in equation \eqref{eq:sf_def}, the force increment is defined $\delta f_L(r) = \big[ \vec{f}(\vec{x}+\vec{r},t) - \vec{f}(\vec{x},t) \big]\cdot \vec{\hat{r}}$. Once again, we see that the dissipation rate has been maintained, along with a `correction'. This is then estimated as $\simeq (2/7)\varepsilon k_0^2 r^3$, where $k_0$ is the forcing wavenumber. Indeed, integrating equation \eqref{eq:KHE_Gotoh02} using the form in equation \eqref{eq:Gotoh_approx}, the equivalent correction is found to be $\simeq (2/35)\varepsilon_W K^2 r^3$. In contrast to both of these approaches, we do not approximate the forcing term but calculate it from the full work spectrum.

\subsection{Consequences for the structure functions}\label{subsec:fKHE_SF}
\begin{figure}[tbp!]
 \begin{center}
  \includegraphics[width=0.75\textwidth]{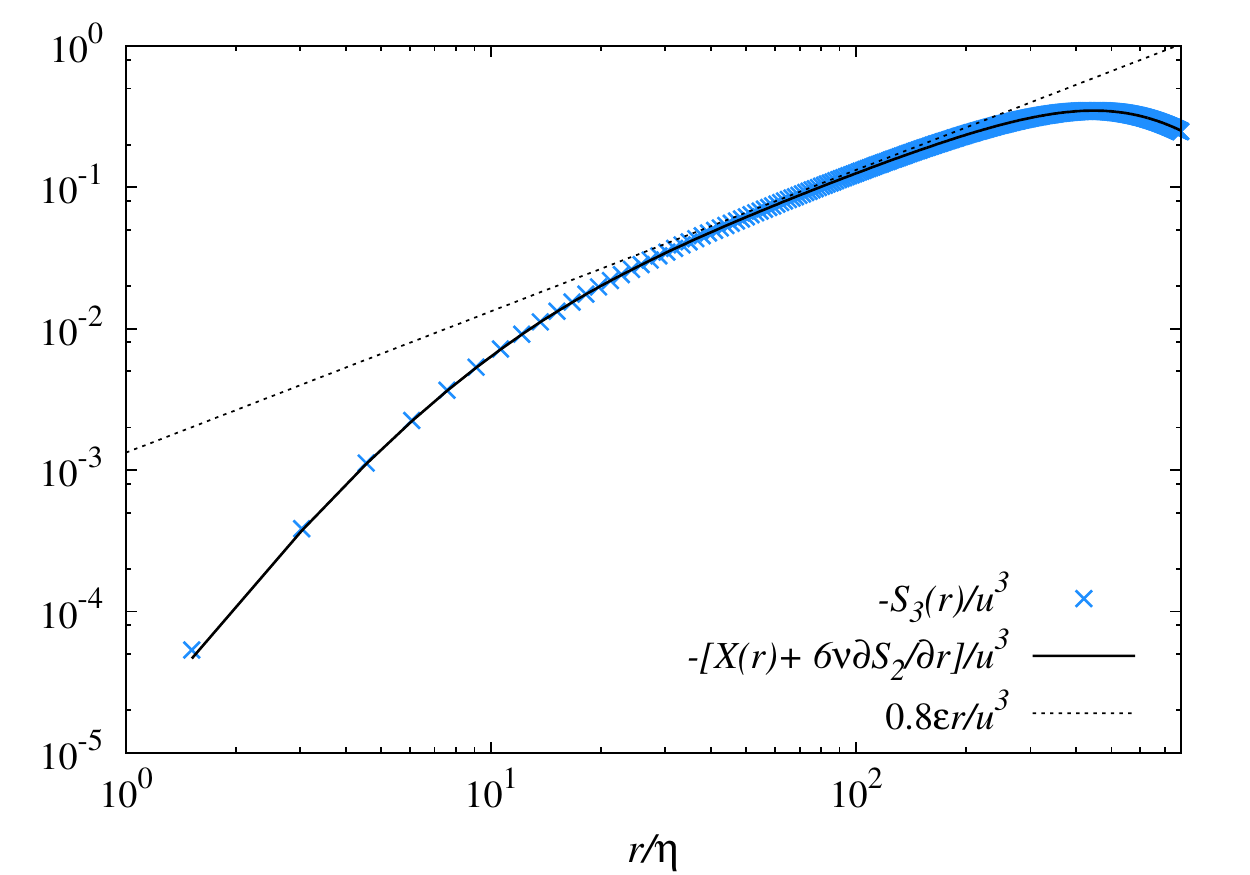}
 \end{center}
 \caption{Viscous and forcing corrections to the third-order structure function.}
 \label{fig:fKHE_SF_correct}
\end{figure}

The implication for the K41 form of the structure functions is immediately clear: Instead of satisfying equation \eqref{eq:S3_with_diss}, the third-order structure function satisfies
\begin{align}
 S_3(r) &= -\frac{4}{r^4} \int_0^r d\xi\ \xi^4 I(\xi) + 6\nu_0 \frac{\partial S_2(r)}{\partial r} \ . 
\end{align}
The deviation from K41 of the third-order structure function at small $r$ was shown in figures \ref{fig:SF_diss_correct} and \ref{ps_sf_1024a} to be accounted for by the viscous term above, and we now see that the deviation at large $r$ is due to the influence of forcing, figure \ref{fig:fKHE_SF_correct}. This is done using the spectral method of the previous chapter, with which we can express the input term as
\begin{align}
 X(r) &= -\frac{4}{r^4} \int_0^r d\xi\ \xi^4 I(\xi) \nonumber \\
 &= -12r \int dk\ W(k)\ \left[ \frac{3\sin{kr} - 3kr\cos{kr} - (kr)^2\sin{kr}}{(kr)^5} \right] \ .
\end{align}
The expression for the viscous term was given in equation \eqref{eq:visc_term_spectral}. Rather than the K41 result, the third-order structure function is expected to follow the solid black line, which it now does for all scales. From the dimensionless structure functions in figure \ref{fig:ps_sf_comp}, we see that for all Reynolds numbers studied here there is still no region which satisfies K41 for the third-order structure function and as such there is no reason why a measurement of the scaling exponent, either directly or using ESS, should match K41.
Note that if we limit ourselves to a region where the input term is constant, $I(r) = \varepsilon_W = \varepsilon$, then we trivially perform the integral to obtain the K41 result, since
\begin{align}
 -\frac{4\varepsilon}{r^4} \int_0^r d\xi\ \xi^4 = -\frac{4}{5} \varepsilon r \ .
\end{align}

\subsection{Spectral computation of the real-space energy balance}
We now extend the idea of the spectral technique to study the local energy balance expressed by the forced KHE given by equation \eqref{eq:fKHE_stationary}. To do this, we first write the equation as
\begin{equation}
 I(r) = M(r) + N(r,\nu_0) \ ,
\end{equation}
with the functions $M(r)$ and $N(r,\nu_0)$ discussed in turn. We note that $M(r)$ may also display Reynolds number dependence. First, the non-linear term is evaluated by inserting the spectral form for the structure function and performing the derivative analytically to be
\begin{align}
 M(r) &\equiv -\frac{1}{4r^4} \frac{\partial}{\partial r} \Big( r^4 S_3(r) \Big) \nonumber \\
 &= -3 \int dk\ T(k)\ \left[ \frac{\sin{kr} - kr\cos{kr}}{(kr)^3} \right] \ .
\end{align}
In the limit $r \to 0$ we see that
\begin{equation}
 M(0) = -\int dk\ T(k) = 0 \ ,
\end{equation}
since the transfer spectrum does no work on the system.

Next, the viscous term is evaluated in a similar manner to be
\begin{align}
 N(r,\nu_0) &\equiv \frac{3\nu_0}{2r^4}\frac{\partial}{\partial r} \left( r^4 \frac{\partial S_2(r)}{\partial r} \right) \nonumber \\
 &= 3 \int dk\ 2\nu_0 k^2\ E(k,t)\ \left[ \frac{\sin{kr} - kr\cos{kr}}{(kr)^3} \right] \ .
\end{align}
This time, in the limit $r \to 0$ we find
\begin{equation}
 N(0,\nu_0) = \int dk\ 2\nu_0 k^2\ E(k) = \varepsilon \ .
\end{equation}

\begin{figure}[tbp]
 \begin{center}
  \includegraphics[width=0.7\textwidth]{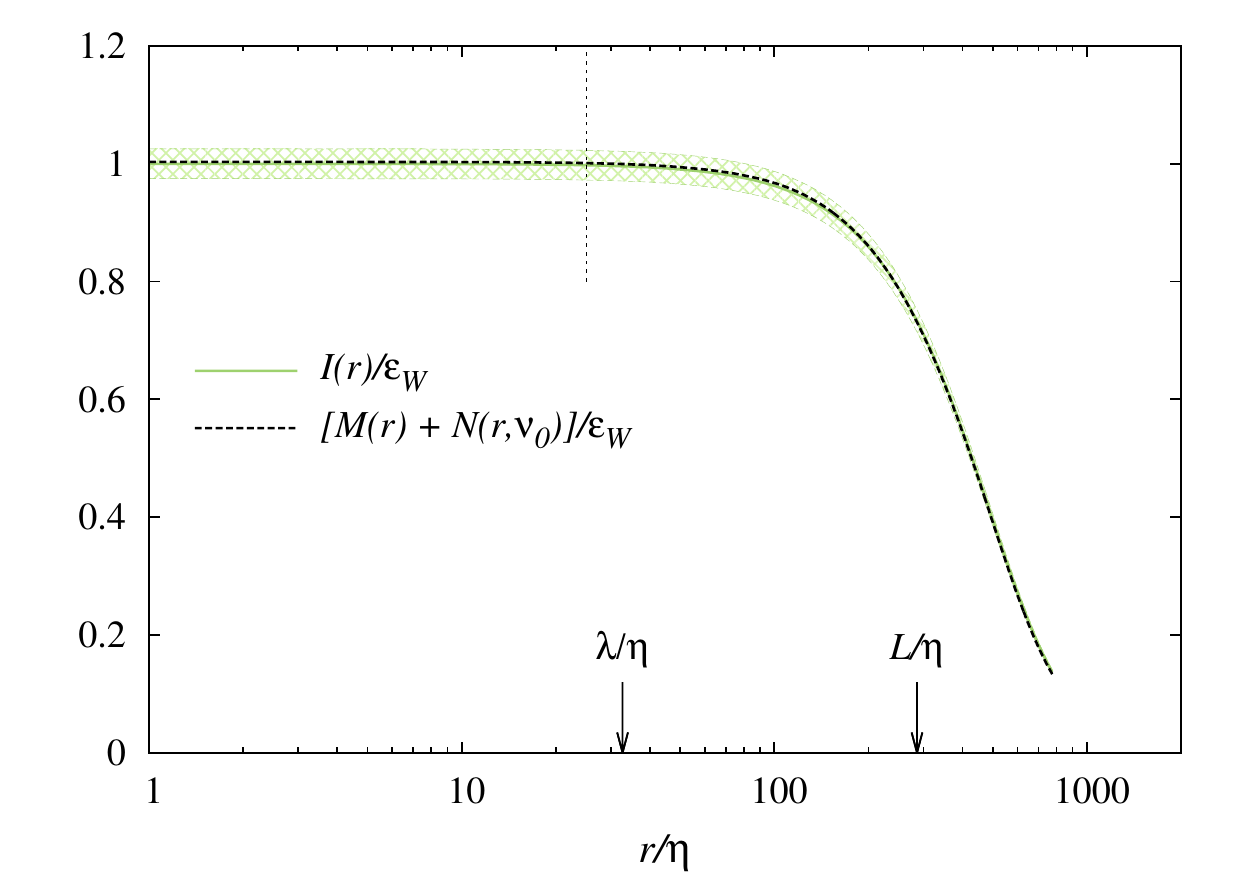}
 \end{center}
 \caption{Numerical verification of the forced KHE. Green line is the measured input term and the dashed black line the sum of the inertial and viscous terms. Vertical dotted line indicates $r_I$, with $\lambda$ and $L$ shown for comparison. Presented for run \texttt{f1024a} with $R_\lambda = 276$.}
 \label{fig:PS_KHE}
\end{figure}
\begin{figure}[tbp]
 \begin{center}
  \includegraphics[width=0.7\textwidth]{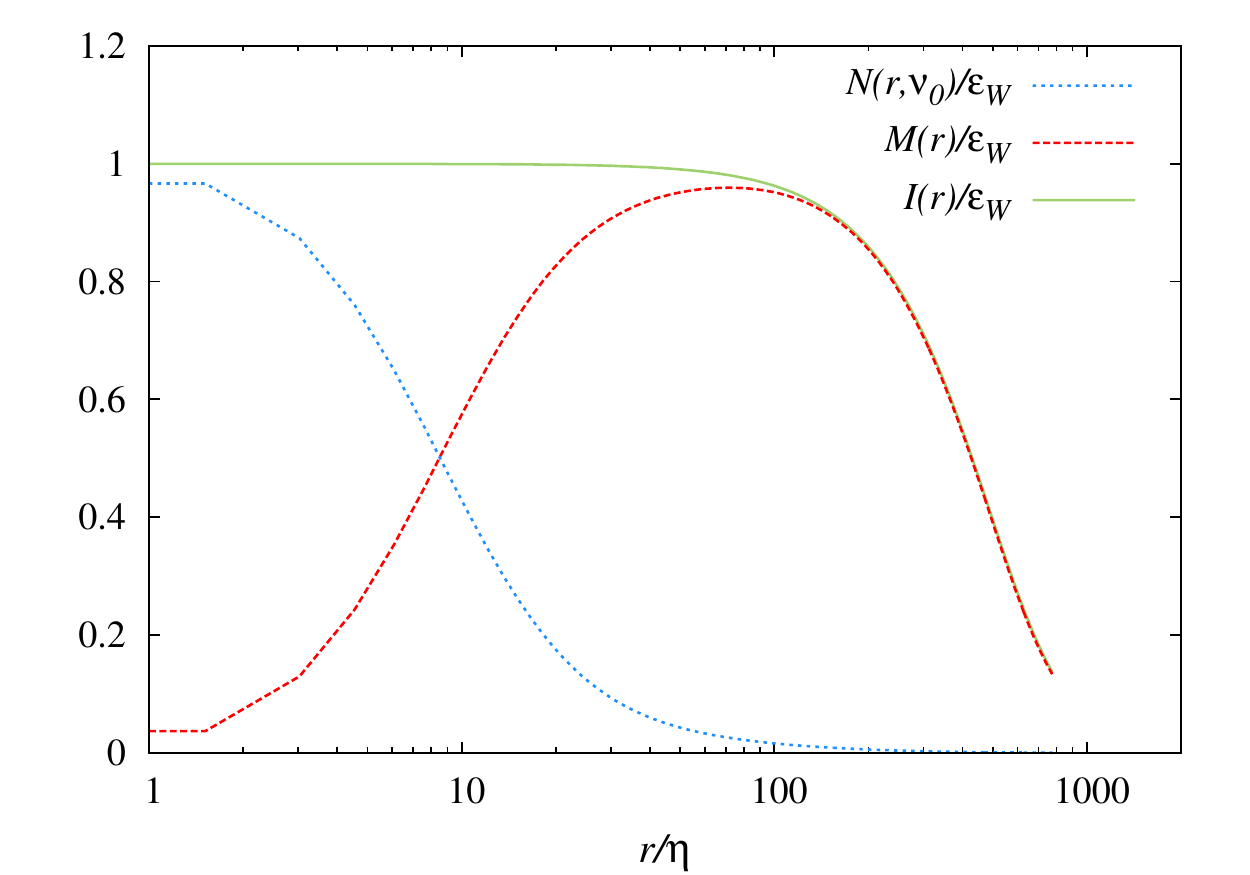}
 \end{center}
 \caption{Local energy balance between input, inertial transfer and viscous loss for the forced KHE. Presented for run \texttt{f1024a} with $R_\lambda = 276$.}
 \label{fig:PS_KHE_terms}
\end{figure}

The local energy balance is shown in figures \ref{fig:PS_KHE} and \ref{fig:PS_KHE_terms}. The first of these figures shows the balance between the input term on the LHS and the measured combination of the non-linear and dissipative terms on the RHS. The agreement is excellent, confirming that the simulation is stationary. We have estimated the error in the calculation of the input term based on the statistical error on the energy spectrum, from which it is derived. We discuss the three curves of figure \ref{fig:PS_KHE_terms} individually:
\begin{itemize}
 \item \textbf{Input term, $I(r)$}\newline
 The input term can be seen to take the value $I(r) = \varepsilon_W$ at small $r$ with the plateau extending to around $r_I = 25\eta$. This indicates the range of scales which are not directly influenced by the large-scale forcing mechanism and may be described by equation \eqref{eq:fKHE_orig}. At small scales, the input is balanced almost entirely by the viscous term, while at large $r$ the input is matched by non-linear term.
 $I(r)$ is interpreted as measuring the total energy injection rate into scales $> r$.

 \item \textbf{Non-linear term, $M(r)$}\newline
 The non-linear term can be seen to closely follow the input term down to $r \sim 100\eta$, below which it tails off to zero.
 $M(r)$ is interpreted as measuring the energy (rate) available to transfer to scales $< r$.

 \item \textbf{Viscous term, $N(r,\nu_0)$}\newline
 The viscous term shows how dissipation is confined to the small scales. As $r$ is increased, the amount of dissipation drops off and above $r = 100\eta$ there is essentially no dissipation.
 $N(r,\nu_0)$ is interpreted as measuring the total rate of energy dissipation by scales $> r$.

The development of an inertial subrange relies on there being no dependence on the forcing mechanism (a plateau for the input term) \emph{and} negligible viscous dissipation. Even for this Reynolds number, we see that the largest scale one could associate with the plateau still coincides with the order of 15\% of the dissipation and cannot really be considered part of an inertial subrange. We therefore conclude that an inertial subrange is not present at this Reynolds number.

\end{itemize}

\subsubsection{Compensating for the measured dissipation rate}\label{subsubsec:compensating}
\begin{figure}[tb!]
 \begin{center}
  \subfigure[Run \texttt{f256a}]{
   \label{sfig:PS_KHE_f256a}
   \includegraphics[width=0.475\textwidth,trim=10px 0 10px 0,clip]{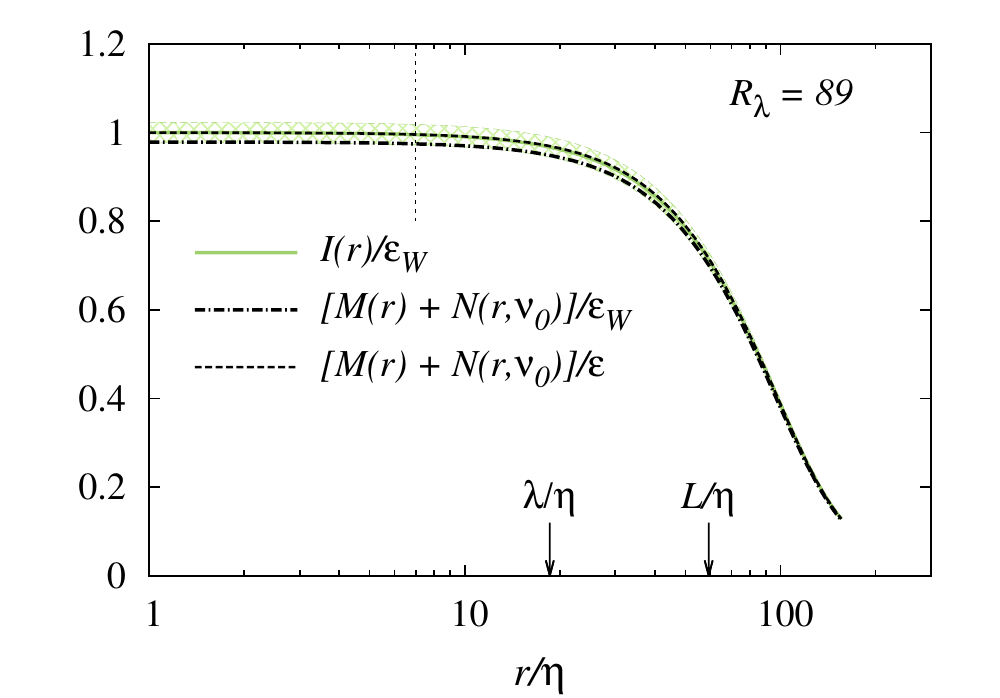}
  }
  \subfigure[Run \texttt{f1024b}]{
   \label{sfig:PS_KHE_f1024b}
   \includegraphics[width=0.475\textwidth,trim=10px 0 10px 0,clip]{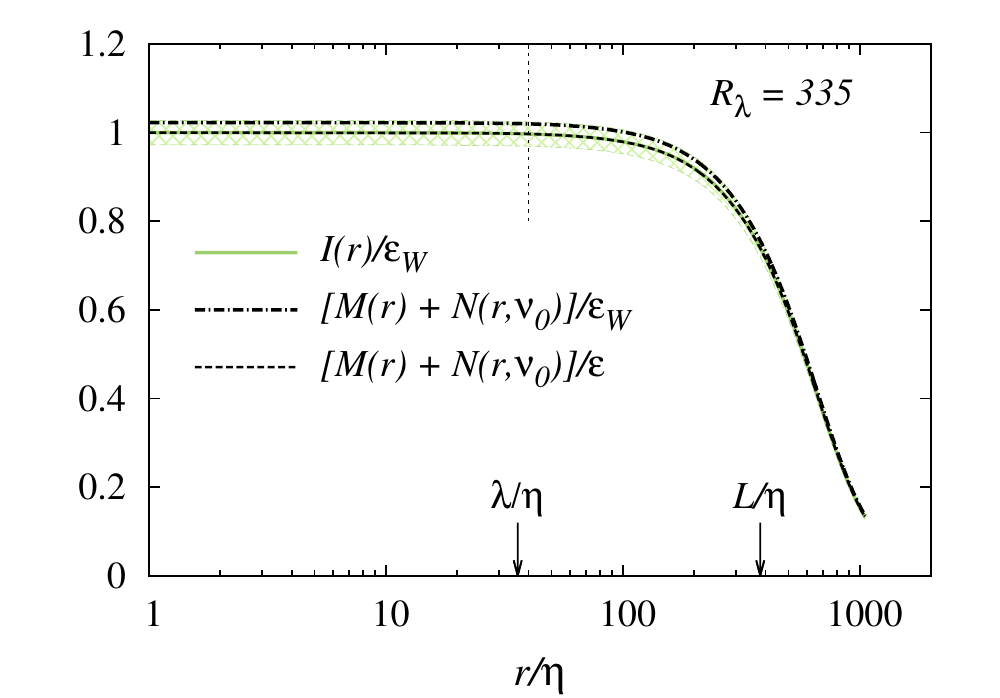}
  }
 \end{center}
 \caption{Comparison of using scaling against $\varepsilon$ and $\varepsilon_W$ when measured dissipation rate is not exactly equal to input rate. Shading represents one standard error on the mean ($\sigma/\sqrt{N}$) calculated for $I(r)$, based on the standard deviation of the energy spectrum. Vertical dotted line indicates $r_I$.}
 \label{fig:PS_KHE_not_eq}
\end{figure}

For stationary turbulence, the measured dissipation rate should be equal to the energy input rate, $\varepsilon = \varepsilon_W$. However, the dissipation rate fluctuates in time about a mean. The sampling window used to calculate $\varepsilon$ can therefore modify the mean value. For example, in the case of the $R_\lambda \sim 89$ and $R_\lambda \sim 335$ simulations, we find
\begin{align}
 \varepsilon(R_\lambda \sim 89)/\varepsilon_W &= 0.98 \nonumber \\
 \varepsilon(R_\lambda \sim 335)/\varepsilon_W &= 1.02 \ .
\end{align}
This has an undesired effect on the terms on the RHS of the modified KHE, since $M(0) + N(0,\nu_0) = \varepsilon$, and the curves sit shifted. This is seen in figure \ref{fig:PS_KHE_not_eq}. This can be rectified when the terms are scaled against the work rate (as they are in the figures) by instead scaling $M(r)$ and $N(r,\nu_0)$ with the measured dissipation rate. This can also be seen in figure \ref{fig:PS_KHE_not_eq}. This has the same result as using $\varepsilon$ instead of $\varepsilon_W$ in the calculation of $I(r)$.

\section{A model for the behaviour of $\Ceps$}\label{sec:model_DA}
We saw in figure \ref{fig:PS_KHE_terms} that the work term $I(r) = \textrm{constant}$ for a region of $r < r_I$. It is in this region that K41 could hold provided that there was negligible viscous dissipation, which was not the case. In this section we look at the effect of varying the work spectrum on the input function and try to develop a model for the inertial subrange.

\subsection{Limit of $\delta$-function forcing}
\begin{figure}[tbp]
 \begin{center}
  \includegraphics[width=0.7\textwidth]{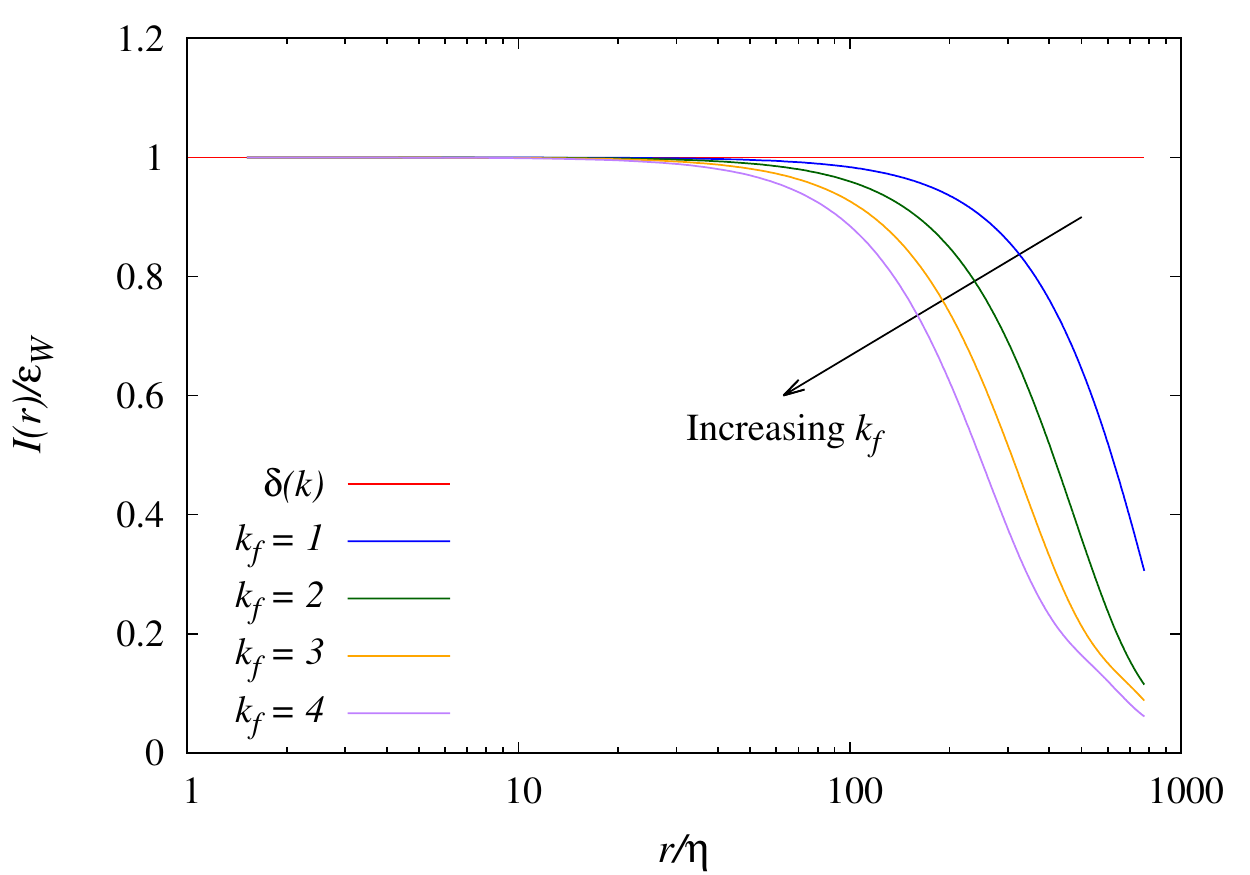}
 \end{center}
 \caption{Effect of varying the width of the forcing shell on the input term, $I(r)$. As we take $k_f \to 0$ we consider the limit of $\delta$-function forcing. Shown for run \texttt{f1024a}.}
 \label{fig:delta_forcing}
\end{figure}

In the above analysis we were restricted to a forcing mechanism based on negative damping of the lowest two wavenumber shells. To model the behaviour of the input term as we vary the \emph{thickness} of this forcing band, we consider the `flat' or top hat work spectrum defined as
\begin{equation}
 W(k) = \left\{ \begin{array}{ll}
  \frac{\varepsilon_W}{k_f} & k \leq k_f + \tfrac{1}{2} \\
  & \\
  0 & \textrm{otherwise}
  \end{array} \right. \ ,
\end{equation}
where the addition of $1/2$ is due to using $\Delta k = 1$ in the shell average. We then vary $k_f \in \mathbb{N}$ and study the change this causes to $I(r)$. From this work spectrum, the resultant input term is found to be
\begin{equation}
 I(r) = \frac{3 \varepsilon_W}{k_f} \sum_{n = 1}^{k_f} \frac{\sin{nr} - nr\cos{nr}}{(nr)^3} \ ,
\end{equation}
which is shown in figure \ref{fig:delta_forcing}.
It shows that as we increase $k_f$ the input term rolls off at lower $r/\eta$, as expected since we are including larger wavenumbers (small length-scales) in our forcing. Decreasing the width of the forcing band leads to the limit of $\delta$-function forcing at the origin, which gives
\begin{align}
 I(r) &= 3 \int dk\ \varepsilon_W \delta(k) \ \left[ \frac{\sin{kr} - kr\cos{kr}}{(kr)^3} \right] \\
 &= \varepsilon_W \ ,
\end{align}
which is a constant for all $r$. A similar result for the force autocorrelation in the limit that the forcing band is taken to zero was noted by Machiels and Deville \cite{Machiels:1998p1017}. The case of $\delta$-function forcing is not realisable in a DNS, since the mode $k = 0$ is not coupled to any other mode by the non-linear interaction and as a consequence the energy cannot be transferred. However, we can still mathematically consider the limit, in which case we would have
\begin{align}
 \varepsilon_W = M(r) + N(r,\nu_0) \ ,
\end{align}
where the $\varepsilon_W$ highlights that we are considering constant input rate, not dissipation. Of course, for stationary turbulence we must have $\varepsilon_W = \varepsilon$.

We now consider the dimensionless dissipation rate by scaling the above equation with the Taylor surrogate, $u^3/L$, thus:
\begin{align}
 \label{eq:fKHE_dim}
 \Ceps(R_L) = A_3(r) + \frac{A_2(r)}{R_L} \ ,
\end{align}
which will be referred to as the \emph{dimensionless KHE}, where the functions
\begin{align}
 A_3(r) &\equiv \frac{L}{u^3} M(r) \nonumber \\
 \frac{A_2(r)}{R_L} &\equiv \frac{L}{u^3} N(r,\nu_0) \ .
\end{align}
These can be seen on dimensional grounds, since $M(r) \sim u^3/L$ and $N(r,\nu_0) \sim \nu_0 u^2/L^2$. Once we divide by the surrogate we have $A_3(r)$ dimensionless and $N(r,\nu_0)L/u^3 \sim 1 / R_L$, which is dimensionless but we extract the explicit Reynolds number dependence.

\subsection{Functional form for the model}\label{subsec:model_func}
Let us take a moment to consider the limits which a functional form for $A_3(r)$ and $A_2(r)$ must satisfy:
\begin{enumerate}
 \item Since we are considering the limit of $\delta$-function forcing, the input term and hence $\Ceps$ are constant with respect to scale, $r$. This means that for equation \eqref{eq:fKHE_dim} to hold, the $r$ dependence must cancel on the RHS.

 \item As we take $r \to 0$, the non-linear term goes to zero and we must have $A_2(r)/R_L \to \Ceps$.

 \item As we take $r \to \infty$, we approach $k = 0$ where there is no dissipation. As such, $A_2(r)/R_L \to 0$ and $A_3(r) \to \Ceps$ for all Reynolds numbers.

 \item As $R_L \to \infty$, the viscous term must vanish such that $\Ceps = \Ceps(\infty)$ cannot be a function of $r$.
\end{enumerate}

\noindent With these limits in mind, we introduce the functional forms
\begin{align}
 \label{eq:fit_A3}
 A_3(r) &= D_3 \Big[ 1 - H(r^*) \Big] \\
 \label{eq:fit_A2}
 A_2(r)&= D_3 H(r^*) R_L + D_2 \ ,
\end{align}
where $r^* = r/\eta$ and $D_3(R_L), D_2(R_L)$ are to be found. The profile function $H(r^*)$ satisfies
\begin{equation}
 \lim_{r^*\to 0} H(r^*) = 1 \qquad\text{and}\qquad \lim_{r^*\to\infty} H(r^*) = 0 \ ,
\end{equation}
and here we use the suitable function
\begin{equation}
 H(r^*) = \frac{1}{\beta {r^*}^\alpha} \left( 1 - \exp \Big[- \beta {r^*}^{\alpha} \Big] \right)
\end{equation}
where $\alpha = \alpha(R_L), \beta = \beta(R_L)$ are fit parameters (both positive real numbers).

\noindent We address the first three limits above in turn:
\begin{enumerate}
 \item Using these functional forms, we obtain
  \begin{equation}
   \Ceps = D_3 + \frac{D_2}{R_L} \ ,
  \end{equation}
  which is independent of $r$, as required, but maintains Reynolds number dependence both explicitly and through $D_3$, $D_2$.

 \item In the limit $r \to 0$, we have
  \begin{align}
   \lim_{r \to 0} A_3(r) = 0 \qquad\text{and}\qquad \lim_{r \to 0} \frac{A_2(r)}{R_L} = D_3 + \frac{D_2}{R_L} = \Ceps \ .
  \end{align}
  The last equality following from point 1. Thus the second limit is satisfied.

  \item The limit $r \to \infty$,
  \begin{align}
   \lim_{r \to \infty} A_3(r) = D_3 = \Ceps \qquad\text{and}\qquad \lim_{r \to \infty} \frac{A_2(r)}{R_L} = \frac{D_2}{R_L} \ .
  \end{align}
  Thus we must have $D_2(R_L) = 0$.

\end{enumerate}

We now introduce a model for the dissipation rate at low Reynolds numbers (large viscosities) where not all of the dissipation passes through the cascade. In this case, energy is lost directly from the large scales. Thus, we split the dissipation rate into that which travelled through the cascade $\varepsilon_T < \varepsilon$ and that which was lost by the large scales directly, $\varepsilon_L$, such that
\begin{equation}
 \varepsilon_W = \varepsilon = \varepsilon_T + \varepsilon_L \ .
\end{equation}
By dimensional analysis, we see that $\varepsilon_T$ has dimensions of $u^3/L$ since it comes from the non-linear transfer, while $\varepsilon_L$ has dimensions $\nu_0 u^2/L^2$ since it comes from dissipation \cite{TennekesLumley:1972}. Note that both of these estimates use the characteristic length and velocity of the large scale motion. Thus we make the assumption
\begin{align}
 \varepsilon_T = C_\Pi \frac{u^3}{L} \qquad\text{and}\qquad \varepsilon_L = C_L \nu_0 \frac{u^2}{L^2} \ ,
\end{align}
where $C_\Pi, C_L$ are constants. In terms of the dimensionless dissipation rate, we have
\begin{equation}
 \label{eq:model_diss}
 \Ceps = C_\Pi + \frac{C_L}{R_L} \ .
\end{equation}
Considering the limit $r \to \infty$ once again with the above relationship in mind, we now take
\begin{equation}
 D_3(R_L) = \Ceps = C_\Pi + \frac{C_L}{R_L}
\end{equation}
as our model equation.

The final limit now gives:
\begin{enumerate}
 \setcounter{enumi}{3}
 \item In the limit of infinite Reynolds number, we see that
 \begin{equation}
  \Ceps(\infty) = C_\Pi \ ,
 \end{equation}
 which is a constant independent of Reynolds number.
\end{enumerate}
Thus, the functional forms given in equations \eqref{eq:fit_A3} and \eqref{eq:fit_A2}, with $D_2(R_L) = 0$ and our model equation $\Ceps(R_L) = \Ceps(\infty) + C_L/R_L$, satisfy all of the constraints set by the limits above. We now consider the implications this model has on the dissipation rate and behaviour of the dimensionless KHE, equation \eqref{eq:fKHE_dim}.

Note that this model is not inconsistent with the interpretation of $M(r)$ being the amount of energy available to transfer, since at large scales this \emph{can} be greater than $\varepsilon_T$ when the Reynolds number is such that not all the dissipated energy passes through the cascade. As we take $r \to \infty$ in the limit of $\delta$-function forcing, all the energy entering is available to transfer, so in this case $M(r) \to \varepsilon_W = \varepsilon_T + \varepsilon_L$.

\subsection{Consequences for the dissipation anomaly}\label{subsec:model_DA}

\begin{figure}[tb!]
 \begin{center}
   \includegraphics[width=0.75\textwidth,trim=0 10px 0 0, clip]{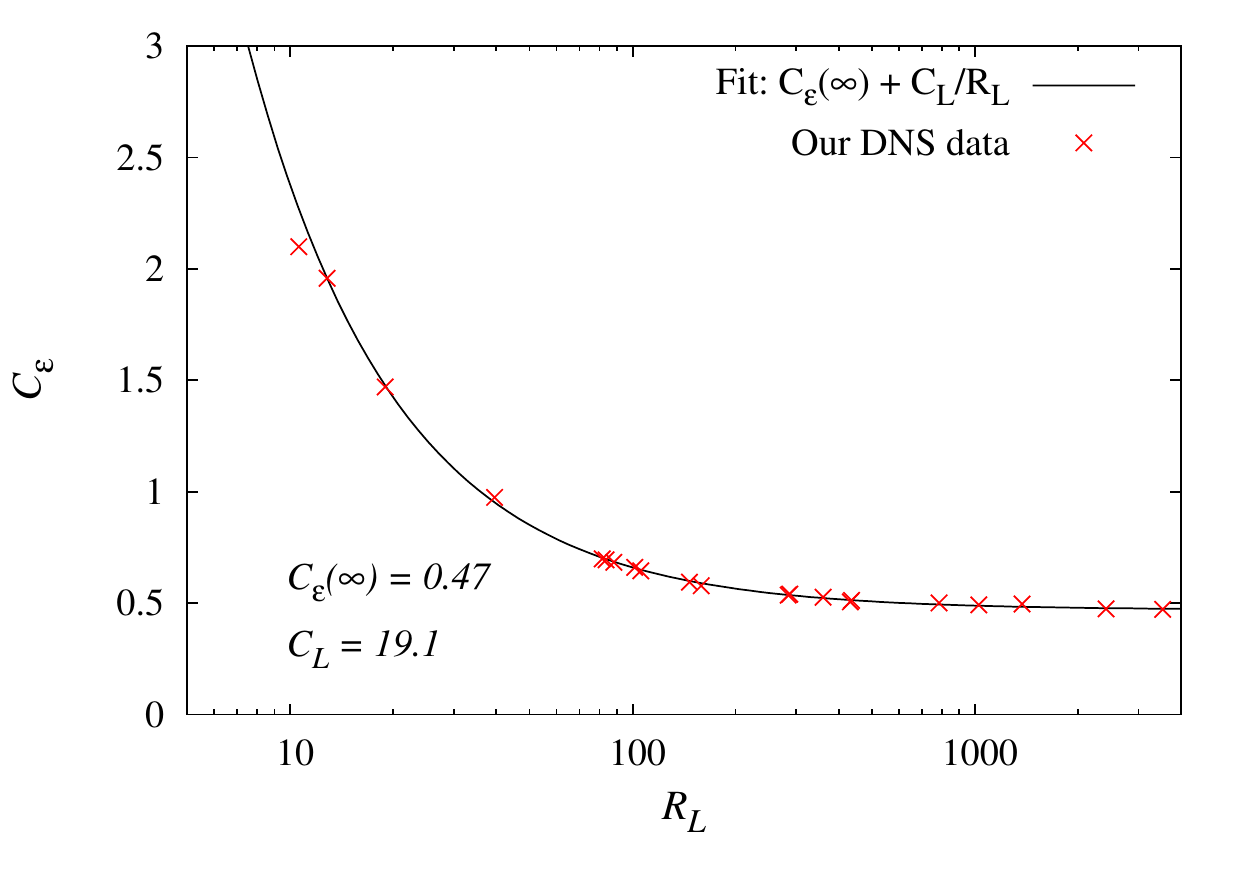}
 \end{center}
 \caption{Fit of the model equation $\Ceps = \Ceps(\infty) + C_L/R_L$ to DNS data.}
 \label{fig:Ceps_RL}
\end{figure}

The model for the dissipation rate introduced above has direct implications for the study of the dissipation anomaly. For the limit of $\delta$-function forcing, the expression
\begin{equation}
 \label{eq:model}
 \Ceps(R_L) = \Ceps(\infty) + \frac{C_L}{R_L} \ ,
\end{equation}
suggests that plotting the measured values of $\Ceps$ against integral scale Reynolds number, rather than Taylor-Reynolds number, is more appropriate. Indeed, this was noted by Batchelor \cite{Batchelor:1953-book} in a footnote as being the most appropriate abscissa, only the data were not available to him. The above equation has been fitted to the DNS data for $\Ceps$ using the parameters $\Ceps(\infty)$ and $C_L$, and figure \ref{fig:Ceps_RL} shows the fit. The data is seen to agree with the model equation very well, over a large range of $R_L$. The original formulation of this relationship based on dimensional analysis was presented in McComb, Berera, Salewski and Yoffe \cite{McComb:2010p1601}.

This should be compared to other analytic works which tend to try to fit the data when plotted against $R_\lambda$. Doering and Foias \cite{Doering:2002p135} derived an upper bound on the dimensionless dissipation coefficient of the form
\begin{equation}
 \Ceps \leq \frac{b}{2} \left( 1 + \sqrt{1 + \frac{4a}{b^2} \frac{1}{R_\lambda^2}} \right) \ ,
\end{equation}
which motivated the fit $A[1+\sqrt{1 + (B/R_\lambda)^2}]$ used by Donzis \etal\ \cite{Donzis:2005p855}. A mean-field closure of the K\'arm\'an-Howarth equation was used by Lohse \cite{Lohse:1994p149} to obtain an approximate form for $\Ceps$, again in terms of $R_\lambda$.

\begin{figure}[tb!]
 \begin{center}
  \includegraphics[width=0.75\textwidth]{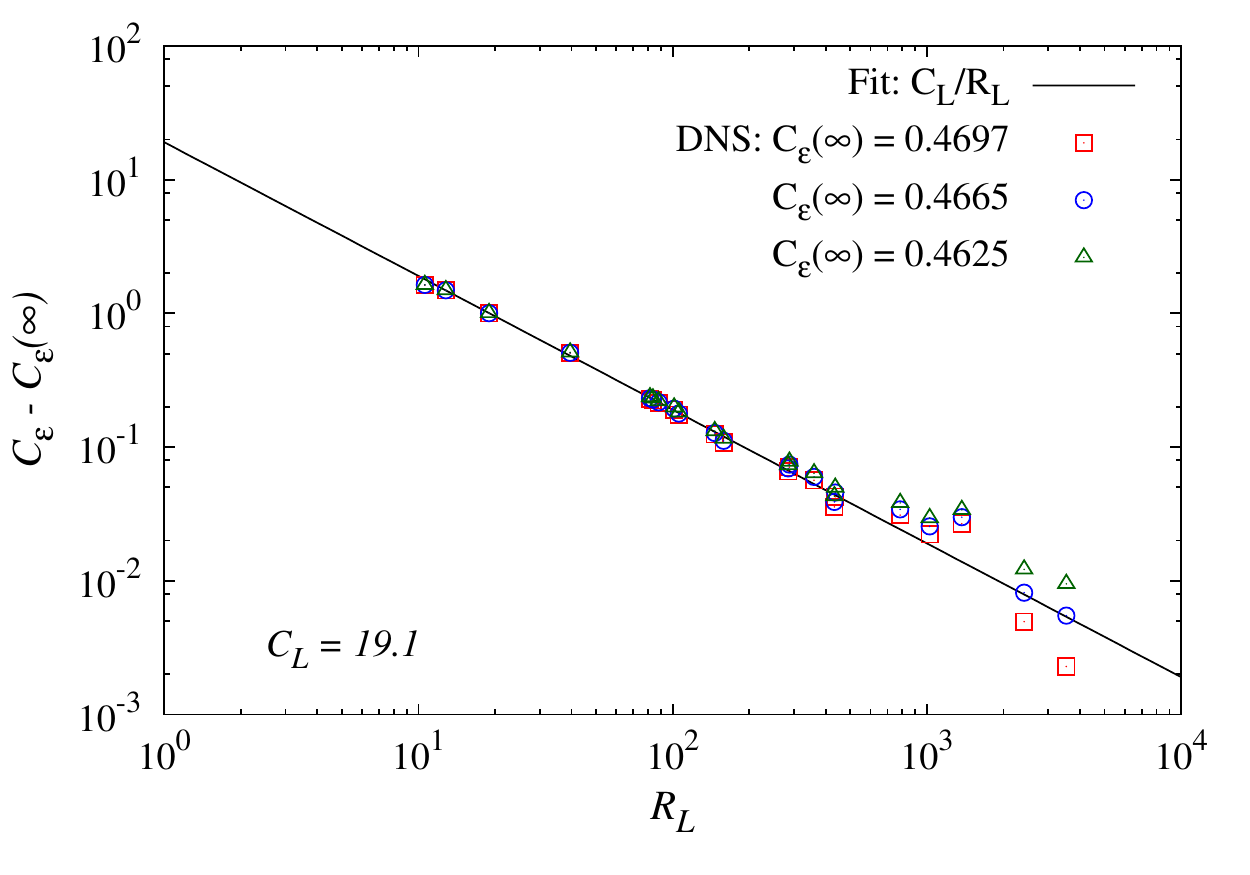}
 \end{center}
 \caption{Log-log plot of the DNS data for $\Ceps$ with the plateau value subtracted to highlight the $1/R_L$ behaviour. The effect of variation of the value of the asymptote $\Ceps(\infty)$ by a small amount is shown. $\Ceps(\infty) = 0.4697$ is the value measured by the fit.}
 \label{fig:Ceps_RL_log}
\end{figure}

To confirm the $1/R_L$ behaviour, we perform two tests: First, the more general form $A + B R^{-n}_L$ was fitted to the DNS data and found $n = 0.9997$. Second, we consider subtracting the value of the asymptote from the DNS data and plotting on logarithmic scales. This is shown in figure \ref{fig:Ceps_RL_log}. For lower Reynolds numbers, the data for $\Ceps - \Ceps(\infty)$ clearly obeys $R_L^{-1}$ power-law behaviour. The cause of the fluctuations at larger $R_L$ is currently unknown. Variation of $\Ceps(\infty)$ to slightly lower values than the fit (since the curve cannot turn up) is also shown in the figure. By reducing the value of $\Ceps(\infty)$, the last two, highest Reynolds number points can be made to sit on the power law at the expense of the three data points around $R_L = 1,000$.

\subsection{DNS fit to the functional forms}
In section \ref{subsec:model_func} we introduced a model for the behaviour of the dimensionless dissipation rate and the non-linear and viscous terms. Section \ref{subsec:model_DA} then showed how the model dissipation equation \eqref{eq:model_diss} could be fit to DNS data for the dissipation anomaly with good results. Using the values $\Ceps(\infty) = \CPiVAL,\ C_L = \CLVAL$ found in the dissipation anomaly fit, we now study the behaviour of the functional forms found above,
\begin{align}
 \label{eq:fitA3}
 A_3(r) &= \left(\Ceps(\infty) + \frac{C_L}{R_L} \right) \left[ 1 - \frac{1}{\beta} \left(\frac{r}{\eta}\right)^{-\alpha} \left( 1 - \exp \left[- \beta \left(\frac{r}{\eta}\right)^{\alpha} \right] \right) \right] \\
 \label{eq:fitA2}
 A_2(r)&= \left\{ \left(\Ceps(\infty) + \frac{C_L}{R_L} \right)\frac{1}{\beta} \left(\frac{r}{\eta}\right)^{-\alpha} \left[ 1 - \exp \left(- \beta \left(\frac{r}{\eta}\right)^{\alpha} \right) \right]\right\} R_L \ ,
\end{align}

As mentioned above, DNS cannot explore $\delta$-function forcing since the mode $k = 0$ corresponds to a uniform translation of the entire system and cannot exchange energy with any other mode. The simulations performed here instead use a forcing method with energy injected into the lowest two wavenumber shells. Therefore, we cannot directly compare the functional forms above with our DNS data. However, since the input term $I(r)$ is a constant in the $\delta$-function forcing limit, we consider attempting to fit the functions (via the parameters $\alpha, \beta$; $\Ceps(\infty), C_L$ already fixed) in the $r < r_I$ plateau region seen in figure \ref{fig:PS_KHE_terms}.

\begin{figure}[tbp]
 \begin{center}
  \subfigure[Fits for $A_3(r)$ and $A_2(r)/R_L$. Parameters were fitted in the region $1 \leq r \leq 25\eta$.]{
   \label{sfig:fitA3_f1024a}
   \includegraphics[width=0.9\textwidth,trim=0 55px 40px 50px, clip]{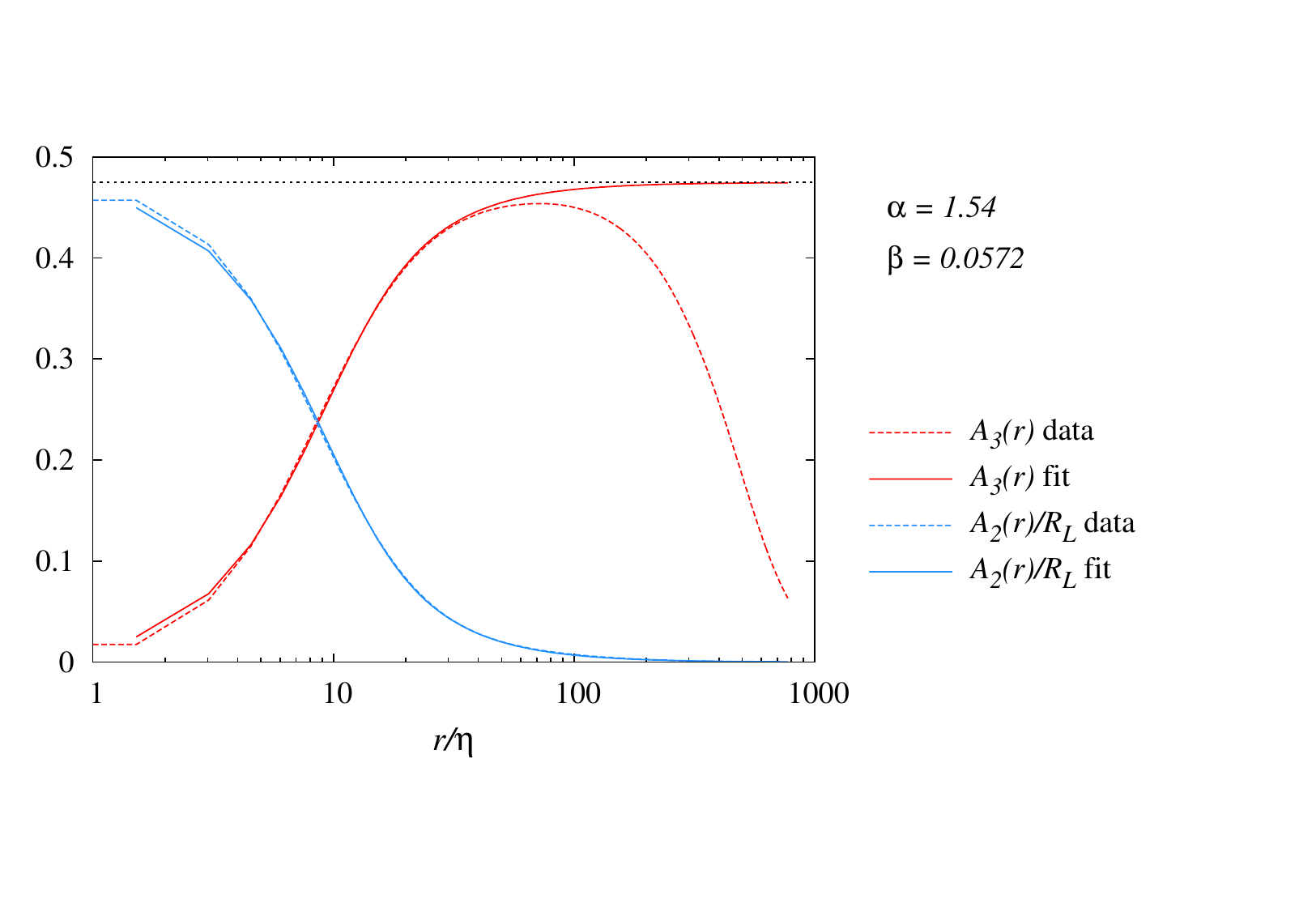}
  }
  \subfigure[Fits for $A_3(r)$ and $A_2(r)/R_L$ as above, scaled using $I(r)/\varepsilon_W$ to compensate for the finite forcing effects.]{
   \label{sfig:fitA3_f1024a_scaled}
   \includegraphics[width=0.9\textwidth,trim=0 55px 40px 20px, clip]{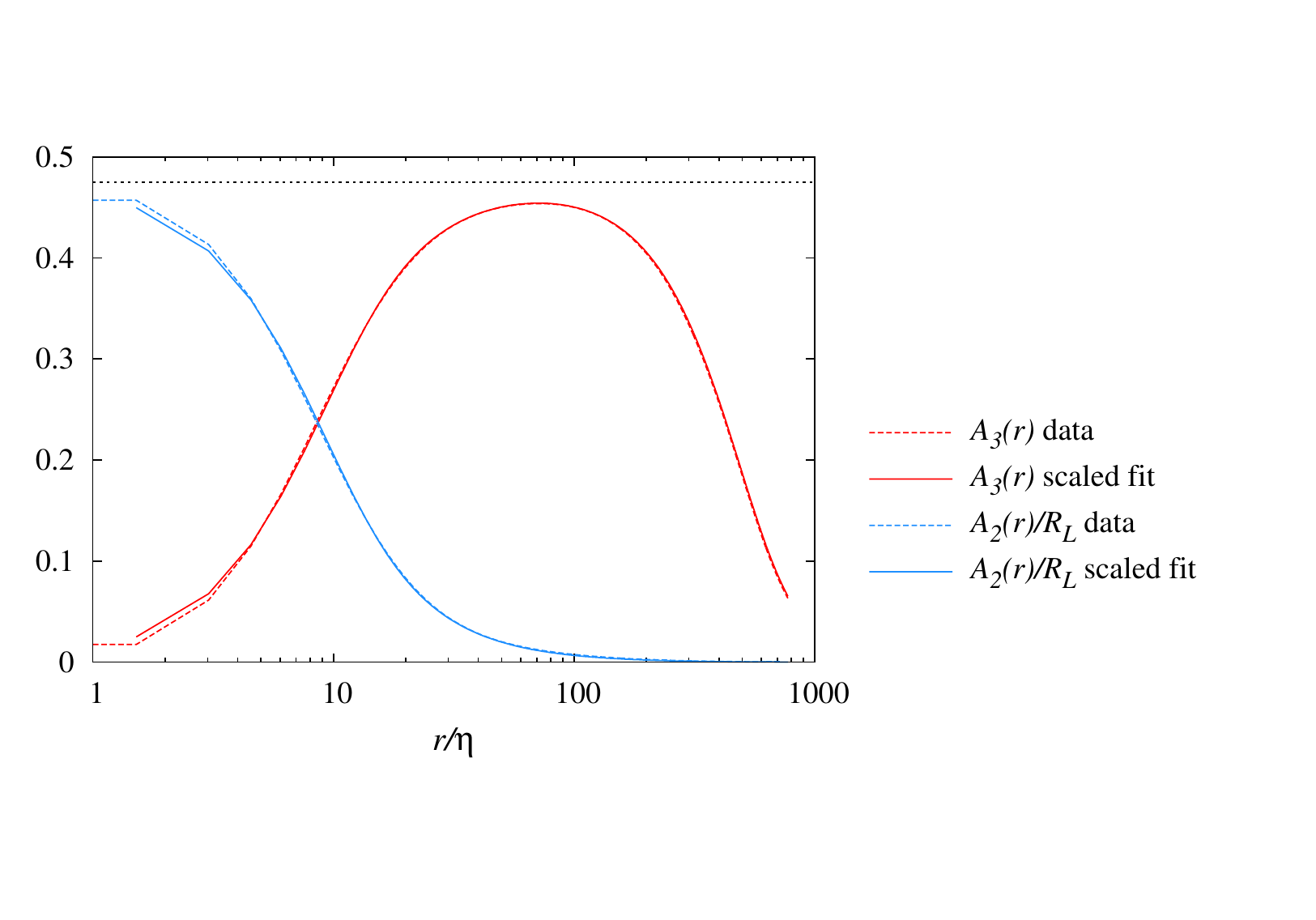}
  }
 \end{center}
 \caption{Fit of the function $A_3(r)$ to DNS data from run \texttt{f1024a} with $R_\lambda \sim 276$. Fit parameters then used to plot $A_2(r)/R_L$. Dotted line represents the measured value of $\Ceps$ from the DNS data.}
 \label{fig:fitA3_f1024a}
\end{figure}

Figure \ref{sfig:fitA3_f1024a} shows the functional forms given in equations \eqref{eq:fit_A3} and \eqref{eq:fit_A2} with the parameters $\alpha$ and $\beta$, fit in the region $1 \leq r \leq 25\eta$ for run \texttt{f1024a}. Note that the functions are not independent and the fit is performed for only one of the two, in this case $A_3(r)$. As can be seen, the fit for $r < 25\eta$ matches the DNS data very well. However, as expected, at larger $r$ the data and fits disagree. Since this is caused by the input term tailing off due to the finite forcing band, we can approximate a correction to this effect by scaling with
\begin{equation}
 A_3(r) \longrightarrow \frac{I(r)}{\varepsilon_W} A_3(r) \qquad\quad\textrm{and}\qquad\quad
 A_2(r) \longrightarrow \frac{I(r)}{\varepsilon_W} A_2(r) \ .
\end{equation}
This will not change the asymptotic value in the fit range, where $I(r) = \varepsilon_W = \varepsilon$, but it is hoped that its affect on the larger scales compensates for the effects of finite forcing. This is shown in figure \ref{sfig:fitA3_f1024a_scaled}. The agreement for all scales is very good.

\begin{figure}[tbp]
 \begin{center}
  \subfigure[Fits for $A_3(r)$ and $A_2(r)/R_L$. Parameters were fitted in the region $1 \leq r \leq 14\eta$.]{
   \label{sfig:fitA3_f512a}
   \includegraphics[width=0.9\textwidth,trim=0 55px 40px 50px, clip]{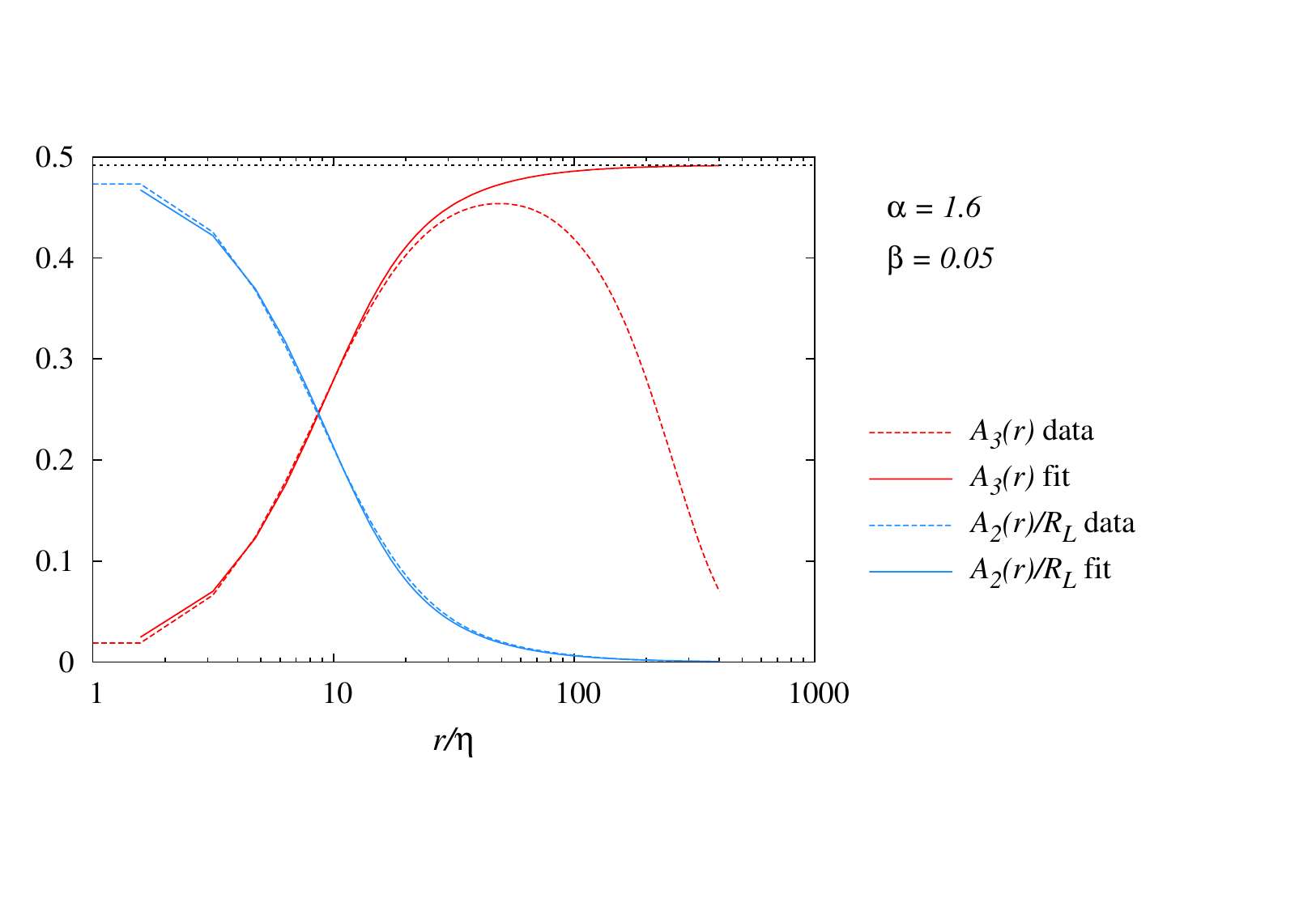}
  }
  \subfigure[Fits for $A_3(r)$ and $A_2(r)/R_L$ as above, scaled using $I(r)/\varepsilon_W$ to compensate for the finite forcing effects.]{
   \label{sfig:fitA3_f512a_scaled}
   \includegraphics[width=0.9\textwidth,trim=0 55px 40px 20px, clip]{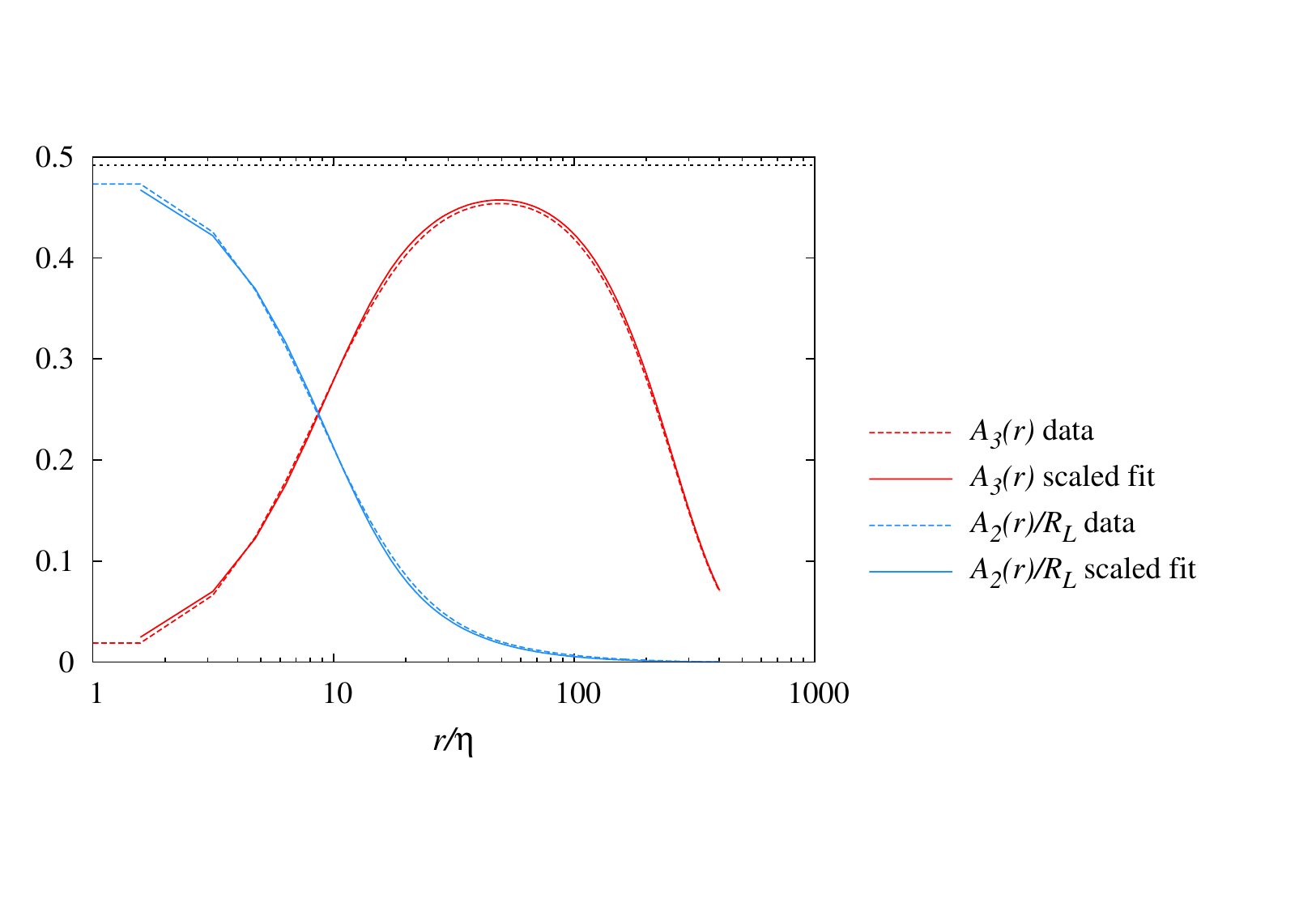}
  }
 \end{center}
 \caption{Fit of the function $A_3(r)$ to DNS data from run \texttt{f512a} with $R_\lambda \sim 177$. Fit parameters then used to plot $A_2(r)/R_L$. Dotted line represents the measured value of $\Ceps$ from the DNS data.}
 \label{fig:fitA3_f512a}
\end{figure}

\begin{figure}[tbp]
 \begin{center}
  \subfigure[Fits for $A_3(r)$ and $A_2(r)/R_L$. Parameters were fitted in the region $1 \leq r \leq 40\eta$.]{
   \label{sfig:fitA3_f1024b}
   \includegraphics[width=0.9\textwidth,trim=0 55px 40px 50px, clip]{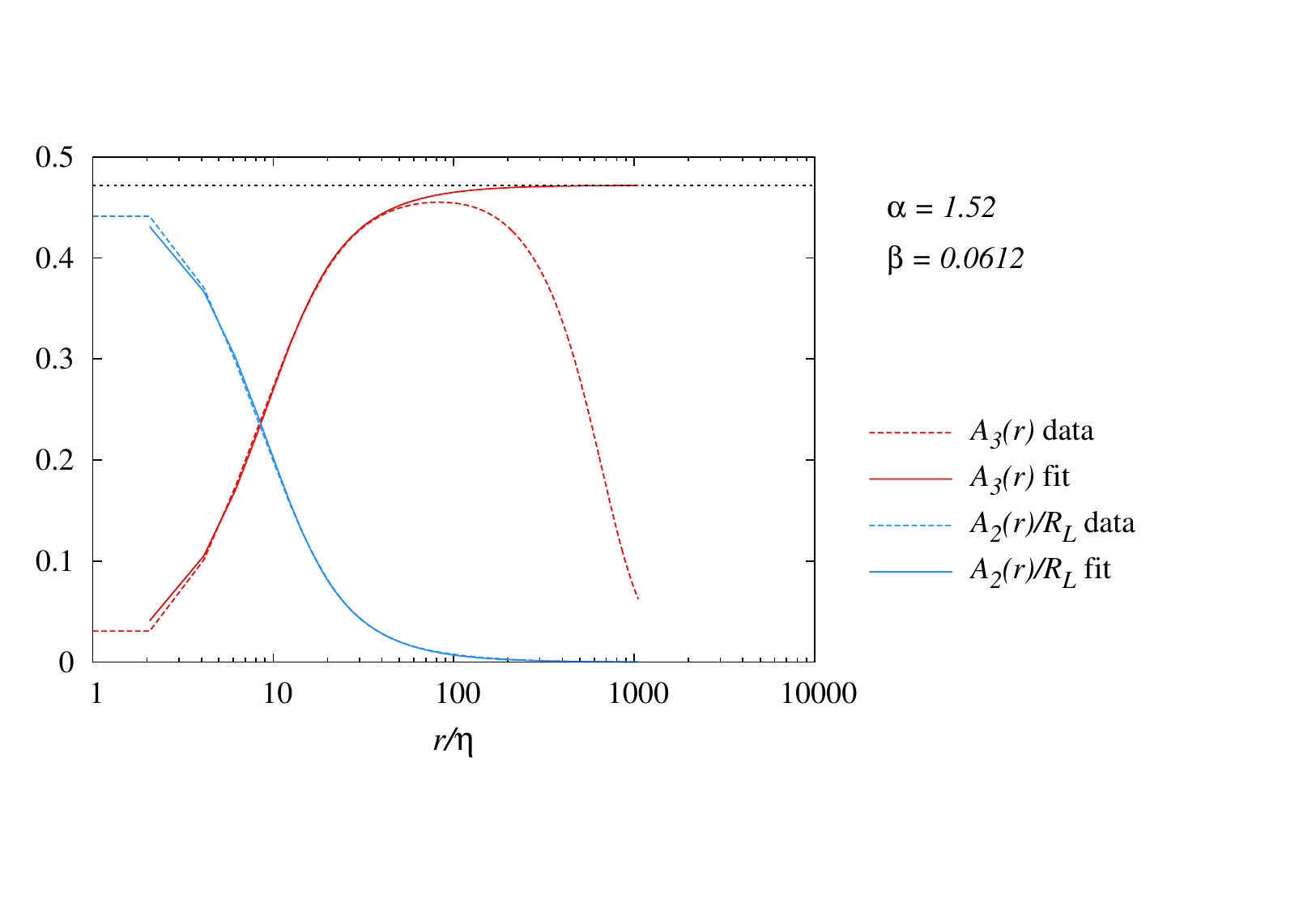}
  }
  \subfigure[Fits for $A_3(r)$ and $A_2(r)/R_L$ as above, scaled using $I(r)/\varepsilon_W$ to compensate for the finite forcing effects.]{
   \label{sfig:fitA3_f1024b_scaled}
   \includegraphics[width=0.9\textwidth,trim=0 55px 40px 20px, clip]{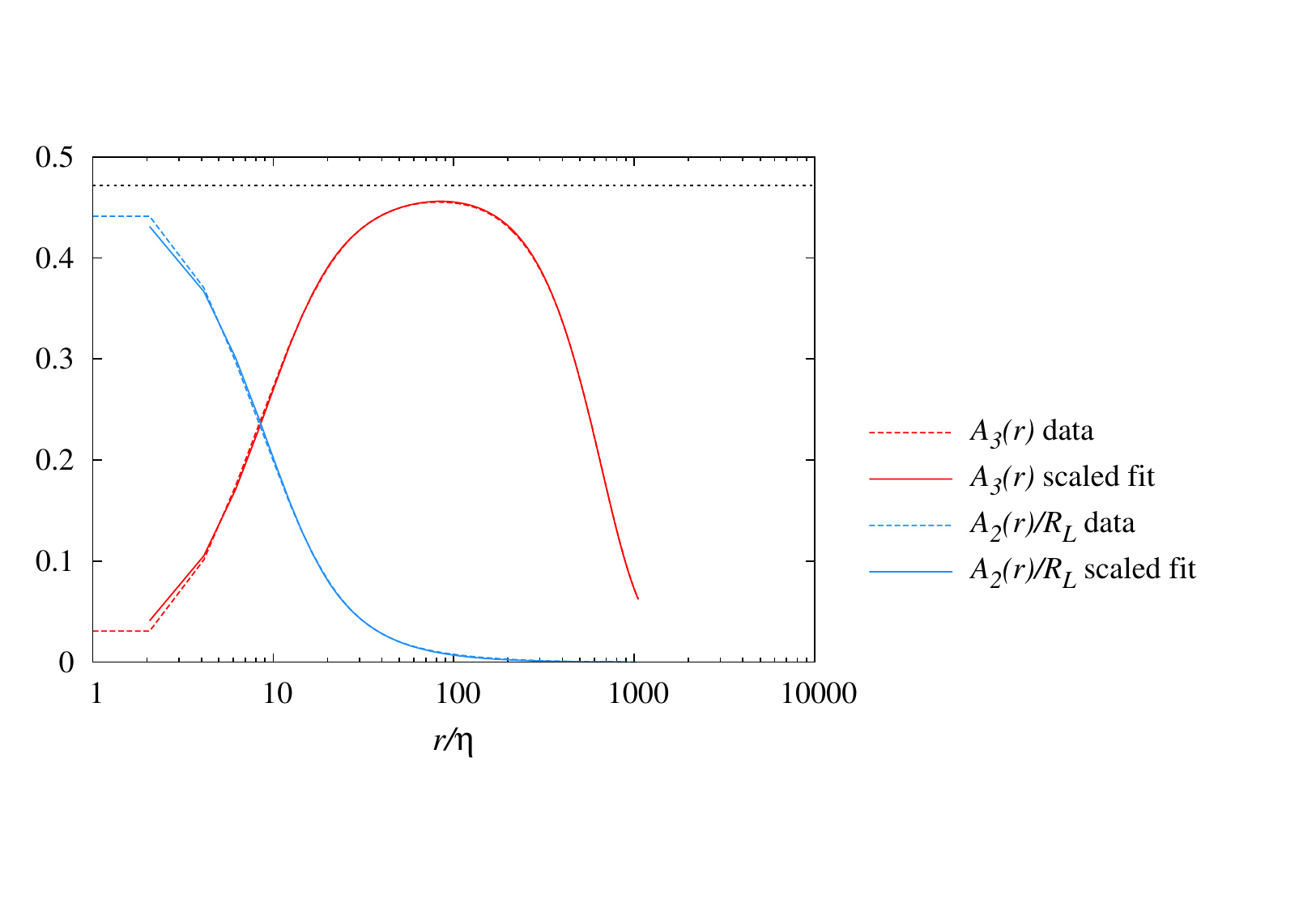}
  }
 \end{center}
 \caption{Fit of the function $A_3(r)$ to DNS data from run \texttt{f1024b} with $R_\lambda \sim 335$. Fit parameters then used to plot $A_2(r)/R_L$. Dotted line represents the measured value of $\Ceps$ from the DNS data.}
 \label{fig:fitA3_f1024b}
\end{figure}

The same analysis has been done for a lower Reynolds number, run \texttt{f512a}, and our highest Reynolds number, run \texttt{f1024b}. These are shown in figures \ref{fig:fitA3_f512a} and \ref{fig:fitA3_f1024b}, respectively. It is difficult to perform this analysis for even lower Reynolds numbers since the input term does not have an extended plateau large enough to perform reliable fits, for example see figure \ref{sfig:PS_KHE_f256a} for run \texttt{f256a}. We also see that the exponent $\alpha$ appears to decrease with increasing Reynolds number, while $\beta$ increases.

\section{Discussion}
Results from direct numerical simulation have been used to study the Reynolds number dependence of the dimensionless dissipation coefficient for both forced and decaying isotropic turbulence. In the case of stationary turbulence, we find results very much consistent with the literature, with a plateau value of $\Ceps(\infty) \simeq 0.5$. We also highlight, by maintaining a constant energy input rate, how the variation of $\Ceps$ for forced turbulence is due to the variation of $u^3/L$, in contrast to decaying turbulence. Future work using variation of the initial energy spectrum and forcing scheme would be required to determine the universality of $\Ceps(\infty)$ for forced turbulence.

For decaying turbulence, we have shown how the choice of measurement time has a big impact on the behaviour of $\Ceps$. This was shown using the time series, which develops a plateau for sufficiently large Reynolds number. The implications of measurement at early evolved time and those which lie on the plateau were discussed. We conclude that $\Ceps(\infty)$ can take a range of values, depending on the choice of evolved time criteria. Measurement using the peak dissipation rate is shown to give good agreement with the $\Ceps$ curve obtained from forced turbulence, while for later evolved times the value of $\Ceps(\infty)$ increases to a maximum (corresponding to the beginning of the plateau observed in the time-series) closer to unity, before decreasing one more. Measurement at significantly later times would be interesting, as we expect $\Ceps(\infty)$ would remain above the result for forced turbulence.

A review of the K\'arm\'an-Howarth equation is presented and its use for stationary turbulence questioned. We then derive a new form of the KHE, which describes both decaying and forced turbulence, using a transformation of the Lin equation. The input term is shown to reproduce the standard KHE for stationary turbulence in the limit of large Reynolds number. The interpretation of the new input term is discussed, before its consequence for the third-order structure function is considered. This is done using the spectral method developed in the previous chapter, where the structure functions (and their derivatives with respect to $r$) are computed from the energy and transfer spectra. We then show how using the new input term accounts for the deviation of $S_3(r)$ from K41 at the large scales.

The spectral method is then used to study the individual terms in the new KHE and the local energy balance they represent. Noting the plateau $I(r) = \varepsilon_W$ at small scales $r < r_I$, we develop an analytic model for the behaviour of these terms in the limit of $\delta$-function forcing, or maximum separation of scales. Fit to DNS data in the plateau region (where available) shows good agreement for the small scales. Scaling the fit functions with the measured input term, $I(r)/\varepsilon_W$, leads to excellent agreement of the model and DNS data for all length-scales. This is then used to model the behaviour of $\Ceps(R_L)$. The development of a similar model for decaying turbulence, able to take into account $\partial S_2(r)/\partial t$, would be of direct interest and could help identify the cause of the difference between the behaviour of $\Ceps$ for forced and decaying turbulence.

In contrast to current analytical work, which attempt to describe $\Ceps = \Ceps(R_\lambda)$, our model equation $\Ceps(R_L) = \Ceps(\infty) + C_L/R_L$ uses the integral-scale Reynolds number. This is then fitted to the DNS data obtained for forced turbulence with excellent agreement. The fit finds a value of $\Ceps(\infty) = 0.47$, consistent with the literature. Development of a model equation for $\Ceps$ in decaying turbulence would be useful for comparison to experimental work.

To our knowledge, the interpretation of equation \eqref{eq:fKHE} is unique and its consequences are of interest to the study of structure functions. The input term could potentially be evaluated in physical-space, allowing comparison to the spectral evaluation to be made. Furthermore, it could possibly be measured in experimental flows; although, realising stationary isotropic turbulence in a laboratory would present a major challenge.

\chapter{Renormalization methods and their application to turbulence}\label{sec:RG}

\section{The renormalization group}

\begin{figure}[tb]
 \begin{center}
  \includegraphics[width=0.85\textwidth]{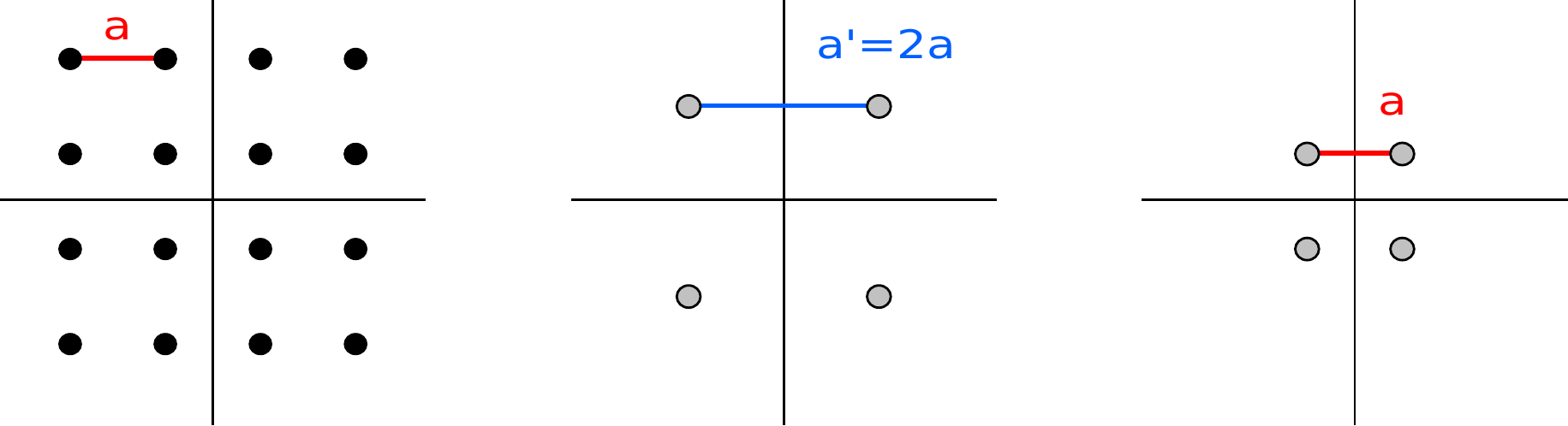}
 \end{center}
 \caption{Illustration of the steps of RG procedure. (Left) Original system; (centre) after removal of scales; and (right) after rescaling.}
 \label{fig:RG_illus}
\end{figure}

Renormalization can be considered as a coarse-graining procedure, where we probe the system at progressively lower resolution. In doing so, we remove small length-scales so that only the behaviour of the large scales is retained. The goal is to develop an effective theory for these large scales which takes into account the action of the small, removed scales by modifying the parameters of the theory. The original or `bare' parameters become renormalized or `dressed' to account for the lost scales. The system is then rescaled and if the resulting system looks the same (that is, the governing equations have the same form) then it is said to be invariant under the renormalization group. This is summarised in figure \ref{fig:RG_illus}.
In momentum space, renormalization corresponds to removing or integrating out a band of high wavenumber modes.

The renormalization group (RG) has enjoyed huge success in condensed matter systems, the theory of critical phenomena and quantum field theory. For example, in the theories of quantum electrodynamics (QED) and quantum chromodynamics (QCD), RG is used to absorb infinities and study the running of the fundamental couplings as the (energy/momentum) scale at which the system is observed is changed.

As we perform this systematic removal of scales, the strength of the couplings between modes can be changed, but new `effective' interactions can also be introduced. If we consider a discrete system with only local, nearest neighbour interactions between sites on the lattice, then in one dimension the effective coupling after the renormalization is essentially the coupling to the next nearest neighbour of the original lattice, modified in some way to account for the removed site:
\begin{center}
 {\includegraphics[width=0.296\textwidth]{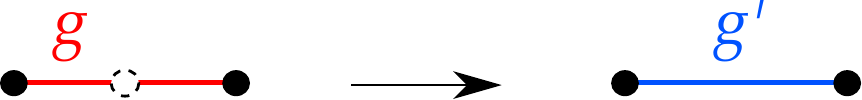}}\ .
\end{center}
In more than one dimension, while our original theory only possesses nearest-neighbour interactions, we can introduce new interactions:
\begin{center}
 {\includegraphics[width=0.6\textwidth]{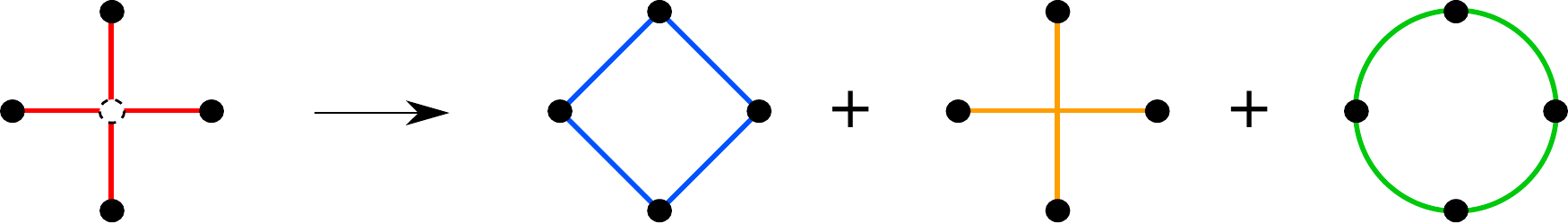}}\ .
\end{center}
We have introduced nearest diagonal neighbour, next nearest neighbour and a plaquette coupling. This is an example of \emph{proliferation} of couplings, where the RG process has generated additional interactions between the remaining degrees of freedom which are required such that our new effective theory can describe the behaviour of the retained scales. As the RG is iterated, it is possible that the values of the couplings stop depending on the cut-off scale, in which case we have reached a \emph{fixed point} of the theory. This is highlighted by the Callan-Symanzik beta function for the coupling $g_i$,
\begin{equation}
 \beta(g_i) = \frac{\partial g_i(\mu)}{\partial \log{\mu}} \ , \hspace{0.5in} \beta(g_i^*) = 0 \ ,
\end{equation}
where $\mu$ is the renormalization scale (UV cut-off) being considered and $g_i^*$ is the fixed point value.

The operators corresponding to the new interactions can be assessed to determine what effect they have on the effective theory as the renormalization progresses; in other words, whether the theory is renormalizable or not. This is basically done by studying the dependence of the operators on the RG scale using power counting. If the operator diverges as we move to the infrared (low wavenumbers), it is said to be a relevant operator since it becomes a dominant contribution; whereas, if it vanishes it is irrelevant. These are identified by couplings where the momentum scale dependence is through a negative power. For a theory to be renormalizable, it should contain no relevant effective couplings. If the operator does not present any dependence on the scale, it is marginal.

A good introduction to renormalization and the RG is given in McComb \cite{mccomb:2004-book}, Le~Bellac \cite{LeBellac:1991-book} and Zinn-Justin \cite{zinn-justin:2002-book}.

\subsection{Application to turbulence}
The running of the couplings allow us to identify where perturbation theory is valid and where it breaks down. This is because a perturbation series involves a truncated expansion in increasing powers of the coupling. When the coupling is small, higher-order corrections become smaller and smaller and the system is well described by retaining just a few terms. On the other hand, if the problem is strongly coupled, the series diverges and perturbation theory cannot represent the system. It has been shown that the QED coupling, $\alpha$, becomes small in the IR while for QCD the coupling $\alpha_S$ becomes small in the UV. Thus perturbation theory describes low-energy QED (extremely) well but fails for QCD.

\begin{figure}[tb]
 \begin{center}
  \includegraphics[width=0.85\textwidth]{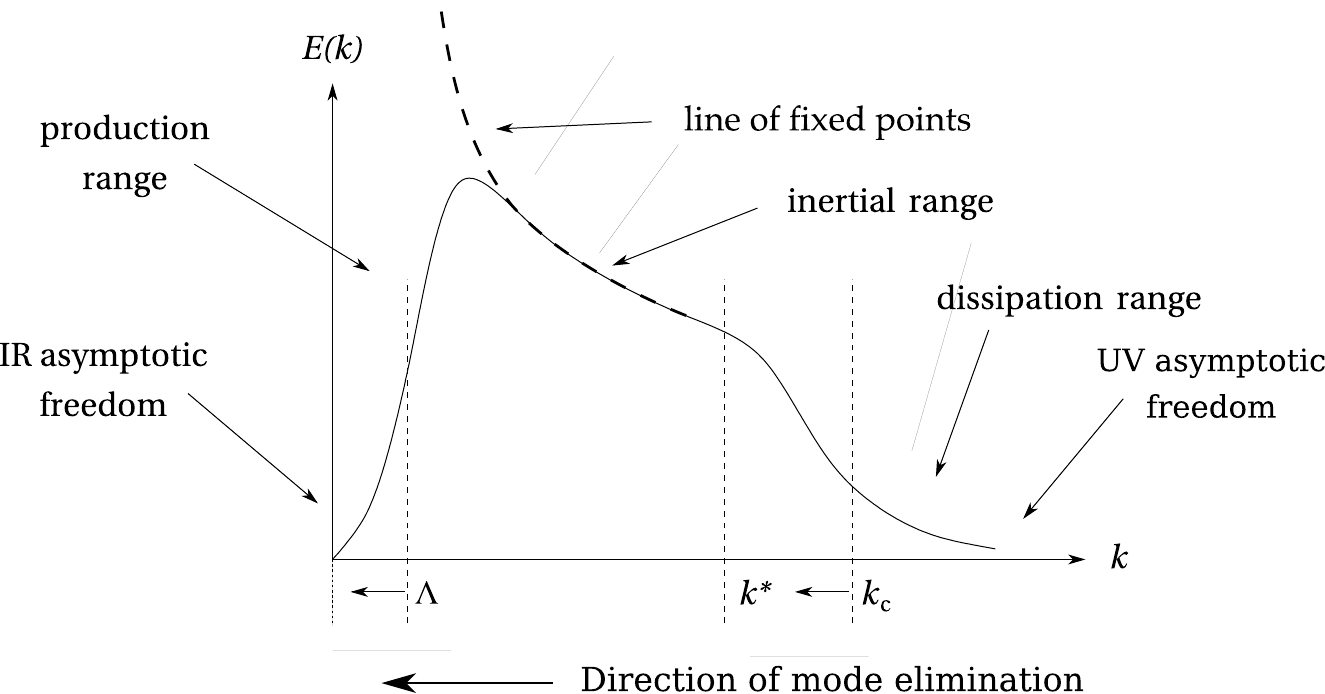}
 \end{center}
 \caption{Illustration of the energy spectrum for a turbulent flow. Renormalization from $\Lambda << k_d$ is towards the Gaussian fixed point $k = 0$. Iterative averaging from $k_c \simeq 0.1 k_d$ is towards the non-Gaussian fixed point at $k = k^*$.}
 \label{fig:RG_energy_spec}
\end{figure}

For turbulence, the expansion parameter is essentially a Reynolds number. We face the problem that, for systems of interest, this is not small and turbulence is a strongly coupled problem. However, we can construct a local (in $k$) Reynolds number based on the energy spectrum,
\begin{equation}
 R(k) = \sqrt{\frac{E(k)}{\nu^2 k}} \ .
\end{equation}
By considering the general form of the energy spectrum, as shown in figure \ref{fig:RG_energy_spec}, we see that this becomes small for both $k \to 0$ and $k \to \infty$ and turbulence offers both IR and UV asymptotic freedom. Thus we have two options: (1) start at low $k$ and go towards the Gaussian fixed point at $k = 0$; or (2) start at some high wavenumber above the inertial range and work our way down.

In the first case, we have excluded the effects of the inertial range from the start and, due the location at low $k$, are always sensitive to the forcing term. For the latter, due to the inertial range satisfying $k^{-5/3}$ scaling, we reach a non-Gaussian fixed point $k^*$ (different to $k_*$, the zero crossing of the transfer spectrum) and continue to follow the $k^{-5/3}$ line. As such, we have no access to, or dependence on, the large scale forcing; only the rate at which energy passes through the inertial range. An example of this approach is the method of iterative averaging by McComb (see McComb \cite{McComb:2006p214} for a discussion and further references), which uses a conditional average to remove the high wavenumber modes and indeed locates a non-trivial fixed point.

We now turn our attention to an alternate, dynamic renormalization procedure applied as $k \to 0$ and, as such, is not strictly a description of turbulence.

\section{Infra-red properties of stirred hydrodynamics}

The first application of the dynamic renormalization group procedure to turbulence was performed by Forster, Nelson and Stephen \cite{Forster:1976p426,Forster:1977p448}. They considered the large scale, long time behaviour of a randomly stirred fluid with the progressive removal of the small length-scales. As they note, their study is only valid at the smallest wavenumbers and as such well below the inertial range \cite{McComb:2006p214}. For several different types of (Gaussian-distributed) forcing, Forster, Nelson and Stephen (hereafter FNS) found the change to the viscosity and force coefficient induced by a single step of the renormalization scheme, from which they determined differential recursion relations and a form for the renormalized viscosity and noise coefficient.

Later, the analysis was extended by Yakhot and Orszag \cite{PhysRevLett.57.1722,Yakhot:1986p357} to a general form of the forcing (of which the studies by FNS were special cases). Despite the limit $k \to 0$, the results were then used to calculate inertial range properties by introducing the authors' \emph{correspondence principle}. This issue has been contested in the literature \cite{Teodorovich:1994p376,Eyink:1994p389,McComb:1990-book}, but is not what we pursue here.

Instead, we focus on another disagreement related to the methodology used by FNS and Yakhot and Orszag (hereafter YO). This stems from a change of variables used by FNS to perform the required integrals. Wang and Wu \cite{Wang:1993p371} and Teodorovich \cite{Teodorovich:1994p376} argued that the substitution shifts the shell of integration such that the identities used to perform the angular parts are no longer valid. The authors then showed how evaluation without using the change of variables led to an inconsistent result.

In an attempt to find which result was right, Nandy \cite{Nandy:1997p375} used a symmetrisation of the method used by Wang and Wu. This produced a third result which, on averaging with that found by Wang and Wu (and Teodorovich), recovered the original result of FNS. It therefore appeared that this symmetrisation was necessary to obtain the FNS result.

Since the approach of FNS has found wide-ranging application in soft-matter systems \cite{Kardar:1986p292,Medina:1989p291,Frey:1994p283} and magnetohydrodynamics \cite{Fournier:1982p400}, it is unsatisfactory to have a dispute over the basic methodology. As was shown in Berera and Yoffe \cite{Berera:2010p789}, the presence of an additional constraint on the internal (or loop) momenta, which was neglected in other approaches, actually prevents the integration shell from breaking the required symmetries for the angular identities to be used. Careful consideration of this extra constraint found that there was a correction to the result found in \cite{Wang:1993p371,Teodorovich:1994p376} which exactly compensates for the difference to the original result of FNS and YO. In addition, the symmetrisation procedure of Nandy \cite{Nandy:1997p375} was shown to be unnecessary with all results actually agreeing with FNS.

In this section we present the analysis of Berera and Yoffe \cite{Berera:2010p789} for the calculation of the renormalized viscosity and the evaluation of the corrections introduced by the additional momentum constraint. The renormalization of the force is then considered in section \ref{sec:RG_force}.

\subsection{General setup of the problem}
We start from the Navier-Stokes equation in physical space, as presented in equation \eqref{eq:NSE_x_iso}, only this time for a $d$-dimensional incompressible fluid (such that $\alpha \in \{1, \cdots, d\}$) subject to stochastic forcing. To move to spectral space we introduce the Fourier transform
\begin{equation}
 u_\alpha(\vec{x},t) = \int \frac{d^dk}{(2\pi)^d} \frac{d\omega}{(2\pi)}\ u_\alpha(\vec{k},\omega)\ e^{i(\vec{k}\cdot\vec{x} + \omega t)} \ ,
\end{equation}
where we note that the time argument has also been transformed using the angular frequency $\omega$ and the normalisation convention has been changed. Using this transformation, the NSE can be expressed in Fourier space as
\begin{align}
 \label{eq:def:nse_fourier}
 \left(i\omega + \nu_0 k^2\right) u_\alpha(\vec{k},\omega) = f_\alpha(\vec{k},\omega) + \lambda_0 & M_{\alpha\beta\gamma}(\vec{k}) \int \measure{j}{\Omega} \int\measure{p}{\Omega'} u_\beta(\vec{j},\Omega) u_\gamma(\vec{p},\Omega') \nonumber \\
 &\times(2\pi)^{d+1} \delta(\vec{j}+\vec{p}-\vec{k}) \delta(\Omega + \Omega' - \omega) \ ,
\end{align}
supplemented with the incompressibility condition $k_\alpha u_\alpha(\vec{k},w) = 0$. The operators $P_{\alpha\beta}(\vec{k})$ and $M_{\alpha\beta\gamma}(\vec{k})$ were defined in equations \eqref{eq:def:projection} and \eqref{eq:def:vertex}, respectively and we have introduced a book-keeping parameter $\lambda_0 = 1$ to the non-linear term. The integral over $\vec{p},\Omega'$ could be trivially done using the $\delta$-function (as was done in equation \eqref{eq:do_delta_int} to give the previous form of the NSE in Fourier space) but we maintain its presence here for comparison to Nandy \cite{Nandy:1997p375}.

The stochastic forcing is defined through its autocorrelation
\begin{equation}
\label{eq:def:stirring_forces}
 \langle f_\alpha(\vec{k},\omega) f_\beta(\vec{k'},\omega') \rangle = 2 F(k) P_{\alpha\beta}(\vec{k}) (2\pi)^{d+1} \delta(\vec{k} + \vec{k'}) \delta(\omega + \omega')\ ,
\end{equation}
where $F(k) = F_0 k^{-y}$ is the forcing spectral density and the projection operator ensures that the forcing is solenoidal, such that the incompressible nature of the fluid is not disturbed. Note that since the RHS of this correlation is real and symmetric under $\vec{k} \to -\vec{k}$, the real-space force autocorrelation must also be real.

A hard UV cut-off is imposed at some wavenumber $\Lambda \ll k_d$, with $k_d$ the dissipation number defined in equation \eqref{eq:k_d}. With the cut-off wavenumber below the inertial range, only the largest length-scale behaviour is accounted for and the theory cannot describe inertial range properties. Since we are at such low wavenumbers, we are very much influenced by the large scale forcing that is driving the system, which in this case is Gaussian distributed.

\subsubsection{Decomposing the velocity field}\label{subsubsec:decompose}

The velocity field is decomposed into contributions to its high and low frequency modes
\begin{equation}
\label{eq:def:decomposition}
u_\alpha(\fvect{k}) = \left\{ \begin{array}{c}
u^{\low}_\alpha(\fvect{k}) \qquad 0 < \lvert \vec{k}\rvert < e^{-\ell}\Lambda \\
\\
u^{\high}_\alpha(\fvect{k}) \qquad e^{-\ell}\Lambda < \lvert \vec{k}\rvert < \Lambda
\end{array} \right. \ ; \quad \ell > 0\ ,
\end{equation}
where we introduce a more compact four-vector notation $\fvect{k} = (\vec{k},\omega)$ such that
\begin{equation}
 \int \fmeasure{k}{\omega} = \int \frac{d^dk}{(2\pi)^d)} \frac{d\omega}{2\pi} \quad\qquad\textrm{and} \quad\qquad \delta(\fvect{k}) = (2\pi)^{d+1} \delta(\vec{k})\delta(\omega) \ .
\end{equation}
The parameter $\ell$ therefore controls the width of the high frequency band that we intend to remove. The NSE is written for the high and low frequency modes as
\begin{align}
 \label{eq:nse_high}
 \left(i\omega + \nu_0 k^2\right)u_\alpha^{\high}(\fvect{k}) &= f_\alpha^\high(\fvect{k})
+ \lambda_0 M^\high_{\alpha\beta\gamma}(\vec{k}) \int\fmeasure{j}{\Omega} \int\fmeasure{p}{\Omega'}\ \delta(\fvect{j}+\fvect{p}-\fvect{k}) \\
 &\qquad \times \Big[u_\beta^{\low}(\fvect{j}) u_\gamma^{\low}(\fvect{p}) + 2u_\beta^{\low}(\fvect{j}) u_\gamma^{\high}(\fvect{p}) + u_\beta^{\high}(\fvect{j}) u_\gamma^{\high}(\fvect{p}) \Big] \nonumber \\
 \label{eq:nse_low}
 \left(i\omega + \nu_0 k^2\right)u_\alpha^{\low}(\fvect{k}) &= f_\alpha^\low(\fvect{k})
+ \lambda_0 M^\low_{\alpha\beta\gamma}(\vec{k}) \int\fmeasure{j}{\Omega} \int\fmeasure{p}{\Omega'}\ \delta(\fvect{j}+\fvect{p}-\fvect{k}) \\
 &\qquad \times \Big[u_\beta^{\low}(\fvect{j}) u_\gamma^{\low}(\fvect{p}) + 2u_\beta^{\low}(\fvect{j}) u_\gamma^{\high}(\fvect{p}) + u_\beta^{\high}(\fvect{j}) u_\gamma^{\high}(\fvect{p}) \Big] \nonumber \ .
\end{align}
The filtered vertex operators $M^\low_{\alpha\beta\gamma}(\vec{k})$ and $M^\high_{\alpha\beta\gamma}(\vec{k})$ are understood to restrict $0 < k < e^{-\ell}\Lambda$ and $e^{-\ell}\Lambda < k < \Lambda$, respectively, in the non-linear term. The former will lead to an additional constraint which has been neglected by other authors when dealing with the loop integral. This is the source of the dispute over the validity of the substitution which was used by FNS and YO.

The high frequency modes are then expressed as a perturbation expansion,
\begin{equation}
 u_\alpha^{\high}(\fvect{k}) = u_\alpha^{\high(0)}(\fvect{k}) + \lambda_0 u_\alpha^{\high(1)}(\fvect{k}) + \lambda_0^2 u_\alpha^{\high(2)}(\fvect{k}) + \cdots \ .
\end{equation}
Inserting this expansion into equation \eqref{eq:nse_high} and collecting orders of $\lambda_0$, one can solve for $u_\alpha^{\high(i)}(\fvect{k})$ in terms of the bare response to the Gaussian forcing,
\begin{align}
 u_\alpha^{\high(0)}(\fvect{k}) &= G_0(\fvect{k}) f_\alpha^{\high}(\fvect{k}) \\
 u_\alpha^{\high(1)}(\fvect{k}) &= G_0(\fvect{k}) M^\high_{\alpha\beta\gamma}(\vec{k}) \int\fmeasure{j}{\Omega}\ \Big[u_\beta^{\low}(\fvect{j}) u_\gamma^{\low}(\fvect{k}-\fvect{j}) + 2u_\beta^{\low}(\fvect{j}) u_\gamma^{\high(0)}(\fvect{k}-\fvect{j}) \\
 &\hspace{2.5in}  + u_\beta^{\high(0)}(\fvect{j}) u_\gamma^{\high(0)}(\fvect{k}-\fvect{j}) \Big] \nonumber \ ,
\end{align}
where the zero-order \emph{propagator} is given by
\begin{equation}
 \label{eq:bare_prop}
 G_0(\fvect{k}) = \frac{1}{i\omega + \nu_0 k^2} \ .
\end{equation}
We then substitute the perturbation expansion of the high frequency modes into the equation for the low frequency modes, equation \eqref{eq:nse_low}, and retain terms up to order $\lambda_0^2$. The expressions for $u_\alpha^{\high(0)}$ and $u_\alpha^{\high(1)}$ are then inserted and we have, to order $\lambda_0^2$,
\begin{align}
 \label{eq:ahh!}
\left(i\omega + \nu_0 k^2 \right) u^{\low}_\alpha(\fvect{k})&= f^{\low}_\alpha(\fvect{k}) \\
&\quad + \lambda_0 M^\low_{\alpha\beta\gamma}(\vec{k})\int\fmeasure{j}{\Omega} \int\fmeasure{p}{\Omega'}\ \Big[u_\beta^{\low}(\fvect{j}) u_\gamma^{\low}(\fvect{p}) + 2u_\beta^{\low}(\fvect{j}) G_0(\fvect{p}) f_\gamma^{\high}(\fvect{p}) \nonumber \\
&\qquad\qquad\qquad\qquad\qquad + G_0(\fvect{j}) G_0(\fvect{p}) f_\beta^{\high}(\fvect{j}) f_\gamma^{\high}(\fvect{p}) \Big] \delta(\fvect{p} -\fvect{k}+\fvect{j}) \nonumber \\
&\quad + 2 \lambda_0^2 M^\low_{\alpha\beta\gamma}(\vec{k}) \int\fmeasure{j}{\Omega}\int\fmeasure{p}{\Omega'} \int\fmeasure{q}{\Omega''}\ M^\high_{\gamma\mu\nu}(\vec{p}) G_0(\fvect{p}) \delta(\fvect{p} -\fvect{k}+\fvect{j}) \nonumber \\
&\qquad\times\Big[u_\beta^{\low}(\fvect{j}) u_\mu^{\low}(\fvect{q}) u_\nu^{\low}(\fvect{p}-\fvect{q}) \nonumber \\
&\qquad\qquad + 2 G_0(\fvect{p}-\fvect{q}) u^{\low}_\beta(\fvect{j}) u^{\low}_\mu(\fvect{q}) f^{\high}_\nu(\fvect{p}-\fvect{q}) \nonumber \\
&\qquad\qquad + G_0(\fvect{j}) f^{\high}_\beta(\fvect{j}) u^{\low}_\mu(\fvect{q}) u^{\low}_\nu(\fvect{p}-\fvect{q}) \nonumber \\
&\qquad\qquad + G_0(\fvect{q}) G_0(\fvect{p}-\fvect{q}) u^{\low}_\beta(\fvect{j}) f^{\high}_\mu(\fvect{q}) f^{\high}_\nu(\fvect{p}-\fvect{q}) \nonumber \\
&\qquad\qquad + 2 G_0(\fvect{j}) G_0(\fvect{p}-\fvect{q}) f^{\high}_\beta(\fvect{j}) u^{\low}_\mu(\fvect{q}) f^{\high}_\nu(\fvect{p}-\fvect{q}) \nonumber \\
&\qquad\qquad + G_0(\fvect{j}) G_0(\fvect{q}) G_0(\fvect{p}-\fvect{q}) f^{\high}_\beta(\fvect{j}) f^{\high}_\mu(\fvect{q}) f^{\high}_\nu(\fvect{p}-\fvect{q}) \Big] \nonumber \ .
\end{align}

This intimidating expression can be simplified if we consider the application of a filtered-averaging procedure, denoted $\langle \cdots \rangle_f$, under which:
\begin{enumerate}
 \item The low frequency modes are statistically independent of the high frequency components;
 \item The low frequency components are invariant: $\langle f^\low \rangle_f \simeq f^\low$ and so $\langle u^\low \rangle_f \simeq u^\low$;
 \item The stirring forces are Gaussian with zero mean: $\langle f^\high \rangle_f = \langle f^\high f^\high f^\high \rangle_f = 0$.
\end{enumerate}
Performing this average, we find that numerous terms disappear and we are left with
\begin{align}
 \label{eq:ahh!2}
\left(i\omega + \nu_0 k^2 \right) \Big\langle u^{\low}_\alpha(\fvect{k})\Big\rangle_f &= \Big\langle f^{\low}_\alpha(\fvect{k}) + \Delta f_\alpha^\low(\fvect{k}) \Big\rangle_f \\
&\ + \lambda_0 M^\low_{\alpha\beta\gamma}(\vec{k})\int\fmeasure{j}{\Omega}\int\fmeasure{p}{\Omega'}\int\fmeasure{q}{\Omega''}\ \Big\langle u_\beta^{\low}(\fvect{j}) u_\gamma^{\low}(\fvect{p})\Big\rangle_f \delta(\fvect{p} -\fvect{k}+\fvect{j}) \nonumber \\
&\ + 2 \lambda_0^2 M^\low_{\alpha\beta\gamma}(\vec{k}) \int\fmeasure{j}{\Omega}\int\fmeasure{p}{\Omega'}\int\fmeasure{q}{\Omega''}\ M^\high_{\gamma\mu\nu}(\vec{p}) G_0(\fvect{p}) \nonumber \\
&\ \quad\times \delta(\fvect{p} -\fvect{k}+\fvect{j}) \Bigg[\Big\langle u_\beta^{\low}(\fvect{j}) u_\mu^{\low}(\fvect{q}) u_\nu^{\low}(\fvect{p}-\fvect{q})\Big\rangle_f \nonumber \\
&\ \qquad + G_0(\fvect{q}) G_0(\fvect{p}-\fvect{q}) \Big\langle u^{\low}_\beta(\fvect{j})\Big\rangle_f \Big\langle f^{\high}_\mu(\fvect{q}) f^{\high}_\nu(\fvect{p}-\fvect{q})\Big\rangle_f \nonumber \\
&\ \qquad + 2 G_0(\fvect{j}) G_0(\fvect{p}-\fvect{q}) \Big\langle u^{\low}_\mu(\fvect{q})\Big\rangle_f \Big\langle f^{\high}_\beta(\fvect{j})  f^{\high}_\nu(\fvect{p}-\fvect{q})\Big\rangle_f \Bigg] \nonumber \ .
\end{align}
The penultimate term violates the triangle condition, since the wavevector arguments add to produce $\delta(\vec{p})$ which causes $M_{\gamma\mu\nu}(0) = 0$ and the term cannot contribute. A similar thing happens to the last term at order $\lambda_0$ on the RHS of equation \eqref{eq:ahh!}, only this term is associated with the induced random force $\Delta f_\alpha^\low$. It has zero mean but modifies the autocorrelation of the forcing to compensate for the eliminated modes. This is discussed further in section \ref{sec:RG_force}.

We notice that this procedure has generated a new coupling between three of the low frequency modes and is an example of proliferation. Thus, as the renormalization scheme is iterated, we generate higher and higher order non-linearities. It was shown by Eyink \cite{Eyink:1994p389} that this operator is not irrelevant but marginal by power counting, see also appendix A of \cite{Berera:2010p789}. Despite this and following FNS and YO, we neglect the new coupling (and all higher orders generated from it). This is justified as it simply represents the order of the approximation with which we are working \cite{Zhou:1997p208}. In any case, the higher-order operators become irrelevant as we take $k \to 0$ \cite{McComb:2006p214}, as we will do.

With this in mind, we write our new equation for the slow modes:
\begin{align}
 \Big(i\omega + \nu_0 k^2 \Big) u^{\low}_\alpha(\fvect{k}) &= f^{\low}_\alpha(\fvect{k}) + \lambda_0 M^\low_{\alpha\beta\gamma}(\vec{k})\int\fmeasure{j}{\Omega}\int\fmeasure{p}{\Omega'}\ u_\beta^{\low}(\fvect{j}) u_\gamma^{\low}(\fvect{p})\ \delta(\fvect{p} -\fvect{k}+\fvect{j}) \nonumber \\
 &\quad + 4 \lambda_0^2 M^\low_{\alpha\beta\gamma}(\vec{k}) \int\fmeasure{j}{\Omega} \int\fmeasure{p}{\Omega'} \int\fmeasure{q}{\Omega''}\ M^\high_{\gamma\mu\nu}(\vec{p})\ u^{\low}_\mu(\fvect{q}) \nonumber \\
 &\qquad \times G_0(\fvect{j}) G_0(\fvect{p}) G_0(\fvect{p}-\fvect{q}) \left\langle f^{\high}_\beta(\fvect{j}) f^{\high}_\nu(\fvect{p}-\fvect{q})\right\rangle \delta(\fvect{p} -\fvect{k}+\fvect{j}) \nonumber \\
 &= f^{\low}_\alpha(\fvect{k}) + \lambda_0 M^\low_{\alpha\beta\gamma}(\vec{k})\int\fmeasure{j}{\Omega}\int\fmeasure{p}{\Omega'}\ u_\beta^{\low}(\fvect{j}) u_\gamma^{\low}(\fvect{p})\ \delta(\fvect{p} -\fvect{k}+\fvect{j}) \nonumber \\
 \label{eq:EOM_current}
 &\qquad+ \Sigma^\low_\alpha(\fvect{k}) \ .
\end{align}
Inserting the autocorrelation of the forcing, given in equation \eqref{eq:def:stirring_forces}, and using the $\delta$-function it introduces to perform the integral over $\fvect{q}$, the \emph{current} has been defined as
\begin{align}
 \Sigma^\low_\alpha(\fvect{k}) = 8 \lambda_0^2 M^\low_{\alpha\beta\gamma}(\vec{k}) \int\fmeasure{j}{\Omega} \int & \fmeasure{p}{\Omega'} \ M^\high_{\gamma\mu\nu}(\vec{p})\ u^{\low}_\mu(\fvect{p}+\fvect{j}) \lvert G_0(\fvect{j}) \rvert^2 G_0(\fvect{p}) P^\high_{\beta\nu}(\vec{j}) F(j) \nonumber \\
 \label{eq:current}
 &\times\delta(\fvect{p} -\fvect{k}+\fvect{j}) \ ,
\end{align}
where we have used the definition of the bare propagator, equation \eqref{eq:bare_prop}, to rewrite $G_0(\fvect{j}) G_0(-\fvect{j}) = \lvert G_0(\fvect{j}) \rvert^2$.

\subsubsection{Graphical representation}
\begin{figure}[tb]
 \begin{center}
 \subfigure[Feynman rules]{
  \label{sfig:RG_frules}
  \includegraphics[width=0.7\textwidth]{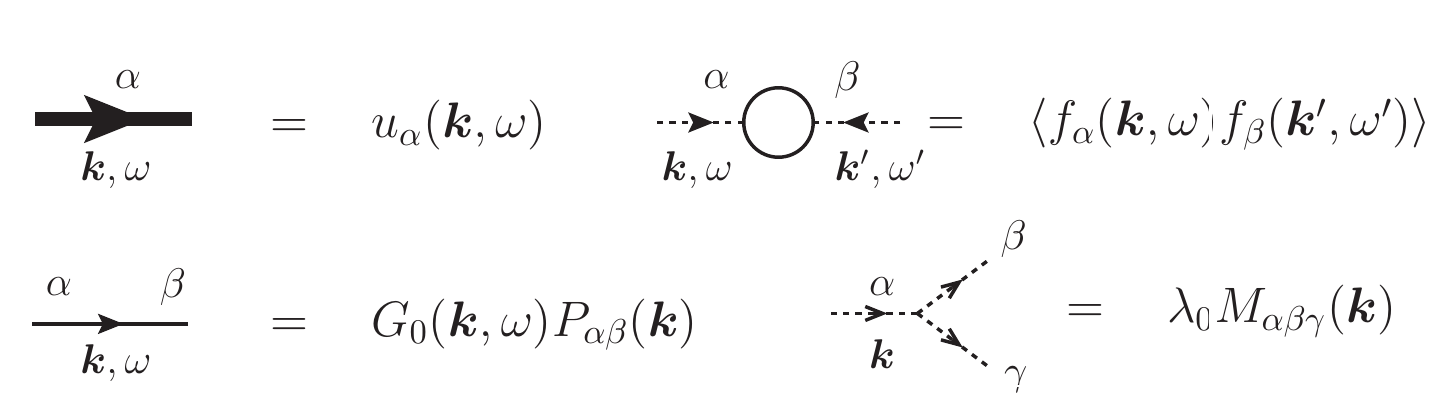}
 }
 \subfigure[Renormalization of the velocity field]{
  \label{sfig:graph_visc}
  \includegraphics[width=0.7\textwidth]{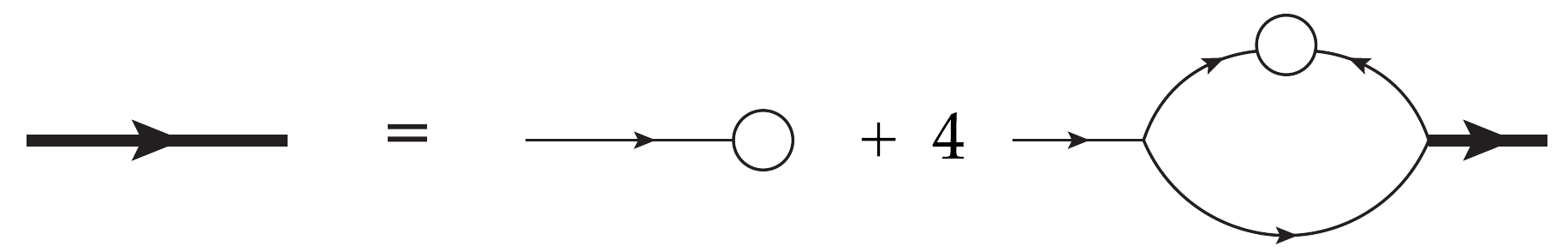}
 }
 \end{center}
 \caption{Feynman rules for generating diagrammatic representations and the Feynman diagram for the renormalization of the velocity field, leading to the one-loop correction of the viscosity.}
 \label{fig:graph_visc}
\end{figure}
We briefly pause to consider the graphical representation of the calculation we are about to embark on. From the first line of equation \eqref{eq:EOM_current}, we see that if we were to multiply through by $G_0(\fvect{k})$ and neglect the non-linear term, we could write
\begin{align}
 u^\low_\alpha(\fvect{k}) &= G_0(\fvect{k})f^\low_\alpha(\fvect{k}) + 4 G_0(\fvect{k}) \lambda_0^2 M^\low_{\alpha\beta\gamma}(\vec{k}) \int\fmeasure{j}{\Omega} \int\fmeasure{q}{\Omega''}\ G_0(\fvect{k}-\fvect{j}) M^\high_{\gamma\mu\nu}(\vec{k}-\vec{j}) \nonumber \\
 &\qquad \times G_0(\fvect{j}) \left\langle f^{\high}_\beta(\fvect{j}) f^{\high}_\nu(\fvect{k}-\fvect{j}-\fvect{q})\right\rangle G_0(\fvect{k}-\fvect{j}-\fvect{q})\ u^{\low}_\mu(\fvect{q}) \ ,
\end{align}
where we performed the integral over $\fvect{p}$ using the $\delta$-function. Using the Feynman rules given in figure \ref{sfig:RG_frules}, this can be represented graphically as shown in figure \ref{sfig:graph_visc}. The lower propagator in the loop carries four-momentum $\fvect{k}-\fvect{j}$ into the right hand vertex and $\fvect{q}$ leaves with the velocity. Since the autocorrelation of the force contains $\delta(\fvect{k}-\fvect{q})$, we assert that this renormalization of the velocity field can be expressed as
\begin{equation}
 u^\low_\alpha(\fvect{k}) = G_0(\fvect{k})f^\low_\alpha(\fvect{k}) + 4 G_0(\fvect{k}) \Sigma^\low_{\alpha\mu}(\fvect{k}) u^\low_\mu(\fvect{k}) \ ,
\end{equation}
where $\Sigma^\low_{\alpha\mu}(\fvect{k})$ is referred to as the self-energy tensor. This stems from high-energy physics, where the renormalized or `dressed' propagator may be written using the Dyson equation (see Wyld \cite{Wyld:1961-formulation}) as $G = G_0 + G_0 \Sigma G + \cdots$, where $\Sigma$ represents the self-energy operator. Thus, whenever we refer to `self-energy' we are talking about evaluation of the current. The graphical treatment of turbulence was originally formulated by Wyld \cite{Wyld:1961-formulation} and a comprehensive discussion is presented in Salewski \cite{thesis:msalewski}. For the symmetry factor of 4 associated with the graph, 2 comes from exchanging which of the two outgoing legs from the left hand vertex connects to the noise correlation and another 2 from the velocity (thick line) instead being incident on the left. Once again, see Wyld.

\subsubsection{Controlling the integration using $\theta$-functions}
We have been careful to ensure that the filtered velocity and vertex operators kept an indication of where the variable is non-zero. This burden is now placed in the hands of a pair of $\theta$-functions, which we define as
\begin{align}
 \label{eq:def:theta}
 \theta^{\low}(\vec{k}) = \theta(\Lambda e^{-\ell}-\lvert\vec{k}\rvert) &= \left\{ \begin{array}{ll}
1 & \lvert\vec{k}\rvert < \Lambda e^{-\ell} \\
1/2 & \lvert\vec{k}\rvert = \Lambda e^{-\ell} \\
0 & \textrm{otherwise}
\end{array} \right.\ \nonumber\\
\theta^\high(\vec{k}) = \theta(\lvert\vec{k}\rvert-\Lambda e^{-\ell}) \theta(\Lambda-\lvert\vec{k}\rvert) &= \left\{ \begin{array}{ll}
1 & \Lambda e^{-\ell} < \lvert\vec{k}\rvert < \Lambda \\
1/2 & \lvert\vec{k}\rvert = \Lambda e^{-\ell} \\
0 & \textrm{otherwise}
\end{array} \right.\ .
\end{align}
These are used to control the domain of integration; for example, $u_\mu^\low(\fvect{k}) = u_\mu(\fvect{k}) \theta^\low(\vec{k})$. Integrals over the momenta now run over $0 < \lvert\vec{\kappa}\rvert < \infty$. We insert the $\theta$-functions and restore full notation such that the current is given by
\begin{align}
 \Sigma^\low_\alpha &= 8 \lambda_0^2 M_{\alpha\beta\gamma}(\vec{k})\theta^\low(\vec{k}) \int \measure{j}{\Omega} \int d^dp\ d\Omega'\ M_{\gamma\mu\nu}(\vec{p})\ u_\mu(\vec{p}+\vec{j},\Omega+\Omega') \ F(j) \nonumber \\
 &\quad\times\lvert G_0(\vec{j},\Omega) \rvert^2 G_0(\vec{p},\Omega') P_{\beta\nu}(\vec{j}) \theta^\high(\vec{p}) \theta^\high(\vec{j}) \theta^\low(\vec{p}+\vec{j}) \delta(\vec{p}-\vec{k}+\vec{j}) \delta(\Omega-\omega+\Omega') \ .
\end{align}
One of the factors $1/(2\pi)^{d+1}$ has been cancelled by the corresponding factor from the definition of the $\delta$-function in the reduced notation. The shell of modes which we are going to eliminate is seen not to just require that $\vec{j}$ is a high frequency mode, but also $\vec{p} = \vec{k}-\vec{j}$ (due to the $\delta$-function). Therefore, we require that all internal (or loop) momenta be in the elimination band. The constraint on $\vec{p}$, which originated with the filtered vertex operator, is neglected in other approaches to this work, as will be discussed in section \ref{subsec:RG_others}.

\subsubsection{Performing the frequency integrals}\label{subsubsec:frequency_int}
\begin{figure}[tb]
 \begin{center}
  \includegraphics[width=0.6\textwidth]{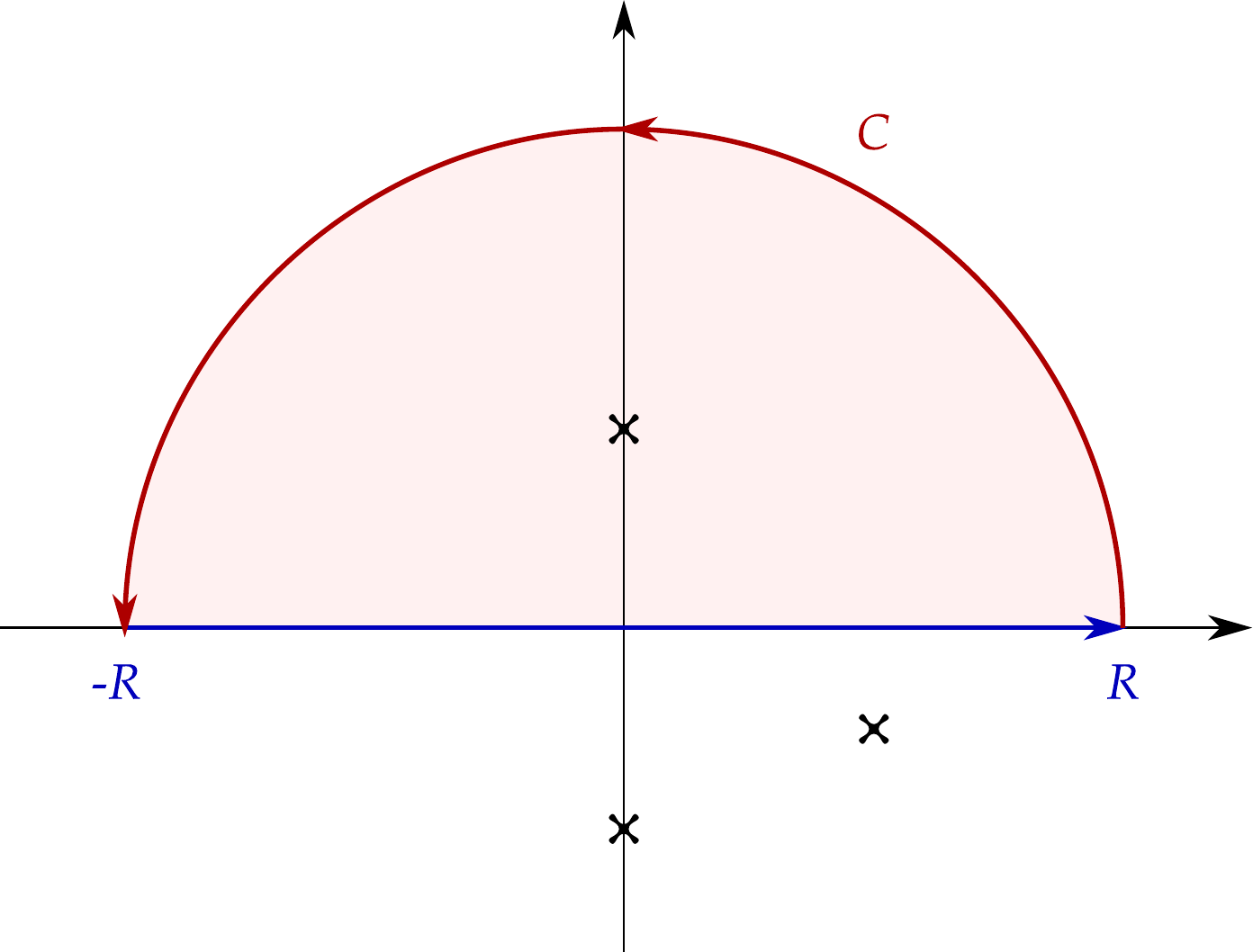}
 \end{center}
 \caption{Evaluation of the frequency integral using the residue from the single pole located at $\Omega = i\nu_0 j^2$ by closing the contour in the upper half-plane and taking $R \to \infty$.}
 \label{fig:frequency_contour}
\end{figure}
\definecolor{dred}{HTML}{AD0000}
\definecolor{dblue}{HTML}{0000BB}

The integral over $\Omega'$ can be trivially done using the term $\delta(\Omega-\omega+\Omega')$. The current can then be written
\begin{align}
 \Sigma^\low_\alpha &= 8 \lambda_0^2 M_{\alpha\beta\gamma}(\vec{k})\theta^\low(\vec{k}) \int \frac{d^dj}{(2\pi)^d} \int d^dp\ M_{\gamma\mu\nu}(\vec{p})\ u_\mu(\vec{p}+\vec{j},\omega)  F(j) P_{\beta\nu}(\vec{j}) \theta^\high(\vec{p}) \theta^\high(\vec{j})  \nonumber \\
 &\quad\times I_\Omega(\omega;\vec{j},\vec{p})\ \theta^\low(\vec{p}+\vec{j}) \delta(\vec{p}-\vec{k}+\vec{j}) \ ,
\end{align}
where the frequency integral
\begin{align}
 I_\Omega(\omega;\vec{j},\vec{p}) &= \int \frac{d\Omega}{2\pi}\ \lvert G_0(\vec{j},\Omega) \rvert^2 G_0(\vec{p},\Omega') \nonumber \\
 &= \int_{-\infty}^\infty \frac{d\Omega}{2\pi}\ \left(\frac{1}{i\Omega + \nu_0 j^2}\right) \left(\frac{1}{-i\Omega + \nu_0 j^2}\right) \left(\frac{1}{i(\omega - \Omega) + \nu_0 p^2}\right) \ .
\end{align}
The poles in $\Omega$ occur at $\Omega = \pm i\nu_0 j^2$ and $\Omega = w - i\nu_0 p^2$. These are shown in figure \ref{fig:frequency_contour}, indicated by $\times$. To perform the integral, we consider first the integration from $-R$ to $R$ and close the contour in the upper half-plane, enclosing only the pole at $\Omega = i\nu_0 j^2$, as shown in the figure. The contour is traversed in an anti-clockwise sense, such that the residue theorem allows us to evaluate the closed path as
\begin{equation}
 \oint d\Omega\ f(\Omega) = {\color{dblue}\int_{-R}^{R}} d\Omega\ f(\Omega) + {\color{dred}\int\limits_{C}} d\Omega\ f(\Omega) = 2\pi i \sum\limits_{a \in \mathbb{S}} \textrm{Res}_{\Omega = a} f(\Omega) \ ,
\end{equation}
where the set $\mathbb{S}$ contains all the poles enclosed by the contour and $f(\Omega)$ is the integrand of $I_\Omega(\omega;\vec{j},\vec{p})$. When we take the limit $R \to \infty$, the contribution from the semi-circle ({\color{dred}$C$}) vanishes due to the overall $1/R^2$ dependence of the integral. For the single simple pole enclosed, the above reduces to
\begin{align}
 I_\Omega(\omega;\vec{j},\vec{p}) &= 2\pi i \sum\limits_{a \in \mathbb{S}} \textrm{Res}_{\Omega = a} f(\Omega) \nonumber \\
 &= \frac{2\pi i}{2\pi} (\Omega - i\nu_0 j^2) \left. \left(\frac{1}{i\Omega + \nu_0 j^2}\right) \left(\frac{1}{-i\Omega + \nu_0 j^2}\right) \left(\frac{1}{i(\omega - \Omega) + \nu_0 p^2}\right) \right\rvert_{\Omega = i\nu_0 j^2} \nonumber \\
 &= \frac{1}{2\nu_0 j^2} \left( \frac{1}{i\omega + \nu_0 j^2 + \nu_0 p^2} \right) \ .
\end{align}
Note that performing the frequency integrals in the opposite order (by using the $\delta$-function to perform the integral over $\Omega$ first) yields the same result. The poles instead occur at $\Omega' = i\nu_0 p^2$ and $\Omega' = \omega \pm i\nu_0 j^2$ and we close the contour in the lower half-plane, enclosing only the pole $\Omega' = \omega-i\nu_0 j^2$. The contour is then traversed in the opposite sense, from $R$ to $-R$, meaning that the contribution picks up a minus sign.

Inserting the frequency integral back into the expression for the current, we have
\begin{align}
 \Sigma^\low_\alpha = \frac{4\lambda_0^2}{\nu_0} M_{\alpha\beta\gamma}(\vec{k}) \theta^\low(\vec{k})\int \frac{d^dj}{(2\pi)^d} \frac{F(j)}{j^2} & \int d^dp\ M_{\gamma\mu\nu}(\vec{p}) \frac{P_{\beta\nu}(\vec{j}) u_\mu(\vec{p}+\vec{j},\omega)}{i\omega + \nu_0 j^2 + \nu_0 p^2}\ \\
 &\times\theta^{\high}(\vec{j}) \theta^{\high}(\vec{p}) \theta^\low(\vec{p}+\vec{j}) \delta(\vec{p}-\vec{k}+\vec{j}) \nonumber \ .
\end{align}
At this point, we can quite simply take the limit $\omega \to 0$ (since we are considering the long time properties) to obtain
\begin{align}
 \label{eq:all_agree}
 \Sigma^\low_\alpha(\vec{k},0) = \frac{4\lambda_0^2}{\nu_0^2} M_{\alpha\beta\gamma}(\vec{k}) \theta^\low(\vec{k}) & \int \frac{d^dj}{(2\pi)^d} \frac{F(j)}{j^2} \int d^dp\ M_{\gamma\mu\nu}(\vec{p}) \frac{P_{\beta\nu}(\vec{j}) u_\mu(\vec{p}+\vec{j},0)}{j^2 + p^2}\ \nonumber \\
 &\times\theta^{\high}(\vec{j}) \theta^{\high}(\vec{p}) \theta^\low(\vec{p}+\vec{j}) \delta(\vec{p}-\vec{k}+\vec{j}) \ .
\end{align}
This expression for the current is the point at which different approaches to evaluating the self-energy diverge. This is basically due to the constraints which are placed on the internal momenta. We now consider the different approaches to calculating the remaining momenta integrals and the disagreement they lead to.

\subsection{Approach of FNS and YO}
To consider the evaluation of FNS \cite{Forster:1977p448} (and later YO \cite{Yakhot:1986p357}), we first perform the trivial integral over one of the momenta, in this case $\vec{p}$, and insert the autocorrelation spectral density $F(k) = F_0 k^{-y}$. This means that our expression of interest becomes
\begin{align}
 \Sigma^\low_\alpha(\vec{k},0) = \frac{4\lambda_0^2}{\nu_0^2} u_\mu(\vec{k},0) M_{\alpha\beta\gamma}(\vec{k}) \theta^\low(\vec{k})\int \frac{d^dj}{(2\pi)^d} \frac{F_0 M_{\gamma\mu\nu}(\vec{k}-\vec{j}) P_{\beta\nu}(\vec{j}) \theta^{\high}(\vec{j}) \theta^{\high}(\vec{k}-\vec{j})}{j^{y+2}(j^2 + \lvert\vec{k}-\vec{j}\rvert^2)} \ .
\end{align}
In appendix A of their paper, FNS explicitly state that the momentum integrals are restricted such that \emph{both} loop momenta, $\vec{j}$ and $\vec{k}-\vec{j}$, lie in the band of fast modes which are to be removed. We have kept both these restrictions through the use of $\theta$-functions. Inserting the definition of the vertex operator and using the properties of the projection operators and incompressibility condition, we have
\begin{align}
 u_\mu(\vec{k},0) M_{\gamma\mu\nu}(\vec{k}-\vec{j}) P_{\beta\nu}(\vec{j}) &= \frac{u_\mu(\vec{k},0)}{2i}\Big[ k_\nu P_{\gamma\mu}(\vec{k}-\vec{j}) - j_\mu P_{\gamma\nu}(\vec{k}-\vec{j}) \Big] P_{\beta\gamma}(\vec{j}) \ ,
\end{align}
and as such
\begin{align}
 \label{eq:current_pre_CoV}
 \Sigma^\low_\alpha(\vec{k},0) = \frac{2F_0\lambda_0^2}{i\nu_0^2} \frac{u^\low_\mu(\vec{k},0) M_{\alpha\beta\gamma}(\vec{k})}{(2\pi)^d} & \int d^dj\ \frac{j^{-y-2} P_{\beta\nu}(\vec{j}) }{j^2 + \lvert\vec{k}-\vec{j}\rvert^2} \ \theta^{\high}(\vec{j}) \theta^{\high}(\vec{k}-\vec{j})\nonumber \\
 &\times\Big[ k_\nu P_{\gamma\mu}(\vec{k}-\vec{j}) - j_\mu P_{\gamma\nu}(\vec{k}-\vec{j}) \Big] \ .
\end{align}
Since we are looking for the (additive) correction to the viscosity, which appears as $\nu_0 k^2 u^\low_\alpha$ in equation \eqref{eq:EOM_current}, we want to consider contributions at order $k^2$. Since there is a factor of $k$ hidden inside $M_{\alpha\beta\gamma}(\vec{k})$ outside the integral, we note that we need only keep terms up to order $k$ in the integrand.

To simplify the problem, FNS (followed by YO) introduce the change of variables $\vec{j} \to \tfrac{1}{2}\vec{k} + \vec{j}$ with the claim that the restriction on the resulting integral is reduced to $\Lambda e^{-\ell} < \lvert\vec{j}\rvert < \Lambda$. With this substitution, and recalling that we only need to consider up to $\order{k}$ in the integrand, we have
\begin{align}
 \Sigma^\low_\alpha(\vec{k},0) = \frac{2F_0\lambda_0^2}{i\nu_0^2} & \frac{u^\low_\mu(\vec{k},0) M_{\alpha\beta\gamma}(\vec{k})}{(2\pi)^d} \int d^dj\  \frac{\lvert \tfrac{1}{2}\vec{k}+\vec{j}\rvert^{-y-2} }{\lvert\tfrac{1}{2}\vec{k}+\vec{j}\rvert^2 + \lvert\tfrac{1}{2}\vec{k}-\vec{j}\rvert^2}\ \theta^{\high}(\vec{j}) \\
 &\hspace{0.5in}\times \Big[ k_\nu P_{\gamma\mu}(\vec{j}) P_{\beta\nu}(\vec{j}) - j_\mu P_{\gamma\nu}(\tfrac{1}{2}\vec{k}-\vec{j}) P_{\beta\nu}(\tfrac{1}{2}\vec{k}+\vec{j}) \Big] \nonumber \ .
\end{align}
Since we are considering $k \to 0$, Taylor expansion of $\lvert \tfrac{1}{2}\vec{k}+\vec{j}\rvert^{-y-2}$ becomes
\begin{align}
 \lvert \tfrac{1}{2}\vec{k}\pm\vec{j}\rvert^{-y-2} &= j^{-y-2} \left( 1 \pm \frac{\vec{k}\cdot\vec{j}}{j^2} \right)^{-(y+2)/2} + \order{k^2} \nonumber \\
 &= j^{-y-2} \left( 1 \mp \left(\frac{y+2}{2}\right) \frac{\vec{k}\cdot\vec{j}}{j^2} \right) + \order{k^2} \ ,
\end{align}
which can also be used to expand the product of projection operators,
\begin{equation}
 P_{\gamma\nu}(\tfrac{1}{2}\vec{k}-\vec{j}) P_{\beta\nu}(\tfrac{1}{2}\vec{k}+\vec{j}) = P_{\beta\gamma}(\vec{j}) - \left(\frac{k_\gamma j_\beta - k_\beta j_\gamma}{2j^2} \right) + \order{k^2} \ .
\end{equation}
Inserting these relations back into the expression for the current, $\Sigma^\low_\alpha$, we find
\begin{align}
 \label{eq:current_pre_int}
 \Sigma^\low_\alpha(\vec{k},0) &= \frac{F_0\lambda_0^2}{i\nu_0^2} \frac{u^\low_\mu(\vec{k},0) M_{\alpha\beta\gamma}(\vec{k})}{(2\pi)^d} \int d^dj\ j^{-y-4}\ \theta^{\high}(\vec{j})\ \bigg[ k_\nu P_{\gamma\mu}(\vec{j}) P_{\beta\nu}(\vec{j}) \\
 &\qquad\qquad\qquad+ j_\mu P_{\beta\gamma}(\vec{j}) \left(\frac{y+2}{2}\right)\frac{k_\nu j_\nu}{j^2} - j_\mu P_{\beta\gamma}(\vec{j}) + j_\mu \left(\frac{k_\gamma j_\beta - k_\beta j_\gamma}{2j^2} \right) \bigg] \nonumber \ .
\end{align}
We now insert the definition of the projection operator, $P_{\beta\nu}(\vec{j}) = \delta_{\beta\nu} - j_\beta j_\nu/j^2$, and use the identities:
\begin{align}
 \int d^dj\ f(j) &= S_d \int dj\ j^{d-1}\ f(j)\ , \nonumber \\
 \label{eq:ang_ident}
 \int d^dj\ j_\alpha\ j_\beta\ f(j) &= \frac{S_d}{d} \delta_{\alpha\beta} \int dj\ j^{d+1}\ f(j) \ ,\quad\text{and} \\
 \int d^dj\ j_\alpha\ j_\beta\ j_\gamma\ j_\nu\ f(j) &= \frac{S_d}{d(d+2)} \Big(\delta_{\alpha\beta}\delta_{\gamma\nu} + \delta_{\alpha\gamma}\delta_{\beta\nu} + \delta_{\alpha\nu}\delta_{\beta\gamma}\Big) \int dj\ j^{d+3}\ f(j) \nonumber \ ;
\end{align}
where $S_d = 2\pi^{d/2}/\Gamma(d/2)$ is the surface area of the unit sphere in $d$-dimensions and $f(r)$ is an arbitrary test function. We therefore find, noting that the penultimate term in equation \eqref{eq:current_pre_int} integrates to zero and the last term vanishes once contracted with $M_{\alpha\beta\gamma}(\vec{k}) u^\low_\mu(\vec{k},0)$,
\begin{align}
 \Sigma^\low_\alpha &= \frac{F_0 \lambda_0^2}{i\nu^2_0} M_{\alpha\beta\gamma}(\vec{k}) k_\nu u^\low_\mu(\vec{k},0) \frac{S_d}{(2\pi)^d} \frac{1}{d(d+2)} \Bigg[ \left(d^2 - 3 - \frac{y+2}{2}\right)\delta_{\mu\gamma} \delta_{\beta\nu} \nonumber \\
 &\qquad +\left(1+(d+1)\frac{y+2}{2}\right) \delta_{\beta\gamma} \delta_{\mu\nu} + \delta_{\mu\beta}\delta_{\gamma\nu}\left(1 - \frac{y+2}{2}\right) \Bigg] \int_{\Lambda e^{-\ell}}^\Lambda dj\ j^{-y+d-5} \ .
\end{align}
Substituting the definition of the vertex operator, $M_{\alpha\beta\gamma}(\vec{k}) = (1/2i)[k_\beta P_{\alpha\gamma}(\vec{k}) + k_\gamma P_{\alpha\beta}(\vec{k})]$, contracting all Kronecker-$\delta$s and collecting terms then gives
\begin{align}
 \Sigma^\low_\alpha &= -\frac{F_0 \lambda_0^2}{\nu^2_0} \frac{S_d}{(2\pi)^d} \frac{1}{2d(d+2)} \left[\left(d^2 - 3 - \frac{y+2}{2}\right) + \left(1 - \frac{y+2}{2}\right) \right] k^2 P_{\alpha\mu}(\vec{k}) u^\low_\mu(\vec{k},0) \nonumber \\
 &\qquad\times\int^{\Lambda}_{\Lambda e^{-\ell}} dj\ j^{-\epsilon-1} \nonumber \\
 &= -\frac{F_0 \lambda_0^2}{\nu^2_0} \frac{S_d}{(2\pi)^d} \frac{d^2 - d -\epsilon}{2d(d+2)} \left(\frac{e^{\epsilon\ell} - 1}{\epsilon\Lambda^\epsilon}\right) k^2 P_{\alpha\mu}(\vec{k}) u^\low_\mu(\vec{k},0) \ ,
\end{align}
where we have introduced $\epsilon = 4-d+y$ and performed the final integral over the magnitude $j$. The projection operator hits the remaining velocity component to change its index to $\alpha$ and we see that the current has the form
\begin{align}
 \Sigma^\low_\alpha(\vec{k},0) &= -\Delta\nu_0(\vec{k},0)\ k^2 u^\low_\mu(\vec{k},0) \ ,
\end{align}
where the viscosity increment has been defined as
\begin{equation}
 \Delta\nu_0(\vec{k},0) = \frac{S_d}{(2\pi)^d} \frac{d^2 - d -\epsilon}{2d(d+2)} \frac{F_0 \lambda_0^2}{\nu^2_0 \Lambda^\epsilon} \left(\frac{e^{\epsilon\ell} - 1}{\epsilon}\right) \ .
\end{equation}

Returning to equation \eqref{eq:EOM_current}, we see the motivation for calling this a viscosity increment. If we insert the deduced form of the current $\Sigma^\low_\alpha$, we can rearrange such that our renormalization procedure has modified the viscosity,
\begin{align}
 \Big(i\omega + \big(\nu_0 + \Delta\nu_0\big) k^2 \Big) u^{\low}_\alpha(\fvect{k}) &= f^{\low}_\alpha(\fvect{k}) + \lambda_0 M^\low_{\alpha\beta\gamma}(\vec{k})\int\fmeasure{j}{\Omega}\int\fmeasure{p}{\Omega'}\ u_\beta^{\low}(\fvect{j}) u_\gamma^{\low}(\fvect{k}-\fvect{j}) \ .
\end{align}
We finally take the limit $k \to 0$ in our result for the viscosity increment, giving the YO (a generalised version of FNS) result
\begin{equation}
 \label{eq:delta_nu_fns}
 \Delta\nu(\vec{0},0) = \nu_0 A_d(\epsilon) \overline{\lambda}^2(0) \left(\frac{e^{\epsilon\ell} - 1}{\epsilon} \right)\ ,
\end{equation}
with the prefactor $A_d(\epsilon)$ and reduced coupling $\overline{\lambda}$ (which is a Reynolds number \cite{PhysRevLett.57.1722}) defined through the relations
\begin{equation}
 \label{eq:result:fns_compact}
 A_d(\epsilon) = \frac{S_d}{(2\pi)^d} \tilde{A}_d(\epsilon)\ , \quad \tilde{A}_d(\epsilon) = \frac{d^2-d-\epsilon}{2d(d+2)}\ , \quad \overline{\lambda}^2(0) = \frac{F_0 \lambda_0^2}{\nu_0^3 \Lambda^\epsilon} \ \quad\textrm{and}\ \quad \epsilon = 4-d+y \ .
\end{equation}

\subsection{Analysis of other authors}\label{subsec:RG_others}
The approach used by FNS and YO to perform the self-energy integral by introducing the change of variables has been criticised in the literature, most notably by Wang and Wu \cite{Wang:1993p371} and Teodorovich \cite{Teodorovich:1994p376}. This is because the authors did not maintain the condition represented by $\theta^\high(\vec{k}-\vec{j})$, which results from performing the integral over $\vec{p}$ in equation \eqref{eq:all_agree}. Only the condition $\theta^\high(\vec{j})$ is retained and, as such, the change of variables breaks the symmetry of the shell of integration required for the identities in equation \eqref{eq:ang_ident} to be valid. The authors therefore adopted alternative methods for evaluating the integrals, but with the consequence that that the resulting expression for the viscosity increment was different. In fact, it worked out to be equation \eqref{eq:delta_nu_fns} only with $A_d(\epsilon)$ replaced by $A_d(0)$.

Having evaluated the viscosity increment with $A_d(\epsilon)$, Yakhot and Orszag \cite{Yakhot:1986p357} went on to calculate various quantities in the inertial range, such as the Kolmogorov constant, using their correspondence principle and were able to demonstrate Kolmogorov $k^{-5/3}$ using $\epsilon = 4$ (or $y = d$). However, in doing so they had to use $\epsilon = 0$ in the evaluation of the prefactor. This was required for a self-consistent asymptotic expansion at each step \cite{Smith:1998p359} and appeared to favour the $\epsilon$-free result. The prefactor $A_d(0)$ is also used in the more field-theoretic approach of Adzhemyan, Antonov and Vasiliev \cite{AAV:1999-book}.

Later, in an attempt to reconcile the differences between the $A_d(\epsilon)$ and $A_d(0)$ results, Nandy \cite{Nandy:1997p375} argued that the result had been biased by the choice of performing one integral using the $\delta$-function over the other. A symmetrisation of the procedure without a substitution was used, after which one recovers the original result of FNS and YO.

\section{Resolution of the conflict}
Since the method of FNS and YO has found successful application in other areas \cite{Kardar:1986p292,Medina:1989p291,Frey:1994p283,Fournier:1982p400}, a fundamental disagreement about the methodology and result should be investigated. We now show how correct treatment of the additional momentum constraint causes the integration shell to remain symmetric under the change of variables and actually leads to a non-zero correction when it is not performed. Inclusion of this contribution recovers the FNS/YO result. For the case of Nandy, the corrections from each evaluation are equal and opposite, explaining why the symmetrisation recovered the result of FNS/YO but highlighting that it is not needed.

\subsection{Expansion of $\theta$-functions}
The $\theta^\high$-function which has been used to control the shell of integration was defined in equation \eqref{eq:def:theta}. The product under consideration here is given by $\theta^\high(\vec{j}) \theta^\high(\vec{k}-\vec{j})$ and we look at the small $k$ behaviour by developing a Taylor expansion for $\theta^\high(\vec{k}-\vec{j})$. The $\theta^\high$ function is made up of two pieces, which we consider separately:
\begin{align}
 \theta(\lvert\vec{k}-\vec{j}\rvert - \Lambda e^{-\ell}) &= \theta(\lvert\vec{k}-\vec{j}\rvert-\Lambda e^{-\ell})\Big\rvert_{\vec{k}=0} + \vec{k}\cdot\Big(\vec{\nabla}\theta(\lvert\vec{k}-\vec{j}\rvert-\Lambda e^{-\ell})\Big)\Big\rvert_{\vec{k}=0} + \ldots \nonumber \\
 &= \theta(\lvert\vec{j}\rvert-\Lambda e^{-\ell}) + \vec{k}\cdot\left.\left[\left(\frac{\vec{k}-\vec{j}}{\lvert\vec{k}-\vec{j}\rvert}\right) \Big(\delta(\lvert\vec{k}-\vec{j}\rvert-\Lambda e^{-\ell})\Big)\right]\right\rvert_{\vec{k}=0} + \ldots \nonumber \\
 \label{eq:theta+exp}
 &= \theta(\lvert\vec{j}\rvert-\Lambda e^{-\ell}) - \frac{\vec{k}\cdot\vec{j}}{\lvert\vec{j}\rvert} \delta(\lvert\vec{j}\rvert-\Lambda e^{-\ell}) + \order{k^2}
\ ;
\end{align}
and
\begin{align}
 \theta(\Lambda - \lvert\vec{k}-\vec{j}\rvert) &= \theta(\Lambda - \lvert\vec{k}-\vec{j}\rvert)\Big\rvert_{\vec{k}=0} + \vec{k}\cdot\Big(\vec{\nabla}\theta(\Lambda - \lvert\vec{k}-\vec{j}\rvert)\Big)\Big\rvert_{\vec{k}=0} + \ldots \nonumber \\
 &= \theta(\Lambda - \lvert\vec{j}\rvert) - \vec{k}\cdot\left.\left[\left(\frac{\vec{k}-\vec{j}}{\lvert\vec{k}-\vec{j}\rvert}\right) \Big(\delta(\Lambda - \lvert\vec{k}-\vec{j}\rvert)\Big)\right]\right\rvert_{\vec{k}=0} + \ldots \nonumber \\
 \label{eq:theta-exp}
 &= \theta(\Lambda - \lvert\vec{j}\rvert) + \frac{\vec{k}\cdot\vec{j}}{\lvert\vec{j}\rvert} \delta(\Lambda-\lvert\vec{j}\rvert) + \order{k^2}
\ .
\end{align}
Multiplying these pieces together, our expansion is therefore
\begin{equation}
 \label{eq:theta+}
 \theta^\high(\vec{k}-\vec{j}) = \theta^\high(\vec{j}) - \frac{\vec{k}\cdot\vec{j}}{j} \Big[ \theta(\Lambda - j) \delta(j - \Lambda e^{-\ell}) - \theta(j - \Lambda e^{-\ell}) \delta(\Lambda - j) \Big] + \order{k^2} \ .
\end{equation}
Thus the elimination shell has been clearly shifted.

\begin{figure}[tb]
 \begin{center}
  \subfigure[Single constraint $\theta^\high(\tfrac{1}{2}\vec{k}+\vec{j})$]{\label{sfig:shell_shift_single} \includegraphics[width=0.475\textwidth]{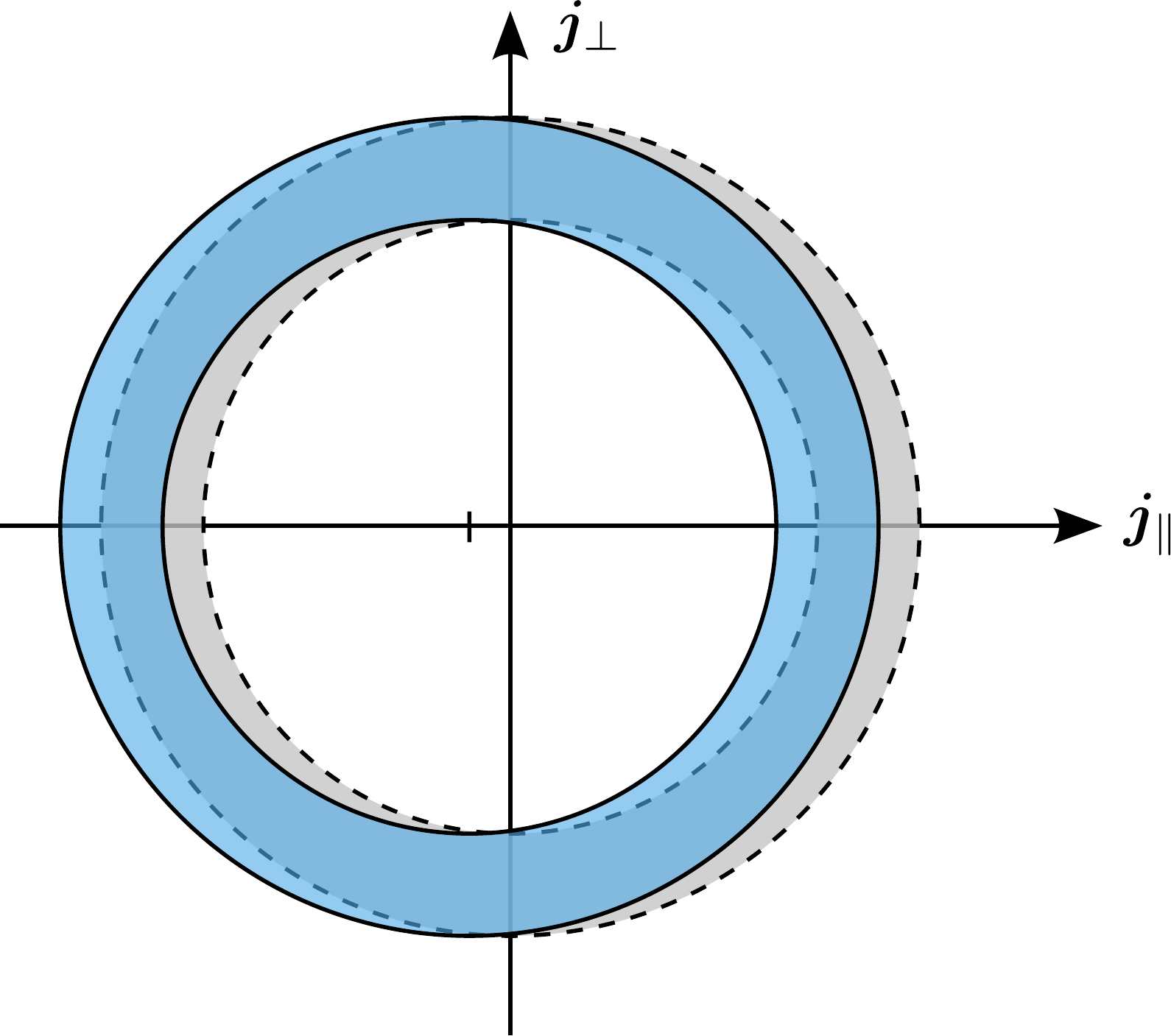} }
  \subfigure[Double constraint $\theta^\high(\tfrac{1}{2}\vec{k}+\vec{j}) \theta^\high(\tfrac{1}{2}\vec{k}-\vec{j})$]{\label{sfig:shell_shift_double} \includegraphics[width=0.475\textwidth]{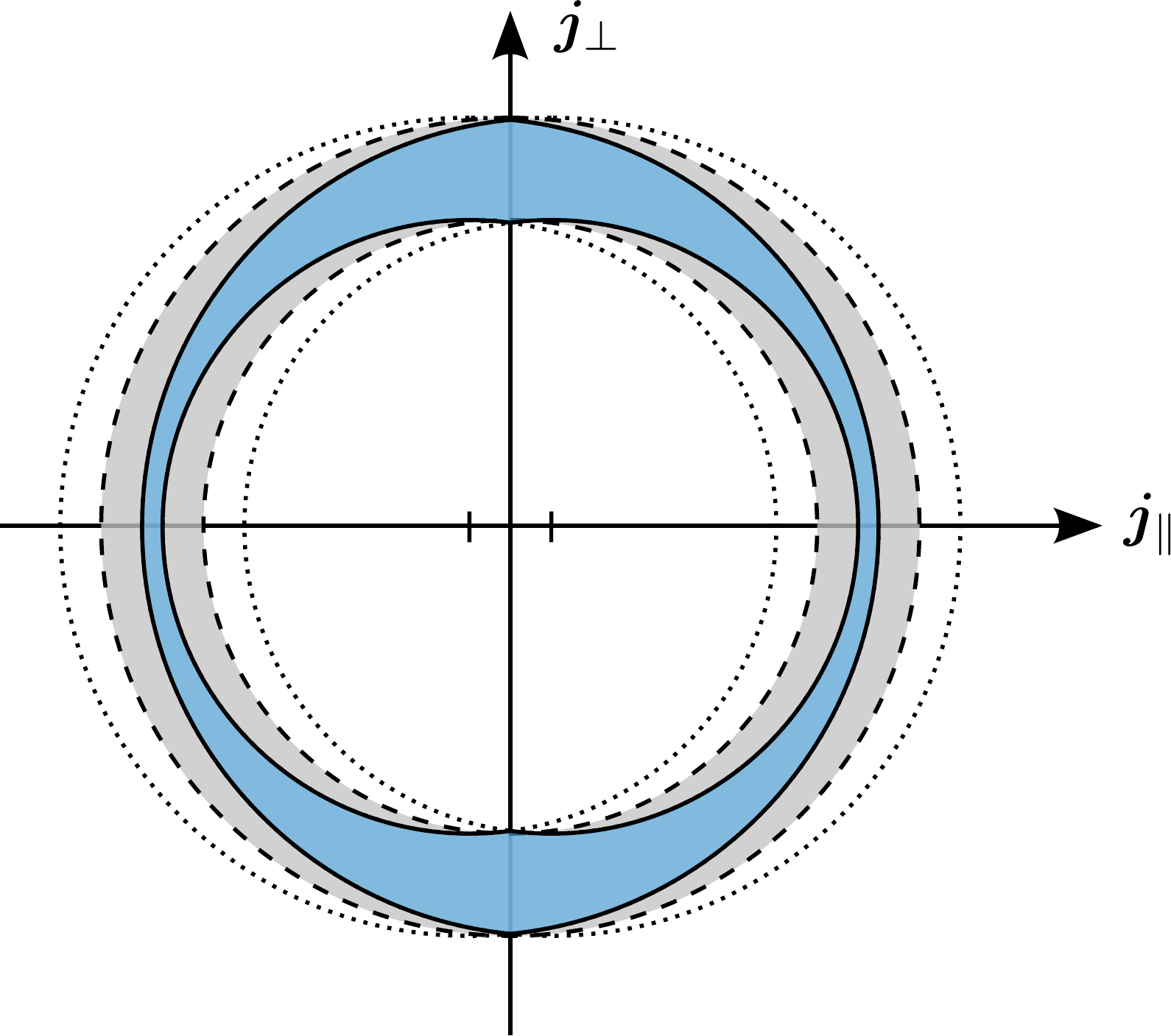} }
 \end{center}
 \caption{Shift of the integration shell due to the constraints. Grey shows the $\theta^\high(\vec{j})$ shell and blue the resultant shell. Tick marks show $\pm\tfrac{1}{2}\lvert\vec{k}\rvert$, exaggerated for effect. In (b) the dotted lines show the two individual constraints. We see that for (a) the shell is clearly not symmetric, whereas for (b) the resultant shell remains symmetric under $\vec{j} \to -\vec{j}$.}
 \label{fig:shell_shift}
\end{figure}

We now consider the change of variables $\vec{j} \to \tfrac{1}{2}\vec{k}+\vec{j}$. With just the requirement $\theta^\high(\vec{j})$, the shell of integration becomes shifted and is no longer centred at the origin but rather $-\tfrac{1}{2}\vec{k}$, as shown in figure \ref{sfig:shell_shift_single}. Since it is no longer symmetric, the required identities given in equation \eqref{eq:ang_ident} are no longer valid. On the other hand, when we include the additional constraint we have both $\theta^\high(\tfrac{1}{2}\vec{k}+\vec{j})$ and $\theta^\high(\tfrac{1}{2}\vec{k}-\vec{j})$. Taylor expansion of these functions gives
\begin{equation}
 \theta^\high(\tfrac{1}{2}\vec{k}\pm\vec{j}) = \theta^\high(\vec{j}) \pm x(\vec{k},\vec{j}) + \order{k^2} \ ,
\end{equation}
where
\begin{equation}
 x(\vec{k},\vec{j}) = \frac{\vec{k}\cdot\vec{j}}{2j} \Big[ \theta(\Lambda - j) \delta(j - \Lambda e^{-\ell}) - \theta(j - \Lambda e^{-\ell}) \delta(\Lambda - j) \Big] \ .
\end{equation}
As such, their product gives
\begin{equation}
 \theta^\high(\tfrac{1}{2}\vec{k}+\vec{j}) \theta^\high(\tfrac{1}{2}\vec{k}-\vec{j}) = \theta^\high(\vec{j}) + \order{k^2} \ ,
\end{equation}
and the contributions at order $k$ cancel exactly. The shell remains symmetric and the identities in equation \eqref{eq:ang_ident} remain valid. This is illustrated in figure \ref{sfig:shell_shift_double}.

The expansion of the $\theta^\high$-function in equation \eqref{eq:theta+} has highlighted a key point: when the change of variables is not used, there is a correction at order $k$ to the integration shell. Since we are required to keep terms up to $\order{k}$ in the integrand, if there is a term which is $\order{k^0}$ then there is potentially an additional contribution.

\subsection*{Expansion of $\theta$-functions as a limiting procedure}

\begin{figure}[tb]
 \begin{center}
  \subfigure[$\theta(x)$]{\label{sfig:theta_limit} \includegraphics[width=0.475\textwidth]{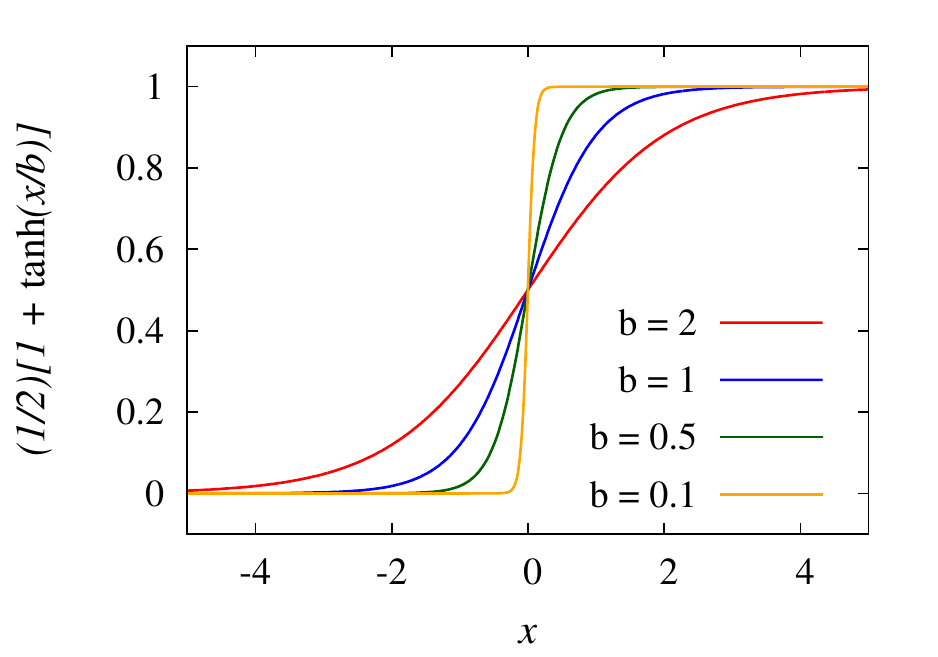} }
  \subfigure[$\delta(x)$]{\label{sfig:delta_limit} \includegraphics[width=0.475\textwidth]{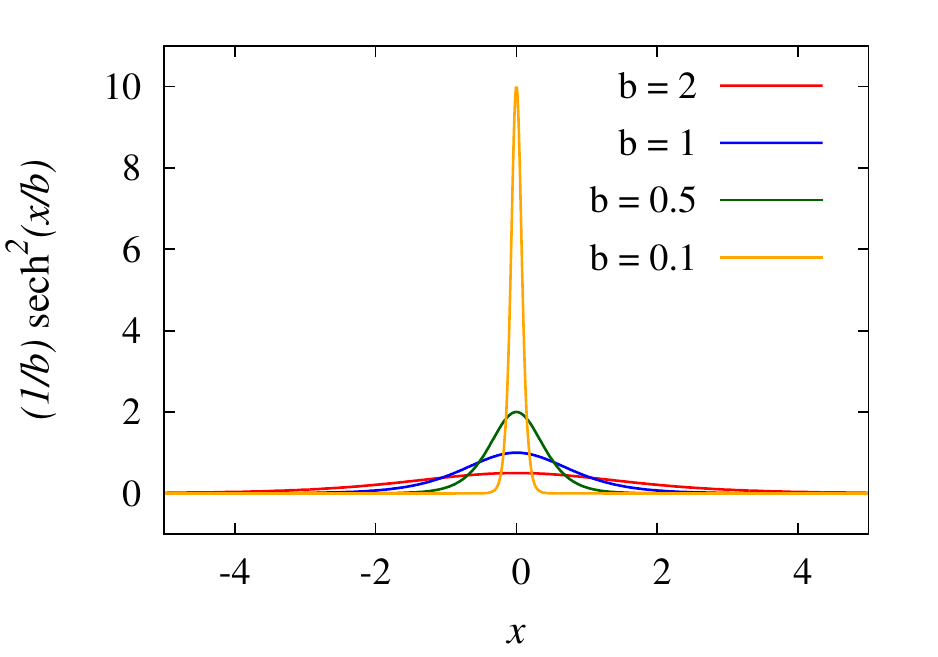} }
 \end{center}
 \caption{$\theta$- and $\delta$-functions as a limit of smooth functions, as given by equation \eqref{eq:theta_limit}.}
 \label{fig:thetadelta_limit}
\end{figure}

The previous subsection showed how the Taylor expansion of the $\theta^\high(\vec{k}-\vec{j})$ function contains an order $k$ contribution which can potentially contribute to the evaluation of the integral in equation \eqref{eq:current_pre_int}. To support the expansions given in equations \eqref{eq:theta+exp} and \eqref{eq:theta-exp}, we consider the $\theta$- and $\delta$-functions as limits of smooth functions,
\begin{equation}
 \label{eq:theta_limit}
 \theta(x) = \lim_{b \to 0} \left[ \frac{1}{2} + \frac{1}{2}\tanh\left(\frac{x}{b}\right) \right] \qquad\text{and}\qquad \delta(x) = \lim_{b\to 0} \left[ \frac{1}{b} \text{sech}^2\left(\frac{x}{b}\right) \right]\ ,
\end{equation}
as shown in figures \ref{sfig:theta_limit} and \ref{sfig:delta_limit}, respectively.
Expansion of $\theta(\lvert\vec{k}-\vec{j}\rvert - A)$, where $A = \Lambda e^{-\ell}$ is the cutoff, is then achieved by considering the Taylor expansion of the $\tanh$ function about $\vec{k} = 0$:
\begin{align}
 \label{eq:tanh_expansion}
 \tanh\left(\frac{\lvert\vec{k}-\vec{j}\rvert - A}{b}\right) = \tanh\left(\frac{\lvert\vec{j}\rvert - A}{b}\right) + \left.k_\alpha \frac{\partial}{\partial k_\alpha}\tanh\left(\frac{\lvert\vec{k}-\vec{j}\rvert - A}{b}\right) \right\vert_{\vec{k}=0} + \cdots \ .
\end{align}
We find the derivative, using the chain rule, as:
\begin{align}
 \frac{\partial}{\partial k_\alpha}\tanh\left(\frac{\lvert\vec{k}-\vec{j}\rvert - A}{b}\right) &= \frac{\partial \lvert\vec{k}-\vec{j}\rvert}{\partial k_\alpha} \frac{\partial}{\partial \lvert\vec{k}-\vec{j}\rvert}\tanh\left(\frac{\lvert\vec{k}-\vec{j}\rvert-A}{b}\right) \nonumber\\
 &= \frac{k_\alpha - j_\alpha}{\lvert\vec{k}-\vec{j}\rvert} \frac{1}{b} \text{sech}^2\left(\frac{\lvert\vec{k}-\vec{j}\rvert-A}{b}\right) \ ,
\end{align}
and evaluate at $\vec{k}=0$ to obtain
\begin{align}
 \left. \frac{\partial}{\partial k_\alpha}\tanh\left(\frac{\lvert\vec{k}-\vec{j}\rvert - A}{b}\right) \right\vert_{\vec{k}=0} &= -\frac{j_\alpha}{\lvert\vec{j}\rvert} \frac{1}{b} \text{sech}^2\left(\frac{\lvert\vec{j}\rvert-A}{b}\right) \ .
\end{align}
Inserting this into the expression for the expansion of $\tanh$ in equation \eqref{eq:tanh_expansion}, we find
\begin{align}
 \theta(\lvert\vec{k}-\vec{j}\rvert - A) &= \lim_{b \to 0} \left[ {\frac{1}{2} + \frac{1}{2}\tanh\left(\frac{\lvert\vec{j}\rvert-A}{b}\right)} - \frac{\vec{k}\cdot\vec{j}}{\lvert\vec{j}\rvert} {\frac{1}{b} \text{sech}^2\left(\frac{\lvert\vec{j}\rvert-A}{b}\right)}  + \cdots \right] \nonumber \\
	       &= \theta(\lvert\vec{j}\rvert-A) - \frac{\vec{k}\cdot\vec{j}}{\lvert\vec{j}\rvert} \delta(\lvert\vec{j}\rvert-A)  + \cdots \ ,
\end{align}
which, recalling that $A = \Lambda e^{-\ell}$, recovers the result of equation \eqref{eq:theta+exp}. A similar analysis may be performed to recover equation \eqref{eq:theta-exp}.

\subsection{Approach of Wang and Wu}
Wang and Wu \cite{Wang:1993p371} did not have a constraint on the loop momentum $\vec{k}-\vec{j}$. This is clear from equation (4) where they explicitly declare that their integral is over the shell $\Lambda e^{-\ell} < \lvert\vec{j}\rvert < \Lambda$ (in our notation). The change of variables used by FNS would therefore cause the shell to be shifted and invalidate the result. Instead, they continue from equation \eqref{eq:current_pre_CoV} without it. Since the first term in the square brackets already has a factor $k_\nu$, the projection operator is simply replaced by $P_{\gamma\mu}(\vec{j})$. We do, however, have to expand the second product
\begin{align}
 P_{\beta\nu}(\vec{j}) P_{\gamma\nu}(\vec{k}-\vec{j}) &= \left( \delta_{\gamma\nu} + \frac{k_\nu j_\gamma}{\lvert\vec{k}-\vec{j}\rvert^2} + \order{k^2} \right) P_{\beta\nu}(\vec{j}) \nonumber \\
 &= \left( \delta_{\gamma\nu} + \frac{k_\nu j_\gamma}{j^2} \left(1 + \frac{2\vec{k}\cdot\vec{j}}{j^2}\right) + \order{k^3} \right) P_{\beta\nu}(\vec{j}) \\
 &= \left( \delta_{\gamma\nu} + \frac{k_\nu j_\gamma}{j^2} \right) P_{\beta\nu}(\vec{j}) + \order{k^2} \ ,
\end{align}
where we made use of the Taylor expansion of $\lvert\vec{k}-\vec{j}\rvert^{-2}$. The current, as investigated by Wang and Wu and denoted by $\hat{\Sigma}^\low_\alpha$ with a hat to distinguish it from FNS and YO, is then given by
\begin{align}
 \label{eq:WW_pre_correct}
 \hat{\Sigma}^\low_\alpha(\vec{k},0) &= \frac{F_0\lambda_0^2}{i\nu^2_0} \frac{M_{\alpha\beta\gamma}(\vec{k}) u^\low_\mu(\vec{k},0)}{(2\pi)^d} \int d^dj\ j^{-y-4}\ \theta^{\high}(\vec{j}) \\
&\qquad\qquad\times \left[k_\nu P_{\gamma\mu}(\vec{j})P_{\beta\nu}(\vec{j}) - k_\nu \frac{j_\mu j_\gamma}{j^2}P_{\beta\nu}(\vec{j}) - k_\nu \frac{j_\mu j_\nu}{j^2}P_{\beta\gamma}(\vec{j}) - j_\mu P_{\beta\gamma}(\vec{j})\right] \nonumber \ ,
\end{align}
where we have explicitly dropped the constraint on $\vec{k}-\vec{j}$ and Taylor expanded the denominator of equation \eqref{eq:current_pre_CoV} as $(j^2 + \lvert\vec{k}-\vec{j}\rvert^2)^{-1} = (1/2j^2) (1 + \vec{k}\cdot\vec{j}/j^2) + \order{k^2}$.

The last term was seen previously to integrate to zero, since it is an odd function of $\vec{j}$ integrated over a symmetric domain. Using the identities in equation \eqref{eq:ang_ident}, this is evaluated to be
\begin{equation}
 \hat{\Sigma}^\low_\alpha(\vec{k},0) = -\left[\frac{F_0\lambda_0^2}{\nu^2_0} \frac{S_d}{(2\pi)^d} \frac{d-1}{2(d+2)} \left(\frac{e^{\epsilon\ell}-1}{\epsilon\Lambda^\epsilon}\right)\right] k^2 u^\low_\alpha(\vec{k},0) \ ,
\end{equation}
from which we find the viscosity increment
\begin{align}
 \label{eq:delta_nu_WW}
 \Delta\hat{\nu}_0(\vec{0},0) &= \frac{F_0\lambda_0^2}{\nu^2_0} \frac{S_d}{(2\pi)^d} \frac{d-1}{2(d+2)} \left(\frac{e^{\epsilon\ell}-1}{\epsilon\Lambda^\epsilon}\right) \nonumber \\
 &= \nu_0 A_d(0) \overline{\lambda}^2(0) \left(\frac{e^{\epsilon\ell} - 1}{\epsilon} \right) \ ,
\end{align}
with the various factors defined in equation \eqref{eq:result:fns_compact}.

\subsubsection{Finding the correction}
As noted above, we dropped the constraint on $\vec{k}-\vec{j}$ and the last term in equation \eqref{eq:WW_pre_correct}, which is order $k^0$, integrated to zero. We now restore $\theta^\high(\vec{k}-\vec{j})$ and consider its expansion as given in equation \eqref{eq:theta+}. Since all other terms in the integrand are already $\order{k}$, only the last term can generate a correction when coupled with the order $k$ contribution from equation \eqref{eq:theta+}, so we may write this correction as
\begin{align}
 \delta\hat{\Sigma}^\low_\alpha(\vec{k},0) &= \frac{F_0\lambda_0^2}{i\nu^2_0} \frac{k_\nu M_{\alpha\beta\gamma}(\vec{k}) u^\low_\mu(\vec{k},0)}{(2\pi)^d} \int d^dj\ j^{-y-5}\ j_\nu j_\mu P_{\beta\gamma}(\vec{j})\ \theta^{\high}(\vec{j}) \nonumber \\
&\qquad\qquad\times \Big[ \theta(\Lambda - j) \delta(j - \Lambda e^{-\ell}) - \theta(j - \Lambda e^{-\ell}) \delta(\Lambda - j) \Big] \ .
\end{align}
Inserting the definition of the projection operator and using the identities in equation \eqref{eq:ang_ident}, this is evaluated to be
\begin{align}
 \label{eq:current_correction_WW}
 \delta\hat{\Sigma}^\low_\alpha(\vec{k},0) &= \frac{F_0\lambda_0^2}{i\nu^2_0} \frac{k_\nu M_{\alpha\beta\gamma}(\vec{k}) u^\low_\mu(\vec{k},0)}{(2\pi)^d} \frac{S_d}{d(d+2)} \nonumber \\
 &\hspace{0.5in}\times\Big[(d+1)\delta_{\mu\nu} \delta_{\beta\gamma} - \delta_{\mu\beta} \delta_{\nu\gamma} - \delta_{\mu\gamma} \delta_{\nu\beta} \Big] \Big[(\Lambda e^{-\ell})^{-\epsilon} - \Lambda^{-\epsilon}\Big] \theta(0) \nonumber \\
 &= -\frac{F_0\lambda_0^2}{2\nu^2_0} \frac{S_d}{(2\pi)^d} \frac{1}{2d(d+2)} \left(\frac{e^{\epsilon\ell}-1}{\Lambda^{\epsilon}}\right) \Big( -2 k^2 u^\low_\alpha(\vec{k},0) \Big) \ .
\end{align}
In evaluating the constraints on $j$ we used the standard convention $\theta(0) = 1/2$, for example see \cite{Gozzi:1983p686,Hochberg:2000p688} or the limiting procedure given in equation \eqref{eq:theta_limit}. Thus the additional constraint has generated a correction to the viscosity increment calculated by Wang and Wu, given in equation \eqref{eq:delta_nu_WW}, of
\begin{equation}
 \delta\hat{\nu}_0(\vec{0},0) = -\frac{F_0\lambda_0^2}{\nu^2_0} \frac{S_d}{(2\pi)^d} \frac{1}{2d(d+2)} \left(\frac{e^{\epsilon\ell}-1}{\Lambda^{\epsilon}}\right) \ .
\end{equation}
When this correction is added to the result found by Wang and Wu, we recover the original result of FNS and YO, presented in equation \eqref{eq:delta_nu_fns}; that is,
\begin{align}
 \Delta\hat{\nu}_0(\vec{0},0) + \delta\hat{\nu}_0(\vec{0},0) &= \frac{F_0\lambda_0^2}{\nu^2_0} \frac{S_d}{(2\pi)^d} \frac{1}{2(d+2)} \left[ \frac{d-1}{\epsilon} - \frac{1}{d} \right] \left(\frac{e^{\epsilon\ell}-1}{\Lambda^{\epsilon}}\right) \ .
\end{align}

Therefore, the method used by Wang and Wu does not produce a different result to FNS and YO, once the constraint that \emph{all} loop momenta are contained in the eliminated shell is properly dealt with. Teodorovich \cite{Teodorovich:1994p376} used a different method to evaluate the angular part of the self-energy integral. However, the author also neglects the additional momentum constraint and arrives at the same result as Wang and Wu, equation \eqref{eq:delta_nu_WW}.

\subsection{Approach of Nandy}
In an attempt to settle the dispute, Nandy \cite{Nandy:1997p375} showed how symmetrising the self-energy integral could be used, along with no change of variables, to recover the result of FNS and YO. Referring to equation \eqref{eq:all_agree}, he points out that there is no reason to do one integral in place of the other, in which case one must do both and average the results. Of course, this should not lead to a different result, but due to the missing momentum constraint it does. As we will show, the corrections caused by more careful consideration of the domain of integration cancel one another, giving the appearance that this symmetrisation procedure was necessary.

We have dealt with half of the problem in the discussion above of Wang and Wu. It remains to return to equation \eqref{eq:all_agree} and instead perform the $\vec{j}$ integral,
\begin{align}
 \tilde{\Sigma}^\low_\alpha(\vec{k},0) = \frac{4\lambda_0^2}{\nu_0^2} M_{\alpha\beta\gamma}(\vec{k}) u^\low_\mu(\vec{k},0)\ \theta^\low(\vec{k}) & \int \frac{d^dp}{(2\pi)^d} \frac{F(\lvert\vec{k}-\vec{p}\rvert)}{\lvert\vec{k}-\vec{p}\rvert^2}\ M_{\gamma\mu\nu}(\vec{p}) \frac{P_{\beta\nu}(\vec{k}-\vec{p})}{\lvert\vec{k}-\vec{p}\rvert^2 + p^2}\ \nonumber \\
 &\times\theta^{\high}(\vec{k}-\vec{p}) \theta^{\high}(\vec{p})  \ .
\end{align}
We use a tilde to denote Nandy's results. In this case, the constraint on $\vec{k}-\vec{p}$ is dropped and Nandy only ensures that $\vec{p}$ is a high frequency mode. The calculation follows in a similar manner to the previous section: We Taylor expand the function $\lvert\vec{k}-\vec{p}\rvert^{-y-2}$ and the denominator; insert the definition of the projection and vertex operators; and collect terms to $\order{k}$ in the integrand to find
\begin{align}
 \label{eq:Nandy_pre_correct}
  \tilde{\Sigma}^\low_\alpha(\fvect{k}) &= \frac{F_0\lambda_0^2}{i\nu^2_0} \frac{k_\nu M_{\alpha\beta\gamma}(\vec{k}) u^\low_\mu(\vec{k},0)}{(2\pi)^d} \int d^dp\ p^{-y-4}\ \theta^{\high}(\vec{p})\ \bigg[ k_\nu P_{\gamma\nu}(\vec{p}) \frac{p_\mu p_\beta}{p^2}  \\
  &\hspace{0.5in} - k_\nu P_{\gamma\mu}(\vec{p}) \frac{p_\beta p_\nu}{p^2} + k_\beta P_{\gamma\mu}(\vec{p}) + (y+3)k_\nu P_{\beta\gamma}(\vec{p})\frac{p_\mu p_\nu}{p^2} + p_\mu P_{\beta\gamma}(\vec{p}) \bigg] \nonumber \ .
\end{align}
Note that, once again, the last term is the only term $\order{k^0}$ in the integrand and integrates to zero here due to symmetry. Continuing with the evaluation by inserting the definition of the remaining projection and vertex operators and performing the integrals using the identities in equation \eqref{eq:ang_ident}, one obtains
\begin{align}
 \label{eq:delta_nu_N}
 \Delta\tilde{\nu}_0(\vec{0},0) &= \frac{S_d}{(2\pi)^d} \frac{d^2 - d -2\epsilon}{2d(d+2)} \frac{F_0 \lambda_0^2}{\nu^2_0 \Lambda^\epsilon} \left(\frac{e^{\epsilon\ell} - 1}{\epsilon}\right) \ .
\end{align}
Thus we see that, under the symmetrisation procedure, we recover the FNS/YO result,
\begin{align}
 \Delta\overline{\nu}_0(\vec{0},0) = \frac{1}{2}\Big( \Delta\hat{\nu}_0(\vec{0},0) + \Delta\tilde{\nu}_0(\vec{0},0) \Big) = \Delta\nu_0(\vec{0},0) \ .
\end{align}

\subsubsection{Finding the correction}

We now return to equation \eqref{eq:Nandy_pre_correct} and restore the additional constraint, $\theta^\high(\vec{k}-\vec{p})$. Since the last term is the only term which can generate a correction at $\order{k}$ once we expand the $\theta^\high$-function, the correction can be written
\begin{align}
 \delta\tilde{\Sigma}^\low_\alpha(\vec{k},0) &= -\frac{F_0\lambda_0^2}{i\nu^2_0} \frac{k_\nu M_{\alpha\beta\gamma}(\vec{k}) u^\low_\mu(\vec{k},0)}{(2\pi)^d} \int d^dp\ p^{-y-5}\ p_\nu p_\mu P_{\beta\gamma}(\vec{p})\ \theta^{\high}(\vec{p}) \nonumber \\
&\hspace{1.25in}\times \Big[ \theta(\Lambda - p) \delta(p - \Lambda e^{-\ell}) - \theta(p - \Lambda e^{-\ell}) \delta(\Lambda - p) \Big] \nonumber \\
&= - \delta\hat{\Sigma}^\low_\alpha(\vec{k},0) \ .
\end{align}
We notice that, under the relabelling of the (dummy) loop momentum $\vec{p} \to \vec{j}$, the correction is identical to that found for Wang and Wu in equation \eqref{eq:current_correction_WW}, only with opposite sign.

The resulting correction to the viscosity increment can, therefore, be directly written:
\begin{align}
 \delta\tilde{\nu}_0(\vec{0},0) &= -\delta\hat{\nu}_0(\vec{0},0) \nonumber \\
 &= \frac{F_0\lambda_0^2}{\nu^2_0} \frac{S_d}{(2\pi)^d} \frac{1}{2d(d+2)} \left(\frac{e^{\epsilon\ell}-1}{\Lambda^{\epsilon}}\right) \ .
\end{align}
Once again, adding this contribution to the calculated viscosity increment given by $\Delta\tilde{\nu}_0(\vec{0},0)$ in equation \eqref{eq:delta_nu_N} recovers the FNS/YO result,
\begin{align}
 \Delta\tilde{\nu}_0(\vec{0},0) + \delta\tilde{\nu}_0(\vec{0},0) &= \frac{F_0\lambda_0^2}{\nu^2_0} \frac{S_d}{(2\pi)^d} \frac{1}{2d(d+2)} \left[ \frac{d^2-d-2\epsilon}{\epsilon} + 1 \right] \left(\frac{e^{\epsilon\ell}-1}{\Lambda^{\epsilon}}\right) \ .
\end{align}
We conclude that, however we choose to evaluate the self-energy integral, as long as all momentum constraints are carefully implemented we arrive at the same result.

\section{Renormalization of the stirring force}\label{sec:RG_force}
In addition to the renormalization of the viscosity, the contribution of the eliminated band of modes to the forcing has also been considered. This is presented in Berera and Yoffe \cite{Berera:2010p789}. Referring back to equations \eqref{eq:ahh!} and \eqref{eq:ahh!2}, the induced random force has the form
\begin{equation}
 \label{eq:induced_force}
 \Delta f^\low_\alpha(\fvect{k}) = \lambda_0 M^\low_{\alpha\beta\gamma}(\vec{k})  \int \fmeasure{j}{\Omega} G_0(\fvect{j})\ f_\beta(\fvect{j}) f_\gamma(\fvect{k}-\fvect{j}) G_0(\fvect{k}-\fvect{j}) \theta^\high(\vec{j}) \theta^\high(\vec{k}-\vec{j}) \ .
\end{equation}
A graphical representation of this expression is given in figure \ref{sfig:RG_force_bubble}. We see that it has zero mean, since the term $\langle f_\beta(\fvect{j}) f_\gamma(\fvect{k}-\fvect{j}) \rangle$ gives a $\delta(\fvect{k})$, which hits the vertex operator outside and forces $\langle \Delta f^\low_\alpha(\fvect{k}) \rangle = 0$. This can also be seen from figure \ref{sfig:RG_force_bubble}, since the averaging procedure forms a closed loop and so there cannot be a momentum flow through the graph.

\subsection{Setting up the calculation}

The induced force can, however, contribute to the autocorrelation. We see this by considering the resultant force $\tilde{f}^\low_\alpha(\fvect{k}) = f^\low_\alpha(\fvect{k}) + \Delta f^\low_\alpha(\fvect{k})$ and looking at
\begin{equation}
 \langle \tilde{f}_\alpha(\fvect{k}) \tilde{f}_\rho(\fvect{k}') \rangle = \langle f_\alpha(\fvect{k}) f_\rho(\fvect{k}') \rangle + \langle \Delta f_\alpha(\fvect{k}) \Delta f_\rho(\fvect{k}') \rangle\ .
\end{equation}
We see that an order $\lambda_0^2$ contribution to the autocorrelation has been created,
\begin{align}
  \label{eq:app:delta_f_cor}
   \langle \Delta f_\alpha(\fvect{k}) \Delta f_{\rho}(\fvect{k}') \rangle &= \lambda_0^2 M_{\alpha\beta\gamma}(\vec{k}) M_{\rho\mu\nu}(\vec{k}') \int\int \fmeasure{j}{\Omega}\ \fmeasure{p}{\Omega'}\ \theta^\high(\vec{j}) \theta^\high(\vec{p}) \theta^\high(\vec{k}-\vec{j})  \theta^\high(\vec{k}'-\vec{p}) \nonumber \\
   &\quad\times G_0(\fvect{j}) G_0(\fvect{k} - \fvect{j}) G_0(\fvect{p}) G_0(\fvect{k}' - \fvect{p}) \langle f_\beta(\fvect{j}) f_\gamma(\fvect{k}-\fvect{j}) f_\mu(\fvect{p}) f_\nu(\fvect{k}'-\fvect{p}) \rangle  \ .
 \end{align}
Since the forcing has been assumed to be Gaussian, we may split the fourth order moment into products of second order moments,
\begin{equation}
 \label{eq:split_4th}
  \langle f_\beta^{\fvect{j}} f_\gamma^{\fvect{k}-\fvect{j}} f_\mu^{\fvect{p}} f_\nu^{\fvect{k}'-\fvect{p}} \rangle =
  \langle f_\beta^{\fvect{j}} f_\gamma^{\fvect{k}-\fvect{j}} \rangle \langle f_\mu^{\fvect{p}} f_\nu^{\fvect{k}'-\fvect{p}} \rangle + \langle f_\beta^{\fvect{j}} f_\mu^{\fvect{p}} \rangle \langle f_\gamma^{\fvect{k}-\fvect{j}} f_\nu^{\fvect{k}'-\fvect{p}} \rangle 
  + \langle f_\beta^{\fvect{j}} f_\nu^{\fvect{k}'-\fvect{p}} \rangle \langle f_\mu^{\fvect{p}} f_\gamma^{\fvect{k}-\fvect{j}} \rangle \ ,
\end{equation}
where we use a symbolic notation for brevity. Considering once again the autocorrelation of our original force, we see that the first term generates two disconnected loops or, equivalently, vanishes due to $\delta(\fvect{k})$ and $\delta(\fvect{k}')$ acting on the vertex operators. Using the invariance of the integral under the change of variables $\fvect{j} \to \fvect{k}-\fvect{j}$ and $\fvect{p} \to \fvect{k}'-\fvect{p}$ (the integration region is maintained) and the symmetry of the vertex operator $M_{\alpha\beta\gamma} = M_{\alpha\gamma\beta}$, we see that the surviving two terms are actually equivalent and we have
\begin{align}
 \label{eq:induced_corr}
  \langle \Delta f_\alpha(\fvect{k}) \Delta f_{\rho}(\fvect{k}') \rangle &= 2 \lambda_0^2 M_{\alpha\beta\gamma}(\vec{k}) M_{\rho\mu\nu}(\vec{k}') \int\int \fmeasure{j}{\Omega}\ \fmeasure{p}{\Omega'}\ \theta^\high(\vec{j}) \theta^\high(\vec{p}) \theta^\high(\vec{k}-\vec{j})  \theta^\high(\vec{k}'-\vec{p}) \nonumber \\
  &\ \times G_0(\fvect{j}) G_0(\fvect{k} - \fvect{j}) G_0(\fvect{p}) G_0(\fvect{k}' - \fvect{p}) \langle f_\beta(\fvect{j}) f_\mu(\fvect{p}) \rangle \langle f_\gamma(\fvect{k}-\fvect{j}) f_\nu(\fvect{k}'-\fvect{p}) \rangle \ .
\end{align}

\begin{figure}[tb]
 \begin{center}
  \subfigure[Graphical representation of the induced force]{
   \label{sfig:RG_force_bubble}
   \includegraphics[width=0.65\textwidth]{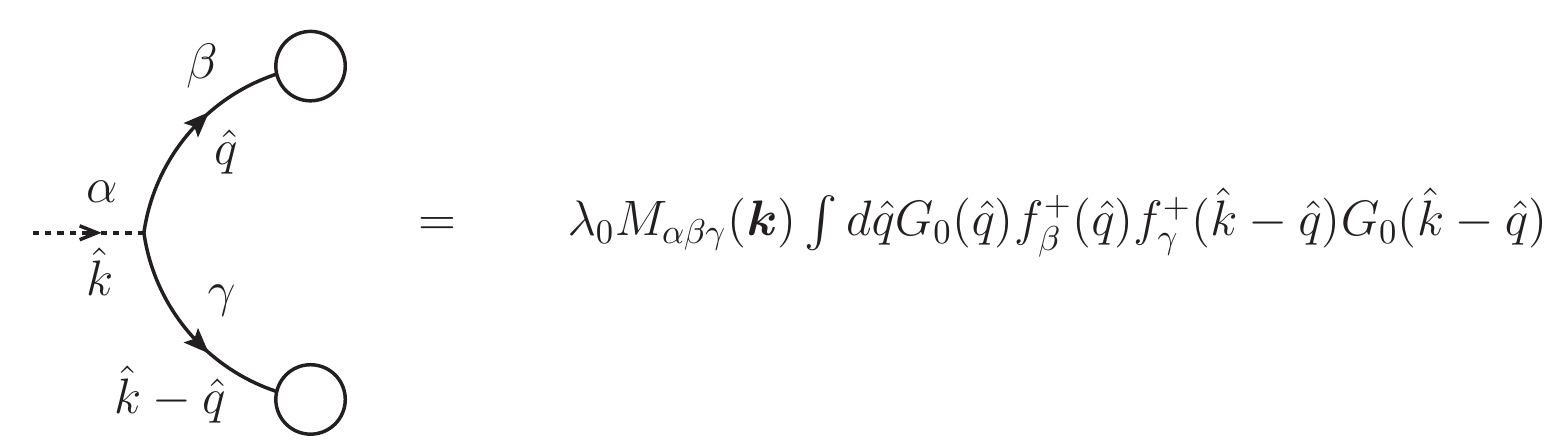}
  }
  \subfigure[Renormalization of the force autocorrelation]{
   \label{sfig:RG_force}
   \includegraphics[width=0.7\textwidth,trim=0 31px 0 0,clip]{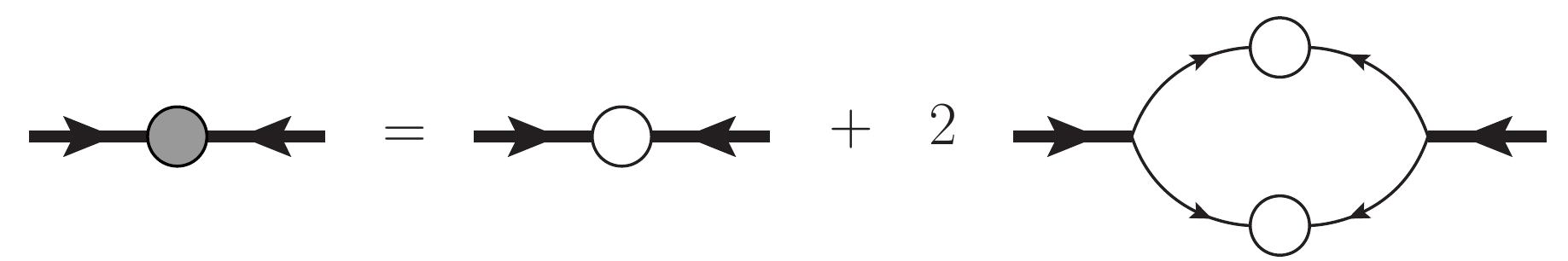}
  }
 \end{center}
 \caption{Feynman diagrams for the induced force and the resultant renormalization of the force autocorrelation. For the case $y = -2$, this leads to the one-loop correction of the force coefficient.}
 \label{fig:RG_force}
\end{figure}
From a graphical point of view, we take two of the induced force shown in figure \ref{sfig:RG_force_bubble} and average them. There are clearly three ways to join them, as we can join any circle with any of the other three. If we connect to the other leg coming from the same vertex, we generate two disconnected loops. The other two contributions, where we connect to legs of the other vertex, lead to the graph shown in figure \ref{sfig:RG_force}. This symmetry factor of two is just associated with relabelling the legs on one of the vertices, and is the origin (or a consequence, depending on how you look at it) of the symmetry of the vertex operator used above. Using the Feynman rules presented in figure \ref{sfig:RG_frules}, we see that the expression represented by the graph in figure \ref{sfig:RG_force} is identical to equation \eqref{eq:induced_corr}, once we restrict all momenta in the loop to be high frequency modes.

\subsection{Evaluation of the induced correlation}

We now continue with the evaluation of equation \eqref{eq:induced_corr}, which proceeds in a similar manner to the viscosity increment. This is done in our reduced notation. First, the correlations of the original force are inserted,
\begin{align}
   \langle \Delta f_\alpha^{\fvect{k}} \Delta f_{\rho}^{\fvect{k}'} \rangle &= 2\lambda_0^2 M_{\alpha\beta\gamma}^{\vec{k}} M_{\rho\mu\nu}^{\vec{k}'} \int \fmeasure{j}{\Omega}\int \fmeasure{p}{\Omega'}\ G_0^{\fvect{j}} G_0^{\fvect{k} - \fvect{j}} G_0^{\fvect{p}} G_0^{\fvect{k}' - \fvect{p}} 
   \langle f_\beta^{\fvect{j}} f_\mu^{\fvect{p}} \rangle \langle f_\gamma^{\fvect{k}-\fvect{j}} f_\nu^{\fvect{k}'-\fvect{p}} \rangle \nonumber \\
   &\hspace{1.25in}\times\theta^\high_{\vec{j}} \theta^\high_{\vec{p}} \theta^\high_{\vec{k}-\vec{j}}  \theta^\high_{\vec{k}'-\vec{p}} \nonumber \\
   &= 8\lambda_0^2 M_{\alpha\beta\gamma}^{\vec{k}} M_{\rho\mu\nu}^{\vec{k}'} \int\fmeasure{j}{\Omega}\int\fmeasure{p}{\Omega'}\ G_0^{\fvect{j}} G_0^{\fvect{k} - \fvect{j}} G_0^{\fvect{p}} G_0^{\fvect{k}' - \fvect{p}} \theta^\high_{\vec{j}} \theta^\high_{\vec{p}} \theta^\high_{\vec{k}-\vec{j}}  \theta^\high_{\vec{k}'-\vec{p}} \nonumber \\ 
   &\hspace{1.25in}\times F(j) P_{\beta\mu}^{\vec{j}} \delta(\fvect{j}+\fvect{p}) F(\lvert\vec{k}-\vec{j}\rvert) P_{\gamma\nu}^{\vec{k}-\vec{j}} \delta(\fvect{k}-\fvect{j}+\fvect{k}'-\fvect{p}) \nonumber \\
   &= -8\lambda_0^2 M_{\alpha\beta\gamma}^{\vec{k}} M_{\rho\mu\nu}^{\vec{k}} \delta(\fvect{k}+\fvect{k}') \int\fmeasure{j}{\Omega}\ \lvert G_0^{\fvect{j}}\rvert^2 \lvert G_0^{\fvect{k} - \fvect{j}}\rvert^2 F(j) F(\lvert\vec{k}-\vec{j}\rvert) P_{\beta\mu}^{\vec{j}} P_{\gamma\nu}^{\vec{k}-\vec{j}}  \nonumber \\
   \label{eq:force_pre_frequency}
   &\hspace{1.25in}\times \theta^\high_{\vec{j}} \theta^\high_{\vec{k}-\vec{j}} \ ,
\end{align}
where in the last line we used $\delta(\fvect{j}+\fvect{p})$ to perform the integral over $\fvect{p}$ and the resultant $\delta(\fvect{k}+\fvect{k}')$ to restrict $\vec{k}'=-\vec{k}$ in the vertex operator (which satisfies $M^{-\vec{k}} = -M^{\vec{k}}$).

The frequency integral is then performed as in section \ref{subsubsec:frequency_int}, where we close the contour in the upper half-plane and collect the residue from two poles at $\Omega = i\nu_0 j^2$ and $\Omega = \omega + i\nu_0 \lvert\vec{k}-\vec{j}\rvert^2$. The result is
\begin{align}
  &\int \frac{d\Omega}{2\pi}\ \lvert G_0^{\fvect{j}} \rvert^2 \lvert G_0^{\fvect{k}-\fvect{j}} \rvert^2 = \nonumber \\
  &\hspace{0.5in}\frac{2\pi}{4\pi\nu_0}\left[\frac{j^2\left(i\omega + \nu_0(j^2 + \lvert\vec{k}-\vec{j}\rvert^2)\right) + \lvert\vec{k}-\vec{j}\rvert^2\left(i\omega - \nu_0(j^2 + \lvert\vec{k}-\vec{j}\rvert^2)\right)}{j^2 \lvert\vec{k}-\vec{j}\rvert^2 \left(\omega^2 + \nu_0^2(j^2 + \lvert\vec{k}-\vec{j}\rvert^2)^2 \right)}\right] \ .
\end{align}
Taking the limit $\omega \to 0$ we see a drastic simplification,
\begin{equation}
 \left. \int d\Omega\ \lvert G_0^{\fvect{j}} \rvert^2 \lvert G_0^{\fvect{k}-\fvect{j}} \rvert^2 \right\vert_{\omega \to 0} = \frac{1}{2\nu_0^3}\left[\frac{1}{j^2 \lvert\vec{k}-\vec{j}\rvert^2 \left(j^2 + \lvert\vec{k}-\vec{j}\rvert^2 \right)}\right] \ ,
\end{equation}
which we insert into equation \eqref{eq:force_pre_frequency} to obtain
\begin{equation}
 \langle \Delta f_\alpha^{\vec{k}} \Delta f_{\rho}^{\vec{k}'} \rangle = \frac{-4\lambda_0^2 F_0^2}{\nu_0^3} \frac{M_{\alpha\beta\gamma}^{\vec{k}} M_{\rho\mu\nu}^{\vec{k}} \delta(\fvect{k}+\fvect{k}')}{(2\pi)^d} \int d^dj\ \frac{\Big(j\lvert\vec{k}-\vec{j}\rvert\Big)^{-y-2} P_{\beta\mu}^{\vec{j}} P_{\gamma\nu}^{\vec{k}-\vec{j}}}{j^2 + \lvert\vec{k}-\vec{j}\rvert^2} \theta^\high_{\vec{j}} \theta^\high_{\vec{k}-\vec{j}} \ .
\end{equation}
Since there is a factor of $k$ associated with each of the vertex operators, we note that the leading order will \emph{always} go as $\order{k^2}$. If we keep terms up to $\order{k^2}$ as we take $k \to 0$, the integrand contains only terms of order $k^0$ and as such we cannot generate any corrections by expanding the constraint $\theta^\high(\vec{k}-\vec{j})$. We are therefore left to evaluate
\begin{equation}
\langle \Delta f_\alpha^{\vec{k}} \Delta f_{\rho}^{\vec{k}'} \rangle = \frac{-2\lambda_0^2 F_0^2}{\nu_0^3} \frac{M_{\alpha\beta\gamma}^{\vec{k}} M_{\rho\mu\nu}^{\vec{k}} \delta(\fvect{k}+\fvect{k}')}{(2\pi)^d} \int d^dj\ j^{-2(y+3)} P_{\beta\mu}^{\vec{j}} P_{\gamma\nu}^{\vec{j}} \theta^\high_{\vec{j}} + \order{k^3} \ .
\end{equation}
Expanding the projection operators and performing the $(d-1)$ angular integrals, we find
\begin{align}
 \langle \Delta f_\alpha^{\vec{k}} \Delta f_{\rho}^{\vec{k}'} \rangle &= \frac{-2\lambda_0^2 F_0^2}{\nu_0^3} \frac{M_{\alpha\beta\gamma}^{\vec{k}} M_{\rho\mu\nu}^{\vec{k}} \delta(\fvect{k}+\fvect{k}')}{(2\pi)^d} \frac{S_d}{d(d+2)}  \\
&\hspace{0.5in}\times\left[ (d^2 - 3) \delta_{\beta\mu}\delta_{\gamma\nu} + \delta_{\beta\gamma}\delta_{\mu\nu} + \delta_{\beta\nu}\delta_{\gamma\mu} \right]\int dj\ j^{-2(y+3)+d-1}\ \theta^\high_{\vec{j}} \nonumber\ .
\end{align}
Finally, we expand the vertex operators, do the remaining integral over $j$ and perform contractions to obtain
\begin{align}
  \langle \Delta f^\low_\alpha(\fvect{k}) \Delta f^\low_{\rho}(\fvect{k}') \rangle &= \frac{\lambda_0^2 F_0^2}{\nu_0^3}  \frac{S_d}{(2\pi)^d} \frac{\delta(\fvect{k}+\fvect{k}')}{2d(d+2)} \left[ 2k^2 P_{\alpha\rho}(\vec{k}) (d^2 - 2) \right] \left(\frac{e^{\ell(\epsilon+y+2)} - 1}{(\epsilon+y+2)\Lambda^{\epsilon+y+2}}\right) \nonumber \\
  \label{eq:delta_ff}
  &= 2F_0 \overline{\lambda}^2(0) B_d  \left(\frac{e^{\ell(\epsilon+y+2)} - 1}{(\epsilon+y+2)\Lambda^{y+2}}\right) k^2 P_{\alpha\rho}(\vec{k}) \delta(\fvect{k}+\fvect{k}')\ , 
\end{align}
where the reduced coupling was defined in equation \eqref{eq:result:fns_compact} and the new prefactor is given by
\begin{equation}
 B_d = \frac{S_d}{(2\pi)^d} \frac{d^2 - 2}{2d(d+2)} \ .
\end{equation}

\section{Discussion}\label{sec:RG_discuss}
We have shown how the calculation of the viscosity increment using a change of variables is not subject to an antisymmetric shell of integration, as claimed by other authors \cite{Wang:1993p371,Teodorovich:1994p376,Nandy:1997p375}. Using $\theta$-functions, we have explicitly shown the preserved symmetry of the shell due to an additional constraint. Instead, calculation when the change of variables is not performed suffers from a shifted integration shell and the order $k^2$ correction was found to recover the original result of FNS.

The correction to the force autocorrelation is seen to always go as $\order{k^2}$ for low $k$. As such, the effect it has on the renormalization of the forcing depends on the choice of $y$ in $F(k) = F_0 k^{-y}$. When $y > -2$, the correction will be sub-leading as we take $k \to 0$ and as such can be safely neglected. However, when $y < -2$ we are presented with a problem since the forcing at large scales becomes dominated by the correction, which is order ${\overline{\lambda}^2(0)}$. This makes the calculation of the viscosity increment order ${\overline{\lambda}^4(0)}$, or \emph{two-loop}. Such a calculation has not been performed; instead analysis has to be restricted to $y \geq -2$.

The expressions for the renormalized viscosity and force autocorrelation, given in equations \eqref{eq:delta_nu_fns} and \eqref{eq:delta_ff}, can be written
\begin{align}
 \nu_I &= \nu_0 \left[ 1 + A_d(\epsilon) \overline{\lambda}^2(0) \left( \frac{e^{\epsilon\ell}-1}{\epsilon}\right) \right] \\
 \langle \tilde{f}^\low_\alpha(\fvect{k}) \tilde{f}^\low_\beta(\fvect{k}') \rangle &= 2F_0 k^{-y} P_{\alpha\beta}(\vec{k}) \delta(\fvect{k}+\fvect{k}')\ \left[ 1 + B_d \overline{\lambda}^2(0) \left(\frac{e^{\ell(\epsilon+y+2)} - 1}{(\epsilon+y+2)\Lambda^{y+2}}\right) k^{y+2} \right] \ .
\end{align}
For the case $y = -2$, which was one of the cases studied by FNS, we see that the renormalized autocorrelation can instead be written
\begin{align}
 \langle \tilde{f}^\low_\alpha(\fvect{k}) \tilde{f}^\low_\beta(\fvect{k}') \rangle = 2F_I k^2 P_{\alpha\beta}(\vec{k}) \delta(\fvect{k}+\fvect{k}') \ ,
\end{align}
where the force coefficient has been renormalized as
\begin{equation}
 F_I = F_0 \left[ 1 + B_d \overline{\lambda}^2(0) \left( \frac{e^{\epsilon\ell}-1}{\epsilon}\right) \right] \ .
\end{equation}
Since $y = -2$ corresponds to $\epsilon = 2-d$, we see $A_d(2-d) = B_d$ and the force and viscosity are renormalized in the same way. We note that the incorrect prefactor $A_d(0)$, obtained by other authors, can only agree with $B_d$ when $d = 2$ (the critical dimension for $y = -2$, where $\epsilon = 0$). Whereas, $A_d(\epsilon)$ agrees with $B_d$ for all $d$, provided $y = -2$.

We now consider the second stage of the RG and rescale the variables. This is done by introducing the scaling factor $s = e^\ell$ such that the spatial coordinates transform as $\vec{x} = s\vec{x}'$ and $t = s^zt'$, where the unprimed variables are the original scale. Therefore, we have $\vec{k}' = s\vec{k}$ and $\omega'=s^z\omega$. The velocity field is taken to scale as $\vec{u}(\vec{k},\omega) = s^\chi \vec{u}'(\vec{k}',\omega')$. In this case, we find (see appendix A of Berera and Yoffe \cite{Berera:2010p789})
\begin{equation}
 \nu(s) = s^{z-2}\nu_I \ , \qquad F(s) = s^{3z-2\chi+d+y} F_I \ , \qquad \lambda(s) = s^{\chi-d-1} \lambda_0 \ ,
\end{equation}
where the exponents $z$, $\chi$ are to be determined. We can then write expressions for the scale dependent renormalized quantities, such as the viscosity,
\begin{equation}
 \nu(s) = s^{z-2} \nu_0 \left[ 1 + A_d(\epsilon) \overline{\lambda}^2(0) \left( \frac{s^\epsilon-1}{\epsilon}\right) \right] \ .
\end{equation}
By taking the limit $s \to 1$ ($\ell \to 0$), we find the differential recursion relation
\begin{align}
 s\frac{\partial \nu(s)}{\partial s} = \nu(s) \Big[ z-2 + A_d(\epsilon) \overline{\lambda}^2(s) \Big] \ ,
\end{align}
which is just the beta-function for the viscosity, since with momentum scale $\mu = \Lambda/s$ we have $\partial/\partial \log{\mu} = -\partial/\partial \log{s}$. Similar expressions can be found for the force and vertex, $\lambda(s)$.

We now have two choices: First, for the case $y = -2$ the noise coefficient is renormalized in the same way as the viscosity, and so we can fix one of the scaling exponents since $3z-2\chi+d-2 = z-2$. This requires $\chi = z + d/2$. With this result, the beta-function for the reduced coupling is independent of $z$ and we have
\begin{equation}
 \beta(\overline{\lambda}) = \frac{\partial \overline{\lambda}}{\partial \log{s}} = \overline{\lambda}(s) \Big[ \epsilon/2 - A_d(\epsilon) \overline{\lambda}^2(s) \Big] \ .
\end{equation}
We can solve $\beta(\overline{\lambda}^*) = 0$ to find a non-trivial fixed point at $\overline{\lambda}^{*2} = \epsilon/2A_d(\epsilon)$. Ensuring that $\beta(\nu^*) = 0$ also requires $z-2+A_d(\epsilon)\overline{\lambda}^{*2}=0$, which simplifies to $z = 2 - \epsilon/2$.

Our second choice is $y > -2$, in which case the noise coefficient should not be renormalized and so $\chi = (3z + d + y)/2$ can be inferred. The reduced coupling then satisfies $\beta(\overline{\lambda}) = \overline{\lambda}(s) [ \epsilon - 3 A_d(\epsilon) \overline{\lambda}^2(s) ]/2$ from which we find the fixed point $\overline{\lambda}^{*2} = \epsilon/3A_d(\epsilon)$. This is the same beta-function and fixed point found using the more field-theoretic method by Teodorovich \cite{Teodorovich:1994p376}. If we require $\beta(\nu^*) = 0$, we then need $z = 2 - \epsilon/3$.

Throughout their work, FNS make use of Galilean invariance (GI) to constrain the vertex. By this, we require that $\lambda(s) = \lambda_0$, which places the condition $\chi = d+1$ on the scaling relations. However, we found that in both cases the constraint on $\chi$ imposed by the renormalization of the force was sufficient to find the fixed point, without the need to call on Galilean invariance. In fact, the consequences of GI are trivial and place no constraint on the renormalization of the vertex \cite{Berera:2007p406,McComb:2005p222,Berera:2005p1606}, apart from at $k = 0$. For $y = -2$, using $\chi = d+1$ with $\chi = z + d/2$ directly gives $z = 2 - \epsilon/2$, as found above. Similarly for $y > -2$ we find $z = 2-\epsilon/3$. So the vertex is not renormalized \emph{at this order}, but not due to GI.

Finally, we comment on the use of this low-$k$ dynamic RG to calculate inertial range quantities, as done by YO \cite{Yakhot:1986p357}. Briefly, the ``correspondence principle'' they invoke states that an unforced system which started from some initial conditions with a developed inertial range is statistically equivalent to a system forced in such a way as to generate the same scaling exponents. Thus, if we generate the right scaling exponent at low $k$ by our choice of forcing, then this artificially generated ``inertial range'' may be used to calculate properties of the inertial range. YO found that a value of $\epsilon = 4$ recovers $k^{-5/3}$ for the energy spectrum (see also Lesieur \cite{lesieur:1990-book}). It should be noted that this Wilson-style $\epsilon$-expansion is strictly only valid for $\epsilon$ small, so there is no evidence that it should describe $\epsilon = 4$. Moreover, the choice $\epsilon = 4$ corresponds to $y = d > 1$ and as such the forcing actually diverges as we take $k \to 0$. Results claiming to describe inertial range properties should therefore be treated with caution.

\chapter{A statistical approach to turbulence}\label{chp:statistical}

In section \ref{sec:intro_stat} we introduced the statistical closure problem. The equation of motion describing the evolution of the $n$th order moment of the velocity field always involves the $(n+1)$th moment, making one more unknown than we have equations. In order to make any prediction, we must introduce some sort of approximation that closes this set of equations. In this chapter, we introduce the application of statistical closure to turbulence. Existing statistical theories and renormalized perturbation theories are discussed before a further development of the recent theory of McComb \cite{McComb:2009p1053} is begun. This involves the introduction of a specific probability density functional for the velocity field, and its properties are studied here. This is very much a work in progress and its future development and application plans are discussed.

\section[Statistical closures and renormalized perturbation theories]{Statistical closures and renormalized perturbation theories in the study of isotropic turbulence}

The study of turbulence as a statistical problem is in no way new. It was started by Osborne Reynolds as early as 1883, and since then there have been numerous attempts (with varying degrees of success) to describe fluid turbulence. Due to the dramatic simplifications offered by homogeneity and isotropy, most of the work in this area concentrates on isotropic turbulence.

\begin{figure}
 \centering
 \includegraphics[width=\textwidth]{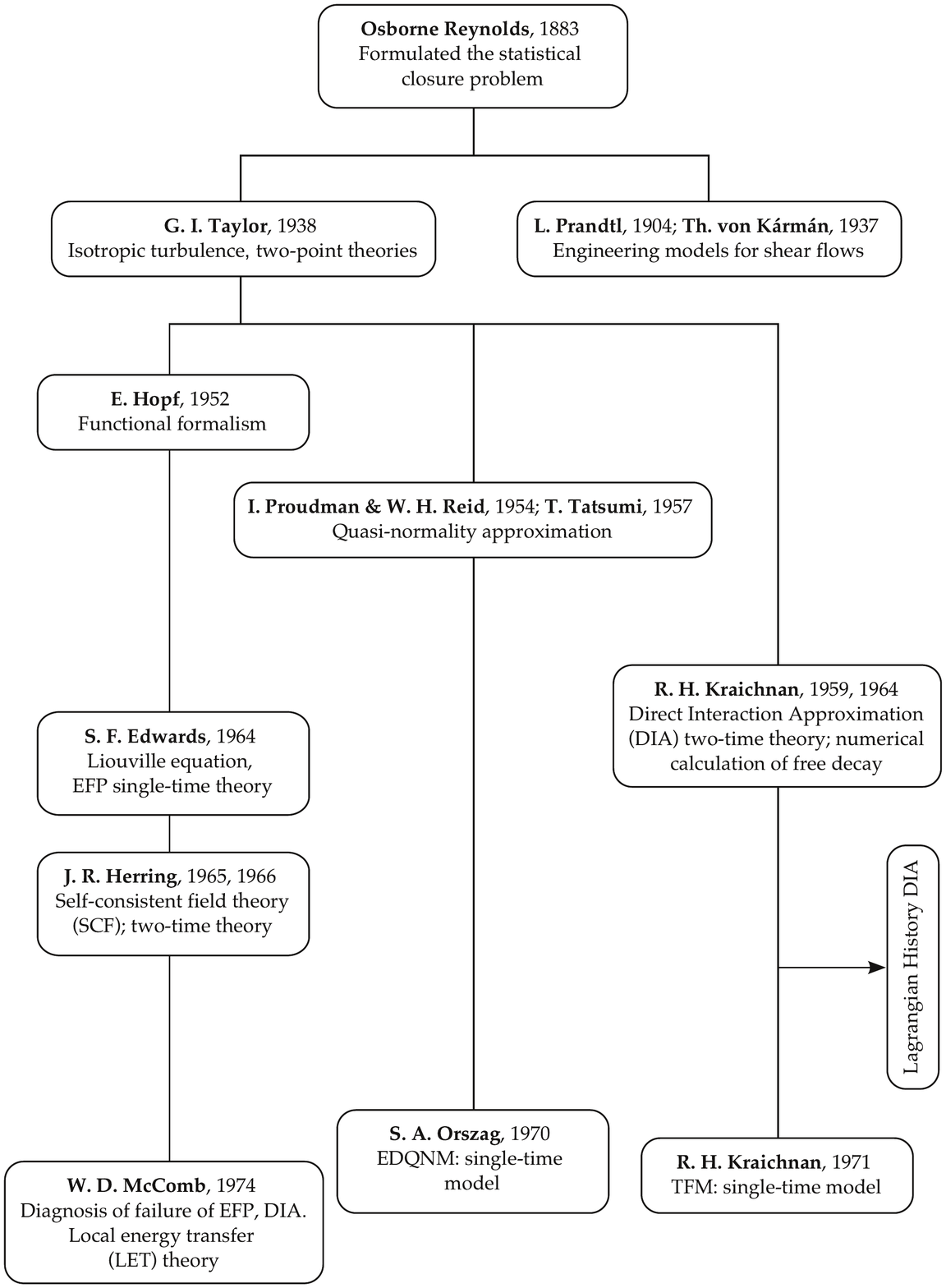}
 \caption{Summary of (Eulerian) statistical theories of turbulence.}
 \label{fig:chp8:summary}
\end{figure}
An outline of the historical development of statistical theories is illustrated in figure \ref{fig:chp8:summary}. The connection between the various theories is discussed by Lesieur \cite{lesieur:1990-book}, along with comprehensive reviews in Leslie \cite{leslie:1973-book} and McComb \cite{McComb:1990-book}. While we do not doubt the importance of all of these models, our attention will be directed to a discussion of quasi-normality and its updated form as the eddy-damped quasi-normal Markovian (EDQNM) approximation. Following this, we discuss the two renormalized perturbation theories (RPTs): the direct interaction approximation (DIA) and the local energy transfer (LET) theory. The connection between DIA and the Edwards-Fokker-Planck (EFP) theory is highlighted.

\subsection{Quasi-normality and EDQNM}\label{subsec:EDQNM}
The quasi-normal hypothesis was put forward by Proudman and Reid \cite{Proudman:1954p1647} and Tatsumi \cite{Tatsumi:1957p1763} in the 1950s. Essentially, it allows the fourth-order moment, introduced by the evolution equation for the third-order moment, to be decomposed into a product of second-order moments. This is a property of normal (or Gaussian) distributions and its assumption at this order does not impose Gaussianity on the velocity field, since we retain the contribution from this third-order moment.

We start from the energy balance equation,
\begin{equation}
 \left( \frac{\partial}{\partial t} + 2\nu_0 k^2 \right) E(k,t) = T(k,t) + W(k,t) \ ,
\end{equation}
and consider the form of the transfer spectrum:
\begin{equation}
 \label{eq:QN_T}
 T(k,t) = 4\pi k^2 \textrm{Re} \left[ M_{\alpha\beta\gamma}(\vec{k}) \int d^3j\ \langle u_\beta(\vec{j},t) u_\gamma(\vec{k}-\vec{j},t) u_\alpha(-\vec{k},t) \rangle \right] \ .
\end{equation}
The evolution equation for the third-order moment can be constructed by writing the Navier-Stokes equation for each of the components $u_\beta(\vec{j},t)$, $u_\gamma(\vec{k}-\vec{j},t)$ and $u_\alpha(-\vec{k},t)$, before multiplying by the remaining two. Expressing the triple moment as
\begin{equation}
 C_{\beta\gamma\alpha}(\vec{j},\vec{k}-\vec{j},-\vec{k};t) = \langle u_\beta(\vec{j},t) u_\gamma(\vec{k}-\vec{j},t) u_\alpha(-\vec{k},t) \rangle \ ,
\end{equation}
we add the three equations and average to find
\begin{align}
 &\left(\frac{\partial}{\partial t} + \nu_0 \Big[j^2 + \lvert\vec{k}-\vec{j}\rvert^2 + k^2\big] \right) C_{\beta\gamma\alpha}(\vec{j},\vec{k}-\vec{j},-\vec{k};t) \nonumber \\
 &\qquad = \int d^3m\ \Big( M_{\beta\rho\delta}(\vec{j}) \left\langle u_\rho(\vec{m},t) u_\delta(\vec{j-m},t) u_\gamma(\vec{k-j},t) u_\alpha(\vec{-k},t) \right\rangle \nonumber \\
&\qquad\qquad+ M_{\gamma\rho\delta}(\vec{k-j}) \left\langle u_\rho(\vec{m},t) u_\delta(\vec{k-j-m},t) u_\beta(\vec{j},t) u_\alpha(\vec{-k},t)\right\rangle \nonumber \\
&\qquad\qquad+ M_{\alpha\rho\delta}(\vec{-k}) \left\langle u_\rho(\vec{m},t) u_\delta(\vec{-k-m},t) u_\beta(\vec{j},t) u_\gamma(\vec{k-j},t)\right\rangle \Big) \nonumber \\
\label{eq:third-mom}
&\qquad = H_{\beta\gamma\alpha}(\vec{j},\vec{k}-\vec{j},-\vec{k};t) \ .
\end{align}
As mentioned previously, this highlights the closure problem, since the equation for the third-order moment requires knowledge of the fourth-order moment. This continues indefinitely, with the fourth-order dependent on the fifth, and so on. Formally, we can construct an expression for the third-order moment using an integrating factor, to obtain
\begin{align}
 C_{\beta\gamma\alpha}(\vec{j},\vec{k}-\vec{j},-\vec{k};t) = \int_0^t ds\ e^{-\nu_0[j^2 + \lvert\vec{k}-\vec{j}\rvert^2 + k^2](t-s)} H_{\beta\gamma\alpha}(\vec{j},\vec{k}-\vec{j},-\vec{k};s) \ .
\end{align}

In order to make the problem tractable, we must close the moment hierarchy in some way. It is at this point that we appeal to the quasi-normal hypothesis and split the fourth-order moments into products of second-order moments; for example
\begin{align}
 &\left\langle u_\rho(\vec{m},s) u_\delta(\vec{j-m},s) u_\gamma(\vec{k-j},s) u_\alpha(\vec{-k},s) \right\rangle \\
 &\hspace{3cm} = P_{\rho\delta}(\vec{m}) P_{\gamma\alpha}(\vec{k}-\vec{j}) C(m;s) C(\lvert\vec{k}-\vec{j}\rvert;s) \delta(\vec{j}) \nonumber \\
 &\hspace{3cm}\quad+ P_{\rho\gamma}(\vec{m}) P_{\delta\alpha}(\vec{j}-\vec{m}) C(m;s) C(\lvert\vec{j}-\vec{m}\rvert;s) \delta(\vec{m} + \vec{k} - \vec{j}) \nonumber \\
 &\hspace{3cm}\quad+ P_{\rho\alpha}(\vec{m}) P_{\delta\gamma}(\vec{j}-\vec{m}) C(m;s) C(\lvert\vec{j}-\vec{m}\rvert;s) \delta(\vec{m}-\vec{k}) \nonumber \ .
\end{align}
We have used isotropy to write $C_{\alpha\beta}(\vec{k};t) = P_{\alpha\beta}(\vec{k}) C(k;t)$. The first term on the RHS cannot contribute, since $\delta(\vec{j})$ forces the wavevector $\vec{j} = 0$, in which case $M_{\beta\rho\delta}(0) = 0$. Collecting the fourth-order moments together and using the $\delta$-functions to trivially perform the integral over $\vec{m}$, we have
\begin{align}
 H_{\beta\gamma\alpha}(\vec{j},\vec{k}-\vec{j},-\vec{k};s) &= 2 M_{\beta\rho\delta}(\vec{j}) P_{\rho\gamma}(\vec{k}-\vec{j}) P_{\delta\alpha}(\vec{k}) C(\lvert\vec{k}-\vec{j}\rvert;s) C(k;s) \nonumber \\
 &\quad+ 2 M_{\gamma\rho\delta}(\vec{k}-\vec{j}) P_{\rho\beta}(\vec{j}) P_{\delta\alpha}(\vec{k}) C(j;s) C(k;s) \nonumber \\
 \label{eq:QN_H}
 &\quad+ 2 M_{\alpha\rho\delta}(\vec{-k}) P_{\rho\beta}(\vec{j}) P_{\delta\gamma}(\vec{k}-\vec{j}) C(j;s) C(\lvert\vec{k}-\vec{j}\rvert;s) \ .
\end{align}
Note that we have used the symmetry $M_{\beta\rho\delta}(\vec{j}) = M_{\beta\delta\rho}(\vec{j})$ and a relabelling to gain the factor of 2. Introducing the coefficient $L(\vec{k},\vec{j})$ --- see appendix \ref{app:Lkj} --- the transfer spectrum can be expressed as
\begin{align}
 T(k,t) = 8\pi k^2 \int d^3j\ L(\vec{k},\vec{j}) \int_0^t ds\ & e^{-\nu_0[j^2 + \lvert\vec{k}-\vec{j}\rvert^2 + k^2](t-s)} \nonumber \\
 \label{eq:QN_T_final}
  &\times C(\lvert\vec{k}-\vec{j}\rvert;s) \Big( C(j;s) - C(k;s) \Big) \ ,
\end{align}
where the coefficient is defined as
\begin{equation}
 L(\vec{k},\vec{j}) = -2 M_{\alpha\beta\gamma}(\vec{k}) M_{\beta\alpha\rho}(\vec{j}) P_{\rho\gamma}(\vec{k}-\vec{j}) \ .
\end{equation}

If we prescribe $C(k;0)$, the evolution equation for the energy spectrum can be integrated forward in time, with the non-linearity computed directly from the the single-time covariance at earlier times. Note that it involves an integral over the complete time history of the system. However, this computation was done in the early 1960s by Ogura \cite{Ogura:1963p1841}, where it was found that, in the case of free decay, the single-time covariance (and hence the energy spectrum) became negative as time progressed. This is completely unphysical. Furthermore, the effect became worse as Reynolds number was increased. This is because at larger viscosities (lower $Re$) the dynamical memory is finite.

The failings of quasi-normality were discussed by Orszag \cite{Orszag:1970p1614}, where the author notes that this unphysical behaviour can be understood on the basis of improper relaxation times. Instead of the time-memory only involving a viscous damping, the history should be cut off because the non-linear interactions destroy coherence. In other words, as time progresses, the correlation of the velocity field with its past should become less important. Improvements to the quasi-normal hypothesis could then be expected by replacing the viscous damping by a modelled eddy damping, $\nu_0 k^2 \to \eta(k)$; for example, Bos, Chevillard, Scott and Rubinstein \cite{Bos:2011p797} use
\begin{equation}
 \eta(k) = \nu_0 k^2 + \lambda \sqrt{\int_0^k dq\ q^2 E(q)} \ ,
\end{equation}
where $\lambda$ is a parameter they choose to be 0.49. A Markovian system is then constructed by using a correlation time to replace the time integral, leading to the EDQNM equation
\begin{equation}
 T(k,t) = 8\pi k^2 \int d^3j\ L(\vec{k},\vec{j})\ \Theta(\vec{k},\vec{j};t)\ C(\lvert\vec{k}-\vec{j}\rvert;t) \Big( C(j;t) - C(k;t) \Big) \ ,
\end{equation}
where the correlation time is given by (for example, Lesieur \cite{lesieur:1990-book})
\begin{equation}
 \Theta(\vec{k},\vec{j};t) = \frac{1 - \exp \Big[ -\big( \eta(k) + \eta(j) + \eta(\lvert\vec{k}-\vec{j}\rvert) \big) t \Big]}{ \eta(k) + \eta(j) + \eta(\lvert\vec{k}-\vec{j}\rvert) } \ .
\end{equation}

\subsection{DIA and LET}
We begin this section with a short interpretation of the perturbation expansion for the velocity field. This is based on the account given in McComb \cite{McComb:2012}. The exact velocity field can be expressed as a perturbation series
\begin{equation}
 \label{eq:vel_pert}
 u_\alpha(\vec{k},t) = u^{(0)}_\alpha(\vec{k},t) + \lambda u^{(1)}_\alpha(\vec{k},t) + \lambda^2 u^{(2)}_\alpha(\vec{k},t) + \order{\lambda^3} \ ,
\end{equation}
where the zero-order field $u^{(0)}_\alpha(\vec{k},t)$ is chosen to be Gaussian-distributed. This can be compared with a direct numerical simulation from a Gaussian initial field. As the simulation is started, the non-linearity gets to work coupling modes and exchanging energy between them and the velocity field at time $t$ develops non-Gaussian corrections. As time progresses, these corrections increase in order, as the modes become coupled in more and more complicated ways. Eventually, once the field has developed, the exact velocity field corresponds to a (potentially infinite) sum of terms representing the multiple interactions of the modes of the Gaussian velocity field from which it is constructed. Of course, the exact velocity field is our observable quantity; we do not have access to $u^{(0)}_\alpha(\vec{k},t)$ in the simulation.

With this in mind, we turn our attention to renormalized perturbation theories (RPTs), which involve the resummation of a subset of the terms generated by the perturbation expansion. A nice treatment is presented in McComb \cite{McComb:2012} and we summarise the main points here. In a symbolic notation, the Navier-Stokes equation can be expressed as
\begin{equation}
 \Big( \partial_t + \nu_0 k^2 \Big) u_{\vec{k}} = f_{\vec{k}} + \lambda \int d^3j\ M_{\vec{k}} u_{\vec{j}} u_{\vec{k}-\vec{j}} \ .
\end{equation}
As in section \ref{subsubsec:decompose}, we introduce the bare propagator or viscous response function as the Green function of the linear operator on the LHS,
\begin{equation}
 \Big( \partial_t + \nu_0 k^2 \Big) R_{\vec{k}}^{(0)}(t,t') = \delta(t-t') \ ,
\end{equation}
such that the velocity field may be written
\begin{equation}
 u_{\vec{k}} = R_{\vec{k}}^{(0)} f_{\vec{k}} + \lambda R_{\vec{k}}^{(0)}\int d^3j\ M_{\vec{k}} u_{\vec{j}} u_{\vec{k}-\vec{j}} \ .
\end{equation}
If the perturbation series given by equation \eqref{eq:vel_pert} is then inserted on both sides, we collect terms at $\order{\lambda^n}$ to find expressions for $u_{\vec{k}}^{(n)}$ in terms of $u_{\vec{k}}^{(0)}$. The exact correlation of the velocity field is then found by inserting these expressions into
\begin{equation}
 C_{\vec{k}} = \big\langle u_{\vec{k}}^{(0)} u_{-\vec{k}}^{(0)} \big\rangle + \big\langle u_{\vec{k}}^{(0)} u_{-\vec{k}}^{(2)} \big\rangle + \big\langle u_{\vec{k}}^{(1)} u_{-\vec{k}}^{(1)} \big\rangle + \big\langle u_{\vec{k}}^{(2)} u_{-\vec{k}}^{(0)} \big\rangle + \order{\lambda^4} \ .
\end{equation}
Note that this is a two-time covariance, $C_{\vec{k}} = C_{\alpha\beta}(\vec{k};t,t')$. The times will be restored when we return to full notation. Using the Gaussian properties of the zero-order field to factorise fourth-order moments into products of second-order moments, the exact covariance may be written as
\begin{align}
 C_{\vec{k}} &= R_{\vec{k}}^{(0)} \Big\langle f_{\vec{k}} f_{-\vec{k}} \Big\rangle \left[ R_{-\vec{k}}^{(0)} + 4\int d^3j\ R_{-\vec{k}}^{(0)}M_{-\vec{k}} M_{-\vec{k}-\vec{j}} R_{-\vec{k}-\vec{j}}^{(0)} C_{\vec{j}}^{(0)} R_{-\vec{k}}^{(0)} + \order{\lambda^4} \right] \nonumber \\
 &\qquad+ 2R_{\vec{k}}^{(0)}\int d^3j\ M_{\vec{k}} C_{\vec{k}-\vec{j}}^{(0)} \Big[ M_{-\vec{k}} C_{\vec{j}}^{(0)} R_{-\vec{k}}^{(0)} + 2 M_{\vec{j}} R_{\vec{j}}^{(0)} C_{-\vec{k}}^{(0)} \Big] \ ,
\end{align}
where $M_{\vec{k}}$ is understood to represent the flow of momentum $\vec{k}$ into a vertex from the left (or the total outgoing to the right) and $C_{\vec{j}}^{(0)}$ the flow of momentum $\vec{j}$ carried by a zero-order covariance from left to right. A negative sign reverses the direction of the flow, which for a vertex implies that the flow is incoming from the right. Note that, since we have truncated our expansion of the velocity field at $\order{\lambda^2}$ we do not consider the renormalization of the vertex. Anything that connects like a vertex can only enter at higher order.

\begin{figure}[tb]
 \begin{center}
  \subfigure[Renormalized response]{ \includegraphics[height=5.05em]{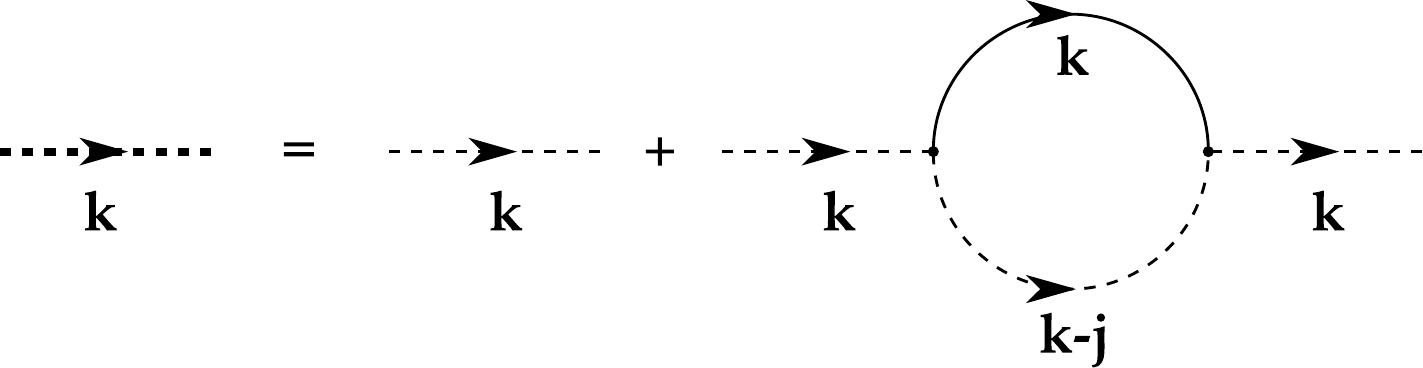} }
  \subfigure[Renormalized covariance]{ \includegraphics[height=5.05em]{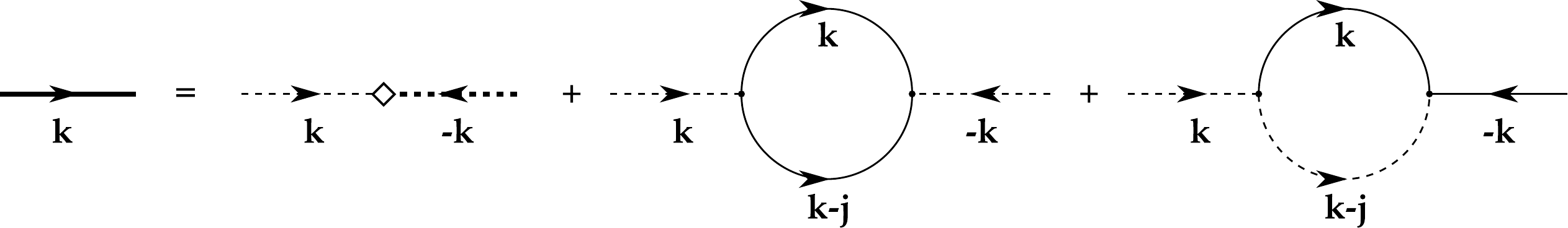} }
 \end{center}
 \caption{Diagrammatic representation of the exact response and covariance. Thick lines denote renormalized quantities, while thin lines represent zero-order. A solid line represents a covariance and a dashed line a response function. The empty diamond is the force autocorrelation. When renormalization is performed, all thin lines on the RHS are replaced by their corresponding thick line.}
 \label{fig:PT_diags}
\end{figure}

The term contained in the square brackets on the first line of the above equation is taken to give the exact response function,
\begin{equation}
 \label{eq:DIA_R_un}
 R_{\vec{k}} = R_{\vec{k}}^{(0)} + 4\int d^3j\ R_{\vec{k}}^{(0)}M_{\vec{k}} M_{\vec{k}-\vec{j}} R_{\vec{k}-\vec{j}}^{(0)} C_{\vec{j}}^{(0)} R_{\vec{k}}^{(0)} + \order{\lambda^4}  \ ,
\end{equation}
such that the covariance can be written
\begin{align}
 \label{eq:DIA_C_un}
 C_{\vec{k}} &= R_{\vec{k}}^{(0)}\Big\langle f_{\vec{k}} f_{-\vec{k}} \Big\rangle R_{-\vec{k}} + 2R_{\vec{k}}^{(0)}\int d^3j\ M_{\vec{k}} C_{\vec{k}-\vec{j}}^{(0)} \Big[ M_{-\vec{k}} C_{\vec{j}}^{(0)} R_{-\vec{k}}^{(0)} + 2 M_{\vec{j}} R_{\vec{j}}^{(0)} C_{-\vec{k}}^{(0)} \Big] \ .
\end{align}
These equations are illustrated by the diagrams in figure \ref{fig:PT_diags}.

From equation \eqref{eq:DIA_C_un}, we write the evolution equation for the exact covariance as
\begin{align}
 \Big(\partial_t + \nu_0 k^2 \Big) C_{\vec{k}} &= \Big\langle f_{\vec{k}} f_{-\vec{k}} \Big\rangle R_{-\vec{k}} + 2\int d^3j\ M_{\vec{k}} C_{\vec{k}-\vec{j}}^{(0)} \Big[ M_{-\vec{k}} C_{\vec{j}}^{(0)} R_{-\vec{k}}^{(0)} + 2 M_{\vec{j}} R_{\vec{j}}^{(0)} C_{-\vec{k}}^{(0)} \Big] \ .
\end{align}
We now do the renormalization and replace all zero-order quantities on the RHS of equations \eqref{eq:DIA_R_un} and \eqref{eq:DIA_C_un} with their exact counterparts, $R_{\vec{\kappa}}^{(0)} \to R_{\vec{\kappa}}$ and $C_{\vec{\kappa}}^{(0)} \to C_{\vec{\kappa}}$. In the diagrams, this is equivalent to replacing thin lines with thick lines. Using isotropy to express $C_{\vec{k}} = P_{\vec{k}} C_k$, where $P_{\vec{k}}$ is the projection operator in this reduced notation, the above equation reduces to
\begin{align}
 \label{eq:DIA_C_symb}
 \Big(\partial_t + \nu_0 k^2 \Big) C_k &= \Big\langle f_{\vec{k}} f_{-\vec{k}} \Big\rangle R_k + \int d^3j\ L(\vec{k},\vec{j})\ C_{\lvert\vec{k}-\vec{j}\rvert} \Big[ C_j R_k - R_j C_k \Big] \ ,
\end{align}
where the factor $L(\vec{k},\vec{j})$ is the same geometric factor encountered in section \ref{subsec:EDQNM} and discussed in appendix \ref{app:Lkj}.

Restoring full notation and dropping the forcing (or at least restricting our attention to $\delta$-correlation in time), equation \eqref{eq:DIA_C_symb} and the equivalent for equation \eqref{eq:DIA_R_un} become:
\begin{align}
 \Big(\partial_t + \nu_0 k^2 \Big) C(k;t,t') &= \int d^3j\ L(\vec{k},\vec{j})\ \left[ \int_0^{t'} ds\ R(k;t',s) C(j;t,s) C(\lvert\vec{k}-\vec{j}\rvert;t,s) \right. \nonumber \\
 &\qquad - \left. \int_0^{t} ds\ R(j;t,s) Ckj;t',s) C(\lvert\vec{k}-\vec{j}\rvert;t,s) \right] \\
 \Big(\partial_t + \nu_0 k^2 \Big) R(k;t,t') &= \delta(t-t') \nonumber \\
 &\qquad- \int d^3j\ L(\vec{k},\vec{j}) \int_{t'}^t ds\ R(j;t,s) R(k;s,t') C(\lvert\vec{k}-\vec{j}\rvert;t,s) \ .
\end{align}
With $R$ relabelled $G$, these are the equations of the direct interaction approximation (DIA) for decaying turbulence derived by Kraichnan \cite{Kraichnan:1958p1038,Kraichnan:1959p244}. Note that, when $t = t'$, the DIA equation for the single-time covariance requires $2\nu_0 k^2$ on the LHS, rather than $\nu_0 k^2$, and a factor 2 on the RHS. In the DIA, the response was taken to be the response of the velocity field to forcing, denoted $G(k;t,t')$,
\begin{equation}
 G_{\alpha\beta}(\vec{k};t,t') = P_{\alpha\beta}(\vec{k}) G(k;t,t') = \left\langle \frac{\delta u_\alpha(\vec{k},t)}{\delta f_\beta(\vec{k},t')} \right\rangle \ , \qquad t > t' \ .
\end{equation}

The DIA equation for the covariance can be connected to the eddy-damped quasi-normal approximation by considering exponential decay for time correlations, such that
\begin{align}
 C(k;t,t') = e^{-\eta(k) (t-t')} C(k;t',t') \ , \qquad G(k;t,t') = e^{-\eta(k) (t-t')} \ ,
\end{align}
where the time $t \geq t'$. If the `memory' of the time integral is taken to be independent of $t$, such that $\Theta(\vec{k},\vec{j}) = 1/\big( \eta(k) + \eta(j) + \eta(\lvert\vec{k}-\vec{j}\rvert) \big)$ in the Markovian approximation, we recover the energy balance derived from the second-order consistency condition for the statistical theory of Edwards \cite{Edwards:1964p277}. However, the equations for the response function do not agree under these assumptions.

While the DIA is a successful theory, it could not support the Kolmogorov spectrum. Instead, DIA predicts an inertial range spectrum which goes as $k^{-3/2}$. This was put down to the general failing of Eulerian spectral closures to satisfy random Galilean invariance, since the two-time moments are not invariant.
This was overcome by reformulating DIA in Lagrangian coordinates (Lagrangian History DIA), which is consistent with a $k^{-5/3}$ inertial range. Furthermore, the DIA and EFP possess an infra-red divergence as we take the limit of infinite Reynolds number, when the inertial range extends over all wavenumbers ($0 < k < \infty$).

The failings of DIA and the Edwards-Fokker-Planck (EFP) approach were discussed in McComb \cite{McComb:1974p316}, where the local energy transfer (LET) theory was introduced. Since the response in DIA is connected to the forcing, the relaxation of the covariance equation puts too much emphasis on the low-$k$, energy containing modes. Instead, the LET introduces the response function through a fluctuation-dissipation relation, such that
\begin{equation}
 C(k;t,t') = R(k;t,t') C(k;t',t') \ .
\end{equation}
The response is therefore associated with correlations of the velocity, or
\begin{equation}
 R_{\alpha\beta}(\vec{k};t,t') = P_{\alpha\beta}(\vec{k}) R(k;t,t') = \left\langle \frac{\delta u_\alpha(\vec{k},t)}{\delta u_\beta(\vec{k},t')} \right\rangle \ , \qquad t > t' \ .
\end{equation}
With this form for the response tensor, the LET theory \emph{is} consistent with a $k^{-5/3}$ inertial range. The covariance equation remains unmodified; however, the evolution equation for the response function contains an extra term which, it turns out, cancels the IR divergence present in the DIA and EFP theories:
\begin{align}
 \Big(\partial_t + & \nu_0 k^2  \Big) C(k;t,t') =  - \int d^3j\ L(\vec{k},\vec{j}) \int_{t'}^t ds\ R(j;t,s) R(k;s,t') C(\lvert\vec{k}-\vec{j}\rvert;t,s) \\
 &+\int d^3j\ L(\vec{k},\vec{j})\int_0^{t'} ds\ \frac{C(\lvert\vec{k}-\vec{j}\rvert;t,s)}{C(k;t',t')} \Big[ R(k;t',s) C(j;t,s) - R(j;t,s) C(k;t',s) \Big] \nonumber \ .
\end{align}
Although, the response function is evaluated from the fluctuation-dissipation rather than solution of this equation. See McComb \cite{McComb:1990-book,mccomb:2004-book,McComb:2012} for a fuller account. We note that both LET and DIA are examples of mean-field theories, with the response treated as being statistically sharp under averaging.

\section{A two-time statistical theory}\label{sec:stat_theory}
We start our discussion of the statistical theory of McComb \cite{McComb:2009p1053} from the Liouville equation describing the evolution of the probability density functional (PDF) of the velocity field, denoted $P[\vec{u},t]$. This is an explicitly time-dependent functional of the velocity field, $\vec{u}(\vec{k},t)$. The equation is presented as
\begin{align}
 \label{eq:Liouville}
 \frac{d P[\vec{u},t]}{d t} = \left[ \frac{\partial}{\partial t} + V(t) + L(t) \right] P[\vec{u},t] = 0 \ ,
\end{align}
where the operators are defined by their action on $P$,
\begin{align}
 V(t) P &= \int d^3p\ \frac{\delta}{\delta u_\rho(\vec{p},t)} \left[ \int d^3q\ M_{\rho\beta\gamma}(\vec{p}) u_\beta(\vec{q},t) u_\gamma(\vec{p}-\vec{q},t) \ P \right] \\
 L(t) P &= - \int d^3p\ \frac{\delta}{\delta u_\rho(\vec{p},t)} \left[ \left( \nu_0 p^2 u_\rho(\vec{p},t) + P_{\rho\sigma}(\vec{p}) F(p) \frac{\delta}{\delta u_\sigma(-\vec{p},t)} \right) P \right] \ .
\end{align}
We note that this equation is linear in the PDF and has been derived elsewhere, such as Edwards \cite{Edwards:1964p277}, Leslie \cite{leslie:1973-book} or McComb \cite{McComb:1990-book}. This can also be derived by consideration of the characteristic (or \emph{generating}) functional, Beran \cite{Beran:1967p749}. Once again, the spectral density of the forcing is denoted $F(k)$, as introduced in equation \eqref{eq:forcing_density}.

\subsection{Model PDF}
In a similar manner to McComb \cite{McComb:2009p1053}, we introduce a Gaussian base distribution $P_0[\vec{u}]$ and consider and expansion of the full PDF in a perturbation series. The base distribution is chosen such that it is normalised to unity and recovers the full covariance,
\begin{align}
 \int \mathscr{D}\vec{u} \ P_0[\vec{u}] &= 1 \\
 \int \mathscr{D}\vec{u} \ P_0[\vec{u}]\ u_\mu(\vec{k},t) u_\nu(-\vec{k},t') &= \langle u_\mu(\vec{k},t) u_\nu(-\vec{k},t') \rangle = C_{\mu\nu}(\vec{k};t,t') \ ,
\end{align}
respectively, and $t \geq t'$. The functional measure $\mathscr{D}\vec{u}$ is understood to represent the variation of the configuration of the field in the continuum limit. Thus the base distribution has the functional form
\begin{equation}
 P_0[\vec{u}] = N \exp\left[ -\frac{1}{2}\int d\tau \int d\tau' \int d^3\kappa\ u_\alpha(-\vec{\kappa},\tau) C^{-1}_{\alpha\beta}(\vec{\kappa};\tau,\tau') u_\beta(\vec{\kappa},\tau') \right] \ ,
\end{equation}
where $N$ is a normalisation constant and the inverse of the correlation tensor is defined through
\begin{equation}
 \int ds\ C_{\alpha\beta}(\vec{k};t,s) C^{-1}_{\beta\gamma}(\vec{k};s,t') = \delta_{\alpha\gamma} \delta(t-t') \ .
\end{equation}
Construction of the generating functional allows one to show that this PDF recovers the full two-time covariance. This is presented in appendix \ref{app:gaussian}. The base distribution is stationary, such that it does not have any explicit time dependence.

We now consider expanding the full PDF about this Gaussian base distribution in a perturbation series, thus
\begin{equation}
 P[\vec{u},t] = P_0[\vec{u}] + \epsilon P_1[\vec{u},t] + \epsilon^2 P_2[\vec{u},t] + \order{\epsilon^3} \ .
\end{equation}
The time-dependence of the PDF enters through the higher-order terms. Since the base distribution is both normalised and recovers the full covariance, constraints have been placed on the higher-order coefficients,
\begin{align}
 \int \mathscr{D}\vec{u}\ \left( \sum_{i \in \mathbb{N}^*} P_{2i}[\vec{u},t] \right) &= 0 \\
 \label{eq:consistency}
 \int \mathscr{D}\vec{u}\ \left( \sum_{i \in \mathbb{N}^*} P_{2i}[\vec{u},t]\ u_\mu(\vec{k},t) u_\nu(-\vec{k},t') \right) &= 0 \ ,
\end{align}
where $\mathbb{N}^*$ is the set of invertible natural numbers and the odd orders cannot contribute since they are antisymmetric in $\vec{u}$.

\subsubsection{Perturbation expansion of the Liouville equation}
In order to proceed, we define the operator $L_0(t)$ which generates our distribution such that $L_0 P_0 = 0$. This has the form
\begin{equation}
 L_0(t) P_0 = \int d^3p\ H_\rho(\vec{p}, t) \left[ \frac{\delta}{\delta u_\rho(-\vec{p},t)} + \int d\tau\ C^{-1}_{\rho\sigma}(\vec{p};t,\tau) u_\nu(\vec{p},\tau) \right] P_0 \ ,
\end{equation}
where the form of $H_\rho(\vec{p},t)$ cannot be fixed by this condition since the square brackets vanish when the operator hits $P_0$. Edwards \cite{Edwards:1964p277} used $H_\rho(\vec{p},t) = \delta/\delta u_\rho(\vec{p},t)$ so that the operator had Fokker-Planck form.

This operator is introduced into the Liouville equation, given in equation \eqref{eq:Liouville}, so that it is written
\begin{equation}
 \left[ \frac{\partial}{\partial t} + L_0 + V + \big(L - L_0\big) \right] P[\vec{u},t] \ .
\end{equation}
Since the operator $V$ stems from the non-linear term and is antisymmetric in $\vec{u}$, we assign it superficial order $\epsilon$. The operator $L-L_0$ generates a correction to the flatness factor of the PDF and is assigned order $\epsilon^2$. The perturbation expansion of the PDF is then inserted, such that
\begin{equation}
 0 = \left[ \frac{\partial}{\partial t} + L_0 + \epsilon V + \epsilon^2\Big(L - L_0\Big) \right] \Big(P_0[\vec{u}] + \epsilon P_1[\vec{u},t] + \epsilon^2 P_2[\vec{u},t] + \cdots \Big) \ .
\end{equation}
Collecting terms with the same order of $\epsilon$, we find
\begin{align}
 \label{eq:order0}
 \epsilon^0 &: \qquad \frac{\partial P_0}{\partial t} + L_0 P_0 = 0 \\
 \label{eq:order1}
 \epsilon^1 &: \qquad \frac{\partial P_1}{\partial t} + L_0 P_1 + V P_0 = 0 \\
 \label{eq:order2}
 \epsilon^2 &: \qquad \frac{\partial P_2}{\partial t} + L_0 P_2 + V P_1 + \Big(L - L_0\Big) P_0 = 0 \ .
\end{align}
The consequences of equation \eqref{eq:order0} at $\order{\epsilon^0}$ are trivial, since we saw above that both terms vanish independently due to the stationarity of the base distribution and the definition of the $L_0$ operator.

Equation \eqref{eq:order1}, on the other hand, is of direct interest, since it allows us to calculate the (leading order) contribution to odd moments of the velocity field; something that the Gaussian $P_0$ cannot tell us anything about. The equation at $\order{\epsilon^n}$ is equivalent to writing
\begin{equation}
 \frac{d P_n}{d t} = 0 \ .
\end{equation}
As a model, we consider the effects of using equation \eqref{eq:order0} for order $n = 0$ to define the total time derivative, such that
\begin{equation}
 \frac{d}{dt} = \left( \frac{\partial}{\partial t} + L_0 \right) \ .
\end{equation}
The consequence of this step is that equation \eqref{eq:order1} is instead written
\begin{equation}
 \frac{d P_1}{d t} = - V P_0 \ ,
\end{equation}
which is then integrated to give a form for the first-order coefficient
\begin{equation}
 \label{eq:P1}
 P_1[\vec{u},t] = -\int_0^t ds\ \int d^3p\ \frac{\delta}{\delta u_\rho(\vec{p},s)} \left( \int d^3q\ M_{\rho\beta\gamma}(\vec{p}) u_{\beta}(\vec{q},s) u_\gamma(\vec{p}-\vec{q},s)\ P_0[\vec{u}]\ \right) \ .
\end{equation}
The full PDF is therefore non-Gaussian, with the first-order antisymmetric piece $P_1[\vec{u},t]$ expressed as an operator acting on the Gaussian base distribution. This has the same result as the approach taken by McComb \cite{McComb:2009p1053}, where the $L_0$ operator is defined such that it also produces $L_0 P_1 = 0$. This could be through the choice of $H_\rho(\vec{p},t)$, but the form of this operator has not yet been identified.

\subsubsection{Second-order consistency}
An important check on our PDF and its expansion is the condition given in equation \eqref{eq:consistency}. It was shown by Edwards \cite{Edwards:1964p277} and Leslie \cite{leslie:1973-book} that, for the EFP theory where $H_\rho(\vec{p},t) = \delta/\delta u_\rho(\vec{p},t)$, the condition
\begin{equation}
 \label{eq:2consist}
 \int \mathscr{D}\vec{u}\ u_\mu(\vec{k},t) u_\nu(-\vec{k},t')\ P_2[\vec{u},t] = 0
\end{equation}
is equivalent to imposing the energy balance equation for $t = t'$. Indeed, for $t > t'$ this condition recovers the two-time covariance equation,
\begin{equation}
 \Big( \partial_t + \nu_0 k^2 \Big) C_{\alpha\sigma}(\vec{k};t,t') = \int d^3q\ M_{\alpha\beta\gamma}(\vec{k}) \big\langle u_\sigma(-\vec{k},t') u_\beta(\vec{q},t) u_\gamma(\vec{k}-\vec{q},t) \big\rangle \ .
\end{equation}
This needs to be shown for general $H_\rho(\vec{p},t)$, or at least the form adopted to ensure that $L_0 P_1 = 0$. We now consider evaluating the two-time covariance equation using the Gaussian base distribution and the derived first-order correction, $P_1$, given by equation \eqref{eq:P1}.

\subsection{Recovering the LET covariance equations}\label{subsec:recover_LET}
The two-time covariance equation is now evaluated using the model PDF introduced in the previous section. The equation we intend to investigate is obtained by multiplying the Navier-Stokes equation for the velocity field $u_\alpha(\vec{k},t)$ by $u_\sigma(-\vec{k},t')$ and averaging against the full PDF, $P$:
\begin{align}
 \Big( \partial_t + \nu_0 k^2 \Big) \int \mathscr{D}\vec{u}\ u_\alpha(\vec{k},t) u_\sigma(-\vec{k},t')\ P &= \int d^3q\ M_{\alpha\beta\gamma}(\vec{k}) \\
 &\quad\times\int \mathscr{D}\vec{u}\ u_\sigma(-\vec{k},t') u_\beta(\vec{j},t) u_\gamma(\vec{k}-\vec{j},t)\ P \nonumber \ .
\end{align}
Since the base distribution gives the full covariance and neither $P_0$ nor $P_2$ can contribute to an odd-order moment, the equation is expressed
\begin{align}
 \Big( \partial_t + \nu_0 k^2 \Big) \big\langle u_\alpha(\vec{k},t) u_\sigma(-\vec{k},t') \big\rangle_0 &= \int d^3q\ M_{\alpha\beta\gamma}(\vec{k}) \big\langle u_\sigma(-\vec{k},t') u_\beta(\vec{j},t) u_\gamma(\vec{k}-\vec{j},t) \big\rangle_1 \ ,
\end{align}
where $\langle \cdots \rangle_n$ denotes that the average has been taken against $P_n$, with $n = 0,1$. The LHS is clearly just the full covariance, and taking the trace along with isotropy we are left to compute
\begin{equation}
 \Big( \partial_t + \nu_0 k^2 \Big) C(k;t,t') = P(k;t,t') \ ,
\end{equation}
where
\begin{equation}
 P(k;t,t') = \tfrac{1}{2}\int d^3q\ M_{\alpha\beta\gamma}(\vec{k})\int \mathscr{D}\vec{u}\ u_\alpha(-\vec{k},t') u_\beta(\vec{j},t) u_\gamma(\vec{k}-\vec{j},t)\ P_1[\vec{u},t] \ .
\end{equation}
Inserting the expression for $P_1$ given in equation \eqref{eq:P1} gives
\begin{align}
 P(k;t,t') &= -\tfrac{1}{2} M_{\alpha\beta\gamma}(\vec{k}) \int\mathscr{D}\vec{u} \int d^3j\ u_{\beta}(\vec{j},t) u_\gamma(\vec{k}-\vec{j},t) u_\alpha(-\vec{k},t') \nonumber \\
 &\qquad\times\int_0^t ds \int d^3p\ \frac{\delta}{\delta u_\rho(\vec{p},s)} \left[ M_{\rho\mu\nu}(\vec{p}) \int d^3q\ u_\mu(\vec{q},s) u_\nu(\vec{p}-\vec{q},s) P_0[\vec{u}] \right] \nonumber \\
 &= \tfrac{1}{2} M_{\alpha\beta\gamma}(\vec{k}) \int\mathscr{D}\vec{u} \int d^3j\ \frac{\delta}{\delta u_\rho(\vec{p},s)} \Big[ u_{\beta}(\vec{j},t) u_\gamma(\vec{k}-\vec{j},t) u_\alpha(-\vec{k},t') \Big] \nonumber \\
 &\qquad\times\int_0^t ds \int d^3p\ M_{\rho\mu\nu}(\vec{p}) \int d^3q\ u_\mu(\vec{q},s) u_\nu(\vec{p}-\vec{q},s) P_0[\vec{u}] \ ,
\end{align}
where in going to the last line we performed an integration by parts with respect to the velocity field and dropped the resulting boundary terms, such that the derivative now acts on velocity components involved in the triple moment. Using the base distribution to perform the average and acting on each of the three velocity components, we find
\begin{align}
 P(k;t,t') &= \tfrac{1}{2} M_{\alpha\beta\gamma}(\vec{k}) \int_0^t ds \int d^3j \int d^3p \int d^3q\ M_{\rho\mu\nu}(\vec{p}) \nonumber\\
 &\qquad\times \left[ \left\langle u_\mu(\vec{q},s) u_\nu(\vec{p}-\vec{q},s) \frac{\delta u_{\beta}(\vec{j},t)}{\delta u_\rho(\vec{p},s)} u_\gamma(\vec{k}-\vec{j},t) u_\alpha(-\vec{k},t') \right\rangle_0 \right. \nonumber\\
 &\qquad\quad+ \left. \left\langle u_\mu(\vec{q},s) u_\nu(\vec{p}-\vec{q},s) u_{\beta}(\vec{j},t) \frac{\delta u_\gamma(\vec{k}-\vec{j},t)}{\delta u_\rho(\vec{p},s)} u_\alpha(-\vec{k},t') \right\rangle_0 \right. \nonumber\\
 &\qquad\quad+ \left. \left\langle u_\mu(\vec{q},s) u_\nu(\vec{p}-\vec{q},s) u_{\beta}(\vec{j},t) u_\gamma(\vec{k}-\vec{j},t) \frac{\delta u_\alpha(-\vec{k},t')}{\delta u_\rho(\vec{p},s)} \right\rangle_0 \right] \nonumber \\
 \label{eq:P_split}
 &= \tfrac{1}{2} M_{\alpha\beta\gamma}(\vec{k}) \int_0^t ds \int d^3j\  \left[I_{\alpha\beta\gamma}+J_{\alpha\beta\gamma}+K_{\alpha\beta\gamma}\right] \ .
\end{align}

The first step is to assume statistical independence of the derivative from the velocity field. This is a mean-field approximation (Leslie \cite{leslie:1973-book}, McComb \cite{McComb:1990-book}) and allows us to split the averages into
\begin{equation}
 \left\langle u_\mu u_\nu \frac{\delta u_{\beta}}{\delta u_\rho} u_\gamma u_\alpha \right\rangle_0 = \left\langle u_\mu u_\nu u_\gamma u_\alpha \right\rangle_0
 \left\langle \frac{\delta u_{\beta}}{\delta u_\rho} \right\rangle_0
\end{equation}
Since the base distribution is Gaussian, we then split the fourth-order moment into products of covariances, as shown in appendix \ref{app:gaussian}. Noting that the correlation tensor is defined through
\begin{equation}
 \left\langle u_\alpha(\vec{k},t) u_\beta(\vec{k}',t')\right\rangle_0 = C_{\alpha\beta}(\vec{k};t,t') \delta(\vec{k}+\vec{k}')
 = C(k;t,t') P_{\alpha\beta}(\vec{k}) \delta(\vec{k}+\vec{k}') \ ,
\end{equation}
pairings of the velocities such as $\vec{q}$ and $\vec{p}-\vec{q}$ violate the triangle condition, since they give rise to $\delta(\vec{p})$ which forces the vertex operator $M(\vec{p}) = M(0) = 0$. We consider the evaluation of $I_{\alpha\beta\gamma}$:
\begin{align}
 I_{\alpha\beta\gamma} &= \int d^3p \int d^3q\ M_{\rho\mu\nu}(\vec{p}) \nonumber \\
 &\qquad\times\Big\langle u_\mu(\vec{q},s) u_\nu(\vec{p}-\vec{q},s) u_\gamma(\vec{k}-\vec{j},t) u_\alpha(-\vec{k},t') \Big\rangle_0 \left\langle\frac{\delta u_{\beta}(\vec{j},t)}{\delta u_\rho(\vec{p},s)}\right\rangle_0 \nonumber \\
&= \int d^3p \int d^3q\ M_{\rho\mu\nu}(\vec{p})\ \nonumber \\
 &\qquad\times\left[ \Big\langle u_\mu(\vec{q},s) u_\gamma(\vec{k}-\vec{j},t) \Big\rangle_0 \Big\langle u_\nu(\vec{p}-\vec{q},s)  u_\alpha(-\vec{k},t') \Big\rangle_0 \left\langle\frac{\delta u_{\beta}(\vec{j},t)}{\delta u_\rho(\vec{p},s)}\right\rangle_0 \right. \nonumber\\
 &\qquad\qquad+\left. \Big\langle u_\mu(\vec{q},s) u_\alpha(-\vec{k},t') \Big\rangle_0 \Big\langle u_\nu(\vec{p}-\vec{q},s) u_\gamma(\vec{k}-\vec{j},t)  \Big\rangle_0 \left\langle\frac{\delta u_{\beta}(\vec{j},t)}{\delta u_\rho(\vec{p},s)}\right\rangle_0 \right] \nonumber \\
 \label{eq:I}
 &= 2M_{\rho\mu\nu}(\vec{j})\ C_{\alpha\mu}(-\vec{k};t',s) C_{\gamma\nu}(\vec{k}-\vec{j};t,s) \ \left\langle\frac{\delta u_{\beta}(\vec{j},t)}{\delta u_\rho(\vec{j},s)}\right\rangle_0 \ , 
\end{align}
where in the last line we gained a factor of 2 due to the symmetry $M_{\rho\mu\nu}(\vec{j}) = M_{\rho\nu\mu}(\vec{j})$. The terms $J_{\alpha\beta\gamma}$ and $K_{\alpha\beta\gamma}$ are evaluated in a similar way to give
\begin{align}
 \label{eq:J}
 J_{\alpha\beta\gamma} &= 2 M_{\rho\mu\nu}(\vec{k}-\vec{j})\ C_{\alpha\mu}(-\vec{k};t',s) C_{\beta\nu}(\vec{j};t,s)\ \left\langle\frac{\delta u_\gamma(\vec{k}-\vec{j},t)}{\delta u_\rho(\vec{k}-\vec{j},s)}\right\rangle_0 \\
 \label{eq:K}
 K_{\alpha\beta\gamma} &= 2 M_{\rho\mu\nu}(-\vec{k})\ C_{\gamma\nu}(\vec{k}-\vec{j};t,s) C_{\beta\mu}(\vec{j};t,s)\ \left\langle\frac{\delta u_\alpha(-\vec{k},t')}{\delta u_\rho(-\vec{k},s)}\right\rangle_0 \ .
\end{align}

We now introduce the response tensor for the LET theory in terms of the functional derivative,
\begin{equation}
 R_{\alpha\beta}(\vec{k};t,t') = \left\langle \frac{\delta u_\alpha(\vec{k},t)}{\delta u_\beta(\vec{k},t')}\right\rangle_0 = P_{\alpha\beta}(\vec{k}) R(k;t,t') \ , \qquad t > t' \ .
\end{equation}
Along with equations (\ref{eq:I}--\ref{eq:K}), we substitute back into equation \eqref{eq:P_split} and use isotropy to write
\begin{align}
 P(k;t,t') &= \tfrac{1}{2}  \int_0^t ds \int d^3j\ 2M_{\alpha\beta\gamma}(\vec{k}) \nonumber \\
 &\qquad\times \bigg[ M_{\rho\alpha\nu}(\vec{j}) P_{\beta\rho}(\vec{j}) P_{\gamma\nu}(\vec{k}-\vec{j})\ C(k;t',s) C(\vert\vec{k}-\vec{j}\rvert;t,s) R(j;t,s) \nonumber \\
 &\qquad\qquad+ M_{\rho\alpha\nu}(\vec{k}-\vec{j}) P_{\beta\nu}(\vec{j}) P_{\gamma\rho}(\vec{k}-\vec{j}) \ C(k;t',s) C(j;t,s) R(\lvert\vec{k}-\vec{j};t,s) \nonumber \\
 &\qquad\qquad+ M_{\alpha\mu\nu}(-\vec{k}) P_{\beta\mu}(\vec{j}) P_{\gamma\nu}(\vec{k}-\vec{j})\ C(j;t,s) C(\lvert\vec{k}-\vec{j}\rvert;t,s) R(k;t',s) \bigg] \nonumber \\
 &= -\tfrac{1}{2} \int_0^t ds \int d^3j\ \Big[ L(\vec{k},\vec{j}) C(k;t',s) C(\lvert\vec{k}-\vec{j}\rvert;t,s) R(j;t,s) \nonumber\\
 &\qquad\qquad\qquad\qquad\qquad+ L(\vec{k},\vec{k}-\vec{j}) C(k;t',s) C(j;t,s) R(\lvert\vec{k}-\vec{j}\rvert;t,s) \nonumber\\
 &\qquad\qquad\qquad\qquad\qquad- A(\vec{k},\vec{j},\vec{k}-\vec{j}) C(\vert\vec{k}-\vec{j}\vert;t,s) C(j;t,s) R(k;t',s) \Big] \ ,
\end{align}
where the coefficients
\begin{align}
 A(\vec{k},\vec{j},\vec{k}-\vec{j}) &= 2 M_{\alpha\beta\gamma}(\vec{k}) M_{\alpha\mu\nu}(-\vec{k}) P_{\beta\mu}(\vec{j}) P_{\gamma\nu}(\vec{k}-\vec{j})\\
 L(\vec{k},\vec{j}) &= -2 M_{\alpha\beta\gamma}(\vec{k}) M_{\beta\alpha\nu}(\vec{j}) P_{\gamma\nu}(\vec{k}-\vec{j})
\end{align}
are discussed in appendix \ref{app:Lkj} and satisfy the relationship
\begin{equation}
 A(\vec{k},\vec{j},\vec{k}-\vec{j}) = L(\vec{k},\vec{j}) + L(\vec{k},\vec{k}-\vec{j}) \ .
\end{equation}
Using this relationship to replace $A(\vec{k},\vec{j},\vec{k}-\vec{j})$, along with the change of variables $\vec{j} \to \vec{k}-\vec{j}$ for the two terms which now contain $L(\vec{k},\vec{k}-\vec{j})$, we arrive at the LET two-time covariance equation,
\begin{align}
 \left(\frac{\partial}{\partial t} + \nu k^2\right) C(k;t,t') &= \int d^3j\ L(\vec{k},\vec{j}) \left[ \int_0^{t'} ds\ R(k;t',s) C(j;t,s) C(\vert\vec{k}-\vec{j}\vert;t,s) \right. \nonumber\\
 &\qquad\qquad\qquad\qquad- \left. \int_0^t ds\ R(j;t,s) C(k;t',s) C(\lvert\vec{k}-\vec{j}\rvert;t,s)\right] \nonumber \ .
\end{align}
Thus, using the first-order coefficient $P_1$ derived from the perturbation expansion of the PDF to evaluate the triple moment in the two-time covariance equation recovers the LET covariance equation. This has assumed the LET form for the response tensor and the statistical independence of the instantaneous response from the velocity field.

\section{Further work}
Further to the work presented by McComb \cite{McComb:2009p1053}, we have determined a form for the base distribution and confirmed, using the introduction of the generating functional, that this form does recover the full covariance for the velocity field. The derivation of the first-order coefficient in the perturbation expansion about this Gaussian base distribution is the same as that derived by McComb.

As mentioned in the introduction to this chapter, this is currently still a work in progress. The planned development and application of this theory are now discussed.

The derivation of $P_1[\vec{u},t]$ was done by assuming, as our approximation, that the total time derivative could be expressed as
\begin{equation}
 \frac{d}{d t} = \left( \frac{\partial}{\partial t} + L_0 \right) \ .
\end{equation}
This was used to arrive at the form of the first-order coefficient in equation \eqref{eq:P1}. An alternative would be to find an operator $H_\rho(\vec{p},t)$, contained within $L_0$, such that $L_0 P_1 = 0$. We intend to find this operator and verify that the second order consistency condition, equation \eqref{eq:2consist}, still recovers the two-time covariance equation.

The use of this PDF to recover the LET covariance equations has already been demonstrated in section \ref{subsec:recover_LET}. Instead, the perturbation expansion of the PDF will be used to perform a renormalization group transformation of the velocity field. Presently, the iterative averaging approach to RG (based at high wavenumbers) relies on the use of a conditional average due to the non-Gaussian nature of the turbulent statistics far from $k = 0$. We expect to re-derive the results of iterative averaging using the PDF. This will unify both the LET theory and iterative averaging into a single theory of turbulence.

\chapter{Conclusions}

This thesis considers both theoretical and numerical approaches to the study of fluid turbulence. While numerical methods are playing an increasingly important role in the field, we feel that analytic approaches have not been exhausted and, despite the non-linear nature of the Navier-Stokes equations, can still offer an understanding of the fundamental properties and difficulties associated with the turbulence problem. Rather, numerical results need to be used to support or disprove theoretical ideas. 

\section{Isotropic turbulence}
Using a combination or direct numerical simulation and theoretical methods, we have studied the properties of isotropic turbulence, in both forced and decaying systems.

\subsection{Decaying isotropic turbulence}
We have discussed the requirement of developing criteria to determine when a decaying system can be considered to have evolved into a solution of the Navier-Stokes equation for which measurements are characteristic of decaying turbulence. Evolved criteria based on dynamical quantities measured from the velocity field are compared to the use of power-law decay for the total energy. In all cases, we have shown that the quantity $u^3/L$ is in fact a better surrogate for the maximum inertial flux than the dissipation rate.

The time evolution of a decay from an evolved velocity field, obtained from forced DNS, showed that the dissipation rate remains constant for a finite time after the forcing has been disabled. This is interpreted as a measurement of the time it takes energy (or information) to pass through the energy cascade to the higher dissipation wavenumbers. Only at this point does the dissipation rate react to the removal of energy input and decay. The onset of this decay is compared to the peak in the dissipation rate observed when starting from a Gaussian initial condition.

\subsection{Forced isotropic turbulence}
The structure functions, as well as the generalised structure functions, have been calculated directly in real space and used to evaluate their scaling exponents when taken to have the form $S_n(r) \propto r^{\zeta_n}$. This was done using extended self-similarity (ESS), with which we obtained exponents consistent with the literature. These show departure from the K41 value of $n/3$, with this disagreement increasing with Reynolds number. This is in contradiction to K41 being an asymptotic theory, although there is no guarantee that ESS exponents are equivalent to the actual scaling exponents of the generalised structure functions.

The calculation of the second- and third-order structure functions directly in physical-space was compared to their evaluation from energy and transfer spectra. The agreement was good for small to intermediate scales, but the constraint that $S_3(r)$ be an odd-function when evaluated in real space, along with the increased anisotropy of the larger scales, causes this agreement not to extend to large scales. While the expressions for $S_2(r)$ and $S_3(r)$ in terms of spectra are present in the literature, we do not know of any direct comparison. Using this method, we have also shown how an alternative measurement of the scaling exponent for $S_2(r)$ instead found that the exponent did seem to be approaching $2/3$, in agreement with K41.

The K\'arm\'an-Howarth equation has been discussed in some detail, with particular attention paid to the origin of the dissipation rate and its presence in the Kolmogorov form for the structure functions, with an emphasis on $S_3(r)$. The K\'arm\'an-Howarth equation for stationary isotropic turbulence, used by Kolmogorov to derive his K41 hypotheses, is ambiguous in its interpretation. Based on theoretical methods, we have derived a forced K\'arm\'an-Howarth equation from the Lin equation and explicitly obtained an energy input term. This new term is \emph{qualitatively} similar to one presented by Gotoh, Fukayama and Nakano \cite{Gotoh:2002p627}, but its interpretation very different.

The new equation reduces to the established form for decaying turbulence. We have also shown how it reduces to the form used by Kolmogorov under the assumption of sufficiently high Reynolds number such that the energy-containing and dissipation scales are well separated. This is also the case when one mathematically considers the limit of $\delta$-function forcing, not accessible to DNS. In this limit, we have found analytic forms for the non-linear and viscous terms of the K\'arm\'an-Howarth equation. Fit to DNS data at small scales, where the input term displays a plateau, gives very good agreement. When scaled by the input term to account for the effects of finite forcing used by DNS, the agreement with data is excellent for all length-scales.

\subsection{Dissipation anomaly}
The model expressions for the terms of the KHE in the limit of $\delta$-function forcing also support the model equation $\Ceps(R_L) = \Ceps(\infty) + C_L/R_L$ for the behaviour of the dimensionless dissipation coefficient, $\Ceps = \varepsilon L/u^3$. The fit of this expression to DNS data is very good and finds a value of $\Ceps(\infty) = 0.47$, in excellent agreement with the literature. This is unique in comparison to other analytic work, which all attempt to describe $\Ceps = \Ceps(R_\lambda)$.

Instead, data from DNS of decaying turbulence has been used to demonstrate the variation of the value of $\Ceps(\infty)$ in decaying turbulence, depending on the choice of evolved time criteria. Our results show that, if measured at the time corresponding to the peak in the dissipation rate, the curve for $\Ceps$ is found to coincide with that for forced turbulence. This is a relatively early time in the decay process; use of power-law decay of the total energy as a criteria is shown to give an increasing value of $\Ceps(\infty)$ until a plateau in the time series of $\Ceps(t)$ develops, after which this value decreases.

\section{Analytic approaches}

\subsection{Dynamic RG and a disagreement over methodology}
The dynamic renormalization group (RG) approach, initially studied by Forster, Nelson and Stephen (FNS), has been carefully investigated. As such, we were able to solve a disagreement which had surfaced in the literature over the methodology and results. This was important, since the original approach used by FNS had found use in other areas of physics. It came down to a substitution used by FNS to simplify the evaluation of a $d$-dimensional integral, which appeared to violate the identities later used to perform the angular parts. Using careful treatment of the shell of integration, we showed how the shell is not shifted due to the change of variables because of an additional momentum constraint which was neglected by other authors. In fact, the absence of this constraint meant that later approaches \emph{should} have had an additional term. These corrections were evaluated and it was shown how all methods recover the original FNS result.

Using the renormalization of the force autocorrelation, which we also evaluated and showed to go as $k^2$ as $k \to 0$, we discussed the constraints on the scaling relations derived in the second stage of the RG procedure. We showed that, in the two cases where the force coefficient was and was not renormalized, a non-trivial fixed point could be identified for the reduced coupling without the need to appeal to Galilean invariance (GI) to prevent the renormalization of the vertex. We then showed how the vertex is not renormalized \emph{at this order of the expansion} but not as a consequence of GI.

\subsection{Development of a statistical theory of turbulence}
A work in progress was presented in the final chapter. This involves the development of a model for the probability distribution functional of the velocity field. A Gaussian base distribution, $P_0$, chosen to recover the exact covariance of the velocity field, is introduced and a perturbation series constructed. The properties of this base distribution, including confirmation that it recovers the full covariance, have been extensive studied. Consideration of the Liouville equation then allowed an expression for the first-order coefficient $P_1$ to be found, as an operator acting on $P_0$. Using $P_0$ and $P_1$ to evaluate the second- and third-order moments, respectively, in the two-time covariance equation we demonstrated how the equivalent equation in the local energy transfer (LET) theory could be recovered.

Future work planned for this theory is to find an explicit form for the $L_0$ operator, designed such that $L_0 P_0 = 0$, which will also support $L_0 P_1 = 0$. The PDF will then be used to eliminate a band of high $k$ modes in an RG transformation. We expect to unify LET and the non-Gaussian iterative averaging approach in a single statistical theory.

\appendix
\chapter{Evaluation of the $L\left(\mathbf{k},\mathbf{j}\right)$ coefficient}\label{app:Lkj}

\section{Derivation of the coefficients}
To derive the geometric $L(\vec{k},\vec{j})$ coefficient, we extend the derivation of the quasi-normal hypothesis in section \ref{subsec:EDQNM}. Referring back to equations \eqref{eq:QN_T}, \eqref{eq:third-mom} and \eqref{eq:QN_H}, we see that the equation for the single-time covariance can be written
\begin{align}
 \label{eq:app:cov}
 \Big(\partial_t + 2\nu_0 k^2 \Big) C(k;t) &= \frac{W(k,t)}{4\pi k^2} + P(k;t) , \
\end{align}
where the non-linearity is contained within
\begin{equation}
 \label{eq:app:cov_P}
 P(k;t) = \int_0^t ds\ e^{-\nu_0(k^2 + j^2 + \lvert\vec{k}-\vec{j}\rvert^2) (t - s)}\ M_{\alpha\beta\gamma}(\vec{k}) \int d^3j\ H_{\beta\gamma\alpha}(\vec{j},\vec{k}-\vec{j},-\vec{k};s) \ .
\end{equation}
The expression for $H_{\beta\gamma\alpha}(\vec{j},\vec{k}-\vec{j},-\vec{k};s)$ was given in equation \eqref{eq:QN_H} and is reproduced here:
\begin{align}
 H_{\beta\gamma\alpha}(\vec{j},\vec{k}-\vec{j},-\vec{k};s) &= 2 M_{\beta\rho\delta}(\vec{j}) P_{\rho\gamma}(\vec{k}-\vec{j}) P_{\delta\alpha}(\vec{k}) C(\lvert\vec{k}-\vec{j}\rvert;s) C(k;s) \nonumber \\
 &\quad+ 2 M_{\gamma\rho\delta}(\vec{k}-\vec{j}) P_{\rho\beta}(\vec{j}) P_{\delta\alpha}(\vec{k}) C(j;s) C(k;s) \nonumber \\
 &\quad+ 2 M_{\alpha\rho\delta}(\vec{-k}) P_{\rho\beta}(\vec{j}) P_{\delta\gamma}(\vec{k}-\vec{j}) C(j;s) C(\lvert\vec{k}-\vec{j}\rvert;s) \ .
\end{align}
To continue, we define the coefficients
\begin{align}
 A(\vec{k},\vec{j},\vec{k}-\vec{j}) &= 2 M_{\alpha\beta\gamma}(\vec{k}) M_{\alpha\rho\delta}(\vec{-k}) P_{\rho\beta}(\vec{j}) P_{\delta\gamma}(\vec{k}-\vec{j}) \\
 \label{eq:def_B}
 B(\vec{j},\vec{k},\vec{k}-\vec{j}) &= -2M_{\alpha\beta\gamma}(\vec{k}) M_{\beta\rho\delta}(\vec{j}) P_{\rho\gamma}(\vec{k}-\vec{j}) P_{\delta\alpha}(\vec{k}) \nonumber \\
 &= -2 M_{\alpha\beta\gamma}(\vec{k}) M_{\beta\alpha\rho}(\vec{j}) P_{\rho\gamma}(\vec{k}-\vec{j}) \\
 B(\vec{k}-\vec{j},\vec{k},\vec{j}) &= -2 M_{\alpha\beta\gamma}(\vec{k}) M_{\gamma\rho\delta}(\vec{k}-\vec{j}) P_{\rho\beta}(\vec{j}) P_{\delta\alpha}(\vec{k}) \nonumber \\
 &= -2 M_{\alpha\beta\gamma}(\vec{k}) M_{\beta\alpha\rho}(\vec{k}-\vec{j}) P_{\rho\gamma}(\vec{j}) \ ,
\end{align}
with which we can express $M_{\alpha\beta\gamma}(\vec{k}) H_{\beta\gamma\alpha}(\vec{j},\vec{k}-\vec{j},-\vec{k};s)$ contained in equation \eqref{eq:app:cov_P} as
\begin{align}
 M_{\alpha\beta\gamma}(\vec{k}) H_{\beta\gamma\alpha}(\vec{j},\vec{k}-\vec{j},-\vec{k};s) &= - B(\vec{j},\vec{k},\vec{k}-\vec{j}) C(\lvert\vec{k}-\vec{j}\rvert;s) C(k;s) \nonumber \\
 &\qquad- B(\vec{k}-\vec{j},\vec{k},\vec{j}) C(j;s) C(k;s) \nonumber \\
 &\qquad+A(\vec{k},\vec{j},\vec{k}-\vec{j}) C(j;s) C(\lvert\vec{k}-\vec{j}\rvert;s) \ .
\end{align}
An identity which originates from the energy-conserving nature of the non-linear term is
\begin{equation}
 \Big[ M_{\alpha\beta\gamma}(\vec{-k}) + M_{\beta\alpha\gamma}(\vec{j}) + M_{\gamma\alpha\beta}(\vec{l}) \Big] P_{\alpha\sigma}(\vec{k}) P_{\beta\rho}(\vec{j}) P_{\gamma\delta}(\vec{l}) = 0 \ ,
\end{equation}
provided that $\vec{k} - \vec{j} - \vec{l} = 0$. From this, it is simple to show that the coefficients satisfy the condition
\begin{equation}
 A(\vec{k},\vec{j},\vec{k}-\vec{j}) - B(\vec{j},\vec{k},\vec{k}-\vec{j}) - B(\vec{k}-\vec{j},\vec{k},\vec{j}) = 0 \ .
\end{equation}
Using this identity to replace $A(\vec{k},\vec{j},\vec{k}-\vec{j})$, we have
\begin{align}
 M_{\alpha\beta\gamma}(\vec{k}) H_{\beta\gamma\alpha}(\vec{j},\vec{k}-\vec{j},-\vec{k};s) &= B(\vec{j},\vec{k},\vec{k}-\vec{j}) C(\lvert\vec{k}-\vec{j}\rvert;s) \Big( C(j;s) - C(k;s) \Big) \nonumber \\
 &\qquad+ B(\vec{k}-\vec{j},\vec{k},\vec{j}) C(j;s) \Big( C(\lvert\vec{k}-\vec{j}\rvert;s) - C(k;s) \Big) \ ,
\end{align}
with which we write $P(k;t)$ from equation \eqref{eq:app:cov_P} as
\begin{align}
 P(k;t) &= \int_0^t ds\ e^{-\nu_0(k^2 + j^2 + \lvert\vec{k}-\vec{j}\rvert^2) (t - s)} \nonumber \\
 &\qquad\times \int d^3j\ \bigg[ B(\vec{j},\vec{k},\vec{k}-\vec{j}) C(\lvert\vec{k}-\vec{j}\rvert;s) \Big( C(j;s) - C(k;s) \Big) \nonumber \\
 \label{eq:app:cov:P}
 &\qquad\qquad+ B(\vec{k}-\vec{j},\vec{k},\vec{j}) C(j;s) \Big( C(\lvert\vec{k}-\vec{j}\rvert;s) - C(k;s) \Big) \bigg] \ .
\end{align}
Since $\vec{j}$ is integrated over all space, we can make the change of variables $\vec{j} \to \vec{k}-\vec{j}$ in \emph{either} term. Doing so to the second term on the RHS, we find
\begin{align}
 P(k;t) &= \int_0^t ds\ e^{-\nu_0(k^2 + j^2 + \lvert\vec{k}-\vec{j}\rvert^2) (t - s)} \nonumber \\
 &\qquad\times \int d^3j\ 2B(\vec{j},\vec{k},\vec{k}-\vec{j}) C(\lvert\vec{k}-\vec{j}\rvert;s) \Big( C(j;s) - C(k;s) \Big) \nonumber \\
&= \int_0^t ds\ e^{-\nu_0(k^2 + j^2 + \lvert\vec{k}-\vec{j}\rvert^2) (t - s)} \nonumber \\
 &\qquad\times 2\int d^3j\ L(\vec{k},\vec{j}) C(\lvert\vec{k}-\vec{j}\rvert;s) \Big( C(j;s) - C(k;s) \Big) \ ,
\end{align}
where the $L(\vec{k},\vec{j})$ coefficient is defined as
\begin{equation}
 \label{eq:def_Lkj}
 L(\vec{k},\vec{j}) = B(\vec{j},\vec{k},\vec{k}-\vec{j}) \ .
\end{equation}
Alternatively, making the change of variable in the first term instead yields
\begin{align}
 P(k;t) = 2 \int d^3j\ L(\vec{k},\vec{k}-\vec{j}) \int_0^t ds\ & e^{-\nu[j^2 + \lvert\vec{k}-\vec{j}\rvert^2 + k^2](t-s)} \nonumber \\
 &\times C(j;s) \Big( C(\lvert\vec{k}-\vec{j}\rvert;s) - C(k;s) \Big) \ .
\end{align}

Multiplying by $4\pi k^2$ on both sides of equation \eqref{eq:app:cov} and using equation \eqref{eq:app:cov:P} for $P(k;t)$, we find the energy equation written as 
\begin{align}
 \Big(\partial_t + 2\nu_0 k^2 \Big) C(k;t) &= W(k,t)+ 4\pi k^2 P(k;t) \ ,
\end{align}
where the transfer spectrum is given by
\begin{align}
 T(k;t) = 4\pi k^2 P(k;t) &= 8\pi k^2 \int d^3j\ L(\vec{k},\vec{j}) \int_0^t ds\ e^{-\nu_0(k^2 + j^2 + \lvert\vec{k}-\vec{j}\rvert^2) (t - s)} \nonumber \\
 &\qquad\times C(\lvert\vec{k}-\vec{j}\rvert;s) \Big( C(j;s) - C(k;s) \Big) \ .
\end{align}
Thus we have arrived at equation \eqref{eq:QN_T_final} for the quasi-normality approximation.

\section{Evaluation}
We start from the definition of the $L(\vec{k},\vec{j})$ coefficient given by equations \eqref{eq:def_Lkj} and \eqref{eq:def_B},
\begin{equation}
 \label{eq:L_def}
 L(\vec{k},\vec{j}) = -2 M_{\alpha\beta\gamma}(\vec{k}) M_{\beta\alpha\delta}(\vec{j}) P_{\gamma\delta}(\vec{k}-\vec{j}) \ .
\end{equation}
Expanding out the projection and vertex operators ($P_{\alpha\beta}(\vec{k})$ and $M_{\alpha\beta\gamma}(\vec{k})$, respectively) using their definitions, we proceed with the evaluation:

\begin{align}
  L(\vec{k},\vec{j}) &= -2\bigg[ \tfrac{1}{2i} \Big(k_\beta P_{\alpha\gamma}(\vec{k}) + k_\gamma P_{\alpha\beta}(\vec{k}) \Big) \tfrac{1}{2i} \Big(j_\alpha P_{\beta\delta}(\vec{j}) + j_\delta P_{\alpha\beta}(\vec{j}) \Big) P_{\gamma\delta}(\vec{k}-\vec{j}) \bigg] \nonumber \\
  &= \tfrac{1}{2} \Bigg[ \bigg( k_\beta \delta_{\alpha\gamma} - 2\frac{k_\alpha k_\beta k_\gamma}{k^2} + k_\gamma \delta_{\alpha\beta} \bigg) \bigg( j_\alpha \delta_{\beta\delta} - 2\frac{j_\alpha j_\beta j_\delta}{j^2} + j_\delta \delta_{\alpha\beta} \bigg) \nonumber \\
  &\qquad\qquad \times \bigg( \delta_{\gamma\delta} - \frac{(k_\gamma - j_\gamma)(k_\delta - j_\delta)}{\lvert\vec{k}-\vec{j}\rvert^2} \bigg) \Bigg] \nonumber \\
  &= \tfrac{1}{2} \Bigg[k_\alpha j_\alpha - 2k_\beta j_\beta + k_\alpha j_\alpha - \frac{k_\beta(k_\beta - j_\beta) j_\alpha(k_\alpha - j_\alpha)}{\lvert\vec{k}-\vec{j}\rvert^2} \nonumber \\
  &\qquad+ 2\frac{k_\beta j_\beta}{j^2}\frac{j_\alpha(k_\alpha - j_\alpha)j_\gamma(k_\gamma - j_\gamma)}{\lvert\vec{k}-\vec{j}\rvert^2} - 4k_\alpha j_\alpha \nonumber \\
  &\qquad- \frac{j_\beta(k_\beta - j_\beta)k_\alpha(k_\alpha - j_\alpha)}{\lvert\vec{k}-\vec{j}\rvert^2} - 4\frac{k_\alpha j_\alpha k_\beta j_\beta}{k^2 j^2}\frac{k_\gamma(k_\gamma - j_\gamma)j_\delta(k_\delta - j_\delta)}{\lvert\vec{k}-\vec{j}\rvert^2} \nonumber \\
  &\qquad+ 2\frac{j_\alpha k_\alpha}{k^2}\frac{k_\beta(k_\beta - j_\beta)k_\gamma(k_\gamma - j_\gamma)}{\lvert\vec{k}-\vec{j}\rvert^2} + 4\frac{k_\alpha j_\alpha k_\beta j_\beta k_\gamma j_\gamma}{k^2 j^2} \nonumber \\
  &\qquad+ 2\frac{k_\alpha(k_\alpha - j_\alpha)j_\beta(k_\beta - j_\beta)}{\lvert\vec{k}-\vec{j}\rvert^2} + 2k_\alpha j_\alpha - \frac{k_\alpha(k_\alpha - j_\alpha)j_\beta(k_\beta - j_\beta)}{\lvert\vec{k}-\vec{j}\rvert^2} \nonumber \\
  &\qquad+ 2\frac{k_\alpha(k_\alpha - j_\alpha)j_\beta(k_\beta - j_\beta)}{\lvert\vec{k}-\vec{j}\rvert^2} - 3\frac{k_\alpha(k_\alpha - j_\alpha)j_\beta(k_\beta - j_\beta)}{\lvert\vec{k}-\vec{j}\rvert^2} \Bigg] \ ,
\end{align}
where the terms with factors of 3 come from $\delta_{\alpha\alpha} = 3$, since we are using the summation convention that repeated indices are summed over. Collecting similar terms, this can be simplified to
\begin{align}
  L(\vec{k},\vec{j}) &= \tfrac{1}{2} \Bigg[ 2k_\alpha j_\alpha \bigg(\frac{j_\beta(k_\beta - j_\beta)j_\gamma(k_\gamma - j_\gamma)}{j^2 \lvert\vec{k}-\vec{j}\rvert^2} + \frac{k_\beta(k_\beta - j_\beta)k_\gamma(k_\gamma - j_\gamma)}{k^2 \lvert\vec{k}-\vec{j}\rvert^2} - 1 \bigg) \nonumber \\
  &\quad+ 4\frac{k_\alpha j_\alpha k_\beta j_\beta k_\gamma j_\gamma}{k^2 j^2} - 2\bigg(1 + \frac{2k_\alpha j_\alpha k_\beta j_\beta}{k^2 j^2}\bigg)\bigg(\frac{k_\gamma(k_\gamma - j_\gamma)j_\delta(k_\delta - j_\delta)}{\lvert\vec{k}-\vec{j}\rvert^2}\bigg) \Bigg] \ ,
\end{align}
which is equivalent to
\begin{align}
  L(\vec{k},\vec{j} &= \Bigg[ \vec{k}\cdot\vec{j} \bigg(\frac{(\vec{k}\cdot\vec{j} - j^2)^2}{j^2 \lvert\vec{k}-\vec{j}\rvert^2} + \frac{(k^2 - \vec{k}\cdot\vec{j})^2}{k^2 \lvert\vec{k}-\vec{j}\rvert^2} - \frac{\lvert\vec{k}-\vec{j}\rvert^2}{\lvert\vec{k}-\vec{j}\rvert^2} \bigg) + 2\frac{(\vec{k}\cdot\vec{j})^3}{k^2 j^2}\frac{\lvert\vec{k}-\vec{j}\rvert^2}{\lvert\vec{k}-\vec{j}\rvert^2} \nonumber \\
  &\qquad- \bigg(1 + \frac{2(\vec{k}\cdot\vec{j})^2}{k^2 j^2}\bigg) \frac{(k^2 - \vec{k}\cdot\vec{j})(\vec{k}\cdot\vec{j} - j^2)}{\lvert\vec{k}-\vec{j}\rvert^2} \Bigg] \ .
\end{align}
The projection $\vec{k}\cdot\vec{j} = kjcos\vartheta = kj\mu$, where $\vartheta$ is the angle between $\vec{k}, \vec{j}$ in the plane spanned by the two vectors. Thus
\begin{align}
  L(\vec{k},\vec{j}) &= \frac{1}{\lvert\vec{k}-\vec{j}\rvert^2} \bigg[ kj\mu \left(\frac{1}{j^2}(kj\mu-j^2)^2 + \frac{1}{k^2}(k^2 - kj\mu)^2 - (k^2 + j^2 - 2kj\mu)\right) \nonumber \\
  &\qquad + 2kj\mu^3(k^2 + j^2 - 2kj\mu) - (1+2\mu^2)(k^2 - kj\mu)(kj\mu - j^2)\bigg] \ .
\end{align}
Expanding the brackets and collecting terms, we arrive at
\begin{align}
  L(k,j,\mu) &= \frac{kj}{k^2 + j^2 - 2\mu kj} \bigg[ \mu\Big(3\mu^2(k^2 + j^2) - 2kj\mu(1+2\mu^2)\Big) \nonumber \\
  &\qquad\qquad\qquad-  (1+2\mu^2)\Big(\mu(k^2 + j^2) - kj(1+\mu^2)\Big) \bigg] \nonumber \\
  &= \frac{kj}{k^2 + j^2 - 2\mu kj} \bigg[ \mu(\mu^2 - 1)(k^2 + j^2) + kj(1+2\mu^2)(1-\mu^2) \bigg] \ ,
\end{align}
which finally yields our expression for the $L(\vec{k},\vec{j})$ coefficient, in terms of the absolute magnitudes $k$, $j$ and the cosine of the enclosed angle, $\mu$:
\begin{equation}
 \label{eq:L_result}
 L(k,j,\mu) = \frac{kj(1-\mu^2)}{k^2 + j^2 - 2\mu kj} \Big[ kj(1+2\mu^2) - \mu(k^2 + j^2)\Big] \ .
\end{equation}

\subsection{A note on numerical evaluation in closures}
It should be noted that, in the evaluation of closures such as LET, it is common to evaluate the momentum integral in spherical polar coordinates (due to isotropy) as
\begin{align}
 \int d^3j\ L(\vec{k},\vec{j}) f(k,j,\lvert\vec{k}-\vec{j}\rvert) &= 2\pi \int dj\ j^2 \int_0^\pi d\theta\ \sin{\theta}\ L(k,j,\cos{\theta}) f(k,j,\cos{\theta}) \nonumber \\
 \label{eq:app:Lkj_int}
 &= -2\pi \int dj\ j^2 \int_1^{-1} d\mu\ L(k,j,\mu) f(k,j,\mu) \ ,
\end{align}
where $\mu = \cos{\theta}$ and so $d\mu = -\sin{\theta} d\theta$. This negative sign is often absorbed into the definition of the $L(k,j,\mu)$ but this is \emph{not} consistent with the original definition. The form above should be used, using the $-$ve sign to switch the limits of the $\mu$ integral.

\chapter{Properties of Gaussian distributions}\label{app:gaussian}

The Gaussian base distribution $P_0[\vec{u}]$ used in section \ref{sec:stat_theory} is now explored in more detail. This is done for a general vector field, $\vec{\phi}(\vec{k},t)$, with a stationary base distribution
\begin{align}
 \label{eq:vector_pdf}
 P_0[\vec{\phi}] = N \exp\left[-\frac{1}{2} \int d\tau \int d\tau' \int d^3\kappa\ \phi^*_\alpha(\vec{\kappa},\tau) C^{-1}_{\alpha\beta}(\vec{\kappa};\tau,\tau') \phi_\beta(\vec{\kappa},\tau') \right] \ ,
\end{align}
where $N$ normalises the PDF to unity and the correlation tensor
\begin{align}
 C_{\alpha\beta}(\vec{k};t,t') = \langle \phi_\alpha(\vec{k},t) \phi_\beta(-\vec{k},t') \rangle = \langle \phi_\alpha(\vec{k},t) \phi^*_\beta(\vec{k},t') \rangle
\end{align}
has an inverse
\begin{align}
 \label{eq:vector_inverse}
 \int ds\ C_{\alpha\beta}(\vec{k};t,s) C^{-1}_{\beta\gamma}(\vec{k};s,t') = \delta_{\alpha\gamma} \delta(t-t')
\end{align}
and satisfies
\begin{align}
 C_{\alpha\beta}(-\vec{k};t,t') &= \langle \phi_\alpha(-\vec{k},t) \phi_\beta(\vec{k},t') \rangle = C_{\beta\alpha}(\vec{k};t',t) \nonumber \\
 &= \langle \phi^*_\alpha(\vec{k},t) \phi^*_\beta(-\vec{k},t') \rangle = C^*_{\alpha\beta}(\vec{k};t,t') \\
 &= \langle \phi^*_\beta(\vec{k},t) \phi^*_\alpha(-\vec{k},t') \rangle^T = C^\dagger_{\beta\alpha}(\vec{k};t,t') \nonumber\ .
\end{align}
The second line is a consequence of the Hermitian symmetry of the field, and the final line just introduces an additional transpose of the tensor indices. The combined result,
\begin{equation}
 \label{eq:vector_C}
 C^*_{\alpha\beta}(\vec{k};t,t') = C_{\beta\alpha}(\vec{k};t',t) \ ,
\end{equation}
is used in the calculation of the generating functional which follows.

\section{The generating functional}
The generating functional for the vector field is found as
\begin{align}
 \label{eq:gen}
 Z_0[\vec{J}] &= \int \mathscr{D}\vec{\phi}\ \exp \left[ \int d^3\kappa \int d\tau\ J^*_\alpha(\vec{\kappa},\tau) \phi_\alpha(\vec{\kappa},\tau) \right]\ P_0[\vec{\phi}] \ ,
\end{align}
where $\vec{J}$ is known as a source and is introduced by noting that
\begin{equation}
 \phi_\alpha(\vec{k},t) = \left. \frac{\delta}{\delta J_\alpha(-\vec{k},t)} \exp \left[ \int d^3\kappa \int d\tau\ J_\mu(-\vec{\kappa},\tau) \phi_\mu(\vec{\kappa},\tau) \right] \right\vert_{\vec{J}=0} \ .
\end{equation}
As such, correlations of the field can be evaluated by instead considering
\begin{align}
 \langle f[\phi_\alpha(\vec{k},t)] \rangle &= \left. f\left( \frac{\delta}{\delta J_\alpha^*(\vec{k},t)} \right) \int \mathscr{D}\vec{\phi}\ \exp \left[ \int d^3\kappa \int d\tau\ J_\mu^*(\vec{\kappa},\tau) \phi_\mu(\vec{\kappa},\tau) \right]\ P_0[\vec{\phi}] \right\vert_{\vec{J}=0} \nonumber \\
 &= \left. f\left( \frac{\delta}{\delta J_\alpha^*(\vec{k},t)} \right) Z_0[\vec{J}] \right\vert_{\vec{J}=0} \ .
\end{align}

To find the generating functional, we insert the form of $P_0[\vec{\phi}]$ into equation \eqref{eq:gen}, upon which we find the exponent to be $-I/2$ with:
\begin{equation}
 I =  \int d^3\kappa \int d\tau\ \left[\int d\tau'\ \phi^*_\alpha(\vec{\kappa},\tau) C^{-1}_{\alpha\beta}(\vec{\kappa};\tau,\tau') \phi_\beta(\vec{\kappa},\tau') - 2 J_\mu^*(\vec{\kappa},\tau) \phi_\mu(\vec{\kappa},\tau) \right] \ .
\end{equation}
We attempt to complete the square using the change of variables
\begin{equation}
 \xi_\alpha(\vec{k},t) = \phi_\alpha(\vec{k},t) - \int ds\ C_{\alpha\mu}(\vec{k};t,s) J_\mu(\vec{k},s) \ , \qquad \mathscr{D}\vec{\xi} = \mathscr{D}\vec{\phi} \ ,
\end{equation}
since we have
\begin{align}
 &\int d\tau \int d\tau'\ \xi^*_\alpha(\vec{k},\tau)  C^{-1}_{\alpha\beta}(\vec{k};\tau,\tau') \xi_\beta(\vec{k},\tau') \nonumber \\
 &\ = \int d\tau \int d\tau'\ \phi^*_\alpha(\vec{k},\tau) C^{-1}_{\alpha\beta}(\vec{k};\tau,\tau') \phi_\beta(\vec{k},\tau') \\
 &\qquad- \int d\tau \int d\tau' \int ds'\ \phi^*_\alpha(\vec{k},\tau) C^{-1}_{\alpha\beta}(\vec{k};\tau,\tau') C_{\beta\nu}(\vec{k};\tau',s') J_\nu(\vec{k},s') \nonumber \\
 &\qquad- \int d\tau \int d\tau' \int ds\ J^*_\mu(\vec{k},s) C^*_{\alpha\mu}(\vec{k};\tau,s) C^{-1}_{\alpha\beta}(\vec{k};\tau,\tau') \phi_\beta(\vec{k},\tau') \nonumber \\
 &\qquad+ \int d\tau \int d\tau' \int ds \int ds'\ J^*_\mu(\vec{k},s) C^*_{\alpha\mu}(\vec{k};\tau,s) C^{-1}_{\alpha\beta}(\vec{k};\tau,\tau') C_{\beta\nu}(\vec{k};\tau',s') J_\nu(\vec{k},s') \nonumber  \\
 &\ = \int d\tau \int d\tau'\ \phi^*_\alpha(\vec{k},\tau) C^{-1}_{\alpha\beta}(\vec{k};\tau,\tau') \phi_\beta(\vec{k},\tau') \\
 &\qquad- \int d\tau \int d\tau' \int ds'\ \phi^*_\alpha(\vec{k},\tau) C^{-1}_{\alpha\beta}(\vec{k};\tau,\tau') C_{\beta\nu}(\vec{k};\tau',s') J_\nu(\vec{k},s') \nonumber \\
 &\qquad- \int d\tau \int d\tau' \int ds\ J^*_\mu(\vec{k},s) C_{\mu\alpha}(\vec{k};s,\tau) C^{-1}_{\alpha\beta}(\vec{k};\tau,\tau') \phi_\beta(\vec{k},\tau') \nonumber \\
 &\qquad+ \int d\tau \int d\tau' \int ds \int ds'\ J^*_\mu(\vec{k},s) C_{\mu\alpha}(\vec{k};s,\tau) C^{-1}_{\alpha\beta}(\vec{k};\tau,\tau') C_{\beta\nu}(\vec{k};\tau',s') J_\nu(\vec{k},s') \nonumber  \ ,
\end{align}
where we used the result equation \eqref{eq:vector_C} in the second term. Using the definition of the inverse, equation \eqref{eq:vector_inverse}, we find
\begin{align}
 \int d\tau & \int d\tau'\ \xi^*_\alpha(\vec{k},\tau) C^{-1}_{\alpha\beta}(\vec{k};\tau,\tau') \xi_\beta(\vec{k},\tau') \nonumber \\
 &= \int d\tau \int d\tau'\ \phi^*_\alpha(\vec{k},\tau) C^{-1}_{\alpha\beta}(\vec{k};\tau,\tau') \phi_\beta(\vec{k},\tau') \\
 &\qquad- \int d\tau \int ds'\ \phi^*_\alpha(\vec{k},\tau) \delta_{\alpha\nu} \delta(\tau - s') J_\nu(\vec{k},s') \nonumber \\
 &\qquad- \int d\tau' \int ds\ J^*_\mu(\vec{k},s) \delta_{\mu\beta} \delta(\tau'-s) \phi_\beta(\vec{k},\tau') \nonumber \\
 &\qquad+ \int d\tau' \int ds \int ds'\ J^*_\mu(\vec{k},s) \delta_{\mu\beta} \delta(\tau'-s) C_{\beta\nu}(\vec{k};\tau',s') J_\nu(\vec{k},s') \nonumber  \\
 &= \int d\tau \int d\tau'\ \phi^*_\alpha(\vec{k},\tau) C^{-1}_{\alpha\beta}(\vec{k};\tau,\tau') \phi_\beta(\vec{k},\tau') - \int d\tau\ \phi^*_\alpha(\vec{k},\tau) J_\alpha(\vec{k},\tau) \\
 &\qquad- \int d\tau'\ \phi^*_\beta(-\vec{k},\tau') J_\beta(-\vec{k},\tau') + \int ds \int ds'\ J^*_\mu(\vec{k},s) C_{\mu\nu}(\vec{k};s,s') J_\nu(\vec{k},s') \nonumber \ . 
\end{align}
This allows us to rewrite the exponent $-I/2$ in the generating functional as
\begin{equation}
 I = \int d^3\kappa \int d\tau \int d\tau'\ \Big[ \xi^*_\alpha(\vec{\kappa},\tau) C^{-1}_{\alpha\beta}(\vec{\kappa};\tau,\tau') \xi_\beta(\vec{\kappa},\tau') - J^*_\mu(\vec{\kappa},\tau) C_{\mu\nu}(\vec{\kappa};\tau,\tau') J_\nu(\vec{\kappa},\tau') \Big] \ .
\end{equation}
Since the latter term does not depend on our integration field, $\vec{\xi}$, we can remove it from the integral and we are left with
\begin{align}
 Z_0[\vec{J}] &= \int \mathscr{D}\vec{\xi}\ \exp \left[ - \frac{I}{2} \right] \nonumber \\
 &= \exp \left[ \frac{1}{2} \int d^3\kappa \int d\tau \int d\tau'\ J^*_\mu(\vec{\kappa},\tau) C_{\mu\nu}(\vec{\kappa};\tau,\tau') J_\nu(\vec{\kappa},\tau') \right] \int \mathscr{D}\vec{\xi}\ P_0[\vec{\xi}] \nonumber \\
 \label{eq:vector_gen}
 &= \exp \left[ \frac{1}{2} \int d^3\kappa \int d\tau \int d\tau'\ J_\mu(-\vec{\kappa},\tau) C_{\mu\nu}(\vec{\kappa};\tau,\tau') J_\nu(\vec{\kappa},\tau') \right] \ ,
\end{align}
where the last line followed from normalisation of the PDF.

\section{Correlations of the field}
From the generating functional, we can calculate a derivative with respect to the source:
\begin{align}
 \label{eq:vector_Z_deriv_1}
 \frac{\delta}{\delta J_\alpha(\vec{k},t)} Z_0[\vec{J}] &= \frac{1}{2} \left[ \int d\tau'\ C_{\alpha\nu}(-\vec{k};t,\tau') J_\nu(-\vec{k},\tau') \right. \nonumber \\
 &\qquad\qquad+ \left. \int d\tau\ J_\mu(-\vec{k},\tau) C_{\mu\alpha}(\vec{k};\tau,t) \right] \cdot Z_0[\vec{J}] \\
 \label{eq:vector_Z_deriv}
 &= \int d\tau\ J^*_\mu(\vec{k},\tau) C_{\mu\alpha}(\vec{k};\tau,t) \cdot Z_0[\vec{J}] \ ,
\end{align}
where in going from equation \eqref{eq:vector_Z_deriv_1} to \eqref{eq:vector_Z_deriv} we relabelled $\tau' \to \tau$, $\nu \to \mu$ in the first term and used equation \eqref{eq:vector_C} to swap the time arguments. We introduce a reduced vector notation for the expressions to come,
\begin{align}
 \phi^{\vec{k},t}_\alpha &= \phi_\alpha(\vec{k},t) \ ,\qquad J^{\vec{k},t}_\alpha = J_\alpha(\vec{k},t) \ , \qquad C^{\vec{k}}_{\alpha\beta}(t,t') = C_{\alpha\beta}(\vec{k};t,t') \ .
\end{align}

\subsection*{Second-order moment}
\begin{align}
 \langle \phi^{\vec{k},t}_\alpha & \phi^{\vec{k}',t'}_\beta \rangle = \left. \frac{\delta}{\delta J^{-\vec{k},t}_\alpha} \frac{\delta}{\delta J^{-\vec{k}',t'}_\beta} \exp \left[ \frac{1}{2} \int d^3\kappa \int d\tau \int d\tau'\ J^{-\vec{\kappa},\tau}_\mu C^{\vec{\kappa}}_{\mu\nu}(\tau,\tau') J^{\vec{\kappa},\tau'}_\nu \right] \right\vert_{\vec{J}=0} \nonumber \\
 &= \left. \frac{\delta}{\delta J^{-\vec{k},t}_\alpha} \left[ \int d\tau\ J^{\vec{k}',\tau}_\mu C^{-\vec{k}'}_{\mu\beta}(\tau,t') \cdot Z_0[\vec{J}] \right] \right\vert_{\vec{J}=0} \nonumber \\
 &= \left. \left[ C^{-\vec{k}'}_{\alpha\beta}(t,t') \delta(\vec{k}+\vec{k}') + \int d\tau\ J^{\vec{k}',\tau}_\mu C^{-\vec{k}'}_{\mu\beta}(\tau,t') \int d\tau'\ J^{\vec{k},\tau'}_\nu C^{-\vec{k}}_{\nu\alpha}(\tau',t) \right] \cdot Z_0[\vec{J}] \right\vert_{\vec{J}=0} \nonumber \\
 &= C^{\vec{k}}_{\alpha\beta}(t,t') \delta(\vec{k}+\vec{k}') \ .
\end{align}
Thus the PDF recovers the full two-time covariance of the (homogeneous) vector field.

\subsection*{Third-order moment}
Using the result for the second-order moment above, we continue to find the third-order moment:
\begin{align}
 \langle \phi^{\vec{k},t}_\alpha \phi^{\vec{k}',t'}_\beta \phi^{\vec{k}'',t''}_\gamma \rangle &= \left. \frac{\delta}{\delta J^{-\vec{k},t}_\alpha} \frac{\delta}{\delta J^{-\vec{k}',t'}_\beta} \frac{\delta}{\delta J^{-\vec{k}'',t''}_\gamma} Z_0[\vec{J}] \right\vert_{\vec{J}=0} \nonumber \\
 &= \frac{\delta}{\delta J^{-\vec{k}'',t''}_\gamma} \bigg[ C^{-\vec{k}'}_{\alpha\beta}(t,t') \delta(\vec{k}+\vec{k}') \nonumber \\
 &\qquad+ \left. \int d\tau\ J^{\vec{k}',\tau}_\mu C^{-\vec{k}'}_{\mu\beta}(\tau,t') \int d\tau'\ J^{\vec{k},\tau'}_\nu C^{-\vec{k}}_{\nu\alpha}(\tau',t) \bigg] \cdot Z_0[\vec{J}] \right\vert_{\vec{J}=0} \nonumber \\
 &= \left[ C^{\vec{k}}_{\alpha\beta}(t,t') \delta(\vec{k}+\vec{k}') \int d\tau\ J^{\vec{k}'',\tau}_\mu C^{-\vec{k}''}_{\mu\gamma}(\tau,t'') \right. \nonumber \\
 &\qquad + C^{\vec{k}'}_{\beta\gamma}(t',t'') \delta(\vec{k}'+\vec{k}'') \int d\tau'\ J^{\vec{k},\tau'}_\nu C^{-\vec{k}}_{\nu\alpha}(\tau',t) \\
 &\qquad + \left. \left. C^{\vec{k}}_{\alpha\gamma}(t,t'') \delta(\vec{k}+\vec{k}'') \int d\tau\ J^{\vec{k}',\tau}_\mu C^{-\vec{k}'}_{\mu\beta}(\tau,t') + \ord{J^3} \right] \cdot Z_0[\vec{J}] \right\vert_{\vec{J}=0} \ . \nonumber
\end{align}
Since this is evaluated at $\vec{J}=0$, the third-order (in fact, any odd-order) moment vanishes, as expected for a Gaussian distribution.

\subsection*{Fourth-order moment}
We continue from the evaluation of the third-order moment above to find the fourth-order moment:
\begin{align}
 \langle \phi^{\vec{k},t}_\alpha \phi^{\vec{k}',t'}_\beta \phi^{\vec{k}'',t''}_\gamma \phi^{\vec{k}''',t'''}_\delta \rangle
 %
 &= \frac{\delta}{\delta J^{-\vec{k}''',t'''}_\delta} \left[ C^{\vec{k}}_{\alpha\beta}(t,t') \delta(\vec{k}+\vec{k}') \int d\tau\ J^{\vec{k}'',\tau}_\mu C^{-\vec{k}''}_{\mu\gamma}(\tau,t'') \right. \nonumber \\
 &\qquad+ C^{\vec{k}'}_{\beta\gamma}(t',t'') \delta(\vec{k}'+\vec{k}'') \int d\tau'\ J^{\vec{k},\tau'}_\nu C^{-\vec{k}}_{\nu\alpha}(\tau',t) \nonumber \\
 &\qquad+ \left. C^{\vec{k}}_{\alpha\gamma}(t,t'') \delta(\vec{k}+\vec{k}'') \int d\tau\ J^{\vec{k}',\tau}_\mu C^{-\vec{k}'}_{\mu\beta}(\tau,t') + \ord{J^3} \right] \nonumber \\
 &\qquad\qquad\times Z_0[\vec{J}] \bigg\vert_{\vec{J}=0} \nonumber \\
 %
 %
 &= C^{\vec{k}}_{\alpha\beta}(t,t') C^{\vec{k}''}_{\gamma\delta}(t'',t''') \delta(\vec{k}+\vec{k}') \delta(\vec{k}''+\vec{k}''') \nonumber \\
 &\qquad+ C^{\vec{k}}_{\alpha\delta}(t,t''') C^{\vec{k}'}_{\beta\gamma}(t',t'') \delta(\vec{k}+\vec{k}''') \delta(\vec{k}'+\vec{k}'') \nonumber \\
 &\qquad+ C^{\vec{k}}_{\alpha\gamma}(t,t'') C^{\vec{k}'}_{\beta\delta}(t',t''') \delta(\vec{k}+\vec{k}'') \delta(\vec{k}'+\vec{k}''') \ .
\end{align}
Thus we find that the fourth-order moment reduces to the three possible pairings of second-order moments.

\section{Comments on isotropy}
The assumption of isotropy significantly alters the formulation of the distribution for the vector field discussed in the previous section. When the system is isotropic, the correlation tensor may be expressed in terms of a scalar function, such as
\begin{equation}
 \label{eq:isotropy_expand_C}
 C_{\alpha\beta}(\vec{k}; t,t') = P_{\alpha\beta}(\vec{k}) C(k;t,t') \ ,
\end{equation}
where $C(k;t,t')$ is a function of $k = \left\lvert\vec{k}\right\rvert$ only. This has important implications for the correlation tensor, since for $-\vec{k}$ we have
\begin{align}
 C_{\alpha\beta}(-\vec{k};t,t') &= P_{\alpha\beta}(\vec{k}) C(k;t,t') = C_{\alpha\beta}(\vec{k};t,t')\ ;
\end{align}
whereas, using the Hermitian symmetry of the field, we also have
\begin{align}
 C_{\alpha\beta}(-\vec{k};t,t') &= C^*_{\alpha\beta}(\vec{k};t,t') = P_{\alpha\beta}(\vec{k}) C^*(k;t,t') \ .
\end{align}
This forces the correlation tensor, and hence the isotropic $C(k;t,t')$, to be real-valued,
\begin{align}
 C(k;t,t') = C^*(k;t,t') = C(k;t',t) \in \mathbb{R} \ .
\end{align}

The inverse of the correlation tensor can be expressed as
\begin{align}
 \label{eq:isotropy_expand_Cinv}
 C^{-1}_{\alpha\beta}(\vec{k};t,t') &= P_{\alpha\beta}(\vec{k}) C^{-1}(k;t,t') \ .
\end{align}
This highlights the fact that the inversion is done in the time coordinates, with the wavevector and tensor parts basically spectating. With this definition, the inverse relation takes the form
\begin{align}
 \int ds\ C_{\alpha\beta}(\vec{k};t,s) C^{-1}_{\beta\gamma}(\vec{k};s,t') &= P_{\alpha\gamma}(\vec{k}) \delta(t-t') \ ,
\end{align}
or for the scalar function
\begin{align}
 \label{eq:isotropy_inverse}
 \int ds\ C(k;t,s) C^{-1}(k;s,t') &= \delta(t-t') \ .
\end{align}
Isotropy also constrains derivatives of the fields,
\begin{align}
 \label{eq:isotropy_derivative}
 \frac{\delta \phi_\alpha(\vec{k},t)}{\delta \phi_\beta(\vec{k'},t')} = \frac{\delta J_\alpha(\vec{k},t)}{\delta J_\beta(\vec{k'},t')} &= P_{\alpha\beta}(\vec{k}) \delta(\vec{k}-\vec{k}') \delta(t-t') \ .
\end{align}

The isotropic distribution can be written
\begin{align}
 \label{eq:isotropy_pdf}
 P_0[\vec{\phi}] = N \exp \left[ -\frac{1}{2} \int d^3\kappa \int d\tau \int d\tau'\ \phi^*_\alpha(\vec{\kappa},\tau) C^{-1}(\kappa;\tau,\tau') \phi_\alpha(\vec{\kappa},\tau') \right] \ .
\end{align}
Using the change of variables
\begin{equation}
 \xi_\alpha(\vec{k},t) = \phi_\alpha(\vec{k},t) - \int ds\ C(k;t,s) J_\alpha(\vec{k},s) \ ,
\end{equation}
the generating functional for the isotropic distribution may be calculated to be
\begin{align}
 \label{eq:isotropy_gen}
 Z_0[\vec{J}] = \exp \left[ \frac{1}{2} \int d^3\kappa \int d\tau \int d\tau'\ J^*_\mu(\vec{\kappa},\tau) C(\kappa;\tau,\tau') J_\mu(\vec{\kappa},\tau') \right] \ ,
\end{align}
from which the second-order moment is evaluated to be
\begin{align}
 \langle \phi_\alpha(\vec{k},t) \phi_\beta(\vec{k}',t') \rangle = P_{\alpha\beta}(\vec{k}) C(k;t,t') \delta(\vec{k}+\vec{k}') \ .
\end{align}

\subsection{The reality of the isotropic two-time correlation function}
Taking the definition $C(k;t,t') = \tfrac{1}{2} \langle u_\alpha(\vec{k},t) u^*_\alpha(\vec{k},t') \rangle$, we see that it is essentially the average of the product of two \emph{different} complex numbers, $A B^*$. When $t' = t$, we have $AA^* \in \mathbb{R}$, and so $C(k;t,t)$ is a real-valued function. However, when $t \neq t'$, it is not immediate clear why this quantity must be real.

For the isotropic system, we can consider the averaging procedure to include all modes on a shell with wavenumber $k$,
\begin{align}
 \langle u_\alpha(\vec{k},t) u^*_\alpha(\vec{k},t') \rangle &= \frac{1}{P(k)} \sum_{\lvert\vec{k}\rvert = k} u_\alpha(\vec{k},t) u^*_\alpha(\vec{k},t') \ ,
\end{align}
where $P(k)$ is the number of points with $\lvert\vec{k}\rvert = k$. For every mode $\vec{k}$ included in the sum, the corresponding mode $-\vec{k}$ is also included. Thus, we can write
\begin{align}
 \langle u_\alpha(\vec{k},t) u^*_\alpha(\vec{k},t') \rangle &= \frac{1}{P(k)} \sum_{\lvert\vec{k}\rvert = k} \tfrac{1}{2} \Big[ u_\alpha(\vec{k},t) u^*_\alpha(\vec{k},t') + u_\alpha(-\vec{k},t) u^*_\alpha(-\vec{k},t') \Big] \nonumber \\
 &= \frac{1}{P(k)} \sum_{\lvert\vec{k}\rvert = k} \textrm{Re} \Big[ u_\alpha(\vec{k},t) u^*_\alpha(\vec{k},t') \Big] \ .
\end{align}
Thus the isotropic tensor must be real. This can also be seen from equation \eqref{eq:second_corr}
\begin{equation}
 C_{\alpha\alpha}(\vec{k};t,t') = \frac{1}{\pi^2 k^2} \int dr\ C(r;t,t')\ kr \sin{kr} \ , \qquad C(r;t,t') \in \mathbb{R} \ .
\end{equation}

\end{document}